\documentclass[epj]{svjour}
\usepackage{graphics, slashed}
\usepackage{amssymb}
\usepackage{amsmath}
\usepackage{slashed}
\usepackage{chngcntr}
\usepackage[normalem]{ulem} 
\usepackage {ulem} 
\usepackage{color}
\usepackage{colordvi}
\usepackage[bookmarks=false]{hyperref}
\counterwithin{figure}{section}
\numberwithin{equation}{subsection}

\def\ee{\end{equation}}
\def\be{\begin{equation}}
\def\eea{\end{eqnarray}}
\def\bea{\begin{eqnarray}}
\def\bal{\begin{align}}
\def\eal{\end{align}}
\def\pp{$pp$}
\def\pbarp{${\bar p}p$}
\def\pbarpmath{{\bar p}p}
\def\spbarps{$S{\bar p}pS$}

\def\sigtotpp{\sigma_{tot}^{pp}}

\def\sigel{\sigma_{elastic}}
\def\siginel{\sigma_{inel}}
\def\sigdiff{\sigma_{diffractive}}
\def\dsigdt{d\sigma_{el}/dt}
\def\sigpair{\sigma^{p-air}}
\def\sigtotep{\sigma^{ep}}
\def\sigtotgamp{\sigma_{tot}^{\gamma p}}
\def\sigtotgampistar{\sigma_{tot}^{\gamma* p}}
\def\sigtotgamgam{\sigma_{tot}^{\gamma \gamma}}
\def\sigtotpn{\sigma_{tot}^{pn}}

\def\x{cross-section}

\def\shat{{\hat s}}
\def\vecr{{\vec r}}
\def\veck{{\vec k}}
\def\vecb{{\vec b}}
\def\vecq{{\vec q}}
\def\vecs{{\vec s}}
\def\vecK{{\vec K}}
\def\vecbp{{\vec b'}}
\def\iqb{i\vecq\cdot\vecb}
\def\sigtot{\sigma_{total}}
\def\sigeltosigtot{$\sigma_{elastic}/\sigma_{total}$}
\def\Imo{\Im m}
\def\Reo{\Re e}

\def\SpbarpS{$S\bar{p}pS$}
\def\Imo{\Im m}
\def\Reo{\Re e}
\def\BC{ \cite{Block:1984ru}}
\def\dsigdtzero{(d\sigma_{el}/dt)_{t=0}}
\def\im{{\cal I}m}
\def\re{{\cal R}e}
\def\energy8{$\sqrt{s}=8\ TeV$}
\def\energy14{$\sqrt{s}=14\ TeV$}
\def\sigmaT{\sigma_T}
\def\sigELA{\sigma_{el}}
\def\eiqb{e^{i\vecq\cdot\vecb}}
\def\emiqb{e^{-i\vecq\cdot\vecb}}

\def\.[{\big{[}}
\def\.]{\big{]}}
\def\.({\big{(}}
\def\.){\big{)}}
\def\eiqb{e^{i\vecq\cdot\vecb}}
\def\shat{{\hat s}}
\def\sqrts{\sqrt{s}}
\def\reltot{\sigel/\sigtot}
\def\relinel{\sigel/\siginel}
\begin{document}
%
\title{Introduction to the physics of the total cross-section at LHC}
\subtitle{A Review of  Data and Models  }

\author{Giulia Pancheri\inst{1}  \and Yogendra N. Srivastava\inst{2}
}                     
%

\institute{INFN Frascati National Laboratory, Via E. Fermi 40, I00044 Frascati, Italy\\
and \\
Center for Theoretical Physics, Massachusetts Institute of Technology, Cambridge, MASS USA 
\and Physics Department, U. of Perugia, Via A. Pascoli 6, Perugia 06123, Italy\\
and\\
Physics Department, Northeastern University, Boston, MASS 02115, USA 
}

%
\date{}

\abstract{
This review describes the  development of the physics of hadronic cross sections up to recent   LHC results  and cosmic ray experiments. We present here  a comprehensive review - written with a historical perspective - about total cross-sections from medium to the highest energies explored experimentally and studied through a variety of methods and theoretical models for over sixty years. We begin by  recalling the analytic properties of the elastic amplitude and the theorems about the  asymptotic behavior of the total cross-section. A  discussion of how proton-proton cross-sections are  extracted  from cosmic rays at higher than accelerator energies and  help the study  of these  asymptotic limits, is presented.  This is followed by a description of the advent of    particle colliders,   through which high energies and unmatched experimental precisions have been  attained. Thus the measured   hadronic elastic and total cross-sections have become crucial  instruments to probe  the so called {\it soft} part of QCD physics, where  quarks and gluons are confined, and  have led to test and refine   Regge behavior and a number of diffractive models.  As the c.m. energy increases,  the total cross-section  also probes the transition into hard scattering describable with perturbative QCD, the so-called mini-jet region. Further tests are provided by cross-section measurements of $\gamma p$, $\gamma^* p$ and $\gamma^* \gamma^*$ for models based on vector meson dominance, scaling limits of virtual photons at high $Q^2$ and the BFKL formalism. Models interpolating from  virtual to real photons are also tested.  \\ 
\hspace{12cm}
{\it It seems to us to be a necessary task to explore bit-by-bit the rigorous consequences of analyticity, unitarity and crossing. 
Who knows if someday one will not be able to reassemble the pieces of the puzzle. } - A. Martin and F. Cheung, based on 1967 
A.M. Lectures at Brandeis Summer School and Lectures at SUNY and Stony Brook \cite{martinbook3:1970}.
} 

\maketitle

\setcounter{tocdepth}{5}
\tableofcontents
\setcounter{secnumdepth}{5}
\section*{Introduction}\label{introduction}

This review aims at illustrating  the development of studies of total hadronic  cross-sections in a  historical perspective,
 primarily for hadrons and photons
 at high energies.

The  optical theorem 
relates a total cross-section linearly to the absorptive part of a forward elastic amplitude. Moreover, at very 
high energies, as the imaginary part of the forward amplitude dominates the real part, the elastic differential 
cross-section in the forward direction becomes proportional to the square of the total cross-section. Thus, 
discussions about total cross-sections become entwined with that of elastic cross-sections. Hence, in this review
considerable attention is also paid both to experimental and 
 theoretical aspects of the elastic cross-sections.

As this rather lengthy review discusses many subjects, we provide below a quick overview to help
a reader choose sections of the review that may be of particular interest.   
Serving  mostly as a  guide through the large material we shall deal with, no references are included in this general introduction, 
but they are of course available in the individual sections, at the beginning of which we provide a  description of contents  
and  a brief guide to the subsections.
 
In Section \ref{sec:general}, kinematics and partial wave expansions are obtained for
the elastic amplitude and general principles, such as unitarity, are employed to derive the optical theorem. 
An introduction to the asymptotic behavior is provided via the Regge formalism, the Pomeranchuk theorem and 
finite energy sum rules. Through analyticity, Martin-Froissart rigorous upper bounds are established for the total
cross-sections. For charged particles, the EM (Coulomb) amplitude is mostly real (and large near the forward
direction) and hence measurements of the needed real part of the ``strong'' forward amplitude (and measurement
of 
the ratio of the real to the imaginary part of the forward scattering amplitude, the so-called
$\rho(s)$ parameter) often involve Coulomb interference and soft radiation. We discuss it in some detail and 
supplement it with a  proposal to employ soft radiation as a tool to measure total cross-sections.

In Section \ref{sec:cosmic}, we discuss how cosmic radiation is employed as a non-accelerator method to measure total
cross-sections and provide valuable information at energies substantially larger than those of earth bound accelerators 
such as the Large Hadron Collider (LHC). Along with some history of the subject beginning with Heisenberg, a description 
of the Glauber formalism for nuclei is presented for the extraction of $pp$ cross-sections from data and corresponding 
uncertainties in the models are discussed. We 
follow the historical path which  led to  
the advances in experimental techniques and theoretical methods 
that  
continue 
to provide
a unique window towards fundamental physics and astrophysics, 
at energies otherwise
unreachable through accelerators 
in the foreseeable future. Recent theoretical results about the power law spectra in the cosmic ray energy distribution
both for fermions (electrons/positrons) and bosons (Helium and other nuclei) are briefly discussed and shown to agree
with high precision data from AMS, Auger and other Collaborations.

Section \ref{sec:measure} deals with pre-LHC measurements of 
$\sigtot$,
the total \pp \  and \pbarp \ \x ,  such as those made at the CERN Intersecting Storage Rings (ISR), the CERN  \SpbarpS and the FermiLab  Tevatron. These
pioneering experiments verified the rise with energy  of $\sigtot$  - suspected  from experiments with cosmic rays - and of  the slope
parameter $B(s)$, which defines  the behaviour of the elastic differential cross-section in the forward region,  
 as well as discovered considerable variations in the $\rho$ parameter. The black disk limit and 
how close we may be to it are also discussed herein.  

In Section \ref{sec:models}, we provide 
 theoretical scenarios and phenomenology of the 
elastic amplitude and hadronic cross-sections,
that span
over 80 years, beginning with the original  Moli\`ere theory of multiple scattering,  
followed by  Heisenberg's considerations about  the energy behavior of $\sigtot$, 
and culminating in various QCD inspired
models. Eikonal and Regge models are discussed along with hadronic matter distribution. We recall the development in QED of the need for soft-photon 
re-summation to avoid the infra-red (IR) catastrophe, and the semi-classical, but Lorentz covariant, methods
for soft radiation subsequently developed.  As a corollary, a Regge trajectory for the photon is obtained. These methods are
extended to discuss soft QCD radiation and the divergent nature of of strong coupling constant for small transverse momenta
$\alpha_s(k_t)$. Asymptotic behavior of scattering amplitudes in QCD, the Balitsky, Fadin, Kuraev and Lipatov (BFKL)  
equation and spontaneously broken gauge
theories are discussed along with the Reggeization of the gauge particles therein. 
Next, eikonal mini-jet models
for $\sigtot$ and their phenomenology are developed. A brief description of the AdS/CFT correspondence for $\sigtot$
is presented. Also, some details of the phenomenology of $\sigtot$ by the COMPETE 
and COMPASS
collaborations are provided.

Details of the energy and momentum transfer dependence, the slope, the dip, the real and the imaginary parts 
of the elastic  (and diffractive) amplitude are discussed in Section \ref{sec:elasticdiff}. Early models and their updates 
such as Durham, Tel Aviv, mini-jet and multi-pomeron models, are presented as required by more refined data. A concise summary 
of the model results are also provided.
 
Photon processes are discussed in Section \ref{sec:photons} beginning with kinematics of interest for real versus virtual
photons and the relevant parton model variables. Sakurai's vector meson dominance, Gribov's model and photo-production at HERA
is taken up next, along with $\gamma\gamma$ and  $\gamma \gamma^*$ processes at LEP and factorization. The transition from real to virtual photon processes is discussed and models such as Haidt's  are presented. The results of the Tel Aviv 
and 
mini-jet models with soft gluon resummation
 are discussed.  The Balitsky-Kovchegov (BK) equation and its various applications
such as geometrical scaling are considered and directions beyond into Pomeron loops, explored.

Section \ref{sec:lhcnow} discusses the 
layout of the LHC experimental areas as had been planned before its start. Expectations were to produce total cross-section data 
with $5\%$ accuracy after a 3 year run. It is gratifying to note that forward physics data with $3\%$ accuracy have already been achieved.
The highest energy physics results are shown for the total, elastic and inelastic $pp$ cross-sections at presently reached LHC energies, $\sqrt s = 7,\ 8$ and $13$ TeV,  
obtained 
 by the TOTEM, ATLAS and  CMS groups.  Predictions at $\sqrt s = 14\ {\rm TeV}$ are indicated.


%
%
 \section{The theoretical framework from unitarity and analyticity}
\label{sec:general}


This chapter is  devoted to a  review of the basic formalism pertaining to elastic scattering and to the well-established 
theorems on total, elastic and inelastic cross-sections. 
Here analyticity and unitarity play a crucial role
for the scattering of hadrons, protons and mesons, such as 
pions and kaons, 
while scattering  of their QCD constituents and
their 
contribution to total cross-section dynamics will be introduced 
when dealing with QCD models.


 For this material, there exist  both books and reviews, nonetheless we reproduce most of the relevant material  to 
 introduce, in a modern  language, the necessary notation and  put together all the theorems 
 which are important for our  present understanding of hadronic physics or for optimal fitting of the existing data.
 One  case at hand is  whether the limitations imposed by the Froissart bound are satisfied  and another case is
 the application to very high energy data fitting by Finite Energy Sum Rules (FESR), derived from analyticity and crossing.  
 

We shall discuss  the early formalism of the partial wave expansion of the elastic scattering amplitude, needed to understand the  
Martin-Froissart theorem, and relate it to the Regge pole expansion which played a major role in phenomenological description of 
inclusive and total cross-sections in the '60s and '70s. To accommodate such a description and the rise of $\sigtot$,  the Pomeron trajectory  
corresponding to the exchange of a state with the quantum numbers of the vacuum was introduced. Thus, a picture  of $\sigtot$, 
with a Regge and a Pomeron exchange,  unrelated to the underlying  parton dynamics  of scattering,  was one of the first and still 
very successful descriptions. 
Finally, from the partial wave expansion for the amplitude,  and  through  the optical theorem, we shall introduce the 
eikonal  representation of the total cross-section. This representation is at present the major formalism, into which QCD 
models for the energy behaviour of $\sigtot$ are embedded. 
This chapter  
is divided with section and subsection headings as indicated in the following:
\begin{itemize}
\item General principles behind relativistic scattering amplitudes in \ref{ss:generalprinciples}
\item Kinematics and analyticity of elastic amplitudes in  \ref{ss:kinematicsanalitycity}
\item Probability conservation and unitarity in  \ref{ss:unitarity}
\item The optical theorem and total cross-section in  \ref{ss:opticaltheorem}
\item Partial wave expansion of elastic amplitudes in  \ref{ss:elasticamplitude}
\item Regge expansion and asymptotic behaviour of amplitudes in  \ref{ss:regge}
\item Finite energy sum rules and duality for the elastic amplitudes in  \ref{ss:FESR}
\item Various derivations of the Martin-Froissart bound in  \ref{ss:froissart}
\item The Pomeranchuk theorem in  \ref{ss:pomeranchuk}
\item Determination of $\rho$ through Coulomb interference  in \ref{ss:coulomb}
with considerations about  Coulomb interference and soft radiation in   \ref{sss:interference} and  \ref{sss:softphotons}.
\end{itemize}

\subsection{General principles}\label{ss:generalprinciples}

Strong interactions are presently understood in terms of interactions between quarks and 
gluons. Quantum chromodynamics (QCD) can give   remarkably accurate
results within perturbation theory, when dealing with very high energy collisions and their final products 
in the large momentum transfer processes.
However, the bulk of collisions among high energy particles involves low  momentum partons which escape
the perturbative treatment. For this purpose, we have to resort to some general principles -valid beyond 
perturbation theory- to establish the necessary formalism and derive some general theorems.
Later we shall 
develop some tools to include QCD phenomena in this general picture.

These general principles were established in the late '50s and consist of unitarity, 
analyticity and crossing symmetry. Each of them is related to basic axioms: 
\begin{itemize}
\item unitarity to the conservation of probability in scattering processes; 
\item analyticity to causality and 
\item crossing symmetry to the relativistic nature of the interaction. 
\end{itemize}
These basic principles are also at the foundations of relativistic Quantum Field Theory (QFT) \cite{berestetskii:1982}.

We shall describe in detail how one obtains the so-called Froissart bound, which imposes limits to the 
asymptotic behavior of the total cross-section in two particle scattering. This limit was obtained first by 
Froissart \cite{Froissart:1961ux} and successively reformulated by Martin \cite{Martin:1962rt} and 
Lukaszuk \cite{Lukaszuk:1967zz}. The importance of this limit cannot be understimated, as most efforts 
to describe theoretically the total cross-section behaviour or most fits to present data must contemplate the 
asymptotic satisfaction of the Froissart bound. For this reason we shall describe how this limit is obtained in 
several different derivations, pointing out in all cases the common hypothesis, which is always the presence 
of a finite mass in final state scattering.

The basic quantity to study in particle physics is the probability that a certain 
set of particles in a given initial state $|i>$ undergo a collision and scatter into a 
final state $ |f>$.

To this effect, the process is described by the quantity
\begin{equation}
S_{fi}=<f|S|i>
\end{equation}
where $S$ is called the S-matrix (S for scattering) and $S_{fi}$ are the matrix elements. 
Since the scattering must also include the possibility that nothing occurs,
 the S-matrix is written in terms of the $T$-matrix, namely
\begin{equation}
S_{fi}=
\delta_{fi}+i (2\pi)^4\delta^4(P_f-P_i)T_{fi}
\label{eq:smatrix}
\end{equation}
where the 4-dimensional $\delta$-function imposes 
energy-momen--tum conservation on all particle momenta $p_j$, and,  with obvious notation,
$P_{i,f}=\sum_{all}p_{i,f}$. 
The relevant matrix elements define the scattering and  are functions of
the momenta of the scattering particles, in particular of the various invariants which 
can be constructed with the momenta. 
Let us then turn to the kinematics before going
 further into the dynamics.
\subsection{Kinematics of elastic scattering}\label{ss:kinematicsanalitycity}
Let us consider  the two body process
\begin{equation}
a(p_1)+b(p_2) \rightarrow c(p_3)+d(p_4)
\end{equation}
Usually, two different set-ups are most frequently encountered: center of mass collisions, as in most if not
all present day accelerator experiments at high energies; and fixed target collisions, as is the case for 
cosmic ray proton-air collisions or low energy photo-production experiments. It is usual to call {\it Laboratory frame} 
where fixed target collisions take place. However, there is another frequently encountered possibility, namely the 
kinematic configuration of two collinear particles of different momentum. This situation is found in electron and 
photon proton collisions at HERA and generally speaking is typical of parton-parton collisions. We shall present in 
the following the kinematics of all these three different possibilities.
 
In the c.m. frame of particles $a$ and $b$, we write
\begin{eqnarray}
p_1^\mu=(E_a,0,0,p)\\
p_2^\mu=(E_b,0,0,-p)\\
p_3^\mu=(E_c,q\sin \theta,0,q\cos \theta)\\
p_4^\mu=(E_d,-q\sin \theta,0,-q\cos \theta)
\end{eqnarray}
which can be described by two independent  variables, to be chosen among  three  relativistic invariants, 
the so called Mandelstam variables, i.e.
\begin{eqnarray}
s=(p_1+p_2)^2=s=(p_3+p_4)^2 \\
t=(p_1-p_3)^2 s=(p_2-p_4)^2\\
u=(p_1-p_4)^2s=(p_2-p_3)^2
\end{eqnarray}
For general processes, we have
\begin{equation}
s+t+u=m_a^2+ m_b^2+m_c^2+m_d^2=h
\label{eq:stu}
\end{equation}
 and thus for 
 elastic scattering, namely
\begin{equation}
a+b\rightarrow a+b
\end{equation}
we have $p\equiv q$ with
\begin{eqnarray}
p^2=\frac {s^2+(m^2_a-m^2_b)^2 -2s(m^2_a+m^2_b)} {4s}\\
=\frac {[s-(m_a+m_b)^2][s-(m_a-m_b)^2 } {4s}
\end{eqnarray}
and 
\begin{eqnarray}
s= m_a^2+m_b^2+2p^2+2\sqrt{m_b^2+p^2}\sqrt{m_b^2+p^2}\\
t=-2p^2[1-\cos\theta]
\end{eqnarray}
where $\theta$ is the scattering angle in the c.m. frame. 

For collisions not taking place in the center of mass, the kinematics reads differently. 
While \pp\ and \pbarp\ scattering in present day accelerators take place through center
 of mass collisions, this was not true for early experiments, where typically a  proton 
or antiproton was directed to a fixed hydrogen target, and it is also not true for meson 
proton scattering, such as $\pi p$,  $Kp$ or $\gamma p$, where pions, kaons or photons are 
directed to a fixed hydrogen target. In such cases, for the kinematics in the 
laboratory frame we get
\begin{equation}
s=m_a^2+m_b^2+2m_aE_b
\end{equation}
so that 
\begin{eqnarray}
s_{pp}=2m_p^2+2m_pE_{lab}\\
s_{\pi p}=m_p^2+m_\pi^2+2m_pE)\pi\\
s_{\gamma p}=m_p^2+2m_pE_\gamma
\end{eqnarray}
In all the above cases, the proton is at rest in the laboratory. A  different case is
 the one encountered at HERA, where the two beams, photons and protons, collide with 
different momenta. For real photons of momentum $q$ colliding with a proton of energy 
$E_p$, one has 
\begin{equation}
q=\frac {s-m^2} {2m^2} E_p (1-\sqrt{1-\frac{m^2} {E^2_p}})
\end{equation}
Kinematics is still different for virtual photon scattering and will
 be described in Sect.~\ref{sec:photons}.

\begin{figure}
\label{fig:physicalregion} 
\resizebox{0.50\textwidth}{!}{%
  \includegraphics{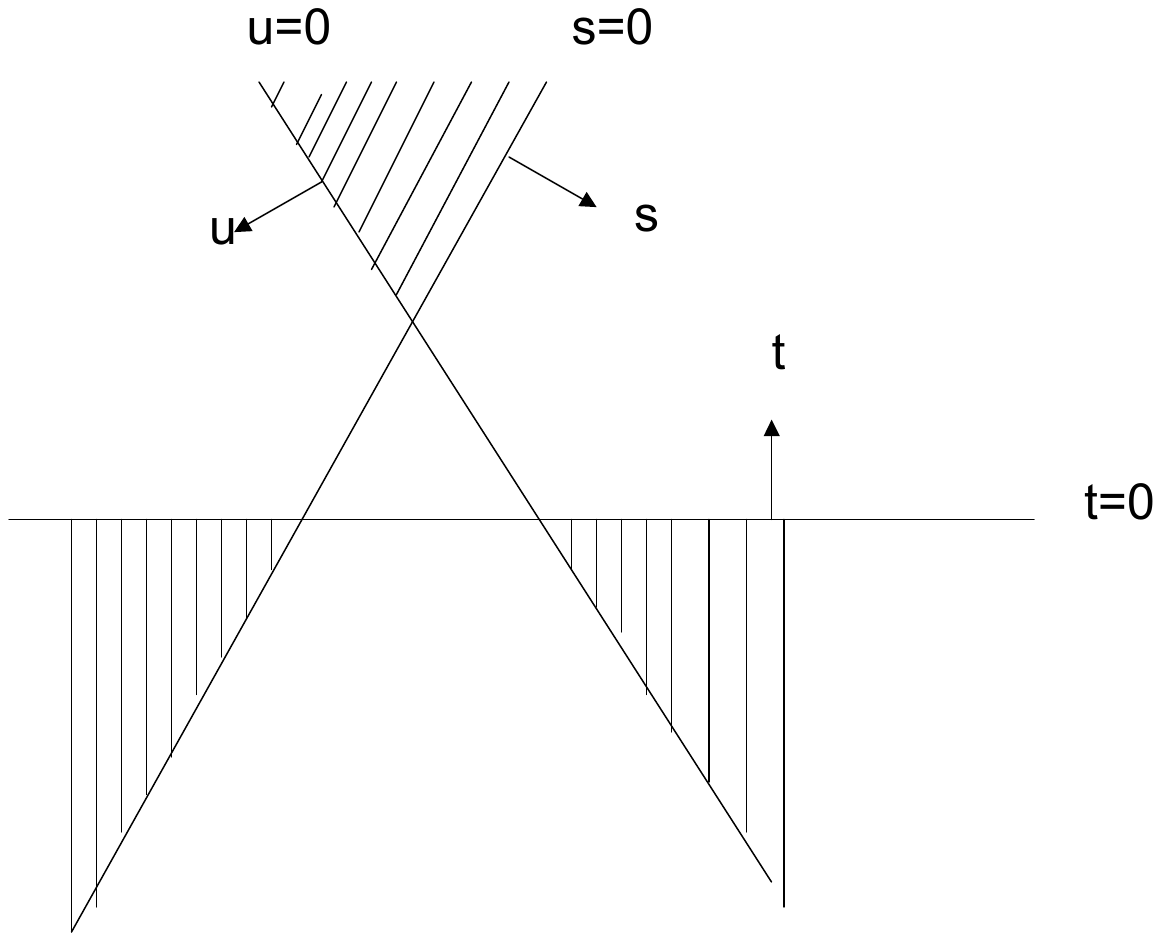}
}
\caption{Physical region for equal mass elastic scattering.}
\end{figure}

Because of energy momentum conservation and of the condition imposed 
 by Eq.~(\ref{eq:stu}), physical processes can take place only for those values of the 
variables  s, t and u which lie in the so called {\it physical region}. Such a region is 
defined as \cite{berestetskii:1982}
\begin{equation}
stu\le as+bt+cu
\end{equation} 
where 
\begin{eqnarray}
ah=(m_1^2m_2^2-m_3^2m_4^2)(m_1^2+m_2^2-m_3^2-m_4^2)\\
bh=(m_1^2m_3^2-m_2^2m_4^2)(m_1^2+m_3^2-m_2^2-m_4^2)\\
ch=(m_1^2m_4^2-m_2^2m_3^2)(m_1^2+m_4^2-m_2^2-m_3^2)
\end{eqnarray}
For the equal mass case, this reduces to the condition $stu\le 0$ and the allowed 
regions are shown in the dashed areas of Fig. ~\ref{fig:physicalregion}.

\subsection{Unitarity and the scattering amplitude}\label{ss:unitarity}
The measurement of the total cross-section is based on two complementary methods: 
counting the number of collisions and, measuring the very forward scattering probability. 
The second method is based on a fundamental physical property i.e., the conservation 
of probability, 
which is embedded in the unitarity property of the S-matrix, namely
\begin{equation}
SS^\dagger={\bf 1}
\end{equation} 
In terms of the matrix elements, we have 
\begin{equation}
(SS^\dagger)_{fi}=\sum_n S_{fn} S_{ni}^*=\delta_{fi}
\label{eq:Sunitarity}
\end{equation}
where $n$ runs on all possible intermediate states. This condition ensures the normalization and 
orthogonality of  states in the reaction. In particular, for the $i=f$ case,  Eq.~\ref{eq:Sunitarity} 
ensures that the sum over all allowed transitions from a given state $|i>$ to any possible final state, 
is one, namely
\begin{equation}
\sum_n |S_{ni}|^2=1
\label{eq:probability}
\end{equation}
Eq.~(\ref{eq:probability}) is the statement of conservation of probability in the scattering.

We can now proceed to derive the optical theorem, by using Eqs.~(\ref{eq:Sunitarity}) and  (\ref{eq:smatrix})
to obtain
\begin{equation}
T_{fi}-T^*_{if}=(2\pi)^4\sum_n\delta^4(P_f-P_n)T_{fn}T^*_{in}
\label{eq:Tunitarity}
\end{equation}
Because  the left hand side of this equation is linear in T, while the right hand side is quadratic,  if the T-matrix 
can be expanded in a small parameter (say a coupling constant), then unitarity ensures that the T-matrix elements 
are hermitian. In the general case, one uses Eq.~(\ref{eq:Tunitarity}) to obtain the optical theorem, namely
\begin{equation}
2 Im T_{ii}=(2\pi)^4\sum_n\delta^4(P_i-P_n)|T_{in}|^2
\label{eq:optical}
\end{equation}
where the amplitude $T_{ii}$ indicates  elastic scattering in the forward direction and where the right hand side, 
a part from a normalization factor, gives the total cross-section for scattering from an initial state $|i>$ into any 
possible final state, as shown in the following subsection. The reader is warned that different authors use
different normalizations for the elastic scattering amplitudes and hence due care must be taken in using 
various unitarity expressions.

 
\subsection{The optical theorem and the total cross-section}\label{ss:opticaltheorem}
We follow here the definitions and normalizations as in \cite{perl:1974}. Let us start with the general definition of total cross-section, 
by first introducing the probability that a given two particle initial state $|i>$ scatters into all possible  final states $|f>$, namely
\begin{equation}
\sum_f P_{fi}=\sum_{f\alpha}\int [ \prod_{n=1}^{N_f} \frac{d^3p_n} {(2\pi)^32E_n}(S^\dagger_{if}S_{fi})]
\end{equation}
where the sun runs over all final states and all possible quantum numbers $\alpha$ of all possible final states. 
Next we use the S-matrix definition in terms of the T-matrix 
\begin{eqnarray}
\sum_f P_{fi} =\sum_{f\alpha}\int  \prod_{n=1}^{N_f} \frac{d^3p_n} {(2\pi)^32E_n}
\nonumber \\ 
\times |T_{fi}|^2
(2\pi)^4\delta^4(P_f-P_i) (2\pi)^4\delta^4(P_f-P_i)
\eea
and define the probability of the scattering per unit volume and unit time, by using the conventional way to 
interpret $(2\pi)^4\delta^4(P_f-P_i)$ as  the four-dmensional scattering volume VT. Using the language of the 
laboratory frame, where the initial state consists of a target particle (T) and a projectile
(P), a further step is taken by considering the scattering per target particle, dividing by the target particle density 
$2E_{T}$, and obtaining the cross-section by further dividing this probability by  the  flux of incoming particles, 
$2E_{P}v_{P, lab}$. We then have
\begin{eqnarray}
\sigma_{tot}\equiv \ \ \  \ \ \  \ \ \ \  \ \ \  \   \  \ \ \ \  \ \ \ \  \ \ \  \ \ \ \  \ \ \  \   \  \ \ \ \  \  \ \ \  \ \  \ \ \ \nonumber \\
 \sum_{f\alpha} \frac{(Probability \ per\  target \ particle\ per \ unit \ time )}{ \ flux \ of \ incoming\ particles}=\ \ \ \nonumber 
\\  \sum_{f\alpha} \frac{ (Prob.\ per\ target\ particle\ per\ unit\ time)}{2E_{P} v_{P,lab}}= \ \ \ \nonumber \\
\frac{(2\pi)^4}{4E_{T}E_{P}v_{P,lab}}\sum_{f\alpha}\int  \prod_{n=1}^{N_f} \frac{d^3p_n} {(2\pi)^32E_n} |T_{fi}|^2\delta^4(P_f-P_i)\ \ \  \ \   \eea 
The next step is to use Eq.~(\ref{eq:optical}) to relate the total cross-section to the imaginary part of  the 
forward scattering amplitude so as to obtain, in the cm frame, 
\begin{equation}
\sigma_{total}=\frac{Im T_{ii}}{2k\sqrt{s}}
\label{eq:opticaltheorem}
\end{equation}
where $k$ is the center of mass momentum of the incoming particles and $\sqrt{s}$ the c.m. energy.
We then see that the total cross-section can be measured in two different ways, either through the total count of all 
the final states hitting the detector or through the imaginary part of  the forward elastic amplitude. In the next section, 
we will establish some definitions and properties of the elastic scattering amplitude.
\subsection{The  elastic scattering amplitude and its partial wave expansion}\label{ss:elasticamplitude}
For two equal-mass particle scattering in the c.m. system, the Mandelstam invariants $s,t,u$ take a particularly simple form 
and the physical region for the s-channel is defined as
\bea
q^2_s=\frac{s-4m^2} {4}>0, \ \  \cos\theta_s=1+\frac{t}{2q^2_s}<1; \\
or\ \ \ \  s>4m^2, t\le 0 , \,u \le 0
\eea
Let then the elastic scattering amplitude $A(s,cos\theta_s)$ be expanded in a series of Legendre polynomials
\be
A^{F}(s,\cos\theta_s) =\frac {\sqrt{s}}{\pi q_s}\sum_{l=0}^\infty (2l+1)P_l(\cos\theta_s)a_l^F(s),
\label{Froissart1}
\ee
where the subscript $F$ refers to the normalization used by Froissart. Martin's normalization differs by a factor $\pi$, namely
\be
A^{M}(s,\cos\theta_s) =\frac {\sqrt{s}}{ q_s}\sum_{l=0}^\infty (2l+1)P_l(\cos\theta_s)a_l^M(s)
\ee
For simplicity, we shall now use $\theta_s\equiv\theta$. Using elastic unitarity, it is rather simple to obtain some limits 
on the partial wave amplitudes $a_l$ .
\subsection{Asymptotic behaviour and Regge theory}\label{ss:regge}
We present here a brief description of the Regge expansion which has been very useful in molding our ideas about 
the behavior of elastic and total cross-sections as a function of energy. The Regge picture forms the backbone of high 
energy phenomenology of cross-sections. To illustrate its central theme, let us consider the partial wave expansion of 
an elastic scattering between two equal mass spinless particles of mass $m$
\be
A(s,\cos\theta_s) = \sum_{l=0}^\infty (2l+1)P_l(z_s)a(l,s), 
\label{Regge1}
\ee
where the partial wave amplitude $a(l,s) = (\frac {\sqrt{s}}{\pi q_s}) a_l^F(s)$ and $z_s = \cos \theta_s$. 
This expansion, for physical $s$-channel scattering ($s > 4 m^2$) certainly converges for $|z_s|\leq 1$. 
The Regge expansion consists in obtaining a representation valid for large $z_s$ through a continuation from 
integral values of $l$ to continuous (complex) values of $l$ via the Sommerfeld-Watson (W-S) transformation. 
In non-relativistic potential scattering, Regge was able to prove that for a superposition of Yukawa potentials, 
the amplitude $a(l,s)$ is an analytic function of $l$ and its only singularities are poles [the famous Regge poles, 
$l = \alpha(s)$)] and that bound states and resonances are simply related to them. The situation in the relativistic case 
is less clear and technically more involved\cite{Collins:1970kn,martin1970elementary}
For integral values of $l$, 
Eq.~(\ref{Regge1}) can be inverted to give
\be
a(l,s) = \frac{1}{2} \int_{-1}^{1} A(s,z) P_l(z)
\label{Regge2}
\ee
While the above equation permits an analytic continuation of the function $a(l,s)$ to complex values of $l$, 
it is not suitable for completing the W-S transformation due to the bad asymptotic behavior of $P(l,z)$ for complex 
$l$\cite{Collins:1970kn}. Hence, a technical nicety, the Froissart-Gribov projection, is required. Assume that 
$A(s,z)$ is polynomially bounded so that that a fixed $s$-dispersion relation (with N subtractions) can be written 
down in the variable $z$:
\begin{eqnarray}
A(s,z) = \sum_{n = 0}^{N - 1} \gamma_n z^n +\ \nonumber\\
 \frac{z^N}{\pi} \int_{z_r}^\infty \frac{dz^{'} D_t(s,z^{'})}{z^{'N}(z^{'}-z)} +\ \nonumber\\
 \frac {z^N}{\pi}\int_{-z_l}^{-\infty} \frac{dz^{'} D_u(s,z^{'})}{z^{'N}(z^{'}-z)}\ \nonumber\\,
\label{Regge3}
\end{eqnarray}   
where $D_t$ and $D_u$ are the $t$ and $u$ channel discontinuities of the amplitude. Substituting the above 
in Eq.~(\ref{Regge2}), we find that
\be
a(l,s) = \frac{1}{\pi}[\int_{z_r}^{\infty}dx D_t(s,x) Q_l(x) + \int_{-z_l}^{-\infty}dx D_u(s,x) Q_l(x)],
\label{Regge4}
\ee
obtained upon using the identity
\be
Q_l(x) = \frac{1}{2} \int_{-1}^{1}(dx) \frac{P_l(z)}{z - x}.
\label{Regge5}
\ee
Since for positive integral values of $l$, $Q_l(-z) = (-1)^{l +1} Q_l(z)$, we may rewrite Eq.~(\ref{Regge4}) as
\be
a(l,s) = \frac{1}{\pi}\int_{z_o}^{\infty}dx [D_t(s,x)+(-1)^l D_u(s,x)] Q_l(x),
\label{Regge6}
\ee
where $z_o$ is the smaller of $z_l$ and $z_r$. To avoid obtaining dangerous factors such as $e^{i\pi l}$ for complex
$l$ when we analytically continue Eq.~(\ref{Regge6}), it is useful to define the ``signatured'' Froissart-Gribov amplitudes $a^{\pm}(l,s)$
\be
a^{\pm}(l,s) = \frac{1}{\pi}\int_{z_o}^{\infty}dx [D_t(s,x) \pm D_u(s,x)] Q_l(x),
\label{Regge7}
\ee 
which can be continued for all $\Re e\ l > N$, since in this region, the above integrals converge. The positive signature amplitude 
$a^+(l,s) = a(l,s)$ for even $l$ and the negative signature $a^-(l,s) = a(l,s)$ for odd $l$. Thus, the W-S transformation is to be 
performed on the signatured total amplitudes
\be
A^{\pm}(s,z) = \sum_{l = 0}^{l = \infty} (2l + 1) a^{\pm}(l,s) P_l(z),
\label{Regge8} 
\ee
separately. The physical amplitude is then given by the combination
\be
A(s,z) = \frac{1}{2}[ A^+(s,z) + A^+(s,-z) + A^-(s,z) - A^-(s,-z)].
\label{Regge9} 
\ee
For each of the amplitudes in Eq.~(\ref{Regge8}), one first replaces the sum by a contour C which encircles all the integers in the 
sum
\be
A^{\pm}(s,z) = \frac{i}{2} \int_{C} \frac{(2l + 1) a^{\pm}(l,s) P_l(-z)}{\sin \pi l},
\label{Regge10} 
\ee  
since the function $\sin \pi l$ has poles at all the integers with residue $(-1)^l/\pi$ and use has been made of the
property that for integers $P_l(-z) = (-1)^l P_l(z)$. The next step is to open the contour as
in  Fig.~\ref{fig:reggeplane}
\begin{figure}
\resizebox{0.5\textwidth}{!}{
\includegraphics{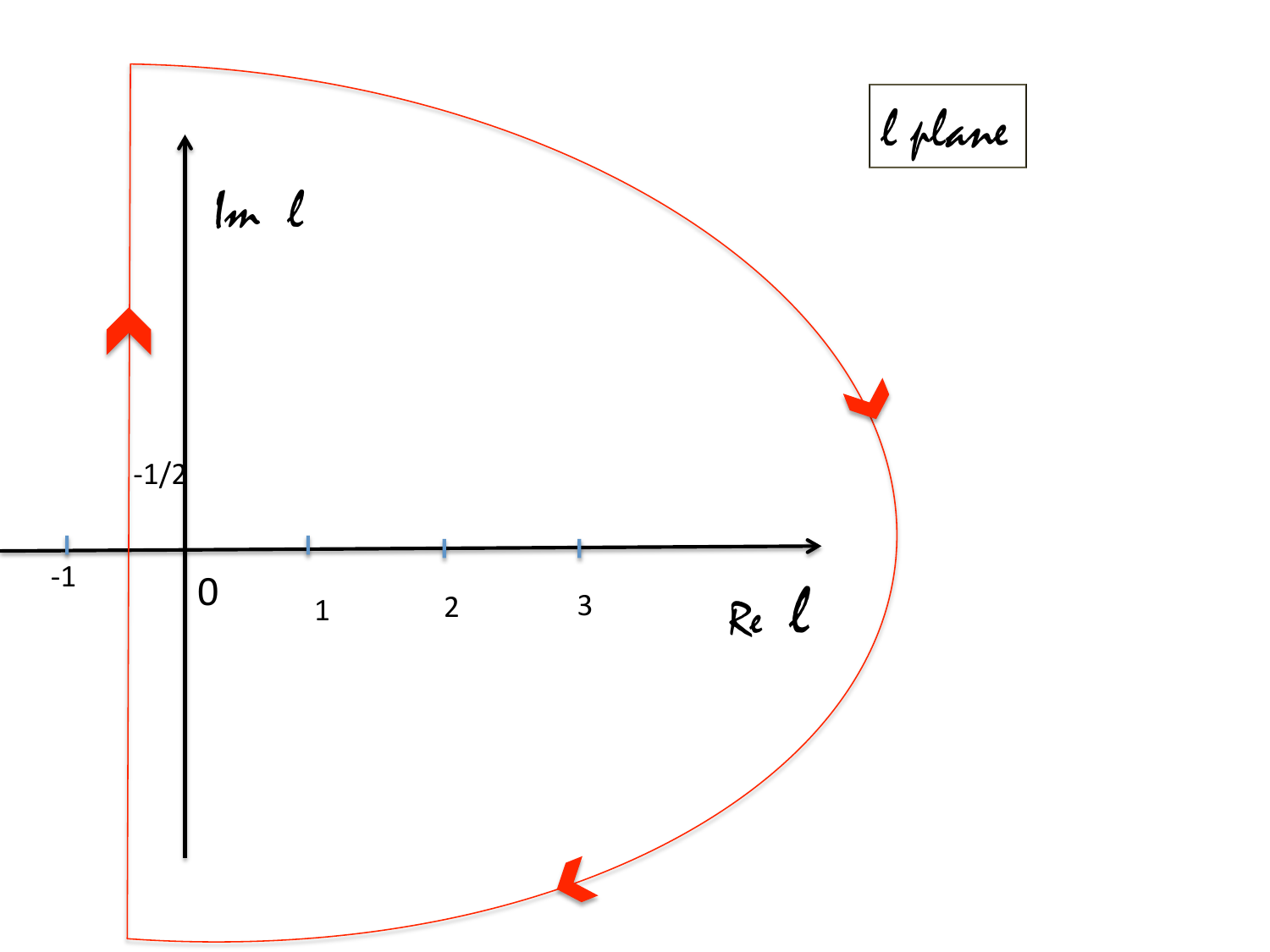}}
\caption{
Graphical representation of the Sommerfeld-Watson transformation contour in the angular momentum plane.}
\label{fig:reggeplane}
\end{figure}
  and one finds 
a large semi-circle in the positive quadrant, a background integral running vertically at $\Re e\ l = -1/2$ and the contributions 
from any singularities in $a^{\pm}(l,s)$. We expect Regge poles and perhaps Regge cuts. Ignoring the cuts for the moment, we may write
\begin{eqnarray}
A^{\pm}(s,z) = \frac{i}{2}\int_{C^{'}} \frac{(2l + 1) a^{\pm}(l,s) P_l(-z)}{\sin \pi l}\ \nonumber\\
- \sum_j \frac{\pi\beta_j^{\pm} (2\alpha_j^{\pm} + 1)P_{\alpha_j^{\pm}}(-z)}{\sin \pi \alpha_j^{\pm}},
\label{Regge11}
\end{eqnarray}   
where the sum includes all poles with $\Re e\ \alpha_j(s) > -1/2$ and the corresponding $\beta_j^{\pm}(s)$ denote their residue.
For large $z$ (which is synonymous with large $t$ for fixed $s$), $P_\alpha(z) \rightarrow\ z^\alpha$ and thus, in this limit 
$A^{\pm}(s,z)$ would be controlled by the Regge pole to the farthest right (called the leading Regge pole). Hence, one 
arrives at the Regge asymptotic behavior result that
\be
A^{\pm}(s,t) \rightarrow\ \chi^{\pm}(s) t^{\alpha_{\pm}(s)},
\label{Regge12}
\ee 
in the limit $t \rightarrow\ \infty$ for fixed $s$. Of course, had we made the Regge expansion in the $t$ channel, 
we would have obtained the result
\be
A^{\pm}(s,t) \rightarrow\ \chi^{\pm}(t) s^{\alpha_{\pm}(t)},
\label{Regge13}
\ee  
in the limit $s \rightarrow\ \infty$ for $t$ fixed. Putting in the proper phases, we obtain for the pole
contribution to the amplitude (which dominates the background integral for all Regge poles with $\Re e\ \alpha(t) > -1/2$) to be of the form
\begin{eqnarray}
A(s,t) \rightarrow\ -\gamma^+(t)\frac{e^{-i\pi\alpha^+(t)} + 1}{\sin \pi \alpha^+(t)}(\frac{s}{s_0})^{\alpha^+(t)}\
 \nonumber\\  
 -\gamma^-(t)\frac{e^{-i\pi\alpha^-(t)} - 1}{\sin \pi \alpha^-(t)}(\frac{s}{s_0})^{\alpha^-(t)}\ \nonumber\\.
\label{Regge14}
\end{eqnarray} 
Using the form -valid for large $s$ (i.e. ignoring masses)-
\be
\sigma_{tot}(s) = (\frac{16 \pi}{s})\Im m\ A(s,0),
\label{Regge15}
\ee
we have
\be
\sigma_{tot}(s) \rightarrow \frac{16 \pi}{s_0}[\gamma^+(0)(\frac{s}{s_0})^{\alpha^+(0) - 1}
+ \gamma^-(0)(\frac{s}{s_0})^{\alpha^-(0) - 1}].
\label{Regge16}
\ee
If $\alpha^+(0) = 1$, then the total cross-section would go to a constant value. This is the celebrated
Pomeron pole. It has the added virtue that the ratio of the real part to the imaginary part of the forward
elastic amplitude would be strictly zero, i.e.
\be
\frac{\Re e\ A(s,0)}{\Im m\ A(s, 0)}|_{\alpha^+(0) = 1}\rightarrow\ 0,
\label{Regge17}
\ee
exhibiting the limiting feature of diffraction scattering. Hence, the early excitement about the Pomeron.

By contrast, were $\alpha^-(0) = 1$, not only would the relative roles of the real and the imaginary parts be 
reversed but there would be a genuine spin 1 massless physical particle pole (analogous to the photon) in the 
elastic amplitude. Since in the hadronic spectrum we have have no massless particles -of any spin- we would 
conclude that $\gamma^-(0) =0$ if $\alpha^-(0) = 1$. Hence, there would be no contribution to the total 
cross-section from an $\alpha^-(0) = 1$ Regge pole (since it would have a vanishing residue). However, the 
real part may be finite then.

Experimental data clearly indicate that (i) all total cross-sections increase at high energies and that (ii) the ``rho'' parameter
\be
\rho (s,o)| = \frac{\Re e\ A(s,0)}{\Im m\ A(s, 0)} < < 1.
\label{Regge18}
\ee
Question then arises as to how to implement these facts phenomenologically in a Regge picture. 
Some theoretical progress has been made regarding the imaginary part in QCD. In the BFKL 
Pomeron\cite{Bartels:2008ce} model, one finds that the Pomeron intercept is slightly greater than 1, 
i.e., $\alpha^+(0) = 1 + \epsilon$, where $\epsilon = (4 \alpha_s N_c/\pi) ln 2$, where $\alpha_s$ is the 
QCD coupling constant and $N_c$ is the number of colours ($3$ for QCD). Thus, 
$\sigma_{tot} \approx\ (s/s_0)^\epsilon$ would rise with energy. While for small enough $\epsilon$, 
this may work for some energy band, it would eventually be in conflict with the Froissart bound discussed 
at length in the subsequent sections. The Froissart upper bound only permits a maximum increase 
$\sigma_{tot} \leq\ \sigma_P\ ln^2(s/s_0)$. 

Powers of logarithms can arise due to the confluence of two (or more) pole singularities. For example, if in the 
angular momentum plane, there occurs a double pole at $l = \alpha(t)$, its contribution to the W-S integral would be 
through a derivative (in $l$ evaluated at $l = \alpha$) \cite{newton1964complex}.
Asymptotically then, if a simple pole 
gave $A(s,t)\approx\ (s/s_0)^\alpha$, a double pole would give $A(s,t)\approx\ (s/s_0)^\alpha ln(s/s_0)$. To saturate 
the Froissart bound, we need two derivatives in $\alpha$ i.e., a third order pole, with of course $\alpha(0) =1$. 
On the other hand, the more general case, i.e., generation of a fractal power such as 
\be
\Im m A(s.t) \rightarrow\ (s/s_0) [ln (s/s_0)]^{1/p}\ (with\ 1/2<p<1) 
\label{Regge19}
\ee
(as found in a 
phenomenologically 
successful model for total cross-section to be discussed later in \ref{sss:BN} of this review) would require a  
confluence of an indefinite number of pole trajectories all converging at $\alpha(0) = 1$. We remark here in passing 
that, near a threshold, due to unitarity, a confluence of an infinite number of trajectories (the ``threshold poles'') does 
occur and it has been well studied \cite{newton1964complex}.
It is an open problem to deduce what happens in the 
vacuum channel of QCD with (almost massless) quarks and gluons. This problem is particularly difficult in QCD because 
it can only be answered satisfactorily after unitarity is imposed -a daunting task indeed.

The spectrum of mesonic masses leads one to conclude that there are four almost-degenerate Regge trajectories 
with intercepts close to $1/2$ \cite{martin1970elementary}
 Hence, in a total cross-section for the scattering of particle $a$ with $b$, 
these terms provide the next to the leading contribution (about half a unit lower than the Pomeron) of the form 
$\sigma(ab)_{Regge}(s) = \sum_{i=1}^4 \sigma_i(ab) (s/s_0)^{(\alpha_{i} -1)}$ with the sum running over the 
$\rho$, $\omega$, $f$ and $A_2$ Regge trajectories. This nomenclature recalls the lowest spin resonance associated 
with a given Regge trajectory. As discussed in the FESR  and duality section, the approximate degeneracy 
$\alpha_i \approx\ 1/2$ is deduced from the absence of resonances in ``exotic'' channels. 

Thus, a phenomenological parametrization based on the Regge picure for the high energy total cross-section of particles 
$a$ and $b$ may be formulated as \cite{Grau:2009qx}
\begin{eqnarray}
\sigma_{tot}(ab) = \sigma_P(ab)[ln (s/s_0)]^{1/p} + \sigma_o(ab)\ \nonumber\\ 
+ \sum_i \sigma_i(ab)(s/s_0)^{(\alpha_{i} -1)} ,\ \nonumber\\ 
\label{Regge20}
\end{eqnarray} 
where the constants $\sigma_P(ab)$, $\sigma_o(ab)$, and $\sigma_i(ab)$ are the respective coefficients of the ``Pomeron'', 
an overall constant and the various Regge terms for the scattering process $a$ on $b$. The constant $p$ obeys the condition 
($1/2\leq p \leq 1$) and $\alpha_i \approx\ 0.5$.

Regarding the asymptotic behavior of the $\rho$ parameter, defined in Eq.(\ref{Regge18}), let us use the generic fractal amplitude 
as given in Eq.(\ref{Regge19}) 
\be
A_{fractal}(s,0) =  A_0 (\frac{s e^{-i\pi/2}}{s_0}) [\ln \frac{s e^{-i\pi/2}}{s_0}]^{1/p}, 
\label{Regge21}
\ee
where $A_0$ is a real constant and we have employed the phase rule $s\rightarrow s e^{-i\pi/2}$ for crossing-even 
amplitudes\cite{Block:1984ru}. This would give for the asymptotic form for $\rho$
\be 
\rho_{fractal}(s,0)\rightarrow\ \frac{ \pi}{2 p\ ln(s/s_0)}.
\label{Regge22}
\ee
This generalizes for arbtrary $p$ a rigorous result\cite{Khuri:1965zz}, valid for an amplitude saturating the Froissart bound 
(here achieved for $p = 1/2$).

While it may be difficult to distinguish between a total cross-section increasing as ($[ln(s/s_0)]^2$) or  [$ln(s/s_0)$] 
[or some power in-between for $1/2<p<1$], it may be easier to use experimental measurements of $\rho$ and employ 
Eq.(\ref{Regge22}) to decipher the value of $p$, since $\rho$ depends on ($1/p$) linearly. In any event, one has two 
consistency conditions provided by Eqs.(\ref{Regge20}) and (\ref{Regge22}) for the parameter $p$.

In the next section, we discuss an important off-shoot from the Regge expansion which goes under the names of 
finite energy sum rules and duality.

\subsection{Constraints from FESR and Duality for the total cross-sections}\label{ss:FESR}

Analyticity in the complex (energy) plane for a function (say a form factor or an elastic scattering amplitude) 
quite generally implies that its values in the ``low'' and ``high'' parts of the complex plane must be intricately related. 
This obvious fact has been used successfully to relate integrals over the low energy parts of amplitudes to those 
over their asymptotic high energy (Regge) parts. 

To illustrate what is involved, consider the simplest but physically quite important example of the charge form-factor 
$F(s)$ of the proton normalized as $F(0) = 1$. Under the usual hypothesis that for space-like values $s = - Q^2 < 0$, 
the function is real and that it has a right hand cut beginning at the physical charged particle-antiparticle thresholds, 
$s_o = 4m_\pi^2, s_1 = 4m_K^2, s_2 = 4m_p^2,...$, we may write a dispersion relation 
\be
F(s) = 1 + \frac{s}{\pi} \int_{s_o}^\infty \frac{ds^{'} \Im m F(s^{'})}{s^{'} (s^{'} - s - i\epsilon)}
\label{F1}
\ee        
Let us use the extra (experimental) information that for large (space-like) $Q^2 \rightarrow \infty$, 
$F(Q^2)\rightarrow 0$. Then, Eq.(\ref{F1}) gives us a sum rule
\be
\frac{1}{\pi} \int_{s_o}^\infty \frac{ds \Im m F(s)}{s} = 1,
\label{F2}
\ee 
which provides a relationship between the integrals over the low and high energy parts of (the imaginary parts) of the 
form factor. Since also, the neutron charge form facor goes to zero for large $Q^2$, we would obtain an expression 
analogous to Eq.(\ref{F2}) also for the neutron except that the right hand side would be zero. In vector meson 
dominance (VMD) models, the couplings of the $\rho$, $\omega$ and the $\phi$ to the nucleons get constrained accordingly.

Actually experimental data regarding form factors are much more stringent: it appears that the fall off of the proton 
form factor is of the ``dipole'' type. For purposes of illustration, let us assume that $Q^2 F(Q^2) \rightarrow 0$ as 
$Q^2 \rightarrow \infty$. Then, we can derive a ``superconvergence'' relation
\be
\int_{s_o}^\infty (ds) \Im m F(s) = 0.
\label{F3}
\ee
Eq.(\ref{F3}) tells us that $\Im m F(s)$ must change sign at least once. To meet this exigency then, a generalized 
vector meson (GVMD) model with other vector mesons $\rho^{'}, \omega^{'}, \phi^{'}$ etc. with their couplings 
(of reversed signs) to the nucleons have to be introduced. It is not our purpose here to advocate GVMD models 
but to illustrate very simply that dispersion relations with some knowledge -be it experimental or theoretical- about 
the behaviour of an amplitude at some value, allows us to put constraints at other values.
\footnote{Sergio Fubini, the discoverer of superconvergence relations, made
an analogy between the knowledge of an amplitude locally  and some knowledge about the amplitude through 
sum rules (e.g., superconvergence integrals) to that between Coulomb's law giving the local value of a field and Gauss' law 
providing an integrated statement about the field.  }

Let us now turn to a specific case that of the elastic meson-baryon amplitudes, with an eye towards their later 
applications to photon-nucleon total cross sections. For fixed $t$, in order to exploit the crossing symmetry between 
the $s$ and the $u$ channels, one defines the variable\cite{martin1970elementary}
\be
\nu = \frac{s - u}{4m} = \omega + \frac{t}{4m},
\label{F4}
\ee
so that $\omega$ denotes the energy of the meson in the rest frame of the baryon and $m$ denotes the mass of the baryon. 
For a crossing-odd scattering amplitude\cite{martin1970elementary}
$T(\nu,t) = T^*(-\nu,t)$, we may write a fixed-$t$ dispersion relation
\be
\label{F5}
T(\nu, t) =\ (possible\ poles)\ +\ \frac{2\nu}{\pi} \int_o^\infty (d\nu^{'}) \frac{\Im m T (\nu^{'},t)}{(\nu^{'2} - \nu^2)},
\ee
Let us assume generic asymptotic Regge terms of the form 
\be
\label{F6}
T_{Regge}(\nu, t) = \sum_i \beta_i(t) [\frac{\pm 1 - e^{-i\pi\alpha_i(t)}}{\Gamma(\alpha_i(t)+1) sin\pi\alpha_i(t)}]\nu^{\alpha_i(t)}, 
\ee
Using arguments previously given, if all the $\alpha(t) < -1$, we would obtain a superconvergence relation
\be
\label{F7}
\int_o^\infty (d\nu) \Im m T(\nu, t) = 0.
\ee
Instead, we can subtract the contributions from all $\alpha_i(t) > -1$ to obtain a superconvergence relation
of the form
\be
\label{F8}
\int_o^\infty (d\nu) [ \Im m T(\nu, t) - \sum_{\alpha_i(t)>-1}\frac{\beta_i(t)}{\Gamma(\alpha_i(t)+1)} \nu^{\alpha_i(t)}] = 0
\ee
Since asymptotically -by construction- the integrand in Eq.(\ref{F8}) goes to zero, we may replace the upper limit of the 
integration to be $\nu = N$ and include the left over Regge terms with $\alpha < -1$, and find
\begin{eqnarray}
\label{F9}
\int_o^N (d\nu) [ \Im m T(\nu, t) - \sum_{\alpha_i(t)>-1}\frac{\beta_i(t)}{\Gamma(\alpha_i(t)+1)} \nu^{\alpha_i(t)}]\ \nonumber\\
+ \sum_{\alpha_j<-1}\frac{\beta_j(t)}{\Gamma(\alpha_j(t) +1)}\int_N^\infty (d\nu) \nu^{\alpha_j(t)}  = 0.\nonumber\\
\end{eqnarray} 
Doing the integral, we have the finite energy sum rule (FESR)
\be
\label{F10}
{\cal S}_o = \frac{1}{N} \int_o^N (d\nu) \Im m T(\nu, t) = \sum_{all\ \alpha} \frac{\beta N^\alpha}{\Gamma(\alpha(t) +2)}
\ee
Also. higher moment sum rules may be written. For even integer $n$, we have
\begin{eqnarray}
\label{F11}
{\cal S}_n = \frac{1}{N^{n+1}} \int_o^N (d\nu)\nu^n \Im m T(\nu, t)\ \nonumber\\ 
= \sum_{all\ \alpha} \frac{\beta N^\alpha}{(\alpha(t) + n +1)\Gamma(\alpha(t) +1)}
\end{eqnarray}

FESR can also be constructed for crossing even amplitudes and we shall return to them later. 

As emphasized in \cite{martin1970elementary},
the relative importance of successive terms in a FESR is the same as 
in the usual Regge expansion: if a secondary pole is unimportant at a high energy above $\nu = N$ then this 
term would be unimportant to exactly the same instant in the sum rule. For $\pi N$ elastic scattering in the 
$t$-channel iso-spin $I_t = 1$, FESR have been exploited with much success to obtain information about the 
$\rho$ and the $\rho^{'}$ trajectories \cite{Igi:1962zz}.
Different variants of the idea have been used, see for example
\cite{Gatto:1967zz}



 In FESR, the scattering amplitude is multiplied by an integral power of the laboratory energy. 
 This was generalized to continuous moment sum rules(CMSR) \cite{Dellaselva68}.  
 In contrast to FESR, in CMSR, the multiplicative energy factor is non-integral. However, 
 CMSR turn out to be simply a superposition of FESR, if the real part of the amplitude is 
 calculated using dispersion relations \cite{Ferrari69}.
For a review of the applications of these ideas to specific processes, see \cite{Violini72}.

An interesting fall out from FESR was the concept of duality\cite{Dolen:1967jr} which in its final form 
may be phrased as follows. Consider a generic amplitude $A(s,t)$ and decompose its imaginary part 
(in the $s$-channel) in terms of the $s$-channel resonances and a smooth background. Then, 
the assertion is that ``direct''($s$) channel resonances are ``dual'' to the crossed ($t$) channel Regge 
trajectories and the Pomeron term(s) is(are) dual to the background. Explicitly, it means that in 
Eq.(\ref{F11}) the integral over the left hand side would contain contributions from $s$ (and $u$) 
channel (baryonic)resonances whereas the right hand side would contain contributions from mesonic Regge trajectories. 

Let us give a practical example of FESR for total cross-section. Suppose experimental data are available 
for a certain total cross-section $\sigma_{tot}$ within a given energy range. Optical theorem then allows us 
to convert this into a knowledge about the imaginary part of the forward elastic amplitude in the same energy range. 
Integrals of this amplitude over the available energy range must match a similar integral for a model describing 
the same asymptotic amplitude (ergo the asymtotic total cross-section). Thus, unknown parameters in the model, 
usually Regge residues and intercepts, can be fixed.  

For the phenomenolgy of high energy $pp$ and $p\bar{p}$ total cross-sections of interest at the Tevatron and LHC, 
one forms combinations of the sum and difference of the two cross-sections, thus focusing attention on crossing-even 
$A_+(\nu)$ and crossing-odd $A_-(\nu)$ forward amplitudes. For the odd amplitude $A_-(\nu)$, the procedure 
described above is applicable. For the even amplitude $A_+(\nu)$, one constructs an odd amplitude $\nu A_+(\nu)$, 
to which the above arguments again apply. We shall discuss how it works in practice when we discuss models 
for total cross-sections.  

\subsection{The Froissart-Martin  bound}\label{ss:froissart}
We shall now derive the Froissart bound following three slightly different 
methods, the original one by Froissart \cite{Froissart:1961ux},   the one 
by Martin in \cite{Martin:1962rt} 
and \cite{martinbook3:1970}, and Gribov's derivation in  
\cite{Gribovbook}.
These different derivations expose the different assumptions underlying them.

\subsubsection{Froissart's derivation of the asymptotic behaviour of the 
scattering amplitude}\label{sss:froissart1}
In \cite{Froissart:1961ux}, the bound on the total cross-section is given an 
intuitive explanation. It must be noted (in hindsight) that this intuitive explanation 
relies upon the existence of confinement. Indeed, the whole description 
applies not to  parton scatterings but to hadronic scattering. 
Let us go through Froissart's intuitive explanation. Let the two particles 
(hadrons) see each other at large distances through a 
Yukawa-type  potential, namely $ge^{-\kappa r}/r$, 
where $\kappa$ is some momentum cut-off.
Let $a$ be the impact parameter, then  the total interaction seen by 
a particle for large $a$ is proportional to $ge^{-\kappa a}$. When 
 $ge^{-\kappa a}$ is  very small, there will be 
practically no interaction, while, when $ge^{-\kappa a}$ is close to $1$,
 there will be maximal probability for the interaction. For such values of $a$,
$\kappa a=\ln |g|$ one then can write for the cross-section 
$\sigma\simeq ( \pi/\kappa^2) \ln^2|g|$. If $g$ is a function of energy and 
we assume that it can grow with energy at most like a power of $s$, then one 
immediately obtains that the large energy behaviour of the total cross-section
is bound by 
$\ln^2 s$. What $\kappa$ is remains undefined for the time being, except that 
it has dimensions of a mass.

Following this heuristic argument, Froissart's paper proceeds to the actual derivation of the bound. 
The derivation is based on the validity of the Mandelstam 
representation and the optical theorem. From the validity of the Mandelstam 
representation for the scattering amplitude and the convergence of the partial
 wave expansion, he derives an upper limit on each partial wave, which depends 
on the value ${\bf L}$ of angular momentum, after which the partial 
wave amplitudes become negligible. 
All the $a_l$ are then put equal to their maximum value $a_l=1$ and,  
then, in the forward direction, one has
\begin{equation}
\sum_0^\infty (2l+1) a_l=L^2+ negligible\ terms \le L^2
\end{equation}
The value of $L$ is determined as being such that for $l\le L$
\be
|a_l| \le
\frac {q_sB(s)}{\sqrt{s}(L-N)} \{ \frac{1}{x_0+(x_0-1)^{1/2}} \}^{L-N}=1
\label{eq:froissartL}
\ee
where $N-1$ is the minimum number of subtractions needed for the validity of the Mandelstam 
fixed-$s$ dispersion relations and $B(s)$ behaves at most like a polynomial in $s$, q being the 
c.m.momentum. Eq.~(\ref{eq:froissartL}) leads to 
\be
L\simeq (q_s/\kappa) \ln (B(s)))
\ee
and from this through the optical theorem to the  bound
\be
\sigma_{total}\le \ln^2s 
\ee
\subsubsection{Andr\'e Martin's 
 derivation }
 \label{sss:martinfroissart}
Martin's derivation does not require the existence of the Mandelstam representation and is thus
more general. Also, it provides an estimate of the constant pre factor to the maximum square
of the logarthimic growth.
We shall write $s =4k^2$ ignoring all particle masses except when necessary and use his 
normalization of the elastic amplitude. 
 
\begin{equation}
\label{1}
\sigma_{TOT}(s) = (\frac{16\pi}{s})\sum_{l=0}^\infty(2l+1) Im f_l(s) = (\frac{16\pi}{s}) A_s(s,0),
\end{equation}
wherein 
\begin{equation}
\label{2}
f_l(s) = \frac{\eta_l(s)e^{2\delta_l(s)} - 1}{2i};\ Im f_l(s) = (\frac{1}{2}) [1 - \eta_l(s) cos(2 \delta_l(s))]
\end{equation}
$0<\eta_l(s)<1$ is the inelasticity and $\delta_l(s)$ is the real part of the phase shift and
\begin{equation}
\label{3}
A_s(s,x) = \sum_{l=0}^\infty(2l+1) Im f_l(s) P_l(x);\ x = (1 + \frac{2t}{s}),
\end{equation} 
denotes the $s$-channel absorptive part of the elastic amplitude. This partial wave series should converge
upto $t > 4m^2_\pi$.

For the Froissart bound, Martin uses the majorization scheme
\begin{eqnarray}
\label{4}
Im f_l(s) =1\ for\ l\leq L_T;\\
\ Im f_l(s) = \epsilon\ for\ l=L_T +1;\\
\ Im f_l(s) =0\ for\ l\geq L_T+2.
\end{eqnarray} 
Few remarks: 
\begin{itemize}
\item (i) The first statement, Eq.(\ref{4}), assumes that even for large $s$, the partial wave amplitude 
is 
elastic and a maximum
i.e., $\eta_l(s)\ =1$ and $\delta_l(s)\ =\pi/2$. This is a gross overestimate since 
we expect that at large $s$, $\eta_l(s)\rightarrow 0$, so realistically we should take $1/2$ and not $1$ for low $l$. 
This then would get the heuristic result Martin obtains towards the end of a recent paper
\cite{Martin:2009pt} improving the total cross-section 
bound by a factor $2$.
\item (ii) Let us also note that Eq.(\ref{4}) assumes that the partial wave amplitudes have a sharp cutoff, i.e., 
its value is exactly $1$ for all $l$ up to $L_T$, then brusquely it drops to $\epsilon$ for $l = L_T +1$ and then identically 
to $0$ for all higher $l$. Clearly, this is a very unphysical assumption for a partial wave amplitude and can not be true 
in any theory which enjoys analyticity in the variable $l$. 
\item (iii) The more reasonable behavior for large $l$, through the convergence of the partial expansion in the
Lehmann ellipse leads to $Im f_l(s)\rightarrow e^{[-l/(s/s_o)]}$ times a very smooth function of $l$ and $s$. 
[Eq.(3.4) {\it et sec} in Martin's book \cite{martinbook3:1970}]. This is also the $p\geq (1/2)$ discussed in \cite{Grau:2009qx} in the context of our BN (Bloch and Nordsieck) inspired model discussed later in \ref{sss:BN},
 and is the minimum realistic
dropoff. However, in obtaining the upper bound, Martin assumes it is identically zero beyond a certain $l$ which is 
certainly true but again unrealistic.
\end{itemize}
\par\noindent
Now to a derivation of the upper bound. Clearly from Eq.(\ref{3}) and Eq.(\ref{4}), we have that 
\begin{equation}
\label{5}
A_s(s,x) > \sum_0^{L_T}(2l+1)P_l(x) = P_{L_T +1}^{'}(x) + P_{L_T}^{'}(x)
\end{equation}
To prove the last identity in Eq.(\ref{5}), use the recursion identity $(2l+1)P_l(x) = P_{l+1}^{'}(x)- P_{l-1}^{'}(x)$ 
and then write the sum to be performed in the opposite order (beginning from the end)
\begin{eqnarray}
\sum_0^{L_T}(2l+1)P_l(x) = 
\left[ P_{L_T +1}^{'}(x)- P_{L_T -1}^{'}(x)\right]\nonumber\\
+ \left[P_{L_T}^{'}(x) - P_{L_T -2}^{'}(x)\right]\nonumber\\
+ \left[P_{L_T -1}^{'}(x) - P_{L_T -3}^{'}(x)\right] +.....
\end{eqnarray} 
All terms cancel, leaving only two terms
\begin{equation}
\label{7}
\sum_0^{L_T}(2l+1)P_l(x) = P_{L_T +1}^{'}(x)+ P_{L_T}^{'}(x).\ \ \ Q.E.D. 
\end{equation} 
For large $L_T$, using Eq.(\ref{3}) and Eq.(\ref{7}), we have
\begin{equation}
\label{8}
A_s(s,x) >  2P_{L_T}^{'}(x) 
\end{equation}
Use the Laplace integral for the Legendre function to bound the right hand side:
\begin{equation}
\label{9}
P_l(x) = \frac{1}{\pi}\int_0^\pi (d\chi)[x + \sqrt{x^2-1} cos\chi]^l,  
\end{equation}
so that we can write for the derivative in a useful form
\begin{eqnarray}
\label{10}
P_l'(x) = 
\frac{l x}{\pi (x^2 -1)}  \int_0^\pi (d\chi)\left[x + \sqrt{x^2-1} 
cos\chi\right]^{l-1} \times \nonumber\\
\left[x -\frac{1}{x} +
\sqrt{x^2 -1} cos\chi\right] \nonumber\\ 
\end{eqnarray} 
Since $x>1$, we can bound the above
\begin{equation}
\label{11}
P_l^{'}(x) > \frac{l x}{\pi (x^2 -1)}\int_0^\pi (d\chi)[x + \sqrt{x^2-1} cos\chi]^l 
\end{equation} 
Using the mean value theorem, we can impose the bound
\begin{equation}
\label{12}
P_l^{'}(x) > \frac{l x \phi_o}{\pi (x^2 -1)}[x + \sqrt{x^2-1} cos\phi_o]^l, 
\end{equation} 
for any $0<\phi_o<\pi$. Since $x\rightarrow 1$ and $(x^2-1)\rightarrow (4t/s)$,
\begin{equation}
\label{13}
2 P_{L_T}^{'}(x) > (Constant)L_T (\frac{s}{4t})
[x + \sqrt{x^2-1} cos\phi_o]^{L_T}\ \ 
\end{equation}
and hence using Eq.(\ref{8}), we have
\begin{eqnarray}
\label{14}
\left[(Constant)(\frac{t}{s}) A_s(s,t)\right] > \nonumber\\ 
L_T \left[x + 
\sqrt{x^2-1} cos\phi_o\right]^{L_T} > 
\left[1 + \sqrt{x^2-1} cos\phi_o\right]^{L_T}. \nonumber\\
\end{eqnarray}
Taking logarithms of both sides we have
\begin{eqnarray}
\label{15}
ln\left[(Constant)(\frac{t}{s}) A_s(s,t)\right] > \nonumber\\ 
L_T ln\left[1 + \sqrt{x^2-1} cos\phi_o\right]
\rightarrow L_T \sqrt{4t/s}\ cos\phi_o \nonumber\\
\end{eqnarray}
We need only two subtractions in $A_s(s,t)$ and so $A_s(s,t)< (s/s_o)^2/ln(s/s_o)$. 
Using it in the above, we  arrive finally to the maximum value allowed to $L_T$
\begin{equation}
\label{16}
L_T < \sqrt{(s/4t)}\ [\frac{ln(s/s_o)}{cos\phi_o}],
\end{equation} 
Now the Froissart bound for the total cross-section follows from
\begin{equation}
\label{17}
\sigma_{TOT} < \frac{16 \pi}{s}L_T^2 = \frac{4\pi[ln(s/s_o)]^2}{t cos^2\phi_o}.
\end{equation} 
Letting $t = 4 m_\pi^2$ and $\phi_o = \pi$, we have the Froissart-Martin result
\begin{equation}
\label{18}
\sigma_{TOT} <  [\frac{\pi}{m_\pi^2}][ln(s/s_o)]^2.
\end{equation} 
All of this can be duplicated in the eikonal scheme and of course much more simply as shown below.
\subsubsection{Eikonal Picture derivation}
\label{sss:eikonal}
In the limit of large $s$ and fixed $t>0$, the eikonal picture emerges under the hypothesis
of identifying the impact parameter $b$ formally as $(l + 1/2)\rightarrow b\sqrt{s}/2$, so that
\begin{equation}
\label{E1}
P_l(1 + \frac{2t}{s}) \rightarrow\ I_o(b\sqrt{t}),
\end{equation}
where $I_o$ is the Bessel function of the ``imaginary argument''. With this identification,
Eq.(\ref{3}) reads
\begin{equation}
\label{E2}
A_s(s,x) = (\frac{s}{2}) \int_0^\infty(bdb)\ Im F(b,s) I_o(b\sqrt{t}),
\end{equation} 
where the ``b-wave amplitude'' reads
\begin{equation}
\label{E3}
F(b,s)\ = \frac{\eta(b,s)e^{2\delta(b,s)} - 1}{2i}
\end{equation}
We may now impose a similar majorization scheme as before
\begin{equation}
\label{E4}
Im F(b,s)\ = 1\ for\ b\leq b_T;\ and\ Im F(b,s)\ = 0\ for\ b> b_T,
\end{equation}
whence
\begin{equation}
\label{E5}
A_s(s,t) > (\frac{s}{2})\int_0^{b_T}(b db)I_o(b\sqrt{t}).
\end{equation}
The last integral can be done. Changing variables $Y_T\ =\ b_T \sqrt{t}$, we have
\begin{equation}
\label{E51}
[\frac{2t}{s} A_s(s,t)] > \int_0^{Y_T}(y dy)I_o(y) = Y_T I_1(Y_T).
\end{equation}
For large $Y_T$,
\begin{equation}
\label{E6}
I_1(Y_T)\rightarrow\ \frac{e^{Y_T}}{\sqrt{2\pi Y_T}},
\end{equation}
so that, for large $Y_T$
\begin{equation}
\label{E7}
[\frac{2t}{s} A_s(s,t)] > \frac{\sqrt{Y_T}e^{Y_T}}{\sqrt{2\pi}} > e^{Y_T},
\end{equation}
upon which by taking the logarthims of both sides, and remembering that $A_s(s,t)<\ (s/s_o)^2/ln(s/s_o)$, we obtain from
\begin{equation}
\label{E8}
[\frac{2t}{s} A_s(s,t)] > \frac{\sqrt{Y_T}e^{Y_T}}{\sqrt{2\pi}} > e^{Y_T},
\end{equation}
that
\begin{equation}
\label{E9}
Y_T < ln(s/s_1),
\end{equation}
and thus that
\begin{equation}
\label{E10}
\sigma_{TOT} < (\frac{s}{2})\int_0^{b_T} (bdb) = [\frac{\pi}{m_\pi^2}][ln(s/s_1)]^2,
\end{equation}
upon imposing $t=4m_\pi^2$.
\subsubsection{Gribov's derivation}\label{gribovfroissart}
What follows is almost {\it verbatim } from Sec. (1.4) of \cite{Gribovbook}.
To show that asymptotically
\begin{equation}
\label{19}
Im A(s,t)|_{t=0}\le const \cdot s \log^2 {{s}\over{s_0}} \ \ \ \ s\rightarrow \infty, 
\end{equation}
Gribov proceeds as follows. His notation differs slightly from the one in the previous section. Defining
\begin{equation}
\label{20}
A(s,t)=\sum_{l=0}^\infty (2l+1)f_l(s)P_l(z),
\end{equation}
with the partial wave amplitudes defined as
\begin{equation}
\label{21}
f_l(s)=8 \pi i [1-\eta_le^{2i\delta_l(s)}]
\end{equation}
[With respect to Martin, the difference is a factor of $16 \pi$].

Using the fact that the singularity of $Im_s A(s,t)$ closest to the physical region of the s-channel is situated 
at $t=4\mu^2$, one tries to estimate $f_l(s)$ at large s. At large $l$, the partial wave amplitude 
must fall exponentially in order to ensure convergence for $t>0$.  This is a consequence of
\begin{eqnarray}
\label{22}
P_l(cosh \alpha) \simeq{{e^{l\alpha}}\over{
\sqrt{2\pi l sinh\alpha}
}}; {\rm for}\  l\rightarrow \infty\nonumber\\ 
cosh \alpha=1+{{2t}\over{s}}
\end{eqnarray}
To ensure convergence for $t<4\mu^2$, the partial wave amplitudes must then decrease as
\begin{equation}
\label{23}
f_l \approx e^{-l\alpha_0}, \ \ \ \ \ \cosh \alpha_0=1+{{8\mu^2}\over{s}}
\end{equation}
Now, in the limit $s>>t$, $\cosh \alpha \approx 1+ \alpha^2/2$, hence $\alpha_0={{\sqrt{4\mu^2}}\over{k_s}}$ and one can write
\begin{equation}
\label{24}
f_l(s )\approx c(s,l) e^
{
-{{l}\over{k_s}} \sqrt{4\mu^2}
},
 \ \ \ \ \ l\rightarrow \infty, s\rightarrow \infty
\end{equation}
where 
$k_s=\sqrt{s-4\mu^2}/2$.
The function $c(s,l)$ may be a slowly varying (non-exponential) function of $l$.

To establish the Froissart bound, Gribov now assumes that the scattering amplitude grows no faster than a power of $s$, 
in the vicinity of  the $t=4\mu^2$ pole in the t channel. This 
condition 
is analogous to the one about subtraction in Martin's derivation, just before  Eq.(\ref{16}). If $A(s,t)<(s/s_0)^N$, 
one can then see that this also valid for  $Im \ c(s,l)$. Let us see how.
\begin{equation}
\label{25}
({{s}\over{s_0}})^N>Im (A(s,t) =\sum_{l=0}^\infty (2l+1) Im f_l(s) P_l(1+{{t}\over{2k_s^2}})
\end{equation}  
Since all the $Im f_l(s)$ are positive due to the unitarity condition as well as the $P_l$ for $t>0$, it 
must  also be true for each term on the sum, namely
\begin{equation}
\label{26}
({{s}\over{s_0}})^N>Im \ c(l,s)  (2 \pi l {{\sqrt{t}}\over{k_s}})^{-1/2}e^{{{l}\over{k_s}}(\sqrt{t}-\sqrt{4\mu^2})           }  
\end{equation}
and for $t<4\mu^2$ it will also be
\begin{equation}
\label{27}
Im \ c(l,s)  < ({{s}\over{s_0}})^N
\end{equation}
and finally we have
\begin{equation}
\label{28}
Im\  f_l(s)\leq ({{s}\over{s_0}})^N e^{-{{2\mu}\over{k_s}} l }
\end{equation}
With the bound on $Im\ f_l(s)$, we can now derive the bound on the imaginary part of the forward scattering 
amplitude and hence on the total cross-section.
\begin{eqnarray}
\label{29}
Im A(s,t=0)=\sum_{l=0}^{\infty} (2l+1)Im\ f_l(s)\\
\le 8\pi \sum_{l=0}^{L} (2l+1) + \sum_{l=L}^{\infty} (2l+1)Im\ f_l(s)
\end{eqnarray}
where one has divided the sum into a term where the partial waves are large and for which
the partial waves take the maximum value allowed by unitarity,  and one which contains all the higher partial waves. 
To  estimate the value of L after which the partial waves are small, consider that they will become less than 1 when
\begin{equation}
\label{30}
({{s}\over{s_0}})^N e^{-{{2\mu}\over{k_s}} L }\leq 1 \ \ \  or\ \  L \leq {{k_s}\over{2\mu}} \log {{s}\over{s_0}}
\end{equation} 
Now using
\begin{equation}
\label{31}
\sum_{l=0}^{L} (2l+1)=L^2
\end{equation}
we immediately obtain
\begin{equation}
\label{32}
Im A(s,t=0) \propto L^2 \propto s\log ^2 {{s}\over{s_0}}
\end{equation}
The question arises as to how large are the neglected terms. We can estimate them by using 
$f_{L+n}\sim f_L e^{-{{2\mu}\over{k_s}}n}$ and then sum the second series as 
\begin{equation}
\label{33}
\sum_{n=0}^{\infty} 2(L+n)e^{-{{2\mu}\over{k_s}}n}\sim L{{k_s}\over{\mu}} +{{k_s^2}\over{2\mu^2}}<< L^2
\end{equation}
These terms are at most of order$ L^2/\log (s/s_0)$ and are subdominant.

Now, using the optical theorem, $Im A(s,0)=s \sigma{tot}(s)$ one obtains the bound
\begin{equation}
\label{34}
\sigma_{tot}(s)\le \sigma_0 \log ^2{{s}\over{s_0}}
\end{equation}

Thus this demonstration uses 
\begin{itemize}
\item position of the t-channel singularity closest to the s-channel physical region, at $t=4\mu^2$
\item convergence of the partial wave series for $t>0$ (and at most up to the singularity)
\item large $l$-behaviour of the Legendre polynomials for $z>1$
\item that the amplitude does not grow with $s$ faster than a fixed power
\item unitarity condition to ensure that $Im \ f_l(s)$ is positive.
\end{itemize}

\subsection{The Pomeranchuk theorem}\label{ss:pomeranchuk}
Here again we  follow Gribov.
The Pomeranchuk theorem 
\cite{Pomeranchuk:1958ll} says that, if total hadronic cross-sections go to a constant 
at very high energy, then asymptotically particle-particle or particle-antiparticle total cross-sections should be equal.
It was derived using the property of crossing symmetry of the elastic scattering amplitude.

We know,  since the early '70s, that  total cross-section grow with energy \cite{Amaldi:1973yv}, and therefore the 
Pomeranchuk theorem could be considered obsolete. However, there are two reasons to discuss it, one of them being 
that  our understanding of high energy particle collisions if dominated by   gluon-gluon scattering in QCD framework,   
would give the same result as the Pomeranchuk theorem, as discussed later in this section. On the other hand, since 
the total cross-section is not a constant at high energy, there is space for the existence of the called {\it odderon}, 
whose exchange may be relevant at very high energies. We shall discuss more about this point in later sections.

The Pomeranchuk theorem is derived very simply. First of all, let us move from the discrete representation of the 
scattering amplitude in angular momentum $l$ to the impact parameter space. To do this, one notices that for the 
total cross-section not to decrease at very high energy, one needs contributions from higher and higher partial waves. 
At high energy, and in the forward direction, the main contribution to the total cross-section comes from higher partial waves. 
For large $l$ then we can use the asymptotic expression for the Legendre polynomials,
\begin{equation}
P_l(\cos\theta)\approx J_0((2l_1)\frac{\theta}{2})\ \ \ \ \ \ \ \ \  l>>1\ \ \ \ \ \ \theta \approx 0 
\end{equation}
We can then substitute the sum over the partial waves with an integral and introduce the impact parameter variable $l k_s=b$. 
Then the partial wave expansion becomes
 \begin{equation}
 A(s,t)\approx k^2_s\int d^2b f(b,s)J_0(b\sqrt{-t})
 \end{equation}
 To obtain a constant total cross-section, assume  the  integral to be  dominated by  values of the impact parameter 
 whose distribution is independent of the energy :  in such a case, the $s$ and $t$ dependences can be factorized. 
 One puts $f(b,s)\approx a(s) B(b)$, to obtain
 \begin{equation}
 A(s,t)\approx s a(s) F(t)
 \end{equation}
 Assuming the further possibility that $a(s)$ has no residual $s$-dependence, from the optical theorem there follows 
 the constant high energy behaviour of the total cross-section. To summarize, constant total cross-sections can be seen 
to arise if one can  factorize the  $s$ and $t$ dependence in the scattering amplitude, and if the dominant partial wave 
amplitudes  also become constant at very high energy. Both assumptions are not necessarily satisfied. These two assumptions 
however allow demonstration of the Pomeranchuk theorem, when we use crossing symmetry to relate two processes, 
in which one particle in the initial  state is substituted with an antiparticle from the final state. This amounts 
to relating the analyticity properties in the $s$-channel to those in the $u$-channel. The argument goes as follows.

Let us consider the two processes
\begin{eqnarray}
a+b \rightarrow c+d\\
a+{\bar d}\rightarrow  c+{\bar b} 
\end{eqnarray}
and let  $A(s,t)$ be the amplitude which describes the first of the above processes. The s-channel physical region for process  
$a+b\to c+d$ is shown in Fig. \ref{fig:s-plane}, i.e.
\begin{figure}
\resizebox{0.50\textwidth}{!}{%
  \includegraphics{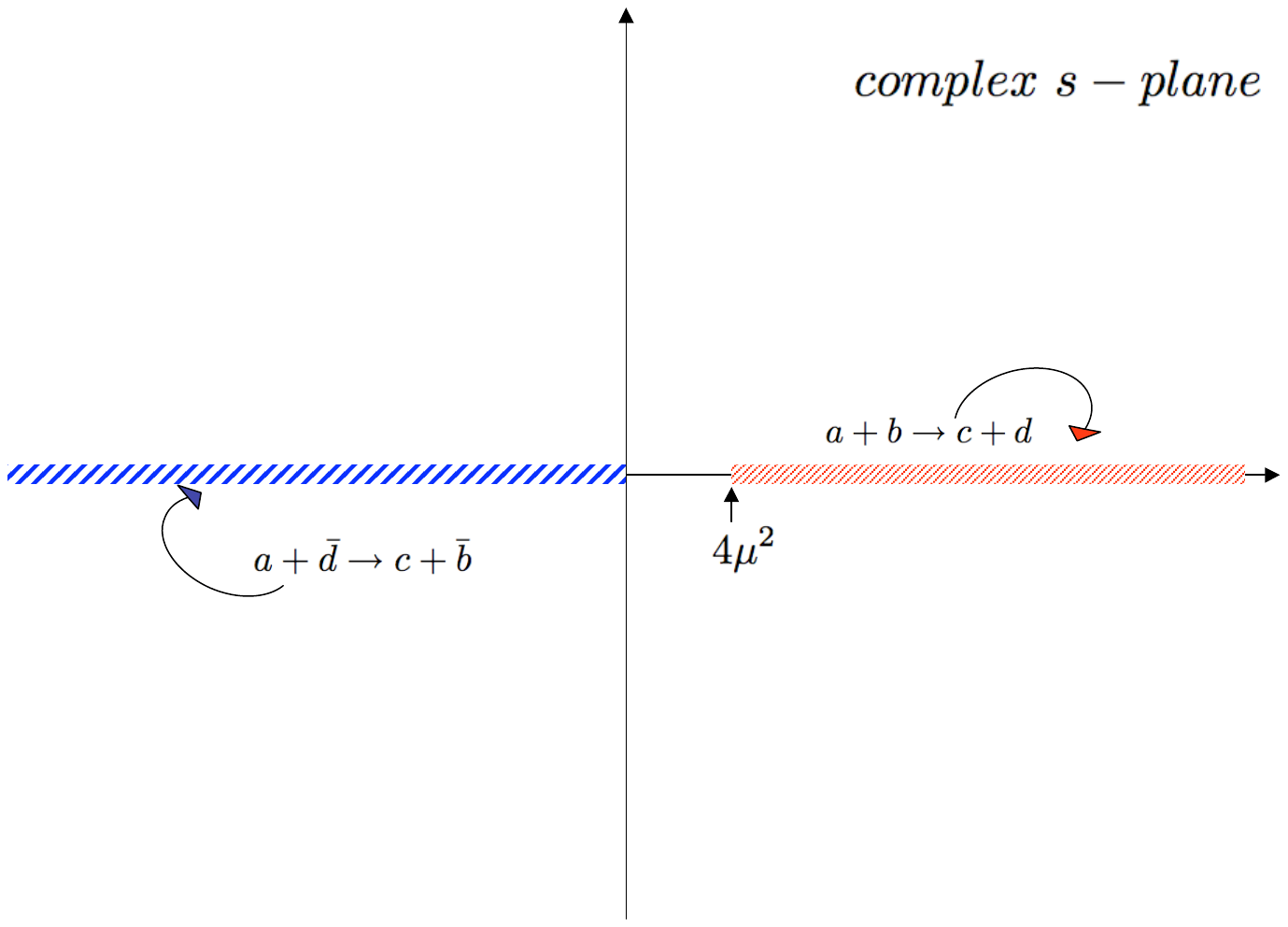}
}
\caption{Physical region for equal mass elastic scattering}
\label{fig:s-plane}       
\end{figure}
with the usual right-hand cut  starting at $s=4\mu^2$ and the left-hand cut starting at $s=0$. It is convenient to assume 
that the amplitude for $a+b\to c+d$ corresponds to the value {\it above} the right-hand cut, namely, to be given by 
$\lim_{\epsilon \to 0} A(s+i\epsilon,t)$, whereas the 
 crossed process $a+{\bar d}\rightarrow  c+{\bar b}$  corresponds to exchange
 $s=(p_a+p_b)^2=(p_c+p_d)^2$ with $u=(p_a-p_d)^2=(p_c-p_b)^2$.  For the crossed process,  the physical region 
 corresponds to $\lim_{\epsilon\to 0}A(u+i\epsilon,t)=\lim_{\epsilon\to 0}A(-s+i\epsilon -t -4\mu^2,t)=\lim_{\epsilon\to 0}A(-(s-i\epsilon)-t -4 \mu^2,t)$, and therefore the physical region for this process, in the $s-$channel is obtained by approaching the real axis, on the left hand side, 
 from below, as indicated in Fig.~\ref{fig:s-plane}. Because the amplitude $A(s,t )$ is real for $0 <s<4\mu^2, t<0$, its value at the edge 
 of the cuts complex conjugates, namely
 \begin{equation}
A(s-i\epsilon,t<0)=A^*(s+i\epsilon,t<0),
\end{equation}
but the amplitude at the left-hand side is for process $a+{\bar d}\to c+{\bar b}$, whereas the one at the right hand side 
is the amplitude for  $a+b\to c+d$  and one then can write
\begin{equation}
A_{a+{\bar d}\to c+{\bar b}}(s)\rightarrow [A_{a+b\to c+d}(s)]^*\ \ \  \ \ \ for \ s\approx -u
\end{equation}
Now apply the above result to the  elastic amplitude for $a+b\to a+b$ and consider the imaginary part of the forward amplitude. 
If the $s$-dependence is of the  simple type which leads to constant total cross-section, namely $A(s,t)\simeq sF(t)$ at large s, 
then optical theorem gives
\begin{equation}
\label{35}
\sigma(a+b) \approx \sigma(a+{\bar b}) \ \ \ \ \ asymptotically
\end{equation} 
For more complicated $s$ and $t$-dependences in the amplitude, i.e. those that do not imply constant total cross-sections, 
Eq.~(\ref{35}) can still be obtained in some simple cases such as $A(s,t)\simeq s \ln{s}^{\beta} F(t)$. 
However, the derivation then  needs the additional hypothesis that the real part of the amplitude does not exceed asymptotically 
the imaginary part, a hypothesis {\it de facto} supported by experimental data. 

 \subsection{Determination of the $\rho$ parameter through Coulomb Interference and soft radiation}
 \label{ss:coulomb}
Here we discuss how 
near the  forward direction, the real part of the hadronic amplitudes is
determined through its interference with the 
Coulomb amplitude. We highlight some of the subtelties associated with the procedure. Also, a proposal is presented for obtaining 
information about the behavior of the purely nuclear amplitude through the measurement of the soft radiation spectrum in a 
{\it quiet event}, i.e., unaccompanied by any other visible particle. 
 
\subsubsection{Coulomb interference}\label{sss:interference}
At high energies, the $\rho$-parameter, which denotes the ratio of the real to the imaginary part of the forward (complex) 
nuclear scattering amplitude $A(s, 0)$
\begin{equation}
\label{c1} 
\rho(s) = \frac{\Re e A(s, 0)}{\Im m A(s, 0)},
\end{equation}
is rather small (about $0.12\div 0.14$). Since, the total (nuclear) cross-section depends only on $\Im m A(s, 0)$ 
and through the optical theorem, the elastic differential cross-section in the forward direction depends on $\rho$ quadratically
\begin{equation}
\label{c2} 
\big{(}\frac{d\sigma_{el}}{dt}\big{)}(t = 0) = \big{(}\frac{\sigma_{tot}^2}{16\pi}\big{)}[1 + \rho^2],
\end{equation}
it is difficult to measure $\rho$ accurately and in any event such a measurement would not determine the sign of the 
real part of the nuclear amplitude.
 
Fortunately, when we augment the nuclear with the Coulomb amplitude (due to one-photon exchange, in the lowest order), 
the interference between the Coulomb and the real part of the nuclear amplitude (for small $t$) allows us to determine both 
the sign and the value of $\rho$. The Rutherford singularity ($\propto\ \alpha/t$) renders the Coulomb amplitude sufficiently 
large to become competitive with the nuclear term, for small $t$. On the other hand, away from very small angles, the 
Coulomb term dies out and one can safely revert to the purely nuclear amplitude. However, to obtain numerically accurate 
information about $\rho$, some care is required to obtain the correct Coulomb phase for the nuclear problem.

To see what is involved, let us consider first Coulomb scattering in non-relativistic potential scattering. The classical 
Rutherford amplitude (or, the Born approximation, quantum mechanically), with a Coulomb $1/r$ potential, for the scattering 
of two charges ($Z_1 e$) and ($Z_2 e$), is given by
\begin{equation}
\label{c3} 
f_C(k, \vartheta) = \frac{2 Z_1 Z_2 \mu \alpha }{t},
\end{equation} 
where $\mu$ denotes the reduced mass, $t\ = - 4 k^2 sin^2 \vartheta/2$ and $\alpha\approx\ 1/137$ is the fine structure constant. 
But, the exact (quantum-mechanical) Coulomb scattering amplitude has an oscillating phase $e^{i\phi_S}$ multiplying the above. 
This phase is given by\cite{Schwinger2001}
\begin{equation}
\label{c4} 
\phi_S = (\frac{Z_1 Z_2 \alpha c}{v})\ ln(sin^2\vartheta/2),
\end{equation} 
where $v$ denotes the relative velocity and the presence of $\alpha$ reminds us of the quantum nature of this phase 
The physical reason for this phase is that the Coulomb potential is infinite range and however far, a charged particle is 
never quite free and hence is never quite a plane wave. For $pp$ or $p\bar{p}$ scattering, in the relativistic limit 
($v\rightarrow\ c$) and for small angles, Eq.(\ref{C4}) reduces to 
\begin{equation}
\label{c5}
\phi_S \approx\ (\mp 2\alpha)\ ln(\frac{2}{\vartheta}).
\end{equation}
Eq.(\ref{C5}) is exactly the small-angle limit of the relativistic Coulomb phase obtained by Solov'ev\cite{Solov'ev66}.
On the other hand, this result was in conflict with an earlier potential theory calculation by Bethe\cite{Bethe58} 
employing a finite range ($R$) nuclear potential in conjunction with the Coulomb potential. According to Bethe, 
the effective Coulomb phase reads
\begin{equation}
\label{c6}
\phi_B \approx\ (\pm  2\alpha)\ ln(k R \vartheta).
\end{equation}   
This discrepancy was clarified by West and Yennie\cite{WestYennie68}. These authors computed the effective Coulomb 
phase through the absorptive part of the interference between the nuclear and the Coulomb amplitude. They found -again 
in the small angle, high energy limit-
\begin{equation}
\label{c7}
\phi_{WY} = (\mp \alpha)[ 2 ln(\frac{2}{\vartheta}) + \int_{-s}^0 \frac{dt^{'}}{|t^{'} - t|}\{1 - \frac{A(s,t')}{A(s,t)}
\}].
\end{equation}  
If one ignores the $t$ dependence of the nuclear amplitude, the integral term above is zero and one obtains 
Solov'ev's result. On the other hand, a result similar to that of Bethe is reproduced, if one assumes the customary 
fall-off $e^{B t/2}$ for the nuclear vertex and a dipole form factor for the EM vertex. Explicitly, if we choose
\begin{equation}
\label{c8}
\frac{A(s,t{'})}{A(s,t)} = e^{B(t^{'} - t)/2} \big{(}\frac{1 - t/\Lambda^2}{1 - t^{'} /\Lambda^2}\big{)}^2,
\end{equation} 
we find
\begin{equation}
\label{c9}
\phi_{WY} \approx\ (\pm \alpha)[\gamma + ln (B|t|/2) + ln (1 + \frac{8}{B \Lambda^2})],
\end{equation}
where $\gamma\approx\ 0.5772$ is the Euler-Mascheroni constant.
This expression for the effective Coulomb phase agrees with Block\cite{Block06}, upto terms proportional to 
($|t|/\Lambda^2$), which are quite small near the forward direction. Hence, Eq.(\ref{c9}) is sufficiently accurate for 
determining  $\rho$ through interference at LHC energies and beyond. 

Block has defined a practically useful parameter $t_o$ for which the interference term is a maximum:
$t_o\ = [8\pi \alpha/\sigma_{tot}]$. For the maximum LHC energy of $14\ TeV$, $t_o\approx\ 7\times\ 10^{-4}\ GeV^2$.
Putting it all together, the Coulomb corrected, differential cross-section for \pp \  or \pbarp\ 
 reads\cite{Block06}
\begin{eqnarray}
\label{c10}
[\frac{d\sigma}{d|t|}]_{o} &=& (\frac{\sigma_{tot}^2}{16\pi}) \big{[} G^4(t)(\frac{t_o}{t})^2\nonumber\\
& \mp& 2\frac{t_o}{|t|} (\rho 
\pm \phi_{WY}) G^2(t) e^{-B|t|/2}\nonumber\\
 & +& (1 + \rho^2) e^{-B|t|}
\big{]},
\end{eqnarray}
where for the magnetic form factor, one may employ $G(t)\ =\ [\frac{1}{1 - t/\Lambda^2}]^2$, with $\Lambda^2\approx\ 0.71 GeV^2$.

One other aspect of the EM radiative corrections needs to be investigated. So far, we have not considered real 
soft-photon emissions in the scattering process. As is well known, contributions due to an infinite number of soft 
photons can be summed via the Bloch-Nordsieck theorem. If $(d\sigma/dt)_o$ denotes the differential cross-section 
without the emitted soft-photons, the inclusion of soft radiation introduces a parameter which is the external energy 
resolution $\Delta E$. A compact expression for the corrected cross-section can be written as follows\cite{Pancheri73} \begin{equation}
\label{c11}
\frac{d\sigma}{dt} = (\frac{\Delta E}{E})^{\hat{\beta}} (\frac{d\sigma}{dt})_o,
\end{equation} 
where the radiative factor $\hat{\beta}$ combines the various combinations of the momenta of the charged emitting 
particles in our equal mass elastic case
\begin{equation}
\label{c11A}
\hat{\beta} = (\frac{2\alpha}{\pi}) [I_{12} + I_{13} + I_{14} - 2],
\end{equation} 
where
\begin{equation}
\label{c11B}
I_{ij} = 2 (p_i\cdot p_j)\int_0^1 \frac{dy}{[m^2 + 2(p_i\cdot p_j - m^2) y(1-y)]}.
\end{equation} 
In the high energy limit, $I_{12}(s)\rightarrow\ 2 ln(s/m^2)$ and $I_{14}(s)\rightarrow\ -2 ln(u/m^2)$, so that the sum
$I_{12} + I_{14}\rightarrow\ 0$ vanishes, leaving us with $I_{13}(t)$. For small angle scattering of interest here, 
we have the correction from real photon emission of the form
\begin{equation}
\label{c11C}
\frac{d\sigma}{dt} = (\frac{\Delta E}{E})^{\beta(t)} (\frac{d\sigma}{dt})_o,
\end{equation} 
where
\begin{equation}
\label{c11D}
\beta(t) \approx\  (\frac{4\alpha}{\pi})(\frac{-t}{3m^2});\ (-t<<m^2),
\end{equation} 
which is rather small and vanishes as $t = 0$. The physical reason is that the amount of radiation is small for low velocities. 
For the CM elastic amplitude, the energy loss $\Delta E$ due to real soft bremmstrahlung is estimated by the lack of collinearity 
in the outgoing particles. Thus, $(\Delta E/E)\approx \Delta \vartheta$). Even for $\Delta \vartheta \sim\ 10^{-4}$, the real radiative 
correction is extremely small and thus can be ignored.

\subsubsection{ Soft photon radiation as a possible tool 
for measurements of the total cross-section}
\label{sss:softphotons} 
Through the above expressions, we may compute the differential probability for a single soft photon produced 
in association with the near forward elastic process. That is, the differential cross-section for the process 
$p(p_1)\ +\ p(p_2)\ \rightarrow\ p(p_1^{'})\ +\ p(p_2^{'})\ + \gamma(k)$, for small $|t|<< m^2$ and small $k$ is given by
\begin{equation}
\label{C12}
\frac{d\sigma}{dk dt} \approx\  (\frac{4 \alpha}{\pi k}) (\frac{-t}{3 m^2}) (\frac{d\sigma}{dt})_o,
\end{equation} 
where $m$ denotes the nucleon mass.

To obtain a leading order of magnitude estimate, let us insert only the nuclear amplitude in Eq.(\ref{C12}) and integrate over all $t$:
\begin{equation}
\label{C13}
\frac{d\sigma}{dk} \approx\  (\frac{4 \alpha}{3 m^2}) (\frac{1}{k}) [\frac{\sigma_{tot}}{4 \pi B}]^2.
\end{equation} 
Putting in nominal values for the LHC, $\sigma_{tot}\approx\ 100 mb$ and for the diffractive width $B\approx\ 20\ GeV^{-2}$, we estimate
\begin{equation}
\label{C14}
\frac{d\sigma}{dk} \approx\  (\frac{1}{k}) (4.35 \times 10^{-3} mb).
\end{equation} 
This would lead us to a comfortable soft photon production rate [associated with elastic scattering]
\begin{equation}
\label{C15}
{\dot{\cal N}} \approx\  (435\ Hertz)\ ln(\frac{k_{max}}{k_{min}}) (\frac{{\cal L}}{10^{32}/cm^2/sec.}),
\end{equation} 
where ${\cal L}$ is the machine luminosity. Thus, observation  of {\it only} soft photons accompanied by no other 
visible particles (an otherwise quiet event) would be very useful in determining some crucial nuclear high amplitude parameters. 
Transcending the specific model estimates, what the soft radiation spectrum really measures is the mean value of $<-t>$ 
associated with the elastic cross section $\sigma_{el}$.

A simple variant of the procedure outlined above of obtaining information about the background process through the 
spectrum of a single photon in an  otherwise quiet event was utilized earlier at LEP. One way adopted to measure the 
number of neutrinos  into which the $Z^o$ could decay consisted in measuring the rate for the process
\begin{equation}
\label{C16}
e^- + e^+ \rightarrow\ \gamma(k) +\ nothing \ visible.
\end{equation}   
Thus, measuring the initial state photon radiation, allowed one to deduce the correct branching ratio of 
$Z^o$ into all neutrino-antineutrino pairs, which obviously escaped experimental detection. On the other hand, 
at LHC, if indeed a single soft photon spectrum in a quiet event  - up to some very small angle - can be measured, 
one can be reasonably be sure that two protons (to conserve baryon number) must have been produced which escape 
within the very small angular cone on either side of the beam. Relaxing the angular aperture might allow one to learn 
something about single diffraction cross section as well. We shall not pursue this interesting topic here any further. 
\section{ Non-accelerator experiments}
\label{sec:cosmic}


In this section, we discuss  the measurement of  proton-proton scattering as performed   through   cosmic ray experiments, the results 
and their interpretation. 

 Until the advent of particle accelerators in the mid-1950's, information about  high-energy elementary particle scattering, and hence 
 its dynamics,  was obtained through the observation of cosmic ray showers, which resulted  from the interaction  of  primary particles 
 (those arriving from the interstellar space and beyond) 
 with the  earth's atmosphere. 
The energy distribution of the primary particles was measured through the depth and extent of particle showers observed after the 
interaction, following a technique we shall describe in more detail later in this section.
 
 Through these observations, it was  possible to extract  proton-proton total cross-sections. 
To this day, the extraction of proton-proton total cross-sections from cosmic ray 
measurements, reaches  center of mass  energies usually higher than contemporary accelerator data. As we shall see however, 
this extraction procedure is still affected to a  large extent upon modeling.

The energy spectrum of cosmic ray particles is shown in
Fig.~ \ref{fig:cosmicPDG} from the 2014 Review of Particle Physics (PDG)  \cite{Agashe:2014kda}, where  an up-to-date 
review of the subject can be found.

 \begin{figure}
\resizebox{0.5\textwidth}{!}{
\includegraphics{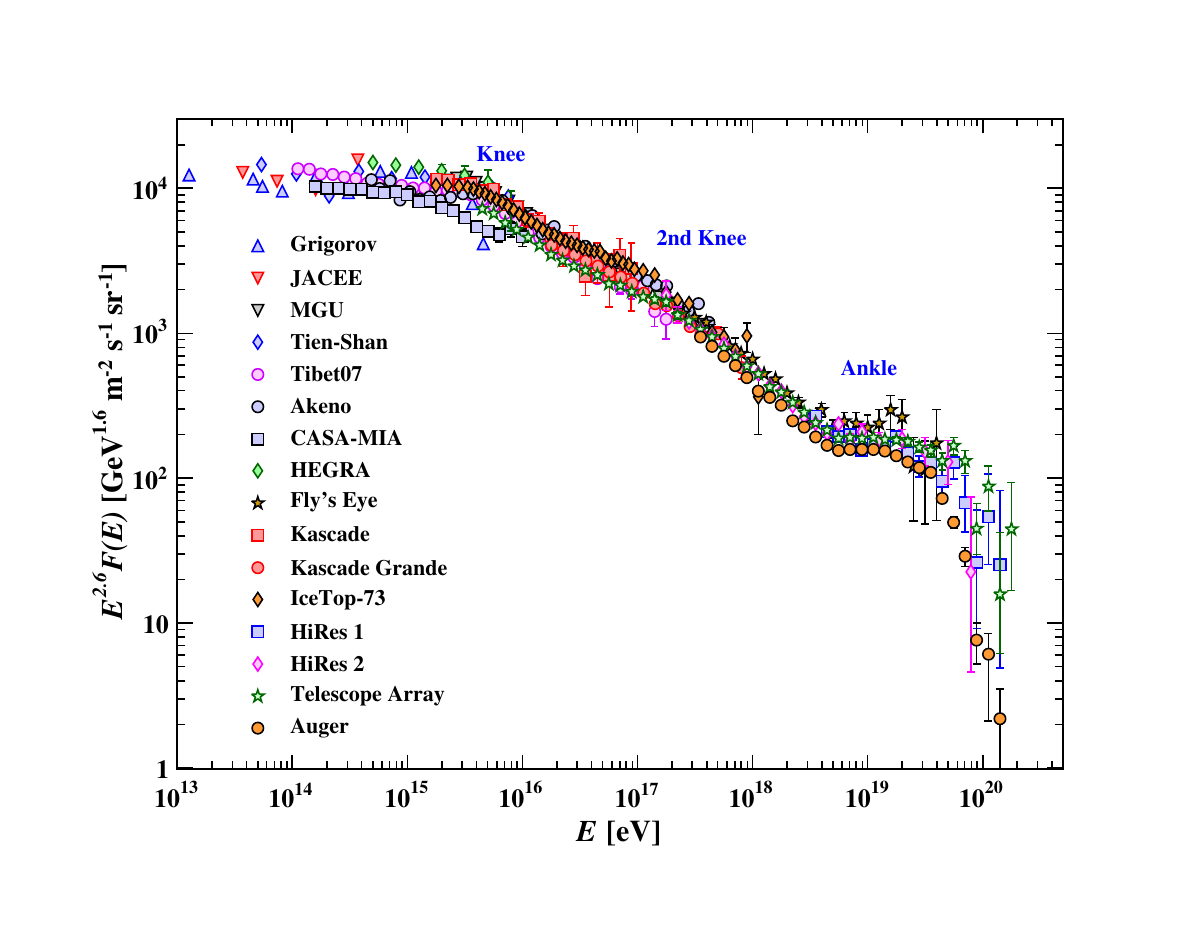}}
\caption{The energy distribution of the cosmic ray flux, multiplied by $E^{2.6}$, from
 \cite{Agashe:2014kda} and 2015 update.}
\label{fig:cosmicPDG}
 \end{figure}

We shall follow 
the historical development of
 the  methods proposed to  
extract  $pp$ cross-sections  from p-air scattering data and present results thus obtained. This section is 
structured as follows:
 \begin{itemize}
 \item Heisenberg's observations  about the effective \x \ for scattering in the Yukawa theory, interesting because 
they involve still currently debated questions 
such as the simultaneous 
occurrence of multiple scattering processes and the range of applicability of perturbation theory, are presented in 
\ref{ss:heisenberg},
\item Glauber's model  for  high energy particle scattering, which is the basis of present methods for  the extraction 
of proton-proton scattering from $p-air$   cross-sections,  is outlined in \ref{ss:glauber},
\item in  \ref{ss:cosmic-pp} developments in the extraction of $pp$ scattering from cosmic ray data  from early 1970's, 
after the appearance of the ISR data, up  to   phenomenological analyses following  1980's accelerator results in the 
TeV energy range, are 
delineated further as follows:
\begin{itemize}
\item early models are  in \ref{sss:earlymodels}
\item   more precise extractions after data from the CERN Super proton-antiproton Synchrotron (\SpbarpS) \  in  \ref{sss:Gaisser1986},
\item    appearance of mini-jet descriptions 
are presented  in \ref{sss:DurandPi},    
\item uncertainties  
in the extraction of the
 $p-air $  data are presented in   \ref{sss:Cosmicuncertainties}, \ref{sss:attenuation},  \ref{sss:air-shower}, 
 with updated analysis of $pp$ data extraction by Bloch, Halzen and Stanev in  \ref{sss:BHS},
 \end{itemize}
\item further clarifications about extraction of $p-air$ cross-section from cosmic rays
in \ref{ss:anchor},
\item  a discussion of critical indices for cosmic ray radiation is  presented in \ref{ss:cosmic-yogi},
\item cosmic ray results  after the new $pp$ total cross-section  measurements  at LHC, from the AUGER  and 
Telescope Array collaborations can be found in \ref{ss:afterLHC}, with associated uncertainties due to diffraction  
discussed in \ref{sss:EU}, 
\item eikonal models as tools for extraction of \pp \ data are discussed in \ref{ss:eikonals}, 
with results from a two-channel model in \ref{sss:GLMcosmic} and a   recent  one-channel analysis  
briefly presented in \ref{sss:ourcosmics}.
\end{itemize}

\subsection{Heisenberg and cosmic radiation }\label{ss:heisenberg}
In a collection of papers  prepared by Heisenberg in 1943 to commemorate A. Sommerfeld's 75th birthday, and later 
translated in 1946 \cite{WH} 
there appear  two important issues. The first concerns the observed power law flux of cosmic ray particles as they appear 
on Earth along with hypotheses regarding the flux of primary cosmic radiation. We shall return to this problem in \ref{sss:critical}. 
The second is a description of the {\it Theory of explosion-like showers}, interesting for the strict analogy 
established by Heisenberg between mesons and light quanta. 

This collection of papers has an interesting history of its own. 
As stated in the foreword  by T. H. Johnson,  the American editor and translator of the book from German to English, the volume was 
published in Berlin in 1943 in commemoration of the 75th birthday of Arnold Sommerfeld. The articles present a general 
view of the state-of-the-art of cosmic ray research. However, and here lies the historical interest of this note, on the very 
day which the book was intended to commemorate, and before many copies had been distributed, bombs fell on Berlin 
destroying the plates and the entire stock of printed volumes. To make the material available to American physicists, 
S. Goudsmith loaned his copy of the German book and T.H. Johnson translated  it. Also interesting are some of the 
comments by Heisenberg in the foreword to the German edition. After mentioning  that investigations on cosmic radiation 
had been sharply curtailed by the misfortunes of the times, Heisenberg  recalls that the papers come from symposia held 
in Berlin during 1941 and 1942, and that the American literature on the subject had been available only up to the summer 
of 1941, so that the present collection could be considered to give a unified representative picture of  the knowledge of 
cosmic radiation at about the end of the year 1941. The book is dedicated to Arnold Sommerfeld,  {\it the teacher of atomic 
physics in Germany}, as Heisenberg says. 

In a brief note, page 124,  Heisenberg is interested in estimating the effective \x \ for scattering in the Yukawa theory, 
to be applied to cosmic ray showers. He objects to what was,  at the time, the  current  interpretation of the Yukawa theory as a 
perturbative one and discusses the presence of multiple simultaneous processes when the energy of the colliding particles 
is above a certain value. Thus, above this value, the assumption that  perturbation theory converges reasonably, i.e. that the 
probability for the simultaneous emission of many particles  be small,  is not valid. According to Heisenberg, there
are two reasons for the occurrence of multiple processes, namely the close relationship of the Yukawa particle (the $\pi$ meson) 
with light quanta, and the peculiarity of the meson-nucleon interaction. Unlike QED, whose convergence depends only on 
the dimensionless constant $\alpha$,  Yukawa's theory depends on a constant with the dimensions of a length (of order 
$10^{-13} \ cm$) and thus perturbation theory will diverge as soon as the wavelength of the particle concerned is smaller 
than this length. Thus in high energy  scattering processes, with very short wavelength of the colliding particles, there occur 
the  possibility of  multiple particle processes.

As for the close analogy between light quanta and meson emission, the similarity lies in the   fact that in the collision of two 
high energy hadrons, several mesons can be created in a way similar to when  an infinite number of light quanta is emitted 
in charged particle collisions. One can describe soft photon emission in a semi-classical way as taking place because in the 
sudden deflection of an electron, the  electromagnetic field surrounding the charged particle becomes detached from the 
particle and moves away like a relatively small wave packet. This process can be described by a delta function in space-time, 
whose Fourier transform is constant. Interpreting this spectrum as the expected spectrum of the radiation, one can calculate 
the mean number  $dn(E)$ of light quanta emitted in a given energy interval $dE$ and thus
\begin{equation}
dn(E)\approx \frac{dE}{E}
\end{equation}   
which is the usual infinite number of emitted soft ($E\rightarrow 0$) photons. 
In complete analogy,  the sudden change in direction of a nucleon will result in multiple meson emission, as the  difference of 
the associated Yukawa fields becomes detached and, as Heisenberg puts it, ``wanders off into space". However there is of 
course a difference, namely that the 
pions are massive and therefore the total number  of emitted 
pions will be finite and 
increase with the logarithm of the collision energy. This effect thus gives in principle the possibility of an increasing multiplicity 
accompanying the high energy collision, but, according to Heisenberg, it is not enough to explain the experiments. Thus a 
second element is introduced, There are non-linear terms in the Yukawa theory which distort the spectrum and give rise to a 
sufficiently large emission to explain experiments. This part however is described only in qualitative terms, at least not in this 
reference. As we shall see later
more developed 
models lead to a \x\  that saturates the Froissart bound.

\subsection{The Glauber model for high energy collisions}\label{ss:glauber}
We shall now discuss Glauber's theory of high energy scattering \cite{Glauber:1959fg}. It derives in part from Moliere's 
theory of multiple scattering \cite{Moliere:1947zz}, whose simpler derivation was obtained by Bethe  \cite{Bethe:1953va}  
in 1955 and which we shall outline in Sect.~\ref{sec:models}. 
Prior to that, Rossi and Greisen had discussed cosmic ray theory \cite{Rossi:1941zz} and many of 
the concepts they used were later elaborated in the theory of high energy scattering.  

Glauber starts by recalling the complexity of high energy collisions, that comes from the large number of final states which open 
up as the energy increases, but  comments on the fact that at high energy it is possible to use a number of approximations to 
deal with this complexity. The inspiration for the treatment of such collisions comes from the diffraction properties of optics, and 
this gives the model its name, i.e.  {\it optical model}. The major difference of course lies in the fact that in optics the obstacles, 
namely the target of the colliding system, is fixed and macroscopic, whereas in nuclei, and of course also in nucleons, the scattering 
is constituted of moving microscopic particles. Thus a quantum mechanical treatment needs to be developed. The model  originally 
deals with elastic scattering alone, treating inelastic scattering as if the particles not scattered elastically had been {\it absorbed} by 
the nucleus. This is the origin of the term  {\it absorption}  still used for inelastic scattering. Glauber explicitly mentions that this work 
can be considered as  an extension and generalization of the Moli\'ere method of multiple scattering \cite{Moliere:1947zz}.
 Notice however a basic difference between Glauber's treatment and Moli\'ere, namely   that Glauber deals with amplitudes, while 
 Moli\'ere with probabilities.  We shall comment on this in  Sect.~\ref{sec:models}.

The scattering amplitude $f(\theta)$ is related to the differential cross-section in the solid angle $d\Omega$  as
\begin{equation}
d\sigma=\frac{Flux\ through\  d\Omega}{Incident\  flux} d\Omega=|f(\theta)|^2d\Omega
\end{equation}
and is related to the potential $V({\vec r})$ through the integral equation
\begin{equation}
f({\vec k},{\vec k'})\equiv f(\theta)=-\frac{m}{2\pi \hbar^2}\int (d^3{\vec r}) V({\vec r}) e^{-i{\vec k'}\cdot {\vec r}} \psi_{\vec k}({\vec r})
\end{equation}
To obtain this expression, a boundary condition has been applied, namely that the potential $V({\vec r})$ is different from zero 
only in a limited region so  that as $r \rightarrow \infty$ the wave function takes the asymptotic form
\begin{equation}
\psi_{\vec k}({\vec r}) \sim e^{i{\vec k}\cdot {\vec r}} + f(\theta) \frac{e^{i kr}}{r}
\end{equation}

Glauber then proceeds to establish some general properties and starts by looking for the consequences of particle conservation. 
For a real potential, he obtains
\begin{equation}
\frac{1}{2i}
\{ 
f(
{\vec k},{\vec k'}
)-f^*(
{\vec k'},{\vec k}
) \}=
\frac{k}{4\pi}
\int f({\vec k_r},{\vec k'})f^*({\vec k}_r,{\vec k}) d\Omega_r
\label{eq:conservationVreal}
\end{equation}
which assumes a particularly simple form for the case ${\vec k}'={\vec k}$, 
\begin{equation}
\Im m f(
{\vec k},{\vec k}
)=\frac{k}{4\pi} \int |f({\vec k}_r,{\vec k}) |^2d\Omega_r=\frac{k}{4 \pi}\sigma_{scatt}
\end{equation}
where $\sigma_{scatt}$ is the total scattering  cross-section. The above relation is also formulated  as the {\it optical theorem}. 
For the case ${\vec k}\ne {\vec k'}$, Eq.~(\ref{eq:conservationVreal}) corresponds to the condition that the operator, which 
yields the final state, is unitary, namely to the so called unitarity condition.

If the final states, as we know to be the case in high energy scattering, will include also inelastic processes, then the potential to 
be considered in such case is a complex potential 
For a complex potential Eq.~(\ref{eq:conservationVreal}) becomes
\begin{eqnarray}
\frac{1}{2i}
\{ 
f(
{\vec k'},{\vec k}
)-f^*(
{\vec k},{\vec k'}
) \} && \nonumber\\
= \frac{k}{4\pi}
\int f({\vec k_r},{\vec k'})f^*({\vec k}_r,{\vec k}) d\Omega_r &&\nonumber \\
- \frac{m}{2\pi \hbar^2}\int (\Im m V({\bf r})) \psi^*_{\vec k'}
\psi_{\vec k} (d^3{\vec r})      &&
\label{eq:conservationVcomplex}
\end{eqnarray}
Again for the case ${\vec k}= {\vec k'}$ one can write the  {\it generalized optical theorem}, namely
\begin{equation}
\Im m f({\vec k},{\vec k})=\frac{k}{4 \pi}(\sigma_{scatt}+\sigma_{abs})=\frac{k}{4 \pi}\sigma_{tot}
\label{eq:opticalglauber}
\end{equation}
where the absorption cross-section $\sigma_{abs}$ has been introduced to  account for particles which have ``disappeared". 
In the optical language, these particles have been absorbed by the scattering material, while  in high energy language 
this cross-section corresponds to the inelastic cross-section, namely to the creation of a final state different from the initial one. 

The three cross-sections defined above, $\sigma_{scatt},\sigma_{abs},\sigma_{tot}$,  can also be expressed using the partial 
wave expansion for the scattering amplitude. Writing
\begin{equation}
f(\theta)=\frac{1}{2ik}\sum_i(2l+1)(C_l-1)P_l(z)
\end{equation}
one obtains
\begin{eqnarray}
\sigma_{scatt}= \int |f(\theta)|^2 d\Omega=\frac{\pi}{k^2}\sum_i (2l+1)|C_l-1|^2\nonumber \\
\sigma_{tot}= \frac{2\pi}{k^2}\sum_i (2l+1)[1-\Re e C_l]\nonumber \\
\sigma_{abs}= \sigma_{tot} -\sigma_{scatt}=\frac{\pi}{k^2}\sum_i(2l+1)[1-|C_l|^2]\ 
\end{eqnarray} 

The expression for the scattering amplitude for an axially symmetric potential is obtained by Glauber, under certain approximations, as 
\begin{equation}
f(\veck ',\veck)=\frac{k}{2\pi i}\int e^{i (\veck -\veck ')\cdot \vecb}\{e^{i\chi(\vecb)}-1  \}d^2\vecb
\end{equation}
where
\begin{equation}
\chi(\vecb)=-\frac{1}{\hbar v}\int_{-\infty}^{+\infty}V(\vecb+{\hat k}z)dz
\end{equation}
He notes that an important test of the self consistency of this approximation is furnished by the unitarity theorem, 
and he proceeds to show that, in the absence of absorption, i.e. for $\chi$ 
purely real, one has
\begin{equation}
\sigma_{scatt}=\int d\Omega_{k'}|f(\veck,\veck')|^2=\sigma_{tot}=\frac{4\pi}{k}\Im m f(\veck,\veck)
\end{equation}
since for $\chi$ purely real
\begin{eqnarray}
\sigma_{scatt}&=&\int |e^{i\chi(\vecb)}-1 |^2d^2\vecb \nonumber \\
&=&2\int (1-\Re e\ e^{i \chi(\vecb)})d^2\vecb\nonumber \\
=\sigma_{tot}
\end{eqnarray}
If there is absorption, namely $\chi$ is complex,  then the conservation of probability implies
for the  inelastic cross-section to be obtained from the difference $\sigma_{tot}-\sigma_{scatt}$, and one has
\begin{equation}
\sigma_{abs}=\int (1-|e^{i\chi(\vecb)}|^2)d^2\vecb
\end{equation}

In what follows in Glauber's paper, various examples are discussed and solved, whenever possible. These are:
\begin{itemize}
\item an absorptive potential (negative imaginary) confined to a sphere of radius $a$ and in such case 
the sphere can be considered to be {\it opaque} in the optical sense
\item a square potential well
\item a Gaussian potential
\item the Coulomb potential
\item a screened Coulomb potential 
\end{itemize}
 In nuclear applications, the incident particle is subject both to  nuclear forces and to the Coulomb field, 
 and   superposition of the phase-shift functions for $\chi(b)$ for the nuclear and Coulomb interactions 
 is suggested.  Thus the nuclear phase shift function needs to be added to the Coulomb one, given by
\begin{equation}
\chi_c(b)=2\frac{Ze^2}{\hslash v}\ln \frac{b}{2a}+\frac{Ze^2}{\hslash v}\frac{b^2}{2a^2}+{\cal O}(\frac{b^4}{a^4})
\end{equation}
which
 represents an expansion in the ratio between the impact parameter distance $b$ and the range $a$ for which the potential is non zero.
According to Glauber, this procedure will take proper account of the two types of effects and of the interference between the two types of scattering.

\subsubsection{Scattering with bound particles}
In the first part of his lectures on high energy collision theory, Glauber discusses scattering of one-on-one  particle.
To study actual scattering experiments of particles on nuclei, one needs to take into account 
that particles are usually in a bound state and thus transitions from one state to another, 
bound or free, can take place. The generalization is done first treating the one-dimensional 
problem and then going to the 3-dimensional one.  
The basic expression for the scattering amplitude in such cases takes the form
\begin{align}
F_{if}(\veck',\veck)&=&\nonumber\\
\frac{k}{2\pi i} \int e^{i(\veck-\veck')\cdot\vecb}\int u^*_f(\vecq)[ e^{i\chi(\vecb-\vecs)}-1] u_i(\vecq)d\vecq \  d\vecb
\end{align}
where $\vecs$ corresponds to the impact parameter coordinate relative to the individual state 
of the target as shown in Fig.~\ref{fig:glaubergeometry}. 
\begin{figure}
\begin{center}
\resizebox{0.6 \textwidth}{!}{
\includegraphics{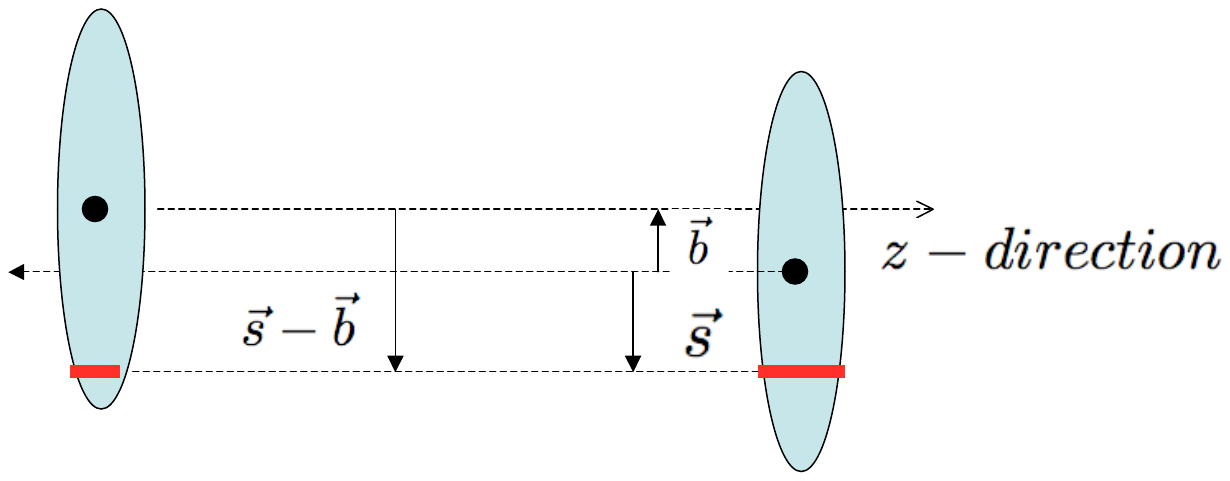}}
\vspace{-2cm}
\caption{
View of the Glauber model geometry in the transverse plane relative to the $z$ axis of scattering direction.}
\label{fig:glaubergeometry}
\end{center}
\end{figure}

The initial states $u_i$ have been defined from the wave function expression
\begin{equation}
\Psi(x,t)=e^{i(kx-\omega t)}\phi(x,t)u_i
\end{equation}
The phase shift function is now generalized from the previous expression so as to include 
the impact parameter of the  target particles, and is thus given as
\begin{equation}
\chi(\vecb-\vecs)=-\frac{1}{\hslash\ v}\int_{-\infty}^{+\infty}V(\vecb-\vecs +{\hat k}z) dz
\end{equation}

\subsubsection{The Glauber model for  high energy scattering of protons by nuclei}
\label{Glauber_nuclei}
In \cite{Glauber:1970jm}, the previously developed theory for multiple scattering is applied 
to describe the results of an experiment by Bellettini {\it et al.}  \cite{Bellettini:1966zz}
for the scattering of $20\ GeV/c$ protons on different nuclei. The starting formula is 
the one for  proton-nucleon collision in the diffraction approximation (small angle), with spin effect neglected, i.e.
\begin{equation}
f(\veck-\veck')=\frac{ik}{2\pi}\int (d^2\vecb)  e^{i(\veck-\veck')\cdot\vecb} \Gamma(\vecb)
\end{equation}
with
\begin{equation}
\Gamma(b)=\frac{1}{2\pi i k}\int d^2\vecq e^{-i\vecb\cdot\vecq}f(\vecq)
\end{equation}
When $\Gamma(\vecb)$ only depends on the scattering angle, one can perform the integration over the azimuthal angle, i.e.
\begin{equation}
f(\veck-\veck')=ik\int bdb J_0(|\veck-\veck'|b)\Gamma(b)
\end{equation}
The proton-proton scattering amplitude  at high energies and small angle is taken to be
\begin{equation}
f(\vecq)=f(0)e^{-\frac{1}{2}\beta^2q^2}
\end{equation}
where $f(0)=(i+\rho)k\sigtot/4\pi$. The values used by Glauber and Matthiae in \cite{Glauber:1970jm} are
$\rho=-0.22$, $\beta^2=10(GeV/c^2)^{-2}$ and $\sigtot=39.0\ mb$. What is needed to compare 
with data (also later in the case of cosmic ray data) are the elastic and the inelastic $proton-nucleon$ 
cross-section. After a number of simplifying approximations, the nuclear elastic scattering amplitude is 
defined by means of  a suitable nuclear phase-shift function $\chi_N(b)$ as
\begin{equation}
F_N(\Delta)=ik\int bdb J_0(\Delta b)[1-e^{i\chi_N(b)}]
\label{eq:fnuclear}
\end{equation}
For large atomic number A, the function $\chi_N(b)$ can take a simple form \cite{Glauber:1959fg}
\begin{equation}
i\chi_N(b)=-\frac{1}{2\pi k}\int e^{-i\vecq\cdot\vecb}f(q)S(q)d^2\vecq
\end{equation}
where $f(q)$ is the proton-proton scattering amplitude and $S(q)$ is the nuclear form factor, 
i.e. the Fourier transform of the nuclear density. The overall density of the nucleon in this case 
is taken as the sum of the single particle densities. Through Eq.~(\ref{eq:fnuclear}), one 
can then  use the optical theorem to calculate the total cross-section. A further approximation 
could be made if the nuclear radius is large compared to the range of the nuclear forces. 
In such a case, the form factor is peaked near zero and
Eq.~(\ref{eq:fnuclear}) is approximated as
\begin{equation}
F_N(\Delta)=ik\int bdb J_0(\Delta b)[1-e^{(2\pi i/k)f(0)T(b) }]
\label{eq:opticalnuclearmodel}
\end{equation}
with the thickness function $T(b)=\int dz \rho(\vecb+{\hat k}z)$. At the time when their paper was written, 
Eq.~(\ref{eq:opticalnuclearmodel}) was referred usually as the optical nuclear model. However, 
according to Glauber and Matthiae, it is not a very good approximation, since the size of the nuclei of 
interest in this study were not much bigger that the range of the nuclear forces. Thus different approximations 
were looked for.

The quantity of interest here is the total inelastic cross-section. This will be obtained as the 
difference between the rate for all possible final states and the elastic differential cross-section. 
For large $\Delta$ values, and for the large nuclear radius approximation, the following expression is proposed
\begin{eqnarray}
\sum_{f\ne i}|F_{fi}(\Delta)|^2
= (\frac{k}{2\pi})^2
\int (d^2\vecb) (d^2{\vec B})
e^{
i{\vec \Delta}\cdot\vecb-\sigtot T(B)}&&\nonumber\\ 
\times \{ exp[\frac{T(B)}{k^2}\int e^{-i\vecq\cdot\vecb}|f(q)|^2d^2\vecq]-1\}\nonumber\\
\end{eqnarray}
Since actually the large radius approximation does not quite hold for light nuclei, one needs to 
use a non approximate expression. Different models for the nuclear density functions and the 
nuclear forces are discussed and the results plotted and compared with the Bellettini {\it et al} data.

The situation is easier for small $\Delta$, namely for small scattering angles, $\Delta\ll R^{-1}$ 
where $R$ is the nuclear radius. When $\Delta$ becomes small,  and  the nuclear radius is large 
compared to the interaction range, for large A one can use
\begin{equation}
\sum_{f\ne i}
|F_{fi}(\Delta)|^2=|f(\Delta)|^2 \{N_1- \frac{1}{A}|\int e^{i{\vec \Delta}\cdot{\vec b}+i\chi_N(\vec b)}
T({\vec b})(d^2{\vecb})|^2 
\},
\end{equation}
where $N_1$ refers to the number of free nucleons.

This model can be used then to extract information about \pp \ total cross-section from cosmic ray experiments, 
given the correlation between the proton-nuclei \x  \ and $\sigtot$. 
However, apart from modeling the nuclear density, which by itself may introduce some degree of theoretical 
uncertainty, there is another problem connected
with the extraction of data from the cosmic ray shower, namely how to extract the energy of the initial hadron 
or proton from the actual measurement of the shower. 
To this we turn in the next sections.

\subsection{Cosmic rays: measurements and extraction of pp data
\label{ss:cosmic-pp}}
Cosmic rays have traditionally afforded information about particle scattering at very high energies. Indeed, 
the rise of the total cross-section, hinted at by the earliest experiments at the  CERN Intersecting Storage Rings (ISR),  could be seen 
clearly in cosmic ray experiments  as shown in \cite{Yodh:1972fv,Yodh:1974cv}.  In  \cite{Yodh:1972fv},
 Yodh and collaborators stress the inconclusiveness of data concerning the energy behavior of the total cross-section. 
 Since the rise, if at all, would be logarithmic, it is pointed out that one needs to go to much higher energies, 
 such as those reached by cosmic ray experiments, where even  logarithmic variations can be appreciable. 
 Fig.~ \ref{fig:cosmic:yodh} from \cite{Yodh:1972fv} shows results from cosmic ray experiments  compared 
 with ISR earliest results \cite{Holder:1971bx}. Subsequent  ISR measurement of $\sigma^{pp}_{tot}$  
then confirmed the early rise and the cosmic ray observations.

\begin{figure}
\resizebox{0.5\textwidth}{!}{
\includegraphics{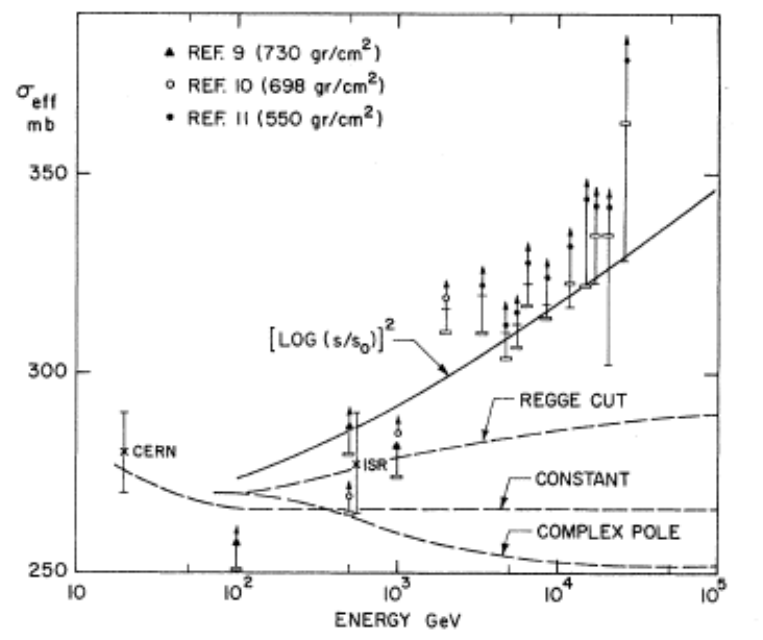}}
\caption{Figure 2 from  \cite{Yodh:1972fv},  in which  measurements for the inelastic p-air cross-section 
from  cosmic ray experiments are compared with various theoretical models and   values extracted from  
then available ISR results \cite{Holder:1971bx} for $\sigtot^{pp} $. Reprinted with permission from 
 \cite{Yodh:1972fv},    \copyright  1972 by the American Physical Society.}
\label{fig:cosmic:yodh}
\end{figure}
 
 In order to relate $pp$ total cross-sections to $p-air$ measurements, Yodh and collaborators  
followed  Glauber multiple scattering theory \cite{Glauber:1970jm} using a method previously discussed in \cite{Newmeyer:1970xq}.
A value for $\sigtotpp $ could be obtained  essentially through the following basic ideas:
 \begin{itemize}
 \item what can be measured in cosmic-rays is the inelastic p-air cross-section, 
 \item once a model for the $p-nucleus$ elastic amplitude is advanced, the  inelastic   $p-nucleus$ cross-section 
 follows from $\siginel=\sigtot - \sigel$,
 \item the $pp$ cross-section corresponds to the mean free path for interaction of a proton in a nucleus and 
 can be input to an effective inelastic cross-section for protons in air.
 \end{itemize}
 Different theoretical models for $pp$  cross-section were then inserted in the calculation and compared 
 with cosmic ray data  as well as an early ISR measurement \cite{Holder:1971bx}. The results were given as a set of different curves, 
 of which the one fitting cosmic rays data followed a behaviour saturating the Froissart bound. With this latter parametrization,  
 the   curve  at  $ \ E_{lab}=
 10^5\  GeV$ gives a most reasonable value of
\be
\sigtotpp=60\ mb \ at \ E_{lab}=
 10^5\  GeV
\ee
corresponding to a c.m. energy $\sqrt{s}=450\ GeV$. We notice that the above value is very close to what subsequent 
fits to $\sigtotpp$ (up to the LHC energies) have given
 \cite{Achilli:2011sw}.

In what follows, after briefly recalling 
the status of  the problem in the early '70s \cite{Barger:1974hi}, we shall   summarize  subsequent developments  following 
\cite{Gaisser:1986haa}, 
for the connection between $p-air$ and \pp \ . Then we shall see how Durand and Pi \cite{Durand:1988cr} applied their   
mini-jet model\cite{Durand:1988ax} to cosmic rays. We shall examine the discussion by Engel {\it et al.} \cite{Engel:1998pw}, 
followed by Block {\it et al.} \cite{Block:1999ub,Block:2000pg}. Subsequently we  update these 
with a discussion from a review\cite{Anchordoqui:2004xb},  
including later results by Block \cite{Block:2006hy,Block:2007rq}
along with  work by Lipari and Lusignoli (LL) \cite{Lipari:2009rm}.

\subsubsection{Cosmic ray experiments and the extraction of energy dependence of $\sigtot^{pp}$ up to 10 TeV
after the ISR data
\label{sss:earlymodels}}
\par\noindent The question of model dependence of the relation between $pp$ total cross-section and 
$p-air$ inelastic cross-section was discussed by Gaisser, Noble  and Yodh in 1974 \cite{Gaisser:1974xk,Gaisser:1974sa}.  
The starting point was of course the ISR confirmation of the rise of the total $pp$ cross-section,  
suggested from  cosmic ray experiments \cite{Yodh:1972fv}. The question being posed in the physics 
community was basically whether a behavior already saturating the  Froissart bound was in action or 
one was observing a transient  behavior due to some sort of threshold, or  to the rising importance of 
parton-parton scattering \cite{Cline:1973kv}.

An answer to this question was considered impossible to obtain from the then available accelerator data alone, 
so the question of reliability of cosmic rays estimates of $\sigtot^{pp}$ naturally arose. The conclusion of this 
paper 
 is that not only  the extraction  of $\sigma_{p-air}$ from air-showers is by itself affected by rather 
large uncertainties, but there is also a large model dependence in the extraction of the $pp$ cross-section from 
the observation of unaccompanied hadrons in the atmosphere. More investigation of the modeling  for both 
crucial steps was needed.

Before proceeding further, we mention the work by Cline, Halzen and Luthe \cite{Cline:1973kv} who interpreted 
the rising cross-sections as receiving a contribution from the 
scattering of  the constituent of the protons, 
the so-called partons, quarks and gluons. 
Parton-parton scattering  would give rise to  jets of particles in the final state.
An early estimate of a jet cross-section 
contribution, given as
\be
\sigma_{tot}=\int_{p_T=(2\div 3)GeV/c}^{p_T=\sqrt{s}/2}\frac{d\sigma}{dp_T} d(phase \ space)
\ee
 was obtained in this paper for jets with final transverse momentum $p_T > (2\div 3)\ GeV/c$ and is shown in Fig. ~\ref{fig:clinehalzen1973}. 
 by the shaded area.  
 \begin{figure}[htb]
\resizebox{0.4\textwidth}{!}{
\includegraphics{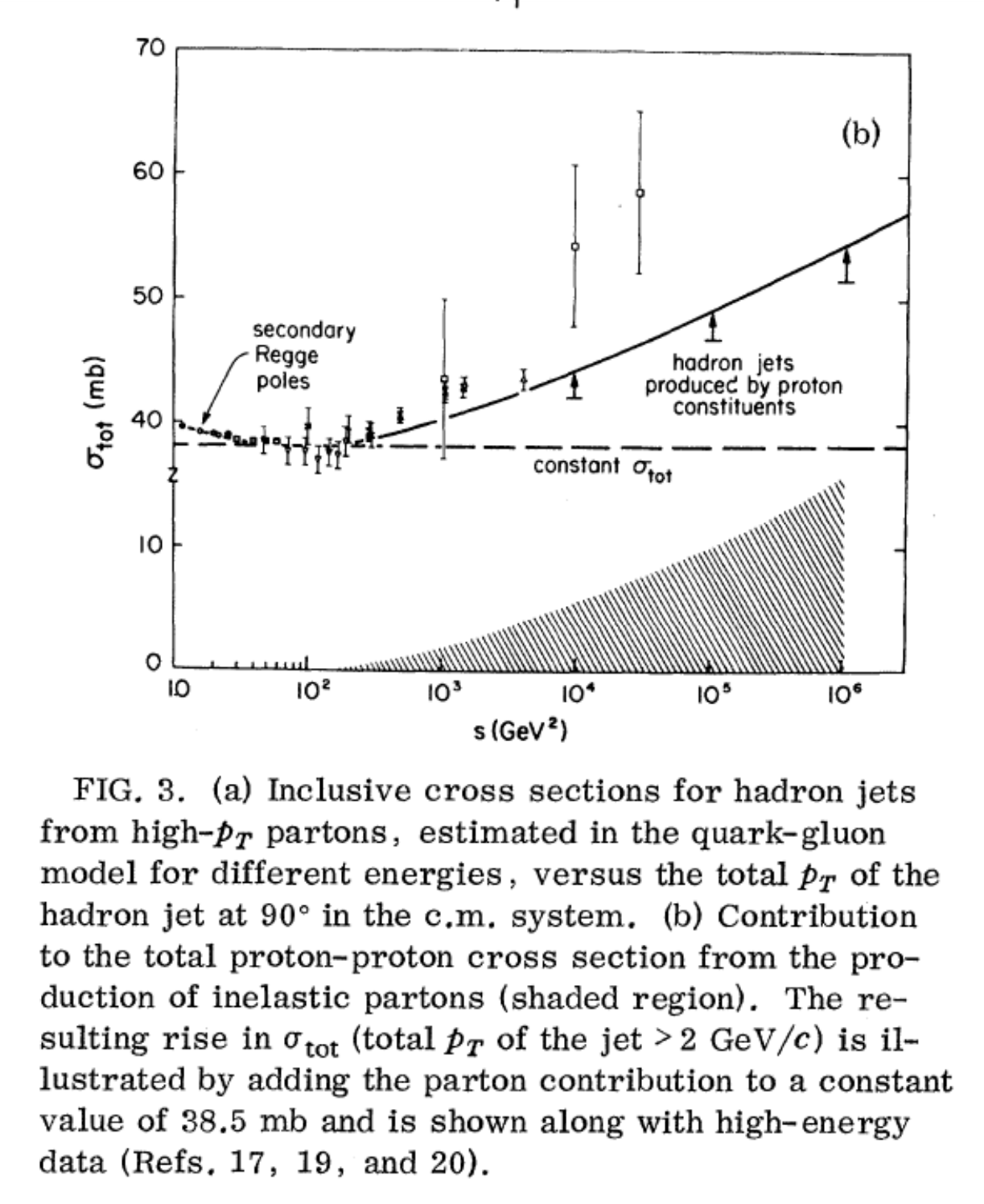}}
\caption{This early attempt to estimate jet contributions to the rising total cross-section shows cosmic ray extracted $
pp$ total cross-section as well as ISR results. Reprinted with permission   from \cite{Cline:1973kv},    \copyright  1973 by the American Physical Society.  }
\label{fig:clinehalzen1973}
\end{figure}
Adding such an estimated contribution to a constant or diffractive term, strongly suggested the (jet) 
phenomenon could to be responsible for the observed rise. It must be noticed that hadron jets had 
not been observed at ISR. Indeed, to get a sizable contribution from such a simple model, sufficient 
to fully explain the cosmic ray excess at high energy, softer  jets, contributing from  a smaller  $p_T \simeq 1\ GeV$, 
are needed: and such small $p_T$ jets would be very hard to observe. 

What appears here for the first  time is the idea of  small transverse momentum cluster of particles, which can be ascribed to 
elementary processes which start appearing in sizable amounts, what will be later called {\it mini-jets}. 
It will be necessary to reach a much higher cm energy,  to actually see particle jets and mini-jets at the 
CERN \spbarps, 
a proton-antiproton accelerator which  would explore energies as high as $\sqrt{s}=540\ GeV$. 

However, in 1974, accelerator data could   give information  on the total hadronic  cross-section  only up to  
$\sqrt{s}\lesssim 60\ GeV$ and new accelerators reaching higher energies, such as near and around   the TeV region,  
were very much in the future. Thus
the question of whether it could be  possible
to extract  the total \pp \ \x\  at c.m. energies  around $10 \ TeV$  and choose among different theoretical models 
for $\sigma^{pp}_{tot}$ was 
of great interest and was further
examined in  \cite{Barger:1974hi}.  
The state-of-the-art for total cross-section studies ({\it circa} 1974) is shown in Fig.~\ref{fig:barger74-new},  where added lines and arrow indicate results in the range of LHC energies, as expected, for instance, in  \cite{Achilli:2011sw,Fagundes:2015vba}. 
\begin{figure}
\hspace{-1cm}
\resizebox{0.6\textwidth}{!}{%
  \includegraphics{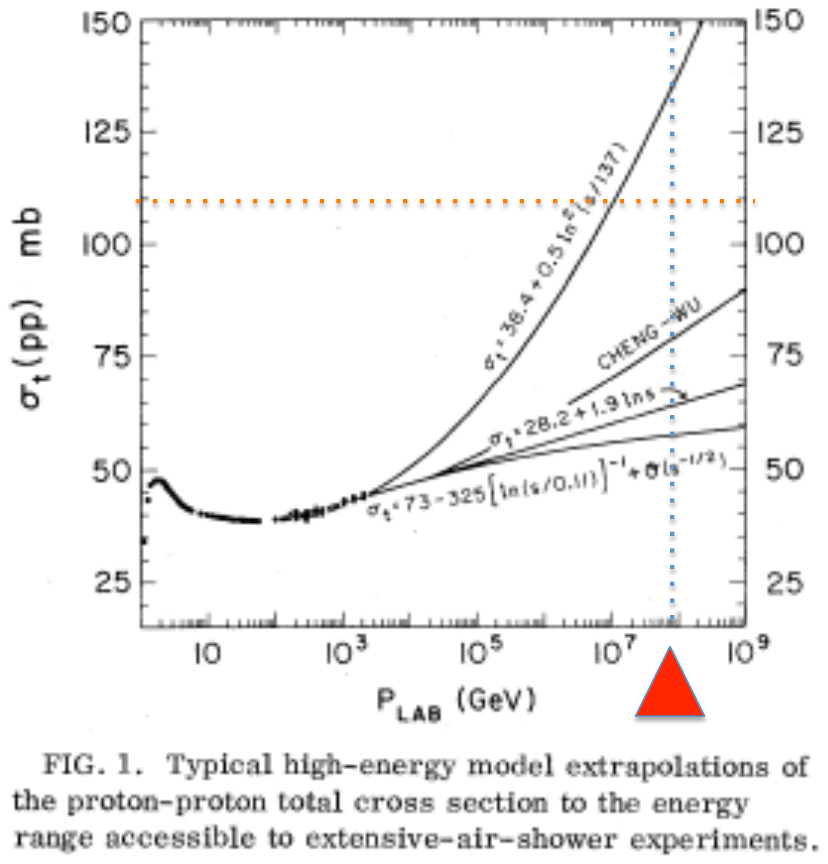}
}
\vspace{-1cm}
\caption{State-of-the-art of total cross-section models  in 1974 from  \protect\cite{Barger:1974hi} and  comparison with  expected results  at LHC13.  Reprinted with permission  from \protect\cite{Barger:1974hi}, \copyright 1974 American Physical Society.}
\label{fig:barger74-new}
\end{figure}

As already known,  the cascade development through which one measures $p-air$ \x\ is 
sensitive only to production processes, and the quantity measured is
 an inelastic cross-section. Then one could write the simplest possible model, which would 
 ascribe the breaking up of the stricken 
nucleus both  to elastic and inelastic scattering from the $nucleon-nucleon$ scattering, using Glauber's formalism, as
\begin{equation}
\siginel^{p-air}\equiv \sigma_{abs}=\int d^2\vecb [1-e^{-\sigtot T(\vecb)}]
\label{eq:airsimple}
\end{equation}
where $\sigtot $  indicates the total nucleon-nucleon (proton-proton) \x. $T(\vecb)$ indicates the profile function of the 
stricken nucleus of atomic number $A$. Two limits of Eq.~(\ref{eq:airsimple}) can be taken as
\begin{eqnarray}
\sigma_{abs}\simeq A \sigtot \ \ \ \ \ \sigtot \ small \\
\sigma_{abs}\simeq \pi R^2_A \simeq C A^{2/3} \ \ \ \ \ \sigtot \ large 
\end{eqnarray}
where $R_A$ is the nuclear radius. 

Such a simple model is unable to  provide enough information on $\sigtot$ at high energy, 
where $\sigtot^{pp}$ can become large, and a more precise expression 
seemed required. Within the Glauber model, convoluting  the nucleon profile function  
with the nuclear density function $\rho(\vecr)$, the following expression was proposed:
\begin{equation}
\sigma_{abs}=\int d^2\vecb [1-{\large |}1-\Gamma_A(\vecb){\large |}^2]
\end{equation}
where 
\begin{equation}
\Gamma_A(\vecb) =1-[1-\int d^2\vecb' dz \Gamma_N(\vecb -\vecb')\rho(z,\vecb')]^A,
\end{equation}
 $\Gamma_N(\vecb)$ is the nucleon profile function. There exist various models for the nuclear density, as already mentioned, 
 depending on the nucleus being light,  heavy or in between. Often, as in \cite{Barger:1974hi}, the gaussian form is used. 
 Next, one needs a  nucleon profile function and the  frequently used expression  is 
 again a gaussian distribution, namely 
 \begin{equation}
\Gamma_N(\vecb)=\sigtot \frac{e^{-b^2/2B}}
{4\pi B}
\end{equation}
This expression is based on the description of the elastic differential cross-section in the small $-t$-region, namely:
\begin{equation}
\frac{d \sigel}{dt}=|\int d^2 \vecb \Gamma_N(\vecb)|^2= [\frac{d \sigel}{dt}]_o\  e^{Bt}
\end{equation}
where $B$ is defined in the usual way, 
\begin{equation}
B(s)=[\frac{d}{dt} (\ln\frac{d\sigel}{dt})]|_{t=0}
\label{eq:B}
\end{equation}
It became clear however that there were other uncertainties to take into account, in particular 
those related to processes which could not easily be classified as elastic or inelastic, but, rather,  were  {\it quasi elastic}.

\subsubsection{Prescriptions for more precise extraction of $\sigtotpp$  after the advent of the CERN $S{p\bar p}S$ data}
\label{sss:Gaisser1986}

\par\noindent

The need to extract from cosmic ray experiments more precise information about the proton-proton cross-section 
led to scrutinize better the original use of the definition of  {\it inelastic} \x\ for nucleon-nucleus collision. Following 
the discussion by Gaisser, Sukhatme and Yodh in \cite{Gaisser:1986haa}, the inelastic $p-air$ expression 
for the cross-section  should include   various  processes which are neither totally elastic nor inelastic. Such are  
\begin{itemize}
\item  $\sigma_{quasi-el}$: quasi-elastic excitations of the air-nucleus, 
\item   $\sigma^*$: processes where there are diffractive excitations of one of the nucleons in the 
stricken nucleus,
\item  $\Delta \sigma$: multiple collisions with 
excited nucleon intermediate states
\end{itemize}
However, the cascade shower which is measured in cosmic rays is not sensitive to these processes. Thus, 
the $p-air$ inelastic \x \  which is being extracted from the cosmic ray cascade  is not  
\begin{equation}
\siginel^{p-air}=\sigtot^{p-air}-\sigel^{p-air}
\end{equation}
but rather 
\begin{eqnarray}
\siginel^{p-air}&=&\sigtot^{p-air}-\sigel^{p-air} \nonumber \\
&-& \sigma_{quasi-el}-\sigma^*-\Delta \sigma
\label{eq:sigprod}
\end{eqnarray}
Hence, in order to be able to extract information 
about \pp \ \x s\  from $p-air$, one needs to have a model for various nuclear excitation processes 
and for the diffractive part of the nucleon-nucleon cross-section which contributes to $\Delta \sigma$. 
These various processes can  be taken into account through the Glauber technique. Eq.~(\ref{eq:sigprod}) 
requires \cite{Gaisser:1986haa}  knowledge of the following quantities:
\begin{itemize}
\item $\sigtot^{pp}$
\item $B^{pp}(t=0)$, the forward elastic slope parameter 
\item $\sigma_{SD}^{pp}$, $\sigma_{DD}^{pp}$,  the total single and double diffractive \pp \ cross-sections
\item $\frac{d^2\sigma}{dt dM_X^2}$ the shape of the diffractive cross-section for $p\ p \rightarrow p+X$ near 
$t_{min} \approx\ - [m^2 (M^2 - m^2)^2]/2 s^2$
\item the nuclear density
\end{itemize}
In \cite{Gaisser:1986haa} use is made of unitarity to either exclude some of the then current models or to restrict the 
various contributions entering Eq.~(\ref{eq:sigprod}). In particular, when inclusion of diffractive processes are taken 
into account, the Pumplin limit \cite{Pumplin:1973cw}, 
to be discussed in Section \ref{sec:elasticdiff}, has to be included:
\be
\sigel+\sigdiff \le \frac{1}{2}\sigtot
\ee
The total, elastic and
quasi-elastic
$p-air$ \x s can be  calculated 
in a straightforward manner using the model parameters for $\sigtot^{pp}$, $\sigel^{pp}$, $B^{pp}$ and $\rho$, 
as for instance in \cite{Block:1984ru}. In brief, after having determined  the values of $\sigtot$ 
as a function of $\sqrt{s}$ as well as that of $B(s)$, the scheme of the calculation is to start with the elastic 
\pp \ scattering amplitude at small angle,
\begin{equation}
f(\vecq)=\frac{k\sigtot}{4\pi}(\rho+i)e^{-Bq^2/2}
\end{equation}
which is now fully determined and insert it 
into the nucleon-nucleon profile function
\begin{equation}
\Gamma_j(\vecb)=1-e^{i\chi_j}=\frac{1}{2\pi k}\int d^2 \vecq e^{-i\vecq\cdot\vecb}f(\vecq)
\end{equation}
which assumes that each nucleon has a static profile, thus automatically identical to a given function, the same for all. 
This nucleon-nucleon profile function is put into the profile factor for the nucleus
\begin{equation}
\Gamma=1-e^{i\chi}=1-e^{i\sum_{j=1,A}\chi_j}
\end{equation}
and then this is 
put into the nucleon-nucleus scattering amplitude, together with the nuclear density of  nucleons in the nucleus $\rho(\vecr)$. 
The nuclear amplitude is thus written as 
\begin{align}
F(\vecq)=\frac{ik}{2\pi}\theta(q^2)\int d^2\vecb e^{i\vecq\cdot\vecb} \times \nonumber \\
\times \int ...\int... \Gamma(\vecb, \vecs_1,..\vecs_A)\prod_{i=1}^A \rho(\vecr_i)d^3\vecr_i 
\end{align}
with $\Gamma(\vecb, \vecs_1,..\vecs_A)$ being the profile function for the nucleus , 
$\vec {s}$ being the component of $\vec {r}$ in the $\vec{b}$ plane. One now 
 uses the basic Glauber hypothesis that the overall phase shift $\chi$ of a nucleon on a nucleus is 
the sum of the  phase-shifts of individual nucleon-nucleon phase-shifts, and the following is now the proposed nucleon-nucleus amplitude:
\be
F(\vecq)=\frac{ik}{2\pi}\theta(q^2) \int d^2\vecb e^{\iqb}[ 1-[1-\int d^ 3\vec r \rho(r) \Gamma_j(\vecb -\vecs)] ^A]
\ee 
where $\Gamma_j$ is the nucleon-nucleon profile function.
One can now see how, by using various models for the nuclear density, and various models for nucleon-nucleon scattering, 
one can estimate $\sigel^{p-air}$, $\sigtot^{p-air}$ through the optical theorem, i.e.
\begin{eqnarray}
\sigtot^{p-air}=\frac{4\pi}{k}\Im mF(0)\\
\sigel^{p-air}=\int d\Omega_k {\large |}F(\vecq){\large |^2}
\end{eqnarray}
Before proceeding further, let us notice that the Block and Cahn model \cite{Block:1984ru}  fits  $B(s)$ and $\sigtot$  from available 
data in a large energy range and the result is a curve  where $B(s)$ and $\sigtot$ can be plotted against each other. 
The effect is that the larger values of $\sigtot^{pp}$ correspond to larger values of $B(s)$, as one can see from  the straight line in 
Fig.~\ref{fig:block-Bsigtot} from \cite{Block:1999ub}. 

The above takes care of the first two terms in Eq.~(\ref{eq:sigprod}), and we now turn to the last three terms. 
While different models were used for the nuclear density function in evaluating the proton-nucleus  amplitude, 
for the  calculation of the other three terms, whose details are in the Appendix of   \cite{Gaisser:1986haa},  
only a gaussian density distribution is used.
 The calculation of $\sigma_{quasi-el}$ is obtained 
from \cite{Glauber:1970jm} by integrating the expression for quasi-elastic scattering and making an expansion, i.e.
\begin{equation}
\sigma_{quasi-el} =\pi R^2\sum_{n=1}^{\infty}\frac{\epsilon^n}{n}
\end{equation}
with
\begin{eqnarray}
R^2=\frac{2}{3}<a^2>\\
\epsilon=\frac{1+\rho^2}{16\pi }\frac{\sigtot}{B}
\end{eqnarray}
$a$ being the rms nuclear radius. Given an input for the nuclear radius, in this model, the quasi-elastic term is again obtained from the 
\pp \ parameters. Next one needs to estimate $\sigma^*$, which represents the correction for diffraction dissociation of the nuclear target. 
This is estimated, in Gaisser's model, from 
\begin{equation}
\sigma^*=\frac{\sigma_{SD}^{pp}}{\siginel^{pp}}(\frac{2\pi}{3}<r^2>)
\end{equation}
where the last term in round bracket represents the cross-section for absorptive $p$-nucleus interaction involving only one nucleon. 
One can use unitarity to put bounds on  the ratio $\frac{\sigma_{SD}^{pp}}{\siginel^{pp}}$ and from the Block and Cahn model, one 
obtains for $\sigma^*$ a  range of values between 28 and 15 mb as the energy changes from $\sqrt{s}=20\ GeV$ to $10\ TeV$.

As stated in \cite{Gaisser:1986haa}, the last term, $\Delta \sigma$, is  a correction to Glauber's model, to include cases when one 
nucleon is excited in one encounter and then returns to the ground state through a subsequent encounter. This correction is 
evaluated to be of the order of $10\ mb$. For its estimate, the differential cross-section for diffractive excitation of the 
nucleon to a mass $M$ is modeled to be
\be
\frac{d\sigma}{dM^2 dt}=A(M) e^{B(M^2)t}
\ee  

Once the five  steps of the calculation have been performed in terms of the given input from $B(s)$ and $\sigtot$ 
for various  energies, one can now try to 
extract the proton-proton cross-section from the measured 
$p-air$ cross-section. 

The procedure consists of two steps. First, one   finds curves of fixed  
value of $\sigma_{p-air}$ in the $(B(s),\sigtot^{pp})$-plane, namely one finds the corresponding points  in the  
$(B,\sigtot)$ plane which give that particular $p-air$ cross-section. This procedure gives  curves such as the 
ones shown  in Fig.~\ref{fig:block-Bsigtot} from a 
later paper by Block, Halzen and Stanev  (BHS) 
\cite{Block:1999ub} to which we shall return next.  
To obtain then $\sigtot$ at a given energy from this procedure depends on the model for $pp$ scattering. 
In a given model let one draw the line which corresponds to predicted values for $B(s)$ at a given energy. 
The same model will also give a value for $\sigtot(s)$ and thus a line can be drawn to join these various point 
in the $(B(s),\sigtot(s))$ plane.  This line will meet the constant $\siginel^{p-air}$ at some points. Now if the 
experiment says that a proton of energy  $\sqrt{s}\simeq 30\  TeV$ has produced  $\sigma_{p-air}=450\ mb$, 
say, all we need to do is to look at which set of $\{B,\sigtot\}$ values the model crosses the constant contour. 
It is possible that the model cannot give such a high value for the $p-air$ cross-section for the $B$ and 
$\sigtot^{pp}$ values input by the model. 
This was for instance the case for one model by Goulianos \cite{Goulianos:1982vk}, in which diffraction 
dominated the cross-section, or a model by Block and Cahn, in which both $\sigtot^ {pp}$ and  $B(s)$ 
were not asymptotically growing.  In such  cases, seen clearly in (Fig. 6 of)  \cite{Gaisser:1986haa}, these 
values can never give a cross-section for $p-air$ as the one observed. Thus, the cosmic ray observation allows 
one to exclude these models, 
 at least within the validity of the given construction based on the Glauber model and on a
correct estimate of the cosmic ray composition.
But other models can give 
values which would be as high as the observed cross-section: in such cases, when the line crosses 
(say a value $450\ mb$) 
of the $p-air$ curve,  one has  correspondence between the experimental point for  $\sigma_{p-air}$ and the 
value of $\sigtot^{pp}$ in that particular model. One can also prepare a different plot, which shows which value of 
$\sigma_{p-air}$ one can get for a given $\sigtot^{pp}$ for different values of the slope parameter.  Depending on 
how different models obtain/parametrize $B$ vs $ \sigtot$, one can then extract 
the relevant information. 
\begin{figure}
\begin{center}
\resizebox{0.5\textwidth}{!}{%
\includegraphics{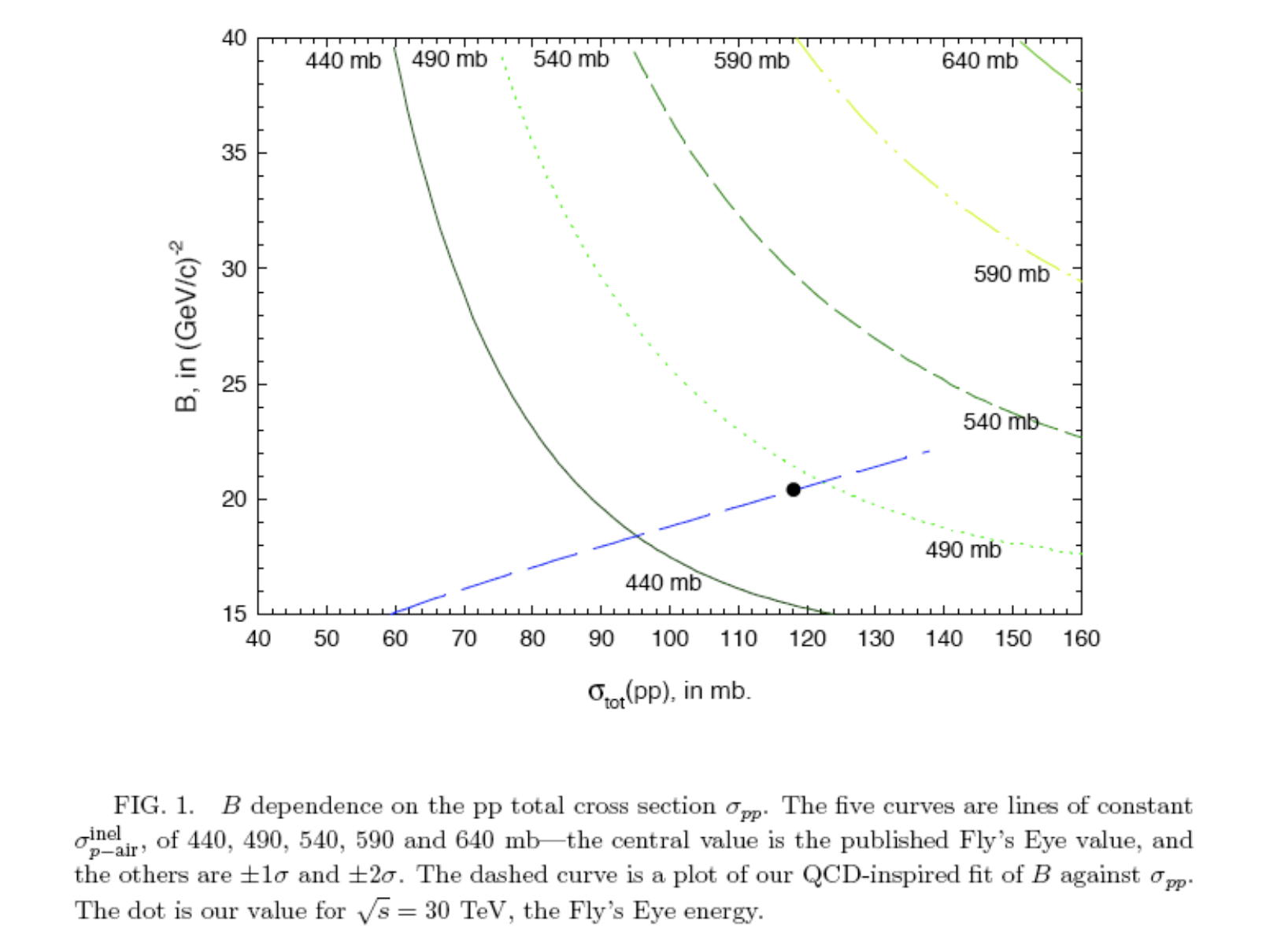}}
\caption{Relations between the total \pp \ \x \ $\sigtot$, the slope parameter $B(s)$ and $\sigpair$ from  
Block, Halzen and Stanev \protect\cite{Block:1999ub}. Reprinted with permission   from \cite{Block:1999ub},    \copyright  1999  by the American Physical Society.  }
\label{fig:block-Bsigtot}
\end{center}
\end{figure}

Conclusions about the validity of a given model, from plots such as the one in Fig.~\ref{fig:block-Bsigtot}, depends also on  the 
uncertainties about the measured values for $\sigpair$. In particular, as emphasized in \cite{Gaisser:1986haa}, fluctuations in 
individual hadronic interactions and on the chemical  composition of the primary particles in the observed showers influence 
the final result. 

\subsubsection{The Durand and Pi mini-jet model for $p-air$  interactions 
}
\label{sss:DurandPi} 
\vspace{0.2cm}

 So far,  we have discussed  approaches  which use the Glauber method to extract  information 
 about $\sigtot^{pp}$ from cosmic ray measurements at very high energy. As we have seen, in some cases, 
 such approach yielded information  not only  on  the energy dependence but also about  the model validity at 
 such very high energies. This was for instance the case of a diffractive dominance model \cite{Goulianos:1982vk} 
 or one of two models by Block and Cahn \cite{Block:1984ru}.
We now  
discuss 
a rather different approach.

The approach followed  by Durand and Pi \cite{Durand:1988cr} employs   their QCD-driven model \cite{Durand:1987yv} 
to predict $\sigma_{abs}(p-Air)$. \footnote{Here, as everywhere else in this review, we adopt the notation used by the 
authors in their articles.}
This model, which is described in detail in Sect.~\ref{sec:models} for the case of $pp$ scattering, uses an eikonal formalism with 
QCD mini-jets as  input for the energy dependence and an  impact parameter distribution (of quarks and gluons in a proton) 
modeled after the proton e.m. form factor.  
In this paper the  model is extended to scattering of protons on nuclei of 
nucleon number A, by basically treating the 
process as the scattering of quarks and gluons from the incoming protons against quarks and gluons in the nucleus.
We shall now follow  their description to see how the model is applied to $p-air$ or more generally to $p-A$.

The authors  start with the proton-nucleus profile function which enters the (proton-nucleus) scattering amplitude $f(t)$ for the given process 
($t$ being the momentum transfer), i.e.
\begin{equation}
f(t)=i\pi \int bdb J_0(b \sqrt{-t})\Gamma(b) \label{eq:DP1}
\end{equation}
Instead of the usual parametrization given as
\begin{eqnarray}
\Gamma(b)=\frac{\sigtot}{4\pi} e^{-b^2/2B}, \\
B=[
\frac{d}{dt}
 \Large{(} 
 \ln \frac{d\sigma}{dt} 
 \Large{)}
]_{t=0}\label{eq:Bslope},
\end{eqnarray}
Eq.~ (\ref{eq:DP1}) is written in terms of an eikonal function whose high energy behavior will entirely be based 
on the mini-jet model, namely
\begin{eqnarray}
f(t)=i\pi \int bdb J_0(b \sqrt{-t})
[1-e^
{- {\tilde \chi}_{pA}}] \label{eq:DP2}\\
{\tilde \chi}_{pA}=\frac{1}{2}  (\sigma_0+ \sigma_{QCD}){\tilde A}(b)
\end{eqnarray}
in close connection with the similar treatment for $pp$. In this, as in other similar models, the 
eikonal function is assumed to have a negligible real part. Corrections for this can be included. The absorption 
cross-section is then given as
\be
\sigma_{abs}(pA)=2 \pi \int b db [1- e^{-2 {\tilde \chi}_{pA}(b,s)}]
\ee
We now sketch the procedure followed by Durand and Pi and postpone  a discussion 
of how this production cross-section differs from the total inelastic cross-section. 
For an eikonal mini-jet model such as the one discussed here one needs to start with the following input:
\begin{enumerate}
\item how partons of given energy, momentum and position $\vecb$ are  distributed  in the nucleus
\item an elementary cross-section for parton-parton scattering $d{\hat \sigma}/d{\hat t}
({\hat s},{\hat t})$
\item density of nucleons in the nucleus
\end{enumerate}

Consider the  $p-A$ scattering process  as built from the uncorrelated scattering of an incoming proton 
with an {\it average} target nucleon $a$, which carries a fraction $1/A$ of the nucleus momentum. 
Quarks and gluons in the incoming proton then scatter against quarks and gluons in the average 
nucleon $a$ inside the nucleus. This model assumes that, at high energy,  the parton distribution 
in  the nucleus A   is given by the  parton  distributions in the nucleon $a$  convoluted with  the 
distribution of nucleons in the nucleus A. We are now dealing with a proton-nucleus scattering 
and, as in the original   mini-jet model for $pp$ scattering,   the impact parameter 
dependence  is  factored out from the energy and transverse moment dependence. Let $\rho_a(\vecb)$ be the 
average impact parameter distribution of partons in nucleon $a$, and $f_{i,a}(x,|{\hat t}|)$ the usual nucleon  
Parton Density Function (PDFs), where $x$ is the fractional longitudinal momentum carried by the parton, 
$\hat t$ the parton-parton momentum-transfer in the scattering.  Next,  parton distributions inside  the nucleus are proposed 
for  a model of $A$ uncorrelated nucleons.
In 
the model,  parton distributions inside a nucleus, $f_{i,A}(x,|{\hat t}|,b)$, are obtained as a   convolution of 
the  distribution of partons in the nucleon $f_{i,a}(x,|{\hat t}|) \rho_a(\vecb)$ with the  distribution of nucleons in 
the nucleus, $\rho_A(\vecr)$, namely 
\begin{eqnarray}
f_{i,A}(x,|{\hat t}|,b)=\frac{1}{A}\sum_{a=1}^{A}\int d^2 \vecr_\perp dz f_{i,a}(x,|{\hat t}|)\times \nonumber\\ 
\rho_a(|\vecb-\vecr_\perp|)\rho_A(\vecr_\perp,z)
\end{eqnarray}
with $\rho_A(\vecr_\perp,z)$ being the nuclear density function, subject to 
the normalization condition
\begin{equation}
\int d^2\vecr_\perp dz \rho_A(\vecr_\perp,z)=A
\end{equation}
Nuclear binding and small differences between protons and neutrons are neglected and one arrives to the following simplified expression
\be
f_{i,A}(x,|{\hat t}|,b)=f_{i,A}(x,|{\hat t}|)\int d^2\vecr_\perp dz \rho_a(|\vecb-\vecr_\perp|)\rho_A(\vecr_\perp,z)
\ee
Folding the above with the elementary parton cross-sections and integrating in the sub-process variables, 
leads to very simple result that the QCD contribution, also called mini-jets, to the eikonal is the {\it same} as the one 
calculable for $pp$ scattering, and  the difference between the hadronic and  the nuclear eikonal function only 
lies in the impact parameter distribution, 
namely the eikonal function for $pA$ scattering is written as 
\begin{equation}
\chi^{QCD}_{pA}(b,s)=\frac{1}{2}{\tilde A}(b)\sigma_{QCD}
\end{equation}
where $\sigma_{QCD}$ is the mini-jet cross-section which will be used to describe (fit) $\sigtot^{pp}$ 
and ${\tilde A}(b)$ is a convolution of the 
nuclear density with the parton density $A(\vecb)$ in the proton, i.e.
\begin{equation}
{\tilde A}(b)=\int d^2\vecr_\perp dz \rho_A(\vecr_\perp,z)A(|\vecb-\vecr_{\perp}|)
\end{equation}
The next step in the calculation is to deal with the non-QCD part, what can be defined as the soft 
scattering contribution, ${\tilde \sigma}_{soft}$. One may ask if
this quantity is the same as in \pp \ scattering. The model  
indicates that the same $\sigma
_{QCD}$ gives the high energy contribution (and hence the rise with energy)
to both nucleon and nuclear proton scattering. But there is no reason 
to expect the soft part to be the same.
Indeed the phenomenology, indicates a smaller ${\tilde \sigma}_{soft}$ in fits to the data. 
The point of view here is that it should be smaller, because soft processes (in mini-jet language -see later-  
final partons with $p_t < p_{t min}$) 
may not be sensitive to such processes as much as the $pp$ cross-section. 
Whatever the reason, the end result is that a good fit yields ${\tilde \sigma}_{soft}=31.2 \ mb$, 
instead of the value $\sigma_{soft}^{pp} = 49.2 \ mb$
used to describe \pp \ scattering.

Folding  the impact parameter distribution of partons of their model with different nuclear  distributions 
according to the nuclear composition of air, gives  the result for $p-air$, and  obtains the expression for  
the absorption cross-section for protons in air
\begin{eqnarray}
\sigma_{abs}(pA)=\int d^2\vecb (1-e^{-2{\tilde \chi}_{pA}(b,s)})\\
{\tilde \chi}_{pA}(b,s)=\frac{1}{2}({\tilde \sigma}_0+\sigma_{QCD}){\tilde A}(b)
\end{eqnarray}
We show their result in Fig.~\ref{fig:durandPicosmic} for two different nuclear density models, and 
${\hat \sigma}_0=31.2\ mb$ fitted to  low energy nuclear data, a value 30\% lower than what enters 
the fit to free $pp$ scattering. The fit to AKENO and Fly's Eye are  quite acceptable. However, the 
inverse procedure, that of trying to extract from the cosmic ray data a value for $\sigtot^{pp}$ poses 
some problems. The Fly's eye value of $\sigma_{abs}(p-Air)=540\pm50\ mb$ is seen to correspond 
to a $\sigtot^{pp}=106 \pm 20 \ mb$ at the cms energy of 30 TeV, a decidedly low value, especially 
after the latest TOTEM results at LHC.  

\begin{figure}
\begin{center}
\resizebox{0.5\textwidth}{!}{%
\includegraphics{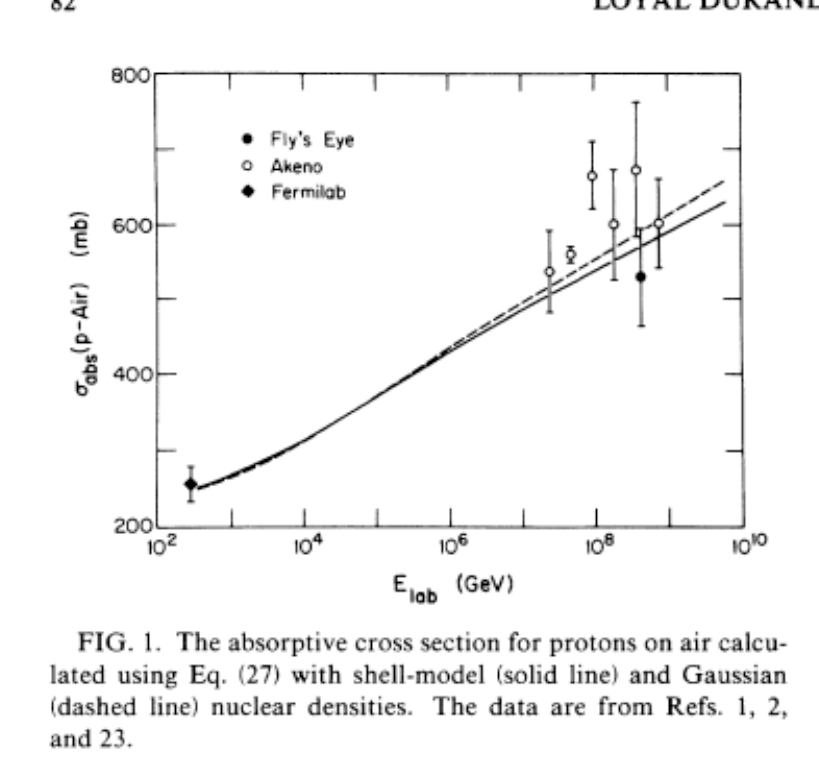}}
\caption{The absorption \x \ for $p-air$  as calculated in the QCD model by Durand and Pi in  \protect\cite{Durand:1988cr} compared with cosmic ray data. Reprinted with permission  from \cite{Durand:1988cr}, \copyright 1988 by the American Physical Society.}
\label{fig:durandPicosmic}
\end{center}
\end{figure}

\subsubsection{ More about uncertainties in extracting 
$\sigma^{pp}_{tot}$
from cosmic ray data, after the Tevatron 
}
\label{sss:Cosmicuncertainties}
While cosmic ray measurements can shed light  on the behaviour of $\sigtotpp$
 at very high energies, the hope to exclude or confirm a given model for $pp$ scattering is clouded by large uncertainties. 
 
 An updated  analysis of such uncertainties was done in 1998 by Engel, Gaisser, Lipari and Stanev \cite{Engel:1998pw}, 
 and a summary of their analysis is presented in the  following.  The starting point is the calculation of the absorption 
 cross-section with the Glauber model \cite{Glauber:1970jm}.   
 First, Engel {\it et al.} \cite{Engel:1998pw} discuss the relation between nucleon-nucleon and nucleon-nucleus cross-sections. 
 This discussion is based  on the procedure  used to extract $\sigtotpp$ from $\sigpair$,  following Gaisser {\it et al.} in  
 \cite{Gaisser:1986haa}, where
the definition of a {\it production} \x\ is adopted, namely
\begin{equation}
\sigma_{p-air}^{prod}=\sigma_{p-air}^{tot} -\sigma_{p-air}^{el}-\sigma_{p-air}^{q-el}
\end{equation}
with $\sigma_{p-air}^{q-el}$  the quasi-elastic $p-air$ \x\ where no particle production takes place, but there are inelastic contributions 
as intermediate states while the nucleon interacts with the nucleus.
  In this procedure  a crucial role is played by the B-parameter, 
 and  the relation between $\sigtot$ and $\sigel$  in 
{\it elementary} hadronic cross-sections.  After this, 
a discussion of how air shower experiments infer the $p-air$ \x \ is presented and the relevant uncertainties summarized.

In this model,  the basic sources of uncertainties arise from modeling of :
\begin{enumerate}
\item $\sigma^{p-air}$, the interaction cross-section between hadrons and the atmosphere, {\it vs.} $\sigma^{pp}$, 
the proton-proton cross-sections (total and elastic)
\item $\sigma^{pp}$ the hadron-hadron cross-section
\item the shower development and the primary cosmic ray composition and the relation with $\lambda_{p-air}$,  
the interaction length of hadrons in the atmosphere. 
\end{enumerate}
   
In the model, the profile function of the stricken nucleus is obtained through a combination of the following inputs:
(i) the elastic differential cross-section in the forward region, (ii) $\sigtot$ and (iii) $B(s)$,  the slope parameter at 
$t=0$, defined as in Eq.~(\ref{eq:Bslope}).  The connection between $B(s)$ and the elementary hadron-hadron \x s, $\sigel$ and $\sigtot$ 
is,  as before,  obtained from the optical theorem and the gaussian approximation for the forward region, namely
\be
\sigma^{el}_{AB}=(1+\rho^2)\frac{(\sigma^{tot}_{AB})^2}{16
\pi B(s)} \label{eq:sigelapp}
\ee
While Eq.~(\ref{eq:sigelapp}) is a good approximation 
to the data,
the high energy behavior of 
$\sigma^{tot}_{AB}$ and $B(s)$ depends on the model used. In \cite{Engel:1998pw}, one of two models 
discussed is  the standard Donnachie and Landshoff (DL)  fit   \cite{Donnachie:1992ny}, i.e.
\be
\sigma^{tot}_{AB}=X_{AB}s^{\epsilon}+Y_{AB}s^{-\eta}
\ee
with $\epsilon \simeq 0.08$ and $\eta\simeq 0.45$. The 
  Regge-Pomeron interpretation of the 
expression for the slope parameter would lead to
\begin{equation}
B(s)=B_0+2\alpha' (0)\ln (\frac{s}{s_0})
\end{equation}
We shall see later, in Sect. ~\ref{sec:models}, that 
a linear extrapolation in ($\ln s$) up to present   LHC results can be challenged  \cite{Schegelsky:2011aa} and that the 
high energy behavior of $B(s)$ is  still an open problem.


In general, once fits to the elastic and total cross-sections have been obtained in a given model,  
the by now familiar plot of $B$ vs $\sigtot$ is used with curves of constant $p-air$ cross-section 
drawn in it. Intersection of the model lines with a given curve allows to extract $pp$ cross-section in a 
given model at the given cosmic ray energy.
One   such plot, from \cite{Engel:1998pw} is  reproduced in Fig.~\ref{fig:engel98-2}. This plot  gives rise 
to large uncertainties: for instance the same low $p-air$ cross-section can be obtained with a small $B$-value 
and a range of $\sigtot^{pp}$ values, and so on.
Given a certain model and its  fit to elastic and inelastic data, and then its input into $p-air$ \x , three observations  
are worth repeating:
\begin{itemize}
\item model for the $pp$ interactions usually show that  at high energy the larger $B(s)$ the larger is $\sigtot$
\item  along a line of constant $\sigpair$, larger $B(s)$ values correspond to smaller $\sigtot$  
\item extrapolations of the slope parameter to higher energy depends on the model and it may lead to large uncertainties in $\sigpair$.
\end{itemize}
Two models are shown in Fig. ~\ref{fig:engel98-2}, the Donnachie and Landshoff  (DL) model \cite{Donnachie:1992ny} and a geometrical scaling model.
Geometrical scaling was a useful approach to the behavior of the elastic differential cross-section up to ISR energies. 
The hypothesis of geometrical scaling is that  the entire energy dependence of the cross-section comes from a 
single source, a radius $R(s)$, thus implying automatically the black disk limit ${\cal R}_{el}=\sigel/\sigtot=1/2$. 
Geometrical scaling however is not observed by experiments  at \spbarps\ energies and beyond. 
We shall return to this point in the section about models and elastic scattering.The dashed area indicates
the region excluded by unitarity, as discussed in subsection \ref{ss:afterLHC}.
\begin{figure}
\resizebox{0.5\textwidth}{!}{
\includegraphics{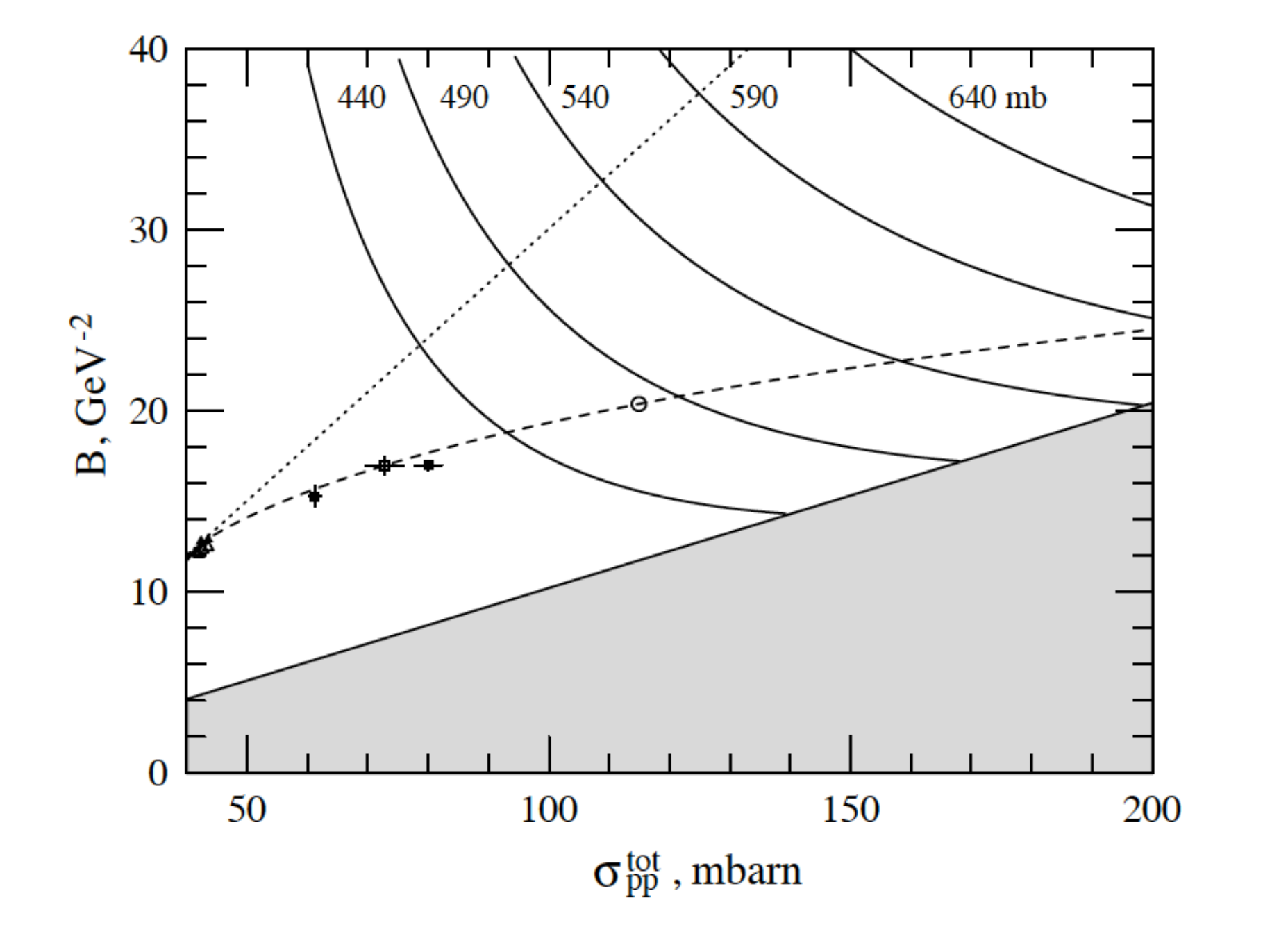}}
\caption{Contour plot of constant $p-air$ cross-section in the  $B(s)$ {\it vs} $\sigtot^{pp}$ plane from \cite{Engel:1998pw}.
Solid symbols are experimental accelerator points, dashed line is the DL model, dotted line a geometrical model. 
Dashed area excluded  by the unitarity constraint, the five curves are the region within one or two standard deviations 
from the central Fly's Eye experiment, $\sigma^
{prod}_{p-air}=540 \pm 50\ ( mb)$ measurement. The open point is the expectation of the DL model at 30 TeV cm energy. Reprinted with permission from \cite{Engel:1998pw}, \copyright  1998 by the American Physical Society. }
\label{fig:engel98-2}

\end{figure}



\subsubsection{Extracting information from cosmic ray showers}\label{sss:attenuation}
The uncertainties encountered in determining values for $\sigtot$ from $\sigpair$ are only one of the problems  
encountered 
in trying to find the asymptotic behavior of  proton-proton scattering. 
 More uncertainties lie 
with the air shower technique which is used  to extract $\sigpair$. 
Such uncertainties  are summarized in a parameter, 
sometimes called $a$, more often $k$, 
which relates the interaction length of protons in air, i.e.
\begin{equation}
\lambda_{p-air}=14.5 \frac{m_p}{\sigma_{p-air}^{prod}}
\end{equation}
where $14.5$ is the mean atomic mass of air, to the attenuation length $\Lambda$
defined as
\begin{equation}
\Lambda=a \lambda_{p-air}
\end{equation}
The {\it attenuation length} $\Lambda$  is a measure of the initial proton energy, and is obtained from the air showers 
generated by the interaction of the primary cosmic ray with the atmosphere. This uncertainty entering  data extraction 
is the rate of energy dissipation by the primary proton. 
The $p-air$ cross-section is thus affected by uncertainty related to the parameter $a$.

But uncertainty also comes from the composition of the most penetrating cosmic rays, 
and this is also related to the question of the origin of cosmic rays. 
Engel {\it et al.} \cite{Engel:1998pw} discuss how different Monte Carlo programs simulate air shower development using the 
same $\Lambda =70\pm 6\  g/cm^2$ and extract different values for $\sigpair$ 
according to the chosen parameter set.

\subsubsection{Air shower modeling}\label{sss:air-shower}
The energy of the particle starting the shower and the interaction length of the proton in air are obtained from the depth 
and extension of the electromagnetic shower they initiated. 
The most important processes of interest for air shower modeling are electron and muon bremsstrahlung and pair production. 
The electromagnetic shower, which develops as an electron or a photon starts losing energy because of EM processes such as  
bremsstrahlung and pair production, can be divided into three phases: 
\begin{itemize}
\item the shower grows as long as  all the particles have energy larger than the typical ionization energy $\epsilon_0$ 
which corresponds to an electron energy  
too small for pair production, and such as to  typically   induce ionization in the nuclei
\item the shower maximum, which has an extension determined by the fluctuations around the point where all the 
particles have energy just about $\epsilon_0$
\item the shower tail where particles lose energy only by ionization or by absorption, or decay.
\end{itemize}
To determine the quantity $X_{max}$ one can use a simple model due to Heitler and summarized in \cite{Anchordoqui:2004xb}. 

These processes are characterized by a typical quantity, the radiation length $X_0$, which represents the mean 
distance after which the high-energy electron 
has lost $1/e$ of its initial energy. Namely, $X_0$ is the constant which defines the energy loss of the electron as it traverses a distance X,
\begin{equation}
\frac{dE}{dX}=-\frac{E}{X_0}
\end{equation}
Then, in a simplified model in which only bremsstrahlung and pair production are responsible for energy losses, the air shower can be modeled, 
as follows. At each step the electron, or photon can split into two branches, each of which will then split into two other branches  as long as the 
energy of each branch is $>\epsilon_0$. 
At a depth $X$, the number of branchings is roughly $n=X/X_0$ and  after $n$ branchings, the total number of particles will be $2^n$. 
The maximum depth of the shower will be reached when the initial energy $E_0$ is equally distributed among the maximum number of secondaries, 
each one of which has energy just above the ionization limit $\epsilon_0$, Thus,   
$E_0=N_{max}\epsilon_0$,  $N_{max}=2^{n_{max}}=2^{X_{max}/X_0}=E_0/\epsilon_0$. This leads to
\begin{equation}
X_{max}=X_0 [\frac{\ln \frac{E_0}{\epsilon_0}}{\ln 2}]
\end{equation}
The matter is further complicated by the fact that not all groups employ the $X_{max}$ method.

The attenuation length $\Lambda$ is obtained from the tail of the function describing the $X_{max}$ distribution.

\subsubsection{Block, Halzen and Stanev: models {\it vs.} measured attenuation length }
\label{sss:BHS}

Shortly after Ref.~\cite{Engel:1998pw}, the uncertainties arising through the  $a$ parameter values 
used by different experiments, were  again discussed  \cite{Block:1999ub,Block:2000pg}  in light 
of the QCD inspired model for $\sigtotpp$ in  \cite{Block:1991yw}, referred to as BHM model. 
This model    incorporates analyticity and unitarity,  in a context 
in which QCD shapes the parameterization of  the terms which contribute to the rise  of  the cross-section. 
We shall discuss it  later  in more detail.

Anticipating  more recent debates, we notice that the cosmic ray   experiments  can give information on  
the interaction $p-air$, and hence on $pp$, but the extraction of data depends not only on the rate at 
which the primary particle dissipates energy in the atmosphere, but  on cosmic ray composition. In this 
 work, Block, Halzen and Stanev choose  to ignore the possibility that the most penetrating particles may 
not be protons, and focus instead on  the consistency of the values extracted by cosmic ray experiments 
with  those extracted by  the Glauber method implemented with their model for $pp$ scattering.
 
 They rename the parameter $a$ as $k$, and use the slightly different nomenclature
\begin{eqnarray}
  \Lambda_m=k\lambda_{p-air}=k \frac{14.5m_p}{\sigma_{inel}^{p-air}}
\label{eq:interactionlength}\\
\sigma_{p-air}^{inel}=\sigma_{p-air} -\sigma_{p-air}^{el}-\sigma_{p-air}^{q-el}
\label{eq:sigmasinair}
\end{eqnarray}
where the subscript $m$ in Eq.~(\ref{eq:interactionlength}) stands for {\it measured}, 
$k$  measuring  the rate at which the primary 
proton dissipates energy into the electromagnetic shower as observed by the experiment, 
and $\sigma_{p-air}$ is the total $p-air$ cross-section.
Once more, here are the steps as described in this paper:
\begin{enumerate}
\item experiments  obtain $\sigma_{inel}^{p-air}$ from Eq.~(\ref{eq:interactionlength})  
at a given energy of the most penetrating primary particle,  
measuring  $\Lambda_m$ and estimating  a value for $k$,
 \item model builders use Eq.~(\ref{eq:sigmasinair}), with   Glauber's theory  and a 
 nuclear density model as described in \cite{Gaisser:1986haa}, and  draw constant contours  
 of  fixed $\sigma_{p-air}^{inel} $  in the  $B(s)$ and $\sigtotpp$ plane,
\item the QCD inspired model \ by   \cite{Block:1991yw} establishes the correspondence between 
$B(s)$ and $\sigtotpp$: i) at any given energy,   extrapolated values for  $\sigtotpp$ and  $B(s) $ 
can be  determined through a fit of the model parameters  of all the $pp$ and \pbarp \ accelerator data.  
Thus a plot such as the one in Fig. ~ \ref{fig:engel98-2} where  the dashed line is for the DL model, 
is now constructed for the BHM model,
\item the intersection of a given curve for $p-air$, which corresponds to the measured $\sigma_{inel}^{p-air}$ 
for primary proton energy, with  the  $B(s)$ {\it vs.}  $\sigtotpp$ line determines the $\sigtotpp$ value at that energy.
\end{enumerate}
   Fig.~\ref{fig:block1999-fig3}, shows the one-to-one correspondence established between 
   $\sigtotpp$ and $\sigma^{inel}_{pair}$ {\it via} the constant contours in the ($B,\sigtotpp$) plane, 
   according to the BHM model.
 \begin{figure}
 \resizebox{0.5\textwidth}{!}{
 \includegraphics{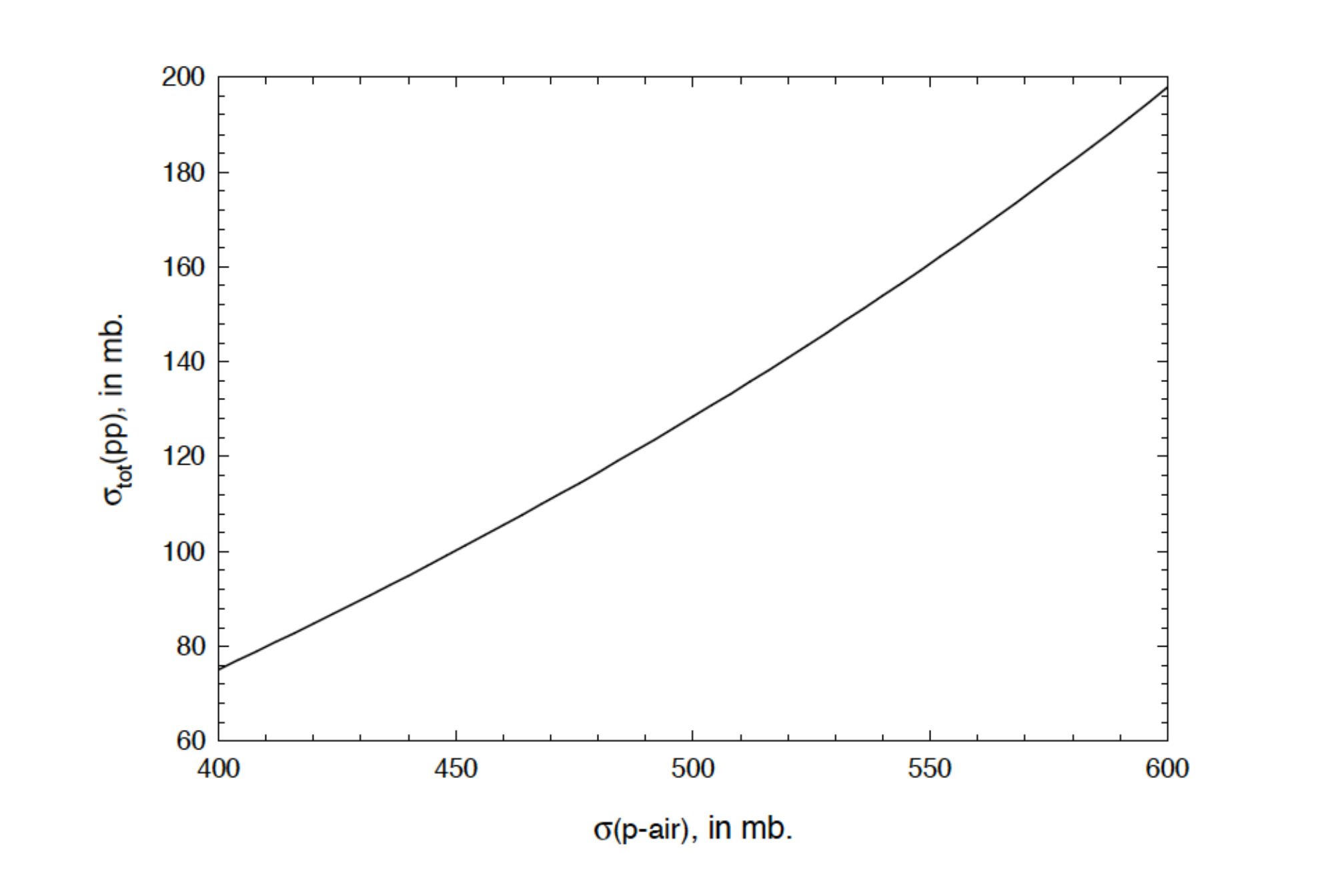}}
 \caption{Plot of the predicted $pp$ cross-section {\it vs.} for any given value of a  measured 
 $\sigma_{p-air}^{inel}$, using the constant contour procedure, from  Fig. 3 of \cite{Block:1999ub}.  Reprinted with permission from \cite{Block:1999ub}, \copyright 1999 by the  American Physical Society. }
 \label{fig:block1999-fig3}
 \end{figure}

It appears that in such a procedure, one  has to first trust two models,  i) the Glauber model   
along with the relation between $B(s)$ and $\sigtotpp$, and ii) the model which extrapolates 
$B(s)$ and $\sigtotpp$ at ultrahigh energies, but then one  has also to trust   the correctness of the 
extraction procedure   of $\sigma_{p-air}^{inel}$ from the air showers.

 In \cite{Block:1999ub}  a contradiction is seen to arise  between the predictions for  $pp$ in the model 
 as obtained from accelerator data and cosmic ray extracted values.

In their conclusion, the authors point a finger to the parameter $k$, as already noted before \cite{Engel:1998pw}. 
Different experiments use different values for $k$, obtained from different analyses of shower simulation. 
The authors thus proceeded to do a $\chi^2$ fit to the cosmic ray data and extract a value for $k$, and obtain 
$k=1.33\pm 0.04\pm0.0026$, which falls in between values used by different Monte Carlo simulations.

Shortly there after, in a subsequent paper \cite{Block:2000pg}, to reduce the dependence on the 
determination of the parameter $k$, a simultaneous fit  to both accelerator data {\it and} cosmic ray data 
was done by the same authors and a reconciliation between $pp$ cross-section between accelerator data 
and cosmic ray data was obtained when a value $k=1.349\pm 0.045\pm 0.028$ was used.  
The resulting agreement is shown in Fig. ~\ref{fig:block-halzen-stanev2000} from \cite{Block:2000pg}. 
This analysis now  gives a value $\sigma^{total}_{pp}(\sqrt{s}=14\ TeV)=107.9\pm1.2\ mb$ and the 
predictions from the model are now in agreement with extracted cosmic ray $pp$ data.
\begin{figure}
\resizebox{0.5\textwidth}{!}{
\includegraphics{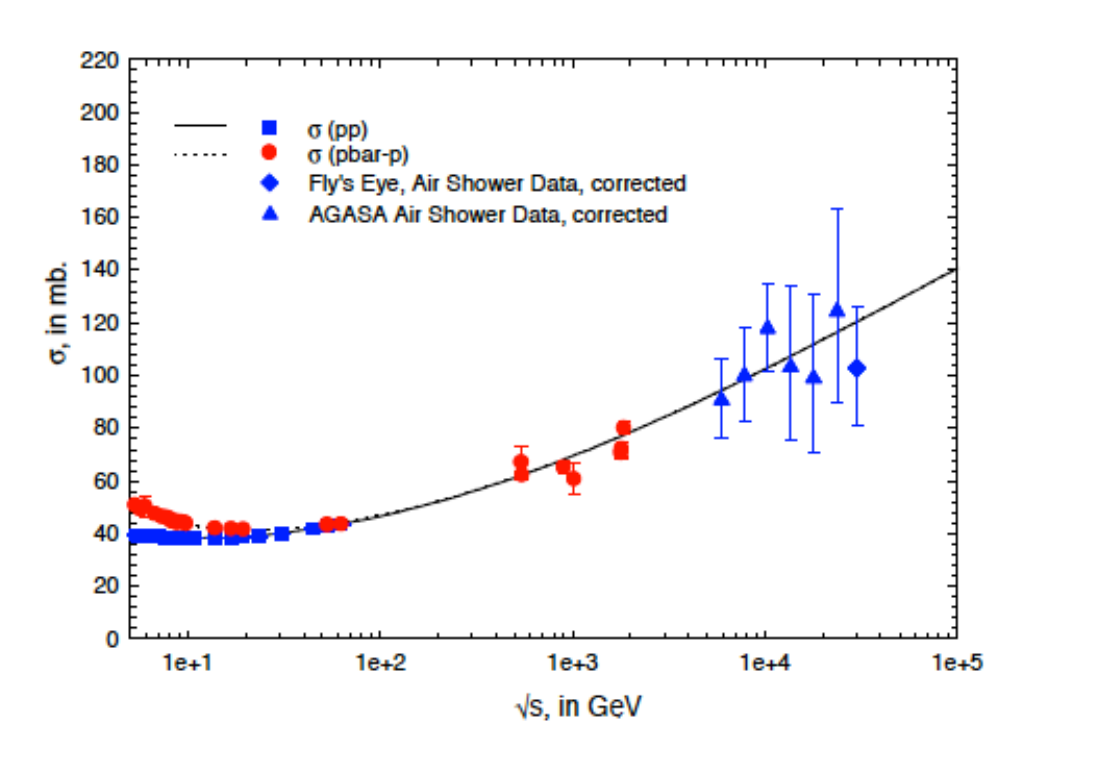}}
\caption{$pp$ scattering total cross-section predictions compared to rescaled Cosmic ray data 
from \cite{Block:2000pg}, as described in the text. Reprinted with permission from \cite{Block:2000pg}, \copyright 2000 by the American Physical Society.} 
\label{fig:block-halzen-stanev2000}
\end{figure}

Other questions 
arise if one were to challenge the assumed primary composition 
or Eq.~(\ref{eq:sigmasinair}), 
as we shall see through a summary of a review paper by Anchordoqui {\it et al.} \cite{Anchordoqui:2004xb}, 
to which we now turn.

\subsection{The extraction of $p-air$ cross-section from cosmic rays }
\label{ss:anchor}
A good review of the experimental techniques used to measure cosmic ray showers and extracting 
information from them can be found in \cite{Anchordoqui:2004xb} and lectures covering many aspects 
of cosmic ray physics can be found in \cite{Anchordoqui:2011}.

This 2004 review focuses on cosmic ray phenomenology from the top of the atmosphere to the earth surface.
For primary cosmic ray energies above $10^5\ GeV$, the flux is so low that direct detection of the primary 
particles above the upper atmosphere is practically impossible. 
In that range, however, the primary particle 
has enough energy to penetrate deeply in the atmosphere and generate  Extensive Air Showers (EAS), 
namely a measurable cascade of detectable products.  Various techniques are used to detect the 
cascade products, and different types of detectors are employed. 

In addition to various ways to extract from air showers  information about primary composition of the incoming 
cosmic rays and energy of the primaries thus selected, one tries to extract $\sigtot^{pp} $from $\sigpair$. 
There are various Monte Carlo simulation programs which do this, the ones mentioned in this review being 
SiBYLL \cite{Fletcher:1994bd}, QGSJET \cite{Ostapchenko:2010vb} and DPMJET \cite{Ranft:1994fd}. 
DPMJET simulates hadronic interactions up to the very high cosmic energies of interest using the Dual Parton 
Model \cite{Capella:1992yb}. The other two are both based on  the eikonal approximation and mini-jet cross-sections, 
but differ in how they introduce the impact parameter distribution of partons in the hadrons. According 
 to this review, in SYBILL the $b-$distribution is the Fourier transform of the proton e.m. form factor, whereas 
 in QGSJET it is taken to be a Gaussian, i.e.
\begin{eqnarray}
A(s,\vec{b})= e^{-b^2/R^2(s)}\\
R^2(s)\simeq 4R^2_0+4\alpha'_{eff}\ln^2 \frac{s}{s_0}
\end{eqnarray}
In this way, they can easily obtain the diffraction peak in agreement with the experimentally observed increase 
with energy. DPMJET  has its name from  the Dual Parton Model and  is based on soft and hard Pomeron exchanges.

We can see now how various models for proton-proton scattering influence the information about the behaviour 
of the total \pp \ \x \ at the highest energy available. In  Fig.~\ref{fig:anchorcosmic}, values   for $\siginel$ from cosmic 
ray data from AGASA and Fly's Eye are plotted against  the two model entries from SYBILL and QGSJET  for the 
inelastic \pp \  \x . At low energy, data come from CERN ISR and the cross-sections are normalized to these values.
\begin{figure}
\resizebox{0.5\textwidth}{!}{%
  \includegraphics{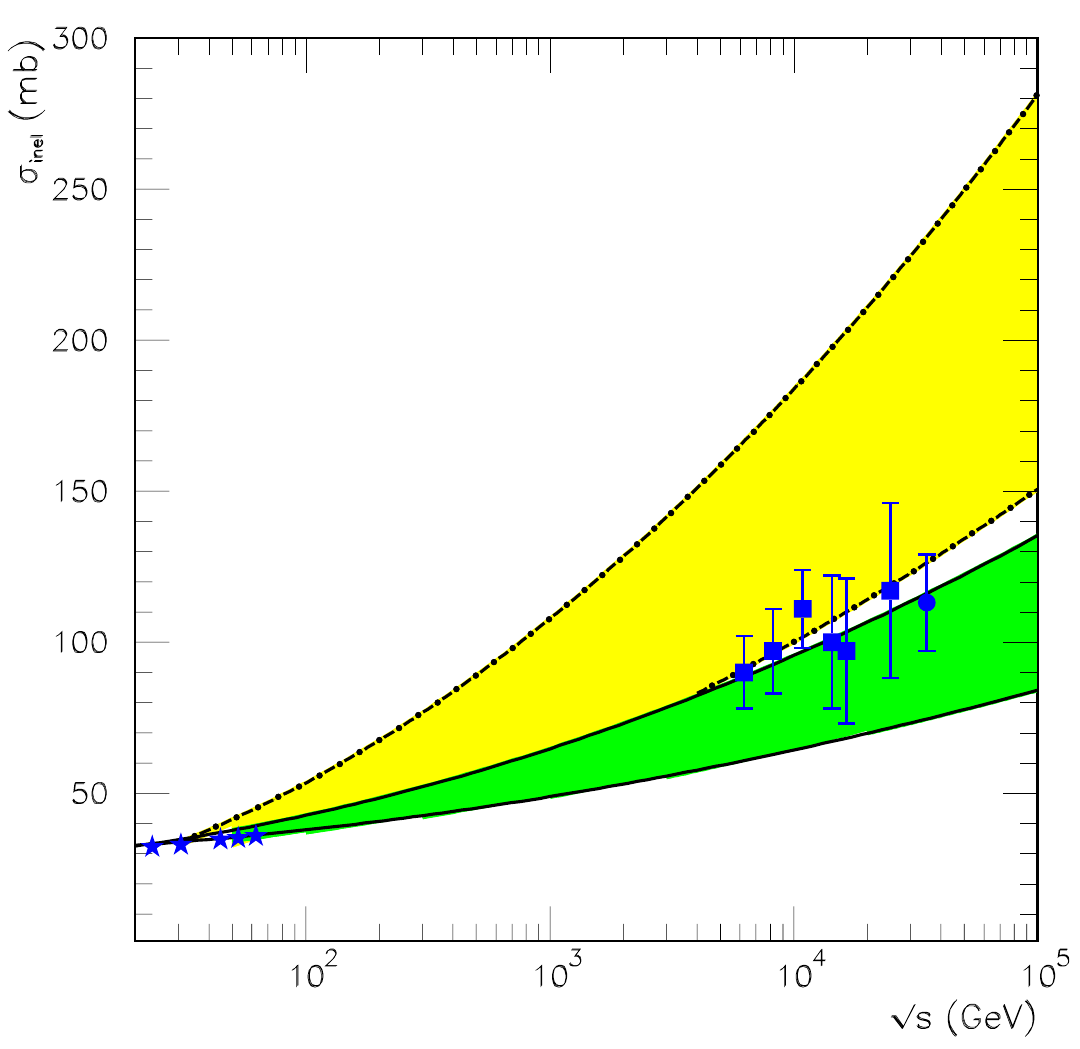}
}
\caption{The above 
 is  Fig. (5) from \protect\cite{Anchordoqui:2004xb} and shows the energy dependence of the \pp \ inelastic cross-section compared with different models for the 
impact parameter distribution inside the protons
This  figure is courtesy of L. Anchordoqui. Reprinted from  \cite{Anchordoqui:2004xb}, \copyright (2004) Elsevier.}
\label{fig:anchorcosmic}       
\end{figure}
The experimental errors indicated in this figure  are mostly due to a limited understanding of the interaction 
of protons  with nuclei and also nuclei with nuclei  at such very high energies, as we shall try to show in what follows. 
An even greater uncertainty seems to come from the modeling of  the 
\pp \ \x \ itself, as indicated by the two different bands. This uncertainty is actually not as large as it appears in this figure, since data at 
$\sqrt{s}\approx  (60\div 1800)\ GeV$ severely limit the high energy behaviour, as one  
can see from the section on models and fits.

Let us now turn to describe how one extracts the data for \pp \ \x \ from $p-nucleus$, basically 
$p-air$ data, always following  \cite{Anchordoqui:2004xb}. In order to 
simulate cosmic ray showers, current event generators  will need first of all to extrapolate the \pp  \  \x\ to very high energy 
(but this is by now provided by the many models and fits described in the next section) but also 
to make a model for the impact parameter distribution of nucleons in  nuclei. Indeed, all the event 
generators or models included in the event generators, use the Glauber formalism, with the nucleon density 
folded into that of the nucleus. 

In ref.~\cite{Anchordoqui:2004xb}, the distinction between a {\it production} and an {\it inelastic} $p-air$ \x\ as related to   
the total vs. the inelastic  $pp$ cross-section is clearly emphasized. We anticipate here that the definition of 
$\sigma^{pp}_{inelastic}$   is model dependent.  Following ref. \cite{Kopeliovich:2003tz}, the following expressions are discussed:
\begin{eqnarray}
{\tilde \sigma}_{inel}\approx \int d^2\vecb \{1-exp[-\sigtot T_A(\vecb)]\}\\
{\tilde \sigma}_{prod}\approx \int d^2\vecb \{1-exp[-\sigma_{inel}T_A(\vecb)]\}
\end{eqnarray}
where $T_A(\vecb)$ gives the impact parameter distribution of the hadronic matter in the target ($air$ for instance) folded 
with that of the projectile particle. $\sigtot$ and $\siginel$  are the relevant quantities for \pp \ scattering or $hadron-hadron $ 
scattering. Here ${\tilde \sigma}_{inel}$ uses the usual eikonal format, with as input the {\it total} \pp \ \x . The physical description  
amounts to  consider all the possible ways in which the proton can interact with another proton, $\sigtot$ and this will then be input 
to the formal expression for the  inelastic $p-air$ cross-section. Then the result, compared with $p-air$ data should allow to extract 
the \pp \ data. The second equation starts with the \pp \ inelastic \x\ and thus corresponds to breaking up single protons in the nucleus. 
What will it give for $p-air$? Clearly it is a scattering process in which at least one proton has broken up, generating a new particle. 
According to Anchordoqui {\it et al.}, $ {\tilde \sigma}_{prod}$ 
gives the cross-section for processes in which at least one new particle is generated. For this to happen, 
one must exclude elastic \pp \ processes, and this is why the input in this case is the  {\it inelastic}  \pp\ \x . 
This latter quantity, ${\tilde \sigma}_{prod}$, is the one which enters the cascade, since this is what will start the cascade. 

Notice that, in the cascade, $\pi-p$ is also playing a role and this \x \ needs  to be entered in the simulations as well. 
The $\pi-p$  \x \ is smaller than the one for \pp, but only by perhaps a factor $2/3$.
The relevant parameters in modeling 
these processes are two, the mean free path, $\lambda= 1/(n{\tilde \sigma}_{prod})$ and the inelasticity $K=1-E_{lead}/E_{proj}$, 
where $n$ is the density of nucleons in the atmospheric target and $E_{lead}$ and $E_{proj}$ are energy of 
the most energetic hadron  with a long life time in the shower and the energy of the projectile particle, respectively.

\subsubsection{Extraction of $\sigtotpp$ in  Block and Halzen  model }\label{sss:bhmodel}
The extraction of $\sigtotpp$ 
from cosmic ray data has been considered once again in \cite{Block:2007rq}. 
 Block summarizes  his description of the connection between $p-air$ and \pp \ data and examines different methods by 
 which data are extracted and analyzed. In addition, a different model for $\sigtotpp$ is used. 
The discussion covers now two different methods by which one can obtain $\sigma_{p-air}^{prod}$ from the 
$X_{max}$ distribution. 

In the first of the methods examined, which is the one used by Fly's Eye, AGASA, Yakutsk and EASTOP, the measured quantity 
is of course $\Lambda_m$, which implies that, in order to extract first of all a value for $\sigma_{p-air}^{prod}$ 
one needs the value of the parameter $k$. The range of values used by different experiments for the parameter $k$ is given in the paper
and it lies within ($1.15\div 1.6$) from EASTOP to Fly's Eye experiment. The smallest value is the one used by EASTOP, 
which is also the most recent and corresponds to more modern shower modeling,

A second method is the one used by the HiRes group, which, according to Block, has developed a quasi model-free method of measuring  
$\sigma_{p-air}^{prod}$. Basically, the shower development is simulated by randomly generating an exponential distribution for the first interaction 
point in the shower. By fitting the distribution thus obtained, one can  obtain $\sigma_{p-air}^{prod}= 460\pm14\pm39\pm11\ mb$ at
$\sqrt{s}=77\ TeV$.

The analysis in this paper differs from the one  in \cite{Block:2000pg} in two respects, one concerning the model 
used for $\sigtotpp$, and the other  the treatment of different cosmic ray experiments. The extraction of 
$\sigma^{prod}_{p-air}$ from the model used for $pp$ description does not use all the results from  the 
QCD inspired model. A   hybrid combination enter into the ($B(s),\ \sigtotpp$) plane, namely: i) $\sigtotpp$ is 
obtained from an analytic amplitude expression, which saturates the Froissart bound \cite{Block:2005pt}, 
i.e.  $\sigtotpp \simeq \ln^2 s$ at asymptotic energies, with parameters fitted to both $pp$ and \pbarp 
\ accelerator data,
ii) $B(s)$ is obtained  via a fit to data from the 
QCD inspired model \cite{Block:2006hy}. Values for $\sigtotpp$ at LHC remain unchanged, but  changes appear in the 
thus determined value for the parameter $k$.

 \subsubsection{The inelastic cross-section and model  uncertainties, including diffraction}\label{sss:inelastic}
A short review by D'Enterria, Engel, Pierog, Ostapchenko and Wener \cite{d'Enterria:2011jc}  deals with 
various  hadronic quantities entering  cosmic rays analyses, 
 such as multiplicity distributions and energy flow.  In addition, extracting an inelastic cross-section from  
 total and elastic scattering  requires, in most models, a  theoretical description  of diffraction, single and double, low and high mass. 
 
In Regge based models, diffraction uses a multichannel formalism along the line of previous analyses by 
Miettinen and Thomas \cite{Miettinen:1979ns}, also discussed  by   Pumplin \cite{Pumplin:1982na,Pumplin:1991ea}, 
and based on the Good and Walker  decomposition of diffractive  scattering \cite{Good:1960ba}.
 Within a  Reggeon Field Theory (RFT) framework,  a QCD based description of diffraction has been applied to 
 cosmic rays by Ostapchenko\cite{Ostapchenko:2010gt} and in the simulation program QGSJET \cite{Ostapchenko:2010vb}. 
 The formalism uses   a multichannel decomposition, and the physics contents include description of non-perturbative effects, 
 such as   gluon saturation, and semi-hard interactions. We shall see in the section dedicated to the elastic differential 
 cross-section  how other authors have introduced diffraction within Regge field theory, in particular Khoze, Martin and Ryskin 
 and Gotsman, Levin and Maor, also using a Good and Walker (GW) formalism with  a Regge model  for the high mass 
 diffraction, through triple Pomeron interactions.
   
Another  approach is found in the work by   by Lipari and Lusignoli  (LL) \cite{Lipari:2009rm},  who have combined the mini-jet approach with the 
GW description of diffractive states.  The approach by LL is  based on the mini-jet description, 
which the authors consider most useful to implement in simulation programs, and on a continuous distribution of diffractive states, 
all contributing to the total diffraction cross-section. Their  work on diffraction is described in the part of our review dedicated to 
the elastic cross-section.

Later, in \ref{sss:EU}, we shall discuss again this point, following a recent analysis by Engel and Ulrich \cite{Engel:2012pa}.

\subsection{Modeling the cosmic ray flux and energy distribution of particles} 
\label{ss:cosmic-yogi}
Of course, a fundamental problem -still unsolved- concerns the origin of the primary cosmic ray flux, specially at high energies, of interest 
for this review. Related issues concern the composition and the energy distribution of the cosmic ray constituents. As some progress 
has been made in this regard, in the following we shall briefly review it. 

\subsubsection{Power law flux and critical indices of cosmic radiation}\label{sss:critical}
As stated at the beginning of this chapter, starting with Heisenberg, many physicists including Landau\cite{Landau:1940l} 
and Fermi\cite{Fermi:1949ff} devoted much time and effort to understand the observed isotropy and a stable power law energy 
spectrum of the cosmic radiation flux. Presently, it is known experimentally
that \cite{Hagiwara:2002fs} the energy distribution law of cosmic ray nuclei in the energy range 
\begin{math}  5 {\rm \ GeV} < E < 100 {\rm \ TeV} \end{math} obtained via the differential flux per unit time per unit area per 
steradian per unit energy obeys
\begin{equation}
\big{[} \frac{d^4 N}{dt dA d\Omega dE}\big{]} \approx 
\frac{(1.8\ {\rm nucleons})}{\rm sec\ cm^2\ sr\ GeV}
\left(\frac{\rm 1\ GeV}{E}\right)^\alpha 
\label{f0}
\end{equation}
wherein the experimental critical index \begin{math} \alpha \approx 2.7  \end{math}. At the ``knee'' of the distribution, i.e. at energy 
\begin{math} E\sim 1\ {\rm PeV} \end{math}, there is a shift in the critical index to the value \begin{math} \alpha \approx 3.1  \end{math}. 

In a recent series of papers \cite{Widom:2014bua},\cite{Widom:2014gza},\cite{Widom:2014xoa}\cite{Swain:2015yoa}
the hypothesis has been made that cosmic rays are emitted from the surfaces of astronomical objects (such as neutron stars) by
a process of evaporation from an internal nuclear liquid to a dilute external gas which constitutes
the ÒvacuumÓ. On this basis, an inverse power in the energy distribution with a power law
exponent of 2.701178 has been obtained in excellent agreement with the experimental value of 2.7. 

The heat of nuclear
matter evaporation via the entropy allows for the computation of the exponent. The evaporation
model employed is based on the entropy considerations of Landau and Fermi that have been applied
to the liquid drop model of evaporation in a heavy nucleus excited by a collision. This model
provides a new means of obtaining power law distributions for cosmic ray energy distributions and,
remarkably, an actual value for the exponent which is in agreement with experiment and explains the
otherwise puzzling smoothness of the cosmic ray energy distribution over a wide range of energies
without discontinuities due to contributions from different sources required by current models. The argument runs as follows.

\subsubsection{Evaporation of fluid particles}
The heat capacity $c$ per nucleon of a Landau-Fermi liquid drop at a non-relativistic low
temperature T is given by
\be
\label{f1}
c = \frac{dE}{dT}= T \frac{ds}{dT} = \gamma T\ {\rm as}\ T \to 0.
\ee
Eq.(\ref{f1}) implies an excitation energy $E = (\gamma/2)T^2$ 
and an entropy $\Delta s = \gamma\ T$ so that
\be
\label{f2}
\Delta s(E) = \sqrt{(2 \gamma E)} = k_B \sqrt{\frac{E}{E_o}}.
\ee
Consider the evaporation of nucleons from such a  droplet
excited say by an external collision. Given the entropy per nucleon $\Delta s$ in
the excited state, the heat of evaporation $q_{vaporization} = T (\Delta s)$, determines
the energy distribution of vaporized nucleons through the activation probability
\be
\label{f3}
P(E) = e^{- \Delta s(E)/k_B} = e^{-\sqrt{E/E_o}},
\ee
using Eq.(\ref{f2}). We now turn to relativistic cosmic ray particle production through evaporation.

\subsubsection{Cosmic ray particle production}
The sources of cosmic rays here are the evaporating stellar winds from gravitationally 
collapsing stellar (such as neutron star) surfaces considered as a big nuclear droplet 
facing a very dilute gas, i.e. the ÒvacuumÓ. Neutron stars differ from being simply very large
nuclei in that most of their binding is gravitational rather than nuclear, but the droplet model of 
large nuclei should still offer a good description of nuclear matter near the surface where it 
can evaporate.

The quantum hadronic dynamical models of nuclear liquids have been a central theoretical feature of nuclear
matter\cite{JDW}. It is basically a collective Boson theory with condensed spin zero bosons (alpha nuclei)
and spin one bosons (deuteron nuclei) embedded about equally in the bulk liquid. In the very high energy
limit, the critical exponent $\alpha$ occurring in Eq.(\ref{f0}) in this model are computed as follows
\cite{Widom:2014bua,Widom:2014gza,Widom:2014xoa}.

\subsubsection{The critical exponent for classical and quantum particles}
The density of states per unit energy per unit volume for ultra-relativistic particles is proportional to the
square of the energy. The mean energy per particle in an ideal gas of particles obeying classical or 
quantum statistics can be succinctly described using, for classical (Boltzmann) statistics $\eta = 0$  
whereas for quantum statistics, $\eta = 1$ for bosons and $\eta = -1$ for fermions, as:
\bea
\label{f4}
E_\eta = \frac{\int_o^\infty [\frac{\epsilon^3 d\epsilon}{e^{\epsilon/k_B T} - \eta}]}
{\int_o^\infty [\frac{\epsilon^2 d\epsilon}{e^{\epsilon/k_B T} - \eta}]}
= \alpha_\eta k_B T\nonumber\\
\eta = 0; \Longrightarrow\ \alpha_{Boltzmann} = 3;\nonumber\\ 
\eta = 1; \Longrightarrow\ \alpha_{bosons} = \frac{3 \zeta(4)}{\zeta(3)} \approx\ 2.701178;\nonumber\\
\eta = -1; \Longrightarrow\ \alpha_{fermions} = \frac{7 \alpha_1}{6} \approx\ 3.151374,
\eea
where 
\be
\zeta(s) = \sum_{n=o}^\infty \frac{1}{n^s}, 
\ee
is the Riemann zeta function.

To establish $\alpha$ as a power law exponent when the energy $E = \alpha (k_B T)$ , one computes (i) the entropy as
\be
\label{f5}
E = \alpha k_B T = \alpha k_B \frac{dE}{dS};\ \Longrightarrow\  S = k_B \alpha\ ln(\frac{E}{E_o})
\ee
and (ii) employs the heat of vaporization to compute the evaporation energy spectrum
\be
\label{f6}
e^{-S/k_B} = \big{(}\frac{E_o}{E}\big{)}^\alpha
\ee
as in Eq.(\ref{f0}).
A more detailed interacting quantum field theoretical calculation of ${\alpha}$ power law exponents involves the 
construction of single particle spectral functions in the context of thermal quantum field theory. While they have here
been computed the critical indices for the free Fermi and free Bose field theories, the results are already in quite satisfactory
agreement with experimental cosmic ray power law exponents.

In a recent paper, the AMS Collaboration\cite{Aguilar:2014fea} has reported detailed and extensive data concerning the
distribution in energy of electron and positron cosmic rays. A central result of the experimental work resides in the energy 
regime $30\ GeV < E < 1\ $ TeV, wherein the power law exponent of the energy distribution is measured to be 
$\alpha_{experiment} = 3.17$. In virtue of the Fermi statistics obeyed by electrons and positrons, the theoretical value
was predicted as $\alpha_{theory} = 3.151374$ in very good agreement with the AMS data. 

The reason for this remarkable agreement would appear to be due to a Feynman ÒpartonÓ structure for the high energy 
asymptotic tails of the single particle spectral functions. In this case that structure would be described by free non-interacting 
particles thanks to asymptotic freedom in QCD. Following Feynman's physical reasoning and employing dispersion 
relations in a finite temperature many body quantum field theory context, in principle, it is possible to compute rather small 
corrections to the renormalized energy dependent power law exponent $\alpha(E)$ for interacting theories. For further details
about a phase transition and the behaviour around the ``knee'' etc. an interested reader may consult\cite{Swain:2015yoa} 
and  references therein.

\subsection{Cosmic ray results after start of the LHC}\label{ss:afterLHC}

Since  2011, accelerator data for  $\sigma^{pp}_{tot}$ at  LHC   c.m. energy of $\sqrt{s}\ge 7$  TeV have provided 
new accurate information on the high energy behavior of the total \pp \  cross-section. The question of whether one 
is now  reaching  a region of saturation of the Froissart bound was posed again. At around the same time,
a new generation of cosmic ray experiments, which  probes ultra-high cosmic ray energies, started to provide data. 
In 2012 the measurement of $\sigma_{p-air}$ was released  by the AUGER Collaboration 
\cite{Collaboration:2012wt}  for an equivalent   cm. energy   $\sqrt{s}_{pp}=57\pm 0.3 \ (stat.)
 \pm 6 \ (sys.)$ TeV. The result  depends on the simulation program, in particular models of hadronic interactions, 
 as described in \cite{Collaboration:2012wt,Ebr:2012dm}, and the model used for extracting  the proton-air production cross-section.  
The spread of  results is shown in Fig.~\ref{fig:AugerFig2} from \cite{Collaboration:2012wt}.
\begin{figure}
\hspace{-1.46cm}
\resizebox{0.7\textwidth}{!}{
\includegraphics{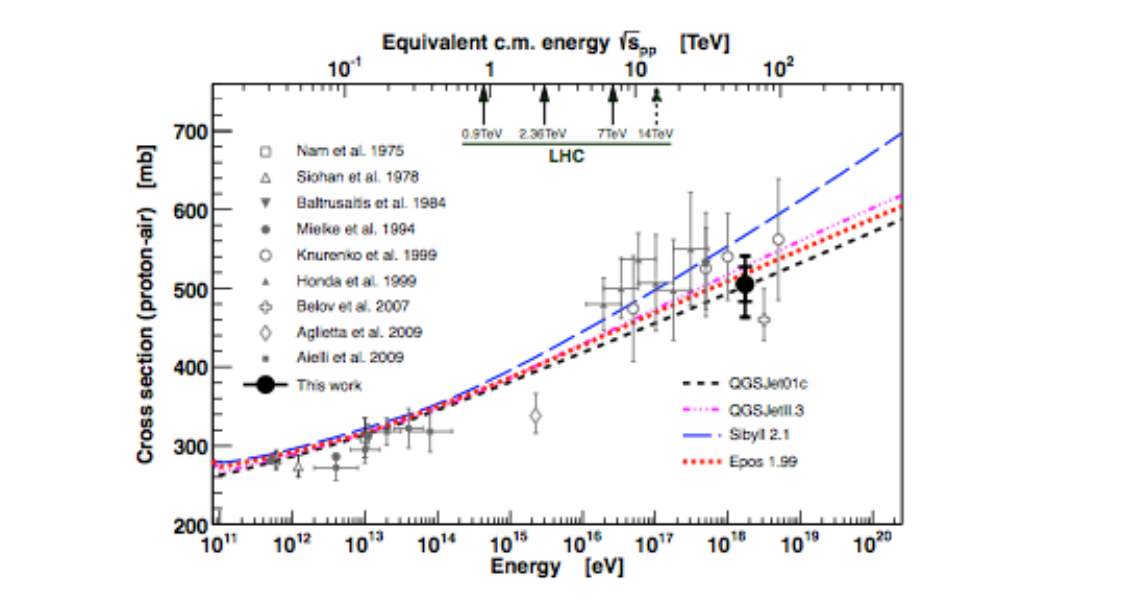}}
\caption{Fig. 2 of \cite{Collaboration:2012wt}, showing the AUGER Collaboration result for $\sigma^{prod}_{p-air}$ 
in comparison with other cosmic ray results, with references as indicated in the figure. Various dotted or 
dashed curves indicate different models used to extract the data through different simulation programs as indicated. Reprinted
 from \cite{Collaboration:2012wt}, \copyright 2012 by the American Physical Society.}
\label{fig:AugerFig2}
\end{figure} 

One important difference between  AUGER result and previous results lies in the assumed primary composition:   
all the measurements in the highest energy group, HiRes, Fly's Eye, Yakutsk and Akeno, assume pure proton composition, 
whereas AUGER opts for a 25\% Helium component. 

Further uncertainties lie in 
how diffraction is  taken into account in the calculation. 
In \cite{Ostapchenko:2014mna}, one can find a 
 recent discussion of  the tension between LHC results  on single and double diffraction as reported by TOTEM, 
 ATLAS and CMS collaborations, and  their impact on the cosmic ray results.
  
 Averaging the result between the different hadronic models leads to
 \be
 \sigma^{prod}_{p-air}=[505 \pm 22 (stat) ^{+28}_{-36}(sys)] mb
 \ee 
at a center of mass energy of $ (57\pm 6) $ TeV. The  correlation between the parameters of the Glauber model 
which converts $p-air$ to $pp$,  $B(s)$ and $\sigma_{inel}(proton-proton)$,  is  shown in Fig.~\ref{fig:AugerFig3}, 
which includes  a comparison with accelerator data, at their respective  energies. The hatched area in the figure 
corresponds to the unitarity limit imposed by the relation between the total, the inelastic and the elastic cross-section,
derived as follows :
\be
\label{Au2}
B \geq [\frac{\sigma_{in}}{4 \pi}].
\ee  
The above inequality may more usefully written in terms of their commonly used units as
\cite{Collaboration:2012wt} 
\be
\label{Au3}
B GeV^2 \geq [\frac{\sigma_{in}}{mb}] [\frac{mb GeV^2}{4 \pi}] \gtrsim [\frac{1}{5}] (\frac{\sigma_{in}}{mb}).
\ee 
[A heuristic derivation of Eq.(\ref{Au3}) proceeds as follows. Assume that there exists an effective B so that
the elastic differential cross-section can be approximated as
\be
\frac{d\sigma}{dt} \approx\ \frac{\sigma_{tot}^2}{16 \pi} e^{Bt}
\label{h1a}
\ee
so that
\be
\label{h2a}
\sigma_{el} = \frac{\sigma_{tot}^2}{16 B \pi}. 
\ee
Since $\sigma_{el} = \sigma_{tot} - \sigma_{in}$, Eq.(\ref{h2a}) may be written as
\be
\label{h3}
B = \frac{\sigma_{tot}^2}{16 \pi (\sigma_{tot} - \sigma_{in})}. 
\ee
Let $x = \sigma_{in}/\sigma_{tot}$, so that Eq.(\ref{h3}) reads
\be
\label{h4}
\frac{4 \pi B}{\sigma_{in}} = \frac{1}{4 x(1 -x)} \ge 1, 
\ee
from which Eq.(\ref{Au3}) follows.] 

The inequality Eq.(\ref{Au3}) is mildly stringent. For example, at LHC [7\ TeV], the left side is $\sim 20$
whereas the right side is $\sim 15$. Incidentally, the lower limit is reached only in the black disk limit
when $\sigma_{in} = \sigma_{el} = \sigma_{tot}/2$. This provides yet another evidence that we are still
nowhere near the black disk limit. But this would be discussed in much more detail in our section
on the elastic cross-section.

Finally, the extraction of the $pp$ total and inelastic cross-sections for $\sqrt{s_{pp}} \simeq 57$ TeV,  leads to the 
quoted results for $pp$ scattering:  
 \begin{align}
 \sigma^{inel}_{pp}&=&[ 92\pm 7 (stat)^{+9}_{-11}sys)\pm 7 (Glauber)] mb \ \ \ \label{eq:auger-inel}\\
 \sigma^{tot}_{pp}&=&[ 133 \pm 13(stat) ^{+17}_{-20}(sys)\pm16 (Glauber)] mb  \label{eq:auger-tot}
 \end{align}
 There is a strong warning in the paper, that the error from the application of the Glauber model may actually 
 be larger than what is quoted here. 
 It is also noted that this error is smaller for the inelastic cross-section, and this can be accounted for by 
 the inelastic cross-section being less dependent on the $B(s)$ parameter than the total.
 We shall return later to the question  of how to estimate the 
 needed inelastic cross-section for cosmic rays. 
\begin{figure}
\resizebox{0.5\textwidth}{!}{
\includegraphics{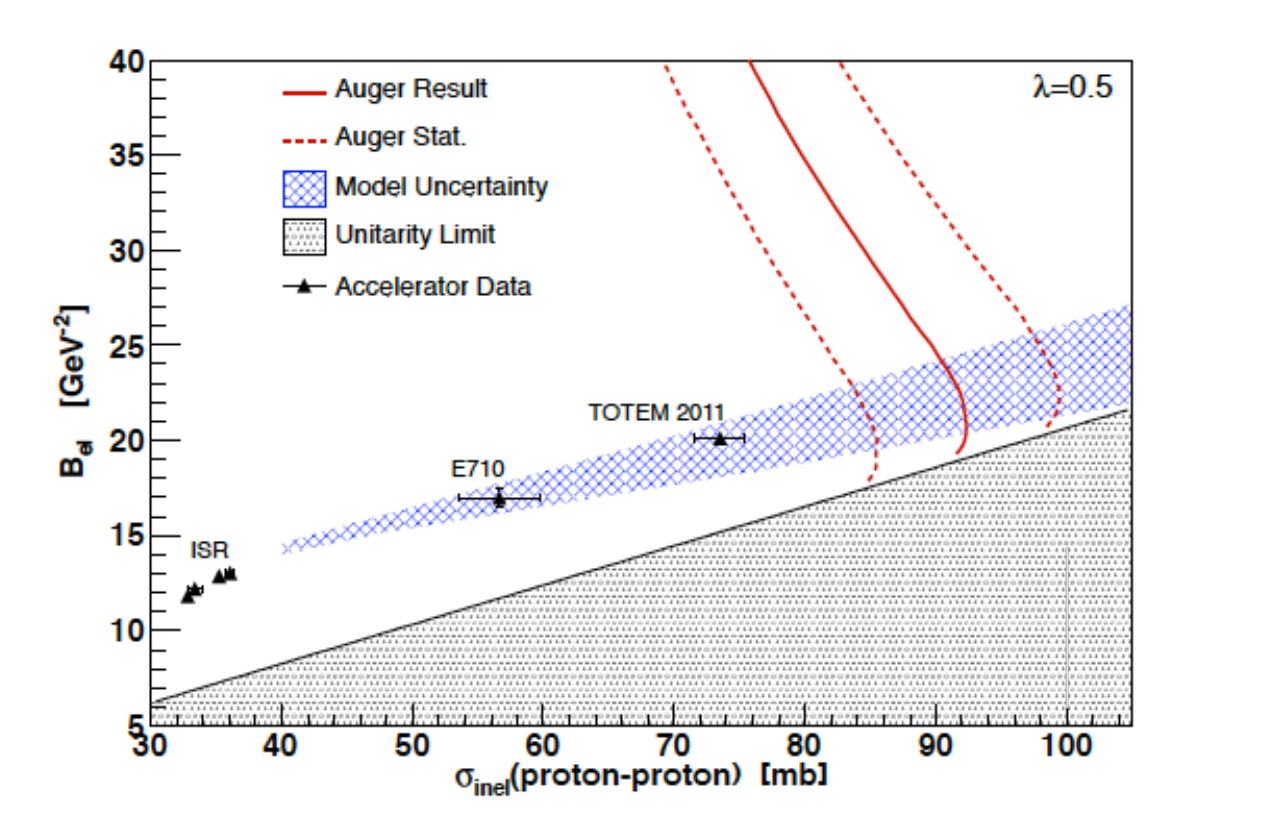}}
\caption{Fig. 3 of \cite{Collaboration:2012wt}, showing the 
AUGER Collaboration result for 
$\sigma^{prod}_{p-air}$ 
 in the $\{B(s),\sigma^{inel}_{pp}(s) \}$ plane.
The  unitarity limit is indicated, as is the uncertainty due to modeling of the $pp$ cross-section.  Reprinted from
 \cite{Collaboration:2012wt}, \copyright (2012) by the American Physical Society.
}
\label{fig:AugerFig3}
\end{figure}
We notice that, very recently, the AUGER collaboration has released  results \cite{Aab:2015bza}  estimated 
in the two energy intervals in $\log(E_{Lab}/eV)$  from 17.8 to 18 and from 18 to 18.5. The corresponding 
values for $\sigma^{p-air}$ are within the errors of the 2012 measurements, the central values lying in a 
curve lower than the one drawn across the central 2012 reported value. These results are shown in the 
right hand plot of Fig.~\ref{fig:cosmic1}, in the context of a mini-jet model whose results are discussed in 
\ref{sss:ourcosmics}.

In the next subsection we shall summarize the method used by the AUGER collaboration to extract the proposed 
values for $ \sigma^{tot}_{pp}$ in \cite{Collaboration:2012wt}.

\subsubsection{A recent analysis of Glauber theory with inelastic scattering \label{sss:EU}}
The details of the actual inputs used in the various Montecarlo simulations used by the AUGER Collaboration to 
extract $\sigma^{prod}_{p-air}$ and thence \sout{to} $pp$ total and inelastic cross-sections can be found in an 
internal report by Engel and Ulrich \cite{Engel:2012pa}. 
Here we shall summarize the salient aspects of this analysis
that includes inelastic screening through a two channel Good-Walker approach in the Glauber theory. 

Considering just two states: $|p>$ and $|p^*>$, where the first is the proton and $p^*$ is an effective state standing
for all inelastic states. A coupling parameter $\lambda$ is introduced so that the 1-channel elastic amplitude 
$\Gamma_{pp}$ becomes a $2 \times 2$ matrix:  
\begin{eqnarray}
\label{EU1}
 |p> = \big{(}\begin{matrix}
 1\\
 0\\
 \end{matrix}\big{)};\  |p^*> = \big{(}\begin{matrix}
 0\\
 1\\
 \end{matrix} \big{)} \nonumber\\
  \Longrightarrow\ \tilde{\Gamma}_{pp} = \big{(} \begin{matrix} 
  1 & \lambda\\
  \lambda & 1 \\
 \end{matrix} )  \Gamma_{pp}          
\end{eqnarray}
The elastic impact parameter amplitude for a hadron $h$ on a nucleus with $A$ nucleons becomes
\begin{eqnarray}
\Gamma_{hA}({\bf b}; {\bf s}_1......., {\bf s}_A)= \nonumber \\ 
= <p| \tilde{\Gamma}_{hA}({\bf b}; {\bf s}_1......., {\bf s}_A)|p>\nonumber\\
= 1 - <p| \prod_{j=1}^A \big{[} 1 - \tilde{\Gamma}_{hN}({\bf b} - {\bf s}_j) \big{]}|p>                
\end{eqnarray}
After diagonalization, it reads
\bea
\label{EU3}
\Gamma_{hA}({\bf b}; {\bf s}_1......., {\bf s}_A) =\nonumber\\
1 - \frac{1}{2}<p| \prod_{j=1}^A \big{[} 1 -(1 + \lambda)\tilde{\Gamma}_{hN}
({\bf b} - {\bf s}_j) \big{]}|p>\nonumber\\                   
- \frac{1}{2}<p| \prod_{j=1}^A \big{[} 1 -(1 - \lambda)\tilde{\Gamma}_{hN}({\bf b} - {\bf s}_j) \big{]}|p>\nonumber\\
\eea
For the Gaussian profile functions, an analytic closed form expression for 
$\Gamma_{hA}({\bf b}; {\bf s}_1......., {\bf s}_A)$
is obtained and, through it, analytic but somewhat long expressions for total, elastic and quasi-elastic cross-sections
for proton-nucleus are obtained and can be found in \cite{Engel:2012pa}.

The parameter $\lambda^2(s)$ 
is  related to the ratio of $\sigma_{SD}(s)$ to $\sigma_{elastic}(s)$ and hence to available accelerator
data on single diffractive dissociation(SD) and elastic  pp-scattering. 
It can be parametrized and extrapolated to higher energies as needed. Thus, accelerator data can in principle 
determine (modulo extrapolation) $\lambda^2(s)$. In practice, empirical
functions such as the following are employed
\bea
\label{EU4}
\sigma_{SD}(s) = \Big{[}\frac{s\ log [10^3\ GeV^{-2} s]}{4s + 400\ GeV^2}\Big{]} (mb);\nonumber\\ 
{\rm valid\ for}\ \zeta_{max} = \frac{M^2_{D,max}}{s} < 0.05.
\eea   
Here $M_{D,max}$ is the maximum invariant mass of the diffractive system expected to be coherently
produced by a nucleon. Typical values $\zeta_{max} = (0.05 \div 0.15)$ are considered to describe SD. The choice of $\zeta_{max}$
controls the scale of $\sigma_{SD}$ and thus $\lambda$. At very high energies $\lambda^2$ should decrease since
we expect $\sigma_{SD} \sim ln(s)$ and $\sigma_{elastic} \sim ln^2(s)$. These authors find that for the AUGER
measurement at $\sqrt{s} = 57\ TeV$, $\lambda(\sqrt{s} = 57\ TeV) = (0.35 \div 0.65)$. This is then folded into the errors
associated with the various cross-section estimates.

The conversion of the proton-air to proton-proton cross-section proceeds along the lines discussed earlier, i.e., 
plots of $B$ vs. $\sigma_{inel}$ are used with constant values of p-air production cross-sections drawn and the intersection giving the
$pp$ inelastic cross-section at that energy. With $\lambda = 0.5$, these authors deduce that at $\sqrt{s} = 57$ TeV:
\bea
\label{EU5}
\sigma_{pp}^{inel}  = [92\pm 7(stat)^{+9}_{-11}(sys) \pm1(slope)\pm 3(\lambda)]\ mb\nonumber\\
\sigma_{pp}^{tot}  = [133\pm 13(stat) ^{+17}_{-20} (sys) \pm13(slope) \pm 6(\lambda)]\ mb,\nonumber\\
\eea   
also shown in Eqs.(\ref{eq:auger-inel}, \ref{eq:auger-tot})
\subsubsection{
The Telescope-Array measurement at 95 TeV c.m. energy }
In 2015, the Telescope Array (TA) collaboration has presented a measurement of the $p-air$ cross-section  at the  never attained 
before c.m. energy  of $\sqrt{s}=95$ TeV.
The method used is that of  {\it K-factor}, in which the attenuation length, and hence the $p-air $ cross-section, is proportional 
to the slope of the tail of the $X_{max}$ distribution. 
As seen before,
the factor K depends on the model used for the shower evolution. In \cite{Abbasi:2015fdr}, averaging over different models, 
a value $K=1.2$ is obtained, with an uncertainty (model dependence) of  $\sim 3\ \%$.  Including a systematic error from 
the uncertainty on the primary cosmic ray composition, the procedure therein described leads to a value
\be
\sigma_{p-air}^{inel}=(567\pm 70.5[Stat]^{+29}_{-25}[Sys]) mb 
\ee
at an energy of $10^{18.68}\ eV$. The proton-air cross-section from this measurement appears to 
lie higher than  the more recent AUGER values, but it is consistent with  the observed trend within all the errors. 
Some of the difference could be ascribed to assuming a different primary composition, a question still not fully resolved. 
Results from a higher statistical sample are expected shortly. 

The TA collaboration has also presented a value for   the total $pp$ cross-section, following the procedures  from 
\cite{Gaisser:1986haa,Engel:1998pw}, which we have described in \ref{sss:Gaisser1986} and \ref{sss:Cosmicuncertainties}. 
Using Glauber theory and the BHS QCD inspired fit \cite{Block:1999ub}, they propose:
\be
\sigma^{tot}_{pp}=170^{+48}_{-44}[Stat.]^{+19}_{-17}[Sys] mb
\ee
While consistent  within the errors with the trend shown by the lower energy measurement by the AUGER collaboration, 
the above value for the $pp$ total cross-section is thus higher 
than the value extracted 
by AUGER. In this respect we should notice that the methods used by the two collaborations to pass from  $p-air$ to 
$proton-proton$ are not the same. In the 
next 
subsection, we 
present 
another model 
about how to extract $pp$ cross-section 
from that of $p-air$.

\subsection{ Eikonal models for  inelastic   $p-air$ scattering. 
}
\label{ss:eikonals}
As we have seen, in order to extract information about the basic $pp$ scattering, cosmic ray measurements require  
models to link  the inelastic $p-air$ cross-section to the total and elastic ones. Glauber theory provides such a connection, 
through an eikonal 
formalism in impact parameter space. However,  one-channel eikonals provide an incomplete picture at high energy, 
as has been noticed since a long time. As we shall see in more detail in the sections dedicated to the total and the elastic 
cross-section, a single channel eikonal model with an approximately real profile function is unable to clearly discriminate 
between elastic and inelastic processes. As a result, various techniques have been developed: we have described in the preceding
a two channel model 
as that due to Engel and Ulrich, and  recalled the analytical model with QCD inspired input by Bloch and Halzen. 
The present uncertainties and difficulties with these extractions lead to 
large errors and hence, in part, to the inability to fully exploit the very high energy data provided 
by the cosmic ray experiments. 

In what follows we shall briefly outline  the results from a multichannel model by  Gotsman, Levin and Maor 
and then show  
corresponding  results 
that can be obtained in a single channel  eikonal model with 
QCD mini-jets and compare them with present data.
\subsubsection{A multichannel model inclusive of diffraction and  triple Pomeron coupling}\label{sss:GLMcosmic}
An example of a QCD model used to extract $p-air$ cross-section can be found  in \cite{Gotsman:2013nya}. This is a 
multi-channel model, to which we shall return in the section dedicated to elastic scattering. It includes diffraction 
contributions and  triple-Pomeron exchanges. The final formula for the inelastic $p-air$ cross-section is given as 
\begin{align}
\sigma_{in}(p+A;Y)=\nonumber\\
\int d^2\vecb[
1-  exp {\large 
(}
 -\{ \sigma^{pp}_{tot}\frac{S_A(b)}{1+\tilde{g}G_{3{\cal P}}G_{enh}(Y)S_A(b)} +\nonumber\\
- (\sigma^{pp}_{el}+\sigma^{pp}_{diff})\frac{S_A(b)}{(1+\tilde{g}G_{3{\cal P}}G_{enh}(Y)S_A(b))^2}    \}     { \large )}
] \label{eq:GLMcosmic}
\end{align}
with $S_A(b)$ the nuclear density, $G_{3{\cal P}}$ the triple Pomeron coupling, $G_{enh}(Y)$ the Green's function of the 
Pomeron exchange, $\tilde{g}$ includes the parameters of the GW diffraction couplings, which are used to determine 
$\sigma_{tot}, \sigma_{el},\ \sigma_{diff}$.

A comparison between data and results from this model is shown in Fig.~\ref{fig:GLMcosmic} from \cite{Gotsman:2013nya}.
\begin{figure}
\resizebox{0.5\textwidth}{!}{
\includegraphics{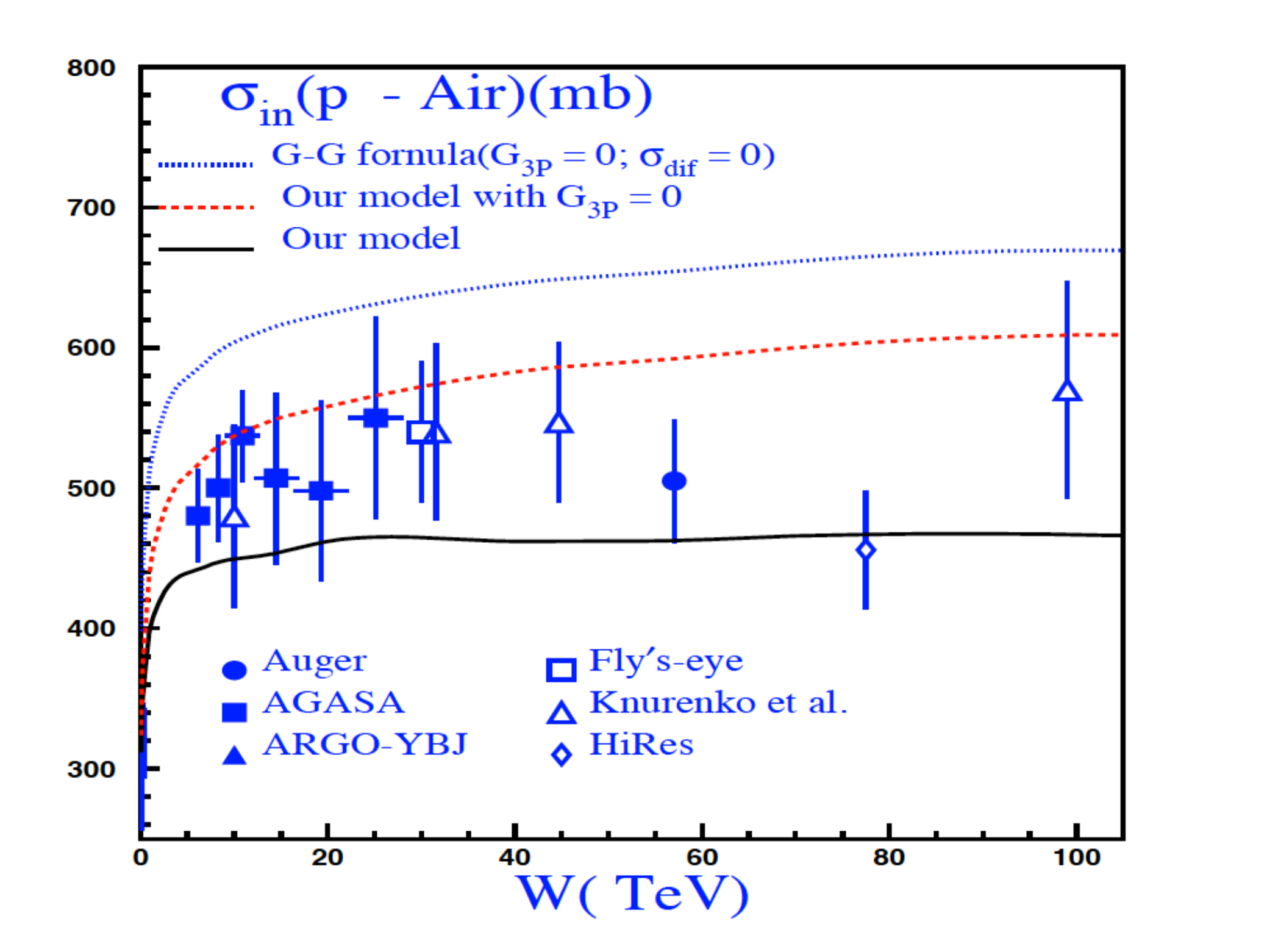}}
\caption{The inelastic $p-air$ cross-section from the two channel model of Gotsman, Maor and Levin, compared 
with cosmic ray data, from \cite{Gotsman:2013nya}.  Reprinted with permission from \cite{Gotsman:2013nya}, \copyright (2013) by the American Physical Society. }
\label{fig:GLMcosmic}
\end{figure}
From Eq. ~(\ref{eq:GLMcosmic}) and the figure, one can notice the following:
\begin{itemize}
\item both recent AUGER and Telescope Array data (within their large errors) can be described by the model. Shown
are 
two curves: for $G_{3{\cal {P}}}=0 $ or  $G_{3{\cal {P}}}=0.03$;
\item the impact parameter (b-)dependence, over which the eikonal is integrated, includes only the nuclear shape. In the 
exponent, it can be factored out of the QCD part. Namely, there is convolution of the nuclear distribution with the inner nucleon 
structure;
\item the case $G_{3{\cal P}}=0$, reduces Eq.~(\ref{eq:GLMcosmic}) to
\be
\sigma_{in}(p+A;Y)=\nonumber\\
\int d^2\vecb[
1-  exp {\large 
(}
 -S_A(b)\{ \sigma^{pp}_{tot}
- (\sigma^{pp}_{el}+\sigma^{pp}_{diff})
 \}     { \large )}
] 
\ee
\item the curve where also  $\sigma^{pp}_{diff}=0$  lies higher than the data. 
\end{itemize}
Thus, the authors conclude that a small triple Pomeron coupling and $\sigma^{pp}_{diff}\neq 0$ can give a good description 
of data, when a two channel GW formalism is employed to describe total and elastic $pp$ \x s. 

An interested reader may also consult some related work in
\cite{Ostapchenko:2014mna} 
that 
is based on QCD and the Regge picture. 
\subsubsection{A single channel model with QCD mini-jets} 
\label{sss:ourcosmics} 
In a recent work \cite{Fagundes:2014fza}, a single channel eikonal formalism has been proposed
for a somewhat different reconstruction of the quantity
$\sigma_{p-air}^{prod}$ measured in cosmic rays, 
from the underlying $pp$ dynamics. 

The starting point of this approach is the 
realization that in single-channel  
mini-jet models, $\sigma_{elastic}$ includes both {\it purely elastic and correlated-inelastic} processes 
\cite{Achilli:2011sw,Fagundes:2015vba}. 

The model we present here  exploits this observation and 
has the virtue of eliminating
the complicated and model dependent untangling of the elastic and diffractive parts to deduce the needed
inelastic 
non-diffractive
contribution, called here   $\sigma^{pp}_{inel-uncorr}$,  that serves as an input in the Glauber reconstruction of
$\sigma_{p-air}^{prod}$. The argument runs as follows.   

 As Eq.(\ref{eq:sigprod}) makes evident, cosmic rays measure and probe the part of the scattering process that is shorn
 of its elastic and quasi-elastic parts. We may thus identify the needed remainder to be the ``inelastic-uncorrelated''
 part of the cross-section. If such is indeed the case, then $\sigma_{inel}$ computed through a single channel 
 mini-jet eikonal formalism based on an exponentiation of the basic parton-parton scattering would indeed
 correspond to the inelastic-uncorrelated 
 cross-section for $pp$ scattering, as we have discussed in \cite{Achilli:2011sw}. 
 This last point will also be discussed in more detail in Sect.~\ref{sec:models}.

 \begin{figure*}
\centering
\resizebox{1.0\textwidth}{!}{
\includegraphics{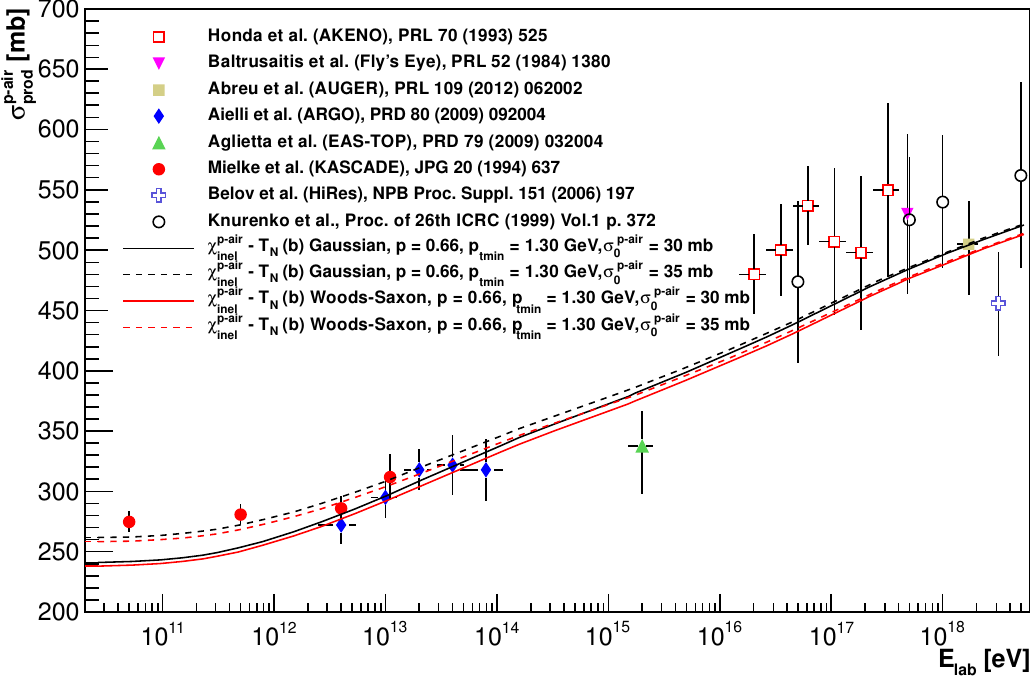}
\includegraphics{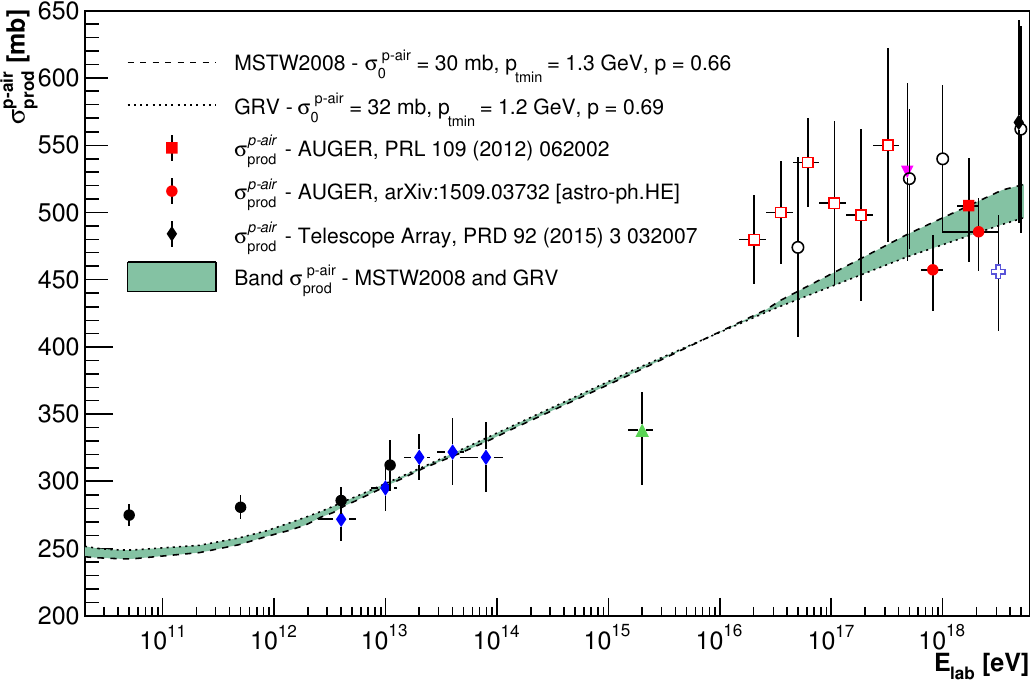}}
\caption{ $p-air$ production cross-section as deduced from the mini-jet model of \cite{Grau:1999em,Godbole:2004kx} and its comparison with experimental 
cosmic ray data. At left we show the model predictions  using MSTW parton densities \cite{Martin:2009iq} and  different nuclear 
density models,  at right using both MSTW and GRV densities \cite{Gluck:1998xa}, and including recent 
AUGER data \cite{Aab:2015bza} as well as those at $\sqrt{s}=95\ GeV$ from the Telescope Array Detector \cite{Abbasi:2015fdr}. Reprinted with permission from  \cite{Fagundes:2014fza}, \copyright 2015 Springer.}  
\label{fig:cosmic1}
\end{figure*}
Under the above hypothesis, the steps relating $pp$ dynamics to the cosmic data in a one channel formalism 
 become rather simple and can be outlined as follows: 
 \begin{itemize}
 \item 1. Neglecting the real part of the scattering amplitude at $t=0$, the same eikonal $\chi_I(s,b)$ can be used to
 describe both $\sigma_{tot}^{pp}(s)$ and $\sigma_{inel-uncorr}(s)$:
 \bea
\sigma_{tot}^{pp}= 4 \pi\int (bdb) [1-e^{-\chi_I(b,s)}]\label{eq:sigtot1}\\
\sigma_{inel-uncorr}^{pp}= 2\pi \int (bdb) [1-e^{-2\chi_I(b,s)}] \label{eq:siginel1}.
\eea
In the mini-jet model of \cite{Grau:1999em,Godbole:2004kx}, 
\be
2\chi_I(b,s)\equiv n^{pp}(b,s)=n^{pp}_{soft} + A(b,s)\sigma^{QCD}_{jet}(p_{tmin},s)
\ee
where the impact parameter function $A(b,s)$ describes the impact parameter space parton distributions in the proton, 
obtained through soft gluon resummation, and  $\sigma^{QCD}_{jet}(p_{tmin},s)$ is calculated through elementary 
parton-parton scattering and library used parton density functions (PDFs), as already discussed in the context of the 
Durand and Pi model in \ref{sss:DurandPi};
\item 2. 
Next, the usual Glauber impact parameter expression is used for the cosmic  ray production cross-section:
\begin{equation}
\sigma_{prod}^{p-air}(E_{lab})= 2 \pi \int (bdb) [1-e^{-n^{p-air}(b,s)}]
\end{equation}
with
\begin{equation}
n^{p-air}(b,s)=T_N(b)\sigma_{inel-uncorr}^{pp}(s)\label{eq:npair}
\end{equation}
where $\sigma_{inel-uncorr}^{pp}(s)$ in Eq.~(\ref{eq:npair}) is obtained from Eqs. ~(\ref{eq:sigtot1}) and (\ref{eq:siginel1}), with the same QCD term, 
but a different parametrization of the low energy part, as also discussed in the Durand and Pi model. 
$T_N(\vecb)$ is the nuclear density, for which the standard gaussian choice is made: 
\begin{eqnarray}
T_{N}(b) = \frac{A}{\pi R_{N}^{2}}\textrm{e}^{-b^{2}/ R_{N}^{2}}\label{eq:nucl_prof}, 
\end{eqnarray}
properly normalized to
\begin{eqnarray}
\int d^{2} \textbf{b} T_{N}(b) = A.
\end{eqnarray}
The parameters used in the profile (\ref{eq:nucl_prof}), namely the average mass number of 
an ``air" nucleus, $A$, and the nuclear radius, $R_{N}$, are 
again standard: 
\begin{eqnarray}
A = 14.5 ,\quad  R_{N} = (1.1 fermi) A^{1/3}.
\end{eqnarray} 
 \end{itemize}
 The authors of \cite{Fagundes:2014fza} have used the above in an eikonal model for the elastic
 amplitude based on gluon resummation (with a singular $\alpha_s$), 
 which we label BN from the Bloch and Nordsieck classical theorem on the infrared catastrophe in QED.
 The BN model
 and the choice of $\chi(b,s)$ are discussed in the elastic cross-section part of this review. 
 
 We show here 
 only the final model  results and their comparison with experimental data in Fig.(\ref{fig:cosmic1}). 
In this figure, the left hand panel shows a comparison of data with the results from two different nuclear density models, 
the gaussian distribution and a Wood-Saxon type potential, as in \cite{Kohara:2014cra}, as well as with  different  low energy 
contributions. The right hand panel shows a comparison of $p-air$ data, including the most recent AUGER  and Telescope 
Array results just discussed, with two different  parton densities used in the BN model, as indicated. The band highlights  
the uncertainty due to the $low-x$ behavior of these two PDFs parametrizations. 

We notice that in this model, the impact parameter distribution of partons is not folded in with the nuclear density distribution, 
rather it is factored out, just as in the model described in \ref{sss:GLMcosmic}. This assumption may  not be valid at low 
energies. In the above mini-jet model this uncertainty is buried in the low energy contribution $n^{p-air}_{soft}$, but it is 
likely to be correct in the very high energies region now being accessed. Thus the model differs from the usual Glauber 
applications. 
  
 The AUGER data at $\sqrt{s} \approx\ 57\ TeV$ are very well reproduced by this model. It is reasonable
 to conclude that a single channel eikonal model that describes well $\sigma_{tot}^{pp}$ indeed generates
 a correct uncorrelated-inelastic $pp$-cross-section. The latter in turn provides the proper input to generate
 $\sigma_{p-air}^{prod}$ as observed in cosmic rays. 
 Work is still in progress to understand the implications of the above description 
 on the construction of multi-channel models.
\subsection{Conclusions}
In this section, we have presented an overview of how, over the past 60 years, cosmic ray experiments have 
provided much needed information (albeit with large errors). They have helped guide the theorists towards more realistic
particle physics models
and make better predictions for their asymptotic betaviour. 
However, many uncertainties still affect the extraction of the more fundamental \pp \ \x \ from the cosmic ray experiments, some of them related to the Glauber formalism and the modeling of quasi-elastic contributions, and others  pertaining to  diffraction in  \pp\ scattering and its relation to $p-air $ processes.  Other uncertainties depend on  understanding the actual composition  of cosmic rays and the relation between   the measured high-energy power law distributions of cosmic electron/positron, proton  and nuclei  and  the origin of high energy cosmic rays.

\section{The measurement of $\sigtot$ before the LHC: description of experiments and their results}
\label{sec:measure}
In this chapter, we give a brief account of crucial hadronic cross-section experiments at particle accelerators, beginning in the 1950's up to the Tevatron, 
including how the measurements  were done at each machine. Relevant experimental data with figures, plots and tables
shall be shown. Whenever it appears useful, there would be a discussion
of the error estimates and the difference between results from different experimental groups.\\ 
The focus of these experiments has been to determine the following 4 basic physical quantities:
\begin{itemize}
\item (i)  total cross-section $\sigtot$; 
\item (ii) elastic cross-section $\sigma_{elastic}$;
\item (iii) slope of the forward elastic amplitude $B$; 
\item(iv) $\rho$-parameter, that is the ratio of the real to the imaginary part of the forward elastic amplitude. 
\end{itemize}

 Before the advent of particle accelerators in the 1950's,  cosmic ray experiments were the only source 
 for  measurements of  total proton  cross-sections.  The situation changed  with  the  first measurements 
 at particle accelerators, which took place   at  fixed target machines, see for instance  \cite{Block:1956zz}.  
 These  measurements were  followed in the 1960's by extensive ones, again at fixed target machines, 
 at the CERN ProtoSynchrotron \cite{Bellettini:1965ab}, at Brookhaven National Laboratory 
 \cite{PhysRev.138.B913}  and in Serpukov \cite{Beznogikh:1969xf}. These earlier measurements
showed decreasing total cross-sections, more so in the case of \pbarp, only slightly in the case of proton-proton. 
These results were in agreement with the picture of scattering as dominated at small momentum transfer by  
exchange of Regge trajectories (leading to a decrease) and multi peripheral production, resulting in the exchange 
of  the Pomeron trajectory, with the quantum numbers of the vacuum, and intercept $\alpha_P(0)=1$. Things 
changed dramatically when the CERN Intersecting Storage Ring (ISR) started operating in 1971. We show 
the scheme of operation of the ISR in Fig.~\ref{fig:ISRgiacomelli} from \cite{Giacomelli:1979nu}.
\begin{figure} 
\begin{center}
\resizebox{0.5\textwidth}{!}{%
  \includegraphics{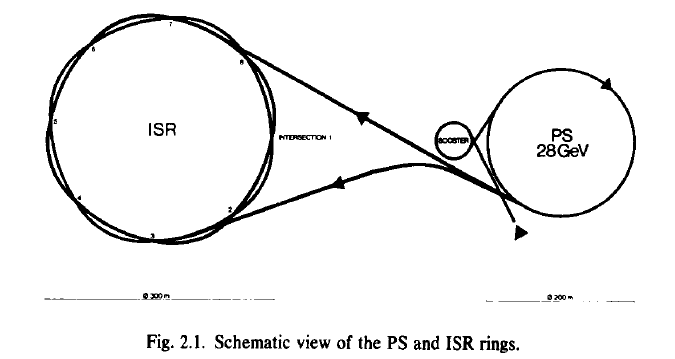}}
\caption{A schematic view of the ISR set up from \cite{Giacomelli:1979nu}. Reprinted from  \cite{Giacomelli:1979nu},
\copyright (1979) with permission  by Elsevier}
\label{fig:ISRgiacomelli}
\end{center}
\end{figure}
Cross-sections were seen to rise. Since then,  accelerator measurements for  the total \x \ for \pp \ 
as well as for \pbarp \ have been performed only at colliders and the \x \ has  continued to increase, 
with the latest measurement released  at  c.m. energies  of $\sqrt{s}=7$ and $8 \ TeV$ at the Large 
Hadron Collider (LHC7 and LHC8) by the TOTEM experiment \cite{Antchev:2013paa}, with predictions 
and measurement in  general  agreement with latest cosmic ray experiments \cite{Collaboration:2012wt}.

The spirit of this section is to show how many different  experiments took place from early 1970's until the 
end of the century, and  established beyond doubt the rising behavior of the total cross-section while  the 
center of mass increased by  almost a factor 100   from the first ISR experiment, at 23 GeV c.m.to the 
1800 GeV at the Tevatron.  

 This section is structured as follows:
\begin{itemize}
\item the description of measurements through fixed target experiments  is given  in \ref{ss:Fixedtarget}, 
\item the measurements at the CERN Intersecting Storage Rings (ISR) are  discussed in   \ref{ss:ISR}, 
including a discussion of measurement of  the $\rho$ parameter and the radiative corrections needed for 
its determination in \ref{sss:rhoRadCorr}, with description of the various methods to measure the total 
cross-section in \ref{sss:fourISRmdethods} and ISR final results in \ref{sss:ISRfinal},
\item experiments confirming  the rise of the total cross-section at the CERN $Sp\bar{p}S$ are 
presented in \ref{ss:spbarps}, with results from UA1 in \ref{sss:UA1}, which includes  a comment on the energy dependence of the slope parameter,
 UA4 and UA4/UA2  results are in 
\ref{sss:UA4}, UA5 and the ramping run in \ref{sss:UA5},
\item measurements at the FermiLab TeVatron  are described in \ref{ss:TeVatron} with results 
from E710, CDF and E811 in \ref{sss:Tevatron-measure} and an 
 overview of the Black Disk model, for a long time a very useful and  commonly held model in \ref{sss:blackDisk},
\item further discussion of the $\rho$ parameter is in \ref{sss:rhoTeV}.
 \end{itemize} 


\subsection{Fixed target experiments}\label{ss:Fixedtarget}
Proton-nuclei cross-sections by Bellettini et al. \cite{Bellettini:1966bg} 
were among the first experiments to measure the expected diffraction pattern from the optical model, for elastic scattering of protons on nuclei.
In this experiment, the momenta of the proton before and after the scattering were measured, the recoil of the nucleus was not measured at all. The experimental technique was the same as the one used for the measurement of proton-proton scattering \cite{Bellettini:1965ab}.

 A system of quadrupoles and bending magnets transported a well collimated (almost monochromatic) beam of protons of average momentum $19\ GeV/c$ from the CERN ProtoSynchrotron to the experimental area.  The incident proton beam was defined by scintillation counters $C_{1,2,3}$ while the scattered protons were detected by another set of counters, $C_{4,5}$, placed after the target. An anti-coincidence counter, placed directly in the path of the beam, was used to reduce the background trigger rate from unscattered particles. The position of the incident and scattered protons were measured by sonic spark chambers, $S_{1,2,3,4,5}$. It is clear from the above description why such experiments received their name: the transmission method.
  
A description of the transmission method can be found in \cite{Citron:1966zz}, where the measurement of pion-proton total cross-section between 2 and 7 $GeV/c$ laboratory momentum is described. 
 The total cross-sections for \pp \ and \pbarp \ were measured \cite{PhysRev.138.B913} along with that for $\pi\ p$ and $K\ p$ on both hydrogen and deuterium targets. In this set of experiments
 total cross sections were  measured between 6 and 22 GeV/c at intervals of 2GeV/c
 and the method utilized was that of a conventional good-geometry transmission experiment with scintillation counters subtending various solid angles at targets of liquid H2 and D2. 
 The results  showed a variation of the cross section with momentum, namely  a small but significant decrease in $\sigma_T(pp)$ [and $\sigma_T(pn)$] in the momentum region above 12GeV/c was found. 
 
 The measurement of total cross-section in  these transmission experiments was done essentially by following the initial and final particle paths through a series of (scintillation ) counters placed at   subsequent intervals and covering different solid angle portions.   For each set of counters, at a given solid angle, a transmission factor was defined to take into consideration signals from the various detecting components  and the total cross-section at a given 
momentum transfer value ($t$) 
 was computed from the expression
\begin{equation}
\sigma(t)=(1/N)\ln (T_E/T_F)
\end{equation}
where $N$ was the number of nuclei per $ cm^{2}$ in the target, $T_E$ and $T_F$ the transmission factors for an empty or a full target respectively. Subsequently partial differential cross-sections measured at different $t$-values were fitted either by a polynomial or preferably by an exponential and extrapolated to zero. 

In Figs.~\ref{fig:Citron1966} and \ref{fig:galbraith}, we show  schematic views of the transmission set up for the measurement of the total cross-section from \cite{PhysRev.138.B913,Citron:1966zz}.
\begin{figure} 
\begin{center}
\resizebox{0.5\textwidth}{!}{%
  \includegraphics{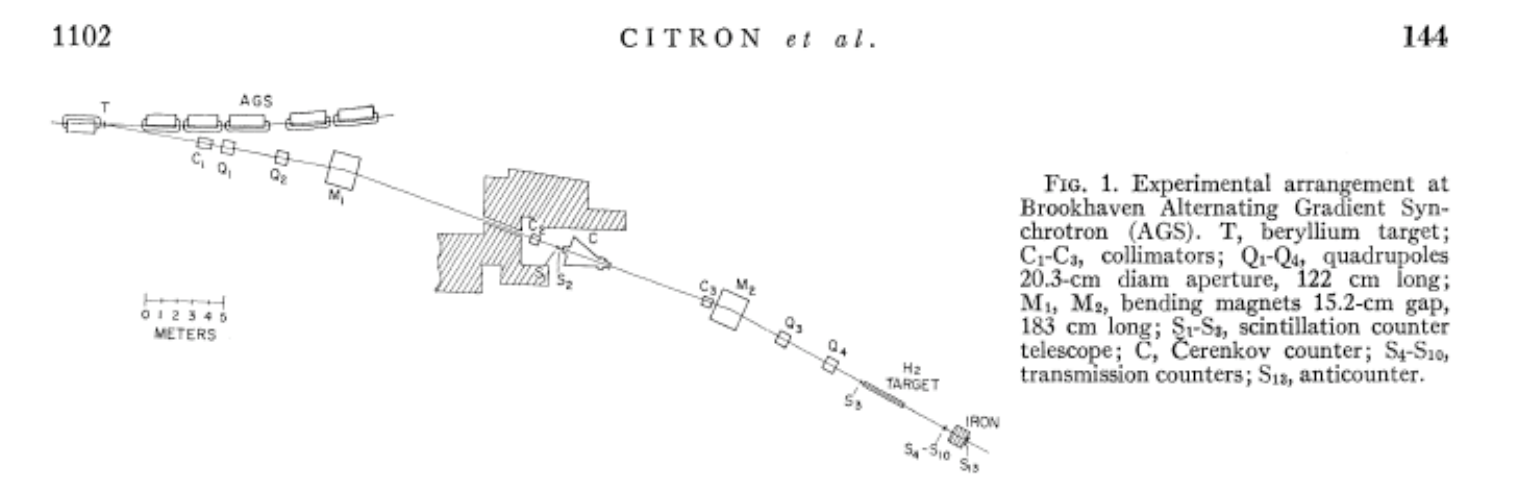}
}
\caption{A schematic view of the transmission type experiment from \protect\cite{Citron:1966zz}.
Reprinted from \cite{Citron:1966zz}, \copyright (1966) by the American Physical Society.}
\label{fig:Citron1966}
\end{center}
\end{figure}

\begin{figure} 
\begin{center}
\resizebox{0.5\textwidth}{!}{%
  \includegraphics{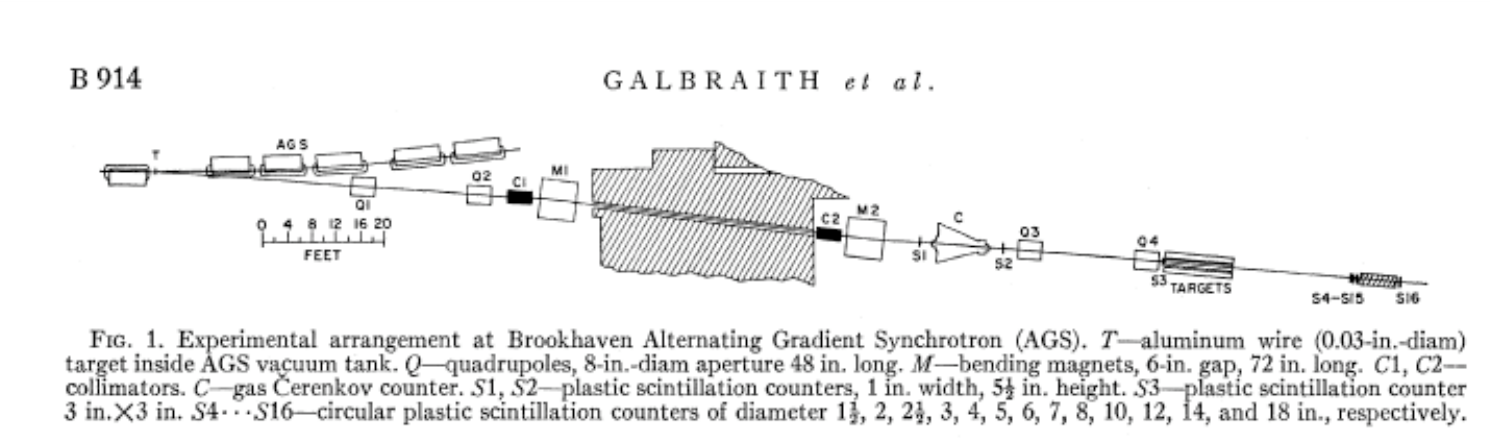}
}
\caption{A schematic view of the transmission type experiment from \protect\cite{PhysRev.138.B913}. 
Reprinted with permission from \cite{PhysRev.138.B913}, \copyright (1965) by the American Physical Society.}
\label{fig:galbraith}
\end{center}
\end{figure}

\subsection{The ISR measurement and the rise of the total cross-section}\label{ss:ISR}
At ISR, in order  to measure physical cross-sections, completely different methods had to be employed.  At least three such methods were used, two of them depending on there so-called luminosity  of the accelerator, one independent. Luminosity is the key parameter for cross-section measurements at intersecting storage rings and is defined as the proportionality factor between  the number of interactions taking place at each beam crossing, R,  and the particle cross-section $\sigma $ to be measured, i.e.  $R=L \sigma$.  The concept of luminosity had been  introduced in the mid-fifties, when storage rings had started being discussed in the community. The name itself is probably due to Bruno Touschek, who used it when proposing the construction of the first electron-positron colliding beam accelerator, AdA, in 1960 \cite{Bonolis:2011wa} and  the process $e^+e^-\rightarrow \gamma \gamma$ was suggested as  the  monitor process for other final states.

At the CERN Intersecting Storage Rings (ISR), $\sigtot$ and/or  $\sigel$ were measured by a number of different experiments, with different methods, in different  $t$-regions.
 A list of all these experiments up to the end of 1978,  can be found in the extensive review of physics at the ISR by Giacomeli and Jacob \cite{Giacomelli:1979nu}. We reproduce information about some of them in Table \ref{tab:isrexp}.
\begin{table*}
\caption{Experiments measuring the total and the elastic \x \ at ISR as of 1977. For complete references see   \cite{Giacomelli:1979nu}, where  C and R refer to Completed and Running experiments, respectively.   }
\label{tab:isrexp}       
\begin{center}
\begin{tabular}{|clclcl}
\hline\noalign{\smallskip}
Observable                      &Experiment                                        & Ref. \\ \hline \hline
\noalign{\smallskip}\hline\noalign{\smallskip}
Elastic scattering   & E601                                                            &\cite{Amaldi:1971kt,Amaldi:1972uw,Amaldi:1973yv}  \\
at small angle         & CERN-Rome                                             & \\ \hline
Elastic scattering   &E602                                                             & \cite{Holder:1971bx,Holder:1971by,Holder:1971pa,Barbiellini:1972ua,Baksay:1978sg} \\
                                  &Aachen-CERN-Genoa-Harvard-Torino& \\ \hline
  $\sigtot$                       &R801-Pisa-Stony Brook  & \cite{Amendolia:1973yw} \\
                                        &CERN-Pisa-Rome-Stony Brook                     &\cite{Amaldi:1976zi,Amaldi:1978vc} \\ \hline
 Small angle scattering&E805                                                        &  \cite{Amaldi:1976yf} \\
                                         & CERN-Rome                                          &  \\ \hline\hline
\noalign{\smallskip}
\end{tabular}
\end{center}
\end{table*}
While early measurements of the  elastic scattering at $\sqrt{s}=30$ and $45\ GeV$ were  not conclusive, 
 a combination of various methods allowed to definitely establish the rise of the $\sigtot$,
 as clearly shown in  Fig.~\ref{fig:ISR} from  
\cite{Amaldi:1973yv}, where  the rise appears  beyond doubt.
\begin{figure} 
\begin{center}
\resizebox{0.5\textwidth}{!}{%
  \includegraphics{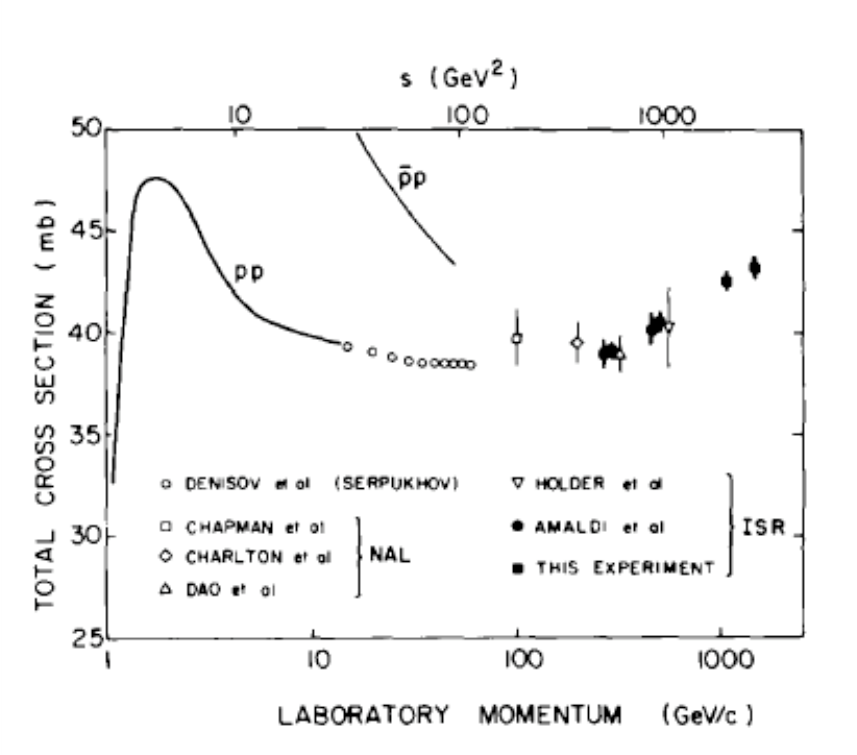}
}
\caption{From Amaldi et al. measurement \cite{Amaldi:1973yv},  one can clearly see the rise of the total cross-section.
Reprinted   from \cite{Amaldi:1973yv}, \copyright (1973) with permission from Elsevier.}
\label{fig:ISR}       
\end{center}
\end{figure}

Unlike fixed target experiments which used the transmission method to measure the total cross-section, storage ring experiments such as those performed at the ISR, needed either a measurement of the total rate, and hence an accurate estimate of the luminosity, or a measurement of the differential elastic cross-section and its extrapolation to the optical point, namely $t=0$. Such measurements are  described in \cite{Amaldi:1972uw} where first results from the operation of the ISR at beam momenta of  $11.8\ GeV$ and $15.4\ GeV$ are reported along with   values for $\sigtot,\rho $ and $\sigma_{elastic}$. These are shown here in Table ~\ref{tab:amaldi}. 
 \begin{table}
\caption{Results of early measurements at   ISR from \protect\cite{Amaldi:1972uw}.}
\label{tab:amaldi}       
\begin{tabular}{|c|c|c|c|}
\hline\noalign{\smallskip}
ISR   beam         & $\sigtot$            & $\rho$                      &$\sigel$ \\
$\sqrt{s}$  &( mb)                  &                                   &(mb) \\
(GeV)        &                            &                                   &         \\
\noalign{\smallskip}\hline\noalign{\smallskip}
11.8           & 38.9$\pm$ 0.7 & +0.02 $\pm$0.05     &6.7  $\pm$ 0.3\\
15.4           & 40.2$\pm$ 0.8 & +0.03$\pm$0.06      &6.9  $\pm$ 0.4\\
\noalign{\smallskip}\hline
\end{tabular}
\end{table}

\subsubsection{ISR measurements for the total cross-section and  the elastic scattering amplitude}

A luminosity dependent measurement of the total cross-section which uses the optical point method, relies on the luminosity and on  extrapolation of the elastic rate down to $t=0$. Through the optical theorem, one has that 
 the total (nuclear) cross-section depends only on $\Im m A(s, 0)$ and the elastic differential cross-section in the forward direction can be written as 
\begin{equation}
\big{(}\frac{d\sigma_{el}}{dt}\big{)}(t = 0) = \big{(}\frac{\sigma_{tot}^2}{16\pi}\big{)}[1 + \rho^2],\label{eq:dsigdt0}
\end{equation}
with only a quadratic dependence on the ratio of the real to the imaginary part of the forward (complex) nuclear scattering amplitude $A(s, 0)$,  
\begin{equation}
\rho(s) = \frac{\Re e A(s, 0)}{\Im m A(s, 0)}.\label{eq:rho}
\end{equation}
However, a method relying only on Eq.~(\ref{eq:dsigdt0}) does not allow a precise determination  of the nuclear amplitude, since
at high energies, from ISR onwards,  the $\rho$-parameter  
is rather small ($\sim 0.1$). It is then difficult to measure $\rho$ accurately and in any event such a measurement would not determine the sign of the real part of the nuclear amplitude.
 
Fortunately, when we augment the nuclear with the Coulomb amplitude (due to one-photon exchange, in the lowest order), the interference between the Coulomb and the real part of the nuclear amplitude (for small $t$) allows us to determine both the sign and the value of $\rho$. The Rutherford singularity ($\propto\ \alpha/t$) renders the Coulomb amplitude sufficiently large to become competitive with the nuclear term, for small $t$. On the other hand, away from very small angles, the Coulomb term dies out and one can safely revert to the purely nuclear amplitude. However, to obtain numerically accurate information about $\rho$, and hence the nuclear amplitude,
 a precise knowledge of the Coulomb amplitude is required and some care has to be applied  to obtain the correct Coulomb phase for the nuclear problem.
 This problem is far from trivial, because of infrared photons present also in the forward direction. A first estimate by Bethe \cite{Bethe:1958zz}, was later  clarified  in \cite{West:1968du}, as  discussed  in subsection \ref{sss:rhoRadCorr}, where we also discuss the question of soft photon emission.

At ISR, experiments using this method
fitted the observed elastic scattering rate  to the expression
\begin{equation}
R(t)\propto \frac{d\sigma}{dt}=\pi |f_c+f_N|^2
\end{equation}
where $f_c$ is the Coulomb scattering amplitude and $f_N$ the nuclear scattering amplitude. The two amplitudes are both complex, with a relative phase to be determined theoretically.  The expression for the Coulomb amplitude is then written as
\begin{equation}
f_C=-2\alpha\frac{G^2(t)}{|t|}e^{i\alpha \phi}
\end{equation}
where the minus sign holds for proton proton scattering, with opposite sign for \pbarp \ scattering, $\alpha$ is of course the fine structure constant and $G(t)$ is the proton electromagnetic form factor. This expression corresponds to "spinless" scattering, a good approximation at high energies. A more complete discussion inclusive of magnetic terms can be found in \cite{Block:2006hy}, part of which will be reproduced in the next subsection. Here we follow the abbreviated discussion in \cite{Amaldi:1972uw}. The expression used for the proton electromagnetic form factor was the usual dipole expression
\begin{equation}
\label{dipole}
G(t)\ =\ [\frac{1}{1 - t/\Lambda^2}]^2,
\end{equation}
with $\Lambda^2\approx\ 0.71 \ GeV^2$.
To complete the parametrization of the Coulomb amplitude, one needs to specify the phase, which was taken to be
\begin{eqnarray}
\alpha \phi=\alpha[\ln (t_0/|t|) -C] \nonumber\\
  t_0=0.08 \ GeV^2, \ \ \  C=Euler's \ constant=0.577\nonumber\\
\end{eqnarray}
For the nuclear, or hadronic amplitude, use was made of the optical theorem, namely
\begin{equation}
f_N=\frac{\sigtot}{4\pi}(\rho +i)e^{Bt/2}\label{eq:fN}
\end{equation}
and of a parametrization of the very small $t$ behaviour described by  a falling  exponential, 
with  $B$  the so-called slope parameter, in fact a function of $s$. One immediately sees that the  rate of elastic events involves all the quantities we are interested in, namely
\begin{eqnarray}
\label{eq:dsigdt}
R(t)=K
[
(
\frac{2\alpha}{t}
)^2 G^4(t)-
(
\rho+\alpha \phi
)
\frac{\alpha}{\pi}
\sigtot
\frac{G^2(t)}{|t|}
e^{Bt/2}\nonumber \\
+(\frac{\sigtot}{4\pi})^2(1+\rho^2)e^{Bt}],\nonumber\\
\end{eqnarray}
where K is a proportionality constant and the  sign in front of $\alpha$ holds to \pp \ scattering and is reversed for \pbarp . 

\subsubsection{Radiative corrections to the determination of the $\rho$ parameter}\label{sss:rhoRadCorr}

Here we discuss further how the real part of hadronic amplitudes near the forward direction is  determined through 
its interference with the Coulomb amplitude and highlight some of the subtleties associated with the procedure.


To see what is involved, let us consider first Coulomb scattering in non-relativistic potential scattering. The classical 
Rutherford amplitude (or, the Born approximation, quantum mechanically), with a Coulomb $1/r$ potential, for the 
scattering of two charges ($Z_1 e$) and ($Z_2 e$), is given by
\begin{equation}
\label{C3} 
f_C(k, \vartheta) = \frac{2 Z_1 Z_2 \mu \alpha }{t},
\end{equation} 
where $\mu$ denotes the reduced mass, $t\ = - 4 k^2 sin^2 \vartheta/2$ and $\alpha\approx\ 1/137$ is the fine structure 
constant. But, the exact Coulomb scattering amplitude has an oscillating phase $e^{i\phi_S}$ multiplying the above. 
This phase is given by\cite{Schwinger2001}
\begin{equation}
\label{C4} 
\phi_S = (\frac{Z_1 Z_2 e^2}{\hbar v})\ ln(sin^2\vartheta/2), 
\end{equation} 
where $v$ denotes the relative velocity and we have restored proper units to exhibit the quantum nature of this phase.
The physical reason for this phase is that the Coulomb potential is infinite range and however far, a charged particle 
is never quite free and hence is never quite a plane wave. For $pp$ or $p\bar{p}$ scattering, in the relativistic limit 
($v\rightarrow\ c$) and for small angles, Eq.(\ref{C4}) reduces to 
\begin{equation}
\label{C5}
\phi_S \approx\ (\mp 2\alpha)\ ln(\frac{2}{\vartheta}).
\end{equation}
Eq.(\ref{C5}) is exactly the small-angle limit of the relativistic Coulomb phase obtained by Solov'ev\cite{Solov'ev66}.
On the other hand, this result was in conflict with an earlier potential theory calculation by Bethe\cite{Bethe58} 
employing a finite range ($R$) nuclear potential in conjunction with the Coulomb potential. According to Bethe, the 
effective Coulomb phase reads
\begin{equation}
\label{C6}
\phi_B \approx\ (\pm  2\alpha)\ ln(k R \vartheta).
\end{equation}   
This discrepancy was clarified by West and Yennie\cite{WestYennie68}. These authors computed the effective Coulomb 
phase through the absorptive part of the interference between the nuclear and the Coulomb amplitude. They found-again 
in the small angle, high energy limit-
\begin{equation}
\label{C7}
\phi_{WY} = (\mp \alpha)[ 2 ln(\frac{2}{\vartheta}) + \int_{-s}^0 \frac{dt^{'}}{|t^{'} - t|}\{1 - \frac{A(s,t'}{A(s,t)}
\}].
\end{equation}  
If one ignores the $t$ dependence of the nuclear amplitude, the integral term above is zero and one obtains Solov'ev's result. 
On the other hand, a result similar to that of Bethe is reproduced, if one assumes the customary fall-off $e^{B t/2}$ for the 
nuclear vertex and a dipole form factor for the EM vertex. Explicitly, if we choose
\begin{equation}
\label{C8}
\frac{A(s,t{'})}{A(s,t)} = e^{B(t^{'} - t)/2} \big{(}\frac{1 - t/\Lambda^2}{1 - t^{'} /\Lambda^2}\big{)}^2,
\end{equation} 
we find
\begin{equation}
\label{C9}
\phi_{WY} \approx\ (\pm \alpha)[\gamma + ln (B|t|/2) + ln (1 + \frac{8}{B \Lambda^2})],
\end{equation}
where $\gamma\approx\ 0.5772..$, is the Euler-Mascheroni constant.
This expression for the effective Coulomb phase agrees with Block\cite{Block06}, up to terms proportional to 
($|t|/\Lambda^2$), which are quite small near the forward direction. Hence, Eq.(\ref{C9}) is sufficiently accurate for 
determining  $\rho$ through interference at LHC energies and beyond. 

As a practical matter, Block has defined a useful parameter $t_o$ for which the interference term is a maximum:
$t_o\ = [8\pi \alpha/\sigma_{tot}]$. For the maximum LHC energy of $14\ TeV$, $t_o\approx\ 7\times\ 10^{-4}\ GeV^2$.
Putting it all together, the Coulomb corrected, differential cross-section reads\cite{Block06}
\begin{eqnarray}
\label{C10}
[\frac{d\sigma}{d|t|}]_{o} = (\frac{\sigma_{tot}^2}{16\pi}) \big{[} G^4(t)(\frac{t_o}{t})^2
 + 2\frac{t_o}{|t|} (\rho + \phi_{WY}) G^2(t) e^{-B|t|/2}\nonumber\\
  + (1 + \rho^2) e^{-B|t|}
\big{]},\nonumber\\
\end{eqnarray}
where for the magnetic form factor $G(t)$, one may employ Eq.(\ref{dipole}).

One other aspect of the EM radiative corrections needs to be investigated. So far, we have not considered real 
soft-photon emissions in the scattering process. As is well known, contributions due to an infinite number of soft 
(IR) photons need to be summed.
If $(d\sigma/dt)_o$ denotes the differential cross-section 
without the emitted soft-photons, the IR corrected cross-section depends upon the external energy resolution $\Delta E$. 
A compact expression for the corrected cross-section can be written as
follows
\begin{equation}
\label{C11}
\frac{d\sigma}{dt} = (\frac{\Delta E}{E})^{\beta(s) - \beta(u) + \beta(t)} (\frac{d\sigma}{dt})_o,
\end{equation} 
where the various radiative factors $\beta(s,t,u)$ are defined and discussed in Sec. \ref{sec:models}.

Recently, there has been a study of the amplitudes for $pp$ and $p\bar{p}$ elastic scattering in the 
Coulomb-Nuclear Interference region based on derivative dispersion relations \cite{Kendi:2009cw}. 
Work on this subject was done early on by Bourrely, Soffer and T.T. Wu  \cite{Bourrely:1984gi} and more 
recently also in collaboration with Khuri \cite{Bourrely:2005qh}. An interested reader may consult these references.

%


\subsubsection{The four methods used at ISR}\label{sss:fourISRmdethods}
Here we present a short discussion of the four methods used at ISR to measure the total cross-section.

The Pisa-Stony Brook method (R801) to measure  the total cross-section at ISR
was based on measuring the luminosity ${\cal L}$ and the inclusive interaction rate $R_{el}+ R_{inel}$, through the definition
\begin{equation}
R(number\ of\ events/second)={\cal L}
(cm^{-2}sec^{-1}) \sigma_{total}(cm^2)
\label{eq:totrate}
\end{equation}
This method had the advantage of making a totally model independent measurement.

Two different approaches were  adopted by the CERN-Rome group. They measured the differential elastic cross-section 
at small angles, but not in the Coulomb region, and then extrapolated it to the optical limit, i.e.  $t=0$. The measurement 
was based on the parametrization of the hadronic part of the cross-section given by Eq.(\ref{eq:dsigdt0}),
which assumed a constant (in t) exponential $t-$dependence of the scattering amplitude, and a parameter 
$\rho$ constant in the range of $t $ of interest. Elastic events at smaller and smaller scattering angles were measured through 
the so-called {\it Roman pots}, which were detectors inserted in containers called the  {\it Roman pots}   
and which could penetrate the beam pipe and get very close to the beam.

The  CERN-Rome group also applied a third complementary method, the one described in the previous subsection, 
 which measured 
  the rate  in the region where the Coulomb and nuclear amplitudes interfere. At ISR this happens in a region $0.001<|t|<0.01\ GeV^2$.
  From Eq.~(\ref{eq:dsigdt}) 
one can fit $d\sigma/dt$ in terms of $\sigtot, \rho, {\cal L}$ and the slope parameter $B$.  

A fourth method was adopted by a combined Pisa-Stony Brook (PBS) and CERN-Rome  collaboration. 
By combining the measurement of the total interaction rate $R_{tot}$, i.e. the PBS  
approach, Eq.~(\ref{eq:totrate}),
with the Cern-Rome method, Eq.~(\ref{eq:dsigdt0}), based on the optical thorem, 
one obtains a {\it luminosity independent } measurement, i.e.
\begin{equation}
\sigtot=\frac{16\pi (dN/dt)_{t=0}}{N_{tot}(1+\rho^2)}
\end{equation}
where $(dN/dt)_{t=0}$ is the elastic  rate measured at $t=0$ and $N_{tot}$ is the total rate. This was a combined Pisa-Stony Brook and CERN-Rome measurement.

For a description of the luminosity measurements at ISR, we refer the reader to \cite{Giacomelli:1979nu}.


These measurements indicated a rising total cross-section. The result was surprising, given the then accepted 
constant cross-section emerging from the simple Pomeron pole model with an intercept $\alpha_P(0)\ =\ 1$. 
The measurement was repeated several times against possible systematic errors. We show in Fig.~ \ref{fig:R801} 
from \cite{Amendolia:1973yw}, the published results for the total cross-section as a function of $s$, the squared 
center of mass energy.
 \begin{figure}
 \begin{center}
 \resizebox{0.5\textwidth}{!}{%
  \includegraphics {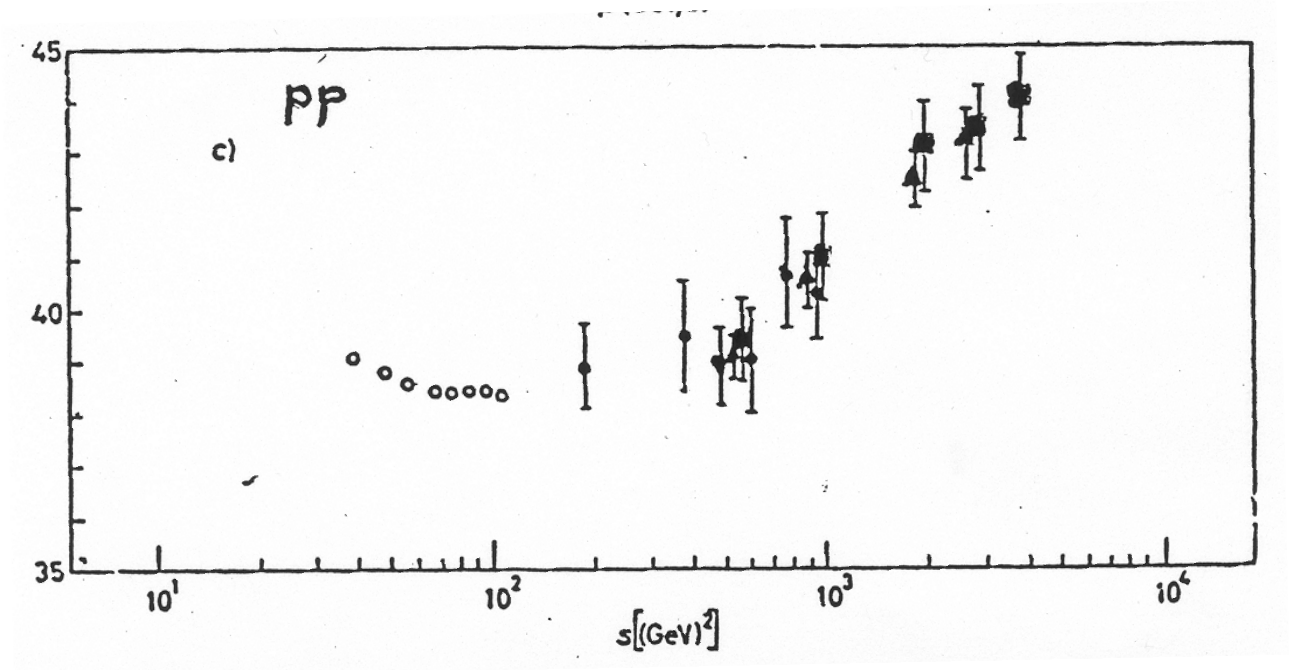}}
\caption{The measurement of the total cross-section by the R801, Pisa-Stonybrook, experiment. This figure is courtesy of G. Bellettini, also published in \cite{Amendolia:1973yw}. Reprinted from \cite{Amendolia:1973yw} \copyright (1973) with permission from Elsevier.}
\label{fig:R801}
\end{center}
\end{figure}
The Pisa-Stonybrook experiment also produced a beautiful pictorial description of single diffraction \cite{Bellettini:1973zu},
 but because of the difficulty in separating the non-diffractive background, these results were not published.


\subsubsection{A final analysis  of ISR results}\label{sss:ISRfinal}
A final analysis from the group from Northwestern, comprehensive of both \pp \ 
 and \pbarp\ scattering in the full range of  ISR energies is given in \cite{Amos:1985wx}. This analysis was published 
 after the CERN $S{\bar p}pS$
  had already been operational for a couple of years and the rise of the total cross-section had been confirmed. We shall now summarize this paper.
 
 In the following we use the notation of  \cite{Amos:1985wx}, where the slope parameter $B$ is indicated by $b$. 
 The method used for measuring $\sigtot$, $\rho$ and  $b$, is luminosity dependent and is based on measuring the 
 differential elastic rate and then making a simultaneous fit of the elastic differential cross-section in and around the 
 Coulomb region, typically $0.5 \times 10^{-3}<|t|<50\times 10^{-3}\  GeV^2$. For $\alpha \phi<< 1$,  from the Rutherford  
 scattering formula  and the optical theorem, the usual expression
 is used to parametrize elastic scattering in this region,  i.e.
 \bea
 \frac{d\sigma}{d|t|}&=&4 \pi\alpha^2 G^4(t)/|t|^2
\mp\sigtot\alpha(\rho\pm\alpha \phi)G^2(t)e^{-b|t|/2}/|t| \nonumber\\
&+&(1+\rho^2)\sigtot^2 e^{-b|t|}/16\pi \label{eq:dsigdtamos}
\eea
For this method, a precise determination of the luminosity is crucial. Notice the known expression for 
Coulomb scattering allows a calibration  of the $|t|$ scale. We show in Fig. \ref{fig:amos1985wx}, one of the many plots 
presented by this collaboration, for \pp, at $\sqrt{s}=52.8\ GeV$.
\begin{figure}
\begin{center}
\resizebox{0.5\textwidth}{!}{%
\includegraphics{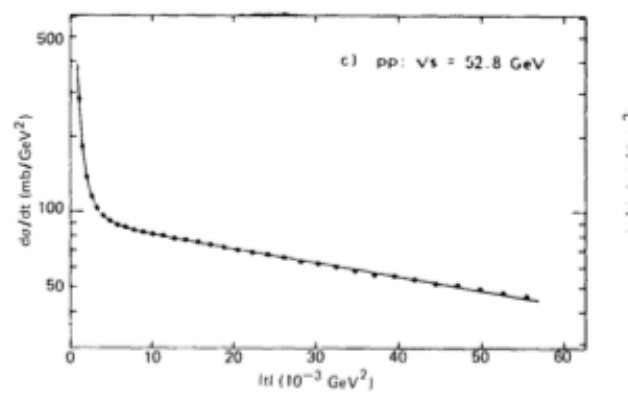}}
\caption{Proton-proton differential elastic scattering cross-section $ \dsigdt$ in the very small  $|t|$ region, at ISR operating 
energy $\sqrt{s}=52.8\ GeV$ from \cite{Amos:1985wx}. Shown are the Coulomb region, the interference and the beginning 
of the nuclear region. Reprinted from \cite{Amos:1985wx} \copyright (1985) with permission by Elsevier.}
\label{fig:amos1985wx}
\end{center}
\end{figure}
This figure clearly  shows the transition between the Coulomb region $-t \lesssim 0.005\ GeV^2$ and the nuclear region , 
$-t > 0.01\ GeV^2$ through the interference region .  Similar distributions are presented in the paper for the full range of ISR 
energies, $\sqrt{s}=23.5-62.3 \ GeV$ for  \pp \ and $\sqrt{s}=30.4-62.3 \ GeV$ for \pbarp. The measured differential cross-sections 
as a function of $|t|$ are presented and a simultaneous fit of Eq.~(\ref{eq:dsigdtamos}) allows to extract the values we 
reproduce in Table~ \ref{tab:amos}.
\begin{table}
\caption{Resulting values for $\sigtot$, $\rho$ and $b$ at ISR from \cite{Amos:1983mh}.}
\label{tab:amos}
\begin{tabular}{|c|c|c|c|c|}
\hline              &$\sqrt{s}$         &$\sigtot$               &$\rho$                    &$b$              \\ 
              & GeV                 &mb                         &                                &$GeV^{-2}$ \\ \hline \hline
 pp         &23.5                 &$39.65\pm 0.22 $ &$0.022\pm0.014  $  &$11.80 \pm 0.30$\\ \hline
 pp         &30.6                 &$40.11\pm0.17 $&$0.034\pm0.008     $ &$12.20\pm 0.30$\\
 \pbarp &30.4                  &$42.13\pm 0.57$ &$0.055\pm0.029   $  &  $12.70\pm 0.50$\\ \hline
 pp        &52.8                  &$42.38\pm 0.15$ &$0.077\pm 0.009    $& $12.87\pm 0.14$\\
 \pbarp &52.6                  &$43.32\pm 0.34 $&$0.106\pm 0.016 $ &$13.03\pm 0.562 $\\ \hline
 pp        &62.3                 &$43.55\pm 0.31 $&$0.095\pm0.011     $  &$13.02\pm0.27$\\
 \pbarp &62.3                  &$44.12\pm 0.39 $&$0.104\pm0.011      $& $13.47\pm 0.52$\\
 \hline
\end{tabular}
\end{table}
\subsubsection{Measurements of $\rho$ and the slope parameter}
The CERN-Rome experiment had measured the ratio of the real to the imaginary part of the forward elastic scattering amplitude. 
The trend with energy of the $\rho$ parameter  was confirmed by other experiments as well and found 
in good agreement with a dispersion relation calculation by Amaldi. These results are shown in Fig.\ref{fig:Amos1983mh} from \cite{Amos:1983mh}. 
 \begin{figure}
 \begin{center}n
 \resizebox{0.5\textwidth}{!}{%
  \includegraphics{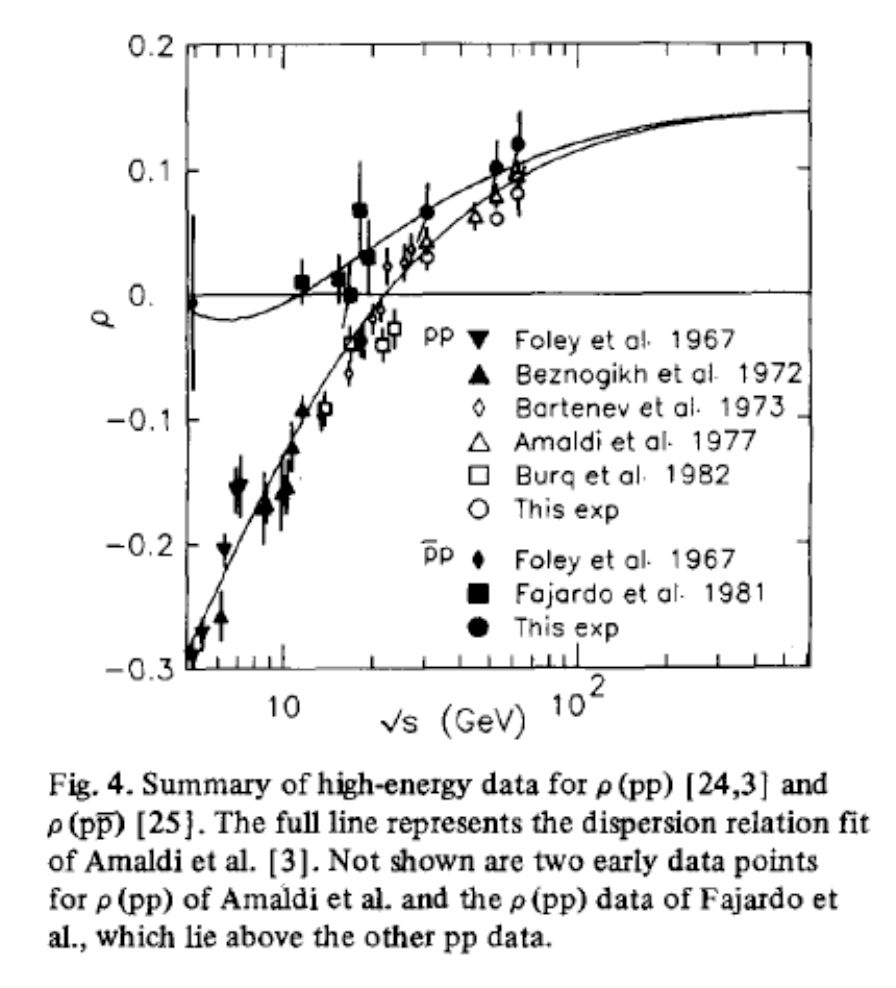}}
\caption{The ratio $\rho$ from $\sqrt{s}=5 \ GeV$ up to the highest ISR energy, from a compilation in \cite{Amos:1983mh}.
Reprinted from \cite{Amos:1983mh} \copyright (1983) with permission by Elsevier.}
\label{fig:Amos1983mh}
\end{center}
\end{figure}

We mention here the question of the $|t|$ dependence of the slope at ISR. In \cite{Barbiellini:1972ua}, the distributions of the differential rates  for \pp \ elastic scattering are presented nominally for 4 different ISR energies, $\sqrt{s}=21.5$, $30.8$, $44.9$ and $ 53\ GeV$. \footnote{However only three such distributions appear in their Fig. 3, the one at 30.8 GeV is apparently absent.} All three sets of data points exhibit a break for $|t|\sim 0.1\ GeV^2$. For all the energies under consideration, the measurement of the slope in the smaller $t$ interval, was found to be larger than the one at larger $t$. At 53 GeV, the two slopes would be $b(0.050<|t|<0.112 \ GeV^2)=12.40\ GeV^{-2}$ and $b(0.168<|t|<0.308 \ GeV^2)=10.80\ GeV^{-2}$. The steepening of the slope at small $|t|$ values increases the forward scattering cross-section above the value extrapolated from larger $|t|$ by about 20 \%.

We now turn to a 1982 paper by a group CERN-Naples together with some members of the Pisa-Stony Brook collaboration \cite{Ambrosio:1982zj}. Measurements were done for \pbarp \ and for  \pp \  at the ISR energy $\sqrt{s}=52.8\ GeV$ for the quantities $\dsigdt$, $\sigel$ and, using the optical theorem $\sigtot$. The measurement of the elastic cross-section was obtained by measuring the elastic rate while simultaneously measuring the luminosity of the colliding beams. The differential cross-section -in the small-$|t|$ region- was then parametrized as usual, i.e. 
\be
\dsigdt=Aexp(bt)
\ee
Fit to the data yielded $\sigel$ and $b$. Elastic scattering was studied in the $t$ range $0.01-1.0\ GeV^2$ using different parts of the detector. The fitting took account of the Coulomb corrections, through Eq.~(\ref{eq:dsigdtamos}).

In Table 1 of this paper, the results for $0.01<|t|<0.05\ GeV^2$ are displayed and the following final values for the slope are given
\be
b(p\bar p)=13.92\pm 0.37\pm0.22\ \ \ \ \ \ \ b(pp)=13.09\pm 0.37\pm 0.21
\ee
where errors quoted are statistical only, and include error on luminosity and uncertainty on determination of the $|t|$ interval (see \cite{Ambrosio:1982zj} for details). When reporting the results for the slope at larger $|t|$ intervals, i.e. $0.09-1.0 \ GeV^2$  for \pp \ and \pbarp, the slope was found to be smaller, $10.34\pm 0.19\pm 0.06 \ GeV^{-2}$ and $10.68\pm 0.20 \pm 0.06\ GeV^{-2}$ respectively, in agreement with \cite{Barbiellini:1972ua}.  Results for the slope in the smaller interval were in agreement not only  with  \cite{Barbiellini:1972ua} but also with \cite{Amos:1981dr} whose measurement covers the interval $|t|=1.0\times10^{-3}$ to $31 \times 10^{-3}\ GeV^2$. The larger $|t|$ interval is not discussed in 
\cite{Amos:1981dr}.
Results for the slope in the larger $|t|$ interval are summarized in  table \ref{tab:blarger}. 
\begin{table}
\caption{\label{tab:blarger}Values of the slope parameter at $\sqrt{s}=52.8\ GeV$ in different $t$ intervals, as shown.}
\begin{tabular}{|c|c|c|c|}
\hline
Experiment                        &reaction   &$|t| interval$    &  b\\ 
                                             &                  &$ GeV^2$        & $(GeV^{-2})$\\     \hline
 \cite{Barbiellini:1972ua} &     \pp       &0.168-0.308     &$10.80 \pm $0.20\\
 \cite{Ambrosio:1982zj}    &\pbarp      &0.09-1               &$10.68\pm 0.20\pm 0.06$\\
 \cite{Ambrosio:1982zj}    &\pp             &0.09-1             &$10.34\pm 0.19\pm 0.06 $\\ \hline
\end{tabular}
\end{table}

Notice that this experiment also gives explicit values for $\sigtot$, the ratio \sigeltosigtot, and the ratio $\sigtot /b$, 
all of which will be discussed in the context of models.

We now look at Amaldi's later work in \cite{Amaldi:1976gr} and  \cite{Amaldi:1979kd}. Ref.~\cite{Amaldi:1976gr} contains a complete 
review of all the data collected at ISR for the usual  slope, total, elastic cross-section and real part of the amplitude in the forward direction, 
while \cite{Amaldi:1979kd} discusses in detail the optical picture and its connection to the Pomeron exchange description.

Before going to illustrate accelerator  measurements at higher energies, we note   a  recent review of ISR results  by Ugo Amaldi, \cite{Amaldi:2015jhq}, where the history of the discovery of the rise of the total cross-section at the ISR is described.

\subsection{Measurements 
at the $Sp\bar{p}S$ \label{ss:spbarps}}
In the early 1980's, an hitherto unimagined energy value in the CM was reached at the CERN $Sp\bar{p}S$, i.e, 
$\sqrt{s}=540\ GeV$. At the $Sp\bar{p}S$, the luminosity \cite{Arnison:1983mm} due to $N_p$ protons and  $N_{\bar p}$ 
antiprotons crossing at zero angle with effective area A is ${\cal L}=fN_p N_{\bar p}/A$ where f is the frequency of revolution of the 
bunches. The effective area is given by $A=wh$ where the effective width ($w$) and height ($h$) of the crossing bunches are :
\begin{equation}
w=\int dx  N_p (x)N_{\bar p}(x)\ \ \ \ \ \ \ \ \ \ \ h=\int dy N_p(y) N_{\bar p}(y)
\end{equation}
where $N_p(x)$ is the normalized proton density profile  along the transverse horizontal axis at the crossing point. At the time of the UA1 early measurement a systematic error of 8\% on the integrated luminosity measurement was reported due to various uncertaintites in the factors entering the luminosity formula.

 At \spbarps \ the total \x \ 
  was measured by experiments UA1 and UA4, later on followed by a combined collaboration UA2/UA4 and by a ramping run measurement by UA5.

\subsubsection{Early     total cross section measurements: UA1 and UA4}\label{sss:UA1}
  
  UA1 experiment made a  measurement of the elastic scattering \x\ \cite{Arnison:1983mm}, with forward detectors covering  angles down to $5\ mrad$, measuring the elastic differential cross-section for  $0.04<-t<0.45\ GeV^2$. 
The data collected by UA1 were fitted by the usual exponential form $dN_{el}/dt\propto e^{Bt}$. A measurement of the integrated luminosity allowed the extraction  of  
values $B\ =\ 13.7\pm 0.2 \pm 0.2\ GeV^{-2}$ for $|t|=0.21-0.45 \ GeV^2$ and $B\ =\ 17.1 \pm 0.1 \ GeV^2$ fot $|t|=0.04-0.18\ GeV^2$. The values $\sigma_{tot}\ =\ 67.6\pm 5.9 \pm 2.7\ mb$ and $\sigma_{el}/\sigtot\ =\ 0.209\pm 0.18\pm 0.08$ were thus obtained. The UA4 experiment  had also made an earlier measurement of $\sigtot$ \cite{Battiston:1982su} and we show both measurements in Fig.~\ref{fig:UA1sigtot} from \cite{Arnison:1983mm}. The two curves shown in this figure correspond to two different fits  by Block and Cahn \cite{Block:1982bv}, where a simultaneous analysis of $\sigtot$ and $\rho$, from $s=25\ GeV^2$ to the ISR data is performed. The two fits follow from the
expression for the even amplitude at t=0 \cite{Bourrely:1973eg} , i.e. 
\begin{equation}
\label{BA}
{\cal M}_+=-is[A+\frac{B(\ln{s/s_0}-i\pi/2)^2}{1+a(\ln{s/s0}-i\pi/2)^2}]+C,
\end{equation}
Let the parameter $a$ take a small value. Then at intermediate-to-high energies such as that at the 
$Sp\bar{p}S$,  one obtains a linear (fit2, lower, dashes) $\ln{s}$  dependence, while obtaining a constant 
behaviour for the total cross-section at really high energies. The other curve  (fit 1,upper, full ) follows from 
an expression for the amplitude which saturates the Froissart bound, namely is quadratic in $\ln{s}$ \cite{Block:1982bv}. 
 Fit 2, with a very small value for the parameter $a$, i.e  $a=0.005\pm 0.0031$, gives a slighly better fit, 
 as the authors themselves point out. The problem with  this fit  is that, at extremely high energies, 
 the total \x \ would go to a constant, a behaviour still not yet observed even at very high cosmic ray 
 energies.
 
In  \cite{Block:1982ew}, the behaviour of the slope parameter $B$ was fitted 
 with the asymptotic form  $\ln^2s$, that is with the same curvature as the total cross-section. 
 Such form follows from the standard parametrization of the forward scattering amplitude as in Eq.~(\ref{eq:fN}).
  The rationale behind this choice of behavior   is that if $B \sim (ln s)$ and $\sigma_{tot} \sim (ln s)^2$, 
  asymptotically
   there is a problem with unitarity, 
 i.e., $\sigma_{el} > \sigma_{tot}$.
But, if $\sigtot $ saturates the Froissart bound, i.e, $\sigtot\sim \ln^2{s}$, and we require that the slope parameter 
$B$ rises as $\sigtot$, then there is a problem with the simple Regge-Pomeron picture, because we would expect
\begin{equation}
\frac{d\sigma}{dt}\propto (\frac{s}{s_0})^{2[\alpha_{Pom}(t)-1]}=e^{-2\alpha' \ln (s)|t|}
\end{equation}
namely the slope parameter to be proportional to $\ln s$. [This argument holds even  with three Pomeron trajectories, 
intersecting at $1$, and conspiring to produce a total cross-section to increase as $(ln s)^2$]. Where, in such picture, 
a $\ln^2{s}$ for the slope parameter $B$  would come from is not clear. On the other hand, using the Block and Cahn 
{\it simple analytic } expression, only a constant or a  $\ln^2{s}$ behaviour results. 

This debate is still of interest, as we discuss in the following sections of this review.


\begin{figure}
\resizebox{0.5\textwidth}{!}{%
\includegraphics{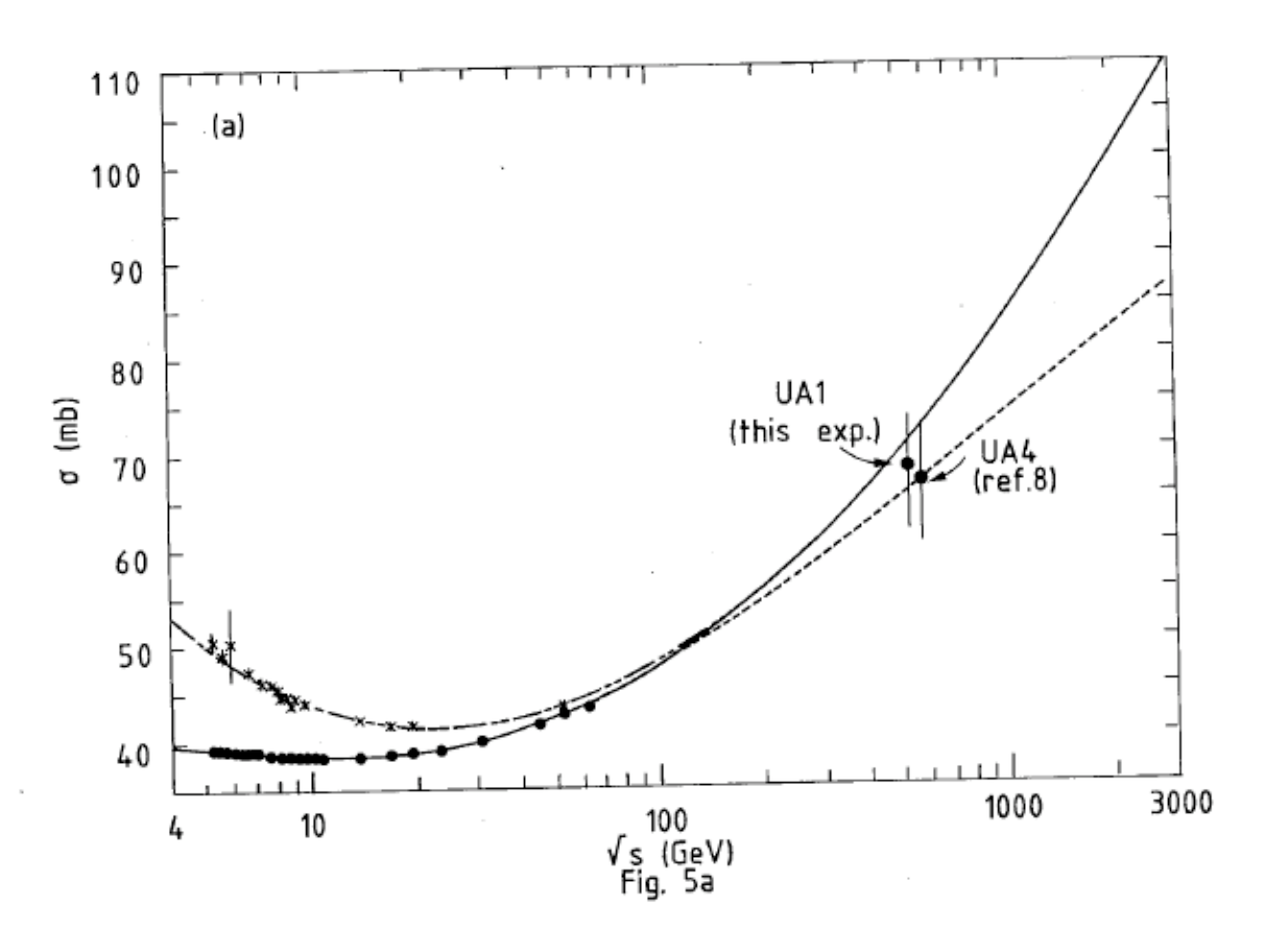}}
\caption{This figure from \cite{Arnison:1983mm} shows the  total cross-section as measured by UA1  at the CERN \spbarps \, compared with the UA4 early measurement \cite{Battiston:1982su}. The two curves correspond to different fits by Block and Cahn. For details see \cite{Arnison:1983mm}, from which this figure is extracted. Reprinted from \cite{Arnison:1983mm} \copyright (1983) with permission by Elsevier. }
\label{fig:UA1sigtot}
\end{figure}

\subsubsection{UA4 and UA2}\label{sss:UA4} UA4 was the experiment dedicated to the measurement of the total and elastic  \x, and of the parameter $\rho$. UA4  measured the total cross-section with the luminosity independent method by comparing the forward differential cross-section with the total elastic and inelastic rate \cite{Battiston:1982su,Bozzo:1984rk}. In \cite{Battiston:1982su} the value $\sigtot=66\ mb$ with a 10\% statistical error was reported. Subsequently, using the same method and through a comparison with a luminosity dependent measurement, the final value of $\sigtot=61.9 \pm 1.5\ mb$ was given \cite{Bozzo:1984rk}. The difference with previous measurements was attributed to a 1.1\% overestimate of the beam momentum above the nominal energy $E=270\ GeV$ and thus to an overestimate of 2.2\% on the total \x.
The inelastic rate at intermediate angles
  was measured by a set of dedicated telescopes , while the total inelastic rate in the central pseudorapidity region was measured by the UA2 detector. 
UA4 was able to measure very small scattering angles, as small as $\sim 1 \  mrad$, down to values 
$0.002<|t|<1.5\ GeV^2$. This was accomplished through the use of the Roman Pot technique, already employed at the ISR.
The result confirmed the shrinking of  the diffraction peak, with a value of the slope, $B$, defined, as customary, from the parametrization $\dsigdt=\dsigdt(t=0)exp(Bt)$. 
We show in Fig.~\ref{fig:UA4sigtot} a plot of the total cross-section from \cite{Bozzo:1984rk}, where the UA4 result is compared  with earlier measurements at ISR and fixed target machines.
\begin{figure}
\begin{center}
\resizebox{0.5\textwidth}{!}{
\includegraphics{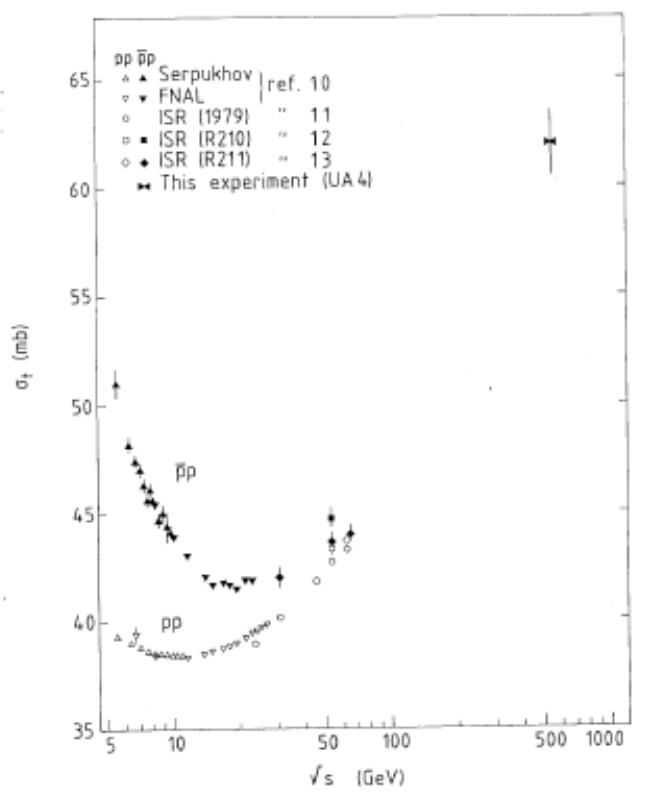}}
\caption{The result of the UA4 experiment for the measurement of the total \x 
 at the CERN $Sp\bar{p}S$ from \cite{Bozzo:1984rk}. Reprinted from \cite{Bozzo:1984rk} \copyright (1984) with permission by Elsevier.}
\label{fig:UA4sigtot}
\end{center}
\end{figure}

The ratio $\rho$ was also measured and found to be surprisingly large, namely $\rho=0.24 \pm 0.4$. 
As the authors themselves note in a subsequent paper \cite{Augier:1993sz}, this measurement was affected by poor beam optics and limited statistics. It was then repeated by a combined UA4/UA2 collaboration under very clean conditions with higher precision and better control of  systematic errors, and found to be
\be
\rho=0.135\pm 0.15
\ee
which supersedes previous measurements and was in agreement with the original theoretical expectations. 
\subsubsection{The ramping run and UA5 measurement}\label{sss:UA5}
The UA5 experiment measured the total $p\bar{p}$ \x \ at $\sqrt{s}=200$ and $900\ GeV$ \cite{Alner:1986iy}. Data were normalized at $\sqrt{900}\ GeV$ from an extrapolation by Amos el al.\cite{Amos:1985wx} and the result, $\sigtot^{900}=65.3\pm 0.7\pm1.5  \ mb$, was found to be consistent with previous measurements. We show in Fig.~\ref{fig:ua5sigtot} the result of UA5 measurement compared with the UA4 measurement \cite{Bozzo:1984rk}, with the extrapolation from \cite{Amos:1985wx}, the expectations from  Donnachie and Landshoff Regge-Pomeron exchange model \cite{Donnachie:1983hf}, as well as a description by  Martin and Bourrely \cite{Bourrely:1973eg,Bourrely:1984pi}, with explicit analytic and crossing symmetric form of the even signature amplitude at $t=0$  as in Eq.(\ref{BA}).
\begin{figure} 
\begin{center}
\resizebox{0.5\textwidth}{!}{
\includegraphics{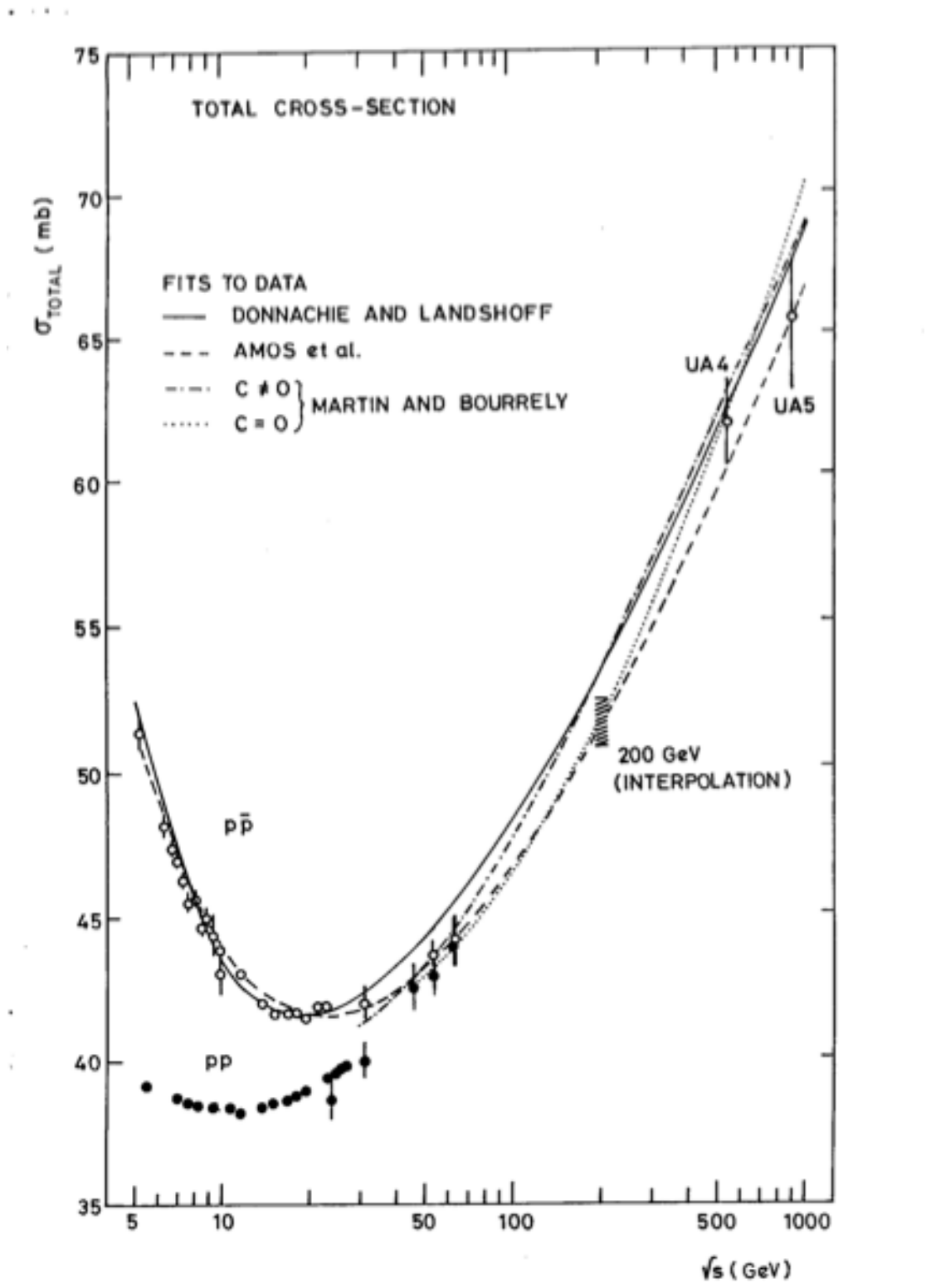}}
\caption{The total \x \ for \pp\ and \pbarp\  from \cite{Alner:1986iy}, including the measurement by the UA5 Collaboration,  compared with models and data, as explained in the text.
Reprinted from \cite{Alner:1986iy}, \copyright (1986) by  Springer.}
\label{fig:ua5sigtot}
\end{center}
\end{figure}
\subsection{Reaching the TeV region}\label{ss:TeVatron}
Starting from 1985, an even higher center of mass energy for proton-antiproton scattering  was reached in 
FermiLab near Chicago through the TeVatron accelerator. 
At  the Tevatron, the total \x \ was measured by three experiments: E710 
\cite{Amos:1989at,Amos:1990jh}, Collider Detector Facility (CDF) \cite{Abe:1993xy} and E811  \cite{Avila:1998ej}.

\subsubsection{Measurements at the TeVatron}\label{sss:Tevatron-measure}
Experiment E710   at the Tevatron was the first to measure the total interaction rate and the forward elastic cross-section \cite{Amos:1990fw}. In the detector for elastic events, two Roman pots were placed one above and one below the beam, with drift chambers and trigger scintillators. The inelastic rate was measured 
at large and intermediate angles through ring shaped scintillators and tracking drift chambers, respectively. The first  measurement  by E710 was based on the optical theorem, i.e. 
\be
\frac{dN_{elastic}}{dt}|_{t=0}=L \frac{d\sigma}{dt}|_{t=0}=L \frac{\sigtot^2}{16 \pi}(1+\rho^2)
\ee
and  extrapolation to zero of the usual exponential behavior of the elastic rate measured in the interval $0.025<|t|<0.08 \ GeV^2$, i.e.
\be
\frac{d\sigma}{dt}=\frac{d\sigma}{dt}|_{t=0} e^{Bt}
\ee
Coulomb effects were included in the analysis, although small in the range used for the extrapolation. A second measurement \cite{Amos:1990jh} used  the luminosity independent method, described in the previous subsection, and the total cross-section was obtained from
\be
\sigtot=\frac{16 \pi}{1+\rho^2}\frac{1}{N_{elastic}+N_{inelastic}}\frac{dN_{elastic}}{dt}|_{t=0}
\ee 

CDF at the FermiLab Tevatron Collider repeated the $\sigtot$ measurement at $\sqrt{s}=546\ GeV$ and extended it to $\sqrt{s}=1.8\ TeV$ \, using the luminosity independent method \cite{Abe:1993xy}. The measurement of the primary particle scattered in the forward direction was possible only on the antiproton side. CDF used the Roman Pot detector technique. Inside the pots, two scintillation counters were used for triggering, with a drift chamber backed by silicon detectors measuring the particle trajectory at 5 points. In Fig.~\ref{fig:cdfrpbellettini} we show the layout of the detector assembly, which also shows the bellows technique employed to move the silicon detectors in and out of the beam.
\begin{figure}
 \begin{center}
 \resizebox{0.5\textwidth}{!}{%
  \includegraphics {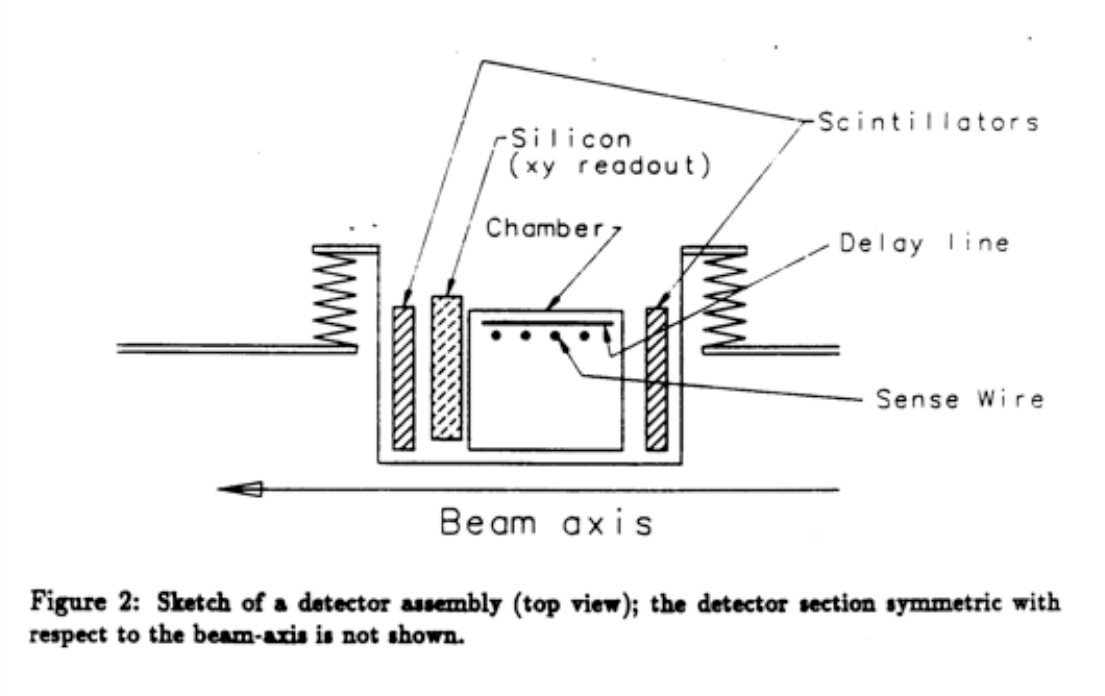}}
\caption{Schematic view of the CDF Roman Pots operation at  the FNAL Tevatron from \cite{Abe:1993xx}.
Reprinted with permission from \cite{Abe:1993xx}, \copyright (1993) by the American Physical Society.}
\label{fig:cdfrpbellettini}
\end{center}
\end{figure}
Elastic events were distinguished by left-right collinearity, while silicon detectors behind the chambers allowed to reach off-line higher angular resolution.

The total \x \ was obtained from a measurement of the  forward elastic and inelastic interaction rate. The rate of inelastic events for scattering at intermediate angles was  measured through two telescopes. On the antiproton side, quasi-elastic antiprotons were detected allowing the measurement of single diffraction with the result
\begin{equation}  
\sigma_{SD}(x>0.85)=7.89\pm 0.33\ mb 
\end{equation}
to be compared with the previous UA4 result of $\sigma_{SD}(x>0.85)=10.4\pm 0.8\ mb $, indicating the difficulty in controlling systematic errors in diffractive event measurements. The measurement of the small angle elastic \x\ was reported in  \cite{Abe:1993xx}.  

The result for the total \x \   measured by CDF
 differs by more than 3 standard deviations from the E710 result, as one can see in Fig.~\ref{fig:CDFE710}. This figure shows a comparison between these two results from \cite{Abe:1993xy}. 
\begin{figure}
 \begin{center}
 \resizebox{0.5\textwidth}{!}{%
  \includegraphics {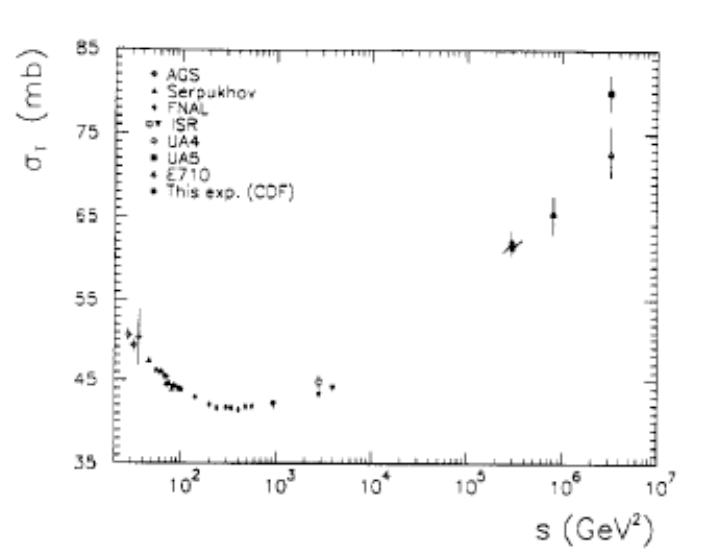}}
\caption{Comparison with lower energy data of  results for the total \x\ from the  two experiments which first measured the total cross-section at the Tevatron, E710 and CDF,  from \cite{Abe:1993xy}. Reprinted with permission  from  \cite{Abe:1993xy} \copyright (1993) by the American Physical Society.}
\label{fig:CDFE710}
\end{center}
\end{figure}
CDF also measured the total cross-section at $\sqrt{s}=546\ GeV$ and the result  with $\sigtot=61.26
\pm 0.93$, obtained assuming $\rho=0.15$,  was consistent with the UA4 result, assuming the same value for $\rho$.

Because of the  discrepancy by more than 3 standard deviations between the CDF and the E710 measurements, E811 analyzed the very small angle data, $0.0045<|t|<0.036\ GeV^2$, using the luminosity independent method in order to measure $\sigtot$ \ \cite{Avila:1998ej} and the $\rho$-parameter \cite{Avila:2002bp} and obtained a result consistent with the one from E710. We show a compilation of all the TeVatron results for $\sigtot$  in Fig.~\ref{fig:811avila1998}.

\begin{figure}
 \begin{center}
 \resizebox{0.5\textwidth}{!}{%
  \includegraphics {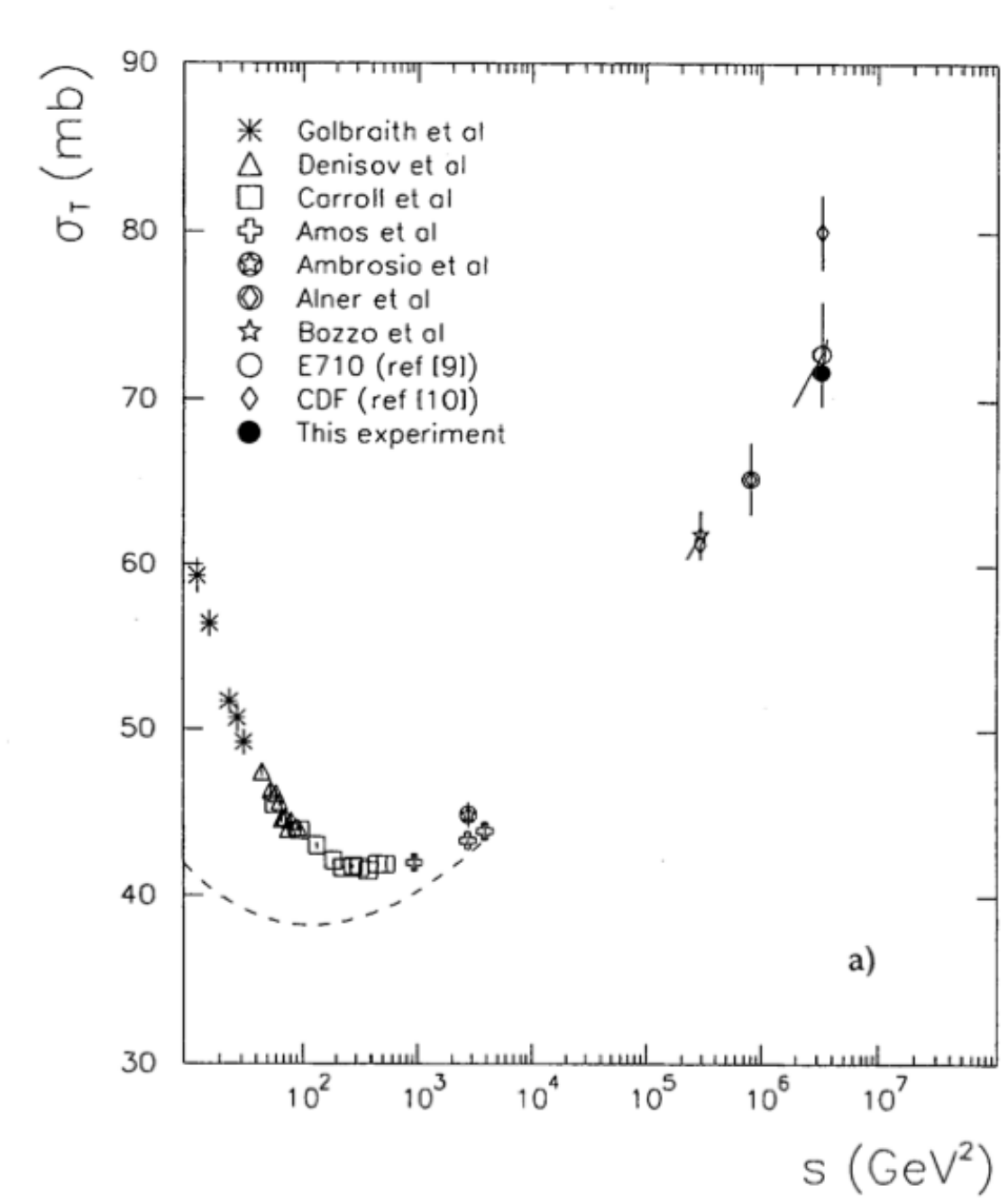}}
  \caption{Results for the total \x\ at the TeVatron, in a compilation by E811,  from \cite{Avila:1998ej}. Reprinted from 
  \cite{Avila:1998ej} \copyright (1998) with permission by Elsevier.}
\label{fig:811avila1998}
\end{center}
\end{figure}

 The final numbers for $\sigma_{tot}(p\bar{p})$ measured at the Tevatron by the  three different experiments, CDF, E710 and E811 are a follows:   
\begin{eqnarray}
\sigtot^{CDF}=80.03\pm 2.24\ mb, \nonumber\\
\sigtot^{E710}=72.8\pm 1.63\ mb, \nonumber \\
 \sigtot^{E811}=71.42\pm 2.41\ mb \nonumber
\end{eqnarray}
Notice that at $\sqrt{s}=1.8 \ TeV$, the CDF result 
\be
\frac{\sigel}{\sigtot}=0.246\pm 0.004
\ee agrees with the E710 value $0.23 \pm 0.012$ \cite{Abe:1993xy}.
\subsubsection{A Comment on the Black Disk Model}
 \label{sss:blackDisk}
Fits to the total \x \  from measurements prior to ISR and up to the latest Tevatron data accomodate  a $\ln^2{s}$ rise. It should however be mentioned that in  models  such as  in Refs. \cite{Grau:1999em,Grau:2009qx} for instance, the rise can be $(ln{s})^{1/p}$ where $1/2<1/p<1$. 

The $\ln^2{s}$ behaviour would reflect a geometrical picture such as that arising from of a black disk with all partial waves to be zero beyond a maximum impact parameter value $b<R$, i.e. angular momentum values $bk=l<L_{max}=kR$. As discussed in more detail in  Section \ref{sec:elasticdiff},  the black disk picture gives
\begin{equation}
\frac{d\sigma}{dt}\sim \pi R^2 \frac{J_1^2(Rq)}{|t|}
\end{equation}
with $\sqrt{-t}\ =\ q\ =\ k\theta$ and $J_1$ the Bessel function of order 1. At small values of $Rq$, $J_1$ can be approximated by an exponential with slope defined by the interaction radius and one can write
\begin{equation}
\frac{d\sigma}{dt}\approx \frac{\pi R^4}{4} e^{R^2t/4} 
\end{equation}
Integrating the above equation, one obtains the elastic and the total cross-sections in the black disk limit, i.e.
\begin{equation}
\sigel=\pi R^2,\ \ \ \ \ \sigtot=2\pi R^2,\ \ \ \ \ \frac{\sigel}{\sigtot}=\frac{1}{2} 
\end{equation} 
As we shall see in later sections, even considering latest LHC and cosmic ray results at $57\ TeV$, such behaviour is not observed yet for the ratio of the two \x s. Indeed what one has so far, before the LHC measurements described in the next sections, is shown in Table ~\ref{isr-tevatron}.
\begin{table}
\centering
\caption{Total and elastic \x \ from ISR to the Tevatron.}
\label{isr-tevatron}       
\begin{tabular}{|clclclc|}
\hline\noalign{\smallskip}
$\sqrt{s} \ (GeV)$ &process& $\sigtot \ (mb)$ & $\sigel/\sigtot$  \\
\noalign{\smallskip}\hline\noalign{\smallskip}
62                           &pp       &    $43.55\pm 0.31 $                         & 0.175$ \pm$ 0.004 \\
546                        & \pbarp& (63.3$\pm$1.5)/(1+$\rho^2$) & 0.213$\pm$ 0.06 \\
1800                     &\pbarp & 71.4 to 80.0                           & 0.246$\pm$ 0.04\\
\noalign{\smallskip}\hline
\end{tabular}
\end{table}
	
\subsubsection{The $\rho$ parameter at the Tevatron}\label{sss:rhoTeV}

In previous subsections we have described the measurement of various quantities related to $\sigtot$, among them the $\rho-$parameter, the ratio of the real to the imaginary part of the forward scattering amplitude.  The behaviour of the $\rho$ parameter  with energy has been discussed in many papers. In most of the literature concerning models for the total \x, $\rho$ is considered to be small and often taken to be $\approx 0$ so as to simplify many analytical calculation, as in most mini-jet models for instance.  The  sudden change of perspective arose in the 1980's, when the experiment UA4 at the CERN $Sp\bar{p}S$ reported a measurement   well above previous theoretical estimates. This results however  was soon superseded by a more precise measurement,  in agreement with theoretical expectations.  At the Tevatron,
 in the range $0.001\le|t|\le 0.14 \ (GeV/c)^2$,
 a 3-parameter least square fit, gave \cite{Amos:1991bp},
 \begin{eqnarray}
 \rho&=&0.140\pm 0.069\\
 B&=&16.99\pm 0.047 \ (GeV/c)^2\\
 \sigtot&=&72.8\pm 3.1 \ mb
 \end{eqnarray} 
 These values are consistent within the quoted errors with the earlier E710 values  \cite{Amos:1989at,Amos:1990jh,Amos:1990fw} and supersede them.
 
 In Fig. ~(\ref{fig:rhoaspen}), we show a compilation from \cite{Block:1998hu}. The full line represents the result from the QCD inspired model by Block and other collaborators, which will be discussed in the section dedicated to models. 
\begin{figure}
\resizebox{0.5\textwidth}{!}{%
  \includegraphics{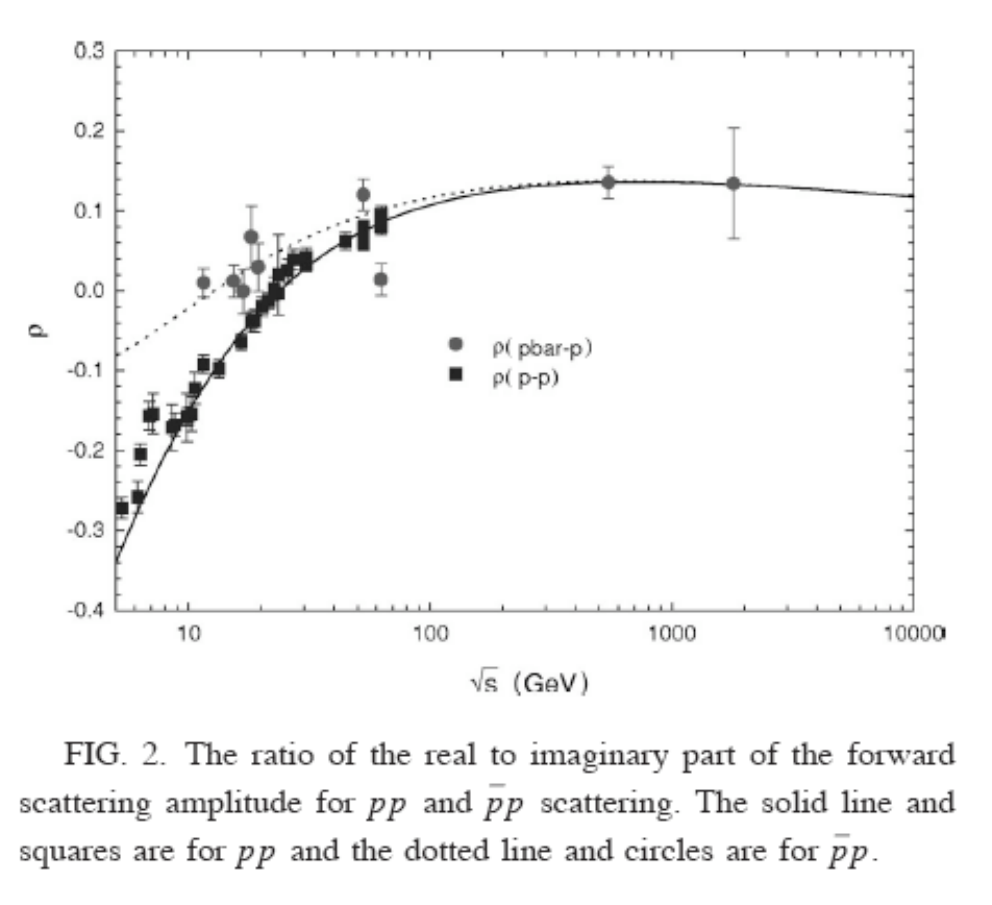}
}
\caption{The $\rho$ parameter as a function of the CM energy and its comparison with the prediction from  \protect\cite{Block:1998hu}.
Reprinted with permission from \cite{Block:1998hu}, \copyright (1999) by the American Physical Society.}
\label{fig:rhoaspen}       
\end{figure}

\subsection{Conclusions}\label{ss:conclusion}
The period of experimentation discussed above led to an enormous change in the view that physicists had held until then, due
to the observed rise in the total cross-section, changes in the value of $\rho$ and the beginning of tension in the 
dependence of the slope parameter with energy. These results would lead to radically different formalisms and models 
for higher energy experiments during the following three decades. This is a subject matter  which we shall discuss at length in the coming  sections of this review.

\section{Theoretical scenarios  and phenomenological applications}

\label{sec:models}

In this section an overview of the state-of-the-art of  theoretical and phenomenological aspects of total cross-sections is presented. 

We show in Fig. ~\ref{fig:sigtot2013} 
a compilation of   total cross-section data, from accelerators and cosmic ray experiments, with photon cross-sections 
normalized at low energy together with proton data  \cite{Godbole:2008ex}.
The dashed and full curves  overimposed to the data are obtained from a mini jet model with soft gluon $k_t-$resummation \cite{Corsetti:1996wg,Grau:1999em,Godbole:2004kx}, which we call  BN model, and which will 
described later in \ref{sss:BN}. 
\begin{figure}[htb]
\centering
\resizebox{0.5\textwidth}{!}{
\includegraphics{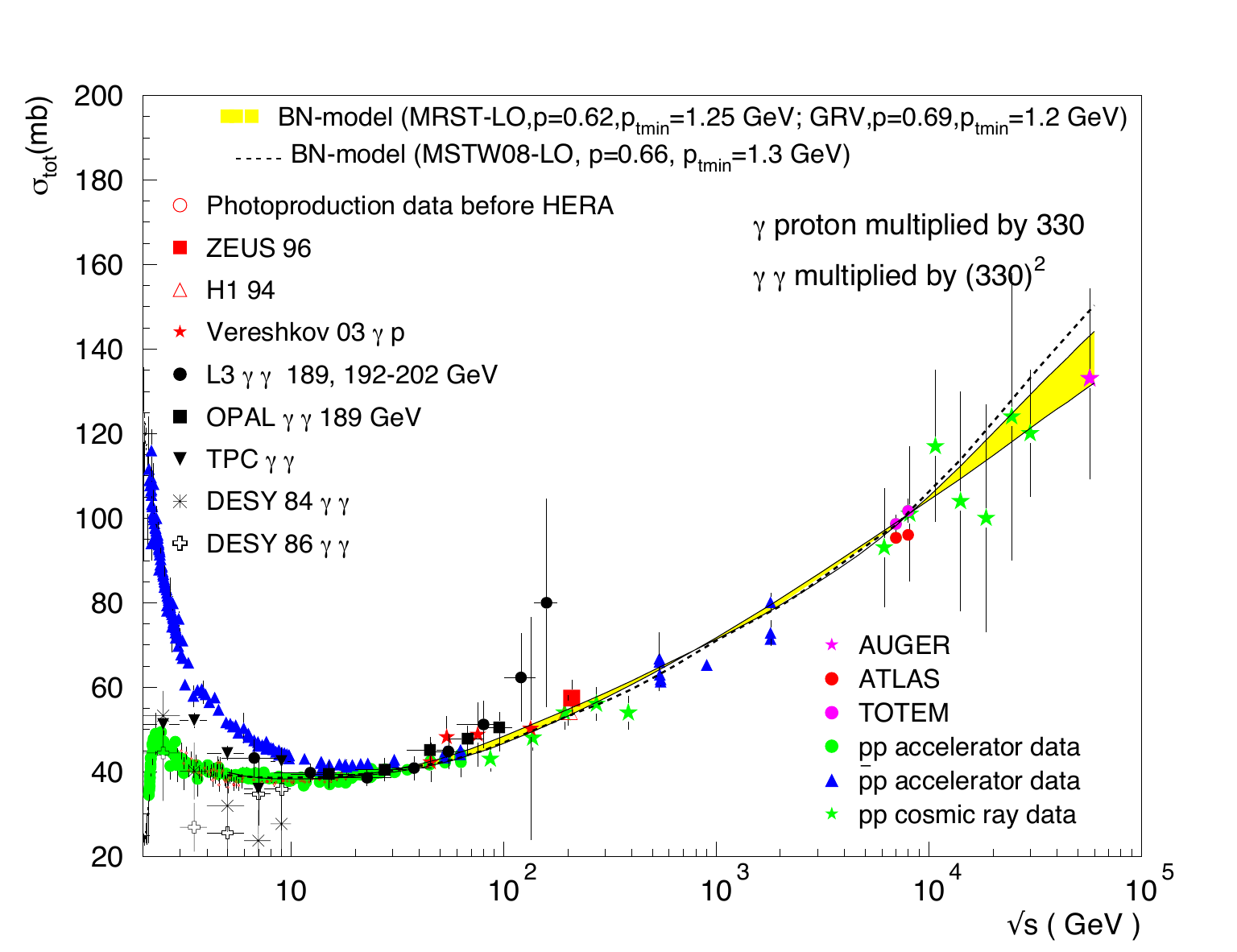}}
\caption{Total cross-section data for  $pp$ and $p{\bar p}$ 
scattering  together with normalized $\gamma p $ and $\gamma \gamma \ $ data. Curves describe predictions from a mini-jet model with 
soft gluon resummation, and 
 has been updated from the corresponding one in \cite{Godbole:2008ex} courtesy of A. Grau, with MSTW08  curve courtesy of D. Fagundes.}
\label{fig:sigtot2013}
\end{figure}

One is often asked what one can learn from  total cross-section measurements. Although the total cross-section is proportional 
to the imaginary part of the elastic scattering amplitude in the forward region, and thus it  can shed only a limited light on   
the dynamics of scattering,  the interest in such a quantity -since more than 60 years-  indicates that  it can give information on 
fundamental questions of particle physics.

Indeed, the total cross-section is the {\it golden observable} as far as QCD confinement dynamics is concerned: its behavior 
is dominated by the large distance behavior of the interaction, and thus by QCD confinement dynamics.  This dominance of 
large distance behavior implies very low-momentum exchanges,  characterized, at high energy, by  gluons with $k_t\rightarrow 0$. 
These very soft quanta need to be resummed, not unlike what happens in QED and,   the problem of the high energy behavior of 
total cross-sections appears  related to the one of radiative corrections to parton-parton scattering. Resummation for such effects, 
and hence integration over the infrared region,  being mandatory,  a knowledge or a model for the behavior of emitted gluons in this 
domain is needed for any parameter free description.
 In our opinion, models which do not access  the infrared region, for instance 
  introducing an infrared cut-off, may provide good 
 phenomenological descriptions and some understanding of the dynamics, but so far fail  to shed light on the essential problem. 

The still unsolved problem of confinement  is presently the reason why  there is no model allowing to   calculate the total \x \  
from first principles  from low energies to high energies. In the context of the total \x , we shall define as {\it low energy} the region 
after the resonances have died out, $\sqrt{s}\approx (5\div 10)$  GeV for protons, and  as {\it high energy} the region where  
$(10\div 20)$ GeV $\le \sqrt{s}\lesssim (10\div 20)$  TeV. We shall refer to higher energies, accessible through cosmic rays, as the very high energy region.

The above  distinction between low and high energy is 
not
purely phenomenological. 
As the c.m. energy rises, partons inside the scattering hadrons  can undergo hard or semi-hard collisions. 
Such collisions, by definition, are  describable with perturbative QCD (pQCD). In this regime,   partons   
of momentum $p=xP_h$ are extracted from  a hadron of momentum $P_h$   with a $1/x$ spectrum and   
scatter into final state partons of transverse moment $p_t$, with a strength calculable through  the asymptotic 
freedom expression for the strong coupling constant, given to lowest order as
\be
\alpha_{AF}(Q^2)= \frac{1}{b_o 
\ln {Q^2/\Lambda^2_{QCD}}
} \label{eq:alphaAF}
\ee
Eq.~(\ref{eq:alphaAF}) is valid for $Q^2>>\Lambda_{QCD}^2$, basically for $Q^2\sim p_t^2 \gtrsim 1 \rm {GeV}^2$.
At the same time, as the c.m. energy rises, parton emission  for given momentum $p$ probes  decreasing 
values of $x$, and, due to the $1/x$ spectrum, 
leading    to an increase of the cross-section, as $x<<1$. Combining the 
spectrum behavior with Eq. ~(\ref{eq:alphaAF}), and calling {\it high energy} the region where pQCD starts taking 
over, we have that the transition to the perturbative region will occur when 
\be
 1/x \geq \sqrt{s}/2\ p_{tmin} >>1\ and \  p_{tmin} \simeq 1\ {\rm GeV}
\ee
For $x\simeq (0.1\div 0.2)$   the turning point where pQCD starts playing a substantial role can be seen to occur when
\begin{eqnarray}
\sqrt{s} \gtrsim (2 / x)\ {\rm  GeV}  \\
 \sqrt{s}\sim (10\div 20) \ {\rm GeV} 
\end{eqnarray}
Indeed, data indicate that, after the  resonances die out,    the \pbarp \  cross-section   keeps on decreasing until reaching a  cm energy 
between 10 and 20 GeV.  It is here that  the cross-section undergoes a relatively fast rise,  easily described by a power law, 
which levels off as the energy keeps on increasing. 
In the case of \pp , the initial decrease is very mild and the rise may start earlier. Notice that for pion  \x s, the   onset of the high energy region may be considered to    start earlier, as one can see 
from a compilation of $\pi \pi$ and $\pi p $ total cross-sections, shown in Fig.~\ref{fig:pions} from \cite{Grau:2010ju}. 
In this figure, the overlaid curves correspond to the same  model as in Fig.~\ref{fig:sigtot2013}, discussed later.
\begin{figure}
\centering
\vspace{-2cm}
\resizebox{0.5\textwidth}{!}{%
\includegraphics{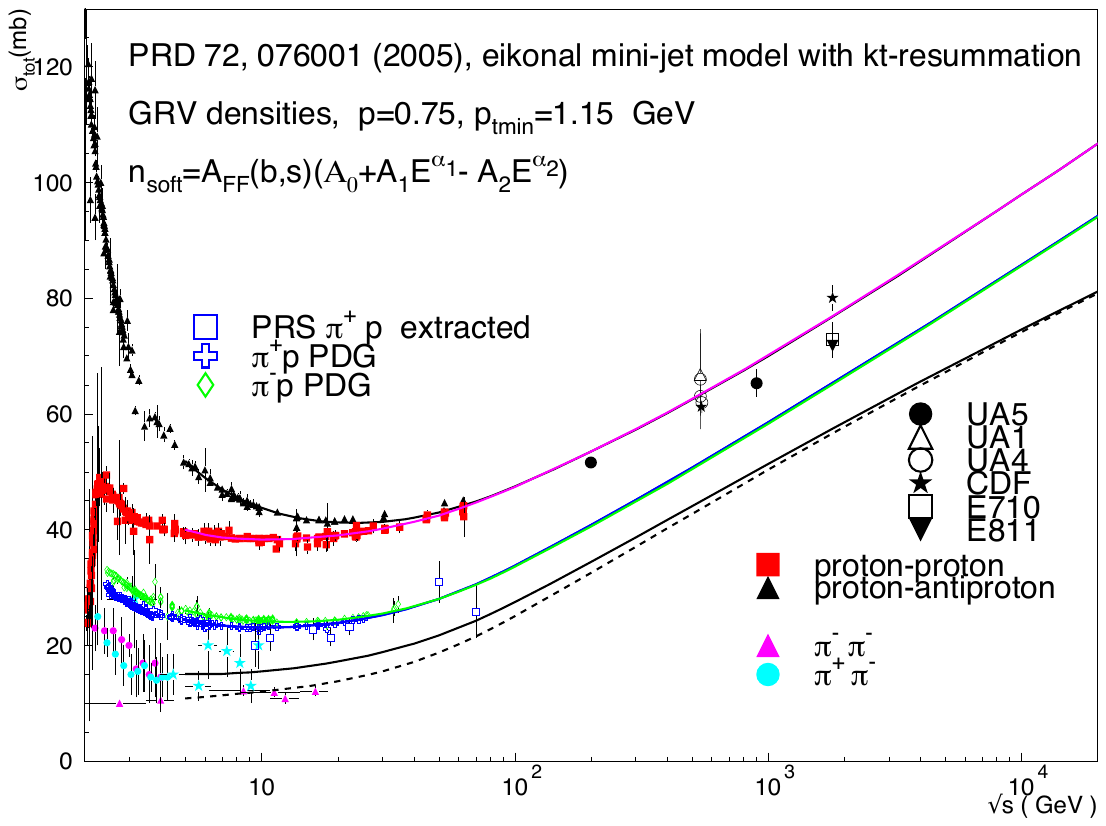}}
\vspace{-3cm}
\caption{Proton and pion total cross-sections, as indicated, from \cite{Grau:2010ju}. Reprinted from 
\cite{Grau:2010ju}, \copyright (2010) with permission from Elsevier.}
\label{fig:pions}
\end{figure}

As for the high energy behaviour of all total \x s, there are two main features which need to be 
properly addressed in any
description of data in the TeV region: (i) 
how to include the 
mechanism  which drives the rise on the one side 
and (ii) what
dynamics
transforms the 
early, almost sudden, power law-like rise into  the smoother observed behaviour, consistent with the Froissart bound, $\sigtot \lesssim \log^2s
$.  A cartoon description of this transformation appears in  Fig.~\ref{fig:cartoonfroissart}.
\begin{figure}
\centering
\resizebox{0.3\textwidth}{!}{%
\includegraphics{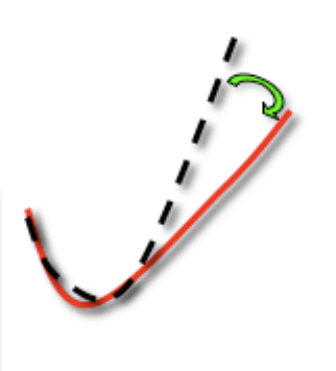}}
\caption{The softening of the total cross-section from early rise to logarithmic type behavior, consistent with the Froissart bound.}
\label{fig:cartoonfroissart}
\end{figure} 
There is a general understanding that the rise is produced by an increasing number of low momentum parton-parton collisions, 
and a similar general understanding that an effect called ``saturation", brings a balance and the Froissart-like behaviour. 
However, models differ in their detailed description. In our  model, such a saturation is a consequence of the 
scheme we propose for   infrared gluon $k_t$-resummation.

 The large number of models available, can be divided in  various groups as follows:
\begin{itemize}
\item geometrical models
\item eikonal models
\item Gauge boson trajectories: the photon and the gluon
\item  Reggeon field theory
\item QCD minijets
\item AdS/CFT approaches
\item fits with Froissart bound asymptotic constraint 
\end{itemize}

 We shall dedicate some space to each of these different approaches, with however particular attention to the 
 QCD models and to efforts on resummation, resulting in what is phenomenologically described as the 
 Pomeron Trajectory. To put this in perspective, we present an extended summary of the problem of infrared 
 radiative corrections in QED, including the question of the photon trajectory, and then a brief overview of 
 the Balitsky, Fadin, Kuraev and Lipatov (BFKL)  approach to hadron scattering.
 
Not all readers are expected to be familiar with all the theoretical backgrounds, and 
it is clearly impossible to render justice to a field which has been active at least for 
70 years, with new data  appearing both from accelerator and cosmic ray experiments. The literature on the subject 
is very large and still increasing and the problem  presents yet unsolved aspects. Because  many books and reviews 
are available on specific models, our choice in this review has been to follow a historical path and highlight some of 
less treated aspects in models. 
The following aspects will be examined:
\begin{itemize}
\renewcommand{\labelitemi}{$\bullet$}
\item Moli\`ere's theory in \ref{subsect:moliere},
\item Heisenberg model in \ref{subsect:heisenberg},
\item on the Froissart bound in \ref{subsec:froissart},
\item the impact picture: Cheng and Wu (also with Walker) and Bourelly, Soffer and T.T. Wu in \ref{subsec:impact},
\item Donnachie and Landshoff Regge-Pomeron description in \ref{subsec:dl},
\item hadronic matter distribution in \ref{subsec:hadronicmatter},
\item role of resummation in QED in  \ref{ss:resum},
 \begin{itemize}
	 \item a digression on the Rutherford  singulartity in \ref{sss:rutherford}
	\item Bloch Nordsieck theorem in \ref{sss:BNQED},
	\item Touschek and Thirring about  covariant formalism of Bloch and Nordsieck theorem in \ref{sss:TT},
	\item Schwinger's exponentiation in \ref{sss:schwinger},
	\item Double logarithms in QED: the Sudakov form factor in \ref{sss:sudakov},
	\item Early 60s and exponentiation in \ref{sixties} ,
	\item Semi-classical approach to resummation in QED in \ref{sss-models:etp},
	\item Reggeization of the photon in \ref{sss:reggeization-photon},
        \end{itemize}
\item Role of resummation in QCD in \ref{ss:resumQCD},
         \begin{itemize}
	 \item BFKL approach in  \ref{sss:BFKL},
	 \item The odderon in \ref{sss:odderontot},
	\item Odderon in QCD in \ref{sss:odderons-QCD},
	\item Gribov, Levin and Ryskin in \ref{sss:GLR},
	\item BFKL inspired models  in \ref{sss:KMRtot}
        \end{itemize}
\item Mini-jet models in \ref{ss:minijets},
	\begin{itemize}
		\item Non-unitary models and the rise of $\sigma_{total}$ in \ref{sss:gaisserhalzen},
		\item QCD inspired eikonal models in \ref{sss:eikonal-mini},
		\item Mini-jets and infrared $k_t$ resummation in \ref{sss:BN} through  \ref{sss-models:froissart},
\end{itemize}
\item AdS/CFT models in \ref{ss:ads},
\item phenomenological fits in \ref{ss:phenofits},
\item the asymptotic behavior of total cross-section models in 
theories with extra-dimensions in 
\ref{ss:High_D}
\end{itemize}
Further discussion of related items can be found in the coming Section \ref{sec:photons}, where  one will also find a presentation of the non-linear Balitsky-Kovchegov (BK) equation.  

\subsection{ Moli\`ere theory of multiple scattering}\label{subsect:moliere}
Most models for the total \x \ are based on  the optical theorem,  and many of them use models for the differential 
elastic cross-section . In this subsection we wish to recall the general features of one such model, 
Moli\'{e}re's theory of multiple scattering \cite{Moliere:1947zz}, developed  for the scattering of electrons on atoms,  
summarized, and compared to other pre-existent models, by Bethe \cite{Bethe:1953va} .

Moli\'{e}re's theory of multiple scattering  is valid for small angle scattering, i.e.  $\sin\theta\approx \theta$, 
and is based on 
the transport equation for $f(\theta,t)$, with $f(\theta,t)d\theta$ the 
number of electrons scattered within an  an angle $d\theta$ after passing through a slab of atoms of thickness $t$, 
and  an ansatz for the probability of small angle single scattering.
The starting point is 
\begin{eqnarray}
\frac{\partial f(\theta,t)}{\partial t}=-Nf(\theta,t) \int \sigma(\chi)\chi d\chi + \nonumber \\
+N\int  f(\theta',t)\sigma({\chi}) d{\vec \chi}
\label{eq:moliere}
\end{eqnarray}
where $N$ is the number of scattering atoms per unit volume, $\sigma(\chi)\chi d\chi$ is the electron-atom cross-section 
into the angular interval $d\chi$ after traversing a thickness $t$.
In Eq.~(\ref{eq:moliere}) the first term corresponds to electrons which were scattered {\it away} from the studied position, 
namely probability of being originally  at angle $\theta$ times the number scattered away in any direction, while the second 
is the probability of electrons scattered {\it into} the observed position from any process, and 
${\vec \theta}'= {\vec \theta} -{\vec \chi}$, $ d{\vec \chi}=\chi d\chi d\phi/2\pi$.  
Taking the Fourier transform of $f(\theta,t)$
\begin{equation}
f(\theta,t)=\int_0^\infty \eta d\eta J_0(\eta\theta)g(\eta,t)
\end{equation}
one can use Eq.~(\ref{eq:moliere}) obtaining, in the notation of Moli\'ere,
\begin{equation}
g(\eta,t)=e^{\Omega(\eta,t)-\Omega_0(t)}
\label{eq:omega}
\end{equation}
where
\begin{equation}
\Omega(\eta,t)=N t \int_0^\infty \sigma(\chi)\chi d\chi J_0(\eta\chi)
\end{equation}
with $\Omega_0(t)\equiv \Omega(\eta=0,t)$ having the physical meaning of the total number of collisions. 
Notice that Eq.~(\ref{eq:omega}) uses the fact that $g(\eta,0)=1$, which follows from the fact that the incident beam is exactly 
in the direction $\theta=0$, i.e. $f(\theta,0)=\delta({\vec \theta})$.
One can then  solve for $f(\theta,t)$ obtaining
\begin{eqnarray}
f(\theta,t)=\int_0^\infty \eta d\eta J_0(\eta\theta) \nonumber \\
\times exp[-Nt\int_0^\infty \sigma(\chi)\chi d\chi\{1- J_0(\eta\chi)\} ]
\label{eq:bethe7}
\end{eqnarray}
Bethe points out that this equation is exact provided the scattering angle is small. He then proceeds to describe  
the approximations used by Moli\'ere to evaluate the integral and, in the remaining sections, to compare these 
results with others.\\
 
Moli\'{e}re's theory of multiple scattering analyzes and proposes models for 
the scattering probability and it is 
an early example of resummation of small angle scattering.

\subsection{The Heisenberg model}\label{subsect:heisenberg}
An early estimate of the total hadronic cross-section was obtained by
 Heisenberg in 1952 \footnote{A translation of this article can be found at http://web.ihep.su/dbserv/compas/src/heisenberg52/engl.pdf} 
\cite{Heisenberg:1952zz},
\begin{equation}
\sigtot \approx \frac{\pi}{m^2_\pi}\ln^2\frac{\sqrt{s}}{<E_0>}
\label{eq:heisenbergfroissart}
\end{equation}
where $<E_0>$ is the average energy per emitted  pion. 
We will 
describe  the argument behind Eq. ~(\ref{eq:heisenbergfroissart} )following the very clear  presentation by 
Kang and Nastase \cite{Kang:2005bj}, who then use it to derive similar results 
in  AdS/CFT.

The description of scattering is the by now familiar picture of two   
hadrons colliding and interacting through their surrounding  cloud of pions. 
Due to Lorentz contraction at high energy, we are dealing with two thin 
pancakes, and the scattering degrees of freedom are in  the transverse plane, 
where the impact parameter $b$ describes the scattering, as in  
Fig.~\ref{fig:kangheisenberg} from \cite{Kang:2005bj}.
 \begin{figure}
\resizebox{0.5\textwidth}{!}{%
  \includegraphics{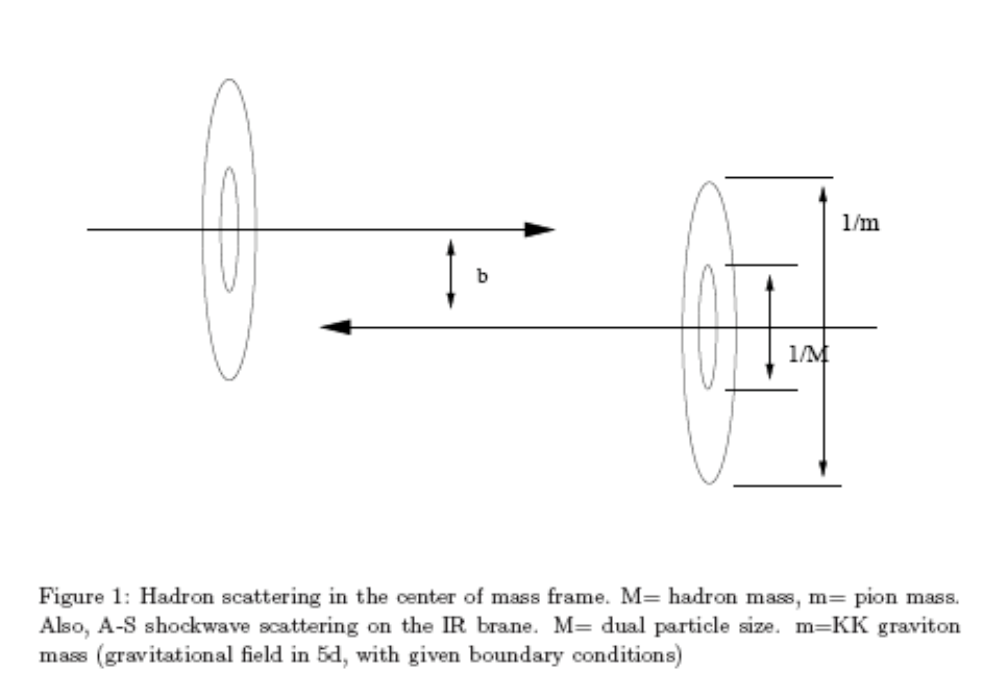}
}
\caption{Description of the scattering in impact parameter space from 
\protect\cite{Kang:2005bj}, leading to Heisenberg total cross-section 
formulation, 
and  to the AdS/CFT derivation, as discussed later. \copyright 2005 Elsevier. Open access under CC BY License.}
\label{fig:kangheisenberg}       
\end{figure}
In such a picture, the total cross-section is written as
\begin{equation}
\label{sigtotmax}
\sigtot = \pi b_{max}^2
\end{equation}
where $b_{max}$ is the largest impact parameter value which still allows pion 
emission, so that the expression at the r.h.s of  Eq~(\ref{sigtotmax}) can 
also be considered 
the maximum value which the total cross-section can take, as the energy 
increases.

The challenge is then to find $b_{max}$ and its energy dependence.
Let 
${\cal E}$ be the energy emitted 
 in the scattering at impact parameter $b$, %
and $<E_0>$  the average pion energy  for scattering at a c.m energy 
$\sqrt{s}$. 
 One can then write
\begin{equation}
{\cal E}=\alpha \ \sqrt{s}\  \ge \ \ 
n <E_0>
\end{equation}
where 
$n$ is the number of pions emitted 
with   $\alpha$ a proportionality constant which reflects the overlap of the 
wave functions of the pions surrounding the 
colliding hadrons. As the transverse distance between the two hadrons 
increases,  the pion wave function can be expected to decrease as an 
exponential, and, at the maximum distance for the scattering to still take 
place, one can write
\begin{equation}
\alpha=e^{-b_{max}m_\pi}
\end{equation}  
since $m_\pi$ is the size of the pion cloud. One then obtains 
Eq.~(\ref{eq:heisenbergfroissart}). This equation however 
does not provide much information, unless one can 
determine how the average pion energy depends on the c.m. energy. 
To obtain this energy dependence,
for instance whether the total \x\  is a constant or increasing with the square of the 
logarithm of the c.m. energy or any other behavior,  
one 
needs a model to calculate $<E_0>$. 
Notice that if $<E_0>$ is proportional to $\sqrt{s}$, then the total 
cross-section would go to a constant, whereas the rise with a logarithmic 
power is only ensured by 
a constant average pion energy. We shall now see, from \cite{Kang:2005bj},
which was the procedure followed by Heisenberg.

One starts with calculating the differential energy radiated away  during the
 collision for the case of a free massive scalar pion of energy $E_0$,
\begin{equation}
\label{eq:spectrum}
\frac{d{\cal E}}{dE_0}=A=constant
\end{equation}
which is 
found to be a constant  up to a maximum 
energy $E_{0,max}=\gamma m_\pi$ with $\gamma\approx \sqrt{s}/ M_H$, 
$M_H$ being the hadron mass.
 
 Apparently a similar argument  was also
pictured by Sommerfeld in  his theory of the production of X-rays (according to Touschek  \cite{Touschek:1968zz}). If the 
collision takes place in a   time interval much shorter than the one 
characterizing the emission, 
 then the process  can be described by a $\delta-$function in time, whose 
Fourier transform is a constant. This would lead to a constant spectrum for 
the energy emitted in the process, namely  Eq.~(\ref{eq:spectrum}).
 
The number of emitted pions per 
unit energy radiated is then obtained as
\begin{equation}
\label{eq:dn}
\frac{dn}{dE_0}=\frac{A}{E_0}
\end{equation}
Now, to get the energy ${\cal E}$ and the number $n$ of emitted mesons, 
we integrate Eqs.~(\ref{eq:spectrum},\ref{eq:dn})
between $m_\pi$ (we need to emit at least one pion) and  $E_{0,max}$. With the maximum energy, which can be emitted, given by 
 $E_{0,max}\approx \sqrt{s}\ m_\pi /M_H$, one gets 
\begin{eqnarray}
{\cal E}=\int_{m_\pi}^{E_{0,max}}
 d{\cal E}
\approx  A (\sqrt{s}\ m_\pi /M_H -m_\pi)
\end{eqnarray}
and
\begin{equation}
n=\int dn =A \int \frac{dE_0} {E_0}=A \ln \frac{E_{0,max}}{m_\pi} 
\end{equation}
which immediately gives the average energy $<E_0>$ from
\begin{equation}
<E_0>=\frac {\cal E}{n}=
\frac{(\sqrt{s} \  m_\pi /M_H -m_\pi)}{\ln \frac{E_{0,max}}{m_\pi} }
\end{equation}
This indicates that the average energy $<E_0>$ increases with the c.m. energy 
$\sqrt{s}$, apart from logarithmic terms. For such a case, then the 
total cross-section would go to a constant. The above result 
follows, according to \cite{Kang:2005bj}, from the equation of motion for a 
free pion. But the pion is not  free at high energy and the equation of 
motion, $[\Box -m^2_\pi]\phi=0$ is not valid. At this point Heisenberg 
took the Dirac-Born-Infeld-like action for the scalar pion 
\begin{equation}
S=l^{-4}
\int d^4 x
\sqrt{
1+l^4[
(\partial_\mu \phi)^2+m^2\phi^2
]
}
\end{equation} 
with a length scale $l$ and obtained
\begin{equation}
\label{eq:infeld}
\frac{d{\cal E}}{dE_0}=\frac{A}{E_0} \to \frac{dn}{dE_0}=\frac{A}{E_0^2}
\end{equation}
Now, using Eq.~(\ref{eq:infeld}), one repeats the above steps, namely
\begin{equation}
{\cal E}=\int_{m_\pi}^{\sqrt{s}\ m_\pi /M_H} d{\cal E}=A \ln \frac{E_{0,max}}
{m_\pi}; 
\end{equation}
and
\begin{equation}
 n=
 \int dn =
 A \int \frac{dE_0} {E_0^2}
\approx  \frac{A}{m_\pi} (1- \frac{M_H}{\sqrt{s}})
\end{equation}
At this point, one can calculate the average energy 
\begin{equation}
<E_0>\equiv \frac{\cal E}{n}= m_\pi
\frac{
\ln 
\frac 
{E_{0,max}}
{m_\pi}
}{
[1-
 \frac{m_\pi}
 {E_{0,max}}
 ]}=m_\pi 
 \frac{\ln\gamma}
 {1-1/\gamma}\approx m_\pi \ln\gamma
\end{equation}
with $\gamma\approx \sqrt{s}/M_H$ .
With this energy distribution for the pion field, the  average energy $E_0\propto m_\pi$ grows only logarithmically.
 This then immediately leads to a maximal behavior consistent with the Froissart limit.

Kang and Nastase comment further on this result. They  write that the minimum energy 
emitted could be mistakenly understood to be $m_\pi$, but it is instead the 
pion energy, which for a free pion would grow linearly with $\sqrt{s}$. On 
the other hand one needs an action with higher power of the derivatives,
such as  the 
Dirac-Born-Infeld action,  to obtain a constant value for $<E_0>$ 
proportional to the pion mass. In their subsequent treatment 
in terms of AdS/CFT, the model by Kang {\it et al.} is applied to the case of pure gauge 
theories and the authors  will talk interchangeably of pions and lightest glueballs.

One can unify the  two derivations from the previous part, by using
\begin{equation}
\label{eq:dep}
\frac{d {\cal E}}{dE_0}=A(\frac{\mu}{E_0})^{p}
\end{equation}
and hence
\begin{equation}
\label{eq:dnp}
\frac{d n}{dE_0}=A\frac{1}{E_0}(\frac{\mu}{E_0})^{p}
\end{equation}
which would give the two previous cases in the limit  $p=0$, i.e. 
constant cross-section and  average pion energy increasing with c.m. 
energy, or  $p=1$ with constant average pion energy and cross-sections 
limited by $\ln^2\sqrt{s}$. For dimensional reasons, we must introduce 
the pion mass already in Eq. ~(\ref{eq:dep}).

Then both the above results can be written in a single expression,  
with $0<p<1$. We write for simplicity $m_\pi=\mu$ 
and then $E_{0,max}\approx \gamma \mu$.
Integrating Eqs.~(\ref{eq:dep}),\ref{eq:dnp}), we get
\begin{eqnarray}
{\cal E}&=&
\frac{A\mu}{1-p}
(\gamma^{1-p}-1)\\
& &\xrightarrow[p \to 1] \ A\mu \ln \gamma\\
& &\xrightarrow[p\to 0] \ A\mu \gamma  
\end{eqnarray}
and 
\begin{eqnarray}
n=\frac{A}{p}\left[1-(\frac{1}{\gamma})^{p}\right]
& &\xrightarrow[p\rightarrow 1]\  A\left[1-\frac{1}{\gamma}\right]\\ 
 & & \xrightarrow[p\rightarrow 0]\ A\gamma \ln{\gamma} 
\end{eqnarray}
so that
\be
<E_0>=\frac{\mu p}{1-p}\frac{\gamma^{1-p}-1}{1-\gamma^{-p}}
\ee
and as a result we have the two limits 
\begin{eqnarray}
<E_0>&\ & \xrightarrow[p\to 1] \ \mu \frac{\ln\gamma}{1-\frac{1}{\gamma}}\\
& &\xrightarrow[p\to 0] \ \mu \frac{\gamma}{\ln\gamma}
\end{eqnarray}
as it should be.

\subsection{ A 
general  observation about the various ways to obtain 
the Froissart bound}\label{subsec:froissart}
Heisenberg's  argument is  geometrical to begin with, but dynamics  enters 
in defining the average pion energy. The geometrical argument is also 
the one used by Froissart \cite{Froissart:1961ux}, and in all the others 
derivations of  the bound, 
including Martin's \cite{Martin:1962rt} and Gribov's \cite{Gribovbook},
 as seen in Section \ref{sec:general}.
These derivations are  all obtained with
\begin{itemize}
\item optical theorem , $\sigtot \propto \Im mA(s,t=0)$ 
\item partial wave expansion truncated at $L_{max}$ so that 
\be
\Im mA(s,t=0)\le\sum_0^{L_{max}}(2l+1)=L^2_{max}
\ee
\end{itemize}
In other words the derivations are related to the partial 
waves falling off at high energy for a finite $L_{max}$ which, in impact 
parameter space,  then becomes proportional to a $b_{max}$. The connection 
between $L_{max}$, alias $b_{max}$, and the energy  comes from the 
high $l-$behaviour of the Legendre functions, and the 
energy dependence 
enters because the scattering  amplitudes are said to grow at most like a 
polynomial in $s$. The difference between Heisenberg's argument and the 
$S$-matrix derivations seems to be that for the latter  the energy dependence comes from 
the hypothesis on the amplitudes  taken to grow 
with energy, whereas for Heisenberg, to obtain the limit one needs an average 
pion energy to be a constant and  total energy emitted 
 proportional to the c.m. energy.  

Let us 
repeat here the heuristic argument given by Froissart, at the 
beginning of his paper, to 
obtain his result.
It must be noted that this intuitive explanation 
relies upon the existence of confinement. Indeed, the whole description 
applies not to  parton scattering but to hadron scattering. 

Let the two  
hadrons see each other at large distances through 
 a Yukawa-type  potential, namely $ge^{-\kappa r}/r$, 
where $\kappa$ is some momentum cut-off.
Let $a$ be the impact parameter, then  the total interaction seen by 
a particle for large $a$ is proportional to $ge^{-\kappa a}$. When 
 $ge^{-\kappa a}$ is  very small, there will be 
practically no interaction, while, when $ge^{-\kappa a}$ is close to $1$,
 there will be maximal probability for the interaction. For such values of $a$,
$\kappa a=\ln |g|$ one can then  write for the cross-section 
$\sigma\simeq ( \pi/\kappa^2) \ln^2|g|$. If $g$ is a function of energy and 
we assume that it can grow with energy at most like a power of $s$, then one 
immediately obtains that the large energy behaviour of the total cross-section
is bound by  $\ln^2 s$. What $\kappa$ is, remains undefined for the time
 being, except that it has dimensions of a mass.

Since Heisenberg's early result, many attempts have been made to reproduce 
it with modern field theory techniques. Leaving aside for the moment, 
the Regge-Pomeron language, one can summarize this result and related
 attempts, including those of Froissart and Martin,  as follows:
\begin{itemize}
\item confinement is input to the derivation as one considers   pions as a cloud around the interacting 
hadrons, represented by a fall-off of the cross-section at large 
impact parameter values. In  \cite{Heisenberg:1952zz} a value $b_{max}$ 
is defined and related to the energy emitted. In Froissart's heuristic 
explanation of his derivation, the potential is of the Yukawa type, 
with the coefficient of the term in the exponential in
 $r$, proportional to a constant $\kappa$, which will later turn out to be 
the pion mass, 
(similar to the coefficient in Heisenberg's exponential);  
\item with such an exponential behaviour, $\sigma_{tot} $, 
by definition  proportional to $b_{max}^2$
, will be proportional to 
\begin{description}
\item $\bullet$ inverse of the scale, which is $m_\pi$
\item $\bullet$ a logarithm of a function of the  energy scales of the collision,  
$f(\sqrt{s},m_\pi,m_H) $
\end{description}  
\item the average emitted pion energy is what determines the function 
$f(\sqrt{s},m_\pi,m_H) $, through the relation $<E_0>=\sqrt{s} e^{-b_{max}m_\pi}$
\item for a free pion field $f(\sqrt{s},m_\pi,m_H) =
\frac {\gamma}{\ln \gamma} $, while for the more realistic case of not free 
pions,  Heisenberg obtains $f(\sqrt{s},m_\pi,m_H)= \ln \gamma$ with 
$\gamma\approx \sqrt{s}/M_H$
\end{itemize}
To obtain the logarithmic dependence in the cross-section, it is then 
necessary to understand the behaviour of the function
 $f(\sqrt{s},m_\pi,m_H) $. In Froissart, 
the elastic amplitude is assumed proportional to a finite power of the energy and this 
brings in the energy term in the logarithm.

Finally, let us notice a  recent paper by Azimov \cite{Azimov:2012tw}, where the fundamentals of the Froissart bound are 
revisited and the possibility of a different asymptotic behaviour of the total cross-section is proposed.



\subsection{The impact picture}\label{subsec:impact}
The impact picture for particle scattering is still at the basis of many of the proposed 
descriptions for the total cross-section. It is often obtained as a direct consequence of 
an optical model for scattering, with direct connection to the optical theorem. 

\subsubsection{Cheng and Wu description of high energy scattering, including work with Walker}
In 1970 Cheng and Wu \cite{Cheng:1971en} (CW)  described the general qualitative 
features of the impact picture for high energy scattering. 

The final  picture had
 arisen through a long series of papers, the first of which  
studied the high energy limit of elastic two body scattering amplitudes in 
(massive)  quantum electrodynamics \cite{Cheng:1969eh}. In this paper, a systematic 
study, at high energy, of all two body elastic scattering amplitudes was 
performed and  the concept of ``impact factor'' was introduced 
for electrons, positrons, nuclei (all point-like) and the (massive) photon. 
 "After 16 months and 2000 pages of 
calculations", as the authors say, it was found that 
the matrix element for 
the elastic scattering process $a+b\to a+b$,  
for small values of  the  momentum transfer ${\bf r}_1\simeq \ 0$, can be stated in the form    \cite{Cheng:1969eh}
\begin{align}
{\cal M}_0^{ab} \simeq 
i (2r_2r_3)(2\pi)^{-2} \nonumber \\
\times \int d^2{\vec q}_\perp 
[({\vec q}_\perp+{\vec r}_1)^2+\lambda^2]^{-1}
[({\vec q}_\perp-{\vec r}_1)^2+\lambda^2]^{-1}\nonumber \\ \times {\slashed S}^a({\vec q}_\perp,
{\vec r}_1){\slashed S}^b({\vec q}_\perp,
{\vec r}_1)\label{eq:CWampl}
\end{align}
This expression 
introduces
the impact factor ${\slashed S}^a$.
In Eq.~(\ref{eq:CWampl}) the amplitude is cast
as an integral over an internal variable, which obtains 
from higher order diagrams, 
with $r_2$ and $r_3$ being averages over initial and final particle momenta.
The results obtained in this paper, which  relies on a non-zero photon mass $\lambda$ to avoid infrared divergences, 
contradicts, according to the authors, previous results on Regge-poles and 
the droplet model for diffractive scattering, both of which rely on potential 
model results. 

In \cite{Cheng:1971en}  the impact picture and the 
eikonal approximation, which will later lead to their numerical prediction 
for high energy scattering (with Walker), are presented and,  in \cite{Cheng:1970bi}, the limiting behaviour of 
cross-sections at infinite energy is stated in the following major 
predictions  for two body scattering: 
\begin{enumerate}
\item
the ratio of the real to the imaginary part of the elastic 
amplitude
\be
\frac{\Re eM(s,0)}{\Im m M(s,0)}=\frac{\pi}{
ln|{\cal S}|
}+{\cal O}
(
\ln|{\cal S}|
)^{-2}\label{eq:rhoCW}
\ee
where ${\mathcal S}$ is obtained asymptotically as
\begin{align}
{\mathcal S}=\big\{  -\frac{(-s)^a}{[\ln(-s)]^2} +\frac{(-u)^a}{[\ln(-u)]^2}\big\}^{1/a}\nonumber\\
\rightarrow \frac{s/s_0}{[\ln (s/s_0)^{2/a}} \label{eq:CWasymptotic}
\end{align}
for $-t\simeq 0$. This expression shows that ${\mathcal S}$ increases at least like a power of $s$, with $a$ a positive constant;
\item
the asymptotic behaviour of the total cross-section
is given as 
\begin{equation}
\sigtot =2\pi^2 R^2+{\cal O}(\ln|{\cal S}|)
\end{equation}
with $R=R_0\ln|{\cal S}|$, and $R_0$ is a constant.
\item 
for the elastic cross-section, the impact picture, extended to $t\neq 0$  
leads to a prediction on the position of the first dip, namely to  geometrical scaling, a result previously obtained 
in \cite{Khuri:1965zz} on very general grounds, i.e. 
\be
-t_{dip} \sigma_{total}=2\pi^3 \beta_1^2+ {\mathcal O}(\ln |{\mathcal S}|)^{-1})
\ee
with $\beta_1$ corresponding to the position of the first zero of the $J_1(\pi \beta)$;
\item the ratio of elastic to the total cross-section  goes to a 
constant, namely
\begin{equation}
\frac{\sigma_{elastic}}{\sigtot}=\frac{1}{2}
\end{equation}
\end{enumerate}
The authors note that, with the exception of the result in Eq. ~(\ref{eq:rhoCW}), all the above is  model independent 
and believed to be firm, whereas the first one could only be valid in QED. We shall return to this point in Sec. \ref{sec:elastic}.

In 1973 Cheng, Walker and Wu proposed a quantitative  model, based on such picture, to 
the study of total cross-sections 
\cite{Cheng:1972ik},  and, later, also applied it to 
study the ratio of the elastic to 
the total cross-section \cite{Cheng:1974dxa}.


The impact picture presented by Cheng, Walker and Wu (CCW) represents the 
collision, as seen by each individual particle, as that of two thin pancakes, 
Lorentz contracted along the direction of motion. The pancakes are seen as 
being made of 
\begin{enumerate}
\item a black core, where essentially  total absorption takes place, 
with a logarithmically expanding radius 
$R(s)\simeq R_0\ln s$, which owes its existence to the production of 
relatively low energy particles in the center-of-mass system 
\item a grey or partially absorptive fringe, roughly independent of the energy.
\end{enumerate}
One of the immediate predictions of this picture is that the ratio of the 
total to the elastic cross-section becomes 1/2 at very high energy. As we discuss later in this 
review in the context  of diffraction,  this  is a  problem with the one channel eikonal representation.\\
Lifting  notation and everything else  from
 \cite{Cheng:1972ik}, for an elastic channel $j$, the amplitude at high energy is written as
\begin{equation}
M_j(s,\Delta)=\frac{is}{2\pi}\int d\vec{x}_\perp exp(-i{\vec\Delta}\cdot{\vec
 x}_\perp)D_j(s,{\vec x}_\perp)
 \label{eq:CW1}
\end{equation}
where ${\vec \Delta}$ is the momentum transfer and $D_j$ is written as
\begin{equation}
D_j(s,{\vec x}_\perp)
=1-exp(
-f_j
(Ee^{i\pi/2})^c
)\times 
exp(-\lambda
(
x_\perp^2+x_{jo}^2
)^{1/2})
 \label{eq:CW2}
\end{equation}
with $E$ the laboratory energy of the projectile (incident particle). While 
$c$ and $\lambda$ are universal constants, $f_j$ and $x_{jo}$ are parameters,
 which however are the same for particles and antiparticles. The factor 
$e^{-i\pi/2}$ is present to allow for crossing symmetry in the amplitudes. 
The normalization of the amplitude is such that the  differential cross-section is  given by
\begin{equation}
\frac{d\sigma}{dt}=|M_j(s,{\vec \Delta})|^2
\end{equation}
where $t=- \Delta^2$. The optical theorem then leads to the expression 
for the total cross-section
\begin{equation}
\sigma_{total}(j)=A_js^{-1/2} +\frac{4.893}{s}\Im m M_j(s,0)
\end{equation}
In the above equation, the authors added a term, defined as a {\it background}
 term, and the factor $4.893$ is a conversion factor to mb.
The asymptotic energy scale is now controlled by the power $c$ in Eq.~(\ref{eq:CW2}). When the paper
 \cite{Cheng:1972ik}
was written, the parameter was fixed through the fit to  hadronic 
data, including results 
 from 
 ISR experiments.
 The value of the parameter    was given as $c=0.082925$.
 
In a subsequent paper \cite{Cheng:1974dxa}, the model is further defined.
 The amplitude is written as the sum of 3 terms, namely a vector and 
a tensor exchange,  and the Pomeron contribution. The authors  write
\begin{equation}
\sigma_{total}(pp/p{\bar p})=
4.893[
\frac{\Im m A_p(s,0)}
{s}+
s^{-1/2}
(
{\bar \gamma}_f\mp
{\bar \gamma}_\omega
)
]\ mb
\end{equation}  
where the subscripts $f$ and $\omega$ refer to the tensor and vector exchange. The Pomeron contribution appears 
through the eikonal formulation, i.e.
\begin{equation}
A_p(s,t)=\frac{is}{2\pi}\int dx_\perp e^{
-i{\vec \Delta}\cdot {\vec x}_\perp
} D(s, x_\perp)
\end{equation}
 with
 \begin{equation}
 D(s, x_\perp)=1-exp[-S(s)F(x_\perp^2)
]
\end{equation}
With this, CWW obtain the fit to the elastic scattering parameter, presently called  the  slope parameter, further discussed in 
Sec. \ref{sec:elasticdiff},   $B(s)$, which is seen to rise slowly with energy. 
It is also seen that the ratio of the 
elastic to the total cross-section should rise by 7\% at ISR.

 \subsubsection{The impact parameter description by  Soffer, Bourrely and   Wu}
A particularly clear description of how the impact picture developed after CW work can be found in a short review paper by Jacques Soffer \cite{Soffer:1987gm} in \cite{Goulianos:1988hk}. It is recalled that QED was the only known relativistic quantum field theory in the late '60s and that CW introduced the small photon mass in order to avoid what Soffer calls {\it unnecessary complications}. It is recalled that the summation of all diagrams for Compton scattering leads \cite{Cheng:1970bi} to the asymptotic expression 
of   Eq. (\ref{eq:CWasymptotic}). 

In the model built by Soffer with Bourrely and Wu (BSW),    the scattering amplitude for proton scattering is written as
\begin{equation}
a(s,t)=a^N(s,t)\pm sa^c(t)
\end{equation}
where the signs refer to \pbarp \ and \pp \ respectively,  the hadronic amplitude is given by $a^N(s,t)$  and the factor $s$ has been factorized out of the  Coulomb amplitude $a^c(t)$. 

The hadronic amplitude is obtained from the impact picture \cite{Bourrely:1978da} as
\begin{equation}
\label{eq:}
a^N(s,t)=is \int_0^\infty b db J_0(b\sqrt{-t})(1-e^{-\Omega(s,b)})
\end{equation}
The eikonal function $\Omega(s,b)$ is split into two terms, reflecting different dynamical inputs, namely
\begin{equation}
  \Omega(s,b)=R_0(s,b)+{\hat S}(s,b)
\end{equation}
where $R_0(s,b)$ includes the Regge contribution important in the low energy region and is different for \pp \ and \pbarp, 
whereas the second term ${\hat S}(s,b)$  is the same for both processes and gives the rising contribution to the total cross-section. 
In this paper, a factorized expression is chosen so that
\begin{equation}
{\hat S}(s,b)=S_0(s) F(b^2)
\end{equation}
with the energy dependence given as in the CW model, namely
\begin{equation}
S_0(s)=
\frac
{
s^c}
{
\ln^{c'}
{s/s_0}
}
 + \frac{u^c}
 {\ln^{c'}{u/u_0}
 }
\end{equation} 
 The impact parameter dependence, and hence the $t$-dependence,  is inspired by  the 
proton electromagnetic form factor, namely it is   the Fourier-transform of 
\begin{equation}
 {\tilde F(t)}=f[ G(t)]^2[\frac{a^2+t}{a^2-t}]
\end{equation}
 where  $G(t)$  is given by a  parameterization inspired by the proton  electromagnetic form factor, i.e. 
\begin{equation}
G(t)=[ (1-t/m_1^2)(1-t/m_2^2) ]^{-1}
\end{equation}

This model had six parameters, which were  fixed from existing data.  As in most models,  the appearance of 
LHC data required some adjustment of the parameters. In Fig.~\ref{fig:bswtotdiff2012} we show some current predictions  
from this model, up to very high energy cosmic rays \cite{Soffer:2013psz}.
\begin{figure}[htbp]
\begin{center}
\resizebox{0.5\textwidth}{!}{
\includegraphics{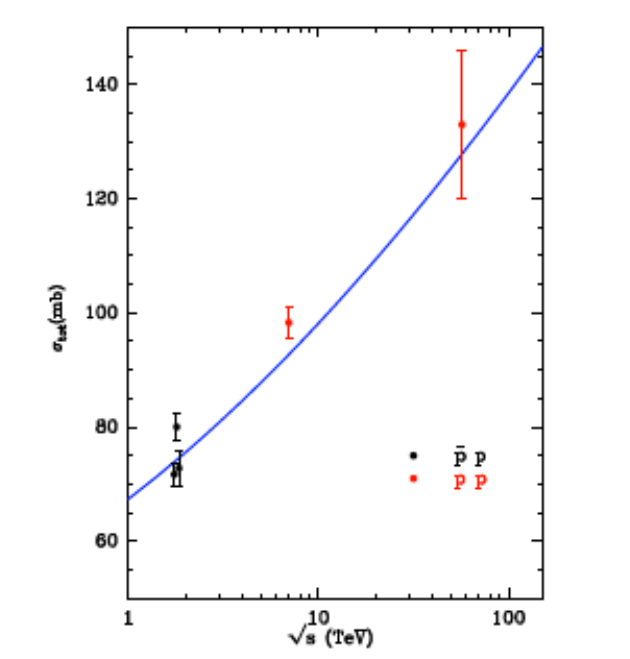}}
\caption{Total cross-section predictions at LHC and beyond, from Soffer's contribution to Diffraction 2012 \cite{Soffer:2013psz}.
Reprinted from \cite{Soffer:2013psz}, \copyright (2013) with permission by AIP Publishing LLC}
\label{fig:bswtotdiff2012}
\end{center}
\end{figure}

The elastic differential cross-section is defined as
\begin{equation}
\frac{\pi}{s^2}|a(s,t)|^2
\end{equation} 
and the total cross-section at $\sqrt{s}=40\ {\rm}TeV$ is predicted to reach a value of $121.2\ mb$ \cite{Bourrely:2002wr}.\\

\subsection{The universal  Regge and Pomeron pole description by Donnachie and Landshoff }
\label{subsec:dl}
In Sect.~\ref{sec:general}, we have seen that the analyticity properties of the 
elastic scattering amplitude
in the complex angular momentum plane, make it   possible to obtain that the 
amplitudes for  large $t$ and small $s$ exhibit a power law behaviour. 
Using crossing, one then obtains the usual large $s$ and small $t$  behaviour to describe 
elastic scattering and different low energy processes.


The  very successful parametrization of all total cross-sections  provided 
in 1992 by Donnachie and Landshoff (DL) \cite{Donnachie:1992ny}  
\begin{equation}
\sigma^{TOT}=Ys^{-\eta}+Xs^\epsilon
\label{eq:dl}
\end{equation}
was  inspired by Regge and Pomeron exchange, and proposed as a universal expression valid for all hadronic total cross-sections. 
In Eq.~(\ref{eq:dl}) 
the first term  is identified as arising from $\rho, \omega, f, a$ ($J=1,2$) exchanges, the second from Pomeron exchange, 
a vacuum trajectory, which, before the observation of the rise of $\sigma_{total}$ had been given a constant intercept $\alpha_P(0)=0$.
Requiring the same values for $X, \epsilon$ and $\eta$, the DL fit to \pp \ and 
\pbarp  \ data gave
\begin{equation}
\epsilon=0.0808\ \ \ \ \ \ \ \ \eta=0.4525
\end{equation}
What is remarkable about this expression is that the same value of $\epsilon$ and 
$\eta$ appeared to fit all available cross-section data, within the 
existing experimental errors, namely $\pi^{\pm} p$, $K^{\pm} p$, $\gamma p$, 
${\bar p} n$, $p n$.The interpretation of this expression for what concerns the 
first  term is that it  correspond to a simple Regge pole with intercept 
$\alpha(0)=1-\eta$. Clearly, if one uses only one such decreasing term for
different sets of data, $\eta$ is understood as being the intercept of an 
{\it effective} trajectory, which actually takes into account different  
Regge terms contributions as well as possible contributions from non-Regge terms, including 
the exchange of more than one Pomeron.
As for  the Pomeron, this is considered as an entity whose distribution and density function were 
actually measured at HERA, 
and to which we shall return in the section dedicated to photon processes, 
Sec.~\ref{sec:photons}.
The fact that the power  is the same for 
different processes finds its justification in that the Pomeron has 
the quantum numbers of the vacuum. Thus, for 
crossing symmetric processes such as 
\pp \ and \pbarp \ not only the 
power is the same, but also the coefficient X. This is borne out by the fit to various processes,
which we reproduce in Fig. \ref{fig:DLx}.
\begin{figure*}[htb]
\centering
\resizebox{1.0\textwidth}{!}{%
\includegraphics{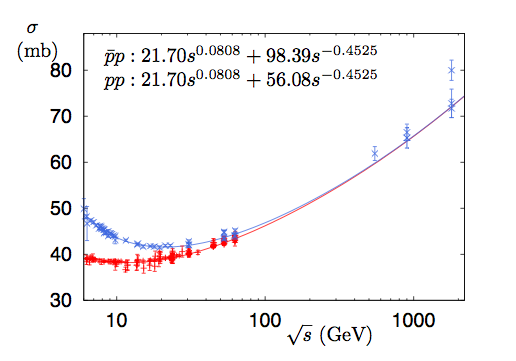}
\includegraphics{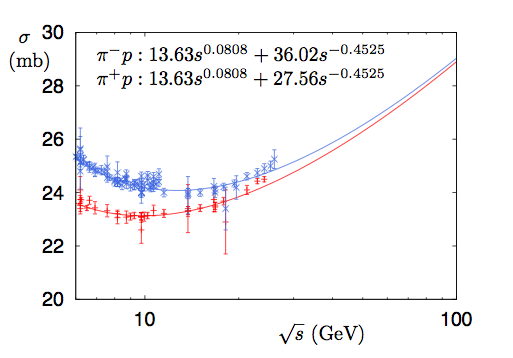}
\includegraphics{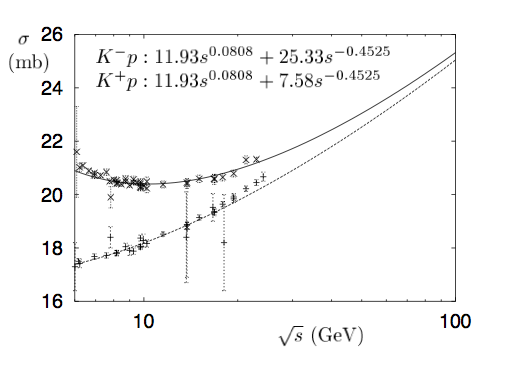}
\includegraphics{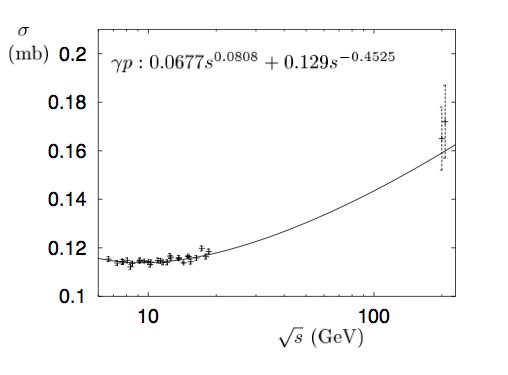}}
\caption{Donnachie and Landshoff universal description of proton, meson and photon total cross-sections with 
Regge and Pomeron pole exchange. The figures are  courtesy of the authors, 
reprinted from \cite{Donnachie:1992ny},
\copyright (1992) with permission by Elsevier.}
\label{fig:DLx}
\end{figure*} 

In the introduction to this section, we have mentioned the interpretation of the rise 
in terms of QCD processes like mini-jets, which will also  be discussed later in more details.   We mention here that in 
\cite{Donnachie:1988ak}  this possibility was considered unlikely because, 
when the contribution  from vector and tensor meson exchange to the total 
cross-section is subtracted off, the rise appears to be present already at 
$\sqrt{s}\le 5\ {\rm GeV}$ and, in some cases, even earlier 
\cite{Donnachie:1985iz}. In the 1992 paper \cite{Donnachie:1992ny}, this statement 
is softened, although it is still said that the form 
$Xs^\epsilon$ is unaffected by the onset at higher energies of new 
production processes, such as charm or minijets. A word of caution is 
however added,
namely that  at the Tevatron one can expect a large number of 
mini-jets.

The question of the rise is also discussed with respect to the impact picture 
by Cheng, Walker and Wu \cite{Cheng:1972ik,Cheng:1974dxa}. DL considered  unhelpful
 to adopt a geometrical approach and to talk of hadrons becoming bigger and 
blacker as the energy increases. 
We note that an early rise with the same power as the one found by DL is also to be found in 
the CWW  \cite{Cheng:1972ik}.

On the other hand,  if the rise is due to mini-jets, at high energy the rise   
should only depend on the gluon densities, and thus be flavour independent. However, 
as in the case of $\gamma p$ and $\gamma \gamma$, there may be other 
ingredients, quark densities, internal structure such as hadronic matter distribution  etc. 
which change the way the cross-section rise. Indeed, as we shall show in the section 
dedicated to $\gamma p$ processes, mini-jet models do not expect the rise to be 
the same for photons as for protons  until such very high asymptotic energies are reached where only gluons play a role.

The advantage of the DL formulation is that it provides a simple and useful parametrization 
of total cross-sections, without the need to introduce the question of the inner structure of protons. 
The disadvantages are  that  it does not connect directly to QCD 
and it violates the limiting behavior imposed by the Froissart bound

\subsection{Hadronic matter distribution}
\label{subsec:hadronicmatter}
A crucial role in many of the models for the total cross-section 
is played by the distribution of matter in the colliding hadrons in the plane perpendicular to 
the initial direction of motion. Practically the only model which does not use an impact parameter 
description is the simple version of the  Regge-Pomeron model by Donnachie and 
Landshoff \cite{Donnachie:1992ny} or phenomenological fits based on analyticity and other 
general properties, 
discussed later in this section.
However, any dynamical description of hadron scattering   uses information 
about the breaking up of the proton, 
which is what is described by the total cross-section.\\

Models in current usage fall into one of the following major categories:
\begin{itemize}
\item matter distributions obtained from the Fourier transform of colliding hadron form factors (FF),
\item parameterisations inspired by  form factor (FF) models, such as the BSW \cite{Bourrely:1978da}, 
or   QCD inspired  model by Block and various collaborators \cite{Margolis:1988ws,Block:1998ge},
\item 
QCD dipole models,
based on the BFKL equation in transverse momentum space \cite{Mueller:1993rr,Mueller:1994jq},
\item Fourier transform of transverse momentum distribution of QCD  radiation emitted during the collision, 
using $k_t$ Soft Gluon Resummation (SGR), 
extended to the infrared region,
as in the so-called BN  model, described later in this section 
\cite{Corsetti:1996wg,Grau:1999em,Godbole:2004kx}.
\end{itemize}
The basic function of these models for  the matter distribution in impact parameter space, is 
to provide a cut-off in space which describes the fact that hadrons have a finite size, namely that 
quarks and gluons are confined into hadrons. From this point of view, models which reflect  this finite 
size, such as  the dipole model for the proton factor, can be a good description, but the problem in 
this case lies in  the energy dependence.

Form factors in fact
are a 
 phenomenological description of the inner structure and need to be placed in a QCD context 
 for a model independent understanding. Also, the problem with form factor models, is that in some 
 cases, such as  the case of photon total cross-sections, the  form factor is not known. One can 
 always use a dipole (for protons) or monopole (for pions) functional form, but then one may have to 
 change phenomenologically the scale parameter. On the other hand, QCD inspired matter distributions, 
 which are based on parton structure,  can be useful only if they are able to access the 
 Infrared Region (IR), since the needed impact parameter description has to describes very large 
 $b$-values. Thus we require a model for ultra-soft, infrared, soft gluon emissions. One such model 
 was proposed  in \cite{Corsetti:1996wg} and its application to total \x \ will be described in \ref{sss:BN}.
In order to approach models based on resummation, we shall dedicate the next subsections to a discussion of 
resummation in  gauge theories, QED first, and then QCD.
\subsection{Role of resummation in QED}\label{ss:resum}
We shall here  first  present a review of resummation in QED, including the Sudakov form factor.


 In this subsection, we  shall provide the reader with a history of resummation beginning with the 
 pioneering work on the infrared catastrophe by Bloch and Nordsieck 
  and its importance for subsequent 
 development both in QED and QCD. The non covariant formulation by 
  Bloch and Nordsieck was substituted by a 
 covariant one by Touschek and Thirring in 1951. Previously, in 1949 Schwinger had anticipated  the need for resummation in perturbative QED. 
 In the fifties and the sixties, there were further refinements  by Touschek, Brown and Feynman, Lomon and 
 Yennie Frautschi and Suura. A different path to resummation was initiated by Sudakov which later 
 played a pivotal role in QCD resummation. For QCD, resummations of soft gluon transverse momentum  
 were developed by Dokshitzer Dyakonov  and Troian, Parisi and Petronzio, Collins and Soper,  Curci Greco and Srivastava, 
 Pancheri and Srivastava. 
 In a subsection to follow, we shall illustrate  the highly influential  work by  by Balitsky, Fadin, Kuraev and Lipatov, 
 based on the evolution equation formalism.
 Finally, 
 their connection with total cross-section computations will be established.

The importance of large distance scattering in models for the total cross-section leads naturally to 
consider the importance of very small momentum interactions, namely the InfraRed (IR) momentum 
region is dual to the  large distance region. When particle  momenta are very small, resummation of 
all processes involving small momenta is necessary. In  QED this has been the subject of interest 
for  almost a hundred years, in QCD for more than  40. In the following we shall present a mini-review of 
the problem of resummation in QED, followed by a discussion of the QCD case. The latter is of course 
still not solved, a possible ansatz to describe the IR region and the connection of resummation with the 
asymptotic behaviour of the total cross-section will be presented later, in the context of the eikonal mini-jet models.  
 
All total cross-sections reflect the effective area defined by the (average) perpendicular distance between 
the two incident particles and hence are controlled by scattering at large distances. Long distances correspond 
to their dual i.e., small (transverse) momenta. Thus, the importance for a theoretical understanding -in any model 
for high energy total cross-section- of scatterings at very small momenta. Unbroken gauge theories, 
abelian[QED] or not [QCD], are plagued by two types of singularities both originating from the masslessness of 
the gauge boson. The exchange of a massless gauge boson in any elastic scattering between (generalized) 
charges gives rise to the Rutherford singularity  as the momentum transfer goes to zero. Thus, a total cross-section 
becomes infinite simply because of the infinite range of the potential. There is a further, infra-red (IR) singularity which 
arises since any charged particle can emit and absorb gauge particles of vanishing momenta leading to all vertices 
and amplitudes be divergent in this limit [The IR catastrophe]. Hence, the need for individual IR singularities 
(in perturbation theory) to be summed and a discussion of the various ways to accomplish this resummation 
is the purpose of this section. But before embarking on it, let us pause to discuss briefly the Rutherford singularity.
\subsubsection{The Rutherford singularity}\label{sss:rutherford}
In QED, the elastic scattering amplitude between say two electrons in the forward direction is indeed infinite. 
However, consider the scattering of a (moderately) high energy electron from a neutral atom considered as 
a collection of $Z$ electrons and a (point-like) nucleus with charge ($Ze$). At momentum transfers sufficiently 
large compared to the binding energy of the electrons, it would not be misleading to compute the elastic ($eA$) 
amplitude in the Born approximation -using the impulse approximation- as an incoherent sum of the incident 
electron scattering from the individual charges of the atom [ an atomic ``parton'' model]. But, this computation 
would hardly suffice for a reasonable estimate of the total cross-section since optical theorem relates the total 
cross-section to the [imaginary part] of the elastic amplitude at zero momentum transfer which in this 
approximation would diverge by virtue of the Rutherford divergence. However, as the momentum transfer goes 
to zero, the incident electron sees the total charge of the atom which is zero and thus there should be no divergence. 
The answer to this problem is of course well known: the coherence between the electrons, which is neglected in 
the impulse approximation, is the culprit and can not be neglected. If $t$ is the momentum transfer, the 
cross-section contains a factor ($[Z - F(t)]^2$), where $F(t)$ denotes the form factor of an electron in the atom
\begin{equation}
\label{5.71}
(\frac{d\sigma}{d\Omega}) = (\frac{2 m \alpha}{t})^2 [Z - F(t)]^2.
\end{equation}  
Since $F(0)\ =\ Z$, there is no Rutherford singularity and the cross-section is indeed finite as it should. 
In fact, a measurement of the forward differential cross-section has been traditionally used to determine the 
charge radius (or the size) of the atom. 

We are recalling these well known facts from atomic scattering for two reasons. One is to remind ourselves that 
elastic scatterings at low momentum transfers depend crucially on the coherence size of the system and hence 
they cannot be neglected even at high energies. The second reason is that for QCD, these facts take on a shade more 
relevant. Individual elastic scattering amplitudes for all coloured particles possess the Rutherford singularity which 
however must disappear for hadronic scattering amplitudes since hadrons are color singlets [``neutral atoms'']. 
By analogy from the atomic case, the final amplitudes must reflect their coherence size i.e., the distribution of color 
charges within the hadron. 

But this is hardly the end of the story for QCD It is 
widely believed (often called IR slavery) that quarks and glue in QCD are confined 
through the IR divergences and that the strong coupling constant $\alpha_s(t)$ becomes very large at small 
momenta $t$.
There are indications for it already in the divergence of the one-loop asymptotic freedom (AF)  formula for $\alpha_s(t)$. 
Hence, a natural paradigm for QCD emerges: If the divergence of the (effective) coupling constant at small $t$ 
is responsible for confinement, the nature and the intensity of its divergence must set the size of a hadron. 
But then, the same must also enter into determining the size of the high energy total cross-sections. We shall return 
to these issues later when we discuss models for total cross-sections. 

After this brief interlude, let us now discuss the underlying issues beneath the IR divergences in QED and QCD.

\subsubsection{Infra-Red catastrophe and the Bloch--Nordsieck cure\label{sss:BNQED}}

The Infra-Red (IR) catastrophe was clearly brought to the fore by Bloch and Nordsieck 
through their two classic papers 
of 1937 \cite{Bloch:1937pw}\cite{Nordsieck:1937zz}. 
Bloch and Nordsieck  observed that previous analyses of radiative corrections to scattering processes were defective in that they 
predicted a divergent low frequency correction to the transition probabilities. This was evident from the soft 
photon emission spectrum in the scattering of an electron  from a Coulomb field (as described by Mott\index{Mott, N.F.} and 
Sommerfeld\index{Sommerfeld, A.} \cite{Sommerfeld:1931az} and by Bethe\index{Bethe, H.} and 
Heitler\index{Heitler, W.} \cite{Bethe:1934za}): as the emitted photon frequency $\omega\rightarrow\ 0$, the spectrum 
takes  the form $d\omega/\omega$.

The two authors  had noticed that, while the ultraviolet [UV] difficulties were already present in the classical theory, 
the IR divergence had no classical counterpart. They anticipated that only the probability for the simultaneous emission 
of infinitely many quanta can be finite and that the probability for the emission of any finite number of them must vanish.

To cure this  ``infrared catastrophe''\glossary{catastrofe infrarossa} phenomenon,  a semi-classical
description was proposed. They noticed that  for emitted photons of  frequencies larger than
a certain $\omega_0$, the  probability for emitting each additional photon is
proportional to
${{e^2}\over{{\hbar}c}} \log{E/{\hbar}\omega_0}$, which becomes large
as $\omega_0 \rightarrow 0$. Thus, the actual expansion is not
${{e^2}\over{{\hbar}c}}$, which would be small, but a larger
number, driven by the logarithm. This led them to analyze the scattering process in terms of what
came to be called Bloch-Nordsieck states, namely states with one electron plus the electromagnetic field, and to substitute the expansion in
${{e^2}\over{ {\hbar}c}}$ with a  more adequate one. The important result
they obtained, in a non covariant formalism, was that, albeit   the probability of emission of any finite number of quanta is zero, when summing on
all possible numbers of emitted quanta,  the total transition probability
was finite. This was so because, by summing on all possible frequencies and numbers of photons, one obtained the result which one would have obtained  by neglecting entirely  the interaction with the electromagnetic
field. Since they could show that the probability for
emission of any finite number of quanta was zero, whereas the total
transition probability was finite and the total radiated energy was finite,
then they anticipated that  the mean total number of quanta
had to be infinite. Thus the idea that any scattering process is always
accompanied by an infinite number of soft photons was introduced and proved
to be true (later, also in a covariant formalism).

In the Bloch and Nordsieck 
 paper we see the emergence of the concept of finite total energy,
with exponentiation of the single photon spectrum which is logarithmically divergent. 
They obtain that the  probability per unit time for a transition in which
$n_{s\lambda}$ light quanta are emitted  always includes a factor
 proportional to
\begin{equation}
exp\{
-\alpha\lim_{\omega_0\rightarrow 0}
\int_{\omega_0}^{\omega_1}
{{d\omega}\over{\omega}} \int d\Omega_k \times \Delta
\end{equation}
where
\begin{equation}
\Delta = [
(
{{{\boldsymbol \mu}}\over
{1-\mu_s}}
-{{{\boldsymbol \nu}}\over
{1-\nu_s}}
)^2-
(
{{\mu_s}\over{1-\mu_s}}-{{\nu_s}\over{1-\nu_s}}
)^2
]
\end{equation}
where ${\boldsymbol \mu}$ and ${\boldsymbol \nu}$ are  the momenta of the incoming and outgoing electron,  $\mu_s$ and $\nu_s$ the projection of $\boldsymbol \mu$ and $\boldsymbol \nu$ along the momentum $\bf k$ of the emitted photon.
Because of the exponentiation of a divergent
term, the transition
probability for a finite number of emitted photons is always zero. On the
other hand, when summation is done over all possible photon numbers and
configurations, the result is finite. Clearly there was still something
missing because the fact that one must emit an infinite number of photons
is obtained by exponentiating an infinite divergent term, and there is no 
hint of how to really cure the IR divergence. In addition the language
used is still non--covariant.

Before going to the covariant formulation, we notice that the 
argument relies on the transition probability being proportional to
\begin{equation}
\Pi_{s\lambda}
e^{-
{\bar n}_{s\lambda}}
{{
{\bar n}_{s\lambda}^{n_{s\lambda}}
}
\over{
n_{s\lambda}!
}}
\end{equation}
namely to a product of Poisson distributions, each of them describing the
independent emission of $n_{s\lambda}$ soft photons, 
and upon neglecting the recoil effects.

\subsubsection{Covariant  formalism  by Touschek and Thirring}
 \label{sss:TT}


Touschek\index{Touschek, B.}  and Thirring\index{Thirring, W.E.}
reformulated the Bloch and Nordsieck  problem in  a covariant formalism \cite{Thirring:1951cz}. 
They proved that 
$|\chi_0^\dagger \chi_0^\prime|^2$, the
probability for a transition from a state $\chi_0$ with no photons  to a state with
an average number ${\bar n}$ of photons, $\chi'_0$,  was given by $e^{-{\bar n}}$, which
  goes to zero as ${\bar n}$ goes to infinity, namely that  the probability of emission of
any finite number of quanta was
zero.

Let us consider the Touschek and Thiring  derivation. 
They point out the importance of the Bloch and Nordsieck  solution and that, although the results 
they obtain are not new and have been discussed by several authors, their  solution being the
only one which admits an accurate solution justifies a general
reformulation of the problem. As already noted in \cite{Bloch:1937pw}
 the simplification
which  enables one to find an accurate solution rests on the neglect
of the recoil of the source particles.

Touschek and Thirring  set out to determine the probability for the production of
a certain number $n$ of quanta in a 4--momentum interval $\Delta$. They obtain
that the probability amplitude for the creation of $n$ particles in
 a state
$r$ is given by
\begin{eqnarray*}
(F^r_n\chi_0)={{1}\over{(2\pi)^{3n}}}\times{{1}\over{\sqrt{n!}}}
\int_{\Delta}dk_1...\int_\Delta dk_n \Pi_i\delta(k^2_i-\mu^2)\\
%
\times u^{r*}_n(k_1....k_n)
(\chi_0'
(\phi(k_1)+\delta \phi(k1))
...(\phi(k_n)+\delta\phi(k_n))\chi_0)
\end{eqnarray*}
where use has been made of a complete set of orthogonal functions $u^r_n$
which satisfy the completeness relation.
$\chi_0$ is the eigenvector describing an incoming state with no quanta at
all in the interval $\Delta$, while $\chi_0'$ is the corresponding one
for the final state.
For the probability to have $n$ photons in the final state they obtain
\begin{equation}
\sum_r |F^r_n\chi_0|^2={{1}\over{n!}} {\bar n}^n |\chi_0' \chi_0|^2
\end{equation}
with
\begin{equation}
{\bar n}={{1}\over{(2\pi)^3 }}
\int_\Delta
dk \delta(k^2-\mu^2)|\delta \phi(k)|^2
\end{equation}
and, by imposing that the total probabiity be 1, they obtain
the Bloch and Nordsieck  
result
\begin{equation}
|\chi_0'\chi_0|^2=e^{-{\bar n}}
\end{equation}
For the derivation, it is necessary that the motion of
the source particles be not affected by the emission of soft quanta, namely that the
wave operator describing the source field be a c--number. Then $\phi^{out}$
differs by $\phi^{in}$ only by a multiple of the unit matrix and,
transforming to
$k$-space, it may be written as
\begin{equation}
\phi^{out}(x)={{1}\over{(2\pi)^3}}\int dk \delta(k^2-\mu^2)
 [\phi^{in}(k)+\delta \phi(k)] e^{ikx}
\end{equation}
where $\delta \phi(k)=-\rho(k)\epsilon(k)$, with $\rho(k)$ the
Fourier-transform of the source density describing the source particles. In
their paper TT first derive their results for a source scalar field, then
they generalize it to a vector source function $j_\mu(x)$ for a point--like
electron, i.e.
\begin{equation}
j_\mu(x)=e\int p_\mu(\tau) \delta(x-\tau p(\tau)) d\tau \label{eq:jmu}
\end{equation}
 where $p_\mu(\tau)=p_\mu$ for $\tau$ less than 0 and $p_\mu(\tau)=p'_\mu$ for
 $\tau$ larger than 0. Notice that the sudden change in momentum imposes the
 restriction that in order to apply the results to a real scattering process,
 the photon frequencies should always be much smaller than $ 1/\tau$,
where $\tau$ is the
 effective time of collision. Otherwise the approximation (of a sudden
 change in momentum) will break down. One then obtains
\begin{equation}
j_\mu(k)= \frac{i e}{(2\pi)^{3/2}}\big( {{p_\mu}\over{(pk)}}-{{p'_\mu}\over{(p'k)}}\big)
\end{equation}
and the average number of quanta ${\bar n}$ now becomes
\begin{equation}
{\bar n}={{e^2}\over{(2\pi)^3}}\int_\Delta dk \delta(k^2-\mu^2)\big[
{{(p\epsilon)}\over{(pk)}}-{{(p'\epsilon)}\over{(p'k)}}\big]^2
\end{equation}
where $\epsilon$ is a polarization vector. Notice that the photon mass $\mu$ remains different from zero, so as
 to ensure convergence of all the integrals. 
However, this is  not necessary when   higher order QED processes are taken into account,  as was shown by explicit Quantum Electrodynamics calculations, starting with Schwinger's work \cite{Schwinger:1949zz}.


\subsubsection{Schwinger's ansatz on the exponentiation of the infrared factor and the appearance of double logarithms}\label{sss:schwinger}

The solution found by Bloch and Nordsieck ,
and later brought into covariant form by Touschek and Thirring,  did not really solve the
problem of electron scattering in an external field and of how to deal with finite energy
losses.
 This problem was  discussed and solved in
QED, where
 the logarithmic divergence attributable to the
IR catastrophe from emission of real light quanta of zero energy
was compensated through the emission and absorption of
virtual quanta. This cancellation took place in the cross--section,
and not between amplitudes. In a short paper in 1949 and, shortly
after, in the third of his  classic QED papers, Julian Schwinger\index{Schwinger, J.} \cite{Schwinger:1949zz}
examined the  radiative
corrections to (essentially elastic) scattering of an electron by a Coulomb
field, computing second order corrections to the first order amplitude and
then cancelling the divergence in the cross--section between these terms and
the cross--section for real photon emission.
The result, expressed as a fractional
decrease $\delta$ in the differential cross--section for scattering
through an angle
$\theta$  in presence of an energy resolution $\Delta E$ of the scattered electron,
is of order
$\alpha$  and given a
\begin{equation}
\delta={{2\alpha}\over{\pi}}log({{E}\over{\Delta E}})\times F(E,m,\theta)
\label{js}
\end{equation}
where $F(E,m,\theta)$ in the extreme relativistic limit is just $\log (2E/m)$. 
Notice here the first appearance of a double logarithm, which will play a crucial role in resummation and exponentiation of radiative corrections.

Schwinger notices that
$\delta$ diverges logarithmically in the limit
$\Delta E\rightarrow 0$ and
points out that this difficulty stems from the neglect of processes with
more than one low frequency quantum. Well aware of the Bloch and Nordsieck
result, he notices that it never happens that  a scattering event
is unaccompanied by the emission of quanta and proposes to replace the
radiative correction factor $1-\delta$ with $e^{-\delta}$, with
further terms in the series expansion of $e^{-\delta}$ expressing the effects
of higher order processes involving  multiple emission of soft photons.

In 1949 however, such refinements, namely the exponentiation of the
radiative correction factor, were still far from being needed, given the
available energies for scattering processes as  Schwinger\index{Schwinger, J.}  points out,
 estimating the actual correction to  then available
experiments, to be about $10\%$. Almost twenty years had to
pass before the exponentiation became an urgent matter for extraction of results from collider experiments,  
such those at SPEAR, ADONE, ACO, VEPP-2M, 
where the double logarithm term 
$\alpha log(E/\Delta E) log (2E/m)$ would climb to $20 \div 30 \%$ and 
beyond \cite{Greco:1975rm}. 

IR radiative corrections were also considered by Brown and Feynman \cite{Brown:1952eu}
some
time after  Schwinger, and the concept of an external parameter ($\Delta E$) 
 which sets the scale of the IR correction was confirmed.
  
It is not clear
whether Touschek\index{Touschek,B.} and Thirring\index{Thirring, W.E.} were aware of
the Schwinger\index{Schwinger, J.}
results when they formulated the covariant version of the
Bloch Nordsieck  method. They do not cite his  results, and their interest
is primarily on obtaining a covariant formulation of the Bloch and Nordsieck  method.
Quite  possibly, at the time they were not interested in the practical
applications of
the problem, which is instead the focus of Schwinger's calculation.

\subsubsection{The Sudakov form factor}\label{sss:sudakov}
The problem of the double logarithms in perturbation theory was investigated  by Sudakov\cite{Sudakov:1954sw} 
who studied  the existence of double logarthims for vertex functions and established 
 their exponentiation.

  Consider the vertex function in QED 
 [$e(p)\rightarrow\ e(q)\ +\ \gamma(l)$] with all three particles off shell in the kinematic limit 
 $l^2>>p^2,q^2>>m^2$. Here the relevant double logarithmic parameter is computed to be 
 ($2\alpha ln(l^2/p^2) ln(l^2/q^2)$) and considerations to all orders show that it exponentiates. 
 Thus, the primitive vertex $\gamma^\mu$ in the stated limit gets replaced by
\begin{equation}
\label{S1}      
\gamma^\mu e^{- [2\alpha ln(l^2/p^2) ln(l^2/q^2)]}.
\end{equation}
These double logarithms imply that as $q^2$ becomes very large, the vertex goes to zero, a rather 
satisfactory result. From a practical point of view, it was shown by Abrikosov {\it et al}
\cite{Landau:1956zr}
that the competing process where a large number of soft real photons are emitted has a far greater 
probability and thus it far overwhelms the Sudakov probability. 

Incidentaly, the double logarithms are completely symmetric in the variables ($p^2, q^2, l^2$). 
Thus, if an electron say $p^2>>q^2,l^2$, a similar result to Eq.(\ref{S1}) holds. The Sudakov limit 
when extended to QCD does become relevant. For example, in the light-quark parton model of QCD, 
one can therefore justify the fact that very far off mass shell quarks are suppressed through the Sudakov 
``form factor''.  

\subsubsection{Status of the field in the early sixties\label{sixties}}
\label{sss:status60}

In the 1950's, with Feynman\index{Feynman, R.} diagram technique available to the
theoretical physics community, many  higher order QED
calculations came to be part of standard theoretical physics handbooks.

Many important contributions to the radiative correction problem appeared
in the '50s and early '60s \cite{Brown:1952eu}, \cite{Erikson:1961er}, \cite{Lomon:1965lo}, with
a major step in the calculation of IR radiative corrections
 done in 1961 by Yennie\index{Yennie, D.R.}, Frautschi\index{Frautschi, S.C.} and Suura\index{Suura, H.} (YFS)  \cite{Yennie:1961ad}.
In their classic paper,
they went though  the cancellation of the IR divergence
 at each  order in perturbation theory in the cross--section and
obtained the final compact expression
for the probability of energy loss. Their result is apparently  disconnected
from the Bloch and Nordsieck result. In their paper, they compute higher and
higher order photon emission in leading order in the low photon
momentum, showing that the leading terms always come from emission from
external legs in a scattering diagram. In parallel, order by order, they
extract the IR divergent term from the virtual diagrams, making the
terms finite through the use of a minimum photon energy. They show that the
result is just as valid using a minimum photon mass, and finally eliminating
the minimum energy, they show the final result to be finite.



\subsubsection{A semi-classical approach to radiative corrections \label{sss-models:etp}}

A semiclassical approach to resummation in QED can be found in  \cite{Touschek:1968zz}, where 
the Bloch and Nordsieck  approach is adopted and an important point made that the picture of an experimentalist as counting
single photons as they emerge from a high energy scattering among charged
particles is unrealistic. It is noted that
 perturbation theory  is
unable to deal with the flood of soft photons which accompany any charged particle
reaction. Then, the question of how light quanta are distributed in momentum is discussed. As discussed 
above, Bloch and Nordsieck  had shown that, by neglecting the recoil of the
emitting electron, the distribution of  any finite number of quanta would
follow a Poisson type distribution, namely
\begin{equation}
P(\{n,{\bar n}\})={{1}\over{n!}}{\bar n}^n e^{-{\bar n}}
\label{poisson}
\end{equation}
and Touschek and Thirring had recast ${\bar n}$ in a covariant form. In \cite{Touschek:1968zz}
the constraint of energy-momentum
conservation is added to this distribution. This is a major improvement, which has sometimes been
neglected in subsequent applications of the method to strong interaction
processes.

Let us repeat the argument through which Touschek obtained the final 4--momentum
probability distribution describing an energy-momentum loss $K_\mu$.
  The final expression is the same as the one proposed
 earlier by YFS, but the derivation is very
different and its physical content more transparent. Touschek  also 
discusses the different energy scales which will become very important later, when dealing with 
resonant states, and in particular with $J/\Psi$\glossary{$J/\Psi$} production.
The derivation is semi-classical and at the end it will be clear that  the quanta considered are 
both real and virtual photons.
The underlying reason for this can be understood from a consideration  by Brown and Feynman
 \cite{Brown:1952eu} in their computation of radiative corrections to Compton
scattering. Brown and Feynman  note that it is difficult to distinguish between real and
virtual quanta of extremely low energy since, by the uncertainty principle, a
measurement made during a finite time interval will introduce an
uncertainty in the energy of the quantum which may enable a virtual quantum
to manifest as a real one.

In \cite{Touschek:1968zz}  the probability of having a total $4$-momentum loss
 $K_\mu$ in a
charged particle scattering process, is obtained by considering all the
possible ways in which $n_k$ photons of momentum $k_\mu$ can give rise to a
 given total energy loss $K_\mu$ and then summing on all the values of
$k_\mu$. That is,  we can get a total  final 4-momentum $K_\mu$
pertaining to the total loss, through emission of $n_{k_1}$ photons of
momentum $k_1$, $n_{k_2}$ photons of momentum $k_2$ and so on.
Since the photons
are all emitted independently (the effect of their
 emission on the source particle being
neglected), each one of these distributions is a Poisson distribution,
and the  probabilty of a 4-momentum loss in the interval $d^4 K$ is written as
\begin{equation}
d^4P(K)=\sum_{n_k}\Pi_{k}P(\{n_k,{\bar n}_k\})
\delta^4(K-\sum_{k} k n_k) d^4K
\end{equation}
where the Block and Nordsieck result of independent emission is introduced through the Poisson
distribution and four momentum conservation
is ensured through the 4-dimensional $\delta$ function, which
selects the distributions $\left\{ n_k,{\bar n_k} \right\}$ with
the right energy momentum loss $K_\mu$.
The final expression
is
\begin{equation}
d^4 P(K)={{d^4K}\over{(2\pi)^4}}\int d^4x \ exp [-h(x)+iK\cdot x]
\label{d4p}
\end{equation}
with
\begin{equation}
h(x)=\int d^3 {\bar n}_k \big(1-exp[-ik\cdot x]\big)
\end{equation}
which is the same as the expression obtained  by YFS through order by order
cancellation of the IR divergence in the cross--section.
In this derivation, which is semiclassical, no mention or no distinction
is made between virtual and real photons as stated above, but it is clear that the
contribution of real photons corresponds  to  the term which is multiplied by the
exponential $e^{-ik\cdot x}$, since this retains the memory that the
total energy--momentum emission is constrained. Thus, single real
photons of momentum ($k$) are all correlated through the Fourier transform
variable $x$. 
The next step was to perform realistic calculations of the
radiative correction factors, using an apparently difficult expression. The
first objective was to obtain the correction factor for the energy, by
integrating Eq. (\ref{d4p}) over the 3--momentum variable. 
Through a very elegant argument based on analyticity, Touschek obtained the probability for a total energy loss
$\omega$ as
\begin{equation}
dP(\omega)=N \beta(E) {{d\omega}\over{\omega}}
\big( {{\omega}\over{E}}\big)^{\beta(E)}
\end{equation}
where N is a normalization factor \cite{Erikson:1961er,Lomon:1965lo}.
In the high energy limit,
\begin{equation}
\beta(E)={{4\alpha}\over{\pi}}\big( \log{{2E}\over{m_e}} -{{1}\over{2}}\big)
\end{equation}

Integrating the 4-dimensional distribution over the energy and longitudinal momentum variables, 
one obtains a transverse momentum distribution of the total emitted radiation \cite{PancheriSrivastava:1976tm}, 
namely
\begin{eqnarray}
{{d^2P({\vec K}_\perp)}\over{d^2{\vec K}_\perp}}=
\int d\omega \ dK_z {{d^4 P(K)}\over{d^4K}}=\\
{{1}\over{(2\pi)^2}}\int d^2{\vec b}\
e^{
i{\vec b}\cdot {\vec K}_\perp -
\int d^3
{\bar n}(k) [
1-e^{-i{\vec b} \cdot {\vec k}_\perp]
}}\\
={{1}\over{(2\pi)^2}}\int d^2{\vec b}\
e^{
i{\vec b}\cdot {\vec K}_\perp -h({\bf b})}\label{eq:d2pk}
\end{eqnarray}
This expression, unlike the energy distribution, does not admit a closed form expression.
 In an Abelian  gauge theory with an energy independent 
 (as in the case of QED),  but not small
 coupling constant,
 the corresponding $\beta$ factor becomes large. A useful approximation for 
 the function $d^2P({\vec K}_\perp)$ was obtained as \cite{PancheriSrivastava:1976tm}
\begin{equation}
\label{trans}  
d^2{\tilde P}(\vecK_\perp)=\frac{\beta (2\pi)^{-1}}{\Gamma(1+\beta/2)} \frac {d^2\vecK_{\perp}}{2E^2A}
 (\frac{K_\perp}{2E\sqrt{A}})^{\beta/2-1}{\cal K}_{1-\beta/2}(\frac{K_\perp}{E\sqrt{A}})
\end{equation}
where this approximate expression is normalized to 1, as is the original distribution, and also admits 
the same average square transverse momentum, given by
\begin{equation}
<K_\perp^2>=2\beta E^2 A
\end{equation}
with E the maximum energy allowed to single gluon emission.  $\beta$ is in general obtained from 
\begin{equation}
\beta=\int d^4 n \theta(n_0)n_0\delta(n_0-1)j_\mu(n)j^{*\mu}(n)\delta(n^2)
\end{equation}
with 
 \begin{equation}
 j_\mu(n)=\frac{ie}{
 (2\pi)^{3/2}
 }
 \sum\epsilon_i
 \frac{p_{i\mu}}{p_i\cdot n}
\end{equation}
The sum runs on all the emitting particles of momenta $p_{i\mu}$, and with $\epsilon_i=\pm$ depending 
on whether the particle is positively or negatively charged, entering or leaving the scattering area.

.
\subsubsection{Reggeization of the photon}
\label{sss:reggeization-photon}

In the previous subsection, we have discussed  photon resummation. Here we turn to the question of Reggeization in QED and, in the next subsection, in QCD. Both play an important role in models for the total cross-section. In fact,  while the Froissart bound regulates the asymptotic behavior of total cross-sections, asymptotic Regge behavior of the scattering amplitudes and the optical theorem allow  direct calculation of the total cross-section. Thus, the question of Reggeization in QED and later in QCD became the center of attention in the 1960s and 1970s.

The question in the early '70s was whether the photon  reggeizes and soon after the discovery of QCD, 
whether the gluon reggeizes. Related questions in 
QED and QCD were regarding the vacuum channel leading singularity which we shall call the Pomeron. 
Reggeization of   gauge bosons in Abelian and non-Abelian gauge theories,  has played an important role 
in models for the total cross-section. These fundamental questions of compelling relevance  for the theories 
as well as for high energy phenomenology were vigorously pursued by several groups and in Subsect.~\ref{sss:BFKL}
we shall discuss 
 the BFKL approach, starting with Lipatov and Fadin and 
Kuraev, then by Lipatov and Balitsky,  on  the gluon Regge trajectory \cite{Lipatov:1976zz,Fadin:1989kf,Fadin:1996tb}. 
These analyses can be followed more easily if a discussion of the photon Regge trajectory is first introduced. To this we now turn.

Reggeization of elementary particles such as the photon or the electron was amply discussed in the 1960's. 
In QED, the infrared divergence and its cancellation 
bore many complications and the conclusion was that  the electron reggeized, not so for the photon
The electron in fact had been shown to  reggeize by Gell-Mann {\it et al.} in a massive photon (Abelian) 
QED up to the fourth order  \cite{GellMann:1964zz} and by Cheng and Wu to the sixth order. 
Such a theory possesses a conserved current but due to the mass of the photon, there are no IR divergences. 
Radiative corrections turned the elementary electron 
inserted into the Lagrangian into a moving Regge trajectory passing through $j = 1/2$ at the mass $m$ of the 
electron. Such a miracle did not occur for the massive 
photon; it did not turn into a Regge pole. Moreover, in the vacuum channel, in the leading 
logarithmic approximation (LLA), no ``Pomeron'' trajectory but a  fixed 
square root branch point at the angular momentum $j = 1 + (11 \pi/32) \alpha^2$ was found  \cite{Cheng:1970bi}.

The question remained open as to what happens in massless QED. The reggeization of a massless 
photon poses a problem in that all charged particle scattering 
amplitudes have IR divergences which are cancelled in the cross-sections. Through a summation of IR 
radiative corrections (not needed in the massive abelian theory) 
and imposition of di-triple Regge behavior, it was found by  Pancheri\cite{PancheriSrivastava:1973je} that 
a photon trajectory does emerge.

In other words, a reggeized behaviour of QED cross-sections can be obtained from the well known 
factorization and exponentiation of infra red corrections 
and, from this, a trajectory for the photon, as was shown in \cite{PancheriSrivastava:1973je}.
We shall describe it here.

In this approach, 
just
as later in the approach to the gluon trajectory of Lipatov and his  co-authors, 
reggeization arises through the exponentiation of single soft photon emission accompanying the scattering. 
Since in QED any reaction is necessarily an inclusive one because of soft photon radiation, 
the process to examine is
\begin{equation}
e(p_1)+e(p_2)\rightarrow e(p_3)+e(p_4) + X\label{eq:qed1}
\end{equation}  
where $X$ stays for any undetermined number of soft photons, hence for which $M_X^2<< s$, where 
$s=(p_1+p_2)^2$. One can now compare the cross-section for  process (\ref{eq:qed1}) with the one 
corresponding to the di-triple Regge limit in hadronic physics \cite{Freedman:1971ae}. Defining the 5 independent invariants 
of process Eq.~(\ref{eq:qed1}) as
\begin{align}
s_1=(p_1+p_2-p_3)^2, \ \ \ \ \ \ \ \ s_2=(p_1+p_2-p_4)^2\\
t_1=(p_2-p_4)^2, \ \ \ \ \ \ \ \ \ \ \ \ \  t_2=(p_1-p_3)^2\\  
M^2_X=(p_1+p_2-p_3-p_4)^2, 
\end{align}
  the limit $t_1\simeq t_2=t$ with $t$ fixed, and $M^2_X<<s,s_1,s_2$, we are within one of the kinematic 
  limits of interest for the inclusive di-triple Regge limit. For $-t<<s, M^2_X<<s\simeq s_1\simeq s_2$, the 
  cross-section of interest becomes
  \begin{equation}
  \frac{d^2\sigma}{dt d(M^2_X/s)}\rightarrow \large{(} \frac{M^2_X}{s}  \large{)}^{1-2\alpha_\gamma(t)} F(t)\label{eq:regge1}
  \end{equation}
  We have seen 
  that for $X$ corresponding to a four-vector $K^2<<s$ soft photon resummation applied to 
  Eq.~(\ref{eq:qed1}) at leading order leads to
  \begin{equation}
  \frac{d^5\sigma}{dtd^4 K}=\large{(} \int d^4 x e^{iK\cdot x -h(x)} \large{)}\frac{d\sigma_0}{dt} 
  \end{equation}
  where $d\sigma_0/dt$ corresponds to the Born cross-section for $e^+e^-\rightarrow e^+e^-$ 
  and the resummed soft photon spectrum is obtained from the regularized  soft photon spectrum
  \begin{equation}
  h(x)=\int d^3{\bar n(k) }[1-e^{-ik
  \cdot x}]
  \end{equation}
 One can rewrite the single photon spectrum 
as
\begin{equation}
d^3{\bar n(k) }=\beta(s,t,u)\frac{d^3k}{2k}f(\Omega_k)
\end{equation}
where  the function $f(\Omega_k)$ is normalized to 1 and gives   the angular distribution of the 
emitted soft photon. With such a definition,  the dependence from the momenta of  emitting particles  is  
specified by   $\beta(s,t,u)$, which is  a relativistic  invariant function of  the Mandelstam variables. 
This compact expression is useful when overall integration over the soft photon momenta is performed. 
From
\begin{equation}
\frac{dP(\omega)}{d\omega}=\int d^3{\vec K} \frac{d^4 P}{d^4K}=\int \frac{dt}{2\pi} e^{i\omega t -h(t)}
\end{equation}
and
\begin{equation}
h(t)=\beta(s,t,u)\int \frac{dk}{k}[1-e^{-ikt}]
\end{equation}
the overall energy dependence from  soft photon emission is now 
\begin{equation}
\frac{dP(\omega)}{d\omega}={\cal N}\beta(s,t,u)\frac{d\omega}{\omega}\large{(} \frac{\omega}{E}\large{)^{\beta(s,t,u)}}\label{eq:qed2}
\end{equation}
where $E$ is a typical scale of the process, and $\cal{N}$ a normalization factor, i.e.
\begin{equation}
{\cal N}=\frac{\gamma^{-\beta(s,t,u)}}{\Gamma(1+\beta(s,t,u))}
\end{equation}
$\gamma$ being the Euler's constant.
 To leading order, $E$ is the upper limit of integration of the soft photon  spectrum, and can 
 be taken to be proportional to the emitter energy.  This choice of the scale makes it easier to 
 modify Eq.~(\ref{eq:qed2} ) when higher order corrections are considered \cite{Touschek:1968zz}.

We can now inspect the function $\beta(s,t,u)$ which will lead us to a phenomenological definition of the 
photon trajectory through Eqs.~(\ref{eq:qed2}) and (\ref{eq:regge1}).
For process (\ref{eq:qed1}), one has 
\begin{equation}
\beta(s,t,u)=\frac{e^2}{2(2\pi)^3}\int d^2 {\hat n}\sum_{i,j=1}^4 \frac{p_{i\mu} \epsilon_i}{p_i\cdot n}\frac{p_j^\mu \epsilon_j}{p_j\cdot n}
\end{equation}
with the four-vector $n^2=0$,  $\epsilon_i=\pm1$, for an electron or a positron in the initial state, or 
positron and electron in the final and the integration is over the angular distribution $d^2 {\hat n}$. 
Performing this integration leads to the following expressions
\begin{align}
\beta(s,t,u)=\frac{2 \alpha}{\pi}\large{\{} I_{12}+I_{1.3}-I_{14}-2\large{\}}\\
I_{ij}=2(p_i\cdot p_j)\int_0^1
\frac{dy}{m^2+2y(1-y)[(p_i\cdot p_j)-m^2]}\end{align}
The soft photon approximation is  the elastic approximation of Eq. ~(\ref{eq:qed1}) and one can now 
take the Regge limit   $s>>-t, -u$ and $s>>m^2$, i.e.
\begin{equation}
I_{12}+I_{14}\rightarrow 0 \ \ \ \ \ \ \  {t<<s, \ \ \ \ \ \ \ s\rightarrow \infty }
\end{equation}
 obtaining, in this limit,
\begin{equation}
\beta({s,t,u})\rightarrow \beta(t)=\frac{2\alpha}{\pi}[(2m^2-t)\int \frac{dy}{m^2-ty(1-y)}-2 ]
\end{equation}
We notice that this function has the correct limit $\beta(t)\rightarrow 0$ for $t=0$, since this is the exact elastic limit and it 
corresponds to no radiation at all at $t = 0$. 

The next step is to integrate Eq.~(\ref{eq:qed2}) up to a maximally observable $\Delta E$
using   as a scale the c.m. energy of the process (\ref{eq:qed1}), i.e. $s=4E^2$. 
Then, one obtains
\begin{equation}
\frac{d\sigma}{dt} = \large{(}\frac{\Delta E}{\sqrt{s}}\large{)}^{\beta(t)} \large{(}\frac{d\sigma_0}{dt}\large{)}
\end{equation}

We can now establish a correspondence between the correction rising from soft photon emission 
and the di-triple Regge limit of Eq. ~(\ref{eq:regge1}). This can be done by integrating the spectrum 
of the inclusive mass $M^2_X$ up to the maximally allowed value, which we can call $\Delta E$ in 
case of no momentum resolution.
We then immediately get
\begin{align}
\frac{d\sigma}{dt}=\int d{M^2_X/s} 
 \frac{d^2\sigma}{dt d(M^2_X/s)}
\\
  \rightarrow \large{(}
\frac{M^2_X}{s}
\large{)}
^{2(1-\alpha_{\gamma}(t))}=\large{(}
\frac{\Delta E}{\sqrt{s}}
\large{)}
^{4(1-\alpha_\gamma(t))}
\end{align}
and hence are led to  the correspondence
\begin{equation}
\alpha_\gamma(t)=1-\frac {\beta (t)}{4}
\end{equation}
and  to \cite{PancheriSrivastava:1973je}
\begin{equation}
\alpha_\gamma(t)=1-\frac{\alpha}{\pi}(\frac{2m^2_e-t}{\sqrt{-t}\sqrt{4m^2_e-t}}\log\frac{\sqrt{4m^2_e-t}+\sqrt{-t}}{\sqrt{4m^2_e-t}-\sqrt{-t}}-1). \label{eq:photontrajectory}
\end{equation}
Notice that to establish the above correspondence one had  to assume  Eq. ~(\ref{eq:regge1}), 
which obtains from the di-triple Regge limit with a vacuum trajectory $\alpha_V(0)=1$, i.e. the cross-section 
for  two reggeized photons into any final state
\begin{equation}
R_\gamma +R_\gamma \rightarrow  X
\end{equation}
is controlled at large energy by a trajectory $\alpha_{vacuum}(0)=1$. This assumption seems reasonable 
[for estimates about the photon trajectory 
computed to order $\alpha$] since the perturbative result for the vacuum trajectory is a branch cut, 
but whose deviation from unity begins at order $\alpha^2$.



To summarize: the proposed  photon {\it trajectory} \cite{PancheriSrivastava:1973je} 
\begin{equation}
\alpha_\gamma(t)=1-\frac{\alpha}{2\pi}[(2m_e^2-t)\int_o^1\frac{dy}{m^2_e-ty(1-y)}-2]\label{eq:gps}
\end{equation}
is an expression obtained in the large $s$, small $t$ limit, from resummation of all soft photons 
emitted in the scattering $e^+e^-\rightarrow e^+e^-$. 
In these expressions, $m_e$ is the mass of the fermion which emits the soft photons and the 
expression exhibits a threshold behaviour with a square root 
singularity (Notice that  being fermions, the threshold behaviour is different from the one required 
for a pion loop, for instance). 
The above expression may also be expressed as a dispersion integral
\begin{align}
\label{eq:gpsy1}
\alpha_\gamma(t)=1-\frac{\alpha}{\pi}[(2m_e^2-t)\times \\
\times \int_{4m_e^2}^\infty \frac{dt^{'}}{(t^{'} - t - i\epsilon)\sqrt{(t^{'}(t^{'} - 4m_e^2))}} - 1]=\nonumber\\
= 1 + (\frac{\alpha}{\pi}) t  \int_{4m_e^2}^\infty \frac{dt^{'} (t^{'} - 2m_e^2)}{t^{'}(t^{'} - t - i\epsilon)\sqrt{(t^{'}(t^{'} - 4m_e^2))}} 
\end{align} 

From Eq.(\ref{eq:gpsy1}), we may directly compute the imaginary part of the photon trajectory, which is positive definite:
\begin{equation}
\Im m\ \alpha_\gamma(t)= \vartheta(t - 4 m_e^2) \alpha \frac{(t - 2m_e^2)}{\sqrt{(t(t - 4m_e^2))}},
\label{eq:gpsy3}
\end{equation} 
and which has the asymptotic limit
\begin{equation}
\Im m\ \alpha_\gamma(t) \to\   \alpha,\ \ \ {\rm for}\  t > > 4 m_e^2,
\label{eq:gpsy4}
\end{equation} 
exactly the same result found by Lipatov for the iso-spin one vector boson trajectory in non-Abelian $SU(2)$ model with a doublet Higgs field \cite{Lipatov:1976zz}. 
We shall discuss it in the next subsection.

We see that the trajectory goes to 1 as $t\rightarrow 0$, in addition to having the threshold singularity corresponding to the fermion loop. 
The actual $t\rightarrow 0$ and $|t|\rightarrow \infty$ limits are also easily taken and lead to
\begin{eqnarray}
\alpha_\gamma(t)&\rightarrow& 1-(\frac{\alpha}{3\pi})
\frac{-t}{m^2_e}\ \ \ \ \ \  \ \ \ \ \ |t|<<m^2_e\\
\alpha_\gamma(t)&\rightarrow& 1-\frac{\alpha}{\pi} \log \frac{-t}{4m^2_e} \ \ \  \ \ \ \ |t|>>4m^2_e\label{eq:gps2}
\end{eqnarray}
namely one recovers the linearity of the trajectory at small $|t|$ and the asymptotic logarithmic limit at large $t$. 

\subsubsection{Comments on the reggeization of the photon}
\par\noindent
In the literature, one finds the statement that in QED the photon does {\it not} reggeize
[See, for example Gell-Mann {\it et al.} \cite{GellMann:1964zz}],
apparently in conflict with the 
Regge trajectory for the photon $\alpha_\gamma(t)$ found and 
discussed in the previous section. Hence, an explanation for the seeming discrepancy is mandatory.

The question of Reggeization in field  theory was begun by Gell-Mann {\it et al.} in the nineteen sixties
and continued in subsequent literature \cite{Mandelstam:1965zz} \cite{Abers:1970wn} \cite{Abers:1967zz}. 
Precisely to avoid IR divergences due to the zero mass of the photon, they and most others, considered
massive fermion QED with a conserved vector current but with a massive photon (vector boson). Then perturbation
theory was used to show that up to the sixth order the fermion reggeizes whereas the vector boson (photon) did not.

On the other hand,  massless QED requires a resummation making it non-perturbative and under the hypotheses
stated in the previous section, the photon does reggeize. Moreover, it was found in \cite{Grisaru:1979rf} that
all gauge vector bosons -including the photon- reggeize in a grand unified theory [GUT] based 
on a semi-simple group with a single coupling constant. However, it is difficult to assign significance to the result obtained 
in \cite{Grisaru:1979rf} in view of the lack of any phenomenological confirmation of GUT.

The problem of reggeization of the gauge bosons in the electro-weak $SU(2)\times U(1)$ theory 
has been discussed in the leading log approximation by Bartels {\it et al.} in \cite{Bartels:2006kr}. Through a
set of bootstrap equations they find that the $W$ boson  does reggeize whereas the $Z^o$ and the photon
do not. 


 \subsection{
 High energy behaviour of QCD scattering amplitudes in the Regge limit \label{ss:resumQCD}}

In the preceding section, we have described in some detail the question of gauge boson trajectories because in Regge theory, the high energy behavior of the scattering amplitude is given  by the exchange of Regge trajectories. We have shown there how  an effective Regge-like behaviour  can be obtained in QED,  from soft photon resummation in charged particle reactions.  In QCD, the role of soft photons is taken on by soft gluons, but, as we know, with enormous differences: not only a running coupling constant, but also un unknown (very likely a singular) behavior in the infrared. On the other hand,  QCD resummation is fundamental to the cross-sections, since, at high energy, the  behavior of the total cross-section is dominated by large distance effects, which correspond to very small momenta, and this immediately leads to the question of resummation of such quanta with very small momenta. 


While at low energy  Reggeons (such as the $\rho$ trajectory) dominate the hadronic scattering amplitude, 
at high energy the leading effect is obtained through a Pomeron exchange, in correspondence with a leading 
vacuum singularity in an even charge conjugation  channel, $C=+1$.   Thus theoretically, it was natural to identify 
the Pomeron as emerging from the exchange of two gluons (accompanied by soft gluon resummation) 
which does not change the quantum numbers of the process.
Phenomenologically, the one-to-one correspondence between the asymptotic total cross-section and the 
Pomeron trajectory, made the Pomeron go from a fixed pole ($\alpha_P= 1$) at low energies,  to a moving pole  
with intercept larger than one, to justify the 
rising cross-section 
observed at the ISR.
As for the $C=-1$ possible partner of the Pomeron, it is called an Odderon, but it is an object {\it so far only seen clearly  in theoretical papers}, as recently quibbed in \cite{Ewerz:2013kda}. The Odderon is also  a QCD effect, and its  trajectory so far seems to correspond to a constant $\alpha_O=1$. 
We shall briefly discuss the odderon in \ref{sss:odderontot} and \ref{sss:odderons-QCD}.


Clearly, for phenomenological applications, the dynamics of scattering among quarks and gluons 
needs to be understood, from  high-$p_t$ jets to that of the infrared gluon emission. In particular, for 
scattering in the soft region, the zero momentum region needs to be incorporated adequately. Since the latter, and 
most important aspect of the problem, has  not yet been completely solved, we can only try to 
give here  some specific examples of how 
one 
approaches the problem of the total cross-section 
in QCD,
starting 
with the preliminary building blocks such as the Balitsky, Fadin, Kuraev, Lipatov (BFKL) 
equation in \ref{sss:BFKL}, followed by the Gribov, Levin and Ryskin (GLR)  treatment 
in  \ref{sss:GLR}
and then the 
Balitsky, Kovchegov,  Peschanski 
equation in \ref{ss:BKgen}. A specific model  that realizes 
many of these QCD notions, the 
Durham-St. Petersburg 
model, will be presented in \ref{sss:KMRtot} and then rediscussed  in more  detail in Section \ref{sec:elasticdiff}. 
Recently an extensive description of  QCD as applied to the high energy scattering amplitudes in the Regge limit 
has appeared \cite{Kovchegov:2012mbw}, with both a theoretical and experimental up-to-date outlook. The field is 
very vast and cannot be covered in depth in this review. In the following, we  shall attempt to outline some of the 
most important physics chapters in the story hoping that our summary would provide a starting point to
a worker interested in the field.

\subsubsection{Non Abelian gauge theory with Higgs symmetry breakdown and the BFKL integral equation}\label{sss:BFKL} 


With the advent of $SU(2)$ Yang Mills  (YM) theory containing a triplet of vector gauge bosons which acquire a mass through an iso-doublet Higgs field, 
investigations turned into answering the reggeization questions for the massive gauge bosons and the Pomeron in the vacuum channel of the theory. 
Such a theory is renormalizable and endows the gauge vector bosons a mass $M$ through the spontaneous symmetry breakdown mechanism.

In a series of papers, Lipatov \cite{Lipatov:1976zz} and co-workers,  Fadin, Kuraev and Lipatov \cite{Fadin:1975cb} \cite{Kuraev:1977fs}; 
Balitsky and Lipatov \cite{Balitsky:1978ic}; \cite{Balitsky:1979ns} and \cite{Lipatov:1985uk}, did fundamental work in this field which goes under
the generic name of BFKL formalism.

Lipatov {\it et al.} found that the gauge vector boson reggeizes, i.e., the elementary iso-vector gauge particle of angular momentum 
$j = 1$ at mass $t = M^2$
turns into a Regge trajectory. Their expression may be written as a dispersion integral
\begin{equation}
\alpha_V(t)=1 + \frac{\alpha_{YM}}{\pi} (t - M^2)\int_{4M^2}^\infty \frac{dt^{'}}{(t^{'} - t - i\epsilon)\sqrt{(t^{'}(t^{'} - 4M^2))}},
\label{y1}
\end{equation} 
where $\alpha_{YM} = g^2/(4\pi)$ and  $g$ is the gauge coupling constant. The above expression verifies explicitly that the gauge vector boson 
trajectory goes to 1 at  $t = M^2$. 

It is instructive to note the remarkable similarity between the expression Eq.(\ref{eq:gpsy1}) for $\alpha_\gamma(t)$ found in QED with the gauge
vector boson trajectory given in Eq.(\ref{y1}). $\alpha_\gamma$ goes to 1 at the physical mass $t = 0$  of the QED gauge boson (the photon), just
as the non-Abelian gauge boson trajectory $\alpha_V$ goes to 1 at its physical mass $t = M^2$. Furthermore, the absorptive part of $\alpha_V(t)$
reads
\begin{equation}
\Im m\ \alpha_V(t)= \vartheta(t - 4 M^2) \alpha_{YM} \frac{(t - M^2)}{\sqrt{(t(t - 4M^2))}},
\label{y2}
\end{equation}   
and it differs from its corresponding expression Eq.(\ref{eq:gpsy3}) for $\Im m\ \alpha_\gamma(t)$ in the replacement $(\alpha, m_e)\ \to\ (\alpha_{YM}, M)$
and in the numerator ($t - 2m_e^2$) to ($t - M^2$). Asymptotically, as stated earlier, the difference vanishes. For large $|t|$, both imaginary parts
go to their respective $\alpha$  or $\alpha_{YM}$.

The situation regarding the vacuum channel or the nature of the Pomeron in this YM theory is still rather obscure. In Lipatov's original paper, it is stated
that if only two particle singularities in the $t$ channel were included, a bare Pomeron trajectory did emerge which may be transcribed in the form
\begin{align}
\alpha_P^{(o)}(t)=1 + \frac{\alpha_{YM}}{\pi} (2t - 5M^2/2)\times \nonumber\\
\times \int_{4M^2}^\infty \frac{dt^{'}}{(t^{'} - t - i\epsilon)\sqrt{(t^{'}(t^{'} - 4M^2))}},
\label{y3}
\end{align} 

But going further and including three particle thresholds in the $t$ channel, led Lipatov to conclude that there is a branch cut in the angular momentum
plane in the vacuum channel due to the exchange of two reggeized vector bosons. This was confirmed in a later paper by Fadin, Kuraev and Lipatov. They obtained
the result that the leading j-plane singularity in the vacuum channel is a branch point at $j =\ 1 + \alpha_{YM} [8 ln(2)/\pi]$ which for an $SU(N)$
theory would read 
\begin{equation}
\label{PomYM}
\alpha_{Pom}^{YM} =\ 1 + \alpha_{YM} [\frac{4 N ln (2)}{\pi}]. 
\end{equation}
Hence, they conclude that in the main LLA, the total cross-sections in a non-abelian gauge theory, 
would violate the Froissart bound. The reason for this violation is that $s$ channel unitarity is not satisfied in the LLA (which assumes $\alpha_{YM}\ ln(s/M^2) \sim\ 1$)
and presumably a proper computation of the vacuum exchange in the $t$ channel would require excursions beyond LLA which must also include
contributions of order $\alpha_{YM}\ ln(s/M^2) > > 1$.

At this juncture, $s$ channel elastic unitarity may be imposed via the eikonal expansion as proposed by Cheng and Wu. 
Such eikonal procedures in various forms have been followed over the years by various groups as discussed in various parts of our review.        


Development of ideas and results from QED and non-abelian $SU(2)$ with a Higgs mechanism  to the theory of interest 
namely QCD [unbroken $SU(3)_{colour}$] 
runs into a host of  well known difficulties. At the level of quarks and glue, one can address the question of the reggeization 
of the gluon trajectory. Unlike the YM theory 
with spontaneous symmetry breakdown which endows the gauge boson with a mass $M$ as discussed in the last subsection, 
gluon remains massless. The difficulty 
is seen immediately as $M \to 0$ in the vector boson trajectory Eq.(\ref{y1}), which diverges  at the lower limit. The reason for 
this divergence is clear in that all 
gauge boson thresholds condense at $t = 0$ as the gauge boson mass vanishes. [Such is not the case for QED due 
to the absence of non-linear couplings of a 
photon to itself].

On the other hand, in the large $t$ limit, we see from Eq.(\ref{y2}) that for SU(N) YM,  $\Im m\ \alpha_V (t) \to (\frac{N}{2}) \alpha_{YM}$. 
Hence, we expect that
\begin{equation}
\label{QCD1}
\Im m\ \alpha_{gluon}(t) \to\  (\frac{3}{2}) \alpha_s\ \ {\rm as}\ t \to \infty,
\end{equation}
[at least for constant $\alpha_s$].This expectation agrees with explicit computations by Lipatov, Balitsky and Fadin in LLA which reads
\begin{align}
\label{QCD2}
\alpha_{gluon}(t) -1 = [(\frac{3 t}{2\pi^2}) \alpha_s] \int \frac{(d^2k)}{k^2 [(q - k)^2]}\nonumber
\\ \approx\ -  (\frac{3}{2\pi}) \alpha_s ln(-t/\lambda^2) ,
\end{align} 
The scale $\lambda$ corresponds to the lower integration cut-off, introduced to avoid the IR region where perturbative QCD cannot be applied.

It is useful to note the similarities and differences betweens BFKL and the approach to the photon trajectory 
obtained through resummation in QED, as described in \ref{sss:reggeization-photon}. In that  approach, and its 
extension  to QCD leading to what we call the BN (for Bloch Nordsieck) model for the total cross-section 
(see \ref{sss:BN}), the IR divergence is cancelled at the level of the observable cross-sections and the asymptotic behavior 
is obtained directly from the cross-section.  On the other hand, the BFKL result is obtained  in terms of matrix elements, 
which are  calculated in different orders in the coupling constant  through dispersion relations and unitarity.

In our  approach \cite{PancheriSrivastava:1973je}  for the photon trajectory, we obtain both a linear term in $t$ for small $t$ 
values as well as a logarithmic behavior for large $t$.
For large $t$, the correspondence between the two trajectories $\alpha_\gamma (t)$ as given by Eq.(\ref{eq:gps2}) and $\alpha_{gluon}(t)$
as given by Eq.(\ref{QCD2}) [written for $SU(N_c)$] is immediate
\begin{equation}
\label{QCD2a}
-\frac{\alpha}{\pi} \ln (\frac{-t}{4m^2}) \mapsto  -\frac{\alpha_s N_c }{2\pi}\ln (\frac{-t}{\lambda^2}),
\end{equation}
from which the substitution $\alpha \rightarrow  (N_c/2)\alpha_s $ maps one into another.
The small $t$-behaviour is much more complicate, because of the unknown infrared behaviour of the strong coupling constant.
In \ref{sss:BN} we shall describe a model for the coupling constant, which allows to  apply resummation in the infrared region, 
and its application to total cross-section studies.

For high energy hadronic (or, photonic) amplitudes, we are primarily interested in the nature of the Pomeron emerging from QCD. 
The physical underlying picture is that a Pomeron
is a bound state of two Reggeized gluons. For this purpose the BFKL approach may be summarized as follows. In LLA, colour singlet 
hadronic/photonic amplitudes are related to their
angular momentum amplitudes through a Mellin transform 
\begin{align}
\label{QCD3}
A(s, t= - q^2) = is \int_{\sigma-i\infty}^{\sigma+i\infty} [\frac{d\omega}{2\pi i}] s^\sigma f_\sigma (q^2)\nonumber\\
f_\sigma(q^2) = \int (d^2k)(d^2k^{'}) {\bf \Phi}^{(1)}(k,q) {\bf \Phi}^{(2)}(k^{'},q) f_\sigma (k,k^{'}; q),
\end{align}
where  $k,k^{'}$ denote the transverse momenta of the exchanged gluons and the function $f_\sigma(k, k^{'}; q)$ can be 
interpreted as the $t$ channel partial wave amplitude for 
gluon-gluon scattering with all gluons off mass shell, with  squared masses: $- k^2$, $-k'^2$, $-(q-k)^2$, $-(q - k^{'})^2$. 
The gluon propagators are included in the function
$f_\sigma(k, k^{'}; q)$. The ${\bf \Phi}^{1,2}$ functions describe the internal structure of the colliding particles 1 \& 2. 
Gauge invariance is then imposed so that
\begin{equation}
\label{QCD4}
\Phi^{1,2}(k, q)|_{k=0} = 0 = \Phi^{1,2}(k, q)|_{k=q}.
\end{equation}
This property is crucial for using a gluon mass $\mu$ in its propagator (as in the SSB YM theory of the last subsection) 
because for colour singlet states there is no IR divergence
and the limit $\mu \to 0$ exists. There is no such IR safety for colour non-singlet amplitudes.  

Armed with the above, for high energy analysis of colour singlet amplitudes, one may imagine to freely employ all the results 
of the SSB YM of the last subsection extended to the
$SU(N=3)$ case. But, such is not the case basically because of asymptotic freedom, i.e., $\alpha_{YM}$ must be replaced 
by $\alpha_s(q^2)$, which for large $q^2$ goes to zero
logarithmically  
\begin{equation}
\label{QCD5}
\alpha_s(q^2) \to\ \frac{4 \pi}{(11  N- 3n_f) ln(q^2/\Lambda^2)};\ {\rm for}\ q^2>> \Lambda^2
\end{equation}
 with $ n_f$  the\ number\ of\ flavours.
On the other hand, for $q^2 \leq\ \Lambda^2$, $\alpha_s$ begins to diverge and hence perturbative QCD becomes inapplicable. 
However, one may derive some useful results in the symptomatic freedom (AF) limit.
As $ln(q^2/\Lambda^2)>>1$, an infinite set of poles  condense to $j \to 1$, whose behaviour may be approximately described through a moving cut
\begin{equation}
\label{QCD6}
\alpha_{Pom}(q^2)|_{ln(q^2/\Lambda^2)>>1} \to 1 + [\frac{4N ln(2)}{\pi}] \alpha_s(q^2).
\end{equation}
The fixed Pomeron branch cut at 
$\alpha_{Pom}^{YM} = 1 + [\frac{4N ln(2)}{\pi}] \alpha_{YM}$ found in Eq.(\ref{PomYM}) is ``made to move'' as $\alpha_s(q^2)$ for large $q^2$. 


In various phenomenological models\cite{Fadin:1996tb}, \cite{Fazio:2011ex}, \cite{Fiore:2008tp} on the other hand, 
the Pomeron trajectory is taken to asymptote to $-\infty$ as $q^2 \to \infty$ 
\begin{equation}
\label{QCD9}
\alpha_P(q^2) = \alpha(0) - \alpha_1 ln (1 + \alpha_2 q^2),
\end{equation}
thus the forward slope of the Pomeron -not directly coupled to $\alpha(0)$- is given by
\begin{equation}
\label{QCD10}
\alpha_P^{'}(0) = \alpha_1 \alpha_2.
\end{equation}


\subsubsection{The odderon}\label{sss:odderontot}

In the following, we shall first give a brief introduction to the odderon.  
The odderon was first introduced on phenomenological grounds  in the early 1970' s by   
Lukaszuk and  Nicolescu  \cite{Lukaszuk:1973nt}  as a $C=-1$ exchange term in the amplitude,  
and further discussed in \cite{Joynson:1975az}.  Since then, there has been no experimental confirmation 
of its existence, although no confirmation of its non-existence has arrived either. One reason for this sort 
of {\it limbo} in which the odderon lives is that this is not a dominant effect and since fitting of the data 
requires a number of terms and concurrent parameters,  it is often possible to mimic its presence by adding some terms. 

Consider the crossing-odd amplitudes in $pp$ \& $p\bar{p}$ scatterings defined as
\be
\label{O1}
F_{-} = \frac{1}{2} [F_{pp} - F_{p\bar{p}}].
\ee
Basically, there are three types of odderons classified  according to their increasing order in energy asymptotic behaviour \cite{Block:2006hy}:
\begin{itemize}
\item Order zero odderons $F_{odd}^{(0)}$ are real and hence only change the real parts of the elastic amplitudes
\item Order one odderons $F_{odd}^{(1)}$ change the cross-section (between a $pp$ \& $p\bar{p}$) by a constant amount
\item Order two odderons $F_{odd}^{(2)}$ [also called maximally singular] lead to a cross-section difference increasing as $ln(s/s_o)$
as well as real parts  that are not equal asymptotically 
\end{itemize} 

Given that the high energy elastic amplitudes are predominantly imaginary, the zero order odderons 
are hard to look for. For obvious reasons, the hunt has been to look for order two or maximally odd amplitudes. 
We shall discuss in various places the results of such searches.

Another quantity which could shed light on the presence of the odderon, is the parameter $\rho(s)$, 
the ratio of the real to the imaginary part of the scattering amplitude for hadrons in the forward direction. 
Possibly, the rather precise measurements of this parameter at LHC could allow to draw some conclusions 
about its presence. From this point of view, we notice that the  somewhat low, preliminary value for the parameter 
$\rho$ at LHC8 could be invoked to be a signal of the odderon.
Also diffractive production of pseudo-scalar and tensor mesons in $ep$ scattering are suggested 
to be a good place were effects from the odderon could be detected.

An extensive review of the status of the odderon appeared in 2003 by Ewerz \cite{Ewerz:2003xi}.
The odderon could be responsible for  the difference between   the elastic differential cross-sections 
for $pp,\ p {\bar p}$, past the forward peak. 
According to Ewerz  \cite{Ewerz:2005rg} for instance,   hadronic exchanges  occurring only  through  
mesonic reggeons are not sufficient to explain the cancellation of the dip in \pbarp. In this respect however, 
one can see that the dip may be slowly reappearing in 
$\ p {\bar p}$, as some analysis of the data show. One can in fact observe that $p {\bar p}$ data  
for the elastic differential cross-section  from ISR to the TeVatron indicates that the observed change in 
curvature becomes more and more pronounced as the energy increases. Whatever the odderon does, 
it would seem that as the energy increases, it may disappear. 

A discussion of the ``missing odderon" can  be found in Donnachie, Dosch and Nachtmann \cite{Donnachie:2005bu} 
where once more the evidence for the  phenomenological odderon is discussed and found lacking. 
 
 A proposal was made to detect the odderon at RHIC and LHC by Avila, Gauron and Nicolescu \cite{Avila:2006wy}.
 The model by Avila, Campos, Menon and Montanha   which incorporates both the Froissart 
 limiting behavior as well as Pomeron and Regge exchanges \cite{Avila:2006ya} is described later
 in the elastic scattering chapter. 

\subsubsection{Odderons in QCD}\label{sss:odderons-QCD}
\par\noindent
In the Regge language, an odderon is defined \cite{Nicolescu:2007ji} as a singularity at 
the angular momentum $J = 1$ at $t = 0$ in the crossing-odd amplitude. Hence, the
obvious question arose as to whether an odderon could be associated with $3-$gluon exchanges in QCD \cite{Bartels:1980pe} 
\cite{Jaroszewicz:1980rw} \cite{Kwiecinski:1980wb} and a definite affirmative answer in pQCD was obtained in \cite{Bartels:1999yt}.

Presently, apart from the phenomenological interest, there has been a strong QCD attention paid to  
NLO corrections to the odderon trajectory \cite{Bartels:2013yga} and on the properties of the odderon in strong coupling 
regime \cite{Brower:2013jga}. 

In   \cite{Ewerz:2013kda}, a  model for soft high-energy scattering, which includes a 
 tensor Pomeron and vector odderon has appeared with detailed description of Feynman-type rules for 
 effective propagators and vertices. 

\par\noindent

\subsubsection{Gribov-Levin-Ryskin (GLR)
model}\label{sss:GLR}
In 1984 V.N. Gribov, Levin and Ryskin  \cite{Gribov:1984tu} wrote a paper meant to  establish the theoretical basis of 
a QCD approach to parton scattering, 
in which the focus was on semi-hard scattering. In the first section they defined {\it semi-hard} scattering. 
The framework is that of deep inelastic scattering  as in Fig.~\ref{fig:GLRfig1.1}.
\begin{figure}
\resizebox{0.5\textwidth}{!}{
\includegraphics{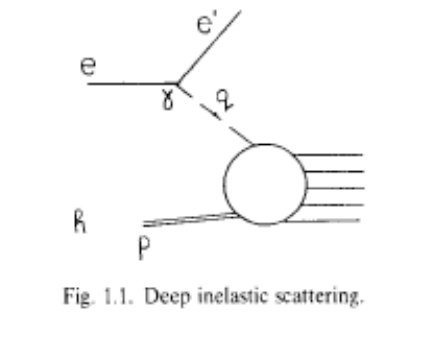}}
\caption{Deep inelastic scattering representative figure from \cite{Gribov:1984tu}.
Reprinted from \cite{Gribov:1984tu}, \copyright (1984) with permission by Elsevier.}
\label{fig:GLRfig1.1}
\end{figure}
The focus on DIS has resulted in introducing an external scale $q^2$ which rendered the extrapolation to 
real photons and, for a while, application to purely hadronic processes difficult. 

GLR argument starting point is that,  as the fractional energy of the scattering partons, $x$, decreases,  the increase of the number of partons 
can be so high as to give a cross-section as large, or even larger,  
as  the actual hadronic cross-section. But this cannot continue indefinitely. When the density of partons becomes 
very large, the partons within a given hadron cannot be any more 
considered to be independent  and they instead start to interact with each other.
GLR define the probability $W(x,q^2)$ that partons interact with each other in a hadron  as
\be
W(x,q^2)=\frac{\alpha_s(q^2)}{q^2 R^2_h}F(x,q^2)
\label{eq:W}
\ee
where $F(x,q^2)$ is  the  parton density function, i.e. gives the number of partons which interact with 
the probing virtual photon,  and $R_h$ the hadron radius. In Eq.~\ref{eq:W}, the ratio
$\frac{\alpha_s(q^2)}{q^2}$ is the  parton-parton cross-section  
and $\Delta b^2_\perp=R^2_h/F(x,q^2)$ is the average perpendicular distance between partons, 
measured through the area where they are and the number density. This is a semi-classical description. 
When the parton-parton \x \ is smaller than the average distance, $W<1$ and for $W \ll 1$ rescattering 
will {\it not } occur, when it is larger, i.e. $W\lesssim 1$ rescattering  will occurr. $W= 1$ is called the {\it unitarity limit}. 
This function allows then to distinguish three regions in the $x$-variable, depending on whether  pQCD applies (or does not apply)  
for the calculation of the parton density function $F(x,q^2)$. Thus, while there is a whole region where nothing can be 
obtained, for $W\le \alpha_s$ the authors  have been able to obtain  interesting results. 

It is worth 
noting that their picture [Fig. 1.9 of their paper] is at the basis 
of the subsequent work by this group.
Consider then the function $F(x,q^2)$ rewritten as
\be
F(x,q^2)=W(x,q^2) \frac{q^2 R^2_h}{\alpha_s(q^2)}
\ee
which is plotted as a function of $x$  in Fig.~\ref{fig:GLR1-9}. The curve denotes the region $W(x,q^2)\le 1$. 
The subsequent behaviour is controlled by the ratio $W/\alpha_s$, both numbers being  less than 1. 

One distinguishes 3 regions, with regions B and region C separated at a value $x_b(q^2)$, where $F(x,q^2)=q^2 R^2_h$. 
For $x<x_b$, $q^2 R^2_h<F\le  \frac{q^2 R^2_h}{\alpha_s(q^2)} $. In region C, $W/\alpha_s>1$. Here the interactions 
are very strong and near the unitarity limit, and the authors have not be able to take them into account.
Region A, is where $W\ll \alpha_s$, region where pQCD applies, it correspond to large $x\sim 1$ and parton interactions 
are negligible.   In region B, $W/\alpha \lesssim 1$ and, for  W not too large,  the authors say that with the Reggeon-type diagram 
technique developed in subsequent chapters they are able to take into account interactions in this region. However, if $W $is 
largish,  $W  \sim 1$, they have not been able to sum all the essential diagrams and only can give qualitative considerations.
We show the three regions in Fig.~\ref{fig:GLR1-9}.
\begin{figure}
\resizebox{0.5\textwidth}{!}{
\includegraphics{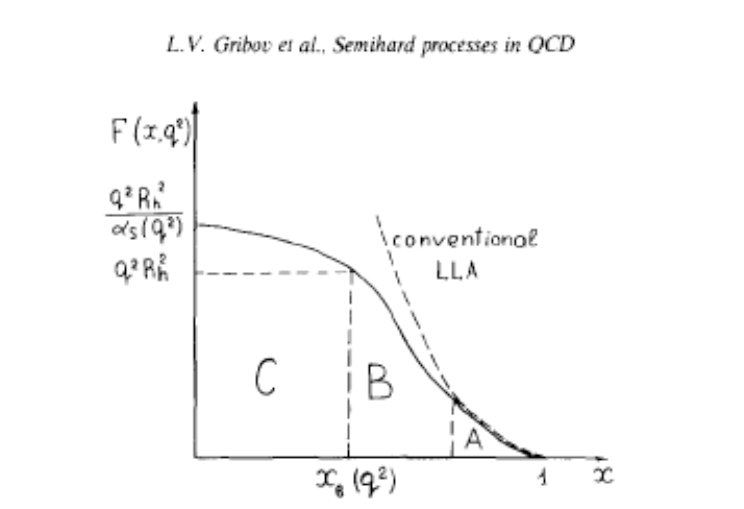}}
\caption{$x$-dependence of the structure function from \cite{Gribov:1984tu}. Reprinted from
\cite{Gribov:1984tu}, \copyright (1984) with permission by Elsevier.}
\label{fig:GLR1-9}
\end{figure}
$A$ is the region where perturbative QCD can be applied, B is the result of this paper, C is not to be obtained yet.

There is a useful, again semi-quantitative, argument to determine the value $x_b(q^2)$, which we reproduce in the following. The conditions are
\bea
F\sim q^2 \ and \  hence \ \ln F &\sim &\ln q^2\\
\frac{\partial F}{\partial x} &\sim& \frac{\alpha_s}{x}F
\eea
the second equation indicating that the origin of the partons as the number of partons increase is due to bremsstrahlung 
(hence the bremsstrahlung spectrum). 
Using the running $\alpha_s$ expression one obtains
\be
\frac{\partial \ln F}{\partial \ln x}\sim \frac{1}{\ln F}
\ee 
from which 
\be
\ln \frac{1}{x_b(q^2)}\sim \ln^2 q^2 
\label{eq:xb}
\ee

After determining this value, the paper goes on to discuss the inclusive jet spectrum, 
the authors introduce the quantity $k_0$ which corresponds 
to the limiting value $x_b(k_0^2)$, i.e. $2k_0/\sqrt{s}=x_b(k_0^2)$ and, using 
Eq.~\ref{eq:xb}, with $q^2=k_0^2$, they obtain
\bea
\ln\frac{1}{x_b(k_0^2)}\sim \ln^2 (x_b^2)\\
k_0\sim e^{c\sqrt{\ln s}}
\eea
The last equation obtains by neglecting a $\ln k_0$ term relative to  $\ln^2k_0$ in Eq.~\ref{eq:xb}.

As a consequence of the  behaviour  thus obtained for the increase in the number of partons below $x_b(q^2)$, 
the authors find that semi-hard processes have a large cross-section and contribute substantially to the average multiplicity as well. 
In particular, since the average multiplicity is proportional to the phase space factor $k_0^2$, one obtains also
\be
{\bar n}\sim e^{2c \sqrt{\ln{s}}}
\ee
The parameter $c$ is determined in subsequent chapters.

The important result of this paper is that semi-hard processes become responsible 
for large part of the cross-section, because of the 
behaviour with energy described through the above equations. Just as the average multiplicity increases, 
the average transverse momentum is also proportional to $k_0$ and thus
\be
<q_t>\sim k_0\propto \Lambda e^{c \sqrt{\ln s}}
\ee
Therefore at very high energy many jets with comparatively large transverse momenta are produced.

The above qualitative description of the region of small $x$ is then further developed in 
chapter 2 of the paper. The authors promise to show 
how to apply the Leading Logarithm Approximation  ( LLA)  to the small $x$ region and to deal 
with the screening effects due to the parton-parton interactions. 
Their aim in the second chapter is to calculate the structure function $F(x,q^2)$ when both 
$\ln q^2/q_0^2$ and $\ln 1/x$ are large. Two problems are encountered: 
the 
necessity to develop LLA in two large logarithms and how to deal with unitarity i.e. the increase of the structure function at very small $x$.

Notice that in the subsection dealing with unitarization in the Double Logarithm Approximation (DLA), a crucial role is played by the quantity
\be
\xi - \xi_0
=b\int_{q^2_0}^{q^2}
\frac{dk^2}{k^2}
\frac{\alpha_s}{4\pi}
\ee
with $b=11N-3N_f$. In their Section 2 .2, the structure of the theory is developed. One of the conclusions is that  it is necessary 
to take into account not just 
corrections of the type $\alpha_s \ln(1/x)\ln (q^2)$ but also $\alpha_s \ln(1/x)$ and $\alpha_s \ln (q^2)$, that multi ladder ``fan'' 
diagrams in the $t$-channel 
are crucial for unitarization. It is asserted  that all other corrections are negligible, 
at least, 
up to the 
normalization of the structure functions. The asymptotic limit of resumming diagrams which 
grow as $(\alpha_s \ln (q^2))^n$ and those which 
grow as $(\alpha_s \ln(1/x))^n$, i.e. the asymptotic behaviour in the DLA, gives
\be
F\propto exp\{ \sqrt{2(\xi -\xi_0)}y\}
\ee
where $y=8N/b\ln 1/x$.

In Section 3, a discussion of the Reggeon Diagram Techique (RTD) in QCD is given. In this version of RTD, the primary object is an LLA ladder, 
which can also be conventionally called a Pomeron. The vertices of interaction between ladders are also calculated in perturbative QCD. 
In Fig.~\ref{fig:GLR3-1} we reproduce Fig. 3.1 of the GLR paper, where the Pomeron and the vertices are schematically indicated.
\begin{figure}
\resizebox{0.5\textwidth}{!}{
\includegraphics{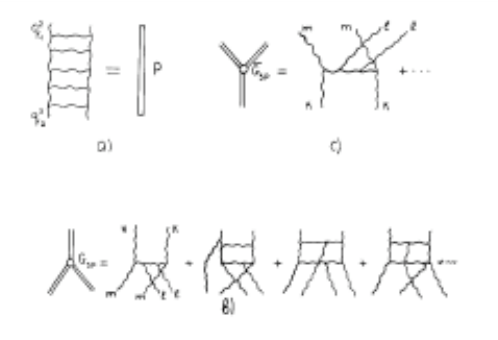}}
\caption{The QCD Pomeron from \cite{Gribov:1984tu} and the triple pomeron vertices. Reprinted from
\cite{Gribov:1984tu}, \copyright (1984) with permission by Elsevier.}
\label{fig:GLR3-1}
\end{figure}
 
 Before proceeding further, one should notice two important differences with respect to  
 an approach 
 based on mini jets and soft gluon resummation, such as the one we shall describe in \ref{sss:BN}:
 \begin{enumerate}
 \item when dealing with parton-parton probability of interaction or even single jet production as in Eq. 4.3 of their paper, i.e.
 \be
 \frac{E d\sigma^{jet}}
 {d^3p}\sim
  \frac{d{\hat \sigma}}{dp_t^2}F(x_1,p_t^2)F(x_2,p_t^2)\sim\frac{\alpha_s}{p_t^4} F_1 F_2
 \ee
 the expression for jet-production is linear in $\alpha_s$, not quadratic; this corresponds to a view in which soft or semi-hard partons are 
 all treated on equal footing, whereas mini-jet models distinguish between hard parton-parton collisions from soft gluon emission, 
 which is separately resummed. 
  \item the other consideration is an observation at the beginning of Sect. 3.3 of their paper about infrared divergence. The authors 
acknowledge that the quarks being coloured, the quark-quark amplitude is divergent, but the divergences are absent when discussing the 
scattering of colourless objects, such as  hadrons. The rationale being that the divergences cancel if one takes into account the interactions 
with the spectator quarks which, together with the interacting quarks constitute the hadron. 
 \end{enumerate}
 
Section 4 of the GLR paper is dedicated to large $p_t$ processes. It is in this section, that the gluon regge trajectory is called in to play 
an important role in the correlations between two large $p_t$ gluon jets. In this case, the double inclusive cross-section in the region
\be
1\ll \Delta \eta\ll \frac{1}{N\alpha_s(p_t^2)} 
\ee
is given by
\begin{align}
\frac{
E_cE_d d\sigma^{2-jet}
}{d^3p_cd^3p_d}=\nonumber\\
= \frac{36}{(16 \pi^2)^2 p_c^2 p_d^2} \int \frac{d^2k_1 d^2k_2 }{\pi^4} \alpha_s^2(p_t^2)\phi(x_1, k_1^2) \phi(x_2, k_2^2)\nonumber\\
\times \delta^2({\bf k}_{1t} + {\bf k}_{2t} - {\bf p}_{ct} - {\bf p}_{dt})
(\frac{{\tilde s}}{4p_{ct}p_{dt}})^{2{\bar \alpha}_G(q_t^2)} T_G^4,\nonumber\\
\label{eq:2jet1},
\end{align}
with the gluon trajectory given as 
\be
\alpha_G(q_t^2)\simeq \frac{3\alpha_s(q_t^2)}{2\pi}\ln\frac{q_t^2}{\mu^2},
\ee
and $\mu$ an infrared cut-off introduced  because the trajectory is infrared divergent \cite{Fadin:1975cb,Kuraev:1977fs}. Of course the infrared 
divergent part of the reggeization is cancelled by the emission of real soft gluons. After this is taken into account, 
it is given as
\be
{\tilde \alpha}_G(p_t^2)=\frac{3\alpha_s(p_t^2)}{2\pi}\ln\frac{p_t^2}{max\{p_{\Sigma^2},k_0^2 \}}+{\cal O}(\alpha_s), 
\ee
with $p_{\Sigma}=p_{ct}-p_{dt}$. Thus
\be
\frac{
E_cE_d d\sigma^{2-jet}
}{
d^3p_cd^3p_d
}\propto (
\frac{{\tilde s}}{4p_{ct}p_{dt}}
)^{2{\bar \alpha}_G(q_t^2)} T_G^4
\label{eq:2jets}
\ee
Finally a form factor $T_G$ is defined as follows:
\be
T_G(p_t^2,(\Delta p)^2) \simeq exp[- \frac{3\alpha_s(p_t^2)}{4\pi}\ln^2\frac{q_t^2}{(\Delta p)^2}]
\ee
which 
incorporates the probability that the global momentum of the emitted (bremsstrahlung) gluons 
which accompany the emission of a hard gluon 
of momentum $p_t$ is smaller than a $\Delta p\ll p_t$.

Electron Positron processes and the properties of produced jets and correlations are both 
examined  in the remaining part of Sect. 4, 
and mostly in Sect.5. After this section, the authors turn to a discussion of the phenomenology 
of semihard processes from the perturbative QCD viewpoint. 


\subsubsection{
KMR model with BFKL Pomeron}
\label{sss:KMRtot}

We shall now examine a specific model in which the theoretical input from the BFKL Pomeron is  included into a 
phenomenological application. Model of this type been developed by various groups, such as the Durham-St Petersburg 
group of Khoze, Martin and Ryskin (KMR), the Telaviv group of    Gotsman, Levin and  Maor (GLM),  Ostapchenko and 
collaborators,  among others, and will be also discussed in Sect. ~\ref{sec:elasticdiff}.
Here we shall describe the model  by Khoze and collaborators \cite{Ryskin:2009tj,Ryskin:2009tk} which has been applied to  
both the elastic and the total cross-section for quite some time. 
A description of the KMR   model for the total cross-section and its extension to elastic scattering 
can be found in \cite{Khoze:2000wk}. In this paper, the discussion is focused on how to take into account the single and double diffractive 
components of the scattering and the following features are discussed
\begin{itemize}
\item an estimate for the diffractive components in a two channel model and comparison with the 
Pumplin bound \cite{Pumplin:1973cw}
\item the t-dependence of the slope parameter $B(t)$, at different energies and how this dependence 
is related to the relative importance 
of pion loops in the calculation of the Pomeron trajectory
\item  survival probabilities of rapidity gaps
\end{itemize}

A simplified version of the model can be found in Appendix A of  \cite{Khoze:2000wk}.
From the expression for the total cross-section in impact parameter space 
\bea
 \sigtot=2\int d^2\vecb_t A_{el}(b_t)\\
 \sigel=\int d^2 \vecb_t |A_{el}(b_t)|^2
\eea
it is clear that $A_{el}$ is purely  imaginary in the model.
A two-channel eikonal is considered, elastic $p\rightarrow p$ and $p\rightarrow N^*\rightarrow p$, 
see Eq. (33) of their Appendix A, 
which will be  discussed in more detail in the section of this review on the elastic cross-section. 
In the case of a single channel they write
\be
\Im m A_{el}=[1-e^{-\Omega(b_t)/2}]
\ee 
With an effective (for illustration)   Pomeron trajectory written as $\alpha_P(t)=\alpha_P(0)+\alpha'_P t=1+\Delta +\alpha'_P t $ and vertex with 
exponential t-dependence $\beta_p exp(B_0 t)$, the {\it opacity } is written as
\be
 \Omega(b,s)=\frac{ 
 \beta_P ^2 (s/s_0)^{\alpha_P(0)-1} 
 }
 {4\pi B_P}
 e^{-b_t^2/4B_P}
\ee
This result is obtained starting  with the usual Regge-Pomeron expression, i.e. 
 \begin{align}
 \Im mA_{el}(s,t)=\beta_P^2(t)\left( \frac{s}{s_0}\right)^{\alpha_P(t)-1}\\
 =\beta_P^2(t)\left( \frac{s}{s_0}\right)^{\alpha_P(0)-1} e^{\alpha'_P\ t\  \log \frac{s}{s_0}}
 \label{eq:KMRampltot}
 \end{align}
  The amplitude in $b-space$ is then obtained  as 
  the Fourier transform of Eq. ~(\ref{eq:KMRampltot})  with $t=-q^2 $
 \bea
 {\cal F}[A(s,t)]&=&\frac{1}{(2\pi)^2}\int d^2\vecq e^{i\vecb \cdot \vecq} A(s,t)\\
& =& \left( \frac{s}{s_0}\right)^{\Delta} e^{(B_0 /2+\alpha'_P\  \log \frac{s}{s_0}) b^2)/4}
 \eea
 and  then eikonalized,   obtaining 
 \begin{align}
 \sigtot=4\pi \Im m A(s,0)=2\int d^2\vecb
  [
 1-e^{
 -\Omega(b,s)/2)}
 ]
 \\
 \Omega(b,s)=\frac{ 
 \beta_P ^2 (s/s_0)^{\alpha_P(0)-1} 
 }
 {4\pi B_P}
 e^{-b^2/4B_P}\\
 B_P=\frac{1}{2}B_0+\alpha'_P \log (s/s_0)
\end{align}

In  two more recent papers \cite{Ryskin:2012ry,Martin:2012mq} the crucial question of the transition 
from soft to hard is examined again. 
We shall first summarize their picture of the transition from \cite{Martin:2012mq} and then, 
in the next section dedicated to the elastic 
differential cross-section,  describe their latest results.
The (QCD) Pomeron is here associated with the BFKL singularity. It is noted that, although 
the BFKL equation should be written for gluons away 
from the infrared region, after resummation and stabilization, the intercept of the BFKL Pomeron 
depends only weakly on the scale for reasonably small 
scales. We reproduce in Fig.~\ref{fig:BFKLpom} their description of the connection between the 
intercept of the BFKL Pomeron and the value for $\alpha_s$.
\begin{figure}
\centering
\resizebox{0.5\textwidth}{!}{
\includegraphics{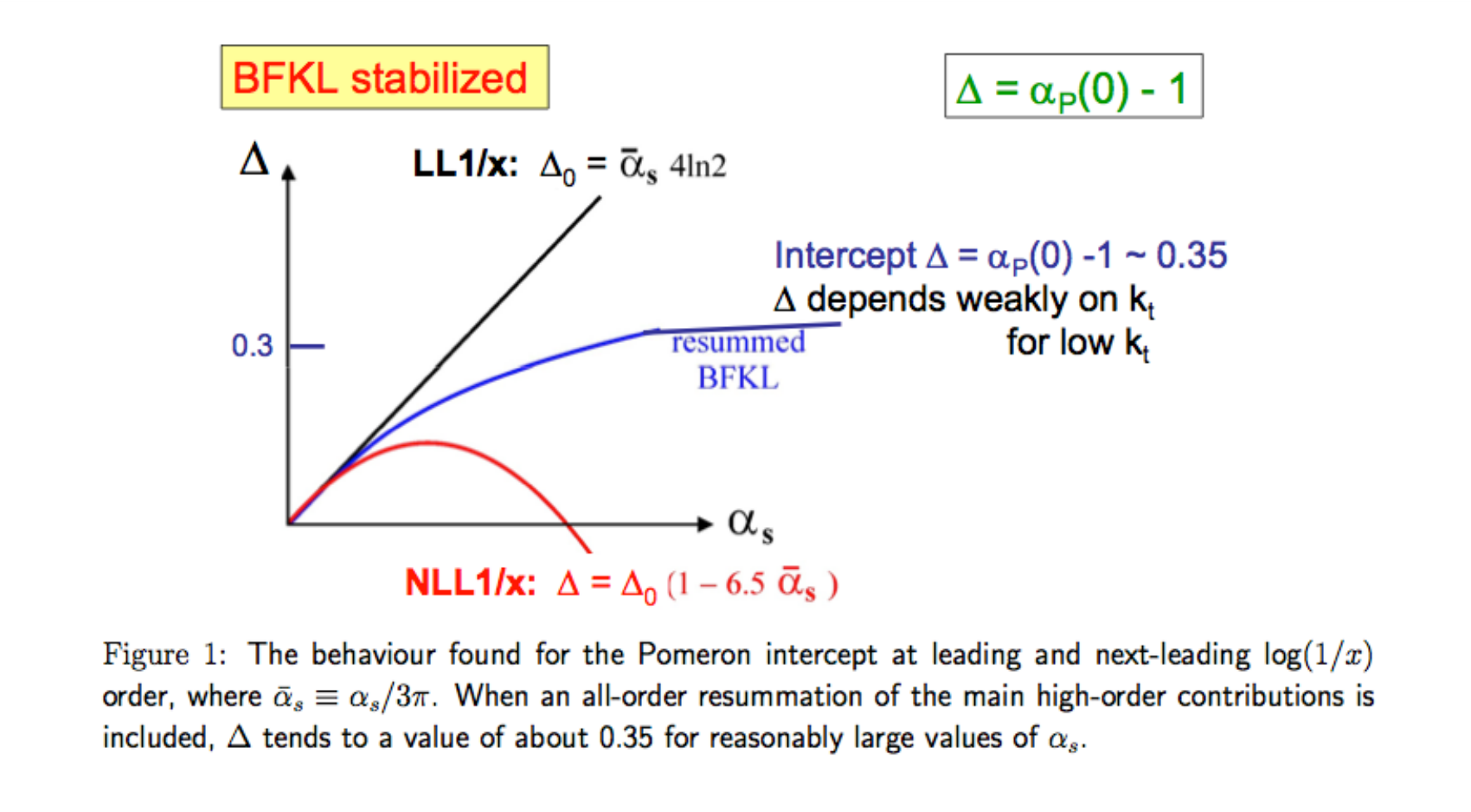}}
\caption{Evolution of the Pomeron intercept from  \cite{Martin:2012mq}. Reprinted with permission from \cite{Martin:2012mq},
 \copyright (2012) INFN  Frascati Physics Series.}
\label{fig:BFKLpom}
\end{figure}
The figure shows how the intercept $\Delta$ goes to a smooth almost constant behaviour 
as $\alpha_s$ increases.

Basically, the Pomeron picture by these authors is summarized as follows \cite{Martin:2012nm}. 
In the soft domain,  Reggeon field theory 
with a phenomenologically soft  Pomeron dominates. In the hard domain, perturbative QCD and a 
partonic approach must be used. In pQCD, 
the Pomeron is associated with the the so called BFKL vacuum singularity. In the perturbative 
domain, there is thus a single hard Pomeron exchanged, 
with $\alpha_P^{bare}=1.3 + \alpha'_{bare} t$ with $ \alpha'_{bare}\lesssim 0.05\ {\rm GeV}^{-2}$. 
In mini-jet language, which we shall describe in the next subsection,  this corresponds to 
having the mini-jet cross-section rising as $\simeq s^{0.3}$. 
The slope  is associated to the size of the Pomeron, i.e. $\alpha'_P\propto 1/<k_t^2>$. 
Thus the bare Pomeron  is associated to the hard scale, 
of the order of a few GeV. In a mini-jet model this hard Pomeron is obtained from  parton-parton scattering 
folded in with the densities and summed over all parton momenta.
This is their perturbative description. But then transition from hard to soft takes places, as one 
moves to smaller $k_t$ values. In KMR 
approach, this 
is due to  multipomeron effects, while in $k_t$-resummation language (see next subsection), this comes about because of 
resummation of soft $k_t$-effects, 
which lower the scale determining $<k_t^2>$ from the hard scale, to the soft one. As a result, 
in the case of the BFKL Pomeron,  the slope increases 
by a factor $\sim 5$, while  at the same time the intercept decreases and one has an effective 
{\it linear} trajectory, $\alpha_P^{eff}\simeq 1.08 +0.25 t$.
A behaviour such as this,  a transition from soft to hard, from a bare to an effective trajectory 
for the Pomeron, was also found by the authors to be 
present in virtual photo production of vector mesons at HERA.

After this general overview of the model, let us see how KMR apply it to elastic scattering. 
The basic building blocks of this model are the following parameters:
\begin{itemize}
\item the bare Pomeron intercept $\Delta=\alpha_P(0)-1, s$-dependent 
\item the bare Pomeron slope $\alpha'\simeq 0$
\item a parameter $d$, which controls the BFKL diffusion in $k_t$
\item the strength $\lambda$ of the triple-Pomeron vertex
\item the relative weight of the diffractive states $\gamma$, determined by low mass diffractive dissociation
\item the absolute value $N$ of the initial gluon density.   
\end{itemize}
KMR have discussed this model in comparison with the recent TOTEM data and 
the values obtained by this program for the total, elastic and diffrative cross-section are given, in this paper, in Tables~\ref{tab:KMRtrento1} 
and ~\ref{tab:KMRtrento2} for two different models, the original KMR \cite{Ryskin:2011qe} and the 3-channel eikonal \cite{Ryskin:2012ry}.
\begin{table}
\caption{Values for various total cross-section components, in the original KMR model  \cite{Ryskin:2011qe}, prior to the LHC data. }
\label{tab:KMRtrento1}
\centering
\begin{tabular}{||c||c|c|c|c||}
\hline
energy& $\sigtot$& $\sigma_{el}$&$\sigma^{SD}_{lowM}$ &$\sigma^{DD}_{lowM}$\\
TeV     &           mb&                     mb&mb                                     &    mb                      
  \\ \hline
1.8 &72.7 &16.6 &4.8 &0.4 
\\ \hline
7    &87.9 &21.8 &6.1 &0.6  \\ \hline
14  &96.5 &24.7 &7.8 &0.8
\\ \hline
100&122.3 &33.3 &9.0 &1.3
\\ \hline
\end{tabular}
\end{table}
\begin{table}
\caption{Values for various total cross-section components in t the KMR 3-channel eikonal 
from \cite{Ryskin:2012ry}, inclusive of LHC TOTEM data at $\sqrt{s}=7\ {\rm TeV}$. }
\label{tab:KMRtrento2}
\centering
\begin{tabular}{||c||c|c|c|c|c|c||}
\hline
energy
&$\sigtot$&$\sigma_{el}$ &$B_{el}$      &$\sigma^{DD}_{lowM}$ &$\sigma^{DD}_{lowM}$ \\
TeV&mb   &       mb             & ${\rm GeV}^{-2}$& mb                                    &                                 
 \\ \hline
1.8 &
79.3 &17.9 &18.0 &5.9 &0.7 \\ \hline
7    &
97.4 &23.8 &20.3 &7.3 &0.9 \\ \hline
14  &
107.5 &27.2 &21.6 &8.1 &1.1 \\ \hline
100&
138.8 &38.1 &25.8 &10.4 &1.6 \\ \hline
\hline
\end{tabular}
\end{table}


\subsection{Mini-jet models}
\label{ss:minijets}

When ISR confirmed the rise of the total cross-section already hinted at by cosmic ray experiments, 
an interpretation was soon put forward that the rise was due to the appearance of partonic interactions 
\cite{Cline:1973kv}.  This early estimate of  jet production contribution 
to the rise of the total cross-section 
and a comparison with existing cosmic ray and accelerator data can be seen from   Fig. ~ \ref{fig:clinehalzen1973}  in \ref{sss:earlymodels}.
In this figure, one could see the appearance of the first parton model for the rise of the total \x  \ and its comparison with data, with 
the shaded area to represents an estimate of the parton contribution.
Subsequently,  models in which the hard component in the rise could be calculated from pQCD or 
could be inspired by pQCD have been put forward, as shall be discussed below.

\subsubsection{Non-unitary mini-jet model by Gaisser and Halzen}\label{sss:gaisserhalzen} 

When data at the CERN \spbarps  \ gave further evidence of the rise of the total cross-section, 
the idea was subsequently 
elaborated by Gaisser and Halzen \cite{Gaisser:1984pg}, who made a model in which the rising 
part of the total cross-section 
was obtained from the  QCD  two jet cross-section, calculated using QCD  parton-parton cross-sections, 
folded in with parton densities. In this calculation,  
\begin{equation}
\sigtot = \sigma_0+\sigma_{jet}(p_{T min})
\end{equation}
with 
\begin{equation}
\sigma_{jet}(p_{Tmin})=\int_{4p_{T\min}^2/s} \frac{d\sigma}{dx_1}dx_1
\end{equation}
and 
\begin{align}
\frac{d\sigma}{dx_1}=\frac{\pi}{18 p_{T\ min}^2}\times \ \ \ \ \ \ \ \ \ \ \ \ \ \ \ \ \ \ \ \ \ \ \ \ \  \ \ \ \ \ \ \ \ \ \ \ \ \ \ \ \ \ \ \ \nonumber \\
\times \int dx_2 F(x_1,Q^2)F(x_2,Q^2)\alpha_s^2(Q^2)H(x_1,x_2,4xp_{T min}^2)
\end{align}
with $H(x_1,x_2,4xp_{T min}^2)$ obtained from the  cross-sections for parton-parton scattering, integrated over all scattering angles 
and parton density functions $F(x_i,Q^2)$. In the above equation, there appear the by-now familiar parameter $p_{Tmin}$, which 
regularizes the parton-parton cross-section, otherwise divergent as $1/p_{T\ min}^2$ for small momenta of the outgoing partons. 
Since these jet cross-sections rise very rapidly with energy, the parameter $p_{Tmin}$ was taken to be energy dependent. 
In Table ~\ref{tab:gasser}, we reproduce  the values of $p_{Tmin}$ needed to describe existing total cross-section   
data from low energy to  high energy values.
\begin{table}
\caption{Table of predicted values for $\sigtot$  by 
Gaisser and Halzen \cite{Gaisser:1984pg}.}
\label{tab:gasser}       
\begin{tabular}{lclclc|}
\hline
$\sqrt{s}$ &$p_{T\ min}$ &  $\sigma_{jet}$ Gaisser Halzen\\
GeV          &GeV                &  mb                                                 \\
\hline
43  & 1.25 & 4 \\
540 & 2 &26 \\ 
4330&3.2  &63 \\
43300&6&127\\
\hline
\end{tabular}
\end{table}
With $\sigma_0=38 \ mb$, Fig.~\ref{fig:gaisshalzen} shows  the corresponding plot,  up to cosmic ray energies.
\begin{figure}
\resizebox{0.5\textwidth}{!}{%
  \includegraphics{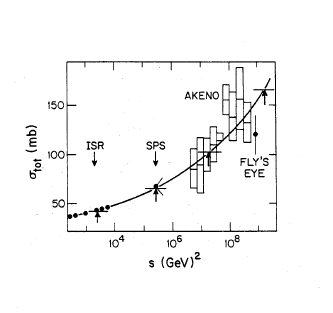}
}
\caption{The first  model for the total cross-section which used QCD calculated 
jet cross-sections to describe the rise, is shown,  from  \protect\cite{Gaisser:1984pg}.
Reprinted with permission  from \cite{Gaisser:1984pg}, Fig.(1), \copyright (1985) by the American Physical Society.}
\label{fig:gaisshalzen}       
\end{figure}
This model had no flexibility to  satisfy unitarity. It was just a simple parametrization, but it already had the 
merit of including soft-hard partonic interactions. 
The importance of such semihard processes had been highlighted in \cite{Gribov:1984tu}, also discussed in 
\cite{Pancheri:1985rm,Pancheri:1985sr,Pancheri:1986qg,Hwa:1988ju}.

\subsubsection{Eikonalization of  mini-jet models}
\label{sss:eikonal-mini}

A subsequent step which would  avoid an energy dependent value of the parameter $p_{T min}$, was the introduction of multiple scattering, 
as  had  been pointed by Durand \cite{Durand:1985qi}. This possibility was realized by Durand and Pi in \cite{Durand:1988ax} who proposed to use 
the mini-jet cross-section as input to the total cross-section through the eikonal representation. Their proposed expression enforced  the idea that 
QCD processes at high energy drive the rise of the total cross-sections, while at the same time satisfying unitarity. The price to pay, as 
always the case when using the eikonal representation,  was the introduction of the impact parameter distribution for the scattering partons. 
In this, as in most other models,  the impact parameter distribution at high energy was taken to be different from the one at low energy. At low energy, 
the distribution was considered to be dominated  mostly by quark scattering,  and, accordingly, taken to be a convolution of the proton electromagnetic 
form factors, while for the gluons, it was a convolution of a proton-like and a pion-like form factor. In the following, we shall reproduce the expressions 
they use and the values of the parameters which give the fits shown in Fig.~\ref{fig:durand1}.
\begin{figure}
\resizebox{0.5\textwidth}{!}{%
  \includegraphics{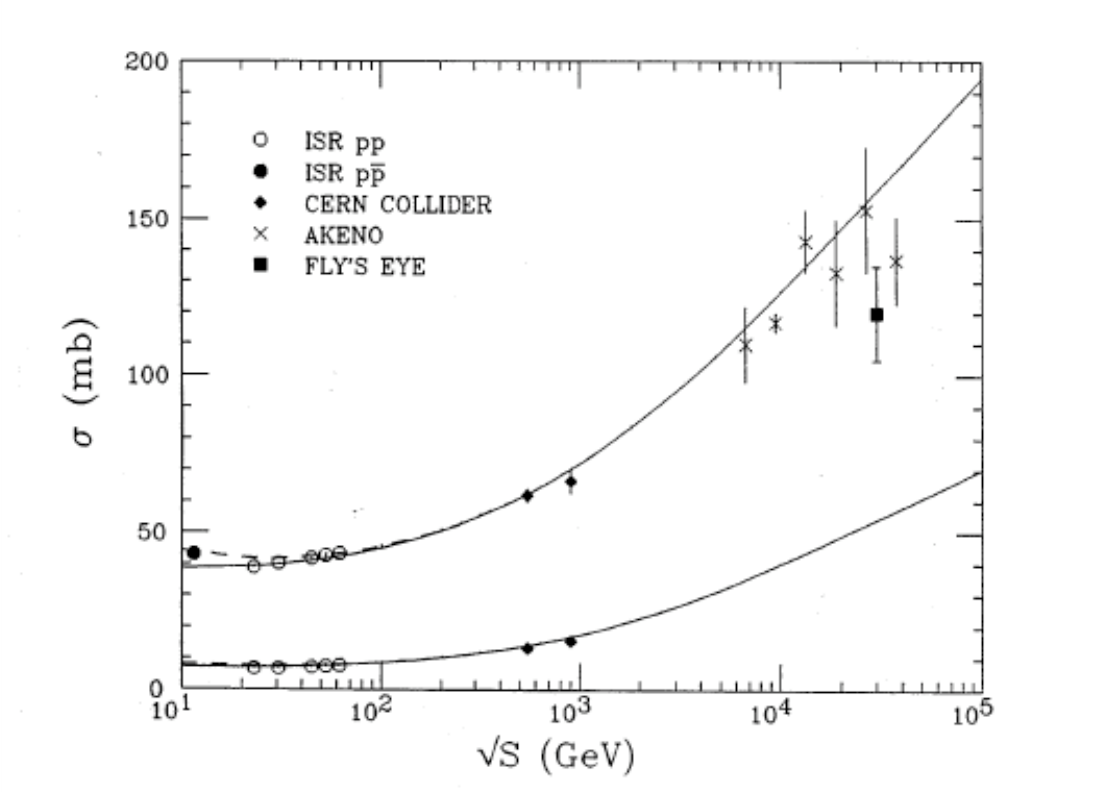}
}
\caption{Description of the total cross-section in the eikonal mini-jet model by Durand and Pi from ref.~\protect\cite{Durand:1988ax}.
Reprinted with permission from \cite{Durand:1988ax}, Fig.(1), \copyright (1989) by the American Physical Society.}
\label{fig:durand1}       
\end{figure}

 The
\pp \ and \pbarp \ elastic scattering amplitudes were written  as
\begin{equation}
f_{pp,\pbarpmath}=i \int b db J_0(b\sqrt{t})[1-e^{-\chi_{pp,\pbarpmath}(b,s)}]
\end{equation}
with  
\begin{equation}
\chi= \chi^R+i \chi^I
\end{equation}
We use here the notation of Ref. \cite{Durand:1988ax}, noting  
 that the definition of $\chi$ has a $-i$ with respect  to the more usual definition, for instance the one  in the  model  described in \ref{sss-models:BN}.
The normalization is such that
\begin{equation}
\frac{d\sigma}{dt}=\pi |f(s,t)|^2
\end{equation}
and then, using  the optical theorem,  one has the usual expressions
\begin{eqnarray}
\sigtot& =&4\pi \Im m f(s,0=)\nonumber\\
&=& 4\pi\int_0^\infty b db [1-\cos\chi^I(b,s)e^{-\chi^R(b,s)}]\\
\sigma_{elastic}&=& 2\pi\int_0^\infty b db |1-e^{-\chi(b,s)}|^2\nonumber \\
&=& 2\pi\int_0^\infty b db [
1-2\cos\chi^I(b,s)e^{-\chi^R(b,s)}\nonumber\\
&+&e^{-2\chi^R(b,s)}]\\
\sigma_{inel}&=&2\pi\int_0^\infty b db [1-e^{-2\chi^R(b,s)}]\
\end{eqnarray}

The authors had  emphasized that the term $e^{-2\chi_R}$  can be interpreted semi-classically as the probability that no collision takes place in which 
particles are produced. The function $\chi^R$ is calculated using parton-parton scattering, impact parameter distributions as mentioned, and, through this 
function, the remaining component $\chi^I$ is obtained from a dispersion relation. This allowed them to obtain both the real and the imaginary part of the 
amplitude and through these, the $\rho$ parameter.The QCD contribution was calculated from mini-jet cross-sections and was input to  the two, 
even and odd, 
eikonals in which the \pp \ and \pbarp \ eikonal functions are   split.  Namely, they write
\begin{equation}
\chi_{\pbarpmath}=\chi_++\chi_-,\ \ \ \ \ \ \ \ \ \ \ \chi_{pp}=\chi_+-\chi_-
\end{equation}
The QCD-like  contribution is input to the even eikonal. At very high energy, if no odderon is present,     the two cross-sections  for \pp \ and \pbarp \  
are equal and one can hope to be able to calculate this part of the eikonal using perturbative QCD. Actually, even at high energy, there will be a 
residual contribution from processes which dominate at low energy, and thus the eikonal is split into a soft and a hard part, namely
\begin{align}
\chi(b,s)=\chi_{soft}(b,s)+\chi_{QCD}(b,s)\\
\chi^I_{QCD}(b,s)=-\frac{2s}{\pi}{\bf P }\int_0^\infty ds'\frac{ \chi^R(b,s')}{(s'^2-s^2)}\\
\chi^R_{QCD}(b,s)=\Re e\ \chi_{QCD}=
\frac{1}{2}\sum_{ij}\frac{1}{1+\delta_{ij}}
\times \int d^2b' dx_1dx_2 \nonumber \\ \times \int_{Q^2_{min}} d|{\hat t }|
\frac{d{\hat \sigma}_{ij}}{d|{\hat t }|}f_i(x_1,|{\hat t }|,|{\vec b}-{\vec b'}|)f_j(x_2,|{\hat t }|,|{\vec b'}|)
\end{align}
One then assumes an approximate factorization between impact space and energy distribution, namely
the probability functions to find a parton of type $i$ with fractional 
momentum $x$ at a distance $\vecb $ from the initial proton direction, are factorized as

\begin{equation}
f_j(x,{\hat t },|{\vec{b}}|) \approx f_j(x,{\hat t })\rho({\vec b})
\end{equation}
where $\rho(b)$ is a function describing matter distribution inside the colliding hadrons. One then can write
\begin{equation}
\chi^R_{QCD}(b,s)=\frac{1}{2} A(b) \sigma^R_{QCD} \label{eq:factorize}
\end{equation}
with 
\begin{equation}
A(b)=\int d^2 b' \rho(b')\rho(|{\vec b }-\vec{b'}|)
\end{equation}
and  $\int d^2b A(b)=1$.

 For the impact parameter distribution, for the soft part the following expressions were used:
 \begin{equation}
A_{\pm}(b)= \frac{\nu_{\pm}^2}{12\pi}\frac{1}{8}(\nu_{\pm}b)^3{\cal K}_3(\nu_{\pm}b)\\
\end{equation}
with ${\cal K}_3(\nu b)$ the special Bessel function of the third kind which comes from the 
convolution of the dipole-type  expression of the proton e.m form factor. 
These functions will enter the even and odd soft eikonals. For the hard part, on the other 
hand, one takes into account that gluons are distributed differently 
from the valence quarks and the expression which is used is
 the convolution of  {\it gluon form factors} given by
\begin{equation}
G(k_\perp^2)=(1+k_\perp^2/\nu_+^2)^{-2}(1+k_\perp^2/\mu^2)^{-1}
\end{equation}
To complete the picture, the authors  set the soft  eikonals as
\begin{align}
2\chi_{+,soft}=A_+(b)\sigma_{soft}=A_+(b)[\sigma_0+\frac{a}{s^\alpha}e^{i\alpha\pi/2}]\\
2\chi_{-,soft}=A_-(b)\frac{R}{\sqrt{s}}e^{-i\pi/4}
\end{align}
with $\sigma_0=\sigma_0^R+ i\sigma_0^I; a,\alpha$ adjustable parameters.

The model, a part from the QCD inputs, namely densities and $Q^2_{min}$, has now 8 parameters, 
$\sigma_0=\sigma_0^R+ i\sigma_0^I, a,\alpha$ and $R$ for the cross-section type terms and 
$ \nu_\pm,\mu$ for the impact parameter distributions. The parameters are then  fixed so as to obtain 
a good fit to the total cross-sections,  elastic and total, to the  elastic differential cross-section and 
to the $\rho$ and slope parameters, given by
\begin{eqnarray}
\rho={\Re}e f(s,0)/{\Im} m f(s,0)\\
B(s)=\frac{d}{dt}[\ln\frac{d\sigma_{elastic}}{dt} ]|_{t=0}
\end{eqnarray}

Within this framework, one obtains the description of the total cross-section shown in Fig.~\ref{fig:durand1}
and  good descriptions of the elastic cross-sections and their energy dependence up to \spbarps \ data.
Notice that the slope parameter $B(s)$ is thus fully determined.

We have dedicated a rather long and detailed exposition to this model since many other models  follow a 
similar  outline  and   models similar to this one  have been used (and still are) in MonteCarlo simulations 
such as PYTHIA \cite{Sjostrand:1987su}. Many of the  features of the Durand and Pi  model are also 
present in the QCD inspired  model which will be described in the next section.

Durand and Pi in \cite{Durand:1988ax} made an effort to obtain a description of   the $\rho$ parameter 
which could accomodate the UA4 measurement \cite {Bernard:1987vq}, namely $\rho=0.24\pm 0.04$ 
and the parameter values for the overall description were influenced by this choice. To explain such a 
large value, it turned out to be quite difficult, in most case it was related to a possible anomalous rise of the total cross-section.
The measurement of the $\rho$ parameter at the Tevatron by  E-710 \cite{Amos:1991bp} and E811 
Collaboration \cite{Avila:2002bp} however did not confirm such a high value for $\rho$, which had in any 
event already been measured again by UA4,s obtaining   a lower value, in line with theoretical expectations.  That the 
$\rho$ parameter could not be this high was pointed out in 1990 by Block {\it et al.} \cite{Block:1989gz} 
who discussed the theoretical implications of such measurements, using a previously developed model 
\cite{Margolis:1988ws}. We now turn to this model.
\subsubsection{ QCD inspired models, Aspen model }
\label{sss:aspen}
Applications of  the  mini-jet idea to the description of the total  and elastic cross-sections  were  developed 
around the 90's by many groups, for instance in  \cite{Innocente:1988jw,Kopeliovich:1989iy}. We shall illustrate 
here the one developed by Block with Fletcher, Halzen, Margolis and Valin \cite{Block:1989gz}, where the 
contribution of semi-hard interactions was fully parametrized, separately indicating quark and glun contributions. 
This model  is also sometimes labelled as the Aspen model \footnote{This name was given by one of us, G.P, as a testimony of the  contribution to the field  from Martin Block, who spent his latest years in Aspen, working further, and until very recently, on the problems of the total cross-section.}
and is at the basis of subsequent developments, where it was applied to photon processes 
\cite{Block:1998hu} and to the extraction of the proton-proton cross-section from cosmic ray experiments \cite{Block:2000pg}.

In this model, a QCD-inspired eikonal parametrization of the data is used. For the total cross-section,  a result similar to 
simple $\ln^2 [s]$ analytic considerations \cite{Block:1984ru} is obtained. One starts with
\begin{eqnarray}
\sigtot=4 \pi \Im m f_N\\
\frac{d\sigma}{dt}=\pi |f_N|^2\\
f_N=i\int bdb J_0(b\sqrt{-t})[1-e^{-P(b,s)/2}]\\
P(b,s)=P_{gg}(b,s)+P_{qg}(b,s)+P_{qq}(b,s)\\
P_{ij}(b,s)=W_{ij}(b,\mu_{ij})\sigma_{ij}(s)
\end{eqnarray}
In the above equations, the probability function $P(b,s)$ is seen to be parametrized in terms of 
three QCD-like terms, corresponding respectively to gluon-gluon, quark-gluon and quark-quark interactions. 
The impact distribution functions for proton-proton scattering are obtained from the convolution of the two proton-like  
form factors, and for $i=j$
\begin{equation}
W_{ii}(b,\mu_{ii})=\frac{\mu^2_{ii}}{96 \pi}(\mu_{ii}b)^3 {\cal K}_{3}(\mu_{ii}b)
\end{equation}
For the gluon-gluon terms, which become more important at high energy and which 
drive the rise of the cross-section, they write
\begin{equation}
P_{gg}(b,s)\simeq W_{gg}(b)s^{J-1}
\end{equation}
where $J$ gives the large $s$ behaviour of the gluon-gluon cross-section integrated over the gluon PDF's in the proton. 
For the probabilities involving quarks, and which are important at low energy, the parametrization is  inspired 
by the x-behaviour of the parton densities and they are written as
\begin{eqnarray}
P_{qq}=W(\mu_{qq}b)[a+b\frac{m_0}{\sqrt{s}}]\\
P_{qg}=W(
\sqrt{\mu_{qq}\mu_{gg}}
b
)[a'+b'\ln \frac{m_0}{\sqrt{s}}]
\end{eqnarray}
The difference at low energy between \pp \ and \pbarp \ cross-sections is obtained by 
first ensuring the correct analyticity properties through the substitution $s\rightarrow s^{-i\pi/2}$ 
and then introducing an {\it ad hoc} odd-crossing amplitude as
\begin{equation}
P_{odd}=W(\mu_{odd}b)a'' \frac{m_0}{\sqrt{s}}e^{-i\pi/4}
\end{equation}
The overall set of parameters can be found in the caption of Figure 1 from  \cite{Block:1989gz}. 
We reproduce this figure for the total cross-section in Fig. ~\ref{fig:blockmargolis}.
\begin{figure}
\resizebox{0.5\textwidth}{!}{%
  \includegraphics{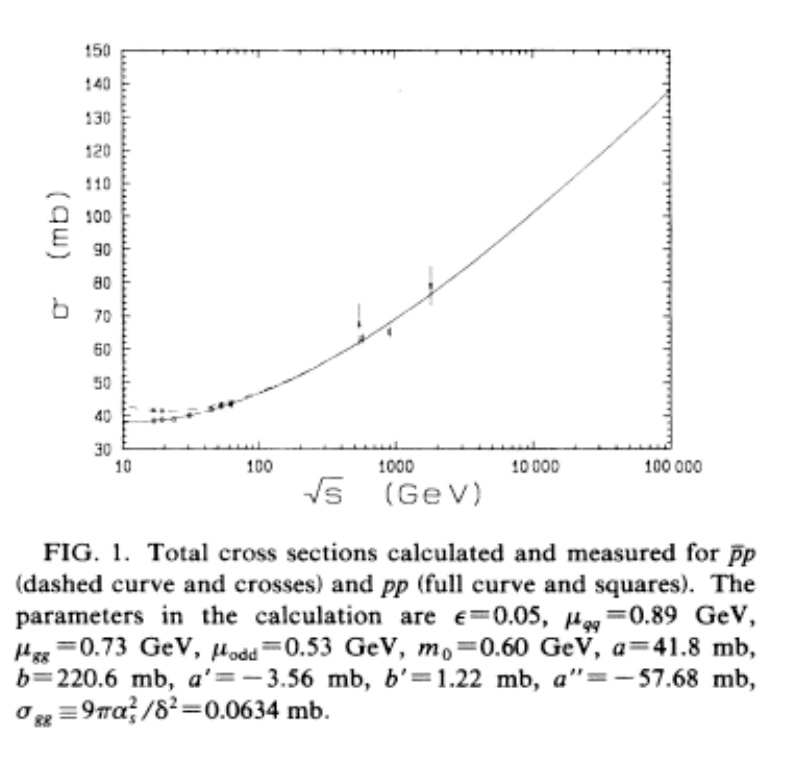}
}
\caption{Description of the total cross-section in the QCD inspired  model by Block {\it et al.},  from \protect\cite{Block:1989gz}.
Reprinted from \cite{Block:1989gz}, Fig.(1), \copyright (1990) by the American Physical Society}
\label{fig:blockmargolis}       
\end{figure}
This model has 11 parameters.

Finally, notice an 
important feature of this model, i.e. how the Froissart-type large energy limit is obtained through the mini-jet contribution in combination with the impact parameter dependence.
The authors search for the critical value of the parameter $b$ for which $P_{gg}\le 1$ and, using 
the large energy behaviour of the mini-jet cross-sections, namely $s^{J-1}$, they find
\begin{equation}
b_c=\frac{J-1}{\mu_{gg}}\ln \frac{s}{s_0} + {\cal O}(\ln\ln \frac{s}{s_0}) 
\end{equation}
Thus, the energy dependence of the QCD cross-sections transforms the  $s^{j-1}$ behaviour at lower energy 
into the black-disk cross-section at high energy, i.e. into 
\begin{equation}
\sigtot=2\pi [
\frac{J-1}{\mu_{gg}}
]^2 
\ln^2 \frac{s}{s_0}
\end{equation}

\subsubsection{Resummation and mini-jets}
\label{sss:BN}

A model which incorporates many  features of QCD is the one developed in \cite{Grau:1999em}, following \cite{Corsetti:1996wg}
 and completed in a number of subsequent papers, in particular in \cite{Godbole:2004kx}.
 The model embodies the idea 
    that resummation down into the infrared region is a crucial  
 component of total cross-section asymptotics and  provides a phenomenological structure linking  
 the infrared behaviour of QCD to the asymptotic limit of the total cross-section. 
 
 The model   
 is labeled following the idea that exponentiation of the spectrum of soft emitted quanta  when 
 $k\rightarrow 0$, first proposed by Bloch and Nordsieck (BN) for QED \cite{Bloch:1937pw}, 
 must be extended to the soft  gluons for the QCD processes and that it must be carried through into  the infrared region. 
 
Thus,  the model is  developed along two basic ideas, that the rise of the total cross-section 
is driven by hard processes, called {\it mini-jets} \cite{Pancheri:1985rm,Pancheri:1986qg} and that 
the softening of the rise into the smooth behaviour consistent with the Froissart bound arises because 
of  soft gluon emission, resummed and 
 extended down into the zero momentum region of the spectrum of the emitted quanta \cite{Grau:2009qx}. 
 To investigate this region, use is made of the ansatz about the infrared behaviour of the effective 
 quark-gluon coupling first introduced in \cite{Nakamura:1983am} for the intrinsic transverse momentum 
 of Drell-Yan pairs, and later implemented to describe the impact parameter distribution of partons in 
 high energy collisions \cite{Corsetti:1996wg}, as we have described previously.
 
 The  basic structure of this model exhibits some of the same features advocated by the work  of   
 GLR, KMR or GLM, described in other parts of this review,  describing implementation of BFKL dynamics, 
 namely soft exchanges (the soft Pomeron) and perturbative QCD (hard Pomeron) for medium energy 
 partons, but the hadronic amplitude at $t=0$ of this model is built through  a probabilistic structure, 
 with the  soft resummation contribution  built as a term factored from the hard parton-parton scattering, 
 not unlike what one does in QED when applying infrared radiative corrections. 
\subsubsection{Hadronic matter distribution and  QCD soft $k_t$ distribution 
 \label{sss-models:kt}}

Because of its obvious relevance to total cross-section estimates through the transverse interaction size of 
hadrons, we shall start by   investigating   the transverse momentum distribution of soft quanta in QCD.  

The expressions for transverse momentum distribution discussed  in the context of QED in  \ref{sss-models:etp} 
cannot be extended simply to QCD\glossary{Quantum ChromoDynamics},
since the coupling constant is momentum dependent. 
When taken into account, the transverse momentum 
distribution due to soft
gluon emissions became extremely interesting. This was first realized by Dokshitzer and collaborators 
\cite{Dokshitzer:1978yd}, who generalized to QCD the Sudakov form factor expression, originally obtained 
in QED \cite{Sudakov:1954sw} and discussed above.
The application of  resummation techniques
to the $K_\perp$-distribution of $\mu$ pairs of mass Q produced in
hadron-hadron collisions  was shortly after studied in  \cite{Parisi:1979se}.
In this paper, it was   argued that
the ``soft limit of QCD\glossary{Quantum ChromoDynamics} could be treated in full
analogy with that of QED with the minor [Author's note : not so minor!] technical
change of $\alpha$ into $\alpha(k_\perp)$''. 

Today  the problem of transverse momentum distributions in QCD\glossary{Quantum ChromoDynamics} 
is still  not completely solved. The reason lies in the lack of our certain knowledge about the momentum 
dependence of the strong coupling constant
$\alpha_s(k_\perp)$ when the gluon momentum goes to zero, as is the case for the soft gluons needed 
in  
 resummation.  Actually, 
 what one needs to know  is  the {\it integral} of $\alpha_s(k_t)$ 
over the infrared region. The IR limit of $\alpha_s$ enters only  when the gluon momenta are close to zero, 
i.e. only in the resummation process, which  implies exponentiation of an integral over gluon's momenta 
with a momentum-dependent $\alpha_s$.
Because the coupling constant 
in QCD\glossary{Quantum ChromoDynamics,}  grows as the momenta become smaller,  a Bloch and Nordsieck  type resummation
 of the  zero momentum modes   of the  soft quanta emitted by coloured quarks
  becomes mandatory.  In this case,
the applicability of the above methods requires a knowledge of (or, an ansatz for) the strong
coupling constant in the IR region.

Following the semiclassical derivation of Eq. ~\ref{eq:d2pk}, the exponent describing the $b$-dependence 
for  QCD now  
reads
\begin{equation}
h( b,E) =  \frac{16}{3}\int^E 
 {{ \alpha_s(k_t^2) }\over{\pi}}{{d
 k_t}\over{k_t}}\ln{{2E}\over{k_t}}[1-J_0(k_tb)].
 \label{hbq}
\end{equation}
Its   use 
is complicated by the  asymptotic  running of the coupling constant  on the one hand and our 
ignorance of the IR behaviour of the theory, on the other.

To overcome the difficulty arising from the infrared region, the function $h(b,E)$, which  describes the 
relative  transverse momentum distribution induced by soft gluon emission from  a pair of,
initially collinear, colliding partons  at LO, 
is split into
\begin{equation}
\label{h1}
h(b,E) = c_0(\mu,b,E)+ \Delta h(b,E),
\end{equation}
where
\begin{equation}
\label{h2}
\Delta h(b,E) =
 \frac{16}{3} \int_\mu^E {\alpha_s(k_t^2)\over{\pi}}[1- J_o(bk_t)]
 {{dk_t}\over{k_t}}
  \ln {
  {{2E}\over{k_t}}}.
\end{equation}
 The integral in $\Delta h(b,E)$
now  extends down to a   scale $\mu \neq 0$, for $\mu> \Lambda_{QCD}$ and one can use the asymptotic freedom 
expression for $\alpha_s(k_t^2)$.  Furthermore, having excluded the zero momentum region from the integration,  
 $J_o(bk_t)$ is  assumed  to oscillate to zero and    neglected.
The  integral of Eq.~(\ref{h2}) is now independent
of $b$ and  can be performed, giving
\bea
\Delta h(b,E) =
\frac{32}{33-2N_f}
\bigg\{& & \ln ( {\frac{{2E}}
{\Lambda }} )\left[ {\ln ({\ln ( {\frac{E} {\Lambda }})}) -
 \ln
({\ln ( {\frac{\mu } {\Lambda }} )})} \right]\nonumber \\
&-& \ln ( {\frac{E}{\mu }} ) \bigg\}. 
\eea
$\Lambda$ being the scale 
in the one-loop expression for $\alpha_s$.
In the range $1/E < b < 1/\Lambda$
an  effective $h_{eff}(b,E)$
is obtained by setting $\mu = 1/b$ 
\cite{Parisi:1979se}. This choice of the scale   introduces a cut-off in impact parameter space which is 
stronger than any power, since
the radiation function, for $N_f=4$,  is now
 \begin{equation}
e^{-h_{eff}(b,E)} =
 \big{[}
{{
\ln(1/b^2\Lambda^2)
}\over{
\ln(E^2/\Lambda^2)
}}
\big{]}^{(16/25)\ln(E^2/\Lambda^2)}
\label{PP}
\end{equation}
which is Equation(3.6) of ref.~ \cite{Parisi:1979se}. The remaining  $b$ dependent term, namely 
$exp[-c_0(\mu,b,E)]$,
is dropped, a reasonable approximation if one  assumes that there is no physical singularity 
in the range of integration
 $0\le k_t \le 1/b$.  This contribution however reappears as an energy independent  smearing function 
 which reproduces phenomenologically 
 the effects of an intrinsic  transverse momentum of partons.
For most applications, this may be a good approximation. 
However, when  the integration in impact parameter space extends to very large-$b$ values, as  is the 
case for the calculation of total cross-sections, the infrared region may be  important and  the possibility 
of  a physical singularity for $\alpha_s$ in the infrared region becomes relevant \cite{Grau:2009qx}. It is this 
possibility,   which we exploit in studying scattering in the very large impact parameter region, $b\rightarrow \infty$.

Our choice for the infrared behaviour of $\alpha_s(Q^2)$  used in obtaining  a
  quantitative description of the  distribution in Eq. (\ref{h2}),
  is a generalization of the   Richardson potential
 for quarkonium bound states \cite{Richardson:1978bt}, which we have proposed and developed in   a number of related
applications \cite{Nakamura:1983xp,Nakamura:1984na}.
Assume a confining potential (in momentum space) given by   the one gluon
exchange term
\begin{equation}
{\tilde V}(Q) = K ({{\alpha_s(Q^2)}\over{Q^2}}),
\end{equation}
where $K$ is a constant calculable from the asymptotic form of $\alpha_s(Q^2)$. Let
us choose for $Q^2<<\Lambda^2$ the simple form
\begin{equation}
\alpha_s(Q^2) = {{B�}\over{ (Q^2/\Lambda^2)^{p}}},
\end{equation}
(with B a constant), so that ${\tilde V}(Q)$ for small Q goes as
\begin{equation}
\label{potential}
{\tilde V}(Q) \rightarrow  Q^{-2(1 +p)}.
\end{equation}
For the potential, in coordinate space, 
\begin{equation}
V(r) =\ \int{{d^3Q}\over{(2\pi)^3}} e^{i {\bf Q.r}} {\tilde V}(Q),
\end{equation}
Eq.(\ref{potential}) implies
\begin{equation}
V(r) \rightarrow (1/r)^3 \cdot r^{(2 + 2p)} \sim  C\ r^{(2p - 1)},
\end{equation}
for large r (C is another constant). A simple check is that for $p$
equal to zero, the usual Coulomb potential is regained. Notice that for
a potential rising with r, one needs $p > 1/2$. Thus, for $1/2< p< 1$,
this corresponds to a confining potential rising less than linearly
with the interquark distance $r $, 
while a  value of $p=1$ coincides with the infrared limit of the Richardson's potential and is also 
found in a number of applications to potential
estimates of quarkonium properties \cite{Yndurain:1997gb}.

Then,  again following Richardson's argument \cite{Richardson:1978bt},  we connect our IR limit for $\alpha_s(Q^2)$ to 
the asymptotic freedom region using the  phenomenological  expression:
\begin{equation}
\label{eq:alphapheno}
\alpha_s(k_t^2)=\frac{1}{b_0}{{p}\over{\ln[1+p({{k_t^2}
\over{\Lambda^2}})^{p}]}}
\end{equation}
with $1/b_0={{12 \pi }\over{(33-2N_f)}}$. The expression of Eq. ~(\ref{eq:alphapheno}) coincides with the usual one-loop formula  for
 values of $k_t>>\Lambda$, while going to a singular
 limit for small $k_t$,  and 
generalizes Richardson's ansatz to values of $p<1$. The range $p < 1$ has an important advantage,
i.e., it allows the integration in Eq.(\ref{h2}) to converge for all
values of $k_t=|k_\perp |$. Some considerations for the case  $p\simeq b_0$ can be found in \cite{Pancheri:2014rga}. 

Using Eq.~(\ref{eq:alphapheno}), one can study the behaviour of $h(b,E)$ for very large-$b$ values  which enter the total cross-section calculation 
and recover the perturbative calculation as well.  {The behaviour of  $h(b,E)$ in   various regions in $b$-space was discussed 
in \cite{Grau:1999em},  both  for a singular and a frozen $\alpha_s$, namely one whose IR limit is a constant. 
 There we saw that, for the singular $\alpha_s$ case,  the following is a good analytical approximation in the very large-$b$ region:
 \begin{eqnarray}
\label{halphas3}
b>{{1}\over{N_p\Lambda}}&>&{{1}\over{M}}\\
h(b,M,\Lambda) &=&{{2c_F}\over{\pi}} \left[ {\bar b} {{b^2\Lambda^{2p}}\over{2}}
\int_0^{{1}\over{b}}{{dk}\over{k^{2p-1}}} \ln {{2M}\over{k}}\right]\nonumber \\
+{{2c_F}\over{\pi}}& &\left[
 2 {\bar b} \Lambda^{2p}\int_{{1}\over{b}}^{N_p\Lambda} {{dk}\over{k^{2p+1}}}
\ln {{M}\over{k}}+{\bar b} \int_{N_p\Lambda}^M {{dk}\over{k}} 
{{\ln{{M}\over{k}}}\over{\ln {{k}\over{\Lambda}}}}\right] \nonumber \\
 &=&{{2c_F}\over{\pi}} \Biggl [ {{{\bar b}}\over{8(1-p)}} (b^2\Lambda^2)^p
\left[ 2\ln(2Mb)+{{1}\over{1-p}}\right] +\nonumber \\
 &\ &{{\bar b}\over{2p}}(b^2\Lambda^2)^p \left[2\ln(Mb)-{{1}\over{p}}\right]
+ \nonumber\\
& \ &{{\bar b}\over{2pN_p^{2p}}}\left[-2\ln{{M}\over{\Lambda N_p}}+{{1}\over{p}}
\right] + \nonumber \\
 &\ & {\bar b} \ln {{M}\over{\Lambda}}\left[\ln {{\ln{{M}\over{\Lambda}}}\over
{\ln{N_p}}}-1+{{\ln{N_p}}\over{\ln{{M}\over{\Lambda}}}} \right] \Biggr ]
\end{eqnarray}
where $N_p=(1/p)^{1/2p}$, $c_F=4/3$ for emission from quark legs and ${\bar b}=12 \pi/(33-2N_f)$. 
The upper limit of integration here is called M, to indicate the maximum  allowed  transverse momentum, 
to be determined, in our approach, by  the kinematics of single gluon emission as in \cite{Chiappetta:1981vv}.  
The above  expression  exhibits the sharp cut-off at large $b$ values which we shall exploit to study the 
very large energy behaviour of our model.

The possibility that $\alpha_s$ becomes constant in the infrared can also be considered. We found that such possibility does not contribute 
anything new  with respect to the already known results. In fact, using the  expression 
\cite{Parisi:1979se,Halzen:1982cb,Altarelli:1984pt}
\begin{equation}
\alpha_s(k_t^2)={{12 \pi}\over{33-2N_f}} {{1}\over{\ln [a^2+ k_t^2/\Lambda^2]}}
\end{equation} 
with $a>1$, in   the same  large $b$-limit as in Eq. (\ref{halphas3}), we have 
\cite{Grau:1999em}  
\begin{align}
b>{{1}\over{a\Lambda}}>{{1}\over{M}} \ \ \ \ \ \ \ \ \ \ \ \ \ \ \ \ \ \ \ \ \ \\
h(b,M,\Lambda)=(constant) \ln(2Mb) + \ double \ \ logs
\end{align}
namely no sharp cut-off in the impact parameter $b$, as expected. More precisely, we have  the following approximate expression:
\begin{eqnarray}
h(b,M,\Lambda)&=&{{2c_F}\over{\pi}}
\Biggl \{
{{
{\bar \alpha_s}
}\over{
8}}
[1+2\ln(2Mb)]
+ \nonumber \\
 &\ & 2{\bar \alpha_s}
  [
  \ln(Mb)\ln(a{\Lambda}b) - {{1}\over{2}} \ln^2{(a \Lambda b)}
]+\nonumber \\
& \ & {\bar b}
 \left[
 \ln{{M}\over{\Lambda}}\ln {{\ln{{M}\over{\Lambda}}}
\over{\ln a }}-\ln {{M}\over{a\Lambda}}
 \right] \Biggr  \}
\end{eqnarray}
with ${\bar \alpha_s}=12\pi/(33-2N_f) \ln(a^2)$.
These approximations are  reasonably accurate, as one
can see from  \cite{Grau:1999em}, where  both the approximate 
and the exact expressions for $h(b,M,\Lambda)$  have been plotted for the singular as well as for the frozen $\alpha_s$ case.  Notice that, in the following sections, we drop   for simplicity the explicit appearance of the $\Lambda$ in the argument of $h(b,M,\Lambda)$.

The above expressions derived for the overall soft gluon emission in a collision, are input for the 
QCD description of the total cross-section \cite{Corsetti:1996wg}. In this description, the impact factor
is defined   as in 
\begin{align}
\sigtot=2\int d^2\vecb [1-e^{-(\Omega_{soft}(s,b)+\Omega_{hard}(b,s)}]\\
\Omega_{hard}(b,s)=A_{BN}(b,s)\sigma_{jet}(s)\\
A_{BN}=N \int d^2 \vecK e^{i\vecb \cdot \vecK } d^2 P(K)
\end{align}
N being a normalization factor such that $\int d^2\vecb A_{BN}=1$ and the subscript $BN$
 indicates that this impact factor is obtain through soft gluon resummation. 
 The detailed application of such model and the phenomenological results are described in the subsection to follow.
 
 \subsubsection{  Bloch and Nordsieck inspired model for  the total cross-section \label{sss-models:BN}}
As discussed above, in the BN model 
soft gluons of momentum $k_t$ are  
resummed up to a maximum value $q_{max}$, 
and partons, mostly gluons at high energy,  of momentum $p_t$. Thus, there are three regions for  the emitted 
parton transverse momentum and hence three scales:
 \begin{description}
 \item (i) $p_t>p_{tmin}$, with $p_{tmin}\simeq 1\ {\rm  GeV}$, one can apply perturbative QCD (pQCD) and 
 calculate the mini-jet contribution to the scattering process,
 \item (ii) $\Lambda <k_t<q_{max}\simeq (10-20)\% p_{tmin}$, where $k_t$ indicates the transverse 
 momentum of a single soft gluon which corresponds to initial state soft radiation from partons with  
 $p_t>p_{tmin}$, and for which one needs to do resummation, and $\Lambda\simeq 100\ MeV\simeq m_\pi$,
 \item (iii) $0<k_t<\Lambda$ for infrared momentum gluons, which require resummation but also an ansatz 
 about the strong coupling in this region. 
 \end{description}
 In the pQCD region, parton-parton scattering and standard Leading Order (LO) parton densities are used to 
 calculate an average mini-jet cross-section as
 \begin{align}
\sigma_{mini-jets}\equiv \sigma^{AB}_{\rm jet} (s) = 
\int_{p_{tmin}}^{\sqrt{s/2}} d p_t \int_{4
p_t^2/s}^1 d x_1 \int_{4 p_t^2/(x_1 s)}^1 d x_2 \nonumber \\
\times \sum_{i,j,k,l}
f_{i|A}(x_1,p_t^2) f_{j|B}(x_2,p_t^2)~~
 \frac { d \hat{\sigma}_{ij}^{ kl}(\hat{s})} {d p_t}.\label{eq:minijets}
\end{align}
 Here   $A$ and $B$ denote particles ($\gamma, \ p,
\dots$), $i, \ j, \ k, \ l$ are parton types and $x_1,x_2$ the fractions
of the parent particle momentum carried by the parton. $\hat{s} =
x_1 x_2 s$  and $\hat{ \sigma}$ are hard parton scattering
cross--sections. 

Eq.~(\ref{eq:minijets}) is a LO parton-parton cross-section averaged over the given parton densities, 
through the phenomenologically determined parton density 
functions\\ 
$f_{i|A}(x,p^2)$, DGLAP evoluted at the scale $p_t$ of the mini-jet produced in the scattering. 
As is well known, however, for a fixed $p_{tmin}$ value, 
and as $p_{tmin}/\sqrt{s} \rightarrow 0$, gluon-gluon processes become more  and more important and, 
since the  LO densities of gluons are phenomenologically 
determined to increase as $x^{1+\epsilon}$, the mini-jet integrated cross-section will increase as 
$s^\epsilon$, with $\epsilon \sim (0.3\div 0.4)$ depending on the densities.
 This rise of the  parton-parton cross-sections, averaged over the parton densities,  has its counterpart in the hard 
 Pomeron of  BFKL models, where a behaviour $s^\Delta$ corresponding to 
  hard Pomeron with very small slope is seen to describe the rise of the  profile function, before saturation 
  starts changing the hard behaviour into a softer one.
Multiplying $\sigma_{hard}$ with the average probability for two colliding partons to see each other (and interact) at a distance $\vecb$ would give the 
average number of collisions at impact parameter \vecb \ when two protons collide.  This number can become very large as the energy increases and so 
proper eikonalization and unitarization is introduced, just as in the QCD mini-jet models we have described earlier.

Implementation of unitarity through the eikonal formulation reduces the rise from this hard term, but unless one has a 
cut-off
in the  impact parameter distribution, the rise will not be adequately quenched. 

We now let 
the mini-jet model morphe into an eikonal mini-jet model for $\sigma_{total}$, i.e.
\be
\label{sigtot}
\sigma_{total}=2 \int d^2 {\vec b} [1-\cos \Reo \chi (b,s)\ e^{-\Imo \chi(b,s)}]
\ee 
where the s-dependence of the imaginary part of the eikonal function $\chi(b,s)$ is driven by the mini-jet cross-section and, in first approximation, 
the eikonal has  been  taken to be purely imaginary. Since the total cross-section is dominated by large impact parameter values, and, at high energy,  
$\rho(s,t=0)\sim 0.1$, this is reasonable approximation at high energy and in the calculation of the total cross-section.
 
 The name {\it mini-jets} was first introduced by M. Jacob and R. Horgan  to describe the flood of small transverse momentum jet-like events 
 expected to dominate at the $S{\bar p}pS$ collider \cite{Horgan:1980fm,Horgan:1980kg}. The importance of mini-jets concerning the rise of the total cross-section 
 was doubted however, as discussed for instance by Jacob and Landshoff in \cite{Jacob:1986xd}. In particular they think impossible for mini-jets to 
 contribute to the total cross-section at $\sqrt{s}\simeq 5\ {\rm GeV}$. This may be true, or, at least we do not know how to incorporate pQCD at such 
 low hadronic c.m. energies, however our phenomenology indicates that the mini-jet  contribution start being noticeable around $\sqrt{s}\gtrsim 10\ {\rm GeV}$. 
 It also appears   that when performing a parametrization of the low energy contribution,  the hard part plays a role to  ameliorate the overall description, 
 from low to hard energies. Examples of these two different procedures can be found in our application of the BN model to pion scattering \cite{Grau:2010ju} 
 and to studies of inelastic cross-section at LHC7 \cite{Achilli:2011sw}.
 In addition, from a microscopic point of view,  mini-jets are the only pQCD phenomenon to which one can ascribe the drive of the rise of the total 
 cross-section. Their contribution can be seen as  the microscopic description    of  the hard Pomeron advocated by Reggeon field models.
 
 We now turn to the question of the impact parameter distribution in eikonal mini-jet models. 
 As discussed at length in \cite{Grau:1999em}, the standard use of hadronic  form factors together with standard library  Parton Density Functions (PDF) in  
 the mini-jet cross-sections does not allow to reproduce 
 {\it both } the initial fast rise of the cross-section as well as well as 
 moderating the rise. 

The difficulty to use the EM form factors with  standard LO pQCD techniques has been one of the problems plaguing the eikonal mini-jet approach. 
We 
thus introduced soft gluon resummation to solve this problem. However, as discussed previously, resummation outside the infrared region 
and in asymptotic freedom region, even at higher orders or beyond the LLA, can hardly touch the basic question: how to introduce a cut-off in impact parameter space  or, 
otherwise stated, how to link the asymptotic behavior of the cross-section with confinement? 

We shall now discuss the model in detail, but we anticipate its outline in graphic form here. We show in Figs.~\ref{fig:cartoon1}, \ref{fig:cartoon2}, 
\ref{fig:cartoon3} and \ref{fig:cartoon4} our cartoon representation of the building of the total cross-section. All these figures are reprinted with permission from  \cite{Pancheri:2007rv},
 with \copyright (2007) by Acta Physica Polonica B.
 \begin{figure*}[htbp]
\begin{center}
\hspace{-2cm}
\resizebox{0.5\textwidth}{!}{
\includegraphics{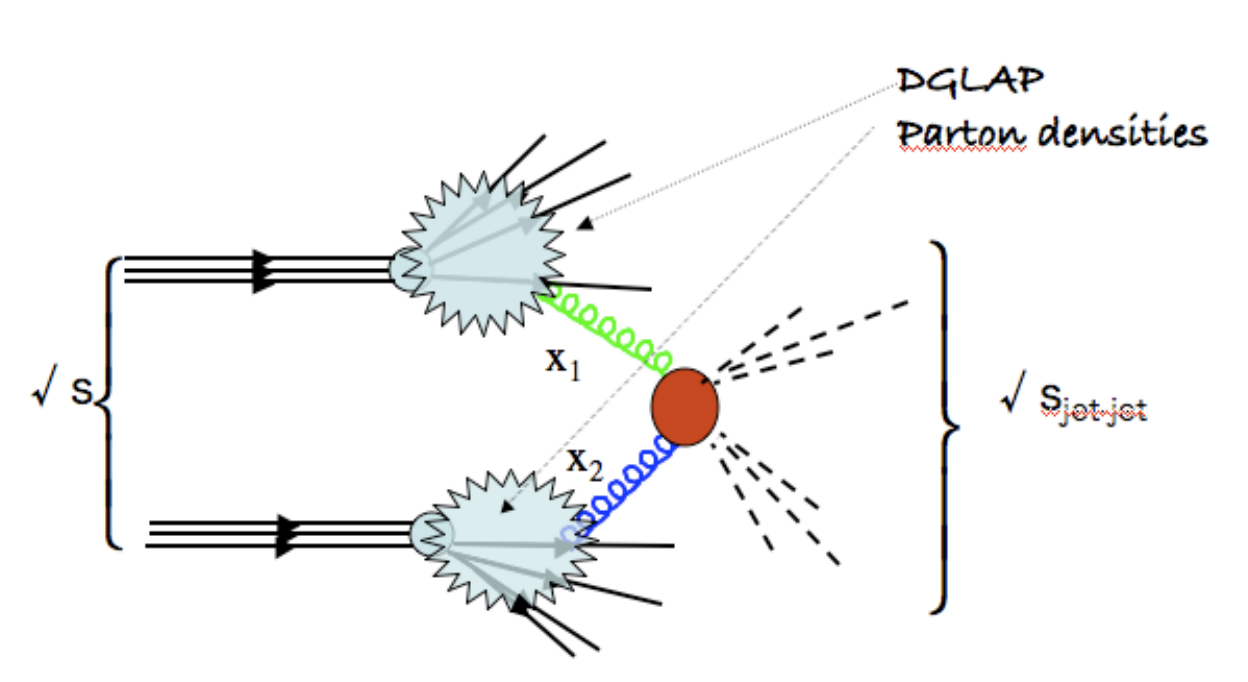}}
\resizebox{0.3\textwidth}{!}{
\includegraphics{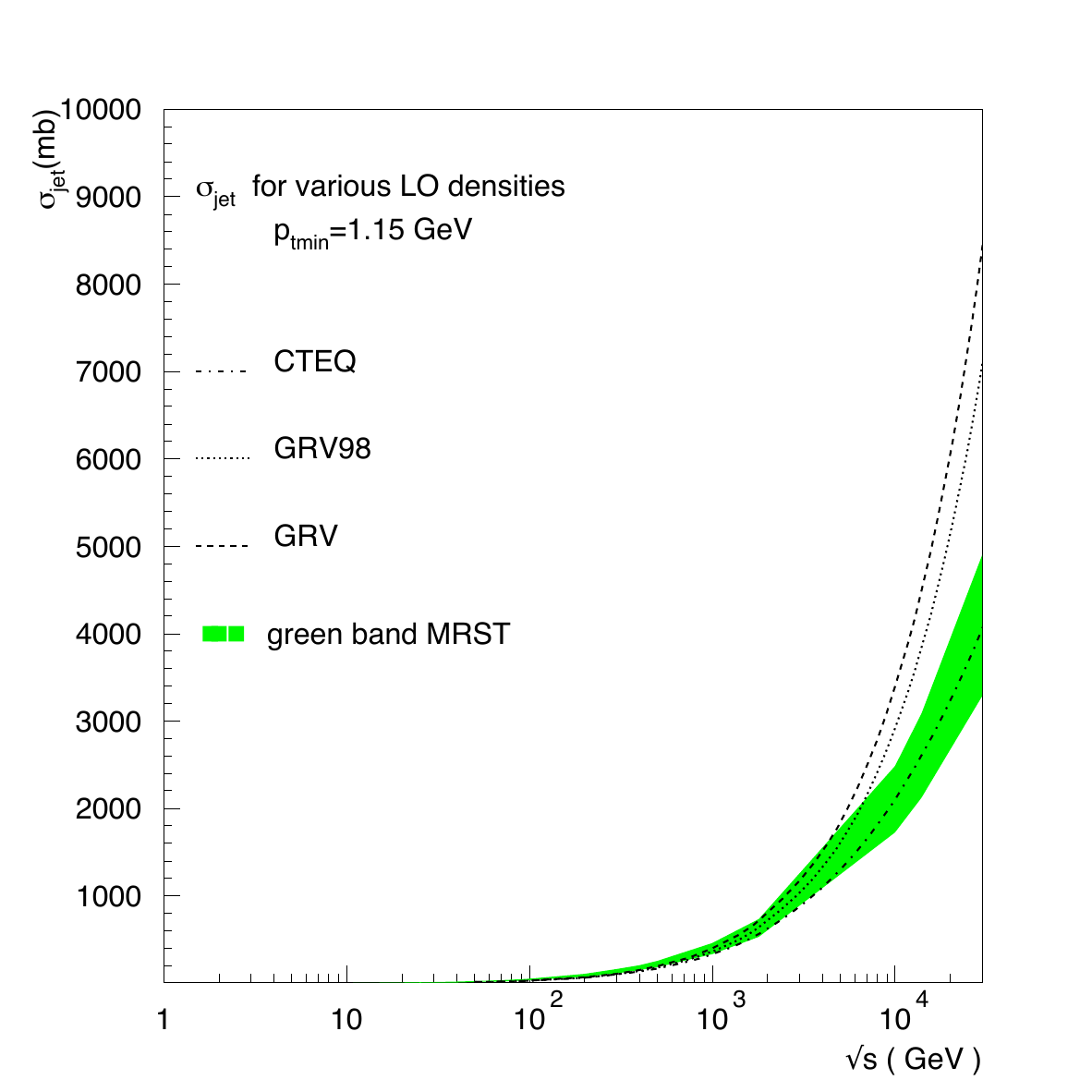}}
\caption{LO Partonic picture of  mini-jet role in hadron-hadron scattering and representative mini-jet calculation from \cite{Pancheri:2007rv}.}
\label{fig:cartoon1}
\hspace{-2cm}
\resizebox{0.5\textwidth}{!}{
\includegraphics{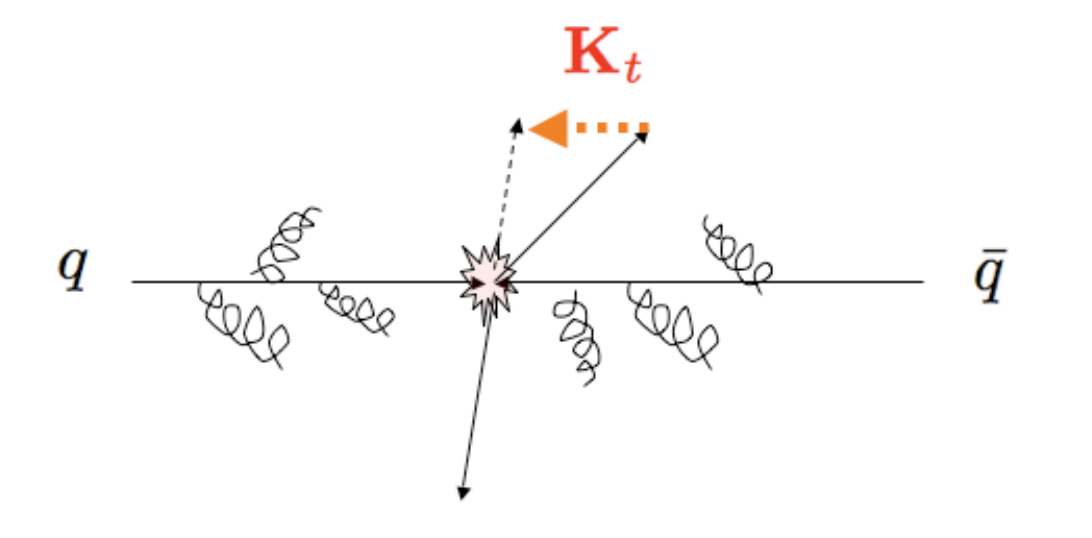}}
\resizebox{0.3\textwidth}{!}{
\includegraphics{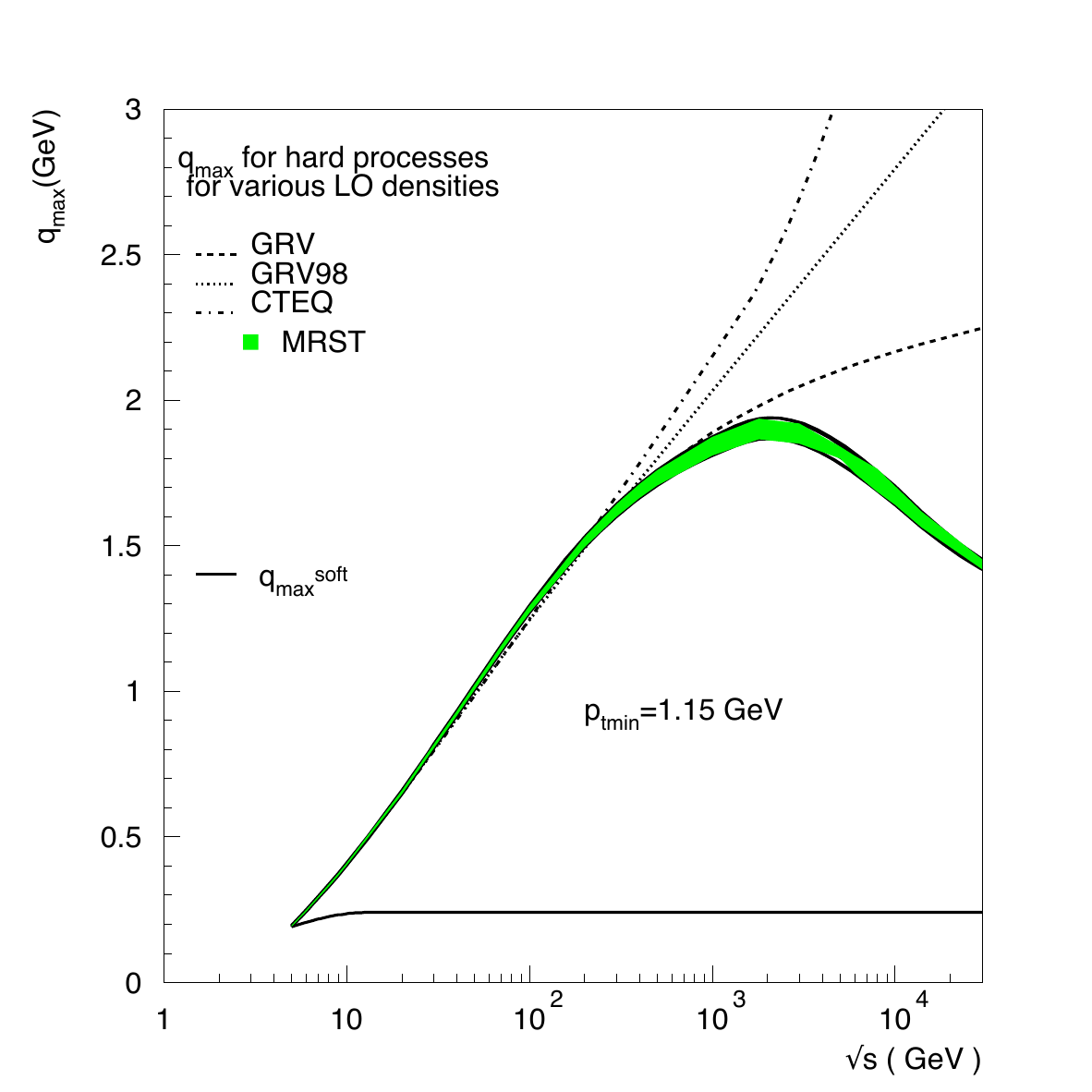}}
\caption{Global soft $K_t$ emission breaks the collinearity of partons.
 At right the maximum momentum. $q_{max}$, allowed to a single gluon, averaged over   
 different PDFs for valence quarks, \cite{Pancheri:2007rv}. 
 Lower curve parametrizes  effects at $\sqrt{s}<10\ {\rm GeV}$. }
\label{fig:cartoon2}
\hspace{-2cm}
\resizebox{0.5\textwidth}{!}{
\includegraphics{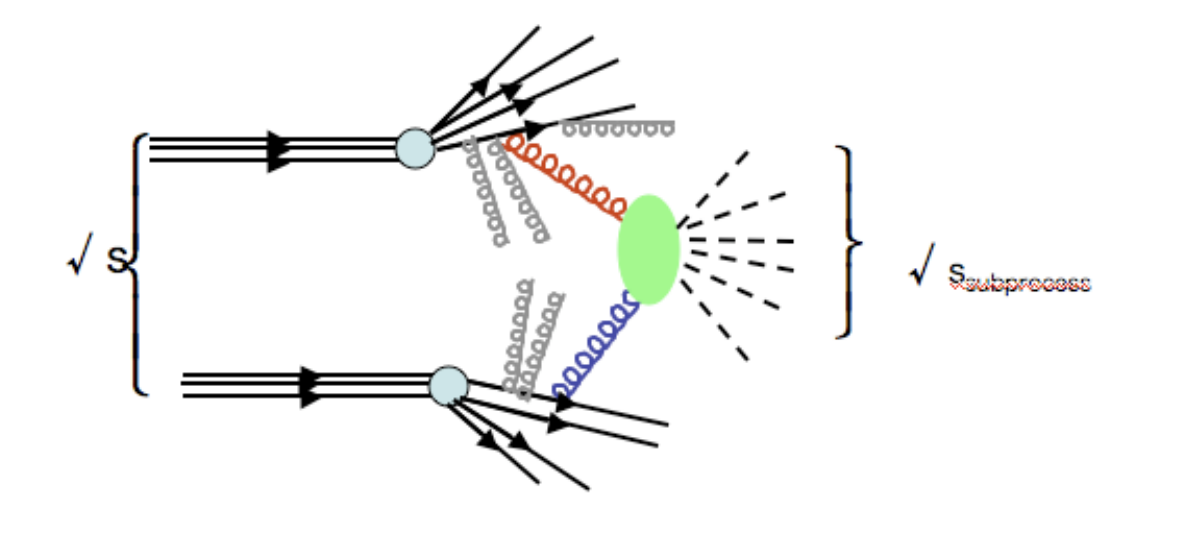}}
\resizebox{0.3\textwidth}{!}{
\includegraphics{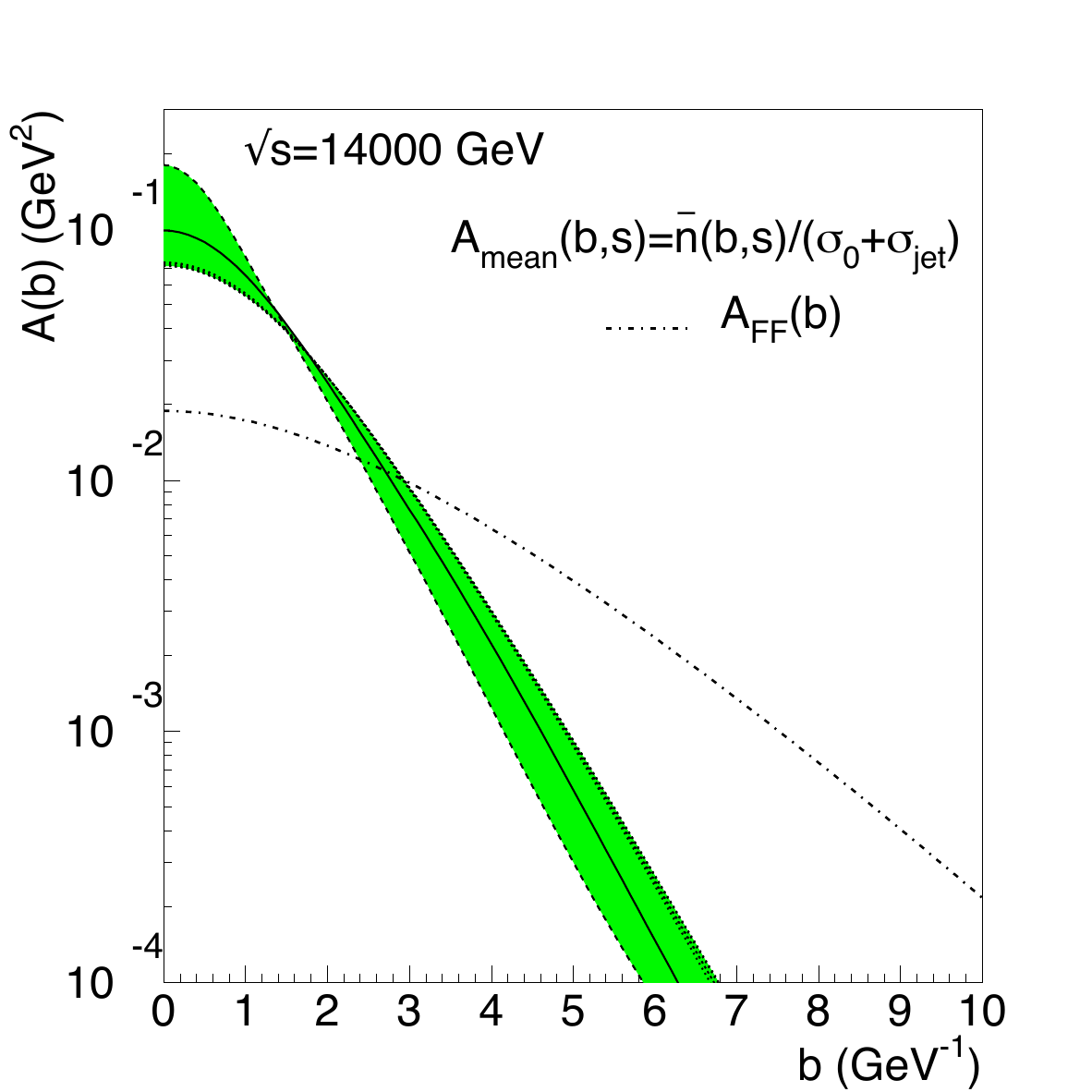}}
\caption{Representative soft gluon radiation from initial quarks in LO picture of hadron-hadron scattering. 
At right, the impact parameter distribution associated to each parton-parton scattering process, 
for different average $q_{max} $ values,  at LHC, from \cite{Pancheri:2007rv}. 
Dotted curve is  convolution of proton e.m. form factors.}
\label{fig:cartoon3}
\hspace{-2cm}
\resizebox{0.5\textwidth}{!}{
\includegraphics{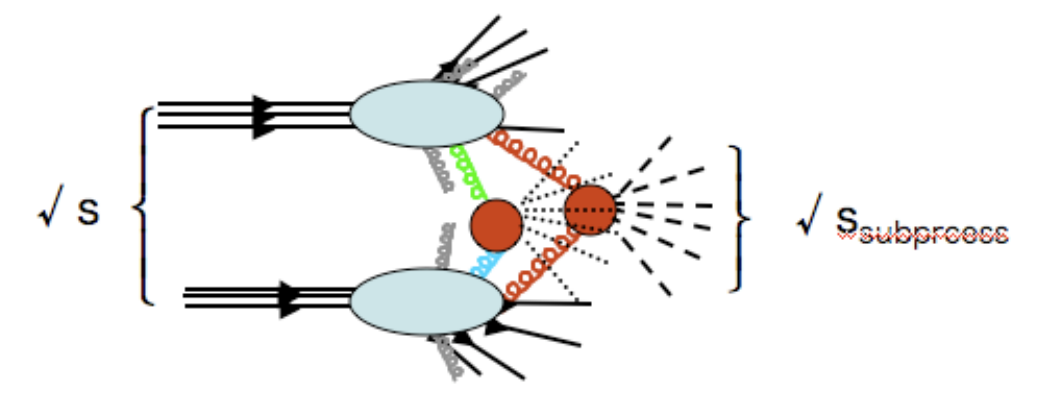}}
\resizebox{0.3\textwidth}{!}{
\includegraphics{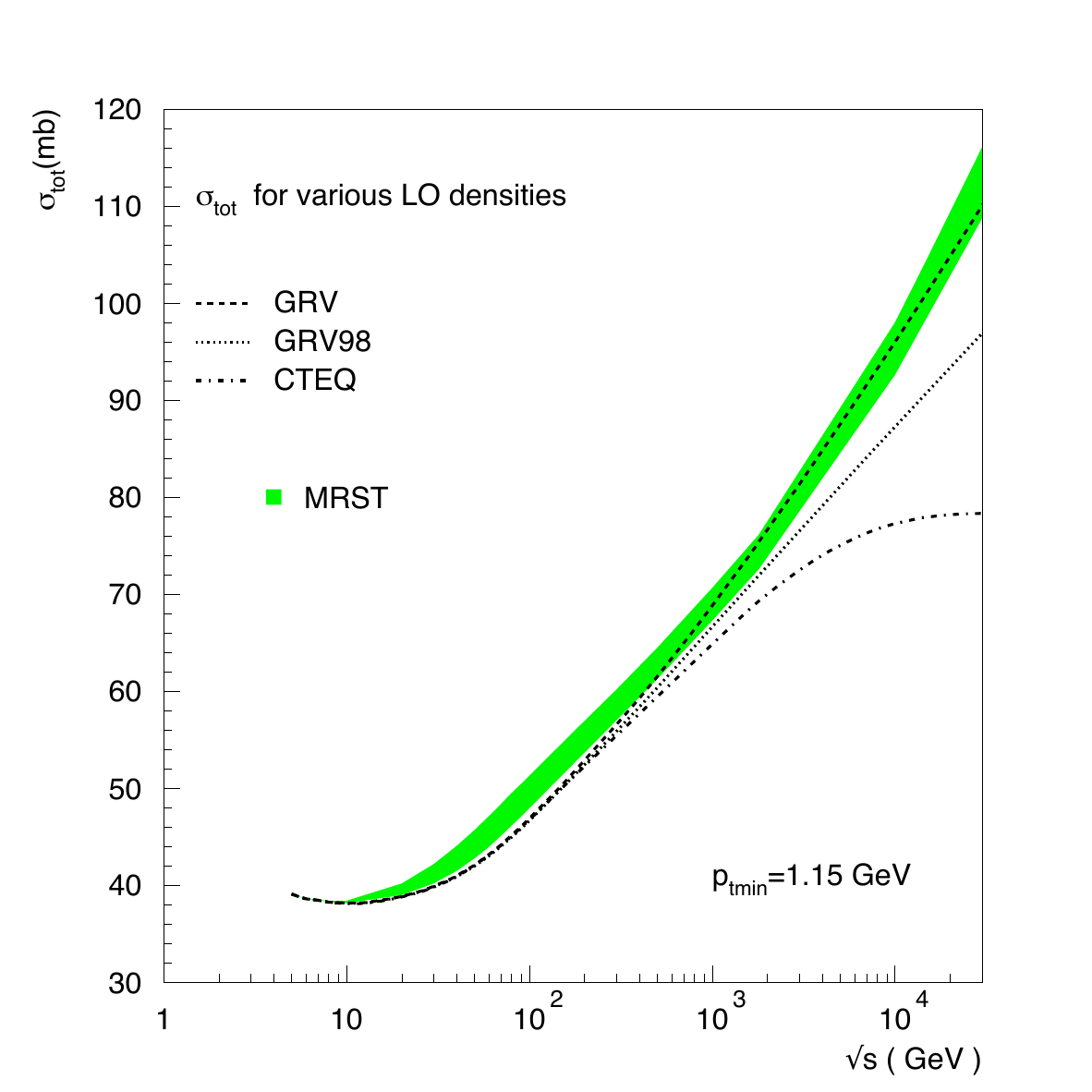}}
\caption{Representative multiple processes, resulting in eikonalization of mini-jet basic process 
corrected for soft gluon emission. At right, the eikonalized total cross-section from different PDFs, 
as from above input, from \cite{Pancheri:2007rv}.This and all the above three figures are  reprinted from \cite{Pancheri:2007rv} \copyright (2007)
with permission by APPB.}
\label{fig:cartoon4}
\end{center}
\end{figure*}
We have discussed in \ref{sss-models:kt}
 our suggestion for the impact parameter distribution. The underlying physics 
is that this distribution is probed in the scattering 
when partons start a brownian like motion inside the protons, each successive change of direction 
generated by emission of soft gluons. Thus, a matter 
distribution $A(b,s)$ can be obtained as  the normalized Fourier transform of the expression for 
soft gluon resummation in transverse momentum space. Through it, the average number of hard collisions is 
calculated to be
\begin{align}
<n(b,s)>=A_{hard}^{BN}(b,q_{max})\sigma_{mini-jets}(s,p_{tmin})\nonumber \\
=
\frac{
e^{-h(b,s)}
}
{\int d^2\vecb e^{
-h(b,s)
}
}
\sigma_{mini-jet}(s,p_{tmin})\label{eq:nhard}
\end{align}
where we have introduced the notation 
 BN to induce that this function is calculated using soft gluon $k_t$-resummation, 
 and the upper limit of integration $q_{max}$ carries energy dependence from the scattering partons.
 We have followed the early work by Chiappetta and Greco about resummation effects 
 in the Drell-Yan process, where   this upper limit is defined by the kinematics of single gluon emission \cite
{Chiappetta:1981bw}. Namely for a process such as
\be
q(x_1)+q(x_2)\rightarrow X(Q^2) + g(k_t)
\ee
with the subenergy of the initial (collinear)  parton pair defined as $\shat=\sqrt{sx_1x_2}$. 
In the no-recoil approximation, $Q^2$ is  the squared  invariant mass of the 
outgoing  parton pair $X$, with parton transverse momentum $p_t>p_{tmin}$.
Kinematics then leads to
\be
q_{max}(\shat,y,Q^2)=\frac{\sqrt{\shat}}{2}
(1-\frac{Q^2}{\shat})
\frac{1}{\sqrt{1+z\sinh^2 y}}
\label{eq:qmax}
\ee
and $y$ is the rapidity 
of the outgoing partons \cite{Chiappetta:1981bw}. 
In our simplified model for the total cross-section, we have approximated $q_{max}(\shat,y,Q^2)$ 
with its maximum value at $y=0$ and averaged its  expression over the parton densities, namely
\begin{align}
q_{max}\equiv <q_{max}(s)>={{\sqrt{s}} 
\over{2}}\times \nonumber\\
{{ \sum_{i,j}\int {{dx_1}\over{ x_1}}
f_{i/a}(x_1)\int {{dx_2}\over{x_2}}f_{j/b}(x_2)\sqrt{x_1x_2}
\int_{z_{min}}^1
 dz (1 - z)}
\over{\sum_{i,j}\int {dx_1\over x_1}
f_{i/a}(x_1)\int {{dx_2}\over{x_2}}f_{j/b}(x_2) \int_{z_{min}}^1 (dz)}}\label{eq:qmaxav}
\end{align}
Furthermore, as discussed in \cite{Achilli:2011sw}, we have made the ansatz that the  
LO contribution to the resummation effect comes from emission from valence quarks. 
Emission from gluons is certainly to be included, and will be be dealt with in further work on the model. 
For a discussion of this point see also \cite{Fagundes:2015vba}.

Following the interpretation by Durand and Pi, as discussed before, and in the spirit of 
Moli\'ere  theory of multiple scattering, we now obtain the pQCD contribution to the imaginary 
part of the scattering amplitude at $t=0$ and hence the total cross-section 
through the identification $\chi(b,s)=<n(b,s)>/2$,
i.e.
\be
\sigma_{total}=2\int d^2\vecb [1-e^{-<n(b,s)>/2}] \label{eq:sigmatotal}
 \ee
 The perturbative calculation we have outlined does not suffice to account for all the process 
 which contribute to the total cross-section. Other partonic processes with momentum 
 $p_t< p_{tmin}$ enter, and, at low energy constitute the dominant contribution. By definition, 
 $p_{tmin}$ separates parton-parton scattering with a pQCD description, 
 from everything  else. Thus one needs to parametrize the low energy part, and in our model 
 we propose  a simple approximate factorization of  the average number of collisions as
 \begin{align}
 <n(b,s)>=<n(b,s)>_{p_t<{p_{ tmin}}}+<n(b,s)>_{p_t>{p_{ tmin}}}=\nonumber \\
 <n(b,s)>_{soft}+<n(b,s)>_{hard}\label{eq:nbsfactorized}
 \end{align}
  We have proposed two different low energy parametrizations:
 \begin{itemize}
 \item $<n>_{soft}^{BN}=A^{BN}_{soft}\sigma_0 [1+\frac{\epsilon}{2\sqrt{s}}]$ with $\epsilon=0,1$ 
 according to the process being \pp \ or \pbarp.
 \item $<n>_{soft}=polynomial\ in \ 1/\sqrt{s}$ 
 \end{itemize} 
 As discussed in Ref. \cite{Godbole:2004kx}, the first of these two low energy parametrizations 
 has the same expression for $A(b,s)$ as in the hard pQCD calculation, 
 except that  the value of $q_{max}$ is chosen  {\it ad hoc } to reproduce the low energy data. 
 The second parametrization is self explaining, and we have started using 
 it when describing $\pi$-p and $\pi \pi$ scattering \cite{Grau:2010ju}. 
Then the elastic and the inelastic total cross-sections follow from  the usual formulae. 
Numerically, the sequence of the calculation is as follows:
\begin{enumerate}
\item choose LO densities (PDFs) for the partons involved in the  process to study, 
such as $\pi, \gamma, protons,\ antiprotons$ and thus calculate the mini-jet cross-section
\item for the given $p_{tmin}$ and chosen LO densities calculate the average value 
for $q_{max}$, the maximum energy carried by a single soft gluon, through the kinematic 
expression given before in Eq.~(\ref{eq:qmax})
\item choose a value for the singularity parameter $p$ in the soft gluon integral and, 
with the $q_{max}$ value just obtained, calculate $A_{hard}^{BN}(b,s)$
\item parametrize the low energy data to obtain $<n(b,s)>_{soft}$
\item eikonalize and integrate 
\end{enumerate}
Different choices of the PDFs call for different values of the parameters $p$ and $p_{tmin}$. 
The sequence of calculations and some typical results for different Parton Densities
can  be found in \cite{Pancheri:2007rv} and are shown in the right hand panels of
 Figs.~\ref{fig:cartoon1}, \ref{fig:cartoon2}, \ref{fig:cartoon3} and \ref{fig:cartoon4}. 
This approach led to the band of predictions shown in Fig. ~\ref{fig:bn} from \cite{Achilli:2007pn}.
\begin{figure}
\resizebox{0.5\textwidth}{!}{%
  \includegraphics{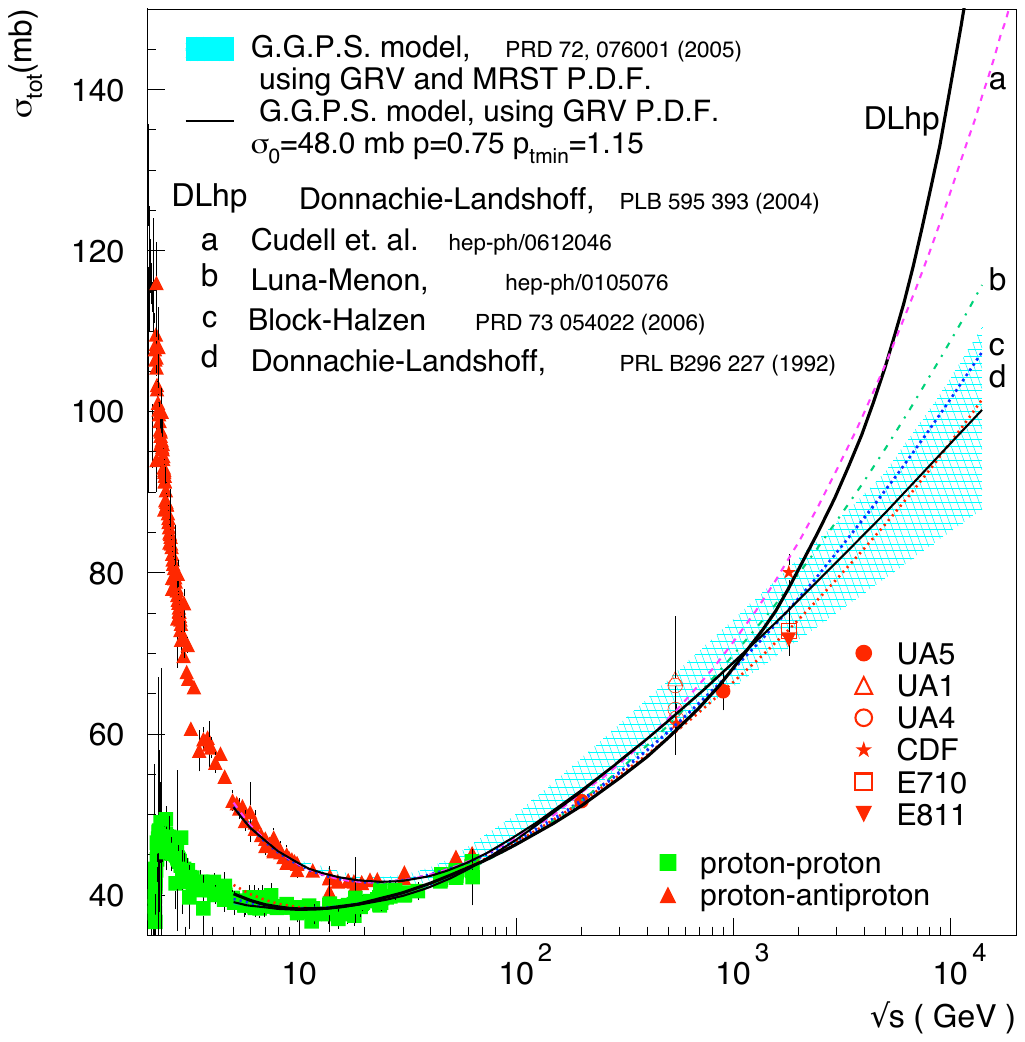}
}
\vspace{-2cm}
\caption{Comparison of the BN model, described in the text, with data and  other model predictions.  
Reprinted from  \protect\cite{Achilli:2007pn}, 
\copyright (2007) with permission by Elsevier.}
\label{fig:bn}       
\end{figure}
In particular we notice that the upper curve of our predicted band nicely accommodates 
the TOTEM result, as we have already shown  in Fig.~\ref{fig:sigtot2013}, 
reproduced at the beginning of this Section.

The above  is a rather  general parametrization of the  total cross-section (of hadrons and photons)  
which we have developed  over the past two decades. 
The central ingredients are pQCD, i.e. mini-jets, and soft resummation in the infrared with a singular 
but integrable effective couling constants for gluons and 
quarks,  and where unitarization is achieved through the impact parameter distribution. 
Not all the details could be specified here, but most can be 
found in \cite{Grau:1999em} and \cite{Godbole:2004kx}.

\subsubsection{Soft gluon $k_t$-resummation and the Froissart bound}\label{sss-models:froissart}

The physics embodied in the phenomenology described in the previous subsection is 
that soft gluon resummation in the infrared region provides a cut-off in 
impact parameter space, which leads to a smooth logarithmic behavior. In our model such behavior  
depends on the ansatz  about limit of the effective {\it quark-gluon} coupling constant when $k_t\rightarrow 0$. To see this, we start with  
taking the very large $s$-limit in Eq. ~(\ref{eq:sigmatotal}). 

At extremely large energy values, we neglect the low energy part, and have
\be
\sigtot \rightarrow 2\int d^2\vecb [1-e^{-n(b,s)_{hard}(b,s)/2}] \label{eq:sighard}
\ee
%
With the QCD jet cross section driving  the rise due to the
increase with energy of the number of partonic collisions
in Eq. (\ref{eq:sighard}), let us recall 
 the energy behaviour of the mini-jet cross-sections.
In the
$\sqrt{s}>>p_{tmin}$ limit, the major contribution to  the mini-jet cross-sections 
comes from collisions of gluons carrying 
 small momentum fractions  
  $x_{1,2}<<1$, a region  where   the relevant 
PDFs   behave approximately as powers of the momentum fraction
$x^{-J}$ with $J \sim 1.3$ \cite{Lomatch:1988uc}.  This leads  to
 the 
 asymptotic high-energy expression for
$\sigma_{jet}$
\begin{equation}
\sigma _{jet}  \propto \frac{1}{{p_{t\min }^2 }}\left[
{\frac{s}{{4p_{t\min }^2 }}} \right]^{J - 1} 
\label{minijetasympt}
\end{equation}
 where the dominant term is   a power of $s$. 
 Fits to the mini-jet cross-sections, obtained with different PDF sets 
\cite{Achilli:2009ej}, confirm the value 
$\varepsilon\equiv J-1\sim0.3$.  

To match  such energy behaviour as in Eq.~(\ref{minijetasympt}) 
with the gentle rise of the total $pp$ and $p {\bar p}$ cross-sections 
at very high energy, 
we inspect  
the impact parameter distribution we have put forward using resummation of soft gluons in the infrared region.
Let us consider the integral for the function $h(b,q_{max})$,
 which is performed up to a value $q_{max}$, corresponding to
the maximum transverse momentum allowed by  kinematics of single gluon 
emission
 \cite{Chiappetta:1981bw}. In principle, this parameter and the
overlap function should be calculated for each partonic
sub-process, but in  the partial factorization of Eq.(\ref{eq:nhard}) we  use the 
average value of
$q_{max}$ obtained by considering all the sub-processes that can
happen for a given energy of the main hadronic
process, as seen before. The energy parameter $q_{max}$ is of the order of 
magnitude of $p_{tmin}$. For present low$-x$ behaviour of the PDFs,  in  the 
high energy limit,  $q_{max}$
is a slowly varying function of $s$, starting as $\ln{s}$, with a limiting 
behaviour which depends on the densities  
\cite{Pancheri:2007rv}.
From Eqs.~(\ref{halphas3}) and (\ref{minijetasympt}) 
one can estimate the very large $s$-limit 
\begin{align}
n_{hard} (b,s)=A_{BN}(b,s)\sigma_{jet}(s,p_{tmin})\sim \nonumber \\
A_0(s)  
e^{-h(b,q_{max})}{\sigma_1}({{s}\over{s_0}})^\varepsilon
\end{align}
and, from this, using  the very large $b$-limit,
\begin{equation}
n_{hard} (b,s)\sim A_0(s) \sigma_1 e^{-(b{\bar \Lambda})^{2p}} 
({{s}\over{s_0}})^\varepsilon
\end{equation}
with $A_0(s)\propto 
\Lambda^2 $ and with a logarithmic dependence on $q_{max}$, i.e.  a 
very slowly varying function of $s$. The large $b$-limit taken above follows from Eq.~~(\ref{halphas3}).  We also have  
\begin{equation}
{\bar \Lambda}\equiv
 {\bar \Lambda(b,s)}=\Lambda 
 \{ 
 {{c_F{\bar b}}\over{4\pi(1-p)}}[
 \ln (2q_{max}(s)b) +{{1}\over{1-p}}]
 \}^{1/2p}
 \end{equation}
It is now straightforward to see  how the   two crucial parameters   of our 
model, namely the power $\varepsilon$ with which the mini-jet cross-section 
increases with energy and the parameter $p$  associated to the infrared   behaviour of 
the effective quark-gluon QCD coupling constant, conjure  to obtain a rise of the total 
cross-section obeying the  limitation imposed by the 
Froissart bound, namely, asymptotically, $\sigtot \lesssim (\ln s)^2$.
Call $\sigma_T(s)$ the asymptotic form of the total cross-section,
 \begin{equation}
  \sigma _T (s) \approx 2\pi \int_0^\infty  {db^2 } 
[1 - e^{ - n_{hard} (b,s)/2} ]
\end{equation}
and insert  the asymptotic expression for $\sigma_{jet}$ at high
energies, which grows as  a power of $s$, 
and the large $b$-behaviour of  $A_{BN}(b,s)$,    obtained
through soft gluon resummation,  and which decreases  in $b$-space at least 
like an exponential ($1<2p<2$).
In such large-$b$, large-$s$ limit, one has
\begin{equation}
  n_{hard}  = 2C(s)e^{ - (b{\bar \Lambda})^{2p} }
\end{equation}
where  $2C(s) = A_0(s) \sigma _1 (s/s_0 )^\varepsilon  $. The resulting
expression for $\sigma_T$ is
\begin{equation}
  \sigma _T (s) \approx 2\pi \int_0^\infty  {db^2 } [1 - e^{ -
C(s)e^{ - (b{\bar \Lambda})^{2p} } } ]
\label{sigT}
\end{equation} 
With the variable transformation
$u=({\bar \Lambda} b)^{2p}$,
and neglecting the logarithmic $b$-dependence in ${\bar \Lambda}$ by 
putting $b=1/\Lambda$,
Eq.~(\ref{sigT}) becomes
\begin{equation}
 \sigma _T (s) \approx {{2\pi}\over{p}}{{1}\over{{\bar \Lambda}^2}} 
\int_0^\infty du u^{1/p -1}
 [1-e^{
 -C(s)
 e^{-u}
 }
 ]  
\end{equation}
Since,  as
$s\rightarrow \infty$, $ C(s)$ grows indefinitely as a power
law,  the quantity between square brackets
$I(u,s)=1-e^{-C(s)e^{-u}}$ has the limits
$I(u,s)\rightarrow 1$ at $u=0$
and $I(u,s)\rightarrow 0$ as $u=\infty$.
Calling $u_0$ the value at which $I(u_0,s)=1/2$ we then put
$I(u,s)\approx 1$ and integrate only up to $u_0$. Thus
\begin{equation}
{\bar \Lambda}^2\sigma_{T}(s)\approx (\frac {2\pi} {p})\int_0^{u_0} du u^{\frac{1-p} {p}}
=2\pi u_0^{1/p}
\end{equation}
and since, by construction
\begin{equation}
u_0=\ln[\frac {C(s)}{\ln 2}]\approx \varepsilon \ln s
\end{equation}
we finally obtain
\begin{equation}
\sigma_{T}\approx \frac {2\pi } {{\bar \Lambda}^2} 
[\varepsilon \ln \frac {s} {s_0}]^{1/p}
\label{froissart1}\end{equation}
to leading terms in $\ln s$. 
 We therefore derive the asymptotic energy dependence 
\begin{equation}
  \sigma _T  \to [\varepsilon \ln (s)]^{(1/p)}
 \label{froissart}
\end{equation}
apart from a  possible   very slow  $s$-dependence  from ${\bar \Lambda}^2$. 
 The same result  is also obtained using the saddle point method.

This indicates that the Froissart bound is saturated if $p=1/2$, but also that we have the two following asymptotic limits
\begin{align}
\sigma_{total} \rightarrow ( \ln s)^2\ \ \ \ \ \ \ \ p \rightarrow 1/2\\
\sigma_{total} \rightarrow \ln s\ \ \ \ \ \ \  \ \ \ \ \ \ p \rightarrow 1
\end{align}
depending on  our approximate singular expression for the strong coupling in the infrared. We notice that the limits $1/2<p<1$ are a  
consequence of our infrared description.  Namely,  from the request for the soft gluon integral  to be finite ($p<1$) follow that the cross-section 
should grow at least like a logarithm,  while  the limitation $p \ge 1/2$ is to ensure the confinement of the partons. Confinement
of partons is essential in creating a ``mass gap'' leading to massive hadrons. Once we have massive hadrons, 
we have a Lehmann ellipse for hadrons 
\cite{Grau:2009qx}. We recall that the existence of a Lehmann ellipse is essential for obtaining the Martin-Froissart bound for total cross-sections.   
Through our model, we have 
delineated the two limits: up to a linear confining potential ($p\rightarrow 1$) or down to a barely confining one ($p\rightarrow1/2$). 

Before closing this subsection and the description of the BN model, in Fig.~\ref{fig:pptot-onech}, from \cite{Fagundes:2015vba}, 
we show how  LHC data up to $\sqrt{s}=8\ 
{\rm TeV}$ can be described by this and other currently used models for the 
total cross-section.

As can be seen from Fig.~\ref{fig:bn} from  \cite{Achilli:2007pn}
it has been the practice  to reproduce total cross-section data  for both $pp$ and ${\bar p} p$, 
up to the TeVatron results.   However,  the large differences among the  Tevatron measurements  
did not allow a precise description at higher energies, such as those explored  at   LHC. Once the LHC data have been  released, and as  it has been the case for all models for the total cross-section,  we have updated 
our analysis. To this aim, we have used   only $pp$ accelerator data,  ISR and the recent LHC measurements, namely ${\bar p} p $ points are shown, but  have  not been used for the phenomenological fit, nor the    
Cosmic ray extracted values for $pp$. A  more    recent  set of LO densities, 
MSTW \cite{Martin:2009iq} has been included in the set of predictions for the BN model. The  values of $p$ and $p_{tmin} $ which better reproduce the LHC result are obtained by varying 
$p_{tmin} \simeq 1\div 1.5\ {\rm GeV}$ and $1/2\lesssim p \lesssim0.8$. The result,  for  the total $pp$ cross sections, 
is shown in   Fig.~\ref{fig:pptot-onech}. 

 The BN results, 
  now  stabilized at $\sqrt{s}=7$ and $8\ TeV$ by tuning the parameters 
  to TOTEM data,   show 
marked differences in the high cosmic ray region. The difference is ascribable to the different low-x behaviour of the PDFs 
used in the mini jet calculation, GRV and MSTW, as discussed  in \cite{Fagundes:2015vba}. In addition, and as we shall briefly discuss at the end of Sec.~\ref{sec:lhcnow}, as of 2016 there is some tension between measurements by the two experiments presently providing values for the total cross-section, TOTEM and ATLAS.
\begin{figure}[h]
\centering
\resizebox{0.5\textwidth}{!}{
\includegraphics{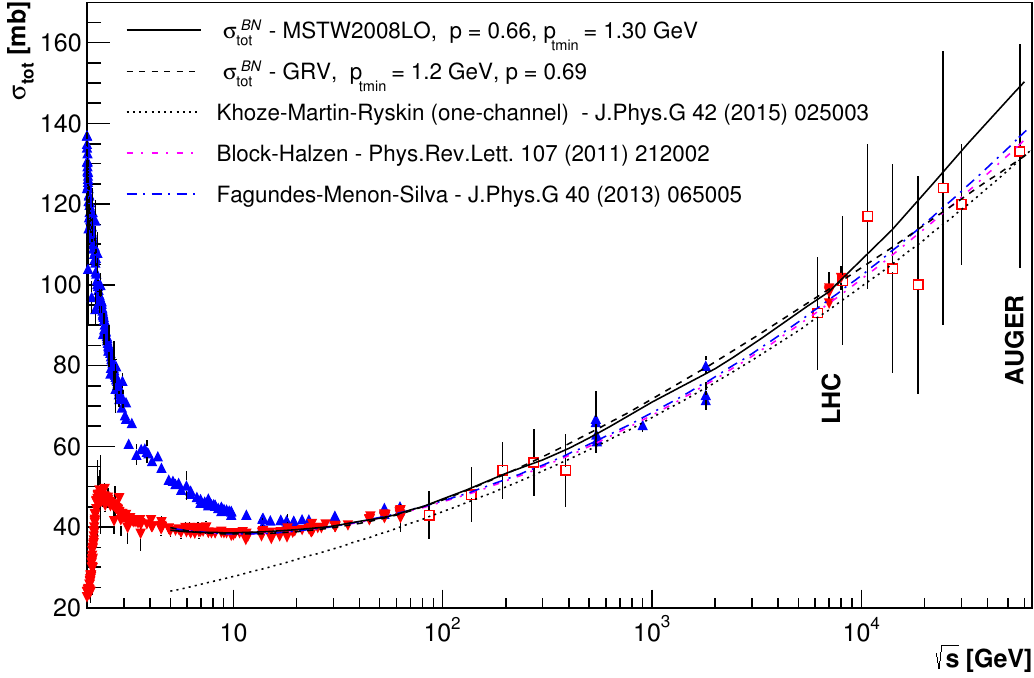}}
\caption{ The one-channel predictions from the one-channel  QCD mini-jet with soft gluon resummation model and $pp$ total cross section (BN model)
are compared with accelerator data at LHC, from TOTEM \cite{Antchev:2013paa,Antchev:2013iaa} and 
ATLAS measurements \cite{Aad:2014dca}. 
   The BN model  results  are compared with  one- channel model 
  from Khoze et al.\cite{Khoze:2014nia}. The  red {\it dot-dashed} curve corresponds to fits to the total cross section 
  by Block and Halzen \cite{Block:2011vz}, the dot-dashed blue line represents the fit by 
  Fagundes-Menon-Silva \cite{Fagundes:2012rr}. The figure is from \cite{Fagundes:2015vba}. 
  Reprinted with permission from  \cite{Fagundes:2015vba}, Fig.(3), \copyright (2015) by the American Physical Society.}
\label{fig:pptot-onech}
\end{figure}
\subsection{AdS/CFT correspondence and 
the total cross-section}\label{ss:ads}
A short, but compact discussion of the Froissart bound in the context of the string/gauge duality can be found in \cite{Tan:2008zz}. 
In \cite{Tan:2008zz}, the Pomeron is defined as 
{\it the leading contribution at large $N_c$ to the vacuum 
exchange at large $s$ and fixed $t$}. It is stated that in both strong coupling and 
pQCD, the Pomeron contribution  grows as $s^{1+\epsilon}$ with $\epsilon>0$, 
so that, in order not to violate the Froissart bound, higher order corrections need to 
be taken into account. 

For the scattering of particles 1 and 2 into particles 3 and 4,  
the standard eikonal representation for the amplitude in the case of large 't Hooft 
coupling is then generalized to the expression
\cite{Brower:2007xg}
\bea
A(s,t)=-2is\int dz dz' P_{13}(z) P_{24}(z') \times\nonumber\\
\times \int d^2b e^{iq_\perp \cdot b}[e^{i\chi(s,b,z,z')}-1]
\eea 
where the wave function $P_{ij}(z)$ refers to the left moving particles, 1 into 3, and right moving 
particles, 2 into 4. The variables $z$ and $z'$ correspond 
to the convolution over moving direction in $AdS_3$
and they are normalized so that, when confinement is implemented,  $\int P_{ij}dz=\delta_{ij}$.

In this description, one obtains the total cross-section through the standard geometrical picture  
$\sigtot\approx b^2_{max}$ and the problem is, as usual, 
that of finding $b_{max}$. 

In this picture, the quantity
\be
\sigtot (s,z,z')=2\Re e \int d^2b[1-e^{i\chi(s,b,z,z')}]
\ee
is the {\it bulk}  \x \  and the physical cross-section 
is obtained after convolution with the wave functions.

The picture in the bulk is split into two regions, respectively called {\it diffractive} and {\it black disk}, as follows:
\begin{description}
\item {\it diffractive}: in this region  $\Im m \chi < \Re e \chi$, and it is $\Re e \chi \approx 1$ which sets the 
limit for contributions to the scattering, with  $b_{max}=b_{diff}$. 
\item {\it black disk}: in this region $\Im m \chi > \Re e \chi$ 
, the interaction is dominated by the weak coupling 
Pomeron and the maximum $b_{black}$  at which scattering still takes place corresponds to where 
absorption is of order 1, namely $\Im m \chi \approx 1$. 

\end{description}
For the example of an even-signature Regge exchange in 4-dimensions, according to \cite{Tan:2008zz}, one can write
\begin{eqnarray}
b_{black}\sim \lambda^{-1/4} m_0^{-1}\ln (\beta s/s_0) 
\\
b_{diff}\sim m_0^{-1}\ln (\beta s/s_0)
\end{eqnarray}
where in the fixed 't Hooft coupling $\lambda=g^2_{YM}N_c$, and  $m_0$ is the scale for  
the trajectory exchanged with  
\be
\alpha_{t}=2+\alpha'(\frac{t}{m_0^2} -1)
\ee

With $\alpha' \sim \lambda^{-1/2}\ll 1$, one obtains $b_{diff}\gg b_{black}$. Thus,  a unique result 
of the strong coupling regime is that the eikonal is predominantly real.

Scattering in the conformal limit, leads to a $\sigtot\sim s^{1/3}$ in the strong coupling regime. 
But with confinement the situation is different and the spectrum has a mass gap which then leads to a 
logarithmic growth. Tan \cite{Tan:2008zz} writes
\begin{equation}
b_{diff} \simeq \frac{1}{m_0}\ln (N^2 s/m_0^2)
\end{equation}
A full discussion of all these regions and the resulting expressions can be found in \cite{Brower:2007xg}.

\subsection{Phenomenogical fits to the total cross-section}\label{ss:phenofits}
We shall now describe two different   phenomenological fits  used to describe $pp$ and ${\bar p} p$ total cross-section, 
published by the  Particle Data Group, one before and the other after the start of LHC.

\subsubsection{ Cudell and COMPETE collaboration}\label{sss:compete}
For the two reactions \pp \ and \pbarp \ of interest here, we show in Fig.~\ref{fig:ppbarp} the data compilation from  
fig. 40.11 of the 2009  Particle Data Group (PDG) \cite{Amsler:2008zzb}. For this and other updates of  mini-reviews and figures, we refer the reader to  the \href{< http://pdg.lbl.gov>}{PDG site}.
\begin{figure}
\resizebox{0.5\textwidth}{!}{%
  \includegraphics{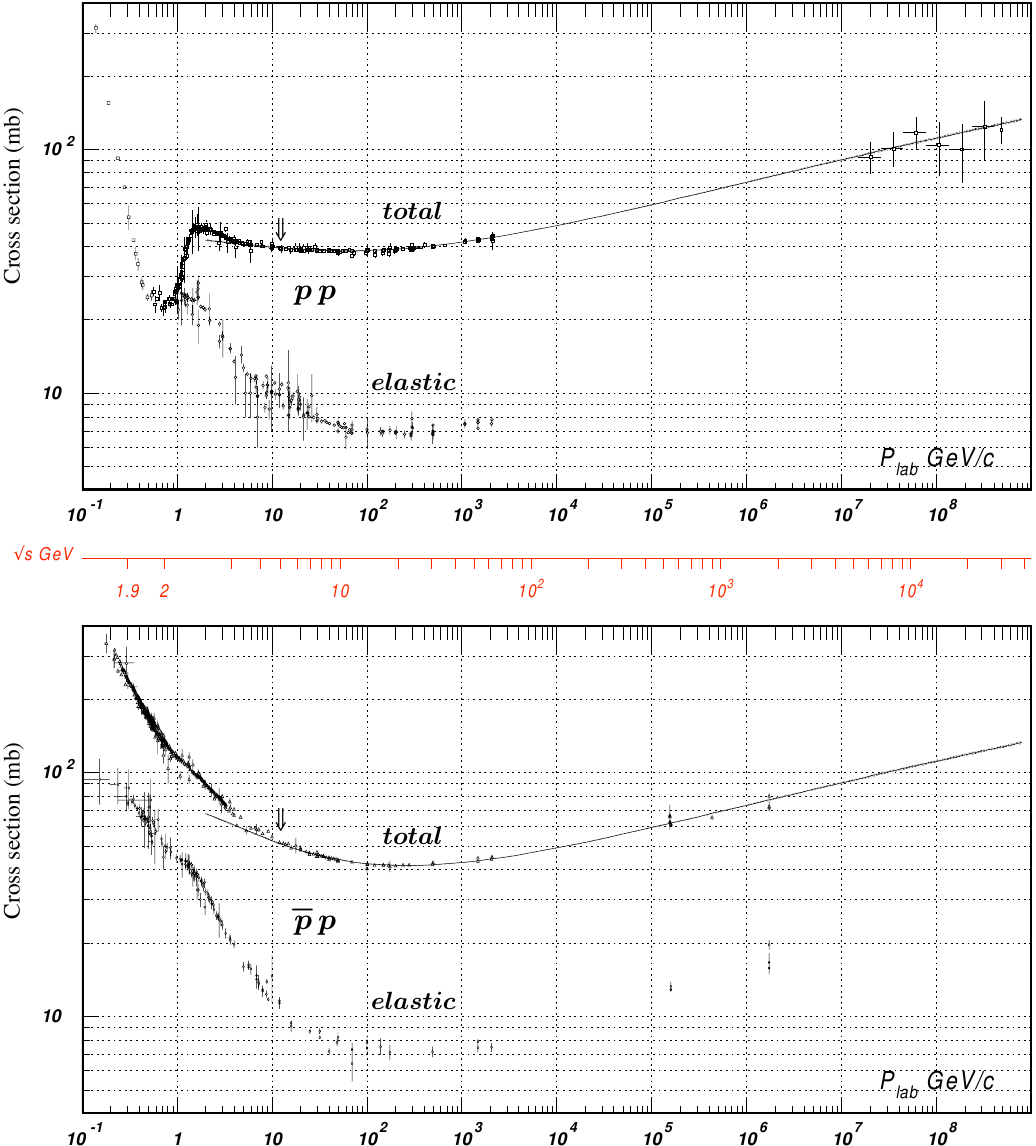}
}
\caption{The total proton
cross-sections, as compiled before the start of the LHC and hence before of the TOTEM and ATLAS measurements,  by the COMPAS collaboration for the 2009 RPP \cite{Amsler:2008zzb} with model input from    \protect\cite{Cudell:2001pn} ( COMPETE Collaboration). Figure downloaded from http://pdg.lbl.gov/2009/figures/figures.html.}
\label{fig:ppbarp}       
\end{figure}
We shall start by summarizing ref. \cite{Ezhela:2002wj} where an overview of the COMPETE program is given. In this paper, 
the authors describe their data base policy and give a number of web access information, which allow to  download and run fitting programs. 

The region to be focused on is the Coulomb-nuclear interference region, and to do this one needs to use, 
for all the experiments, the following common set of theoretical inputs:
\begin{itemize}
\item common parametrization of electromagnetic form factors, where there is a problem with 
the usual VMD term, since a fit to the $\dsigdt$ alone 
gives a better $\chi^2/d.o.f$ than the fit to the combined data set of $\dsigdt$ and   $G_E/G_M$,
\item common procedure to analyze data in the Coulomb interference region,
\item common  set of strong interaction elastic scattering parameters,
\item common study of Regge trajectories : already there is a problem here since the slope of the meson trajectories is different 
depending on the flavour content, although not so much for the baryons.
\end{itemize}
The problem with the above program, of course, is that the Regge description may be only a (albeit good) approximation and while it may 
eliminate the systematic differences, it will still have a model dependence. But more about this later.

The COMPETE collaboration had been cleaning and gathering all total cross-section data \cite{Cudell:2001pn} then  available at the 
PDG site \cite{Amsler:2008zzb}. Data have been fitted with the expression
\begin{align}
\sigma_{{\bar a}b}=Z^{ab}+B\ln^2 (\frac{s}{s_0})+Y_1^{ab}(\frac{s_1}{s})^{\eta_1}+Y_2^{ab}(\frac{s_1}{s})^{\eta_2}  \label{eq:fitcudellpbarp} \\
\sigma_{ ab}=Z^{ab}+\ln^2 (\frac{s}{s_0})+Y_1^{ab}(\frac{s_1}{s})^{\eta_1}-Y_2^{ab}(\frac{s_1}{s})^{\eta_2}
\label{eq:fitcudellpp}
\end{align}
where $Z^{ab}, B, Y_i^{ab}$ are in millibarn, $s,\ s_0,\ s_1$ are in ${\rm GeV}^2$. The scale   
$s_1$ is fixed to be $1/ {\rm GeV}^2$, whereas $\sqrt{s_0}\approx 5\ {\rm  GeV}$. 
The physical interpretation of this fit is that the power law terms reproduce the Regge 
behaviour from the imaginary part of the forward scattering amplitude, 
with two Regge poles if $\eta_1\ne \eta_2$, whereas the constant term and the $\ln^2[s]$ term reflect 
the so-called Pomeron exchange. For a summary of 
the numerical values of the various parameters , we refer the reader to the PDG review. %

 The first two terms in Eqs. ~(\ref{eq:fitcudellpbarp}), (\ref{eq:fitcudellpp})  reflect the bulk of semi-perturbative QCD processes 
 which start dominating the total cross-section as soon as the 
 c.m. energy of the hadronic process goes above 
$\sqrt{s}\approx  (10\div 20) \ {\rm GeV}$. There are various ways to refer to these terms.  
In our QCD model \cite{Grau:1999em,Godbole:2004kx} the 
term which brings in the rise is the one which comes 
from gluon-gluon scattering tempered by soft gluon emission from the initial state, 
as we have described in  \ref{sss:BN}. The constant term is more complicated to 
understand. It is probably due to quark scattering, well past the Regge region. 
Notice that in the simple Donnachie and Landshoff successful original parametrization \cite{Donnachie:1992ny}, 
there is no constant term in the cross-section. 
The cross-sections only apparently go to a constant, which is where the minimum of the cross-section lies, 
just after the Regge descent and just before the 
cross-section picks up for the asymptotic rise. Indeed, whereas the Regge terms can be 
put in correspondence with resonances in the s-channel, the 
constant term is harder to interpret, except as the old Pomeron which was supposed to 
give constant total cross-sections. 
 
Let us now review a   comprehensive discussion of fits to pre-LHC  total cross-sections data  
\cite{Cudell:2009bx}. Cudell notes the difficulty to 
make precise predictons at LHC because of a number of problems with present data on the total cross-section:
\begin{itemize}
\item No data are available between the ISR energy, $\sqrt{s} \approx (60\div70)  \  {\rm GeV}$
 and the \spbarps \ at $\sqrt{s} \approx (500\div 600) \ {\rm GeV}$
 \item at Tevatron energies, $\sqrt{s} \approx 1800\ {\rm GeV}$,  there is a $2\sigma$ discrepancy between 
 the value calculated by two TeVatron collaborations, E710 and CDF
 \item the $t$ dependence of the differential elastic cross-section as $ t \rightarrow 0$ may not
  be a simple exponential  $exp(Bt)$ where 
 B assumed constant in t thus affecting the measurement of the total cross-section. 
 \end{itemize}
 To the above one should  add that total \pp \ cross-section at cosmic ray energies have very large errors, 
 mostly due to the theoretical uncertainty in the  
 procedures adopted to extract \pp \ total cross-section from $p-air$ cross-section, as discussed. 
 Predictions from the COMPETE collaboration 
 from \cite{Cudell:2009bx} are shown in Fig~\ref{fig:cudellpp}.
\begin{figure}
\resizebox{0.5\textwidth}{!}{%
  \includegraphics{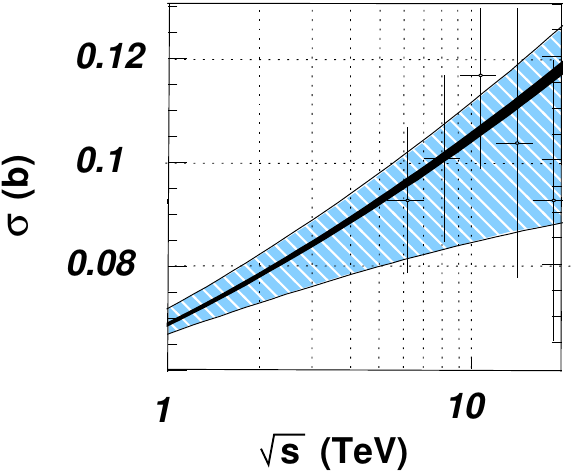}}
\caption{The range of values predicted  for the total proton-proton cross-section 
  by the COMPETE  Collaboration 
  as shown in  \protect\cite{Cudell:2009bx}.  Figure is courtesy of J-R Cudell. Reprinted from
  \cite{Cudell:2009bx}, in  CERN Proceedings CERN-PROCEEDINGS-2010-002.}
\label{fig:cudellpp}       
\end{figure}

Many of the problems discussed by Cudell can be related to the type of unitarization scheme. A new analytic unitarization scheme was 
proposed in \cite{Cudell:2008yb}, but the actual problem is the difficulty of doing a good fit to both elastic and  total cross-section data. 

In \cite{Cudell:2001pn}, it is pointed out that there are problems with the $\rho$ parameter data, 
where $\rho \equiv \rho (s)=\Re e A(s,t=0)/\Im m A(s,t=0)$. Hence, it is said, that the {\it first and safest} 
strategy is to obtain constraints from 
the reproduction of $\sigtot$ only. However, the final result is obtained by fitting the total cross-section and $\rho$. 


In   \cite{Cudell:2001pn} the fits to lower energy
 total cross-sections are parametrized with
\begin{equation}
\sigma^{a\mp b}=\frac{1}{s}(
(R^{+ab}(s)\pm R^{-ab}(s)+P^{ab}(s)+H^{ab}(s)
)
\end{equation}
with
\begin{eqnarray}
R^{+ab}(s)=Y_1^{ab}\cdot (s/s_1)^{\alpha_1}, \  with \ s_1=1\ {\rm GeV}^2\nonumber \\
R^{-ab}(s)=Y_2^{ab}\cdot (s/s_1)^{\alpha_2 }
\end{eqnarray}
wherein
\begin{equation}
\label{Y}
P^{ab}(s)= sC^{ab},  
\end{equation}
describes a simple Pomeron pole  at  $J=1$ 
and $H^{ab}(s)$ is the rising term, which can be 
\begin{itemize}
\item a supplementary simple pole with larger than one intercept
\item  a double pole at $J=1$,  namely $L_{ab}=s(B_{ab}\ln(s/s_1)+A_{ab})$
\item  a triple pole at $J=1$, namely $L_{ab}=s(B_{ab}\ln^2(s/s_1)+A_{ab})$
\end{itemize}
Thus, their  parametrization for $\sigtot$ is a sum of various  $I_n$ terms, with the 
parametrization for $\rho$ 
given below.

For the $\rho$ parameter, in the Appendix, the authors list the following parametrization:
\begin{eqnarray}
R^+_{pole}=-I^+_{pole} \cot[\pi/2 \alpha_+]\\
R^-_{pole}=-I^-_{pole} \tan[\pi/2 \alpha_-]\\
R_L=\frac{\pi}{2} s C_L\\
R_{L2}=\pi s \ln (s/s_0)C_{L2}
\end{eqnarray}
where
\begin{eqnarray}
I^+_{pole}=C^+(s/s_1)^{\alpha^+}\\
I^-_{pole}=\mp C^-(s/s_1)^{\alpha^-}\\
I_L=C_Ls \ln (s/s_1)\\
I_{L2}=C_{L2} \ln^2 (s_0)
\end{eqnarray}
At the end of all this, the result favoured now and found in  2008 PDG \cite{Amsler:2008zzb}, is the one given by Eqs. ~(\ref{eq:fitcudellpbarp}), (\ref{eq:fitcudellpp}).

The paper \cite{Cudell:2001pn} also contains a rather long discussion about  the sign in front of the logarithmic terms, which might hint at the result 
we had originally found in our Pramana paper \cite{Godbole:2006qk}.

Apart from fits to the total cross-section,  Cudell and Selyugin  in \cite{Cudell:2008df} also address the question of the measurement itself, which, 
as described previously in this review, is based on two methods, the optical point and the luminosity based one. The optical point type measurement, 
also called the non-luminosity measurement, is based on the extrapolation of the elastic differential cross-section to the value $t=0$.
The   extrapolation has been usually  done assuming an exponential behaviour $exp[{B(s)t}]$ for the differential elastic cross-section near the $t=0$ point. 
However it is known  the exponent is not stricly linear in $t$. In this paper \cite{Cudell:2008df}  the authors examine the 
possibility that at the LHC expectations based on simple Regge pole models are modified and that the usual expectation of $\sigtot \approx 90\div 125 \ mb$ 
be superceded by the higher values predicted from a number of unitarization schemes, such as hard Pomeron, which would give cross-sections around $150\ mb$ 
or U-matrix unitarization which can give cross-sections as high as 230 mb. The impact of such different expectations is discussed, together with the possibility 
that $\rho$ has a strong $t$-dependence. 
In this paper this t-dependence of the $\rho$ parameter is considered to be a possible reason for the difference 
in the measurement of $\sigtot$ at the Tevatron. 

 Closing this example of a recent fit, we recall the latest results from the TOTEM collaboration, namely $\sigtotpp(8\ TeV)=(101.7\pm 2.9)\ mb $ \cite{Antchev:2013paa} 
 and the preliminary result $\rho(8 \ TeV)=0.104\pm 0.027(stat)\pm 0.01{syst} $ presented at the 2014 Rencontre de physique de La Thuile. As of 2016,  both TOTEM and ATLAS collaboration have released data at $\sqrt{s}=8$ TeV, as we briefly discuss in Sec. \ref{sec:lhcnow}.

 \subsubsection{COMPAS group( IHEP, Protvino)}\label{sss:compas} 
The COMPAS group has presented (in a version of PDG 2012 \cite{Beringer:1900zz}, updated in the first half of 2013)
a phenomenological fit to all total hadronic cross sections and the ratio of the real-to-imaginary parts of the forward elastic 
scattering hadronic amplitudes. New data on total pp collision cross sections from CERN-LHC-TOTEM  and new data from 
cosmic rays experiment have been added. They note -in agreement with we what we also find and as we have discussed 
elsewhere in the present review- that, the models giving the best fit of accelerator data also reproduce the experimental cosmic 
ray nucleonÐnucleon data extracted from nucleon-air data with no need of any extra phenomenological corrections to the data.

COMPAS uses four terms in the total hadronic cross-section for hadron $a^{\pm}$ on hadron $b$:
\bea
\label{Com1}
\sigma_{tot}(a^{\pm} b) = H [\ln(\frac{s}{s^{ab}_M})]^2 + P^{ab}\nonumber\\ 
+ R^{ab}_1(\frac{s}{s^{ab}_M})^{-\eta_1}  \pm R^{ab}_2(\frac{s}{s^{ab}_M})^{-\eta_2}.
\eea
The adjustable parameters are defined as follows: 
\begin{itemize}
\item $H = \pi/M^2$ (in mb) is named after Heisenberg.
\item the scaling parameter $s^{ab}_M = (m_a + m_b + M)^2$.
\item A factorizable set $P^{ab}$ (in mb) stands for a constant Pomeron.
\item Two factorizable sets $R^{ab}_i$ (in mb), (for $i=1,2$)  stand for the two leading Regge-Gribov trajectories.
\item The data for purely hadronic reactions used were $\bar{p}$, $p$, $\pi^{\pm}$ and $K^{\pm}$ on $p, n, d$; $\Sigma^- $ on $p$. 
\item Also used, were fits to $\gamma p$, $\gamma d$ and $\gamma \gamma$.
\end{itemize}
The above parameterizations were used for simultaneous fits to the listed reactions with 35 adjustable parameters. To trace the variation 
of the range of applicability of simultaneous fit results, several fits were produced with lower energy $\sqrt{s}  \ge 5, \ge 6, \ge 7, . . .
{\rm GeV}$ cutoffs, until the uniformity of the fit across different collisions became acceptable with good statistical value. Downloadable
figures are available on the PDG site. Fig. ~\ref{fig:compas-total} reproduces their results for various total cross-sections.
\begin{figure}
\resizebox{0.5\textwidth}{!}{%
  \includegraphics{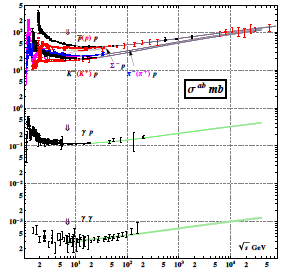}
}
\caption{Total 
cross-sections, as compiled by the Compas Collaboration for PDG \cite{Beringer:1900zz}. 
Reprinted with permission from  \cite{Beringer:1900zz}, Fig.(46.9), \copyright (2012) by the American Physical Society }
\label{fig:compas-total}       
\end{figure} 
\subsection{Asymptotic total cross sections in theories with extra dimensions
\label{ss:High_D}}
The search for asymptotia has been driving many  models, with the question asked as to whether present measurements of the 
total cross-section have reached a stable situation, where one cannot expect new phenomena to be detected in the energy behavior 
of $\sigma_{tot}$. In this subsection, this question will be addressed  by focusing on behavior reflecting extra-dimensions.
 
The rate at which cross sections grow with energy is sensitive to the presence of extra dimensions
in a rather model-independent fashion. In \cite{Swain:2011js}, one can find a review of how rates would be 
expected to grow if there are more spatial dimensions than 3 which appear at some energy scale, 
making connections with black hole physics and string theory. The salient point -as discussed for example in 
\cite{Chaichian:1987zt}- is that the generalization of  the Froissart-Martin bound for space-time dimensions 
$D > 4$ leads generically to a power law growth rather than the maximum square of logarithm growth 
with energy allowed in $D = 4$. 

\subsubsection{Asymptotic relation between cross-section and entropy \label{entropy}}
A clear physical argument for estimating the total cross section at a high energy $s = E^2$ was given by
Eden, a long time ago\cite{Eden}. It runs as follows: 
\par\noindent
(i) If the elastic scattering amplitude at high energy is dominated by the exchange of the lightest particle of mass
$\mu$, then the probability of the exchange at a space-like distance $r$ between the particles, reads
\be
\label{En1}
P(r, E) = e^{-2 \mu r + S(E)},
\ee
where the entropy S(E) (in units the Boltzmann constant $k_B$) determines the density of final states. 
\par\noindent
(ii) The probability becomes of order unity at a distance $R(E) = S(E)/(2\mu)$
\par\noindent

(iii) The total cross section is then given by 
\be
\label{En2}
\sigma_{tot}(E) = 2\pi R(E)^2 = \frac{\pi}{2\mu^2} S(E)^2
\ee
The asymptotic total cross section at large $E$  is thereby determined by the entropy S(E).

A typical entropy estimate may be made via the following reasoning: The equipartition theorem for a gas
of ultra-relativistic particles implies a mean particle energy ($\bar{\epsilon}$) varying linearly with temperature
$\bar{\epsilon} = 3 k_B T$. A Boltzmann gas of such particles has a constant heat capacity. A system with a 
constant heat capacity $C_\infty$ obeys
\be
\label{En3}
E = C_\infty T = C_\infty \frac{dE}{dS};\ \Rightarrow dS = C_\infty (\frac{dE}{E}),
\ee
leading to the following important logarithmic relationship between entropy and energy
\be
\label{En4}
S(E) = C_\infty ln(\frac{E}{E_o}) 
\ee
Hence, the total cross-section for a constant heat-capacity system saturates the Froissart-Martin bound:
\be
\label{En5}
\sigma_{tot}(E) = (\frac{\pi}{2}) (\frac{C_\infty}{\mu})^2 ln^2(\frac{E}{E_o}).
\ee
In a more general thermodynamically stable situation, the entropy S(E) is determined parametrically by the heat capacity
as a function of temperature\cite{Swain:2011js}:
\be
\label{En6}
S(T) = \int_o^\infty C(T') \frac{dT'}{T'}.
\ee
The saturation Eqs.(\ref{En3}) and (\ref{En5}) will then hold true only in the high energy and high temperature limit of a stable
heat capacity $C(T \to \infty) = C_\infty$.

To compute the total high energy cross section for models with extra dimensions, the central theoretical
problem is to understand the entropy implicit in such models. Below we list the entropies and total 
cross-sections for a Hagedorn string and for $n = (D-4)$ extra compact dimensions. 
\subsubsection{Entropies for higher dimensions and string theory \label{Dim_strings}}
A Hagedorn string entropy grows linearly with energy in the asymptotic limit\cite{Hagedorn:1965h1,Hagedorn:1968h2}
\be
\label{En7}
S(E) \to \frac{E}{T_H},
\ee
where $T_H$ is the Hagedorn temperature. The Hagedorn entropy for bosonic and fermionic strings have
a similar linear growth with energy but with different coefficients. Thus, for such theories the total cross-section
is expected to grow as $\sigma_{tot}(s) \sim s$.

On the other hand, above the threshold for the observation of $n = (D - 4)$ extra compact dimensions, 
the total cross-section would grow as\cite{Swain:2011js}
\be
\label{En8}
\sigma_{tot}(s) \sim [\frac{s}{s_o}]^{(n + 2)/2}
\ee
Thus, once even if one such threshold is crossed (that is for $n = 1$ or $D = 5$), $\sigma_{total}(s) \to (s/s_o)^{3/2}$.
It is fair to conclude from the recent LHC total cross-section data that no such extra dimension thresholds have opened up until 
$\sqrt{s} = 8\ TeV$, and notwithstanding large errors with the AUGER cosmic ray data, 
not even until $\sqrt{s} = 57\ TeV$. Such a result is in consonance with the fact that no evidence for beyond
the standard model physics such as that due to extra dimensions has been found in any data from LHC for 
$\sqrt{s} \leq 8\ TeV$. 

The arguments presented in \cite{Swain:2011js} have been accepted in \cite{Block:2012yx} with an aim to extend it
and the latter authors suggest that higher dimensions might be ruled out to arbitrarily high energies via the same arguments. 

\subsection{Concluding remarks}
In this section, we have attempted to give an overview of the existing models for the total cross-section, highlighting the 
chronological order of its long history. The total cross-section, as the imaginary part of the forward scattering amplitude 
describes the very large distance behavior of the interaction, but   understanding of the underlying strong interaction 
dynamics can only be completed by studying the amplitude for $-t\neq 0$.  This,   we shall approach in the next section 
dedicated to the elastic cross-section. 


\section{The Elastic-cross-section }
\label{sec:elasticdiff}

We shall now summarize the state of the art of the  differential elastic cross-section, discuss  some representative models
and present their predictions. We shall try to put in perspective  the phenomenological work developed over more than 50 years, 
up to the latest  measurements made at the LHC running at  $\sqrt{s}=7\ {\rm TeV}$ (LHC7)  and $8\ {\rm TeV}$ (LHC8) 
\cite{Antchev:2011zz,Antchev:2013paa,Aad:2014dca,Aaboud:2016ijx}.

In  Fig.~\ref{fig:totemdiffel-LHC} we reproduce the first  plot  of  the elastic differential 
cross-section measurement  by the  TOTEM  Collaboration at LHC \cite{Antchev:2011zz}. 
It was the first time since almost 40 years,  that the distinctive dip in the \pp \ differential cross-section had been seen again, 
shifting to the left by a factor 3, as the energy increased more than a hundred times.
\begin{figure}
\resizebox{0.5\textwidth}{!}{%
\includegraphics{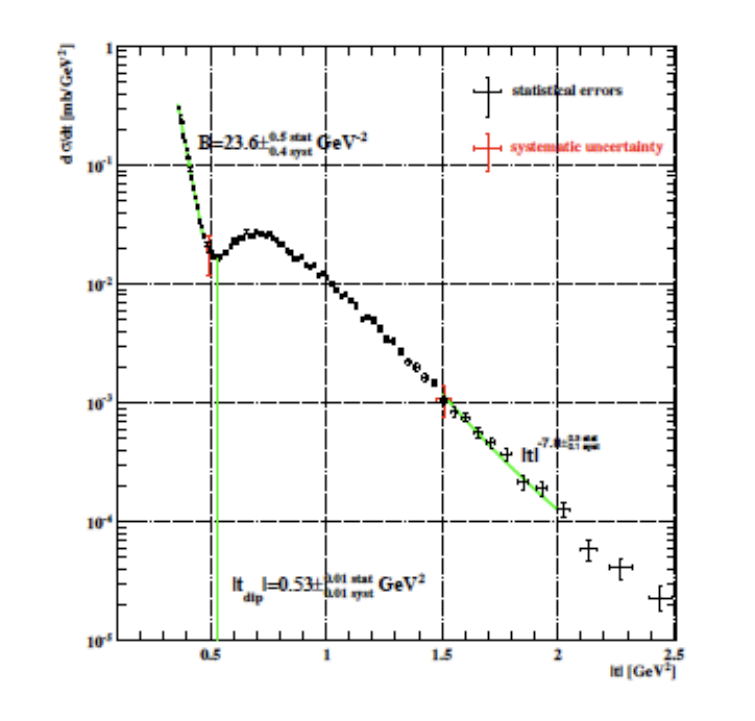}}
\caption{The first TOTEM measurement of the differential elastic cross-section from \cite{Antchev:2011zz}. 
Published
 by IOP under CC BY license.} 
\label{fig:totemdiffel-LHC}       
\end{figure}

As clearly shown in
 Fig.~\ref{fig:fromtotemdiffel-ISR}, 
 \begin{figure*}
\centering
\resizebox{0.8\textwidth}{!}{%
\includegraphics{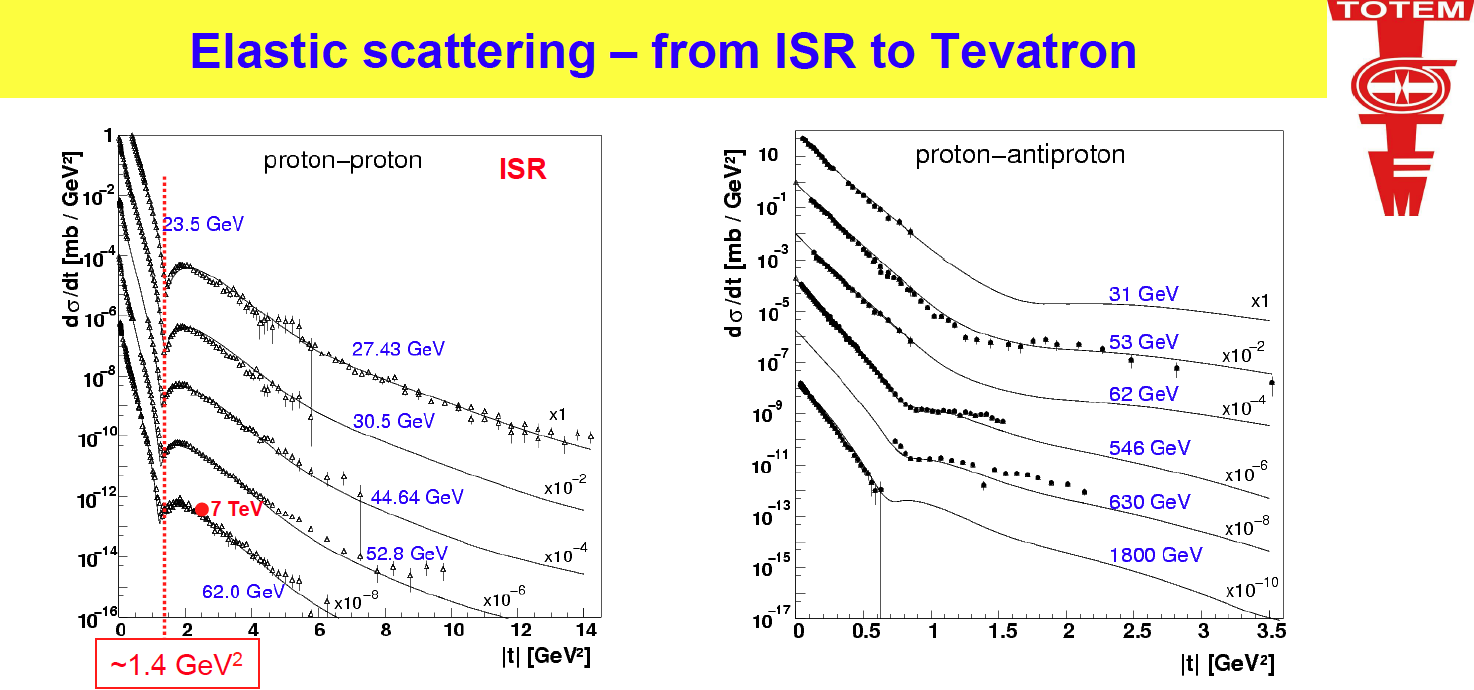}}  
\caption{ISR and Tevatron  elastic differential  cross-sections from  talk by K.Eggert,  at  \href{<https://indico.in2p3.fr/event/6004/session/7/contribution/116/material/slides/0.pdf>}{Hadron Collider Physics Symposium},  November 2011, Paris, France.  Reproduced  with permission.}
\label{fig:fromtotemdiffel-ISR}
\end{figure*} 
 elastic scattering is characterized by the following quantities:  the optical point, i.e. imaginary and real parts of the scattering amplitude at $t=0$;  
the precipitous decrease at small t, related to the slope $B(s,t=0)$; the change in slope  and occurrence of the dip where the imaginary part of 
the amplitude becomes smaller than the real part; the $|t|$ behaviour after the dip and connections with perturbative QCD. 
It is worthy to note  that  at ISR no dip is observed in \pbarp , but only a change in slope, whereas  the dip is 
quite pronounced in \pp, both at lower and higher energies.

The TOTEM experiment has  measured $ \sigtot$, $\siginel$, $ \sigel$, $\dsigdt$ 
at LHC energies $\sqrt{s}=7,\ 8\ {\rm TeV}$
and, from these, has given values for the slope 
parameter $B(s,t)$, at  different $-t$-values, the position of the dip and provided a functional form for the  behaviour of 
$\dsigdt$ after  the dip  \cite{Antchev:2013iaa,Antchev:2013haa,Antchev:2013gaa}.
 From these data, one can  extract the ratio \sigeltosigtot \ and check whether the asymptotic black disk limit has been reached.
At the time of this writing, data for the total and elastic cross-sections at $\sqrts = 8\ {\rm TeV}$ have been published from both the TOTEM \cite {Antchev:2013paa} and ATLAS \cite{Aaboud:2016ijx} Collaborations,  with     
new data  appearing  from the TOTEM Collaboration for   the differential cross-section \cite {Antchev:2015zza}.  In Table \ref{tab:totem} we present the available results from 
TOTEM for some of these quantities.
\begin{table*}
\caption{TOTEM results, with $\siginel=\sigtot - \sigel$,   for 7 TeV \cite{Antchev:2013iaa}  and for 8 TeV 
\cite{Antchev:1495764,Antchev:2013paa},  luminosity independent measurements.}
\label{tab:totem}
\centering
\begin{tabular}{|c|c|c|c|c|c|c|}
\hline
$\sqrts$&$\sigtot $           &B                       &$\sigel $            & $\siginel$      &$\reltot$                   &$\relinel$\\
TeV      &     mb                  & $ {\rm GeV}^{-2}$   & mb                    & mb                  &                                  &                  \\ \hline
7           &  $98.0\pm 2.5$& $19.9\pm 0.3$&$25.1\pm 1.1 $&$72.9\pm 1.5$&$0.257\pm 0.005$&$0.345\pm 0.009$\\
8            & $101.7\pm 2.9$&                     &$27.1\pm 1.4 $&$74.7\pm 1.7$&$0.266\pm 0.006$&$0.362\pm 0.011$ \\ 
\hline\end{tabular}\end{table*}
In addition, most recent data 
at $\sqrt{s}=8\ {\rm TeV}$ (LHC8), have shown that a pure exponential behavior for the slope in the region 
$0.027<-t<0.2\  {\rm GeV}^2$  can be excluded \cite{Antchev:2015zza} with significance greater than 7 standard deviations.


An exhaustive  discussion of all the  quantities defining the elastic differential cross-section can be found in \BC. 
Although the review of Ref. \BC \ pre-dates both the Tevatron and  LHC measurements, most  of its 
content and some of its conclusions are still very much valid. In the following we shall describe these different  quantities, 
and the asymptotic theorems which govern their energy dependence.
Tables and figures  for each of them, $\sigtot$, $B(s,t)$, $\rho(s,t)$, $t_{dip}$ are available at the  \href{http://pdg.lbl.gov/}{ Particle Data Group (PDG)} site,
here we shall reproduce them as encountered in describing models.

We shall define the following  measures of asymptotia: the total cross-section itself $\sigtot$, 
satisfaction of asymptotic sum rules for  the elastic scattering amplitude, the forward slope $B(s)$, the ratio  
\sigeltosigtot.
Then, we shall examine representative 
models for elastic scattering, starting from the simplest possible case, the black and gray disk model. 

Models for the elastic  and  total cross-sections, are based on  two major approaches: the Pomeron-Regge 
road and the  unitarity-Glauber formalism. The Pomeron-Regge theory expresses  the differential  elastic scattering 
amplitude in terms of power laws $s^{\alpha(t)}$, and it has provided more than 50 years of   good phenomenology  
for  both elastic and diffractive scattering. However,   this raises    problems with unitarity and the Froissart bound. 
As well known, the elastic scattering amplitude cannot be just described through a simple pole, since then the 
high energy behavior of the total cross-section  would violate  the Froissart bound; on the other hand, the diffraction 
peak is well represented by a Pomeron pole. Thus the problem is that at $t=0$, the elastic differential 
 cross-section is proportional to $\sigma_{total}^2$ (modulo a small contribution from the real part), hence 
 for  $t\sim 0$, the cross-section at the optical point can increase at most  as 
the fourth power of  logarithm, while, at the same time,  the differential elastic cross-section, as soon as $t\neq 0$ 
does indeed exhibit the exponential behavior  characteristic of the Pomeron pole contribution.

The Glauber-type description is unitary and it can easily embed the Froissart bound, as we have discussed, 
for instance,  in the context of the 
QCD inspired model of Block and collaborators or   our BN mini-jet model, both of them discussed in  Sec.\ref{sec:models}.
 However,  
a one 
channel
eikonal, for  both  elastic and inelastic scattering, fails   in its capacity to describe separately the three 
components of the scattering, elastic, inelastic and 
(single and double) diffractive, as measured up to present energies. This is immediately obvious if one considers 
the expression for the inelastic total cross-section obtained in the one channel eikonal:
\begin{align}
F(s,t)=i\int d^2 \vecb e^{i\vecq \cdot \vecb} [1-e^{i\chi(b,s)}]\\
\sigma_{total}=2\int d^2 \vecb [1-\cos \Reo \chi(b,s) e^{-\Imo \chi(b,s)}]\\
\sigma_{elastic}=\int d^2 \vecb |[1-e^{i\chi(b,s)}]|^2\\
\sigma_{inel}=\sigma_{total}-\sigma_{elastic}=\int d^2 \vecb [1-e^{-2\Imo \chi(b,s)}]\label{eq:siginel}
\end{align}
where $t=-q^2$. 
It must be noted that Eq.~(\ref{eq:siginel}) can be obtained by  summing all possible inelastic collisions 
independently distributed in $\vecb$-space. Assuming in fact that for  every  impact parameter value the 
number of possible collisions  $n(b,s)$ is Poisson distributed around a mean number of collisions $\bar{n} (b,s)$, i.e.
\be
P(\{n, \bar{n}(b,s)\})= \frac{e^{- \bar{n}(b,s)}}{n!} \bar{n}(b,s)^{n}
\ee
it immediately follows that a sum on all possible collisions together with integration on all values of the impact parameter, leads to
\be
\sigma_{independent \ collisions}=\int d^2 \vecb [1-e^{- \bar{n}(b,s)}]\label{eq:sigpoisson}
\ee
Comparing Eqs.(\ref{eq:siginel}) and (\ref{eq:sigpoisson}),  shows that one can obtain $\Imo \chi(b,s) $ from $\bar{n}(b,s)$, 
but also that Eq.~(\ref{eq:siginel}) for the inelastic cross-section only includes independent collisions. Since diffractive processes, 
single, double, central, do exhibit correlations, these processes need to be discussed with a formalism beyond the one-channel eikonal. 

Thus the question arises in eikonal models as to how to include correlated inelastic processes, which  are identified through 
particular final state configurations.   
Among the  models which embed some of these properties, are  those due to Khoze, Martin and Ryskin (KMR), 
Gotsman, Levin and Maor (GLM), Ostapchekp, Lipari and Lusignoli, to be seen 
later in this section.  

The way we choose to present this part of the review is to start   with general definitions and properties of quantities 
defining the scattering and some comparison with  data from LHC at $\sqrt{s}=7\ {\rm TeV}$ (LHC7). Then, models for the 
elastic differential cross-section from the optical point to past the dip will be presented, both in their historical development 
and in their contribution to describe TOTEM data. We shall conclude this section on elastic scattering with a short review 
of models specifically addressing  diffraction 
and a discussion of the inelastic part of the total cross-section.

Each of the items above, is discussed as follows:
\begin{itemize}
\item general features of the elastic cross-section are discussed in \ref{ss:general-el} with
\begin{itemize}
\item the slope parameter in \ref{sss:slope},
\item the real part of the scattering amplitude in \ref{sss:rho} and \ref{sss:real},
\item sum rules for real and imaginary parts of the scattering amplitude at b=0, in \ref{sss:sumrules},
\item asymptotia and the ratio $\sigel/\sigtot$ in \ref{sss:rel},
\item the dip structure and geometrical scaling in \ref{sss:dip} and \ref{sss:GS},
\end{itemize}
\item early models in impact parameter space and their updates are to be found in \ref{ss:early-models} with 
\begin{itemize}
\item the Glauber picture applied to \pp \ scattering in \ref{sss:chouyang} and \ref{sss:durandlipes}, 
\item the black disk picture in \ref{sss:blackdisk},
\end{itemize}
\item models with Regge and Pomeron exchanges are presented in \ref{ss:reggepomeron} with  
\begin{itemize}
\item an early model by Phillips and Barger (PB) and its updates in \ref{sss:PB}, 
\item a model by Donnachie and Landshoff in 
\ref{sss:DLdsigdt} 
\item  a model in which the slope parameter increases with energy at the same rate as   
the total cross-section in \ref{sss:schegelskyryskin},\end{itemize}
\item models including an Odderon exchange can be found in \ref{ss:pomoddregge},
\item eikonal models are discussed in \ref{ss:eikonaldsigdt}
\item selected models including diffraction are to be found in \ref{ss:Diffel} with a comment on single-channel mini jet models with soft gluon radiation in \ref{sss:diffandsoftguons},
\item  a parametrization of  diffraction to obtain the total, elastic and inelastic cross-section  from one-channel eikonal models is presented in \ref{ss:Onechannel}.
\end{itemize}
\subsection{General features of the elastic cross-section}\label{ss:general-el}
As stated earlier, elastic scattering is characterized by the following quantities: 
\begin{itemize}
\item  the optical point, i.e. the  imaginary and real parts  of the amplitude at $t=0$,  
\item the precipitous decrease at small t, related to the forward  slope $B(s,t=0)$,
\item  the change in slope  and occurrence of the dip where the imaginary part of 
the amplitude becomes smaller than the real part, 
\item  the $|t|$ behaviour after the dip and connections with perturbative QCD. 
\end{itemize}
 In the following we shall describe these 
 different 
 quantities, and the asymptotic theorems which govern their energy dependence.

\subsubsection{About the slope parameter}\label{sss:slope}
An earlier rather complete discussion of this issue can be found in \BC. Here we discuss its definition 
and show the present experimental status. 

Although most models do not attribute a single exponential behaviour, and hence a single value for the slope, to the small-$t$ 
behaviour of the elastic differential cross-section, experimentalists usually describe the diffraction peak with a single  slope and a single term, i.e.
\be
\dsigdt(t\simeq 0)=\dsigdtzero\  e^{B(s)t}
\label{eq:slope}
\ee
This expression leads to the approximate result
\be
B(s)\simeq \frac{1}{16 \pi} \frac{\sigtot^2 (1+ \rho^2)}{\sigel} 
\ee
We plot in Fig.~\ref{fig:slope}, the values for $B(s)$, reported by experiments from ISR to LHC7, using  Eq.~(\ref{eq:slope}) definition.  For the TOTEM experiment, 
the value for $B(s)$ at LHC7 corresponds to the measurement in the interval $0.02<|t|<0.33\  {\rm GeV}^{-2}$ \cite{Antchev:2011vs}. 
\begin{figure}
\centering
\resizebox{0.5\textwidth}{!}{
\includegraphics{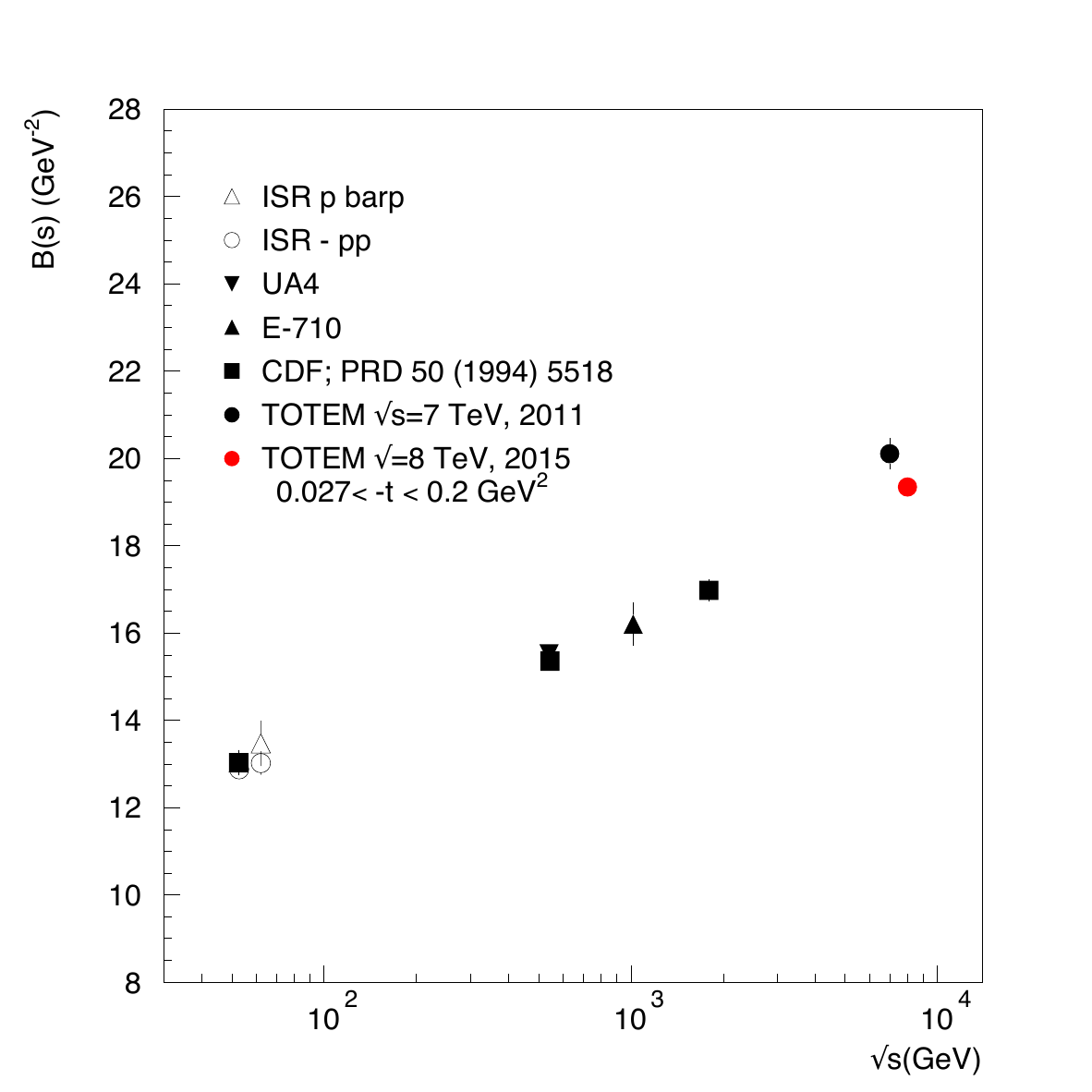}}
\caption{The energy dependence of the slope parameter, Eq.~(\ref{eq:slope}), from ISR to LHC7. Data are from \cite{pdg} for lower energies, 
from \cite{Antchev:2011vs} for LHC7.  We also include the TOTEM result at $\sqrt{s}=8\ {\rm TeV}$ from \cite{Antchev:2015zza}, 
for the case of a single exponential fit in   the $0.027\ {\rm GeV}^2<-t<0.2 \ {\rm GeV}^2$ region.  }
\label{fig:slope}
\end{figure}
There are  general considerations   which relate the asymptotic behaviour of $B(s)$ to that of the total cross-section, in particular one can derive 
the asymptotic relation, so called MacDowell  and Martin bound \cite{MacDowell:1964zz}, 
discussed recently in \cite{Fagundes:2011hv}, given as 
\be
B(s)\ge  \frac{ \sigma_{total}}{18 \pi} \frac{\sigtot}{\sigel}
\ee
Since the ratio $\sigtot/\sigel \geq 1$, 
the above relation implies that the rise with energy of $B(s)$ will at some point catch up with that of the total cross-section. 
We mention here in passing that at LHC7, the right hand side of the above inequality is approximately $18 \ {\rm GeV}^{-2}$ and 
the measured slope on the left hand side is about $20\  {\rm  GeV}^{-2}$.
Hence, the inequality is near to saturation.
 \par\noindent
 To reiterate, if and when the total cross-section will have  reached  an energy such as to saturate the Froissart bound, 
 then  one should expect $B(s)$ to grow with energy as $(ln s)^2$.   Fig. ~\ref{fig:slope} indicates that up to the Tevatron 
 measurements,  data could to be consistent with a $\log s $  type behaviour. 
 After the LHC7 TOTEM data appeared, the possibility of a stronger rise was examined in \cite{Schegelsky:2011aa}. 
 However, the LHC8 result (red dot in Fig. ~\ref{fig:slope}) sheds doubts on the single exponential slope analyses, 
 and it would need to be  rediscussed  when higher LHC data, at 13 and 14 TeV, will be available. 
  
One way to describe the variation in $t$ as one moves away from $t\simeq 0$
has  been  to introduce the curvature parameter $C(s)$ and   parametrize the diffraction peak as 
\be
\dsigdt=\dsigdtzero \ e^{B(s)t +C(s)t^2}
\ee
Such a parametrization  needs a change in sign for $C(s)$ as $t$ moves away from the very forward direction. 
Higher powers, such as a cubic term $t^3$ are also discussed in the recent 2015 TOTEM analysis for the slope 
\cite{Antchev:2015zza}.  More generally, 
away from $t\simeq 0$,  the slope parameter is  a function of both $t$ and $s$, defined as
\be
B_{eff}(s,t)=\frac{d}{dt}\log \dsigdt
\label{eq:bst}
\ee
Since models differ in their parametrization of the forward peak, depending on the extension in $t$, if the dip region has to be included, 
Eq. ~(\ref{eq:bst}) is to be used. One can distinguish the following basic modelings for the forward peak
\begin{itemize}
\item impact parameter models
\item one or more Pomeron pole exchanges
\item di-pole and tri-pole exchanges
\item Pomeron exchanges unitarized via eikonal representation
\item soft gluon resummation and exponential damping (work in progress). 
\end{itemize}
We see in what follows the results from some of these models.

\subsubsection{The real part of the elastic scattering amplitude, at $t=0$,  and the energy dependence of the $\rho(s)$ parameter
}\label{sss:rho}
The elastic scattering amplitude has both a real and an imaginary part. At $t=0$,  the imaginary part is proportional to  the
total cross-section, but there is no such simple way to obtain the real part, 
{ although arguments, based on asymptotic theorems, have been used to extract an asymptotic value for
 $\rho(s)=\Re e F(s,t=0)/\Im m F(s,t=0) \rightarrow \pi/\ln s$ as $s \rightarrow \infty$.} 

In the next paragraph we shall describe how 
one can construct asymptotically a real part for  values of  $t\neq 0$. In this paragraph, we show an 
analysis of 
various high energy data, by Alkin, Cudell and Martynov \cite{Alkin:2011if}, aimed at determining  the parameter 
$\rho(s)=\Re e F(s,t=0)/\Im m F(s,t=0)$ through integral dispersion relations. The authors use various Pomeron 
and Odderon models, employing simple Pomeron or double and triple poles.  Recall that these three different 
cases correspond to an asymptotic behaviour for the total cross-section given by  a power-law, a logarithmic rise, 
or a $(\log s )^2$, while the Odderon term is  always rising less than the Pomeron. Ref.  \cite{Alkin:2011if} 
contains a rather clear description of the phenomenology implied by these different models. 
We show in Fig.~\ref{fig:rho2011} from \cite{Alkin:2011if} a  compilation of data for the $\rho$ parameter for $pp$ scattering,  
 compared with four different models. 
\begin{figure}
\centering
\resizebox{0.5\textwidth}{!}{
\includegraphics{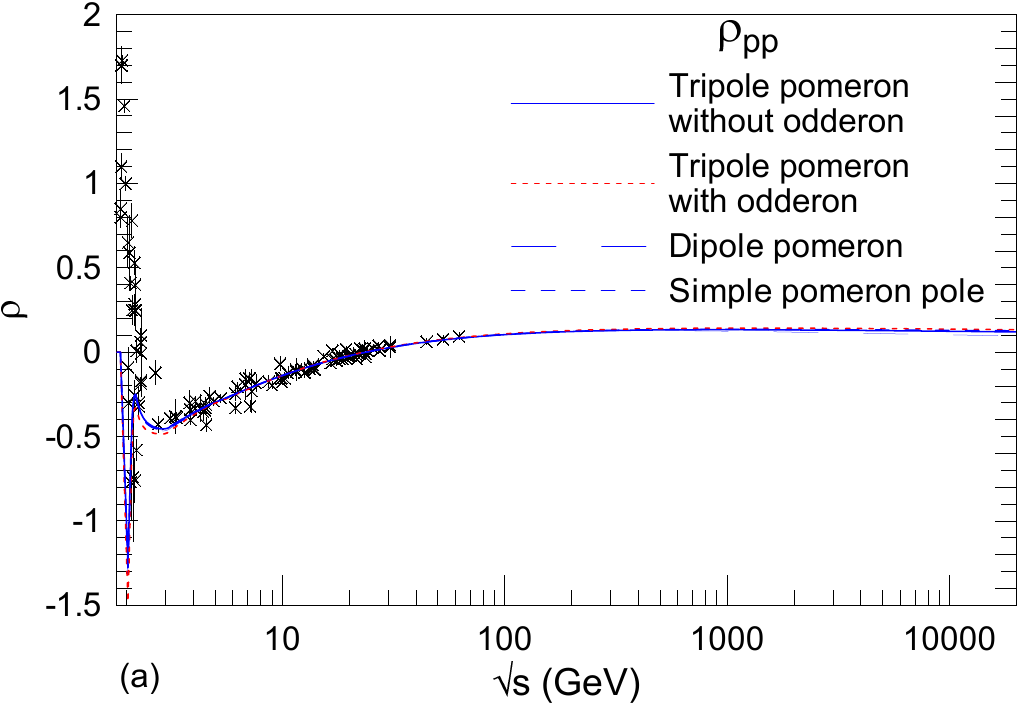}}
\caption{Data for the $\rho$ parameter for $pp$ scattering compared with different models as indicated in Fig.~(10) of  \cite{Alkin:2011if} \copyright (2011) by Springer. 
Reproduced with  permission of Springer. Figure is courtesy of J-R Cudell.}
\label{fig:rho2011}
\end{figure}

If one uses the LHC7 TOTEM result to gauge which model gives the best value for $\rho$, then  the result of this analysis 
seems to select the simple Pomeron model, since this is the model which gives a total cross-section closest to the TOTEM 
measurement, i.e. $\sigtot (7\ TeV)= (94.9 \div 96.4)\ {\rm mb} $ vs. $\sigtot^{TOTEM}=98.3 \pm 0.2^{stat}\pm 2.8^{syst}\ mb$. 
In such case,  $\rho (7\ TeV)= (0.138 \div 0.186)$. The triple-pole  Pomeron model also gives an acceptable total cross-section,
 with $\sigtot (7\ TeV)= (94.1 \div 95.1)\ mb $, corresponding to   $\rho(7\ TeV)= (0.130 \div 0.142)$. We note here that the  
 determination of  the total cross-section initially released,  used the predictions from \cite{Cudell:2002xe} with 
 $\rho(7\ TeV)= 0.138^{+0.01}_{-0.08}$, 
 while the most recent TOTEM analysis, as of this writing, \cite{Antchev:2015zza}, uses a value $\rho(8 \ TeV)=0.140\pm0.007$, using the COMPETE collaboration favoured value \cite{Cudell:2002xe}.
 
Using   the  model by Block and collaborators \cite{Block:1998hu}, we reproduce data and predictions for $\rho$ for both \pp\ and \pbarp\ in Fig.~\ref{fig:rhoaspen}.
Just as in the previous figure, this figure shows that, at high energy, $\rho$ has been measured to be positive, swinging from the negative values  
at energies below the ISR,  to values $\rho\sim 0.12$ at the Tevatron.

{Following \cite{Auberson:1971ru}, the behavior of $\rho(s)$ can be seen to arise rather 
naturally from present phenomenological analyses of the total cross-section alone. 
Consider  the asymptotic terms of a  frequently used parametrization for the even amplitude at $t=0$, }
\bea
\Im m F_+(s,0)&=&\frac{1}{4\pi}{\big [ }  H_1(\ln \frac{s}{s_0})^2  + \nonumber \\
&+&H_2 \ln \frac{s}{s_0}+(\frac{\pi^2}{4}H_1+H_3){\big ]}\\
\Re e  F_+(s,0&=&\frac{1}{4}{\big [ }  H_1\ln \frac{s}{s_0}+\frac{1}{2}H_2 {\big ]}
\eea
 While the coefficient $H_1$ is obviously positive, as it is related to the asymptotic 
 behavior of the obviously positive total cross-section, the sign of $H_2$ is {\it a priori } 
 undefined. Fits to $\sigtotpp$ give $ H_2\le 0$, even when the $\rho$ paramerer is not part of the fits. 
 This immediately gives that  $\rho(s)$ can  go through zero, as it does for 
 $10 \ {\rm GeV}\lesssim  \sqrts \lesssim  30\ {\rm GeV}$. Asymptotically then one has
  \bea
  \rho_+(s)
  &=&\frac{
  \pi
   {\big [ } 
   H_1 \ln  \frac{s}{s_0}+\frac{1}{2} H_2 
   {\big ]}
   }
  {
  {\big [ } 
   H_1(\ln \frac{s}{s_0})^2  +H_2 \ln \frac{s}{s_0}+(\frac{\pi^2}{4}H_1+H_3){\big ]}}\ \ \ \ \ \\
&  =&
\frac{\pi}{(\ln \frac{s}{s_0})}[1+\  non-leading\  in \ \ln s/s_0]
 \eea

Thus, we see that analyticity 
coupled with the assumption of Froissart limit saturation 
predicts that, at sufficiently high energy, $\rho\sim \pi/\log{s}$. While $\rho$ rises at  low energy,  
data from  the $Sp{\bar p}S$ and the Tevatron  region are non-committal and consistent with a constant $\sim 0.12$. 
 In our proposed Eikonal Minijet with soft gluon resummation, the imaginary part of the forward scattering amplitude is 
 proportional to $[\log{s}]^{1/p}$, with $1/2<p<1$. Through the substitution     $s\rightarrow se^{-i\pi/2}$, the expression for the scattering 
amplitude is made analytic and one can see that $\rho\sim \pi/2p \log{s}$.

The behavior of  $\rho$ as the energy 
swung through the ISR region, and $\sigtotpp$ 
began rising, 
became an object of intense scrutiny in the early '70's and led to the first suggestion of the  existence of the Odderon, 
[discussed later] in  \ref{sss:odderon}. 
In \cite{Lukaszuk:1973nt}, the 
observed rise of the total cross-section 
at 
ISR was  the occasion to  
 propose  that both the imaginary {\it and} the real parts of the amplitude could  behave asymptotically  
 as $(\ln s/s_0)^2$ and  fits to both $\sigtotpp$ and $\rho(s)$ in the then available energy range were 
 seen to be compatible with $\rho(s)$ passing through a zero in the ISR region.  We shall mention later, in 
  \ref{sss:odderon} how this  proposal was then applied to study charge exchange 
 reactions and then morphed in an added term named the Odderon. 
 One notices the obvious fact that $\rho(s)$ changes sign in the energy region where the total cross-section 
 changes curvature. The change in curvature is attributed in mini-jet models to the fact that perturbative  QCD 
 processes become observable, which is also the region where an edge-like behavior has been noticed by Block 
 and collaborators in the scattering amplitude \cite{Block:2015sea}.

\subsubsection{The asymptotic behaviour of the real part of the scattering amplitude at $t\neq 0$ \label{sss:real}}
We shall discuss here how one can  construct an asymptotic $ \re F(s,t)$ given the imaginary part $\im F(s,t) $.
{A discussion of some of  the issues presented here can be found in a 1997 paper by Andre' Martin \cite{Martin:1997vy}. }

The construction of the elastic scattering amplitude at asymptotic energies uses a number of asymptotic theorems. The imaginary part at $t=0$ 
is anchored to the optical theorem and its asymptotic value is bound by the Froissart theorem. The real part at $t=0$ is asymptotically obtained 
through the Khuri-Kinoshita theorem 
\cite{Khuri:1965zz}. At $t\neq 0$, models for the imaginary part also allow to obtain an asymptotic value for the real part. 
This is discussed in an early paper by Martin \cite{Martin:1973qm}.
According to Martin, if the total cross-section behaves asymptotically as $\log ^2 s$, then the real part of the even amplitude 
$F_+(s,t)$, again asymptotically, behaves as
\begin{equation}
\Re e F_+(s,t)\simeq \rho(s) \frac{d}{dt} [t \Im m F_+(s,t)]
\label{eq:real}
\end{equation}
where, as usual,  $\rho(s)=\Re eF_+(s,0)/\Im m F_+(s,0)$. The result of Eq.~(\ref{eq:real}) is obviously consistent with 
the expression for the differential cross-section at $t=0$, namely one has
\begin{align}\
(\frac{d\sigma}{dt})_{t=0}\sim (constant)\{ 
(\Im m F(s,0))^2+(\frac{d}{dt} t\Im mF(s,t))_{t=0}^2
\}\\
=(constant) (\Im m F(s,0))^2[1+\rho(s)^2]
\end{align}
It is also interesting to note that the asymptotic expression given by Martin, 
automatically satisfies one of the two asymptotic sum rules for the elastic amplitude 
in impact parameter space, which will be discussed later,  in \ref{sss:sumrules}.

Eq.~(\ref{eq:real}) was derived by Martin in the asymptotic regime  $\sigtot\sim \log^2s$, but it is actually more general and holds also for 
$\sigtot\sim (\log s)^{1/p}$ with $1/2<p<1$. Let us start with the case  $p=1/2$. Defining the even amplitude
\begin{align}
F_+(s,t)=F_+(s,0) f(t)\\
f(0)=1
\end{align}
and assuming the asymptotic behaviour
\be
F_+(s,0)\sim i\beta (\log s/s_0)^2,
\ee
the real part is built using the   amplitude properties of  analyticity and crossing symmetry. Using the 
 additional property of  geometrical scaling obeyed by the asymptotic amplitude  \cite{Auberson:1971ru}, the argument then runs as follows:
\begin{itemize}
\item 
introducing the scaling variable $\tau=t \log^2 s$, geometrical scaling  \cite{Auberson:1971ru} says that 
\be
f(s,t)\equiv f(s,\tau)=f(t (\log s/s_0)^2)
\ee
\item
 for small values of $t$, the even amplitude must be crossing symmetric, i.e. symmetric under the exchange $s\rightarrow se^{-i\pi/2}$, 
and  the Froissart limit  and the geometric scaling variable turn into
\begin{align}
F_+(s,0)\rightarrow i\beta(\log s/s_0-i\pi/2)^2\\
\simeq i\beta (\log s/s_0)^2 +\beta \pi \log s/s_0 \nonumber \\
\tau \rightarrow t (\log s/s_0-i\pi/2)^2\simeq t (\log s/s_0)^2-i\pi t \log s/s_0 \nonumber\\
\rightarrow \tau(1-\frac{i\pi}{\log s/s_0}) \\
f(\tau-i\pi \frac{\tau}{\log s/s_0})\simeq f(\tau) -i\pi  \frac{\tau}{\log s/s_0} \frac{df}{d\tau} \nonumber\\
\simeq  f(\tau) -i\pi  \frac{t}{\log s/s_0} \frac{df}{dt}
\end{align}
so that 
\begin{align}
F_+(s,t)=F_+(s,0) f(\tau)\simeq \nonumber \\
 [i\beta (\log s/s_0)^2 +\beta \pi \log s/s_0][ f(\tau) -i\pi  \frac{t}{\log s/s_0} \frac{df}{dt}] \nonumber\\
\simeq i [\beta (\log s/s_0)^2  f(\tau)-\beta \pi ^2\frac{tdf}{dt}]\nonumber \\
+f(\tau)\beta \pi \log s/s_0+\beta \log s/s_0 \pi t   \frac{df}{dt}\nonumber\\
\simeq i \beta (\log s/s_0)^2  f(\tau) +  \beta \pi \log s/s_0   [f(\tau)+\tau  \frac{d f(\tau)}{d\tau}]
\nonumber \\
\simeq  i\beta (\log s/s_0)^2  f(\tau)
+\frac{\pi}{\log s/s_0}\beta( \log s/s_0)^2\frac{d( t  f(t))}{dt} 
\end{align}
Since asymptotically, the Khuri-Kinoshita theorem says that 
\be
\rho(s)\simeq \frac{\pi}{\log s/s_0}
\ee
we thus have that if
\begin{equation}
\Im m F_+(s,t)\simeq \beta (\log s/s_0)^2  f(t)
\end{equation}
then
\begin{equation}
\Re e  F_+(s,t)\simeq\rho(s)\frac{d}{dt } ( t \Im m F_+(s,t))\nonumber
\end{equation}
\end{itemize}
The demonstration leading to Eq.~(\ref{eq:real}) holds even if the total cross-section 
does not saturate the Froissart bound, namely we can also start with
\be
\Im m F_+(s,0)\simeq \beta (\log s/s_0)^{1/p}  
\ee
with $1/2\le p  \le 1$ and still obtain Eq.~(\ref{eq:real}), with
\be
\rho(s)\simeq \frac{\pi}{2p \log s/s_0}
\label{eq:rhop}
\ee
The limits on $p$ are obtained here from the phenomenological requirement that 
the total cross-section is asymptotically rising at least like a logarithm, 
i.e. the case $p=1$, and that it satisfies the Froissart bound, corresponding to $p=1/2$. 
We have seen, when discussing our mini-jet model in \ref{sss-models:froissart}, how to relate these requirements  
to a phenomenological description of confinement in the infrared region. 

For $p\neq 1$,  we shall now sketch the demonstration, which runs very close to the one just given for $p=1/2$.

Let the  asymptotic behaviour of the even amplitude at $t=0$ be such that     $F_+(s,0)\sim \beta (\log s/s_0)^{1/p}$. Then
\begin{align}
F_+(s,0)\sim i\beta(\log s/s_0 -i\pi/2)^{1/p}\\
\simeq i\beta (\log s/s_0)^{1p}[ 1-\frac{i \pi  }{2p \log s/s_0}]
\end{align}
Now the scaling variable $\tau\sim t F_+(s,0)=t (\log s/s_0)^{1/p}$, and the scaling in the variable $\tau $ gives
\be
f(\tau)\rightarrow f(\tau-\frac{i\pi \tau}{2p \log s/s_0})\simeq f(\tau) -\frac{i\pi \tau}{2p\log s/s_0}(\frac{d f}{d\tau})
\ee
and
\be
F_+(s,t)\simeq F_+(s,0) f(\tau)
\ee
Following steps similar to the $p = 1/2$ case and using 
$\rho(s)$ given by Eq.~(\ref{eq:rhop}) gives the same result as before, i.e.
\be
\Re eF_+(s,t)\simeq \rho(s)\frac{d}{dt}(t \Im m F_+(s,t))
\label{eq:realderivative}
\ee
This expression can be used to obtain a real part in eikonal models with a purely real eikonal function. 
One would obtain, for the full amplitude at $t=-q^2$,
\begin{align}
A(s,q) = i \int bdb (1-e^{-\chi(b,s)})J_0(bq)\nonumber\\
+ \int bdb
(1-e^{-\chi(b,s)})[J_0(bq)\rho(s))-\rho(s)\frac{qb}{2}J_1(qb)]
\end{align}
which leads to
\be
\dsigdt= \pi\{I_0^2+\rho^2[I_0-\frac{\sqrt{-t}}{2}I_1]^2\}
\ee
with
\bea
I_0=\int bdb (1-e^{-\chi(b,s)})J_0(qb)\\
I_1=\int b^2 db (1-e^{-\chi(b,s)})J_1(qb)
\eea
Before leaving this discussion of the real part of the scattering amplitude, we notice that  
the above is valid for the dominant high energy part of the amplitude. Real terms can be 
present at non leading order in the amplitude,  such as the one proposed by Donnachie 
and Landshoff, arising from a three gluon exchange \cite{Donnachie:1979yu}, and described 
later in this section.

\subsubsection{Asymptotic sum rules for the elastic scattering amplitude at impact parameter $b=0$ \label{sss:sumrules}}
Here we shall derive  two asymptotic sum rules which are integrals over momentum transfer for the real and the imaginary parts of the
elastic amplitude \cite{Pancheri:2005jr,Pancheri:2004xc}.

At high energies, ignoring all particle masses, let the complex elastic amplitude $F(s, t)$ be normalized so that
\be
\label{ES1}
\sigma_{tot}(s) = 4 \pi \Im m F(s, 0);\ \frac{d\sigma}{dt} = \pi |F(s, t|^2.
\ee
With this normalization, the elastic amplitude in terms of the complex phase shift $\delta(s, b)$ reads
\be
\label{ES2}
F(s, t) = i \int (bdb) J_o(b\sqrt{-t}) [1 - e^{2i\delta_R(s,b)} e^{-2 \delta_I(s, b)}],
\ee
and its inverse
\be
\label{ES3}
[1 - e^{2i\delta_R(s,b)} e^{-2 \delta_I(s, b)}] = -i \frac{1}{2} \int_{-\infty}^0 dt J_o(b \sqrt{-t})  F(s, t).
\ee

 Rewriting Eq.~\ref{ES3} as
\begin{align}
1 - [\cos(2\delta_R(s,b))+i \sin(2\delta_R(s,b))] e^{-2 \delta_I(s, b)}] =\nonumber\\
 -i \frac{1}{2} \int_{-\infty}^0 dt J_o(b \sqrt{-t}) [\Re e F(s, t)+i \Im m F(s,t)],
  \end{align}
 we have 
 \begin{align}
1 - \cos(2\delta_R(s,b))e^{-2 \delta_I(s, b)}=\frac{1}{2} \int_{-\infty}^0 dt J_o(b \sqrt{-t}) \Im m F(s,t)]\\
\sin(2\delta_R(s,b)) e^{-2 \delta_I(s, b)} = \frac{1}{2} \int_{-\infty}^0 dt J_o(b \sqrt{-t}) \Re e F(s, t)
 \end{align}
Consider the hypothesis of total absorption. 
 This is a stronger hypothesis than the one which leads to the Froissart-Martin bound, namely that there must exist a finite angular momentum value, 
 {\it below} which all partial waves must be absorbed. Under the stronger hypothesis 
 that  in the ultra high energy limit, namely in the central 
region ($b = 0$) a complete absorption occurs, 
and in the black disk limit of $\delta_I(s,0)\to \infty$, 
we 
have the following two asymptotic sum rules
\be
\label{ES4}
S_I = \frac{1}{2} \int_{-\infty}^o dt\  \Im m F(s, t)\ \to\ 1\ {\rm as}\ s\to\infty,
\ee
\be
\label{ES5}
S_R= \frac{1}{2} \int_{-\infty}^o dt\  \Re e F(s, t)\ \to\ 0\ {\rm as}\ s\to\infty.
\ee
Satisfaction of these sum rules is a good measure to gauge whether asymptotia and saturation 
of the Froissart-Martin (FM) bound have been reached.
Notice that the FM bound is obtained under a weaker hypothesis than complete absorption 
and one has $S_I \to 2$ and $S_R \to 0$. Our
phenomenological analysis of TOTEM, presented in \ref{sss:PB}, leads to Eq.(\ref{ES4}) thereby reducing the FM bound by a 
factor $2$. According to the  phenomenology presented in \cite{Grau:2012wy}, at TOTEM(7TeV), 
$S_I \approx\ 0.94$ and $S_R\approx\ 0.05$ bolstering our faith in the sum rules.

If $\Re e F(s,t)$ is constructed through the Martin recipe Eq.(\ref{eq:realderivative}), 
then the second sum rule Eq.(\ref{ES5}) is automatically satisfied.

\subsubsection{Elastic vs. total \x : the ratio
 and
 the unitarity limit \label{sss:rel}}

The ratio of the elastic to the total cross-section plays an important role in all discussions about asymptotic behaviour. We shall start 
our analysis by recalling some general characteristics of this ratio from considerations about total absorption.

Let us write  the expression for the total and the elastic cross-sections  in term of real and imaginary parts of the complex phase shift 
$\delta(s)$ i.e.
\bea 
f_{el}(q)=i\int d^2 \vecb e^{i\vecq\cdot\vecb}[
1-e^{2i \delta(b,s)}
]\\
=2\pi i \int b db J_0(qb)[1-e^{2 i \delta_R(s)-2\delta_I}]\\
\sigmaT=2 \Re e \int d^2\vecb [1-e^{2 i \delta_R-2\delta_I}]\\
\sigELA=\int d^2\vecb |1-e^{2 i \delta_R-2\delta_I}|^2
\eea
Then  examine two limiting cases
\begin{itemize}
\item 
elastic scattering only, $\delta_I=0$, i.e. 
\end{itemize}
\begin{align}
\sigmaT^{(1)}=2\int d^2\vecb [1-\cos 2 \delta_R]\\
\sigELA^{
(1)
}=\int d^2
\vecb |1-e^{i 2  \delta_R}|^2=2\int d^2\vecb [1-\cos 2\delta_R]
\end{align}
Thus $\sigELA\equiv \sigmaT$, and all the scattering is purely elastic.
\begin{itemize}
\item
 a different  limit, $\delta_R=0$, i.e. 
\end{itemize} 
\bea 
\sigmaT^{(2)}=2\int d^2\vecb [1 -e^{-2  \delta_I}]\\
\sigELA^{(2)}=
\int d^2 \vecb 
 (1-e^{ -2  \delta_I} )^2\\
\sigmaT^{(2)}-2\sigELA^{(2)}=2\int d^2\vecb e^{-2  \delta_I}[1-e^{-2  \delta_I}]\ge 0
\eea
i.e. \sigeltosigtot $ \le 1/2$ for $\delta_I>0$. 

 Other important  limits are  examined in  detail in the Block and Cahn 1984 review \BC, and 
 we reproduce here parts of their discussion on the black and grey disk limit.

Using  Block and Cahn convention, with $t=-q^2$, in the c.m. given by $t=-2k^2(1-cos\theta^*)$, one has
\bea
f_{cm}(t)=\frac{k}{\pi}\int d^2\vecb \eiqb a(b,s)\label{eq:BC1}\\
a(b,s)=\frac{1}{4\pi k}\int d^2\vec q \emiqb  f_{cm}(q)\label{eq:BC2}\\
\sigtot=\frac{4 \pi}{k}\Imo f_{cm}(t=0)\label{eq:BC3}\\
\frac{d\sigma_{el}}{dt}=\frac{\pi}{k^2}\frac{d\sigma_{el}}{d\Omega^*}=\frac{\pi}{k^2}|f_{cm}(t)|^2\label{eq:BC4}
\eea
Writing the amplitude $a(b,s)$  as
\be
a(b,s)=\frac{i}{2}[1-e^{2i\chi(b,s)}]\label{eq:BC5}
\ee
one  then  has
\bea
\sigel=4 \int d^2\vecb \ |a(b,s)|^2\\  
\sigtot=4\int d^2\vecb \Imo \ a(b,s)
\eea
The black and grey disk model case corresponds to  a  scattering amplitude  zero outside a finite region in impact parameter space, i.e.
\be
a(b,s)=\frac{iA}{2} \theta(R(s)-b)
\ee
where the radius $R(s)$ of the disk is in general energy dependent.
In this very simple model
\bea
\sigtot=2A\pi R^2(s)\\
\sigel=\pi A^2 R^2(s)
\eea
In the optical analogy, also extensively described in \cite{Amaldi:1976gr},   total absorption corresponds to 
$A=1$,  i.e. $\chi_R=0$ and $\chi_I=\infty$, and $\sigtot=2\sigel$. Purely elastic scattering is $A=2$. Defining the ratio
\be
{\cal R}_{el}=\frac{\sigma_{el}}{\sigma_{total}}=\frac{A}{2}
\ee
there are  the following cases : i) $A=2$, all the scattering is elastic, a possibility close to the model by Troshin and Tyurin, 
characterized by large elastic component,  recently rediscussed in \cite{Troshin:2012nu}, ii) $ A=1$, total absorption, the 
scattering is equally divided between elastic and inelastic scattering, iii) $A\le 1$, the total inelastic scattering contribution, 
which includes also diffraction,  is larger than the elastic. It is common practice to refer to the cases $A=1$ and $A<1$ as 
the black and the gray disk model respectively \cite{Block:1984ru}.

The black disk model  is  rather simple, but it is of present interest to investigate whether at  LHC,
the black disk limit has been reached \cite{Block:2011vz}. 
   Using both the TOTEM data at LHC7 and the cosmic ray data from the Auger Collaboration for the inelastic cross-section
    \cite{PierreAuger:2011aa,Ulrich:2011zz,Mostafa:2011qk}, it is possible to estimate how close one is to the asymptotic black disk limit. 
    We show this in  
    Fig.~\ref{fig:sigeltosigtotnewxifae}, where  the ratio ${\cal R}_{el}$ is obtained from accelerator data (black triangles) \cite{pdg}, including 
    TOTEM's \cite{Antchev:2011vs,Antchev:2011zz}, and the value extracted from Auger. The ratio at 57 TeV is obtained by using 
    Block and Halzen (BH) estimate for the total cross-section ~\cite{Block:2011vz} $\sigma^{\rm BH}_{\rm total}(57\ TeV)=(134.8\pm 1.5)$ mb,  
    which is based on the analytic amplitude method  of {Ref.}~\cite{Block:2005ka}. We then obtained 
$\sigma_{\rm elastic}(57\ TeV)=\sigma^{BH}_{\rm total}-\sigma^{\rm Auger}_{\rm inelastic}=(44.8\pm11.6) \ mb$. We also show the asymptotic 
result  (green dot)
from {Ref.}~\cite{Block:2011vz}. For a recent published reference see \cite{Collaboration:2012wt}.
\begin{figure}
\centering
\resizebox{0.5\textwidth}{!}{
\includegraphics{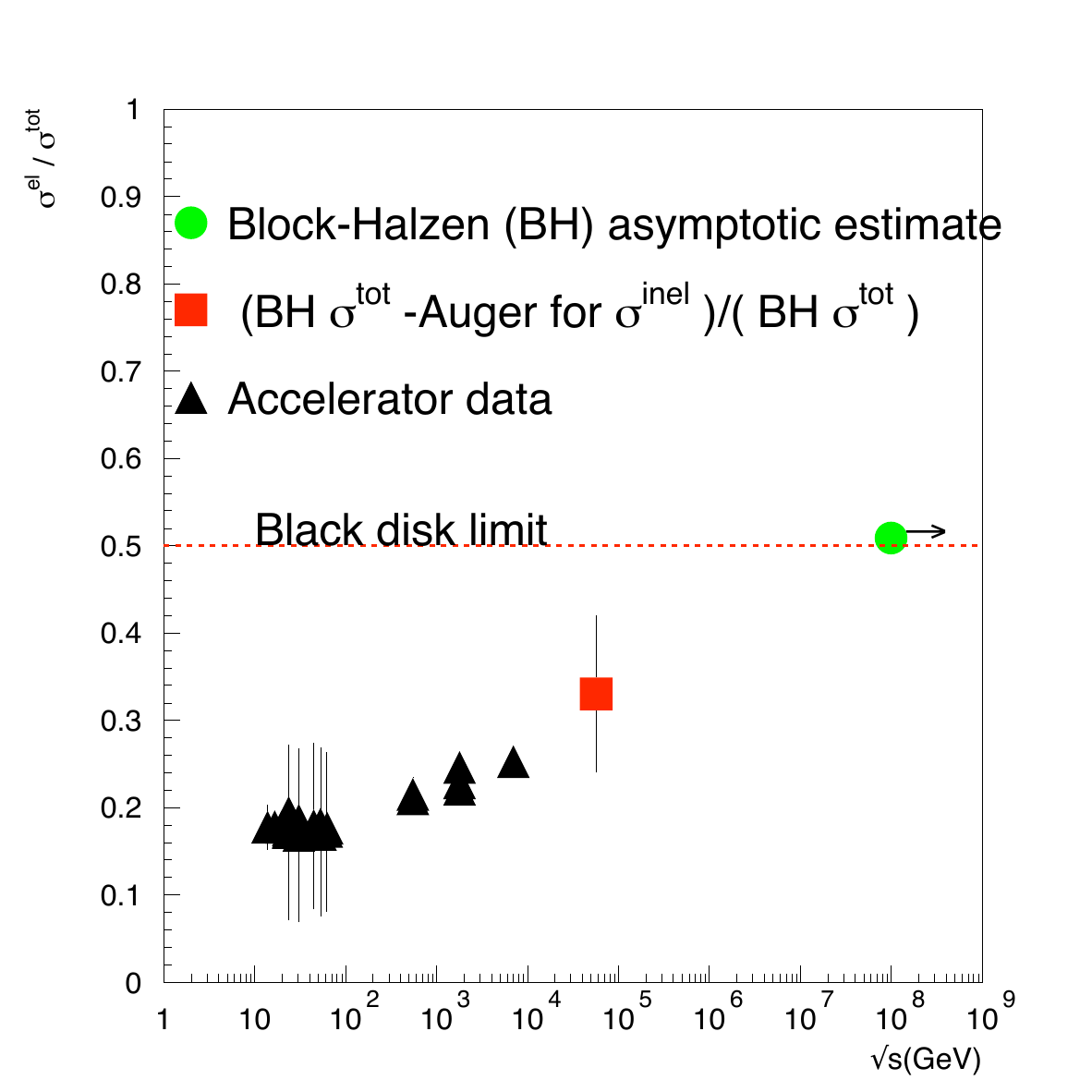}}
\caption{The ratio of the elastic to the total cross-section, using accelerator data, including TOTEM \cite{Antchev:2011zz,Antchev:2011vs} 
and a compilation which uses the Block and Halzen estimate for the total cross-section at $\sqrt{s}=57\ {\rm TeV}$ and extraction of the inelastic 
total cross-section from Auger Collaboration data \cite{PierreAuger:2011aa,Ulrich:2011zz,Mostafa:2011qk}. The green dot represents 
the asymptotic estimate by Block and Halzen.  
Reprinted from \cite{Grau:2012wy}, \copyright (2012), with permission from  Elsevier.}
\label{fig:sigeltosigtotnewxifae}
\end{figure}
From this compilation and a similar one in \cite{Fagundes:2012fa}, \footnote{This analysis indicates a larger error than the one in \cite{Grau:2012wy}.} 
the ${\cal R}_{el}=1/2$ limit does not 
appear to have been reached yet, not even at the highest energy of 57 {\rm TeV}. One may wonder whether it can in fact be reached. 
There are good reasons to expect that a more realistic limit is
\be
{\cal R}_{el}=\frac{\sigel+\sigdiff}{\sigtot}\le  \frac{1}{2} \label{eq:relpumplin}
\ee
where $\sigdiff$ include single and double diffraction. This is the limit advocated by Jon Pumplin \cite{Pumplin:1968bi,Pumplin:1982na,Pumplin:1991ea}. 
In the  subsection dedicated to diffraction  we shall describe Pumplin's model for diffraction and see how the  limit of Eq.~(\ref{eq:relpumplin}) arises.
\subsubsection{The differential cross-section and the  dip structure}\label{sss:dip}
A very interesting characteristic of the data released by the TOTEM
experiment \cite{Antchev:2011zz} is the {\it return of the dip}, namely the 
observation of a very distinctive  dip at $|t|=0.53 \ {\rm GeV}^2$, signalling that the dip, 
observed at the ISR only in $pp$ scattering, and not seen or measured in \pbarp, 
is   now reappearing,   in $pp$.
The dip position has moved from ISR energies, where $-t _{dip}\sim 1.3 \ {\rm GeV}^2$ 
to $-t_{dip}=0.53\ {\rm GeV}^2$ at LHC7. Further shrinkage is expected, but the question of 
how to predict its energy behavior is model dependent, as we show 
in Fig. \ref{fig:tdip}, where   the position of the dip, as measured at various energies, 
  is compared to  a linear logarithmic fit \cite{Fagundes:2013aja} and expectations from  geometrical models \cite{Bautista:2012mq}.
  The uncertainty is related to the difficulty with most models to describe the entire region 
  from the optical point to past the dip, from ISR to LHC energies. Some models which 
  had described this structure at ISR, failed to accurately predict its position and 
  depth at LHC, others describe very well the behavior for small and/or large $|t|$, 
  but not the entire region, as we shall see.

\begin{figure}
\centering
\resizebox{0.5\textwidth}{!}{
\includegraphics{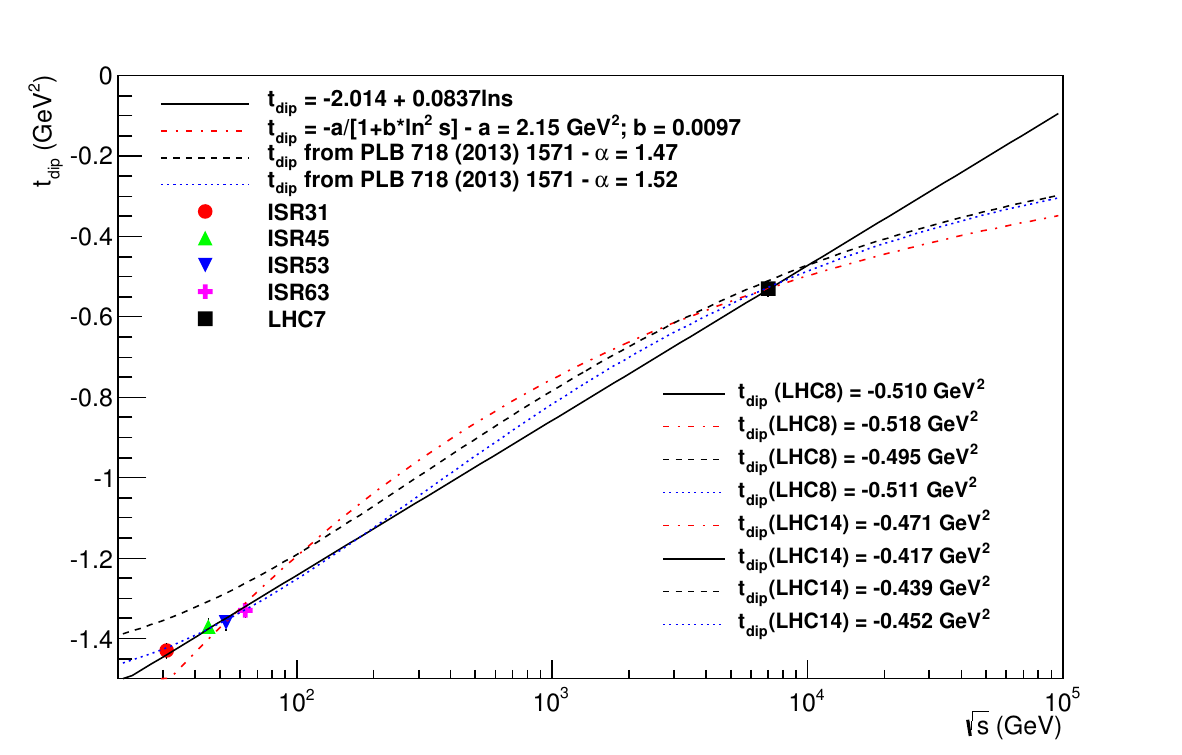}}
\caption{The position of the dip as the energy increases, extracted from geometrical models 
\cite{Bautista:2012mq} and a linear logarithmic fit. The figure is courtesy of D. Fagundes, following  Fig. ~(6) from  \cite{Fagundes:2013aja}, \copyright (2013) by the American Physical Society.}
\label{fig:tdip}
\end{figure}

In Fig. ~\ref{fig:tdip} the dashed and dotted lines (blue) are obtained through models 
for the amplitude in which Geometrical Scaling is partly embedded \cite{Bautista:2012mq}. Clearly 
so long as  
${\cal R}_{el} \neq 1/2$,    two energy scales are present in $pp$ scattering, 
one from the elastic {\it and} the other from the total cross-section. 
In  \ref{sss:GS} we shall discuss this point in more  detail.

\subsubsection{
Geometrical scaling
\label{sss:GS}} 
The idea of geometric scaling is originally due to Dias de Deus and it has been extensively studied in the literature
\cite{DiasDeDeus:1987bf} \cite{Buras:1973km}, \cite{DiasdeDeus:1982ay}, \cite{DiasdeDeus:1974we}, 
\cite{Bautista:2012mq}, \cite{Fagundes:2013aja} both for 
elastic and inelastic amplitudes, for particle multiplicities, etc. For a recent review, see \cite{Praszalowicz:2014bta}.

We shall here limit ourselves only to its application to elastic amplitudes and the position of the dip as a function of
the energy that is of timely relevance for the LHC data on elastic $pp$ data between $\sqrt{s} = (7\div 14)\ {\rm TeV}$. 

As shown in Fig.~\ref{fig:sigeltosigtotnewxifae}, the black disk limit is not reached even until $\sqrt{s} = 57\ {\rm TeV}$ and the geometrical scaling
dip structure being anchored upon it, is hence violated. However, we show in the following that a mean geometrical scaling
based on the two scales works quite well for the position of the dip versus energy.
 
The elastic amplitude $F(s,t)$ has a real and an imaginary part. In the forward direction $t = 0$, the imaginary part
is anchored on the total cross-section
\be
\label{gs1}
\Im m F(s, 0) = \frac{\sigma_{tot}(s)}{4 \pi},
\ee  
and thus is positive-definite and obeys the Froissart-Martin bound. To a certain extent so is real part in the forward direction.
It has an upper bound via the Khuri-Kinoshita theorem
\be
\label{gs2}
\rho(s,0) = \frac{\Re e F(s,0)}{\Im m F(s,0)} \to\ \frac{\pi}{ln(s/s_o)},
\ee
provided the Froissart bound is saturated. By contrast, we have no such general results for $t \neq 0$. As discussed in 
\ref{sss:sumrules}
we  have two sum rules we expect to be satisfied asymptotically
\be
\label{gs3}
S_I = \frac{1}{2} \int_{-\infty}^o dt\  \Im m F(s, t)\ \to\ 1\ {\rm as}\ s\to\infty,
\ee
\be
\label{gs4}
S_R= \frac{1}{2} \int_{-\infty}^o dt\  \Re e F(s, t)\ \to\ 0\ {\rm as}\ s\to\infty.
\ee
Geometrical scaling as applied to the imaginary part of the elastic amplitude -for example- may be stated as follows: that at high energies
\bea
\label{gs5}
\Im m F(s, t) \to\ [\Im m F(s, 0)] \phi(\tau);\ {\rm where}\ \tau = (- t) \sigma_{tot}(s)\nonumber\\
\ {\rm with}\ \phi(\tau = 0) = 1\nonumber\\
\eea 
For $\Re e F(s, t)$, Martin uses analyticity and a saturation of the Froissart-Martin \& the Khuri-Kinoshita  limit, to obtain the form
\be
\label{gs6}
\Re e F(s, t) = \rho(s, 0) \frac{d}{dt} [t \Im m F(s, t)]
\ee
It is easy to see that the sum rule for the real part Eq.(\ref{gs4}) is identically satisfied, i.e., $S_R = 0$, if
Eq.(\ref{gs6}) is obeyed. 

On the other hand, 
\be
\label{gs7}
\int_o^\infty (d\tau) \phi(\tau) = (8 \pi) S_I \to\ (8 \pi)\ {\rm as}\ s\to\ \infty. 
\ee
Now let us focus on the movement of the dip in the elastic cross-section as a function of $s$. Geometric scaling
would imply that
\be
\label{gs8}
t_{dip}(s) \sigma_{tot}(s)\ \to\ {\rm a\ constant\ as}\ s\to \infty.
\ee
Writing the high-energy cross-sections assuming a simple diffraction pattern, we have (for $-t = q^2$)
\bea
\label{gs9}
\sigma_{tot}(s) = 2 \pi b_T^2;\ \sigma_{el}(s) = \pi b_e^2\nonumber\\
\frac{d\sigma}{dt} = [\frac{\sigma_{tot}^2(s)}{16 \pi}] \big{[}\frac{2 J_1(qR)}{qR}\big{]}^2,
\eea 
so that the optical point is correct. In the black disk limit, 
\be
\label{gs10}
b_T = b_e = R;\ \mathcal{R}_{el}^{BD} = \frac{\sigma_{el}}{\sigma_{tot}}\to\ 1/2.
\ee
For the black disk, the dip occurs at the first zero when $q_{dip}^{BD}b_T \approx\ 3.83$. Defining the geometric
scaling variable $\tau_{GS} = q^2 \sigma_{tot}$ and as shown in the left panel of Fig(\ref{fig:geo-mean scaling}), it does not work.

On the other hand, we can define a mean geometric scaling with $b_T \neq b_e$ where the radius $R$ in
Eq.(\ref{gs9}) is taken as the geometric mean and the mean geometric scaling variable $\tau_{GS}^{mean}$ 
reads
\be
\label{gs11}
R = \sqrt{b_T b_e};\  \tau_{GS}^{mean} = q^2 \sqrt{\sigma_{el} \sigma_{tot}}
\ee
The dip is now given by 
\be
\label{gs12}
-t_{dip}(s) = q^2_{dip}(s) \approx\ [\frac{(3.83)^2 \pi \sqrt{2}}{\sqrt{\mathcal{R}_{el}(s)} \sigma_{tot}(s)}] 
\ee
On the right hand panel of Fig(\ref{fig:geo-mean scaling}), we show a comparison of our mean geometrical scaling prediction
with experimental data. The agreement is quite good. We mention in passing that this is yet another example 
that the black disk limit is quite far from being reached even up to $57\ {\rm TeV}$. 
\begin{figure*}
\resizebox{0.9\textwidth}{!}{%
  \includegraphics{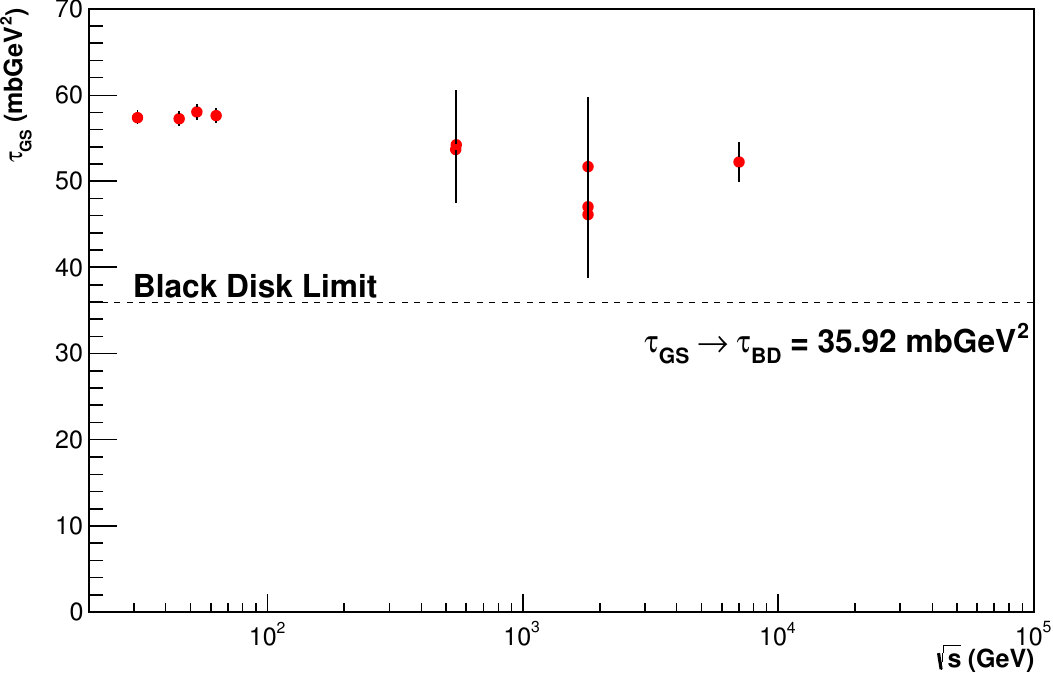}
  \includegraphics{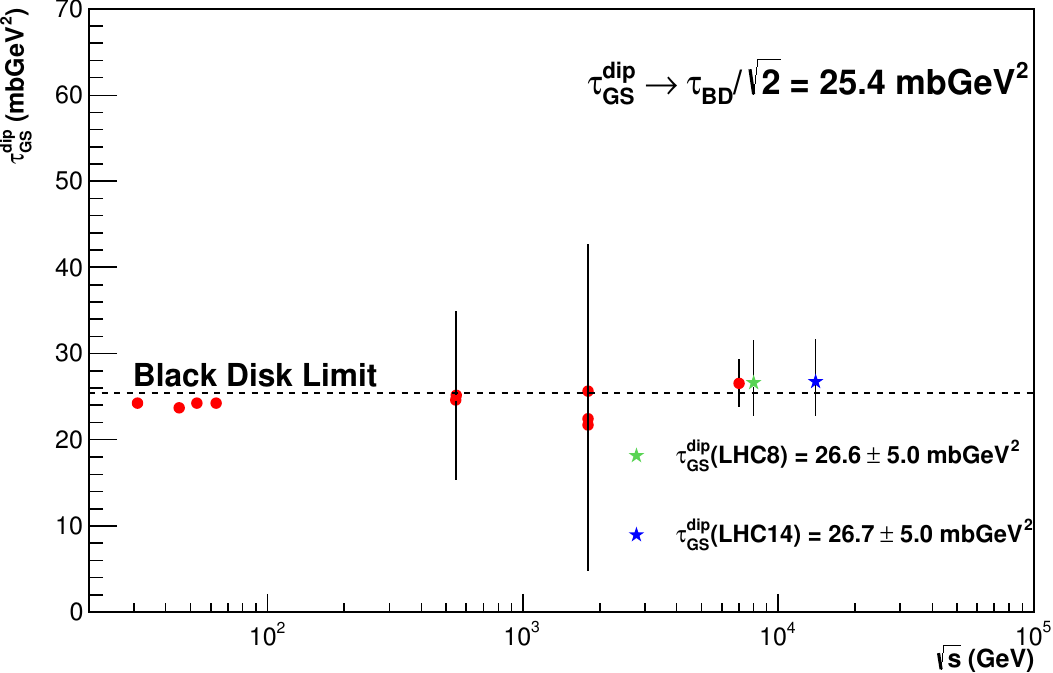}}
\caption{Data up to LHC7 for  energy dependence of the dip $-t_{dip}(s)$ from geometric scaling [left panel] and from mean 
geometric scaling [right panel]. Green and red crosses are predictions at LHC8 and LHC14 from the empirical model of \cite{Fagundes:2013aja}.
Reprinted  from \cite{Pancheri:2014roa} with kind permission from Societ\`a Itaiana di Fisica.}
\label{fig:geo-mean scaling}       
\end{figure*}
On the other hand, we notice a recent work by Block and collaborators \cite{Block:2015sea}, which we shall see in \ref{ss:conclusion-elastic}, 
where  the Black Disk limit is  used  to make predictions at very, extremely, large energies.

\subsection{Early Models in impact parameter space}
\label{ss:early-models}
Let us now examine various models and fits. Empirical fits abound in total and elastic cross-section description. 
They are helpful in developing models, as a guide toward understanding data. To be most useful, of course, 
empirical fits should follow constraints imposed by general theorems on analyticity, crossing symmetry and unitarity, 
all of which (or, more realistically as much as possible) should then be satisfied by the models one 
builds.

Between 1967 and 1968, 
 Chou and Yang  \cite{Chou:1968bc} and Durand and Lipes \cite{Durand:1968ny} presented  a framework for calculation of the elastic scattering amplitude 
 between elementary particles  based on the impact picture and on physical ideas very similar  to those in the Glauber model. The presence of kinks in the elastic differential 
 cross-section was
 discussed.
We shall start with Chou and Yang and then discuss the results 
 by Durand and Lipes.

\subsubsection{The Chou and Yang model}\label{sss:chouyang}
The Chou and Yang model was first discussed in 1967 and fully written in 1968.To discuss particle scattering, rather than that of nucleons on nuclei,  
 Chou and Yang 
 had to put target and projectile  on equal footing, and  comply with existing phenomenology. They started in \cite{Chou:1968bc}, 
 with  the partial wave expansion of the scattering amplitude 
(as we have described in the first section).  From 
\bea
a={\slashed \lambda}^2\sum_l
 (2l+1) P_l(cos\theta)\frac{1}{2}(1-S)
 \eea
 the high energy limit was obtained  by    transforming the sum   into an integral,  through the substitution 
 $P_l(cos\theta)\rightarrow J_0(b\sqrt{-t})$  and the definition $b={\slashed \lambda}(l+1/2)$. 
 This led to the  eikonal expression for the scattering amplitude
\be
 a(t)=\frac{1}{2\pi} \int  (1-S)e^{i\vecq \cdot \vecb} d^2\vecb
\ee
with $t =-q^2$.

 The crucial  assumption of the Chou and Yang  model was that the attenuation of the probability amplitude for two hadrons 
 to go through each other was governed by the local opaqueness of each hadron.  
   In this model, the transmission coefficient $S$ is only a function of the impact parameter $b$, and
Chou and Yang proposed to calculate the transmission coefficient  $S(b)$  through  the Fourier transform of the form factors 
of the colliding particles. 
Subsequently the authors  proceeded to show that their proposal for particle scattering was the same as the Glauber proposal 
for nucleon-nucleus scattering,  with the  two dimensional  Fourier transform of the form factors playing the role of  the nuclear 
density of  the Glauber model. In formulating this connection between the two models, a limit of infinitely many nucleons was postulated, 
namely the nucleus was seen as a droplet of very  finely  granulated nuclear medium, in the spirit of the droplet model formulated 
previously by Byers and Wu \cite{Byers:1966sf}.

The quantity $<s>=-\log S(\vec b)$ is called the {\it opacity}, with the name deriving from the following physical interpretation:
 if an incoming wave hits a slab of material of thickness $g$, the slab partly absorbs and partly transmits the wave and the transmission 
 coefficient through the slab would be 
\be
S=e^{-\alpha g}
\ee
Thus $\log S(b)$ would be proportional to the thickness of the slab and can be considered as the opaqueness of the slab to the wave. 
For particle scattering, the thickness of the slab represents  how much hadronic matter is encountered by the incoming wave, when 
the wave passes through the hadron at impact parameter $\vec b$. Thus $S$ is a  function of the impact parameter $\vec b$, which 
is then integrated over all possible values. $S(b)$ was considered to be asymptotically energy independent, and so would then be 
$\dsigdt$. This was consistent with the fact that, at the time,  data suggested that all cross-section would reach an asymptotic limit, 
independent of energy. 
It should in fact be be noted that  the Chou and Yang model, was  formulated before 1970, i.e, before 
ISR experiments definitely proved that the total cross-section was rising.


To compare with experimental data for $\dsigdt$, two different phenomenological expressions were considered, a single exponential 
in $t$, i.e. a gaussian in $\vec b$-space, and a sum of two exponentials. The resulting fits to the data available at the time, are shown in 
Fig.~\ref{fig:chou-yang}, where they are labeled with $A$ and $B$ respectively. 
A further comparison with data is done by using the 
form factor expression instead of the ansatz on exponential $t$ behaviour of the amplitude, a  comparison which we show in the 
right-hand  plot in  Fig.~\ref{fig:chou-yang}. 
\begin{figure*}
\resizebox{0.9\textwidth}{!}{%
  \includegraphics{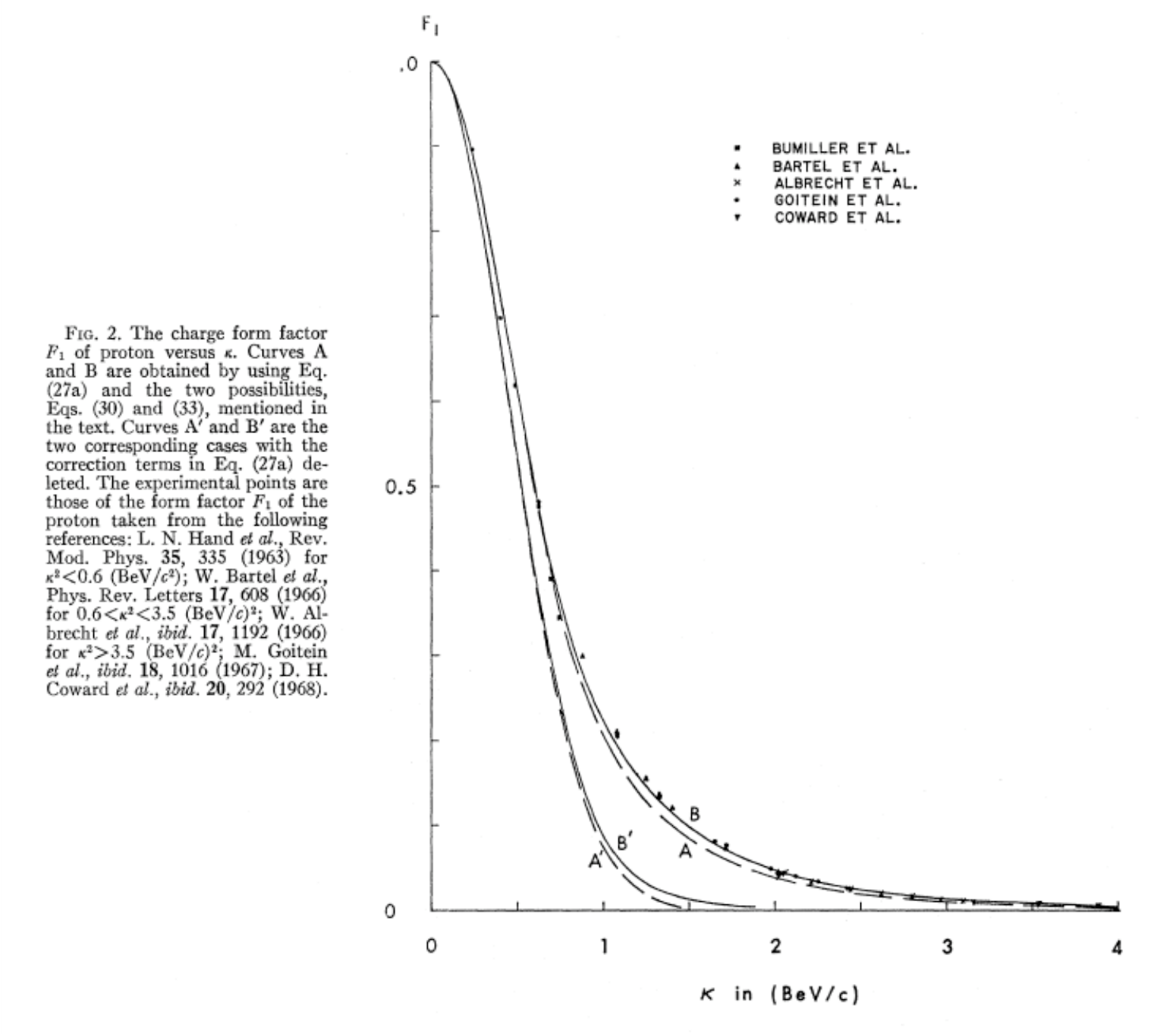}
  \includegraphics{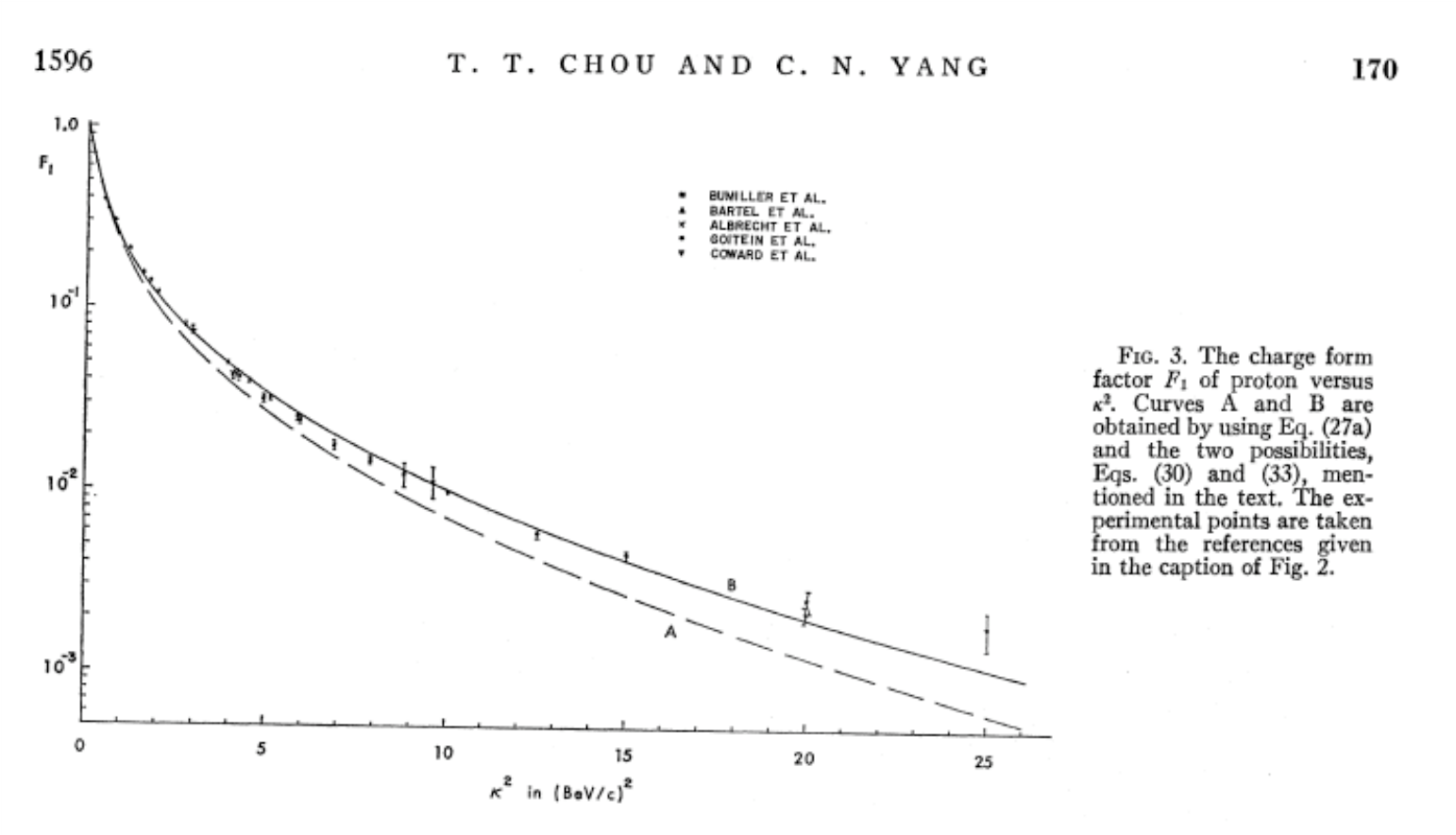}}
\caption{Comparison between the   Chou and Yang model  from \cite{Chou:1968bc}  and existing data for the differential cross-section. 
The two figures correspond to different approximations, left panel, and to a comparison between the form factor expression and their 
model for $\dsigdt$, right panel. 
Reprinted Figs.~(1,2 ) with permission from \cite{Chou:1968bc}.  \copyright (1968) by the American Physical Society.}
\label{fig:chou-yang}       
\end{figure*}

As a conclusion, 
 the  major points  of this model can be summarized as being the following: 
 \begin{enumerate}
\item   the transmission coefficient $S(\vec b)$ can  be obtained from   the convolution of the form factors of the scattering particles,
 \item this model, with the amplitude expressed through the transmission coefficient $S$ includes  the limit of a model previously proposed with 
  T.T. Wu \cite{Wu:1965se}, in which the scattering differential elastic cross-section $d\sigma/dt $ was
    proportional to the 
  fourth power of the proton form factor,
\item correction terms are important, namely one 
needs to use the full $(1- S)$
\item one can expect the  appearance of a distinctive diffraction pattern in the squared momentum transfer in the elastic differential cross-section, 
with dips and bumps.
\end{enumerate}

\subsubsection{The diffractive model of Durand and Lipes}\label{sss:durandlipes}
 A diffractive picture in impact parameter 
space, in a close correspondance to  the  Chou and Yang model, was formulated by Durand and Lipes in 1968 
\cite{Durand:1968ny}. 
 
In  \cite{Durand:1968ny}, it is shown that the  elastic scattering diffractive amplitude should exhibit two diffraction minima, which can be 
filled by the real part of  $pp$ scattering  amplitude and that at large momentum transfer the amplitude is proportional to the product of the 
form factors of the scattering particles. The model uses the impact parameter picture and the paper follows along the lines of   
Wu and Yang \cite{Wu:1965se} and Byers and Yang \cite{Byers:1966sf}.

The suggestion that the $pp$ scattering \x \ at large momentum transfer be proportional to the 
fourth power of the proton form factor, suggested by Wu and Yang \cite{Wu:1965se}, 
was prompted by the  observations of the rapid decrease of the cross-section 
away from $t=0$. Namely,  such a decrease could be seen as the breaking up of 
the proton extended structure as the momentum transfer was 
becoming large. For small momentum transfer values, on the other hand ,  
the coherent  droplet model by Byers and Yang \cite{Byers:1966sf} 
supplied inspiration.

The basic physical assumptions underlying the diffraction model by Durand and Lipes were :
i) elastic scattering at high energy results from the absorption of the incoming wave into the many inelastic channels which are opening up at 
high energy as the extended proton structure (where partons are confined) breaks up, ii) at high energy, the absorption depends on the amount of 
relative interpenetration of the two scattering protons, namely on the distance between the scattering centers, in impact parameter space. With these 
assumptions and the form factor expression for the matter distribution in two hadrons, Durand and Lipes wrote the scattering amplitude in the 
(now)
familiar form 
\begin{align}
f(s,t)=i \int b db J_0(b\sqrt{-t})[1-S(b)]\\
S(b)=e^{-k\rho(b)}\\
G_A(t) G_B(t)=\int_0^\infty b db \rho(b)J_0(b\sqrt{-t})
\end{align} 
where the density of matter inside the scattering region is obtained as the convolution 
of the form factors of the two particles A and B in the initial state. 

The absorption coefficient $\kappa$ was understood to be a function of the initial energy. 
Using the dipole expression for the proton electromagnetic 
form factor, they obtained
\be
S(b)=e^{-\frac{1}{8}A (\nu b)^3 K_3(\nu b)}
\ee
with $A$  proportional to the absorption coefficient $\kappa$. 
Predictions including both a real and imaginary parts for $A$ were given, fitting the real part of A through the total cross-section. 
The results of such a model, with value of the parameter $\nu^2=1 \ {\rm GeV}^2 $, are shown in Fig.~\ref{fig:durandlipes} from \cite{Durand:1968ny}. 
The comparison with data in the very forward region shows a very good agreement with the model.  Notice the prediction of two dips which are 
filled partially through a complex absorption coefficient with a non-zero  imaginary part.
\begin{figure}
\centering
\resizebox{0.5\textwidth}{!}{
\includegraphics{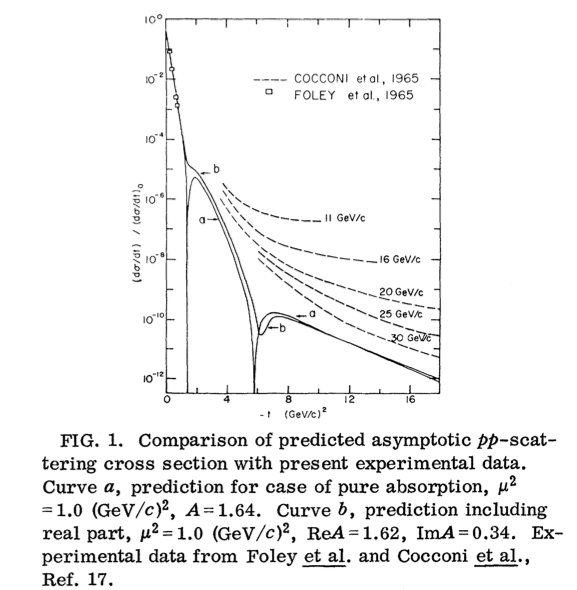}}
\caption{In this figure from \cite{Durand:1968ny} comparison is made with existing data and  the diffraction model by Durand and Lipes. The two curves 
labelled respectively $a$ and $b$ refer to fits made with a purely real or an imaginary  absorption coefficient.
Reprinted with permission from  \cite{Durand:1968ny}, Fig.(1), \copyright (1968) by the American Physical Society.}
\label{fig:durandlipes}
\end{figure}  
Some of the observations  drawn by these authors, still of interest are :
\begin{itemize}
\item asymptotically,  this diffraction model should describe high energy scattering at fixed momentum transfer, although the 
asymptotic limit described  by the model can also be ascribed in a different language to asymptotic contributions from the 
Pomeron trajectory (called Pomeranchuk Regge trajectory in this paper) and its cuts, with other contributions, which will disappear 
at increasing energy, describable by other Regge exchanges,
\item two diffraction minima appear  in the diffraction model.
\end{itemize} 
\subsubsection{The black disk model}\label{sss:blackdisk}
One of the most popular representations of the proton as it emerges from high energy scattering, is that of the proton as a black disk. 
This definition of a black or gray disk  is related to the picture of scattering as one of total or partial absorption. 
To show its making, and before entering into a description of models including diffraction, 
we shall present here the  discussion of some  simple elastic scattering models from  Sect.4 of the Block and Cahn review \cite{Block:1984ru}. 
For definitiveness, these models were illustrated by  fixing  the parameters so as to reproduce the total cross-section and the value 
of the B-parameter at ISR, respectively $\sigma_{tot}=43\ mb$ and $B=13\ ({\rm GeV}/c)^{-2}$. 

With  $a(b ,s)$  the scattering amplitude in impact space space and related quantities as in Eqs.(\ref{eq:BC1}, \ref{eq:BC2}, \ref{eq:BC3}, \ref{eq:BC4})
the following models are considered,
\begin{itemize}
\item an amplitude which is purely imaginary with a constant value $a=iA/2$  inside a  radius R, corresponding to a purely black disk 
for $A=1$, so that near the forward direction one has
\be
\frac{d\sigma}{dt}=\pi A^2 R^4 [\frac{J_1(qR)}{qR}]^2 
\ee
In this model, the real part can be added as shown in Eq.~(\ref{eq:realderivative}), i.e. one has
\begin{align}
\dsigdt |_{Black disk}=\pi R(s)^4A^2\{ [
\frac{J_1(Rq)}{Rq}]^2+\nonumber \\
\frac{\rho^2}{4}J_0(Rq)^2       \}
\end{align}
Variations of this model include an s-dependent radius, as in the model obtained by Ball and Zachariasen \cite{Ball:1972xq} in solving the 
multiperipheral equation for diffractive elastic scattering,
with $R=R_0\log s, A=\kappa[\log(s/s_0)]^{-1}$, and  giving  a total cross-section increasing as $\log s$, and an elastic cross-section which is constant,
\item a parabolic shape, i.e. $a=iA[1-(b/R)^2]$ inside a radius A, and zero outside,
\item a gaussian shape in impact parameter space,
 \be a=1/2A exp[-(b/R)^2]
 \ee which leads to
$\sigma_{tot}=2\pi AR^2, \sigma_{el}=\pi A^2 R^2/2$
\item Chou and Yang \cite {Chou:1968bc,Chou:1983zi} model, or the Durand and Lipes model\cite{Durand:1968ny}.
\end{itemize}
 As in both Chou and Yang and Durand and Lipes model,  the matter distribution inside the hadrons is described through their electromagnetic form factor, i.e.  through the 
 convolution of the Fourier transform of the dipole expression.  Writing
 \bea
 a(b,s)=\frac{e^{2i\delta -1}}{2i}\equiv \frac{1}{2}(1-e^{-\Omega(b)})\\
 \Omega(b)=A\frac{1}{8}(\Lambda b)^3K_3(\Lambda b)
 \eea
 and fitting the results to reproduce the total cross-section at ISR, it is found that $\Lambda^2 \sim 0.71 ({\rm GeV}/c)^2$.
 In  Fig.~\ref{fig:blockcahnfigs7-8}, we show the profiles of  the amplitudes in $b$-space at left, and the differential cross-sections resulting 
 from the different shapes at right, as obtained in \BC.
 \begin{figure*}[htb]
\resizebox{0.9\textwidth}{!}{%
  \includegraphics{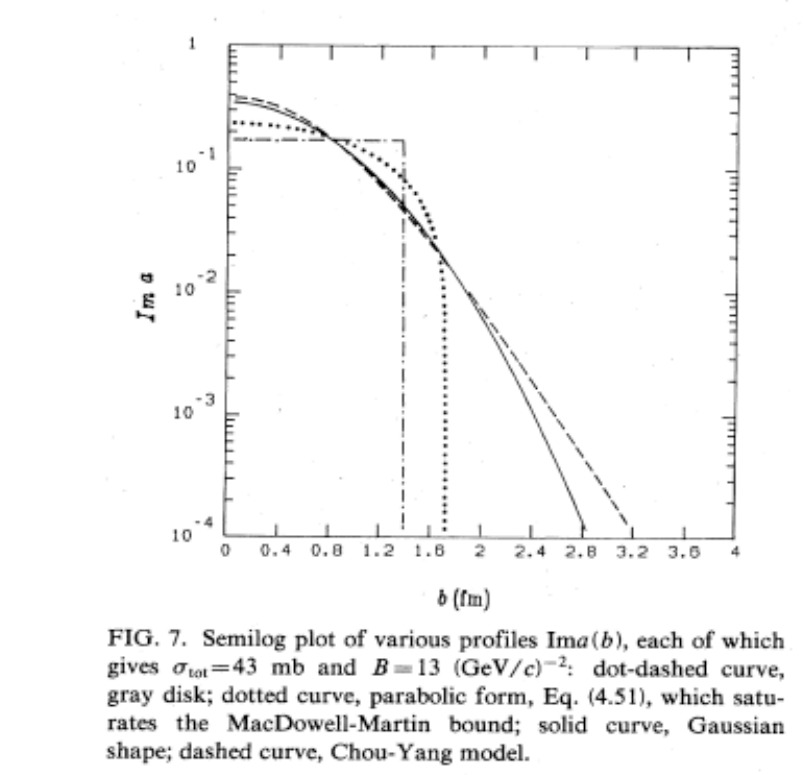}
  \includegraphics{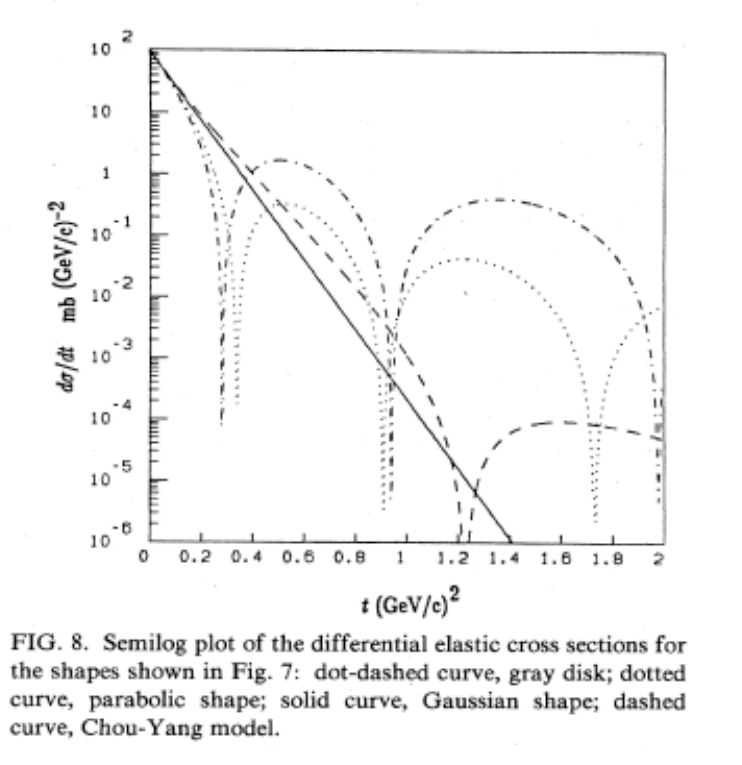}}
\caption{
Shown are     examples of profile 
functions, $\Im m\ a(b,s)$ (at left),  and the resulting elastic differential cross-section (at right): dot-dashed curve, 
grey disk; dotted curve, parabolic form; solid curve, Gaussian shape; dashed curve, Chou-Yang model. 
Reprinted with permission  from \cite{Block:1984ru}, Figs.(7,8), \copyright (1985) by the American Physical Society.}
\label{fig:blockcahnfigs7-8}      
\end{figure*}

  These figures bring to focus  the basic difference between models with a 
 Gaussian matter distribution and the others: the Gaussian shape leads to an exponential decrease in $-t$, with no wiggles or dips and bumps, 
 whereas all the other distributions, grey disk, parabolic shape, or Chou and Yang model, exhibit a diffraction pattern with minima and maxima. 
 Such difference among models persists till today. 



\subsection{ Exponentials and parametrizations through Regge and Pomeron exchanges}\label{ss:reggepomeron}
The models in impact parameter space summarized
so far underline the optical nature of scattering, without  
specific dynamical inputs, except for  the hadron form factors. On the other hand, data on elastic scattering from ISR appeared to 
conform to a picture in which the elastic scattering amplitude obtains from Regge-Pomeron exchange. We shall examine some of 
such models. An additional group of models  embeds  Regge-Pomeron exchanges into the eikonal representation, thus ensuring 
satisfaction of unitarity, in addition to analyticity of the input Born scattering amplitude. Some of these models include    
QCD and  diffractive contributions, 
as we shall discuss in a separate subsection.

\vskip 0.3 cm
\par\noindent
\subsubsection{The model independent analysis by Phillips and Barger (1973) 
}\label{sss:PB}
\par\noindent
In 1973, Barger and Phillips (PB) proposed what they called a model independent analysis of \pp \ scattering \cite{Phillips:1974vt}. 


They propose two slightly different parameterizations, using a  phase 
and two 
exponentials to describe the different slope of the cross-section as a function of  momentum transfer in the range $-t= (0.15\div5.0)\ {\rm GeV}^2 $. 
The first, and more model independent parametrisation, is given as 
\begin{eqnarray}
\frac{d\sigma}{dt}=|{\tilde A}(s,t)+{\tilde C}(s,t)e^{i\phi(s)}|^2 \label{eq:bpeq1-1}\\
{\tilde A}(s,t)=i\sqrt{A}e^{\frac{1}{2}B(s)t }\label{eq:bpeq1-2}\\
{\tilde C}(s,t)=i\sqrt{C}e^{\frac{1}{2}Dt } \label{eq:bpeq1-3}
\end{eqnarray}
Data from $p_{lab}=12\ {\rm GeV}$ to $p_{lab}=1496 \ {\rm GeV}$ ($\sqrt{s}\approx (5\div 53)\ {\rm GeV}$) were fitted with this expression.
The first exponential is seen to have normal Regge shrinking, namely
\be
B=B_0+2\alpha' \log s \ \ \ \ \ \ \ \ \ \alpha'\approx 0.3\ {\rm GeV}^{-2}
\label{eq:bpslope}
\ee
while the second exponential term is almost constant.
The values of the parameters at each energy value are given in Table ~\ref{tab:bp}.\begin{table}
\caption{From \cite{Phillips:1974vt}, parameters obtained from a fit to $pp$ data in the interval $(0.15<|t|<5.0)\ {\rm GeV}^2$ for the model 
given in Eq.~(\ref{eq:bpeq1-1}). }
\label{tab:bp}
\tabcolsep=0.11cm
\begin{tabular}{|c|c|c|c|c|c|c|}\hline
$p_{LAB}$&$\sqrt{s}$&$\sqrt{A}$&B&$\sqrt{C}$&D&$\phi$\\
{\rm GeV}&{\rm GeV}&
$\sqrt{mb}/ {\rm GeV}$
&${\rm GeV}^{-2}$
&
$\sqrt{mb}/ {\rm GeV}$ &${\rm GeV}^{-2}$&rad \\
\hline 
12.0&4.84&7.47&7.33&0.370&1.66&2.06\\
14.2&5.25&7.53&7.79&0.305&1.69&2.12\\
19.2&6.08&7.96&8.00&0.232&1.73&2.12\\
24.0&6.78&7.97&8.07&0.194&1.76&2.16\\
29.7&7.52&7.82&8.41&0.175&1.81&2.16\\
1496.0&52.98&6.55&10.2&0.034&1.70&2.53\\
\hline
\end{tabular}
\end{table}
Notice that, with this parametrisation,  the phase $\phi$ is fitted to be always larger than $\pi/2$, so that the 
interference term is always negative and,
 in  the energy range examined here,   the fitted phase is energy dependent.

 To understand better the role played by the energy dependence, the authors proposed a second parametrisation, 
 clearly inspired by Regge phenomenology,  i.e.
\begin{equation}
\frac{d\sigma}{dt}=|{i\sqrt{A}e^{(\frac{1}{2}B+\alpha'\log s -i\alpha'\pi/2)t}}+(\frac{\sqrt{C_0}}{s}-i\sqrt{C_\infty})e^{\frac{1}{2}Dt}|^2
\label{eq:bpeq3}
\end{equation}
with an explicit energy dependence and the following parameter values, valid for all the energies: 
$\sqrt{A}=6.88, B=5.38, \alpha'=0.306, \sqrt{C_0}=10.3, \sqrt{C_\infty}=0.035, D=1.78$, 
in the units of Table \ref{tab:bp}.The results are presented   
in the two panels of Fig. ~\ref{fig:bp} and the figure shows  that the  quality of the two fits is good and  about the same for both models.
\begin{figure*}[htb]
\hspace{2cm}
\resizebox{0.5\textwidth}{!}{
\includegraphics{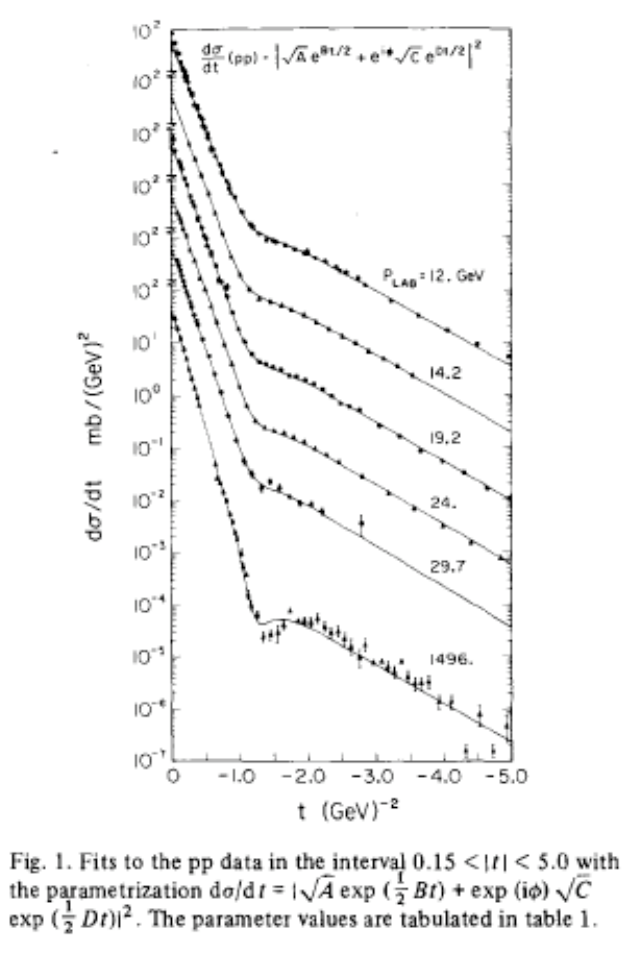}}
\resizebox{0.45\textwidth}{!}{
\includegraphics{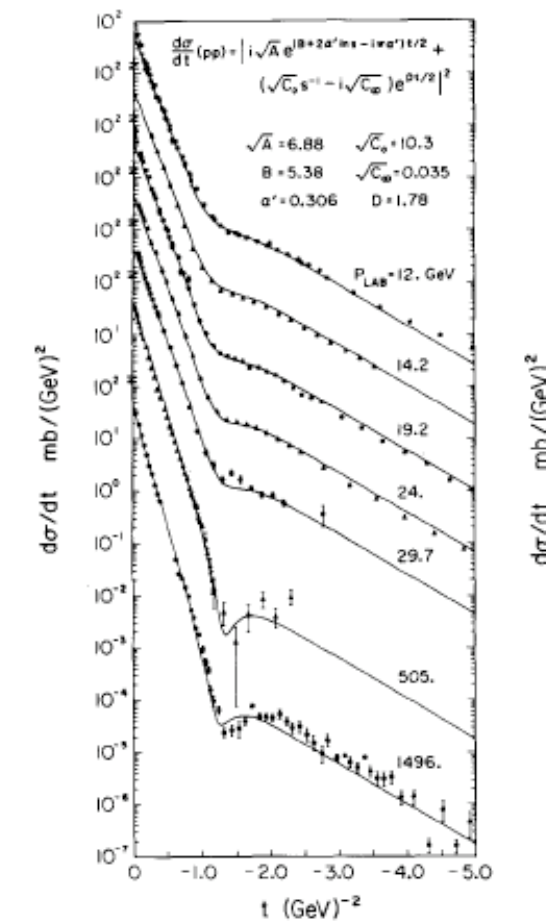}}
\caption{\label{fig:bp}Model independent fits by  Phillips and Barger \cite{Phillips:1974vt}: {parameters are energy dependent at left, constant at right.}
Reprinted from \cite{Phillips:1974vt}, \copyright (1973) with permission by Elsevier.}
\end{figure*}

The PB parametrization applied to the preliminary TOTEM data at LHC7 gave the results discussed in
  \cite{Grau:2012wy}.  This parametrization however misses the optical point by some 5-10\%, but the description of the region $0.2<|t|<2.5\ {\rm GeV}^2$ is quite satisfactory. 
  We shall return to the question of the very small $-t$ behavior  later.


In \cite{Phillips:1974vt}, it is also suggested that the ``second exponential" can be identified with a term proportional 
to $p_T^{-14}$, valid, according to the authors, for all available $pp$ data for $s>15\ {\rm GeV}^2$. Following this suggestion, 
and mindful of the fact that the form factor dependence of the amplitude would contribute as $(-t)^{-8}$, we have tried a 
slight modification of the Barger and Phillips model, which  consists in substituting the second term in Eq.~(\ref{eq:bpeq1-1}) 
with a term proportional to the 4th power of the proton form factor, namely to write
 \bea
 {\cal A}(s,t)=i[\sqrt{A(s)}e^{Bt/2}+\frac{\sqrt{C(s)}}
{(-t+t_0)^4} e^{i\phi}]\label{eq:ffmodel}\\
\frac{d\sigma}{dt}=|\sqrt{A(s)}e^{Bt/2}+\frac{\sqrt{C(s)}}
{(-t+t_0)^4} e^{i\phi}|^2 
\eea which leads to
\begin{equation}
\frac{d\sigma}{dt}=Ae^{Bt}+\frac{C}{(-t+t_0)^8} +2\cos{\phi}\frac{\sqrt{A}\sqrt{C}e^{Bt/2}}{(-t+t_0)^4}
\label{eq:bpours}
\end{equation}
When applied to LHC7 data, this parameterization would comprehend both the proposal 
by Donnachie  \cite{Donnachie:2011aa,Donnachie:2013xia} (further discussed at the 
end of this section), who advocates a power law decrease 
of the type $t^{-8}$, as well as the fit proposed by the TOTEM collaboration, 
$t^{-n}$, with $n=7.8\pm 0.3_{stat}\pm {0.1}_{syst}$ \cite{Antchev:2011zz}. 

A discussion of how these two  models, two exponentials and a phase or one exponential, 
a form factor and a phase,  would be realized at LHC, reading    data from \cite{Antchev:2011zz}, can be found in \cite{Grau:2012wy}.

  That the behaviour past the dip at LHC is consistent with an exponential was  also noticed in \cite{Troshin:2012nu}.
  
 One can see however that the expression
\be
\sigtot=4 \sqrt{\pi} [\sqrt{A}+\sqrt{C} \cos \phi]
\ee
gives a value for $\sigtot$ below the data at all energies, including at LHC7, 
where the fitted value misses TOTEM datum  by some 5\%, but particularly so at low energy and that
 the t-dependence of the t $B_{eff}$ for  $t\sim 0$, while reasonably  well described at LHC, is less  
 so at lower energy values. These two points are connected, since we have already noticed that at very small 
 $|t|$ values at ISR the slope seems to increase.  An interpretation of this effect is given in the Durham model
  \cite{Khoze:2000wk}, with pion loops contributing to the effective Pomeron trajectories at very 
  small $t$-values \cite{Anselm:1972ir} and shall be discussed together with diffraction.

  To overcome this problem and still provide a useful tool, we have proposed an empirical  modification of the original BP model amplitude 
   \cite{Fagundes:2013aja}, i.e.
\be
\mathcal{A}(s,t)=i[G(s,t)\sqrt{A(s)}e^{B(s)t/2}+e^{i\phi(s)} \sqrt{C(s)}e^{D(s)t/2}]. \label{eq:mbp}
\ee
and
  \be
G(s,t)=F^2_P(t)=1/(1-t/t_0)^4 \label{eq:mbp2}
\ee 
With free parameters in Eqs. ~(\ref{eq:mbp}, \ref{eq:mbp2}),  the resulting analysis  of elastic 
$pp$ data at LHC7  is shown in Fig.~\ref{g:lhcbp5}.  This figure  includes a comparison  
with  a parametrization of the tail of the distribution  given  by the TOTEM collaboration (dotted line). Such 
a parametrization,   $(-t)^{-8}$,  was suggested in \cite{Donnachie:1979yu}, 
and recently proposed again  in \cite{Donnachie:2013xia}, where it is shown to describe large $-t$ data from ISR to LHC8, and 
both for $pp$ and \pbarp, as we shall discuss shortly. 
 \begin{figure}
\resizebox{0.48\textwidth}{!}{
\includegraphics{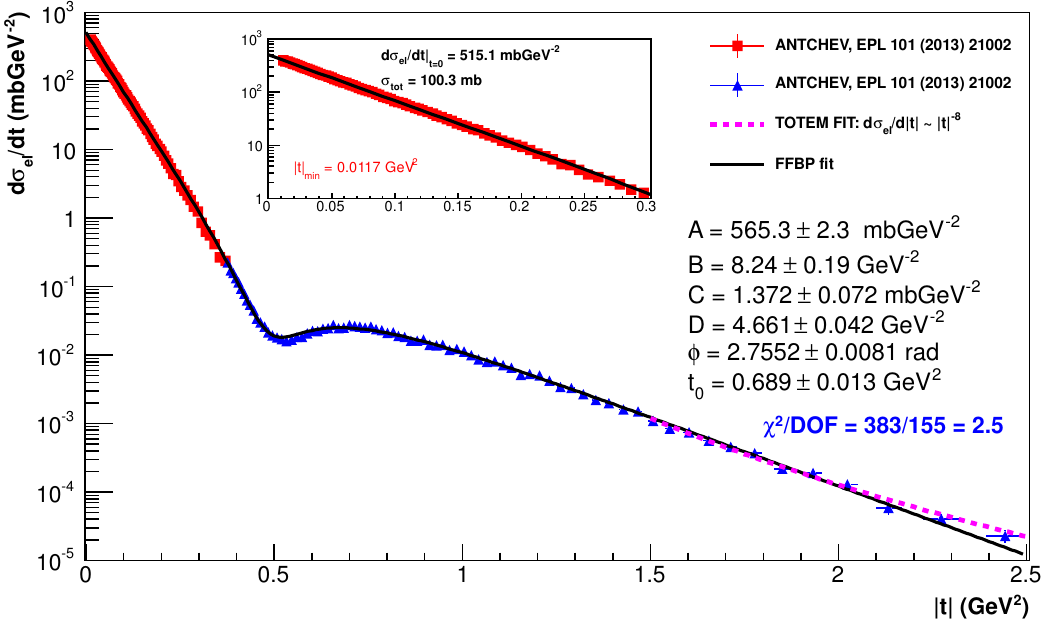}}
\caption{Fits to the  LHC7 data sets with  model of Eqs. (\ref{eq:mbp},\ref{eq:mbp2})  
(labeled { FFBP} in the frame), with $t_0$ a free parameter. Data sets  and parameter 
values  can be found in \cite{Fagundes:2013aja}. {\it Inset:}   LHC7 data near the 
optical point are shown in comparison with the present model, which includes  
the proton form factor modification. 
Reprinted Fig.(2)  from \cite{Fagundes:2013aja},  \copyright (2013) by the American Physical Society.}
\label{g:lhcbp5}
\end{figure}

 Before leaving this model, we remark 
 that  the dip present both at lower and higher energies in $pp$ is only 
a faint appearance in \pbarp. The simple expression for the amplitude with two exponentials and a phase may still be a 
valid phenomenological tool also for \pbarp, but one cannot draw any conclusion about the energy dependence of the parameters from this channel;
as one gets close to the dip, and the first term dies out to zero, non leading terms,     present in \pbarp \ and not in $pp$ fill the dip 
and probably need to be added to the original model, namely one has, as pointed out in many models, a situation such as
\be
A_{p{\bar p}}=A_{pp}+{\cal R}(s,t)
\ee
where $\cal R$ depends on the Regge trajectories exchanged in the $t-$channel, 
 including a possible odderon, 
 as we shall see later.

 Concerning the coefficient $B(s)$, we notice that in the two exponential model just described, this coefficient is related to 
 the slope as defined in Eq. ~(\ref{eq:bst}) through the  relation
 \begin{align}
 B_{\rm eff}(s,t)=\frac{1}{\dsigdt} \times \ \ \ \ \ \ \ \ \ \ \ \ \ \nonumber\\
 [ABe^{Bt}+CDe^{Dt}+ \sqrt{A}\sqrt{C}(B+D)e^{(B+D)t/2} \cos\phi ] 
\label{eq:beff-expression}
 \end{align}



The PB model can be seen as an  attempt to describe the elastic differential cross-section without appealing to unitarity. 
It was defined as {\it model independent analysis}, but it could also be understood as a modeling of elastic scattering 
through Regge and  Pomeron exchanges. Since then many fits to data have appeared with a few Pomerons or many Reggeons and 
Pomerons. We shall next look at   recent contributions, along these lines,  by Donnachie and Landshoff. 

\subsubsection{ Soft and hard Pomeron exchanges in Donnachie and Landshoff model }\label{sss:DLdsigdt}

In \cite{Donnachie:2011aa,Donnachie:2013xia} the authors stress that the data from the TOTEM experiment 
do confirm the results of Regge theory, which 
is in many ways   
 a major success of high energy physics. At high energy however  
such success depends on interpreting a power law as the ultimate solution, and at the end, as we shall see, resulting in using 
models to describe the details of the data. Indeed if one believes that the Froissart bound is actually reflecting the existence of 
confinement, power laws, which contradict the Froissart bound, must at a certain point give up to the true asympotic behaviour. 

In Ref. \cite{Donnachie:2011aa} a hard Pomeron had been invoked to describe LHC7 data, 
but in the  more recent contribution of Ref.~\cite{Donnachie:2013xia},
this is not considered necessary. Instead, a double Pomeron exchange is seen  to provide a better fit, 
as they  show. Of particular interest here is the fact that  in both their modeling of the data, it is triple gluon exchange, 
$ggg$ term, which  describes the decrease  of the elastic differential cross-section 
past the dip in the tail region. This is a purely real contribution, which behaves as $\simeq (-t)^{-8}$ and could be energy independent. 

In \cite{Donnachie:2011aa}   three different types of data are analyzed: 
\begin{itemize}
\item $pp$ elastic scattering at the ISR, $\sqrt{s}=30.54\ {\rm GeV}$
\item HERA data for the proton structure function $F_2(x,Q^2)$ at small $x$, used to determine the hard and soft pomeron powers
\item the TOTEM $pp$ data for both the elastic differential distribution as well as the total cross-section.
\end{itemize}
The expression used to fit the LHC data is
\begin{eqnarray}
& A_{DL}(s,t)=\sum_{i=0,3}Y_ie^{-\frac{1}{2}i\pi \alpha_i(t)}(2\nu \alpha'_i)^{\alpha_i(t)}\\
& 2\nu=\frac{1}{2}(s-u)\ \ \ \ \ Y_i=-X_i\ (i=0,1,2)\ \ \ \ \ \  Y_3=iX_3\nonumber
\end{eqnarray}
with $X_0,X_1,X_2,X_3$ real and positive. The 4 Regge trajectories are parametrized as
\be
\alpha_i(t)=1+\epsilon_i+\alpha'_i t
\ee
with $\epsilon_i$ treated as free parameters and
\begin{itemize}
\item $\alpha_0(t)$ is the hard Pomeron, with \\ $\alpha'_0=0.1\ {\rm GeV}^{-2}$, which, they find, {\it works well}
\item  $\alpha_1(t)$ is the well known soft Pomeron, with\\ $\alpha'_1=0.25\ {\rm GeV}^{-2}$, 
\item $\alpha_2(t)$ is the degenerate $f_2,a_2$ trajectory, with \\ $\alpha'_2=0.8\ {\rm GeV}^{-2}$
\item  $\alpha_3(t)$ corresponds to  $\omega, \rho$ trajectory, with \\ $\alpha'_2=0.92\ {\rm GeV}^{-2}$
\end{itemize}
The normalization of the amplitude is 
\be
\sigma^{TOTAL}=s^{-1}\Im  m A_{DL}(s,t).
\ee
 The parameters $\epsilon_i$ are related to the proton structure function. Thus a simultaneous fit to the HERA data and to 
 LHC7 gives  the figures shown in Figs.~\ref{fig:DL2011-nohardPom} and \ref{fig:DL2011-hardPom}, where the differential 
 elastic cross-section is computed without and with a hard Pomeron, respectively.
\begin{figure}
\centering
\resizebox{0.5\textwidth}{!}{
\includegraphics{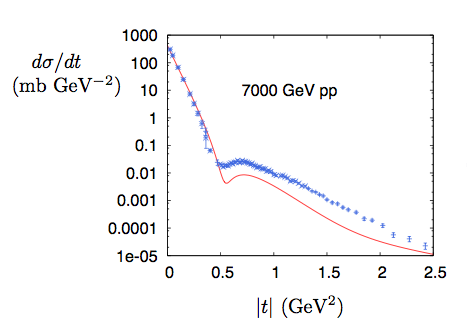}}
\caption{The elastic differential cross-section at LHC7, without inclusion of a  hard Pomeron  \cite{Donnachie:2011aa}. Figure is courtesy of the authors.}
\label{fig:DL2011-nohardPom}
\end{figure}

\begin{figure}
\centering
\resizebox{0.5\textwidth}{!}{
\includegraphics{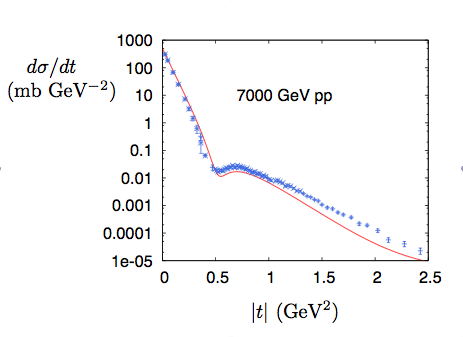}}
\caption{
The elastic differential cross-section at LHC7, with inclusion 
of a  hard Pomeron,  from \cite{Donnachie:2011aa}. Figure is courtesy of the authors.}
\label{fig:DL2011-hardPom}
\end{figure}
The difference between the two figures is due to the fact that, since the total cross-section 
cannot come out right without a hard pomeron, a hard Pomeron is added to obtain 
the total cross-section as given by TOTEM. This gives 
 the parameters for the hard pomeron as $\alpha'_0=0.1\ {\rm GeV}^{-2}$ and $\epsilon_0=0.362$. 

At large $t$, the authors add a real  term to the amplitude of the type $C  st^{-4}$ which 
they understand as due to triple gluon exchange \cite{Donnachie:1996rq,Donnachie:1979yu}. 
In order to make it finite at $t=0$ a possible  expression is proposed as
\be
\frac{C s}{(t_o - t)^4}.
\ee
It is  pointed out  in this paper  that  in order to correctly model the dip, both the real and the imaginary parts must become 
very small at the same value of $t$. 

 In the more recent contribution \cite{Donnachie:2013xia}, where no hard Pomeron is invoked, 
 the amplitude is  given as three single Pomeron exchange terms, one double exchange and a triple gluon term. 
 We show a representative fit of the LHC7 data from this paper in Fig. ~\ref{fig:DL2013}.
\begin{figure}
\resizebox{0.5\textwidth}{!}{
\includegraphics{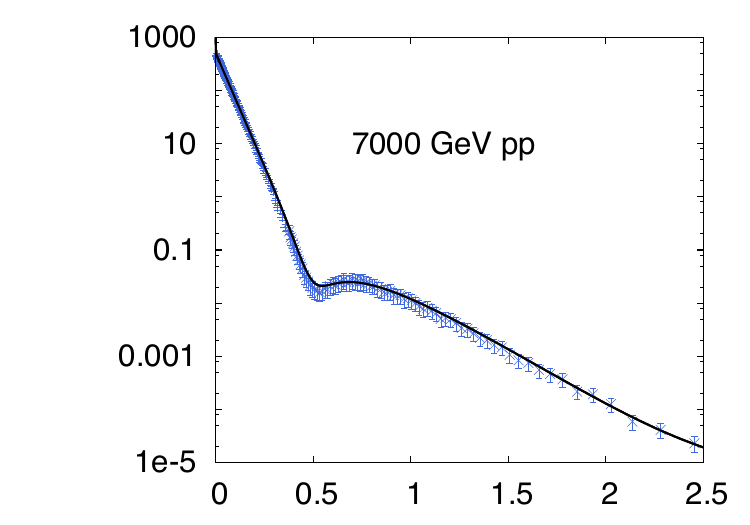}}
\caption{\label{fig:DL2013}The elastic differential cross-section at LHC7
 from \cite{Donnachie:2013xia}, revisited as described in the text. Figure is courtesy of the authors,
reprinted from  \cite{Donnachie:2013xia}, \copyright (2013) with permission from Elsevier.}
\end{figure}

 Each one of the three single-exchange terms is related to  a trajectory  $\alpha_i(t)$, with
 \bea
 A(s,t)=F(t)
 \big{[}
- \frac{
 X_P } {2\nu}
  e^{-\frac{1}{2}i\pi \alpha_{P}(t)}
  (2\nu \alpha^\prime_P)^{\alpha_P(t)} \nonumber \\
  -\frac{
 X_+ } {2\nu}
  e^{-\frac{1}{2}i\pi \alpha_+(t)}
 (2\nu \alpha^\prime_P)^{\alpha_+P(t)} \nonumber\\
 \mp \frac{
 X_- } {2\nu}
  e^{-\frac{1}{2}i\pi \alpha_-(t)}
 (2\nu \alpha^\prime_-)^{\alpha_-(t)}
  \big{]}
 \eea
{ for the $pp$/\pbarp \ amplitude, respectively, $\nu=(s-u)/2m$, and $F(t)$ is a form factor. The three trajectories are parametrized as}
 \be
 \alpha_i(t)=1+\epsilon_i +\alpha_i^\prime t
 \ee
 To these  single-exchange terms, a double Pomeron exchange, $PP$ term,   is then added. It corresponds to a trajectory
 \be
 \alpha_{PP}(t)=1+2 \epsilon_P+\frac{1}{2}\alpha^\prime (t)
 \ee 
but   the corresponding amplitude term is non just a power $s^{\alpha_{PP}(t)}$, and    additional logarithmic terms appear at the denominator. 

The tail of the distribution  is given by  triple gluon exchange \cite{Donnachie:1979yu}, 
and is such that at large $|t|$ one has
\be
g(t)=C\frac{t_0^3}{t^4}
\ee
A joining with the other parts of the amplitude is obtained by {\it trial and error}, as the authors say, and is parametrized with  
\be
\frac{C}{t_0}e^{2(1-t^2/t_
0^2)}
\ee
The small $|t|$ in the Coulomb region is included as 
\be
\mp \frac{\alpha_{EM}}{t}
\ee 
The description of data from ISR to LHC for $pp$ is very good, less so the description of \pbarp \ data, but this 
is common to many present fits. The values for the 12 parameters which give the best fit to total and elastic data 
are given in the paper, with in particular $\epsilon_P=0.110$.

\subsubsection{ The model by Schegelsky and Ryskin}\label{sss:schegelskyryskin}
In \cite{Schegelsky:2011aa}, the emphasis is on the small-$t$ behaviour
and a concern that the  elastic slope may not 
 be just  a simple linear power in $\log s$.
The authors start with   the usual  Regge and Pomeron parametrization for the elastic scattering amplitude, namely
\be
T_{ab}=F_a(t)F_b(t)C_P s^{\alpha_P(t)}+F_R(t)s^{\alpha_R(t)}
\ee 
where the first term corresponds to the Pomeron, the second to a Reggeon and would be negligible at high energy. The 
differential elastic cross-section at high energy thus takes the form
\be
\frac{d\sigma_{ab}}{dt}=\frac{\sigma_0^2}{16\pi}F_a^2(t)F_b^2(t)(\frac{s}{s_0})^{2\epsilon+2\alpha'_P t}
\ee
where the slope of the Pomeron trajectory accounts for the growth of the interaction radius caused by a long chain of 
intermediate (relatively low energy) interactions. 
Agreement with data, with an elastic slope given by
\be
B_{el}=B_0+2\alpha^{' eff}_P \log\frac{s}{s_0}
\label{eq:bel}
\ee
is obtained by assuming a Gaussian type behaviour for the form factors $ F_a^2(t)F_b^2(t)=exp(B_0t)$. 
The second term in Eq.~(\ref{eq:bel}) is supposed to be universal, and the value obtained by examining fixed target experiments,
 i.e. up to $\sqrt{s}=24 \ {\rm GeV}$, is given as $\alpha'=0.14\ {\rm GeV}^{-2}$. On the other hand, the original analysis by Donnachie and 
 Landshoff \cite{Donnachie:1983hf} at $\sqrt{s}=52.8\ {\rm GeV}$ would lead to $\alpha'=0.25\ {\rm GeV}^{-2}$. 

This discrepancy points to the fact that the  energy dependence of the elastic slope may be more complicated than a simple logarithm. 
The first idea is that as the energy increases, multiple interactions take place. In Regge-language, these multiple interactions are 
described by  multiple Pomeron diagrams. 

One important point is that in impact parameter representation, as the energy increases towards the black disk limit,  the imaginary 
part of the elastic scattering amplitude $\rightarrow 1$.  However, while 
asymptotically going to the black disk limit, at the periphery the amplitude is still growing. 
 This leads, according to the authors, to an effective growth of the slope. 
The reason is that the continuing increase of the amplitude at the periphery of the  impact parameter space implies an increasing radius. 
In this way the authors understand the discrepancy between their result for $\alpha'_P$ relative to Donnachie and Landshoff. They also 
provide another way to understand this behaviour, namely the interplay between one and two Pomeron exchanges, which have different 
signs, so that as the two Pomeron effect increases, the amplitude drops more rapidly.

Assuming an  $s$-dependence different  than the simple logarithm of  Eq.~(\ref{eq:bel}), may explain why
$\alpha_P^{'eff} \neq \alpha'_P$. The authors propose
\be
B_{el}=B_0+b_1 \log\frac{s}{s_0} + b_2 \log^2\frac{s}{s_0}
\label{eq:beleff}
\ee
so that,  if one  rewrites Eq.~(\ref{eq:beleff}) as
\be
B_{el}=B_0+\alpha_P^{'eff} (s) \log \frac{s}{s_0}\equiv B_0+ b_1 \log\frac{s}{s_0} + b_2 \log^2\frac{s}{s_0}
\label{eq:b2eleff}
\ee
one would obtain 
 $\alpha_P^{'eff} (s)=b_1+b_2\log s/s_0$. To determine the coefficients, the authors
plot the results from a series of experiments, as we show in the left panel of Fig.~\ref{fig:ryskin}.
 \begin{figure*}
 \resizebox{0.5\textwidth}{!}{
 \includegraphics{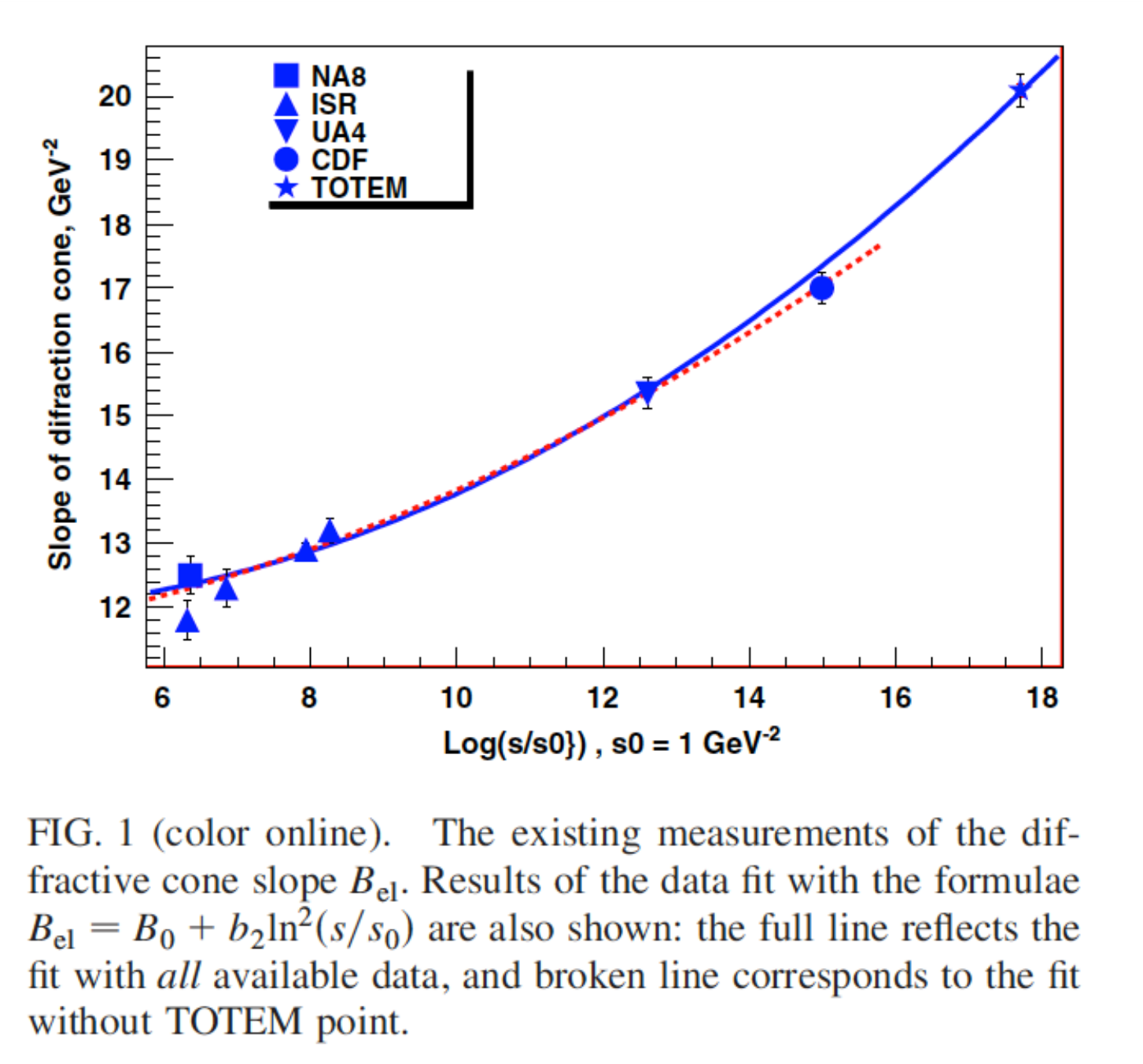}}
 \resizebox{0.5\textwidth}{!}{
  \includegraphics{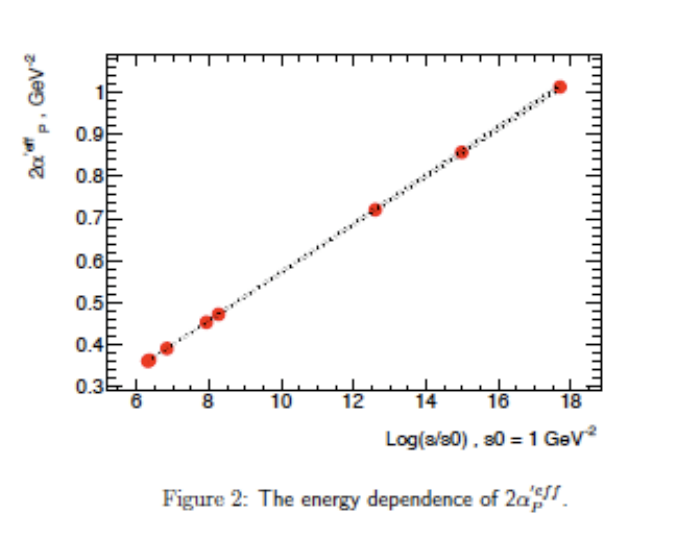}}
 \caption{\label{fig:ryskin} From   \cite{Schegelsky:2011aa}: at left, the effective slope 
 for the elastic differential cross-section from a set of experiments, compared with the 
 proposed effective parametrization of Eq.~(\ref{eq:beleff}). At right the energy dependence of $2\alpha_P^{'eff}$.
 Reprinted with permission from  \cite{Schegelsky:2011aa}, Figs.(1,2), \copyright (2012) by the American Physical Society.}
 \end{figure*}
For $s_0=1\ {\rm GeV}^2$, the coefficient $b_1$ is consistent with zero, while, for $b_2$, the authors obtain 
$b_2=0.037\pm 0.006\ {\rm GeV}^{-2}$. Since $b_1$ is consistent with zero, the authors propose to drop this 
term altogether and refit the data  with only a $(\log s/s_0)^2$ term. This fit gives  $b_2=(0.02860\pm 0.00050)\ {\rm GeV}^{-2}$. 
   
Their result  is then discussed in light of the fits done by Block and Halzen in \cite{Block:2011vz} and the possible 
saturation of the Froissart bound. Unlike Block and Halzen, the authors here do not think that the black disk limit 
is yet reached, and believe  the proton to be  still relatively transparent,
 so that  $\sigma_{total}=2 \pi R^2$, with $B_{el}=R^2/4$,   is not yet at  its geometric value. One word of caution 
 is however put forward by these authors, namely that the non linear logarithmic rise of the elastic slope is 
 basically determined by the recent TOTEM measurement of $B(7 \ {\rm TeV})\sim 20\ {\rm GeV}^2$. Up to and including  
 the {\rm TeV}atron measurement, a linear logarithmic energy rise for  $B_{el}$ is in fact actually still compatible, 
 as one can see from Fig.~\ref{fig:ryskin}.

 \vskip 0.3 cm
\par\noindent
\subsection{ Analyses with Pomeron, 
{Odderon}  and Regge exchanges}\label{ss:pomoddregge}
\par\noindent
The dip structure has been connected to  C-odd  exchanges in the t-channel, phenomenologically referred to as the Odderon. 
The existence of such a state is predicted by QCD and has been advocated  in particular  by Nicolescu, also  
 in collaboration with other authors \cite{Gauron:1993ic,Nicolescu:2007ji}.

 As we have mentioned in Sect.~\ref{sec:models}, the QCD treatment of the Odderon    began in 1980 
 \cite{Bartels:1980pe}  \cite{Kwiecinski:1980wb}. It was  examined  extensively by Bartels and others in non-abelian gauge 
 and color glass condensate theories.  Present QCD studies of  the Odderon trajectory 
focus on  NLO contributions \cite{Bartels:2013yga} and properties in the strong coupling limit\cite{Brower:2013jga}. 
Still, the major question to face is whether and how to detect its presence in LHC experiments and this takes us to the many 
phenomenological analyses which include Reggeons, Pomerons and Odderons..

\vskip 0.3 cm
\par\noindent
\subsubsection{Phenomenological analyses with and without the Odderon contribution} \label{sss:odderon}
\par\noindent 
 To  discuss  the proposal to detect the odderon at RHIC and LHC by Avila, Gauron and Nicolescu (AGN2006) \cite{Avila:2006wy}, 
 we  start with  the model by Avila, Campos, Menon and Montanha  (ACMM2006) which incorporates both the Froissart limiting 
 behavior as well as Pomeron and Regge exchanges \cite{Avila:2006ya}. 

 In   \cite{Avila:2006ya}, the elastic amplitude is normalized so that
 \begin{align}
\frac{d\sigma}{dq^2}=\frac{1}{16\pi s^2}|\Re e F(s,q^2)+i \Im mF(s,q^2)|^2\\
\sigma_{tot}(s)=\frac{\Im m F(s,0)}{s}\\
\end{align}
Although, unlike the total cross-section, both the real and the imaginary parts of the scattering amplitude
enter the elastic differential cross-section, in the small $t$-region, $|t|<0.2 \ {\rm GeV}^2$, the cross-section can be expected to be  
still mostly imaginary.
The  steep decrease in this region  
 region can be empirically  parametrized  in terms of an exponential, i.e. one writes
\be
 \Im mF(s,q^2)\approx \alpha(s) e^{-\beta(s)q^2}
\ee
Experimental data suggest that their energy dependence can be parametrized   in terms of polynomials in $\log s$.  
The difference between $pp$ and \pbarp \ is then introduced as :
\bea
\frac{\Im m F_{pp}(s,q^2)}{s}&=&\sum_{i=1}^{n}\alpha_i (s) e^{-\sum_{i=1}^n \beta_i(s)q^2}\nonumber\\
\frac{\Im m F_{p{\bar p}}(s,q^2)}{s}&=&\sum_{i=1}^{n}{\bar \alpha}_i (s) e^{-\sum_{i=1}^n {\bar \beta}_i(s)q^2}\nonumber\\
\alpha_i(s)&=&A_i+B_i\log s+C_i\log^2 (s)\nonumber\\
{\bar \alpha}_i(s)&=&{\bar A}_i+{\bar B}_i\log s+{\bar C}_i\log^2 (s)\nonumber\\
 \beta_i(s)&=&D_i+E_i\log s\nonumber\\
{\bar \beta}_i(s)&=&{\bar D}_i+{\bar E}_i\log s
\label{Avila1}
\eea
The above parametrization implies $10n -1$ parameters, having imposed the constraint
\be
\sum_{i=1}^n(C_i -{\bar C}_1)=0
\ee
which is valid  
when the Froissart bound is reached and can be considered a  generalized form of the Pomeranchuk theorem \cite{Avila:2006ya}. 
 
Concerning the real part of the amplitude, the authors make use of the first order derivative dispersion relations 
for the even/odd amplitudes, defined as
\bea
F_{pp}(s,q^2)&=&F_+(s,q^2)+F_-(s,q^2)\\
F_{{\bar p}p}(s,q^2)&=&F_+(s,q^2)-F_-(s,q^2)
\eea
which lead to
\bea
\frac{\Re e F_+(s,q^2)}{s}=\frac{K}{s}+\frac{\pi}{2}\frac{d}{d\log s}\frac{\Im m F_+(s,q^2)}{s}\\
\frac{\Re e F_-(s,q^2)}{s}=\frac{\pi}{2}(1+\frac{d}{d\log s}\frac{\Im m F_-(s,q^2)}{s})
\eea
The validity of the above expressions only extends to a region $0<q^2< q_{max}^2$ and depends on the value chosen 
for $q_{max}$. In the subsequent fit to the experimental data, different values of $q_{max}$ are considered, leading to different parameter sets.

The above parametrization, applied to the vast set of lower energy data up to the Tevatron, gives the result shown in 
Fig.~\ref{fig:Avilaparam}. We see that,  for the energy $\sqrt{s}=14\ {\rm TeV}$, the parametrization  does not show existence of a dip.
\begin{figure}
\resizebox{0.5\textwidth}{!}{%
  \includegraphics{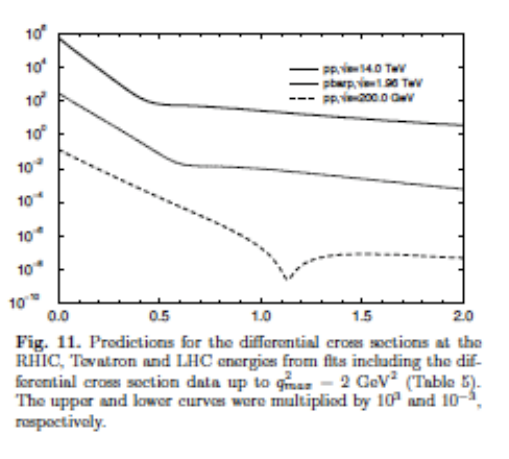}}
\caption{Predictions for the elastic differential cross-section by Avila et al. \cite{Avila:2006ya}
 at LHC and the Tevatron for $pp$ and \pbarp . Reprinted with permission  from \cite{Avila:2006ya} \copyright (2006) Springer.}
\label{fig:Avilaparam}       
\end{figure}

The above approach has been discussed because its {\it no dip at LHC} result in contrast with the TOTEM measurement, 
can falsify some of the assumptions, or lead to the need to introduce the Odderon, as done in a subsequent work with 
Nicolescu, as we discuss next.

 
Avila,  Gauron and Nicolescu  (AGN2006) \cite{Avila:2006wy} 
examine the data in light 
of the possible  existence of the {\it odderon}, a QCD effect corresponding to a singularity in the complex J-plane 
at  $J=1$ at $t=0$ in the amplitude $F_-$, which is odd under crossing. The odderon is considered a non-leading 
QCD effect,  which can only be detected in the elastic differential cross-section. In this analysis, its contribution is 
parametrized according to the properties of a number of  Regge, Pomeron and Odderon trajectory exchanges, 
in the context of total cross-sections saturating the Froissart bound. With such parametrization, at LHC energies a  dip
is expected both for $pp$ and \pbarp \ scattering, something which cannot be proved experimentally at LHC. 
In \cite{Avila:2006wy} the suggestion is advanced to try to check the presence of the dip at RHIC, namely 
at $\sqrt{s}=500 \ {\rm GeV}$.   

The AGN2006  analysis describes the ISR data at $\sqrt{s}=52.8\ {\rm GeV}$ where both $pp$ and \pbarp \ elastic differential 
cross-sections were measured. At this energy, while $pp$ shows a dip, \pbarp \ does not, and  the model with 
the Odderon describes these features well.  At $\sqrt{s}=200\ {\rm GeV}$ and $500\ {\rm GeV}$, where only $p{\bar p}$ scattering 
was  present, the dip has morphed in a break in 
the slope, with not much difference between the two processes, with the \pbarp \ curve remaining slighly higher.  
At $900\ {\rm GeV}$, the break is predicted to be more pronounced and at the Tevatron energies there is the hint of a dip developing, with  
\pbarp \ points remaining higher. Until recently,  experimental data at the Tevatron did not go beyond 
$|t|\simeq 0.6\ {\rm GeV}^2$. Presently published data from the D0 Collaboration \cite{Abazov:2012qb},  
plotted  in Fig. ~\ref{fig:bp-Tevatron}, indicate  a pronouced  break in the slope (as at \SpbarpS\ ), possibly even a  dip. 
This would correspond to the fact that the terms from Regge exchange are becoming increasingly negligible.  

\begin{figure}
\centering
\resizebox{0.5\textwidth}{!}{
\includegraphics{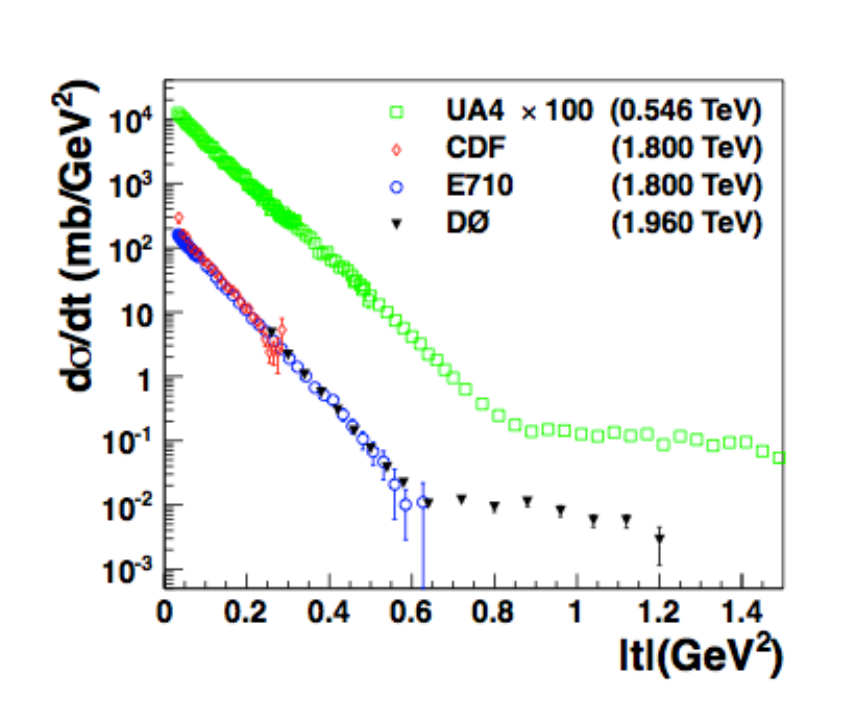}}
\caption{Data for  the elastic differential cross-section for \pbarp\  scattering from the three experiments at the Tevatron,  
compared with UA4 data, from \cite{Abazov:2012qb}.
Reprinted with permission from \cite{Abazov:2012qb}, Fig.(12), \copyright (2012) by the American Physical Society.}
\label{fig:bp-Tevatron}
\end{figure}
 There are no predictions for LHC7, but at $\sqrt{s}=14\ {\rm TeV}$ the dip is now predicted both for $pp$ and \pbarp. 
\begin{figure}
\resizebox{0.5\textwidth}{!}{%
  \includegraphics{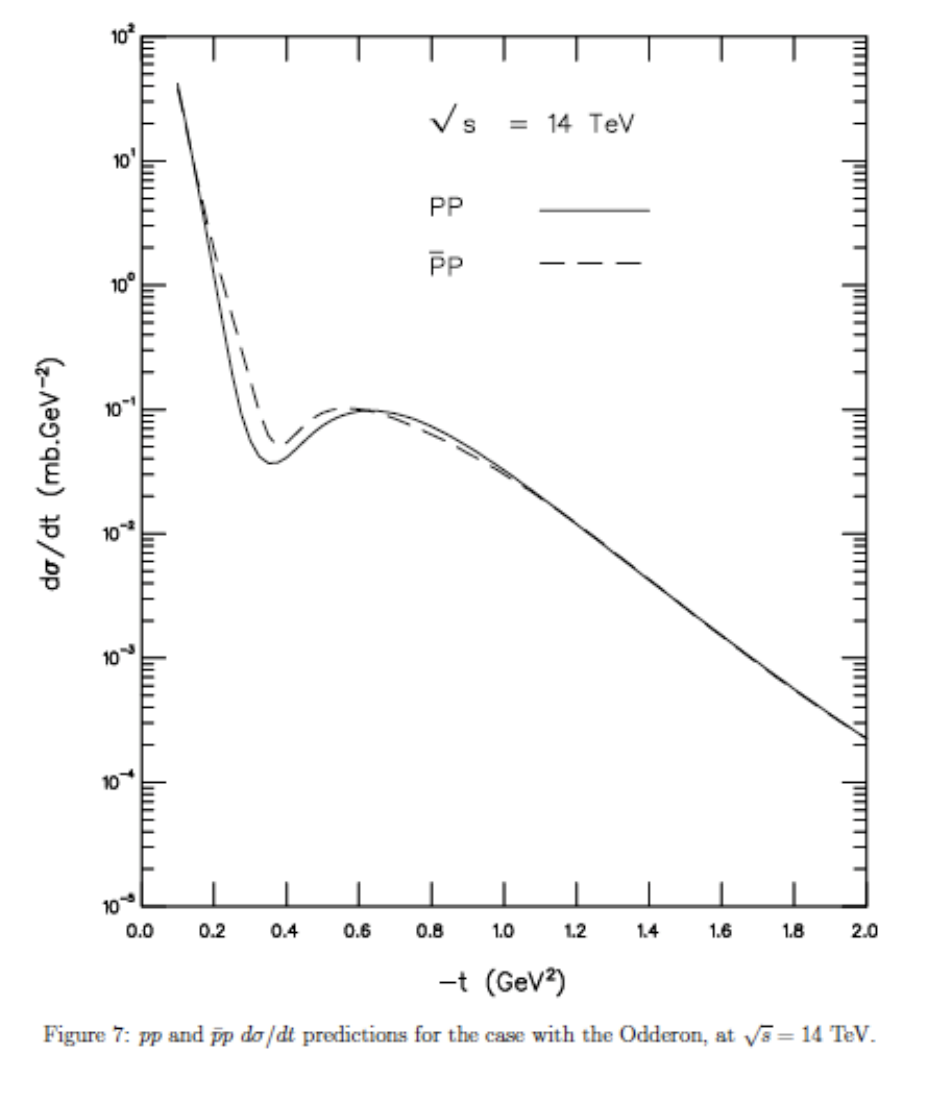}
  
 }
\caption{Predictions for the elastic differential cross-section by Avila et al. \cite{Avila:2006wy} at LHC for $pp$ and \pbarp .
 Reprinted with permission from \cite{Avila:2006wy}, \copyright (2006) Springer. }
\label{fig:odderon-14tev}       
\end{figure}
At LHC running at $\sqrt{s}=14 \ {\rm TeV}$ (LHC14) the position of the dip
is expected at $|t|\sim 0.35 \ {\rm GeV}^2$, to be compared with the position at the ISR at $|t|\simeq 1.3\ {\rm GeV}^2$ 
and the position at LHC7 reported by TOTEM at $|t|\sim 0.53\ {\rm GeV}^2$.

We will discuss now in detail  the parametrization in \cite{Avila:2006wy}. The starting point is the maximal asymptotic 
behaviour of the total cross-sections, 
consistent with data, i.e.
\bea
\sigma_{tot}&\propto & \log^2 s\\
\Delta \sigma(s)&\equiv &\sigma_{tot}^{{\bar p}p}-\sigma_{tot}^{pp}\propto \log s, \  \  \  \  \ as \ \ \ s\rightarrow \infty 
\label{eq:maximal}
\eea
According to \cite{Avila:2006wy}, the choice of a maximal behaviour of the total cross-section is not necessarily 
implying that the imaginary part of $F_-(s,0)$ is also maximal, i.e. that $F_-(s,0)\sim \log s$, but, the authors argue, 
it is the natural choice that strong interactions be as strong as possible. With this assumption, 
a parametrization for the two, even and odd crossing, amplitudes is prepared and then
determined by comparison with the data. 

These authors start with a general form for the hadronic amplitudes  compatible with Eq.~(\ref{eq:maximal}). 
Their strategy is to consider both the existence and the non existence of the odderon, parametrizing all the existing data 
(832 points when the paper was written) for both processes $pp$ and \pbarp \ 
for the quantities  $\sigma_{tot}(s), \ \rho (s)$ and $d\sigma_{el}(s,t)/dt$. After choosing the best description for the existing data, 
the model is  applied to predict future data and to recommend experiments which might measure the difference between 
the two amplitudes $F_+(s,t)$ and $F_-(s,t)$. 
The expressions for these even and odd amplituds are written as sum of Pomeron and Reggeon contributions, 
following general theorems given in 
\cite{Gauron:1986nk,Gauron:1989cs,Auberson:1971ru}, 
namely
\bea
F_+(s,t)&=&F_+^H(s,t)+F_+^P(s,t)+\nonumber \\
& &F_+^{PP}(s,t)+F_+^R(s,t)+F_+^{RP}(s,t)\nonumber \\
F_-(s,t)&=&F_-^{MO}(s,t)+F_-^O(s,t)+\nonumber \\
& & F_-^OP(s,t)+F_-^R(s,t)+F_-^{RP}(s,t)
\eea
where the upperscript $H$ correspond to a polynomial in $\log {\bar s}$, with 
\be
{\bar s}=(\frac{s}{s_0}) e^{-\frac{1}{2}i\pi}
\ee
maximally increasing as  $\log^2(s)$ 
with  exponential $t-$ dependence, $P$ is the contribution from the Pomeron Regge pole, $PP$ the 
Pomeron -Pomeron Regge cut, $R$ corresponds to the secondary Regge trajectories, $RP$ the 
reggeon-Pomeron cut. For the odd under crossing amplitude, $MO$ represents the maximal Odderon contribution, 
also increasing as $\log^2 s$,  $O$ is a minimal Odderon Regge pole contribution with $\alpha(0)=1$, $OP$ 
is a minimal Odderon-Pomeron cut, $R$ a secondary 
Regge trajectory associated with the particles $\rho(770)$ and $\omega (782)$, $RP$ a reggeon-Pomeron Regge cut. 

In 
\cite{Avila:2006wy} the no odderon case is obtained by choosing parameters such that all the three amplitudes with  
contributions from the odderon are zero,
 i.e. $F_-^{MO}(s,t)=F_-^O(s,t)=F_-^OP(s,t)=0$. Because of the contribution from the Regge and the Reggeon-Pomeron cut, with 
\begin{align}
\frac{1}{s}F_-^R(s,t)=-C^-_R\gamma_R^-(t)e^{\beta_-^R t}
[i+\tan(\frac{\pi}{2} \alpha_-^R(t) )]
(\frac{s}{s_0})^{\alpha^-_R(t)-1} \nonumber \\
\frac{1}{s}F_-^{RP}(s,t)=
(\frac{t}{t_0})^2C^-_{RP}e^{\beta_-^{RP} t}
[
\sin(
 \frac{\pi}{2} \alpha_-^{RP}(t)
 )+\nonumber \\
 +i \cos(
\frac{\pi}{2} \alpha_-^{RP}(t)
)]
 \times \frac{(\frac{s}{s_0})^{\alpha^-_{RP}(t)-1}}{\log[(s/s_0)exp(-i\frac{\pi}{2})]} 
 \end{align}
the amplitude $F_-$ is not zero, but it fails to give a good description of the ISR data at $\sqrt{s} = 52.8\ {\rm GeV}$.
The results can be summarized with the 
fact that, starting with the ISR data, the case without odderon does not reproduce the data at $\sqrt{s}=52.8\ {\rm GeV}$ as well as it can do   
with the Odderon. According to the authors  this is due to the fact that the $t$-dependence of these two remaining terms  is   constrained by 
the parameters of the Regge and Pomeron trajectories, which are non-leading and fail to interfere with the even amplitude $F_+$ in the correct way.

Turning to the other case, namely the odderon being present, the model can count on  twelve  more parameters, which add to the 23 parameters 
determining the $F_+$ amplitude.  

Numerically, it is now possible to do a good fit to   ISR data which show the clear  difference between 
\pp  \ and \pbarp, with and without the dip.
At LHC 14 TeV, the 
dip is fully in place, but, unlike intermediate energies,  it would be present in both processes, and the difference between 
the two cross-sections is small.  The predicted dip is 
at $|t|\sim 0.4\ {\rm GeV}^2$ and the cross-section after the dip rises to $\sim 10^{-1}\ {\rm mb/ GeV}^2$.

A pattern of oscillations is observed to develop  in the difference $\Delta (\frac{d\sigma}{dt}(s,t))$ for $pp$ and \pbarp .
with two minima, one around   $|t|=0.1\ {\rm GeV}^2$ and the 
other around $|t|=1.1\ {\rm GeV}^2$.  In  Fig.~\ref{fig:nicolescu-ISR} we show the fit to the differential cross-section at ISR and the predicted difference.
\begin{figure*}
\resizebox{0.9\textwidth}{!}{
\includegraphics{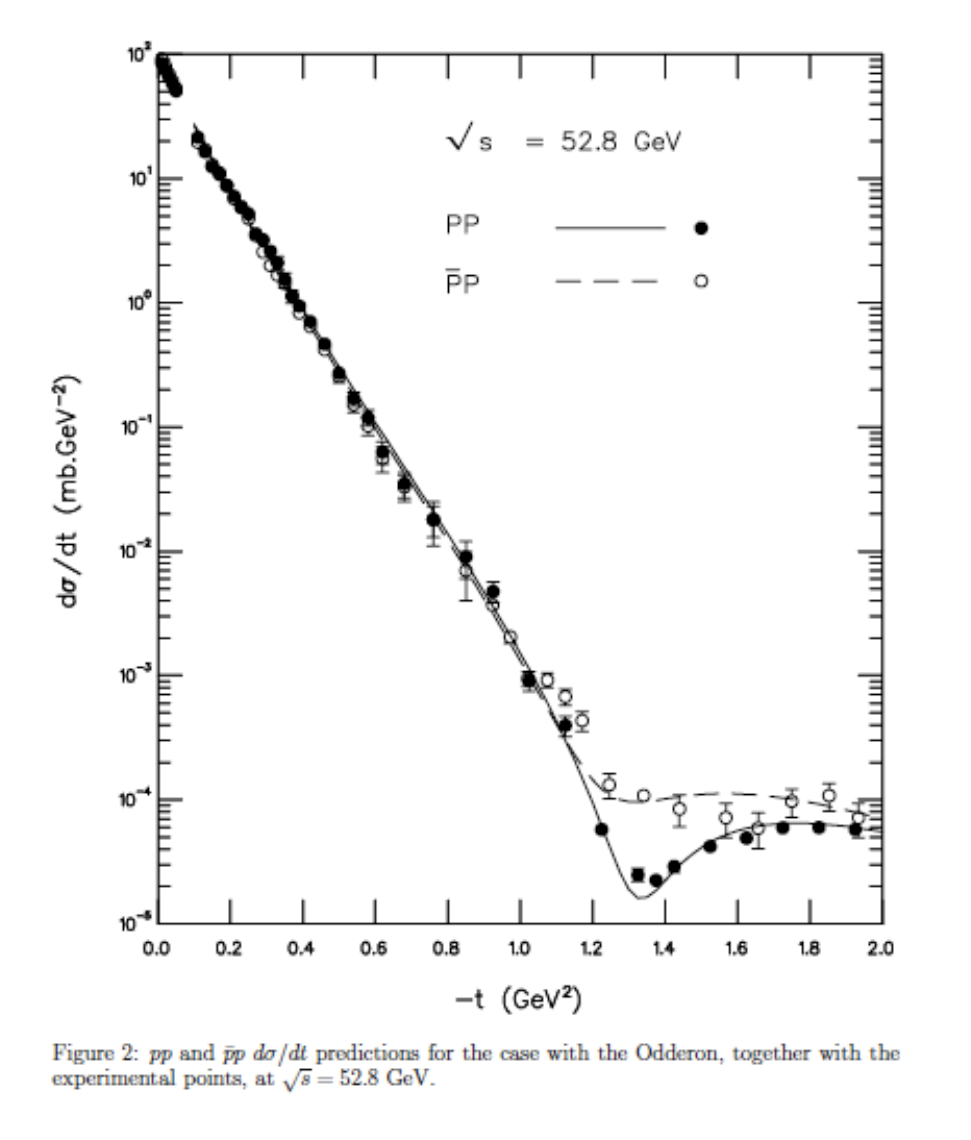}
\includegraphics{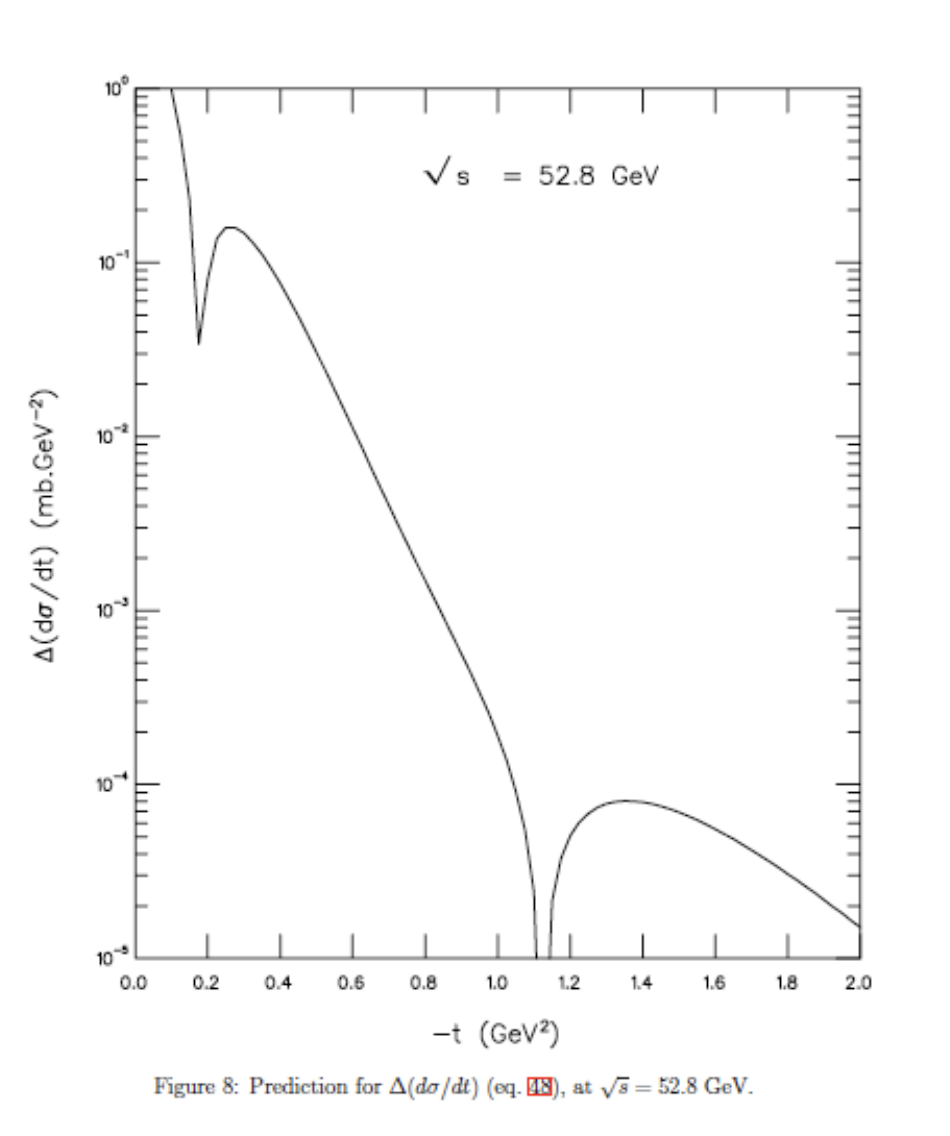}} 
\caption{The fit for the differential cross-sections for $pp$ and \pbarp \ at left, and, at right,  the predicted oscillation in the difference of the cross-sections. 
Both figures are from Avila, Gauron and Nicolescu  \cite{Avila:2006wy}. Reprinted with permission from \cite{Avila:2006wy} \copyright (2006) Springer.}
\label{fig:nicolescu-ISR}
\end{figure*}

In a subsequent paper \cite{Martynov:2007kn} the model is extended to higher values of $0<|t|<16\ {\rm GeV}^2$. 
They define as the {\it standard data set} the data set 
proposed by Cudell, Lengyel and Martynov \cite{Cudell:2005sg}.
To generalize the results beyond the range $|t|=2.6 \ {\rm GeV}^2$ new terms are added to the parametrization originally proposed, namely
\begin{enumerate}
\item $N_{\pm}$ which behave like $t^{-4}$  and growing like $\log s$. The motivation for $N_-$ was advanced to include the 
Odderon  3-gluon exchange, with elementary gluons.
\item a term $Z_{-}$ which has a role as ``cross-over term" for very small $t\sim 0.16\ {\rm GeV}^2$
\item two linear functions of the type $1+A_{MO/O}t$ multiplying the maximal odderon and the Odderon pole  term, 
and which  are necessary to describe the 
smallness of the Odderon forward coupling at present energies. 
\end{enumerate}
 With these many parameters the fit is now very good and the model predicts a dip at 14 TeV around $|t|\sim 0.5\ {\rm GeV}^2$ 
 and a shoulder around $|t| \sim 0.8 \ {\rm GeV}^2$.
 They note that at LHC energies the exponential behaviour of $d\sigma/dt$ is no longer valid.

A different set of parametrization has been considered in \cite{Selyugin:2012np,Selyugin:2015pha,Selyugin:2016lls} that
includes Coulomb interference region as well as large $t$ values, in terms of about 10 parameters.
\vskip 0.3 cm
\par\noindent
\subsubsection{
Jenkovszky's Pomeron/Odderon Dipole model }\label{sss:jenkovszky}
\par\noindent
We now discuss the  contribution by 
Jenkovszky with different collaborators from \cite{Jenkovszky:2012tfa,Jenkovszky:2011hu} 
and references therein.  The  model described in \cite{Jenkovszky:2011hu}  
extends  the Donnachie and Landshoff model,  to include the 
dip-bump structure of the differential elastic cross-section, non-linear Regge trajectories, a  possible Odderon, 
i.e. C-odd asymptotic Regge exchange. The extension should also be such as to be compatible with $s$ and 
$t$-channel unitarity. The authors recall  that the first attempt to reproduce the dip-bump structure was done 
through the Chou and Yang model \cite{Chou:1983zi}, whose major drawback is that it does not contain any explicit 
energy dependence. As mentioned, at the beginning of this section, the Chou and Yang model describes an impact 
parameter distribution obtained from the proton 
electromagnetic form factor. In the Chou and Yang model, just as in the eikonal models in general, the dip-bump structure is obtained because 
of the zeros of the imaginary part of the amplitude, induced by the Fourier transform from $b-$ space. 
In \cite{Jenkovszky:2011hu},    the pattern of the  
  parametrization  proposed by Barger and Phillips in 1973 \cite{Phillips:1974vt}, already discussed, is followed. The amplitude is normalized so that
  \be
\frac{d\sigma}{dt}=\frac{\pi}{s^2}|A(s,t)|^2\ \ \ \ and \ \ \ \ \sigtot=\frac{4\pi}{s}\Im m A(s,t=0)
\ee
with   constants to be determined from the fits, and the amplitudes written such as embodying  the Regge-Pomeron 
description, itself reflecting  the collective processes participating to high energy scattering. 

The  model  in \cite{Jenkovszky:2011hu}  is based on four contributions to the scattering amplitudes for $pp$ and \pbarp, \ 
and the amplitudes are written as a sum of Regge pole amplitudes, namely 
\be
A^{\pbarpmath}_{pp}=A_P(s,t)+A_f(s,t)\pm [A_O(s,t)+ A_\omega(s,t)]
\ee
where $P,O$ stand for the Pomeron and Odderon contribution respectively $C$-even and $C$-odd with intercept 
$\alpha(0)>1$, $f$ and $\omega$, again 
$C$-even and $C$-odd and $\alpha(0)<1$.
The Pomeron and the Odderon are treated on the same footing, their parameters to be determined by the fit to the data. 
For Pomeron and Odderon, a dipole 
expression is chosen, with a non-linear trajectory, i.e.
\begin{align}
A_P(s,t)=i\frac{a_P}{b_P}\frac{s}{s_0}[r_1^2(s)e^{r_1^2[\alpha_P-1]}-\varepsilon_Pr_2^2(s)e^{r_2^2[\alpha_P-1]}]\\
A_O(s,t)=\frac{a_O}{b_O}\frac{s}{s_0}r_O^2(s)e^{r_O^2[\alpha_O-1]}
\end{align}
with $r_1^2=b_P+\log (s/s_0)-i\pi/2, \ r_2^2= \log (s/s_0)-i\pi/2, \ r_O^2=b_O+\log (s/s_0)-i\pi/2$. 
For the Pomeron trajectory, both non linear and linear cases are considered, while the Odderon is taken to lie on a 
linear trajectory, similar to the reggeon trajectories for $f$ and $\omega$. The adjustable parameters are the trajectory 
slopes and intercepts, the constants $a_{P/O}$ and $b_{P/O}$ and $\varepsilon$, which quantify 
the presence of absorption.

The fits to the usual set of observables, differential elastic cross-section, total cross-section and $\rho$
 parameter, lead the authors to conclude that the Odderon contribution is necessary in order to correctly describe the 
 dip-bump structure from ISR to the Tevatron.  
 It should be noticed that in this model there are no wiggles beyond the first dip, since this model does not use the 
 eikonal form for the amplitude, but directly 
 parametrises it in the $(s,t)$ plane. For LHC, we reproduce their predictions in Fig.~\ref{fig:jenk}. 
 \begin{figure}
 \resizebox{0.5\textwidth}{!}{
 \includegraphics{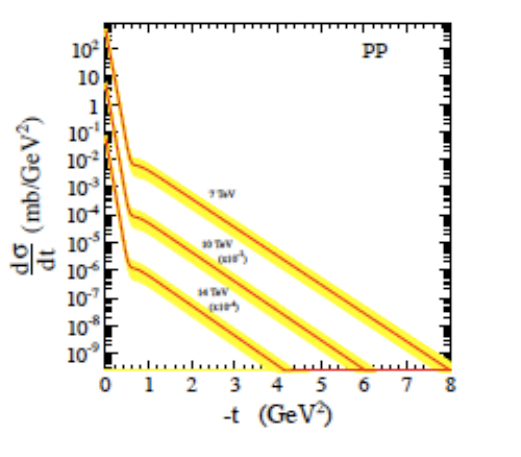}}
 \caption{Prediction from the model by Jenkovszky et al. \cite{Jenkovszky:2011hu} for the elastic differential cross-section 
 at LHC, with linear Pomeron and 
 Odderon trajectories. The band indicates uncertainties in the fit procedure. Republished with permission of L.  Jenkovszky,  from \cite{Jenkovszky:2011hu}, \copyright (2011) World Scientific; permission conveyed through Copyrigt Clearance Center, Inc.}
 \label{fig:jenk} 
 \end{figure}
 More particularly, the conclusions from this model in 
 \cite{Jenkovszky:2011hu} 
 are the following:
 \begin{enumerate}
 \item A single shallow dip is expected at LHC followed by a smooth behaviour
 \item The Odderon, described as a dipole, 
 with a positive intercept and almost flat behaviour in $t$ is indispensable 
 to describe the dip-bump region 
 \item the diffractive minimum in \pp\  at LHC7 is expected at $|t|=0.65\ {\rm GeV}^2$, receding to $0.6\ {\rm GeV}^2$ at LHC14
 \item the contribution from the non-leading Regge trajectories can be neglected in the  LHC  region.
 \end{enumerate}
Recently, the possibility to extract the odderon and the pomeron contribution from \pp\ and \pbarp data \ has 
been revisited in \cite{Ster:2015esa}. After early results from LHC8 have been published,  the model has been 
applied \cite{Fagundes:2015vva} to include a threshold singularity in the Pomeron trajectory, requested by 
$t-$channel unitarity  and related to the pion pole to account for  the non linear slope at very small $-t$-values, 
as  reported by the TOTEM collaboration \cite{Antchev:2015zza}. The possibility that the amplitude reflects a 
square root singularity  was suggested long time ago \cite{Anselm:1972ir,CohenTannoudji:1972gd}, and is 
also present in the work by Khoze, Martin and Ryskin, discussed in \ref{sss:KMRel}. 
\subsection{Eikonal models driven by Pomeron exchanges, parton dynamics and QCD inspired inputs}\label{ss:eikonaldsigdt}
Eikonal models  allow to satisfy unitarity and comply with the asymptotic requirements of the Froissart bound. 
Models using this formulation differ depending on the particular dynamical input determining  the eikonal function. 
We shall briefly illustrate some results from models which include  partonic descriptions of the proton, 
such as the Islam model,   the so-called
 Aspen model, and resummation of many  Pomeron-like exchanges. 
We also include in this subsection the eikonal based model by 
Bourrely, Soffer and TT Wu. In a separate subsection we shall illustrate  QCD models which specifically include 
diffractive processes and derive their input from QCD evolution equations, notably the BFKL approach. 
 
\subsubsection{
Quarks and gluons in the Islam model}\label{sss:islam}
\par\noindent

A  model based on the eikonal representation but reflecting the internal structure of hadrons 
is given by  the Islam model \cite{Islam:2000nc,Islam:2004ke,Islam:2013cba}.

In Islam's original model for elastic scattering, the scattering is described as a two component process: 
the first giving rise to diffractive scattering where the two pion clouds surrounding the 
scattering nucleons interact with each other, whereas the second hard scattering 
process, dominating at large scattering angle, takes place via vector meson 
exchange, while the pion clouds independently interact.

In a more recent paper \cite{Islam:2004ke} the model evolves into  three components, a soft cloud, a hard exchange 
at low c.m. energy and a hard component at high energy, identified with the BFKL Pomeron. Let us see how this picture 
is realized, also based on Luddy's presentation at Blois 2009 \cite{Islam:2010rpa}.
This model is lately referred to as {\it Condensate enclosed Chiral-bag Model}, and can be seen pictorially from a 
drawing of this presentation,  Fig.~\ref{fig:luddy}.
\begin{figure}
\resizebox{0.5\textwidth}{!}{%
  \includegraphics{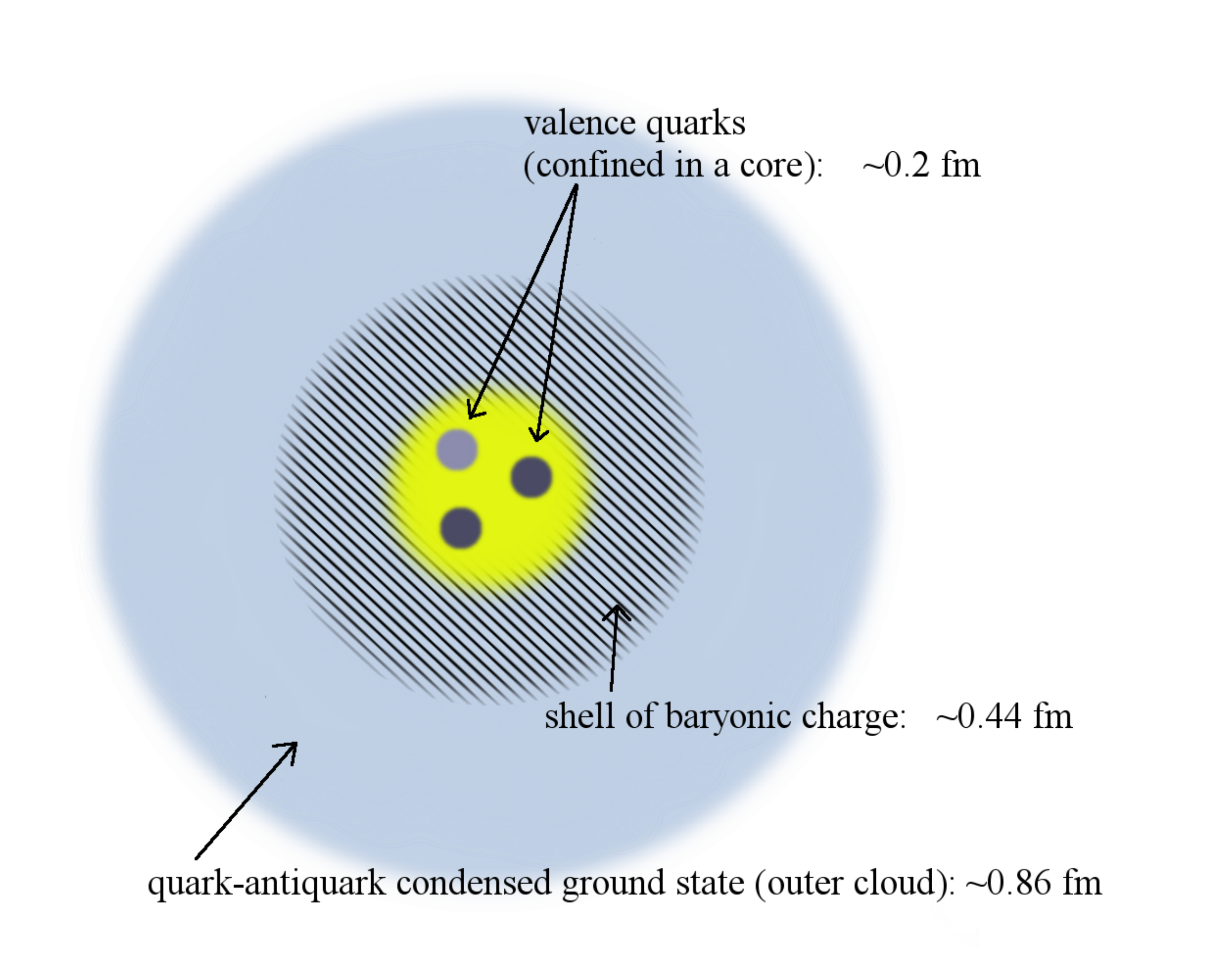}
}
\caption{Pictorial representation of the Condensate enclosed Chiral-bag model from Luddy's presentation at Blois 2009. Reprinted  from  \cite{Islam:2010rpa}, in CERN-PROCEEDINGS-2010-002, \copyright (2010) CERN. Figure is courtesy of the authors.}
\label{fig:luddy}       
\end{figure}
In this  version of the model \cite{Islam:2004ke}, a $qq$ hard scattering term brings in four new parameters, namely the relative strengths 
between the $\omega$ exchange term and the hard $qq$ term, the hard pomeron intercept $\alpha_{BFKL}=1+\omega$, 
the black disk radius $r_0$ and the mass $m_0$ which determines the size of the quark bag.
 In total this model has 17 parameters 
which can be fixed giving a quite satisfactory description of the total and differential cross-section, at various c.m. 
energies and of the $\rho$ parameters. 

In \cite{Islam:2010rpa,Islam:2007nr}, as in the previous papers, the proton  is described through three basic elements: 
an external cloud of  $q {\bar q}$ pairs (sea quarks), 
an inner shell of   baryonic  charge and a central quark-bag containing the valence quarks.  The external cloud and the 
inner shell have an obvious connection to 
QCD phenomenology, a pion cloud or some gluon condensate and the valence quarks, while
 one nucleon probes the  baryonic charge of the other via $\omega$-exchange.   Thus the picture is:

\begin{enumerate}
\item At very small values of the momentum transfer, the scattering is diffractive and we see the interaction of one cloud of one nucleon interacting 
with the  cloud of the other nucleon
\item then the $\omega$-exchange starts dominating
\item at even higher values, it is quark-quark scattering which takes over, i.e. pQCD.
\end{enumerate}

The diffractive contribution 
\be
T_D(s,t)=ipW\int b db \ J_0(bq)\Gamma_D(s,b)
\ee
is built phenomenologically with a diffraction profile function given as
\bea
\Gamma_D(s,b)&\equiv& 1-e^{i\chi(s,b)}\nonumber \\
 &=& g(s)[\frac{1}{1+e^{(b-R(s))/a(s)}}+\frac{1}{1+e^{-(b-R(s))/a(s)}}-1]\nonumber
\label{eq:islamdiffraction} \eea
 The above profile function renders the asymptotic behaviour of the total cross-section through the function 
 $R(s)=R_0+R_1(\log s -\frac{i\pi}{2})$, 
 $a(s)=a_0+a_1(\log s-\frac{i\pi}{2})$with $g(s)$ an even crossing energy dependent coupling constant. 
 Through the assumed energy dependence of the 
 radius $R(s)$, the function $\Gamma_D(s,b)$ 
 gives rise to an asymptotic energy dependence which saturates the Froissart bound. It is thus possible to  obtain the 
 following results, {with corresponding asymptotic theorems in italics} :DE
 \bea
 \sigma_{tot}(s)&\sim &(a_0+a_1\log s)^2 \  Froissart-Martin\  bound \nonumber \\
 \rho(s)& \simeq& \frac{\pi a_1}{a_0+a_1\log s}\ derivative \ dispersion \ relations \nonumber \\
 T_D(s,t)&\sim& i s \log^2 s f(|t|\log^2s) \  AKM \ scaling\nonumber \\
 T_D^{pp}(s,t)&=&T_D^{p\pbarpmath}(s,t)  \ crossing \ even\ property
 \eea
where  AKM  stands for Auberson, Kinoshita and Martin scaling \cite{Auberson:1971ru}.
The contribution from    $\omega$-exchange is written as
  \begin{equation}
  T_\omega(s,t)=\pm i{\hat \gamma}  e^{i{\hat \chi}(s,b)}s \frac{F^2(t)}{m^2_\omega-t}
  \end{equation}
  where the first factor represents the absorptive effect from soft hadronic interactions in $\omega$ exchange , 
  and the $\pm$ refers to \pbarp \ and \pp \ scattering.  
  The squared form factor in the amplitude indicates that one is probing two baryonic charge distributions. Finally the 
  last term, which dominates at larger $|t|$ values, is due to valence quark scattering, with the quarks interacting via  
  reggeized gluon ladders, described through the BFKL Pomeron. This initially single 
  valence quark scattering is then 
  unitarized. 
  We show these predictions in Fig.~\ref{fig:islamdiffel}, where the different contributions are indicated separately.
  \begin{figure}
\resizebox{0.5\textwidth}{!}{
\includegraphics{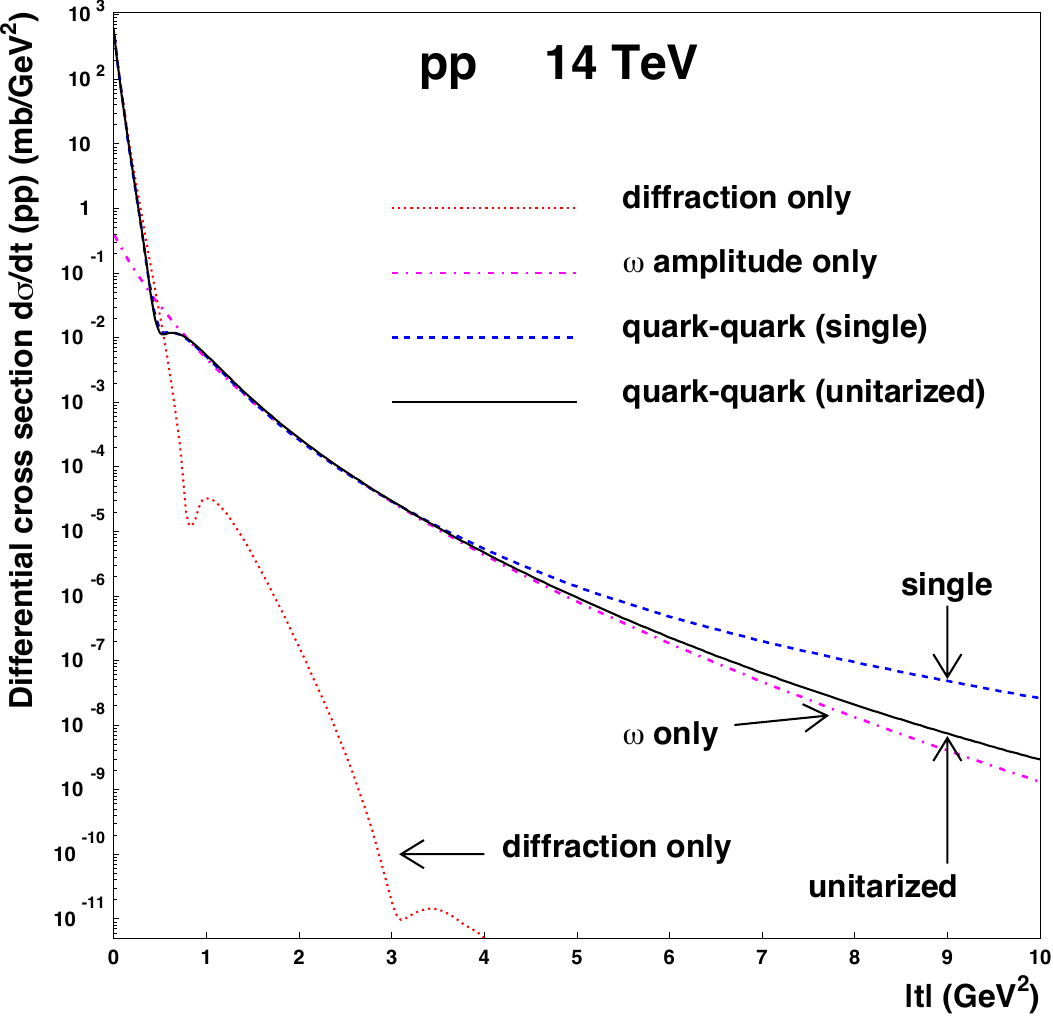}}
\caption{The differential elastic cross-section from Islam model \cite{Islam:2007nr} at $\sqrt{s}=14\ {\rm TeV}$. Reprinted from Fig. (2) of \cite{Islam:2007nr}, in DESY-PROC-2007-02, \copyright (2007) DESY. Figure is courtesy of the authors. }
\label{fig:islamdiffel}
\end{figure}
A recent fit to data, inclusive of  predictions for LHC14,  is shown in Fig.~\ref{fig:islam} from \cite{Islam:2013cba}. 
\begin{figure}
\resizebox{0.5\textwidth}{!}{%
 \includegraphics{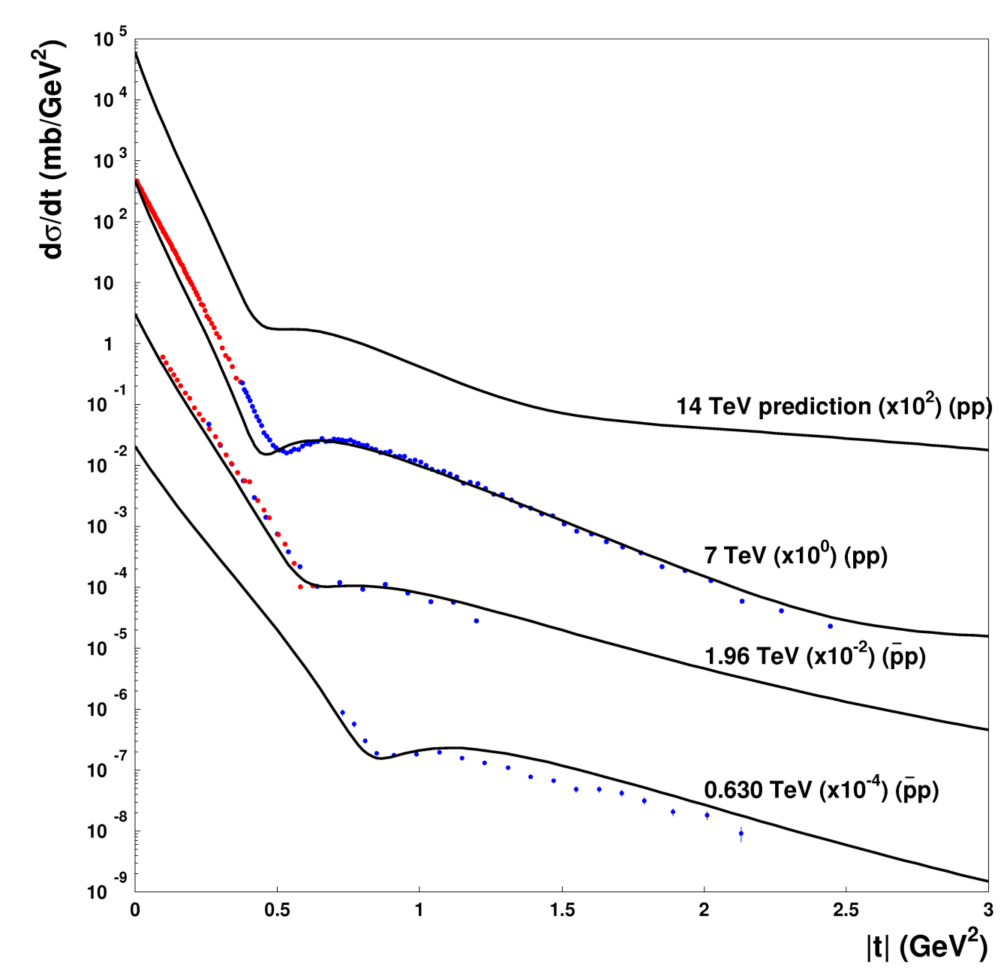}}
\caption{A  2013 study of  the differential elastic cross-section in Islam and Luddy model. Reprinted from \cite{Islam:2013cba}. Figure is courtesy of the authors.  }
\label{fig:islam}       
\end{figure}
In 2015, the model took a significant step forward, with incorporation of  polarization of the outer quark-antiquark cloud region of the
proton by the enclosed baryonic charge as described  in  \cite{Islam:2015fex}. 
\subsubsection{The eikonal  model by Bourrely, Soffer and Wu}\label{sss:BSW}
Another  model of interest is due to  Bourrely, Soffer and Wu (BSW). In  \cite{Bourrely:2002wr}, the total as well as the 
differential cross-sections are discussed. Recent results and a comparison with LHC data can be found in 
\cite{Soffer:2013tha,Bourrely:2012hp}.
We show in Fig.~\ref{fig:BSWdiffel} their pre-LHC  predictions at various energies for the elastic differential cross-section.
    \begin{figure}
   \resizebox{0.5\textwidth}{!}{
\includegraphics{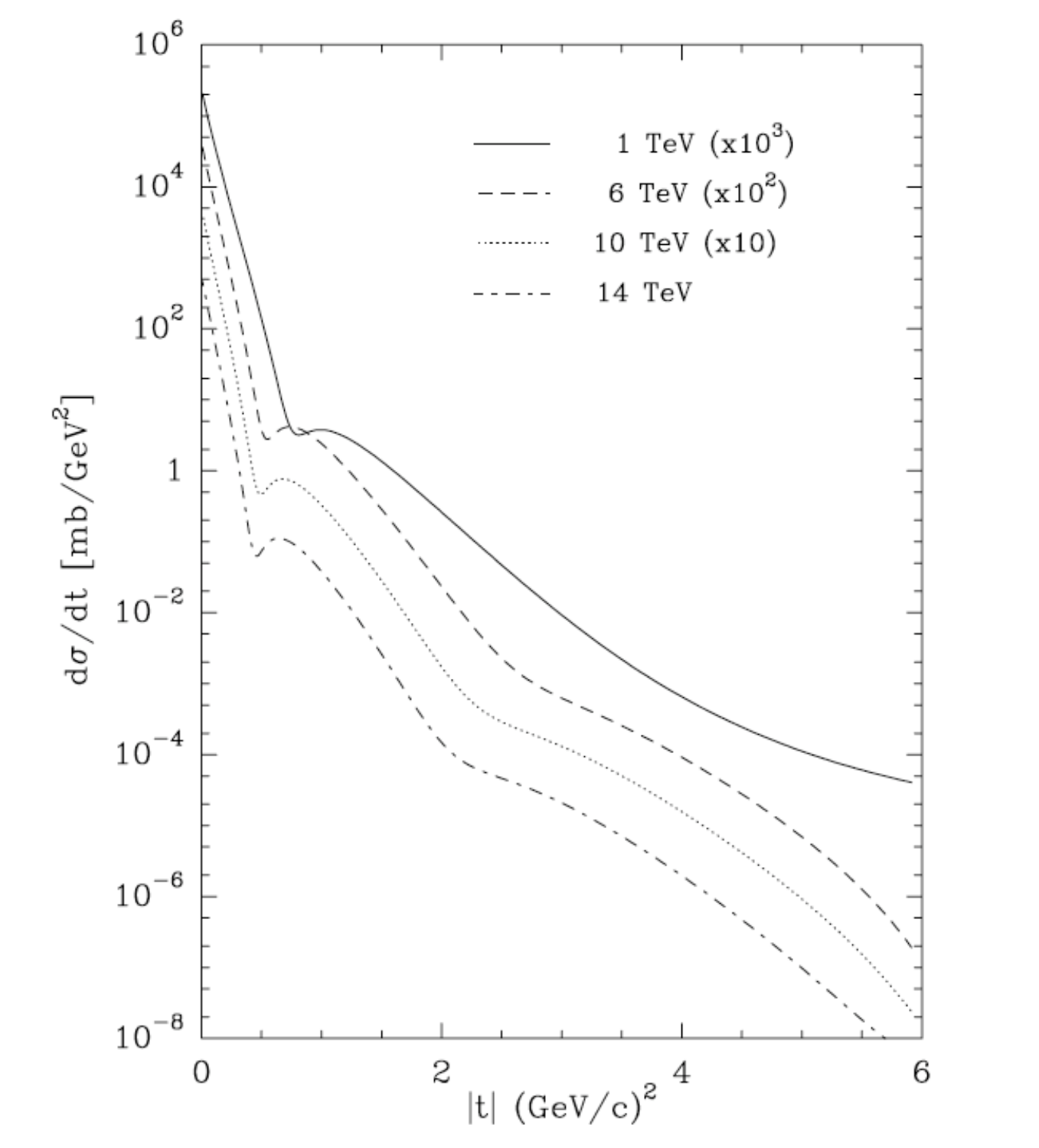}}
\caption{The elastic differential cross-section at various cms energies from BSW model \cite{Bourrely:2002wr}  \copyright (2002) by Springer.
Reprinted with permission  from Springer.}
\label{fig:BSWdiffel}
\end{figure}

This model follows in the steps of the very early work by Cheng and Wu. A particularly clear description of how the impact picture developed 
after the work by Cheng and Wu (CW)  can be found in a short review paper by Jacques 
Soffer \cite{Soffer:1987gm} and also in  \cite{Goulianos:1988hk}. It is recalled that QED was the only known relativistic quantum field theory in 
the late '60s and that CW introduced a small photon mass $\lambda$ in order to avoid what Soffer calls {\it unnecessary complications}. 
After a brief introduction to the CW results, Soffer \cite{Soffer:1987gm} describes the  model developed 
together with Bourrely and  Wu, 
starting with the elastic amplitude for proton scattering namely
\begin{equation}
a(s,t)=a^N(s,t)\pm sa^c(t)
\end{equation}
where the $\pm$ signs refer to \pbarp \ and \pp \ respectively.  The hadronic amplitude is given by $a^N(s,t)$  and 
the factor $s$ has been factorized out of the  Coulomb amplitude $a^c(t)$. For the latter, one has
\begin{equation}
a^c(t)=2\alpha\frac{G^2(t)}{|t|}e^{\mp i \alpha \phi(t)}
\end{equation}
where $\alpha$ is the fine structure constant, $\phi(t)$ is the phase introduced by West and Yennie \cite{West:1968du} as
\begin{equation}
\phi(t)=\log [\frac {t_0}{t}] -\gamma
\end{equation}
$\gamma$ being the Euler's constant and $t_0=0.08\ {\rm GeV}^2$.
$G(t)$ is  the proton  electromagnetic form factor, and the model chosen by Soffer in \cite{Soffer:1987gm} is 
\begin{equation}
G(t)=[ (t-m_1^2)(t-m_2^2) ]^{-1}
\end{equation}

The usual quantities are defined accordingly as
\bea
\sigma_{tot}&=&\frac{4\pi}{s}\Im m a(s,t=0) \nonumber \\
\frac{d\sigma(s,t)}{dt}&=&\frac{\pi}{s^2}|a(s,t)|^2 \nonumber\\
B(s,t)&=&\frac{d}{dt}\log (\frac{d\sigma}{dt})
\eea
and the
hadronic amplitude is obtained from the impact picture \cite{Bourrely:1978da} as
\begin{equation}
\label{eq:Bourrely}
a^N(s,t)=is \int_0^\infty b db J_0(b\sqrt{-t})(1-e^{-\Omega(s,b)})
\end{equation} 
The eikonal function $\Omega(s,b)$ is split into two terms, reflecting different dynamical inputs, namely
\begin{equation}
  \Omega(s,b)=R_0(s,b)+{\hat S}(s,b)
\end{equation}
where $R_0(s,b)$ includes the Regge contribution important in the low energy region and is different for \pp \ and \pbarp, 
whereas the second term ${\hat S}(s,b)$  is the same for both processes and gives the rising contribution to the total cross-section. 
This term  is factorized into energy dependence and impact parameter dependence as
\begin{equation}
{\hat S}(s,b)=S_0(s) F(b^2)
\end{equation}
with the energy dependence given as in the CW model. Thus  the model exhibits a Pomeron energy dependence given by a complex 
crossing symmetric expression
\begin{equation}
S_0(s)=
\frac
{s^c} {(\log \ s/s_0)^{c'}}
 + \frac{u^c}{ (\log u/u_0)^{c'}}
\end{equation} 
where $u$ is the third Mandelstam variable. At high energy and small momentum transfer, the  real and imaginary parts of the amplitude 
can be obtained through the substitution $\log u=\log s -i\pi$.
As for  the essential impact parameter dependence, and hence the $t$-dependence, this   is parametrized  through an expression similar to  the 
proton electromagnetic form factor, namely the Fourier-transform ${\cal F}(t)$ of $F(b^2)$ is
\begin{equation}
{\cal F}(t)=f [ G(t)]^2[\frac{a^2+t}{a^2-t}]
\end{equation}
This model had six parameters, of which two, $c$ and $c'$, related to    the energy dependence, and the other four,  $a,m_1,m_2,$ and $f$  
describing the impact parameter dependence. 



Including the LHC7 data,  the values obtained for the six parameters are given in Table \ref{tab:bsw},  from Soffer's  contribution
to Diffraction 2012 \cite{Soffer:2013tha}. 
\begin{table}[htdp]
\caption{Parameter values for the BSW model from Diffraction 2012 \cite{Soffer:2013tha}}
\begin{center}
\begin{tabular}{|cc|}\hline
c=0.167&c'=0.748\\
$m_1=0.577 \ {\rm GeV}$&$m_2=1.719\ {\rm GeV}$\\
$a=1.858 {\rm GeV}$&$f=6.971 {\rm GeV}^{-2}$\\
\hline
\end{tabular}
\end{center}
\label{tab:bsw}
\end{table}%
The recent discussion of the model  in \cite{Soffer:2013tha} 
gives the description shown in Figs.~\ref{fig:bswdiff2012-realim7tev}, \ref{fig:bswdiffeldiff2012},
respectively for the scattering amplitude and  the  differential cross-section.
Fig.~\ref{fig:bswdiffeldiff2012} shows  that the tail of the distribution after the dip reflects the usual oscillations characteristic of   eikonal models. 
\begin{figure}[htbp]
\begin{center}
\resizebox{0.5\textwidth}{!}{
\includegraphics{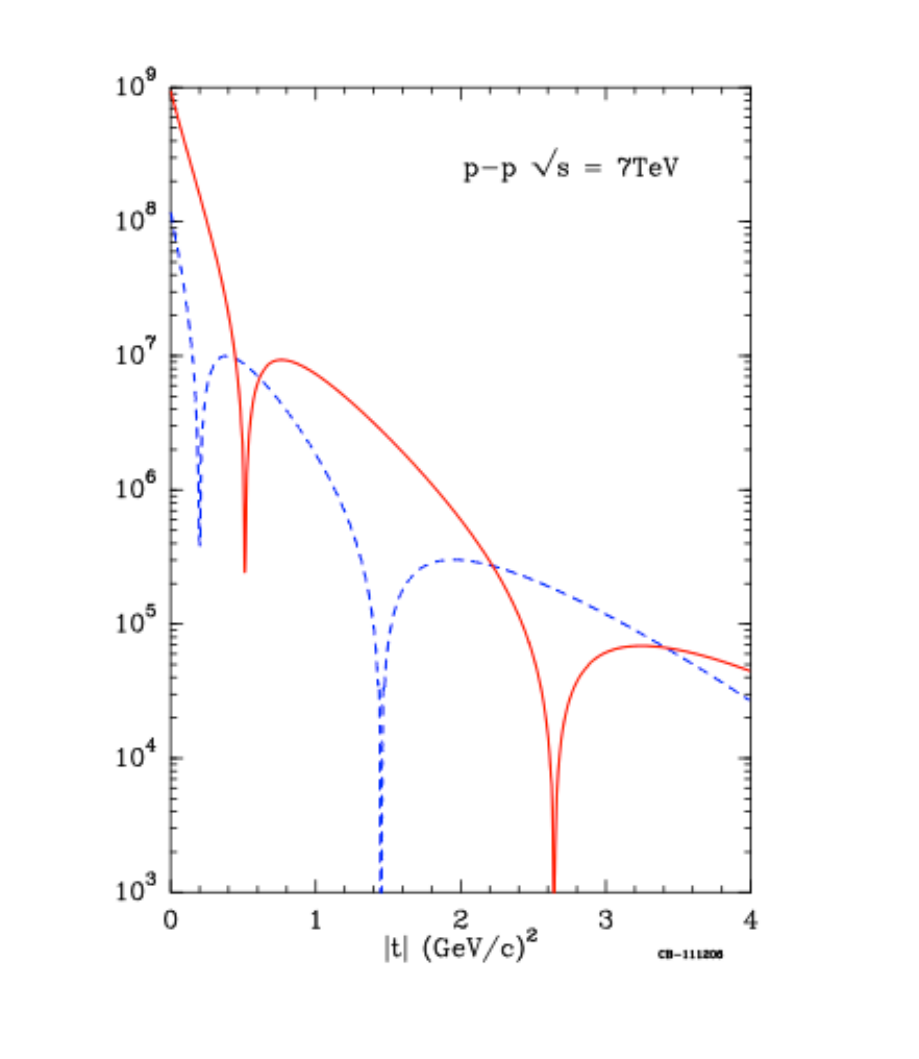}}
\caption{Absolute values of real and imaginary parts of the elastic differential cross-section at 7 TeV from \cite{Soffer:2013tha,Bourrely:2012hp}.
Reprinted with permission from \cite{Soffer:2013tha},
Fig.(2), \copyright (2013) by AIP Publishing LLC. }
\label{fig:bswdiff2012-realim7tev}
\end{center}
\end{figure}

\begin{figure}[htbp]
\begin{center}
\resizebox{0.5\textwidth}{!}{
\includegraphics{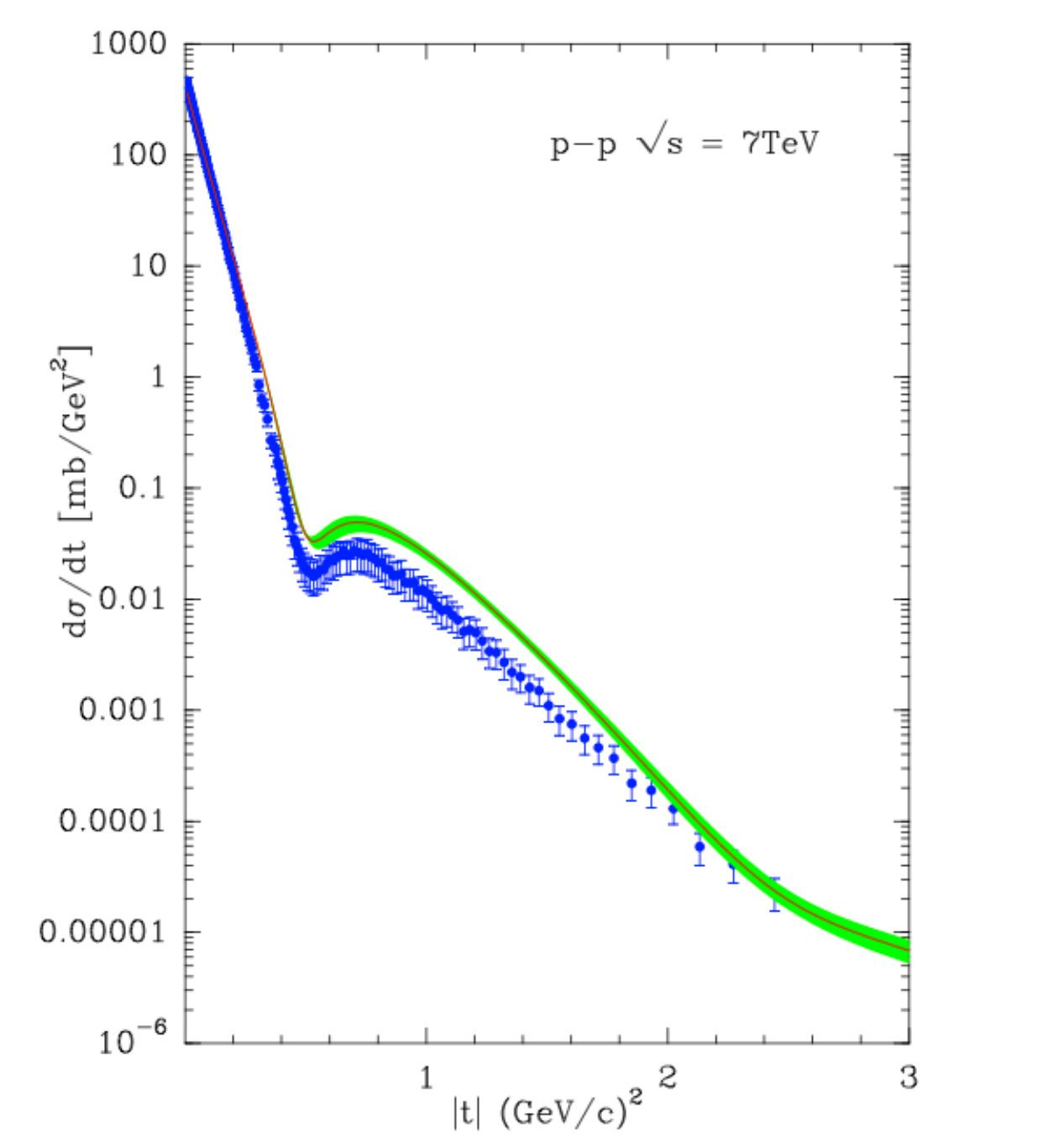}}
\caption{Comparison between  TOTEM data \cite{Antchev:1472948} and  BSW model description of the elastic differential 
cross-section at 7 TeV from \cite{Soffer:2013tha,Bourrely:2012hp}, calculated with 68\% CL. 
Reprinted with permission from \cite{Soffer:2013tha},
\copyright (2013) by AIP Publishing LLC.}
\label{fig:bswdiffeldiff2012}
\end{center}
\end{figure}

We note 
that the $\vecb$-dependence, hence the $t$-dependence, is obtained through the form factor 
$F(\vecb^2)$, which is not the nucleon factor nor a convolution of two nucleon form factors, as it would be in the 
eikonal mini-jet models or the Glauber models. This {\it form factor } is independent of the overall energy. 

\subsubsection{Many Pomeron structures in  eikonal models}\label{sss:manypomerons-eikonal}
 We discuss here four papers by  Desgrolard, Giffon, Martynov, Petrov, Predazzi and 
  Prokudin, who have worked  together in different combinations. These authors have been involved in
a precision analysis of resonances and the forward  region, as in  Desgrolard, Giffon, Martynov and Predazzi  
  \cite{Desgrolard:2000sf}, but they also deal with the elastic differential cross-section \cite{Desgrolard:2000dn}.  
  In   Petrov and Prokudin \cite{Petrov:2001eu}  
  the three  Pomeron model is introduced.  Then the model is again discussed in  \cite{Petrov:2002bm}   and the Coulomb interference 
  problem is picked up by the same authors, this time with Predazzi  in \cite{Petrov:2002nt} and before in \cite{Petrov:2002bm} .
  
  The main point of this approach is the need to go beyond one or two  Pomeron pole description of the elastic 
  differential cross-section, and allow for many Pomerons.
Let us begin with Desgrolard,  Giffon, Martynov and Predazzi \cite{Desgrolard:2000dn}, 
 which contains  in its introduction a good description of the state-of-the-art 
 in the year 2000, at the time LEP was closed and construction for LHC started.
 
This model is based on eikonalization of the input Born amplitude. Namely, the scattering amplitude will be given through the eikonal 
function $\chi(s,b)$, and one has the usual set of equations
\bea
\sigma_{total}&=& \frac{4 \pi}{s}\Im m A(s,t=0)\\
\frac{d\sigma}{dt}&=& \frac{\pi}{s^2}|A(s,t)||^2\\
\rho&=&\frac{\Re e A(s,t=0)}{\Im m A(s,t=0)}
\eea
The eikonalized amplitude in $(s,t)$ space can be written through the Fourier-Bessel transform
\be
A^{\pbarpmath}_{pp}=\frac{1}{2s} \int  H^{\pbarpmath}_{pp}(s,b)J_0(b\sqrt{-t}) bdb 
\ee
 of the  function $H^{\pbarpmath}_{pp}$ which is to be  defined 
  in terms of  an amplitude $h^{\pbarpmath}_{pp}$, which 
 is the Fourier transform of the Born amplitude in $(s,t)$ space, 
\bea
h^{\pbarpmath}_{pp}(s,b)=2s \int   a^{\pbarpmath}_{pp}(s,-q^2)J_0(bq) qdq
\eea
The next problem is the nature of the resummation procedure which takes one from the Born 
amplitude to the full amplitude. 
The Born  input for the crossing-even and crossing-odd  amplitudes 
 \be
 a_{pp}^{\pbarpmath}(s,t)= a_+(s,t)\pm a_-(s,t)
 \ee
   will be determined by the available data on the total cross-section, the differential elastic cross-section and the ratio $\rho(s,t=0)$. 
  The even part is parametrized through the contribution of a Pomeron and an $f$-reggeon, while the odd part is an 
  Odderon and an $\omega$-Reggeon, i.e.
  \bea
  a_+(s,t)=a_P(s,t)+a_f(s,t) \\
   a_-(s,t)=a_O(s,t)+a_\omega(s,t)
  \eea
For the Reggeon, the {\it standard } form is used, namely
\bea
a_R(s,t)=a_R{\tilde s}^{\alpha_R(t)}e^{b_R t}\\
\alpha_R(t)=\alpha_R(0)+\alpha_R't 
\eea  
  where R stands for the $f$- or $\omega$-trajectories.
 For the Pomeron, the authors investigate two possibilities, a monopole (M) or a dipole (D), respectively
 \bea
 a_P^M(s,t)&=&a_P{\tilde s}^{\alpha_P(t)} e^{b_P(\alpha_P(t)}\\
 a_P^D(s,t)&=&a_P{\tilde s}^{\alpha_P(t)}
 [e^{b_P(\alpha_P(t)-1)}
 (
 b_P+\log {\tilde s}
 )\nonumber\\
 +d_P\log {\tilde s}]
 \eea 
 As for the Odderon, the chosen form is 
 \be
 a_O(s,t)=(1-e^{\gamma t})a_O^{(M/D)}(s,t)
 \ee
 with M or D standing for a monopole or a dipole.  It should be noticed that in this paper, the authors state that for the ratio 
 $\rho ( s,t=0)$ to be fitted by the data, it is necessary that the Odderon contribution vanishes at $t=0$.The trajectories are all taken 
 to be linear in $t$, i.e.
 \be
 \alpha_i(t)=\alpha_i(0)+\alpha'_i t\equiv 1+\delta_i +\alpha'_i t
 \ee
There is a relationship between eikonalization and unitarization \cite{Giffon:1996gm,Martynov:1988ts,Finkelstein:1989mf}, 
which gives the following constraints
\be
\delta_P \ge \delta_0, \ \  \alpha'_P\ge \alpha'_0
\ee
Most fits give $\delta_O<0$.   

  Once the Born amplitude is stated, the authors discuss  different eikonalization procedures, one called Ordinary 
  Eikonalization (OE) in which one puts
 \be
  H^{\pbarpmath}_{pp,OE}(s,b)=\frac{1}{2i}(\sum_1^\infty \frac{[2ih^{\pbarpmath}_{pp}(s,b)]^n}{n!})
 \ee
 and the other is the Quasi Eikonal (QE) with
 \be
 H^{\pbarpmath}_{pp,QE}(s,b)=\frac{1}{2i}(\sum_1^\infty \lambda^{n-1}\frac{[2ih^{\pbarpmath}_{pp}(s,b)]^n}{n!})
 \ee
The explicit analytical form is 
\be
H^{\pbarpmath}_{pp,QE}(s,b)=\frac{1}{2i\lambda}(exp[i \lambda \chi^{\pbarpmath}_{pp}]-1)
\ee
When $\lambda=1$, one obtains the OE. Still another form of eikonalization corresponds to the case when the weight $\lambda$
is different for different terms. In the QE case, the various terms in the sum have the same weight, but one can consider the possibility 
that the terms have different weights, and this case is called the Generalized Eikonal (GE). The case with  three  $\lambda$'s is discussed in detail, 
and fits are given. One observation here is the development of structures in the amplitude as a function of  $t$. Apart from the dip, there are oscillations, 
which are a consequence of the properties of the Bessel function from the Fourier transform. It is possible that some special feature of the eikonal  
may eliminate these oscillations, but in this model they are still present. 
   
 The authors discuss fits within the OE and the QE. They have more freedom to do a best fit with the GE with 2 or 3 parameters,  and 
 different cases for fixed or variable parameters for   secondary reggeons are examined. The predictions of this model are shown in Fig.~\ref{fig:desgr} 
 from \cite{Desgrolard:2000dn}.
They emphasize three points:
\begin{itemize}
\item i) the presence of the Odderon contribution is necessary in order to describe the differential cross-section in the dip region and at large 
$t$, the Odderon intercept consistently turns out to be negative from the fits, 
\item ii) the real part of the even amplitude has a zero at small $|t|$-values, which moves  toward $0$ as the energy increases, 
being at $|t|=0.23 \ {\rm GeV}^2$ at $\sqrt{s}=14\  {\rm TeV}$. There are also other secondary zeros.  
\item iii) the eikonalized Odderon contributes to reproduce ``perfectly" the large $|t|$- region.
\end{itemize}

\begin{figure}
\resizebox{0.5\textwidth}{!}{
\includegraphics{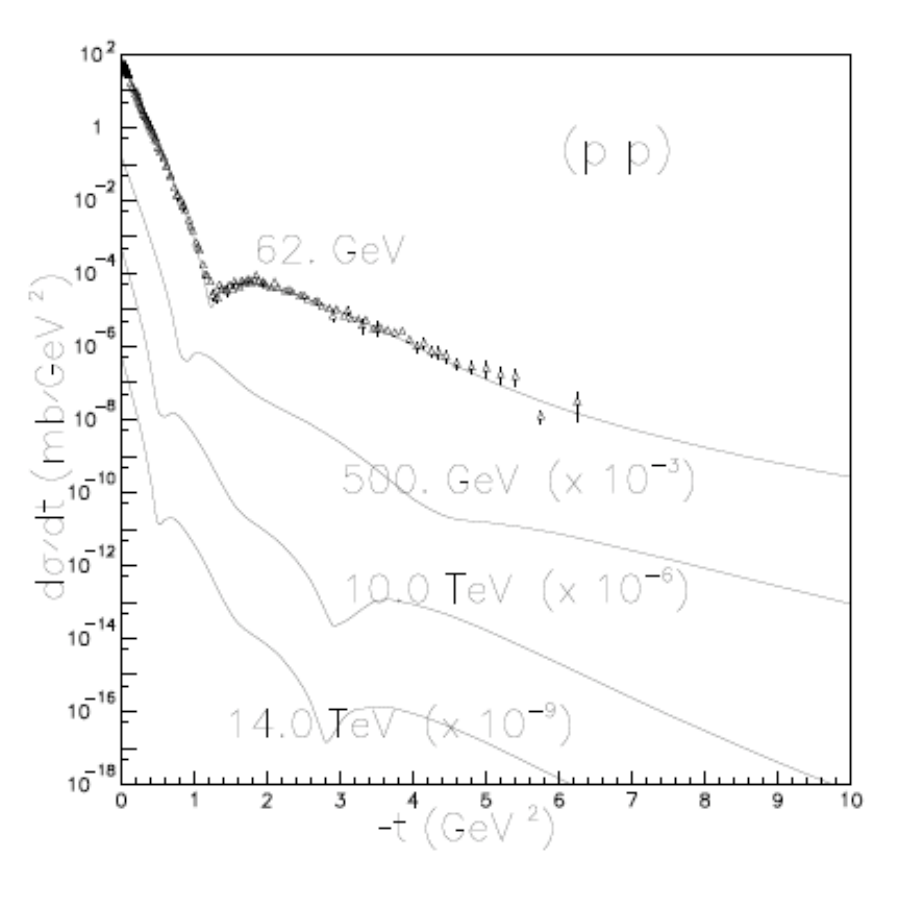}}
\caption{The elastic differential cross-section for $pp$ from Desgrolard, Giffon, Martynov 
and Predazzi \cite{Desgrolard:2000dn}. Reprinted with permission of Springer   from \cite{Desgrolard:2000dn}, \copyright (2000) Springer.}
\label{fig:desgr}
\end{figure}

   We now proceed to Petrov  and Prokudin \cite{Petrov:2001eu}, entitled  ``The first three Pomerons".
This paper contains a short but useful introduction to various models, 
and well summarizes the state-of-the-art at the time.  They 
observe that  the large number of different models describing the scattering 
both at $t=0$ and for small $t$-values hints to the fact that the most general way to deal with the problem is to introduce an arbitrary 
number of Pomerons. In \cite{Petrov:2001eu} a first step is attempted to formulate an  eikonal structure  with many Pomerons. 
They begin with a three Pomeron structure, since, according  to the authors, one and two Pomeron structures are inadequate. 

Writing  the unitarity condition as
\be
\Im m T(s,\vecb)=|T(s,\vecb)|^2 + \eta(s,\vecb)
\ee 
where $T(s,\vecb)$ is the scattering amplitude in the impact parameter representation, $\eta(s,\vecb)$ the contribution of inelastic channels, 
the scattering amplitude in terms of an eikonal function $\delta(s,\vecb)$ as
  \be
  T(s,\vecb) =\frac{e^{2i\delta(s,\vecb)}}{2i}
\ee
  with $\Im m \delta(s,\vecb)\ge 0$ for $s> s_{inel}$. If the eikonal function has only simple poles in the complex J-plane, and the poles are 
  parametrized as linearly rising Regge trajectories, modulo the signature factor, the contribution to the eikonal in $t$ is written as
  \bea
  \hat{\delta}(s,t)=\frac{c}{s_0}(\frac{s}{s_0})^{\alpha(0)} e^{t \rho^2/4}\nonumber \\
  \rho^2=4 \alpha'(0)\log \frac{s}{s_0} +r^2
  \eea
  in $\vecb$-space one then obtains
  \begin{equation}
\delta(s,b)=\frac{c}{s_0}(\frac{s}{s_0})^{\alpha_(0)-1} \frac{e^{-b^2/ \rho^2}}{4\pi \rho^2}
  \end{equation}
  
  Three Pomerons are then introduced describing both $pp$ and \pbarp, as follows:
  \bea
  \delta^{\pbarpmath}_{pp}(s,b)&=&\delta^+_{P_1}(s,b)+\delta^+_{P_2}(s,b)+\delta^+_{P_3}(sb)\mp \nonumber \\
& &  \delta^-_{O}(s,b)+\delta^+_{f}(s,b)\mp\delta^-_\omega(s,b) 
  \eea
  where $P_i$ are the Pomeron contributions, the $\pm $ sign refers to even/odd  trajectories, $O$ referring to  $odderon $,
   $f$ and $\omega$ even and odd trajectories.    To restore analyticity and crossing symmetry, one substitutes $s$ with 
  \be
  {\tilde s}=\frac{s}{s_0}e^{-i\frac{\pi}{2}}
  \ee
  and  obtains the appropriate signature factors for the various terms contributing to the eikonal as
  \bea
 \delta^+(s,b)=i\frac{c}{s_0}(\frac{\tilde s}{s_0})^{\alpha_(0)-1} \frac{e^{-b^2/ \rho^2}}{4\pi \rho^2}\\
 \delta^-(s,b)=\frac{c}{s_0}(\frac{\tilde s}{s_0})^{\alpha_(0)-1} \frac{e^{-b^2/ \rho^2}}{4\pi \rho^2}\\
 \rho^2=4 \alpha'(0)\log {\tilde s}+r^2
    \eea
  The trajectories are dealt with in the linear approximation, with a fit to the meson spectrum determining  the parameters of the secondary 
  Regge trajectories, $f$ and $\omega$. The parameters defining the 3 Pomerons and the odderon contribution (a total of 20) are obtained 
  by a fit to the total cross-sections, the $  \rho$  parameter and the elastic differential cross-section, the latter in the range $0.01\le|t| \le14\ {\rm GeV}^2$.  
  Data for the total elastic cross-section are not included in the fit, but are a result of the model. This is not surprising given the fact that by fitting 
  the total and the differential cross-section, one fixes both the normalization (optical point) and the slope. The  fits require that the three Pomerons 
  and the Odderon as well have  intercept at $t=0$ larger than 1, which also implies that eventually there will be a violation of the Froissart bound, 
  the slope of the odderon being very close to zero.The exercise is repeated with only two Pomerons, but the result is 
  not very good: in this case, unlike the three pomeron case,   the odderon trajectory is fitted to have  intercept less than 1. 
  
  This paper also contains a good discussion of the connection of the model to string theory  models, and to the predictions for BFKL.
  
  Concerning predictions 
   in the small $t$ region, the model indicates a dip at LHC around $0.5\ {\rm GeV}^2$ 
  and a wiggle around $-t \lesssim 3\ {\rm GeV}^2$. We show in Fig.~\ref{fig:3pomerons} the predictions for the elastic 
  differential cross-section at RHIC and at LHC.
    \begin{figure}
\resizebox{0.5\textwidth}{!}{
\includegraphics{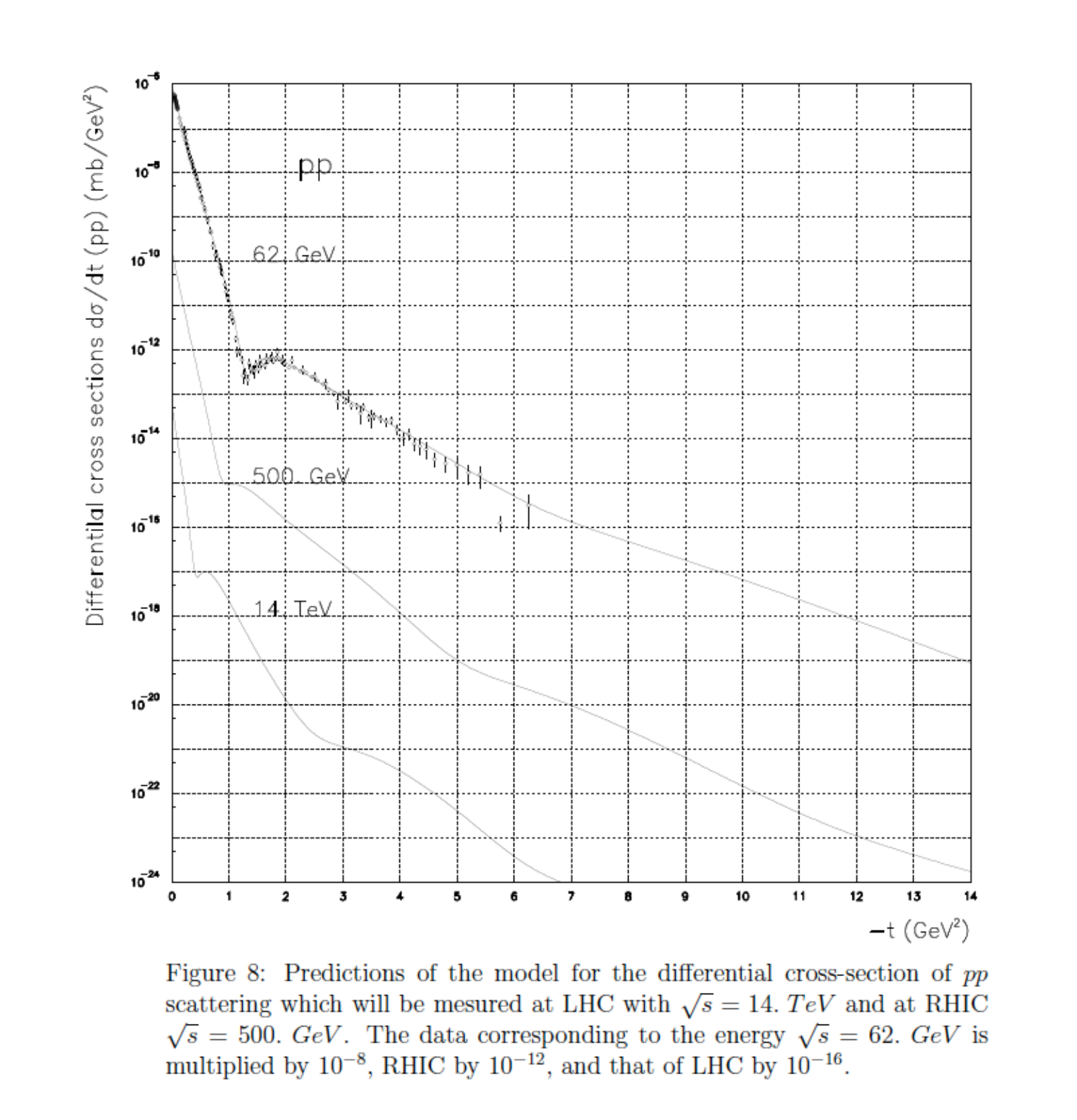}}
\caption{The elastic differential cross-section for $pp$ from \cite{Petrov:2001eu}. Reprinted with permission, \copyright (2002) by Springer. }
\label{fig:3pomerons}
\end{figure}
 \subsubsection{
 The Aspen model}
 
This model \cite{Block:1998hu} 
is a QCD inspired version, applied to both proton and photon processes, of a description of the total 
and elastic differential cross-section based on a large amount of previous work. In the 1984 review, Block and   Cahn \cite{Block:1984ru} 
describe in great detail the constraints arising from unitarity, analyticity and crossing symmetry, introducing their own  proposal for low and 
high energy parametrization.  In the 2006 review, the phenomenological description is updated and an extensive presentation of the subsequent 
work is given \cite {Block:2006hy}. Thus, 
in the following
the term {\it Aspen model} refers explicitly to the description in Appendix A of \cite{Block:2006hy}. 
A description of this model has already been presented in \ref{sss:aspen},
here we recall its main  points.
The Aspen model uses the eikonal representation in order to ensure unitarity. 
It   embeds in addition  
the constraints of analyticity, 
and crossing symmetry. The model includes 
both a crossing odd and a crossing even eikonal, i.e.
\bea
\chi^{\pbarpmath/pp}&=&\chi^{even}\pm\chi^{odd}\\
\chi^{even}&=&\chi_{gg}(s,b)+\chi_{qg}(s,b)+\chi_{qq}(s,b)\\
 &=&i\sum_{ij} [\sigma_{ij}(s)W(b;\mu_{ij})]
\eea
with $ij=gg,qg,gg$, $\mu_{qg}=\sqrt{\mu_{qq} \mu_{gg}}$ and
\bea
W(b;\mu_{ij})=\frac{\mu^2_{ij}}{96 \pi}(\mu_{ij}b)^3K_3(\mu_{ij}b)\\
\Sigma_{gg}=\frac{9\pi \alpha_s^2}{m_0^2}
\eea 
and 
\be
\chi^{odd}=-\sigma_{odd}W(b;\mu_{odd})=-C_{odd}\Sigma_{gg}\frac{m_0}{\sqrt{s}}W(b;\mu_{odd})
\ee
with $W(b;\mu_{odd})$ having the same functional form as the other b-distributions, $W(b;\mu_{ij})$. All the $b$-distributions are normalized 
to 1 and are obtained as the Fourier transform of a dipole. 
As for the cross-sections $\sigma_{ij}(s)$, their QCD inspired parametrization leads to the following large $s$-behaviour:
\be
\sigma_{gg}\sim \log^2 s, \ \ \ \sigma_{qg}\sim \log s, \ \ \ \sigma_{qq}\sim constant 
\ee
For large $s$-values, the even contribution is made analytic through the substitution
\be s\rightarrow se^{-i\pi/2} 
\ee
The two constants $C_{odd}$ and $\mu_{odd}$ are fitted to the data. At high energies,  the odd eikonal vanishes 
like $1/\sqrt{s}$, since this  is the term which accounts for the difference between $pp$ and \pbarp \ interactions, 
and at high energies such a difference should vanish.  We show in Fig.~\ref{fig:fig14martinreport} predictions for 
LHC14TeV and comparison of the model with the Tevatron data \cite{Amos:1990fw,Amos:1991bp}. 
\begin{figure}
\resizebox{0.5\textwidth}{!}{
\includegraphics{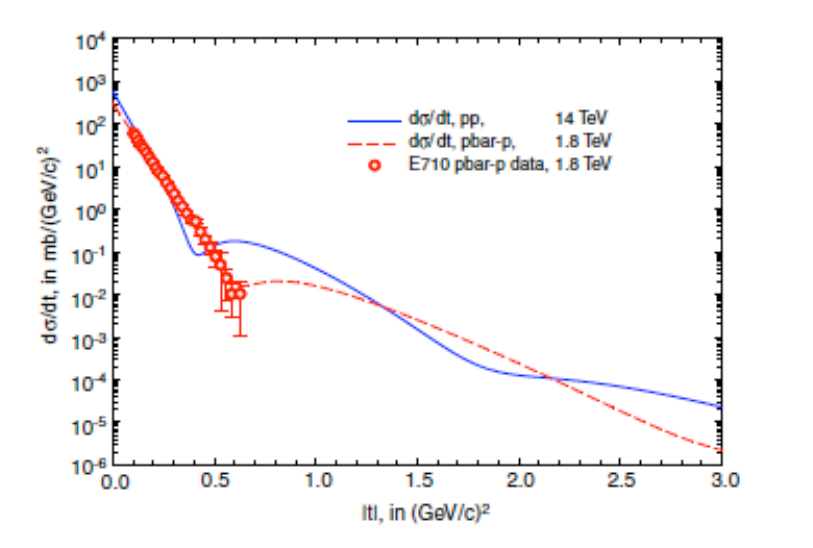}}
\caption{The elastic differential cross-section in the Aspen model from Block's Report \cite{Block:2006hy}. The full line represents expectations at 
LHC at $\sqrt{s}= 14\ {\rm TeV}$,  the dashed line are predictions at the Tevatron collider in comparison with the E710 experiment 
\cite{Amos:1990fw,Amos:1991bp}.
Reprinted from \cite{Block:2006hy}, \copyright (2006), with permission by Elsevier.}
\label{fig:fig14martinreport}
\end{figure}
We note that the predicted curve for LHC has a dip for  $-t\simeq0.5\ {\rm GeV}^2$ and second (slight) dip (more like a wiggle)  around $1.8 \ {\rm GeV}^2$. 
Presently,  data up to $-t=2.5\ {\rm GeV}^2$ at LHC7 show no other structure but  the dip at $-t=0.53\ {\rm GeV}^2$.

The Aspen model has been the inspiration  for the  Dynamical Gluon Mass model \cite{Luna:2005nz} in which an energy dependent  
mass $m_0(s)$  for the gluon is introduced to regulate the low-$p_t$ divergence in the mini-jet like eikonal functions. Recently, the group 
from Campinas   
 has been concerned 
  with the slope and the total  and elastic  cross-sections \cite{Fagundes:2011hv,Fagundes:2012fa}. The 
  ratio ${\cal R}_{el}=\sigma_{elastic}/\sigma_{total}$ is discussed as it  can give information about different models and their 
  asymptotic behavior. The question   whether  
  $B(s)$ is growing linearly with $\log s$ as expected in Regge -Pomeron descriptions, or whether it would grow as 
  $\log^2 s$ \cite{Schegelsky:2011aa} is addressed.  This would  be relevant  in cosmic ray physics, where the ratio 
  $\sigma_{total}/B$ for proton-proton scattering defines the nucleon-nucleon impact parameter amplitude (profile function),  and 
 measurements of the $p-air$ cross-section are thus related to this ratio.


\subsection{Models including 
the diffractive contribution to the scattering amplitude}\label{ss:Diffel}
Diffractive processes contribute to the total cross-section.   They are inelastic processes  with spatial correlations 
to the incoming particles, and their experimental definition depends on the type of experiments as well as on the 
cuts imposed on the final state. Theoretically the description of diffraction  is rendered difficult because no exact 
theorems exist about its energy dependence. We show in Fig. ~\ref{fig:olgainel} a compilation of data for the   
inelastic cross-section, from lower to very high energies, including some  results from   LHC experiments  in 
different rapidity regions, and, in some cases, including extrapolations into the diffractive region.
The spread of results at LHC7 reflects different experimental cuts and different extrapolations into the low mass region.
The blue band gives the results from the soft $k_t-$resummation model, described in \cite{Achilli:2011sw}.  
The blue circle at 14 TeV gives the Block and Halzen (BH) prediction  \cite{Block:2011vz}.
\begin{figure}
\hspace{-1cm}
\resizebox{0.6\textwidth}{!}{
\includegraphics{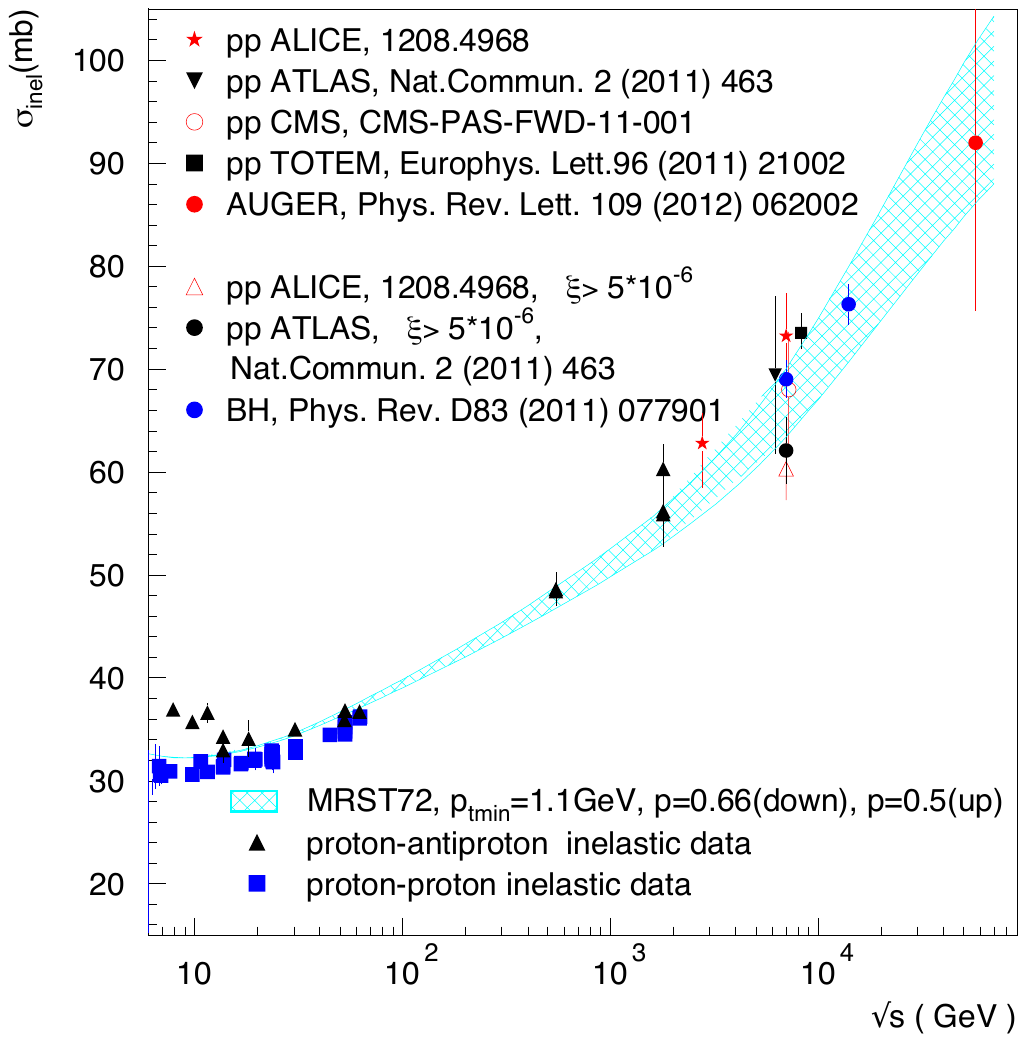}}
\vspace{-2cm}
\caption{Data for inelastic scattering   in different kinematic regions,  with model expectations (blue dot) from Block and 
Halzen (BH) \cite{Block:2011vz} and  from the soft $k_t$-resummation  model \cite{Achilli:2011sw}.
Figure is courtesy of Olga Shekhovtsova.}
\label{fig:olgainel}
\end{figure}

Discussion of diffraction as distinct from both inelastic and elastic processes has a long history. Models for diffraction 
are present in the literature since early observations 
of final state particles appearing in configurations clearly correlated along the incoming projectile. As we shall discuss 
in the following, some of these models use a quantum mechanics formalism, inspired by Good and Walker decomposition 
of diffractive  scattering \cite{Good:1960ba}, while the QCD approaches, such as those by the Durham 
\cite{Khoze:2000wk} and Tel Aviv \cite{Gotsman:2012rm} group, use, in addition,  a triple Regge formalism, 
to account for high mass  as distinct from low mass diffraction. Recently a contribution by Gustafson  to clarify 
the connection between some of these  approaches has appeared \cite{Gustafson:2012hg}.

In what follows  we shall  summarize some representative 
 works  concerned with the description of diffraction.  We shall describe in some detail the contributions by 
 Pumplin and Miettinen, started with some early work  \cite{Pumplin:1973cw} and   \cite{Fialkowski:1975ta} and then   
 followed by Pumplin with Miettinen \cite{Miettinen:1978jb} and by Miettinen and Thomas \cite{Miettinen:1979ns}. 
 A clear description of this approach can be found in \cite{Pumplin:1982na}.  We shall also summarize the application of 
 some of these ideas to mini-jet models by Lipari and Lusignoli \cite{Lipari:2009rm}.  The Regge-Pomeron analysis of 
 diffraction performed by two groups, Durham (KMR) and Tel Aviv (GLM), will follow.  A digression to  the string theory 
 Pomeron description by the Brown  University group \cite{Brower:2007xg}  will be included, as it is central to the  Tel Aviv model.  
 Many phenomenological analyses exist  in the literature, for a recent analysis  up to Tevatron data,  a good summary can 
 be found in the work by Dino Goulianos  
 \cite{Goulianos:2010mm}.


\vskip 0.3cm 
\subsubsection{ The Pumplin limit for diffractive processes}
\label{Pumplin2} 
We now discuss  the      generalization of  the break-up of the  total cross-section into elastic, diffractive  \cite{Fialkowski:1975ta,Miettinen:1978jb}
and inelastic components and how to obtain the so-called {\it Pumplin limit} \cite{Pumplin:1973cw}, i.e. . 
\begin{equation}
{\cal R}_{el+diff}(s) =  [ \frac{\sigma_{el+diff}(s)}{\sigma_{tot}(s)} ]\le \frac{1}{2}\label{P14}
\end{equation}
to be used instead of the black disk limit 
\begin{equation}
{\cal R}_{el}(s) =  [ \frac{\sigma_{el}(s)}{\sigma_{tot}(s)} ]\le \frac{1}{2} \label{eq:blackdisk}
\end{equation}
Let the ``b-wave'' unitary $S(s,b)$ matrix  be decomposed as
\begin{equation}
S^\dagger S = (1 -2i T^\dagger)(1 +2iT) =1; \ \Im m T = T^\dagger T.
\label{P2}
\end{equation} 
If we define twice the imaginary part of the elastic amplitude by ${\tilde M}(s, b) =\ 2 \Im m T_{ii}(s, b) $, 
Eq.(\ref{P2}) leads to the relation
\begin{equation}
2 {\tilde M}(s, b) - {\tilde M}^2 = 4 \sum_{n\neq 1} |T_{ni}|^2 \equiv\ G_{in}(s, b),
\label{P3}
\end{equation} 
valid for large $s$ if we neglect the small real part of the elastic amplitude.

The inelastic scattering sum in $G_{in}$ due to multi-particle production may be written as
\begin{equation}
G_{in}(s, b) = \sum_n \frac{d^2\sigma^{(n)}}{d^2{\bf b}}.
\label{P4}
\end{equation} 
If one assumes a statistically independent production of particles, one is led to the Poisson distribution
\begin{equation}
\frac{d^2\sigma^{(n)}}{d^2{\bf b}} = (\frac{Q(s,b)^n}{n!}) e^{-Q(s,b)}.
\label{P5}
\end{equation} 
Then Eq.(\ref{P3}) leads to the well known eikonal form for the imaginary part of the elastic amplitude
\begin{equation}
\tilde{M}(s, b) = 1 - e^{-(Q/2)} \equiv\ 1 - e^{-\Omega(s,b)},
\label{P6}
\end{equation} 
with $2 \Omega(s,b) = Q(s,b) = <n(s,b)> $ denoting the mean number of collisions occurring at a given
$s$ and $b$.

It is well to note that Eq.(\ref{P3}) has two solutions for ${\tilde M}$: 
\begin{equation}
\tilde{M}(s, b) = 1 \pm e^{-(Q/2)} \equiv\ 1 \pm e^{-\Omega(s,b)} 
\label{P6a}
\end{equation}
The solution chosen in Eq.(\ref{P6}) is the smaller one which corresponds to ${\tilde M} \to\ 0$ as the
mean number of collisions goes to zero. Hence,
\begin{equation}
0 \leq\ \tilde{M}(s, b) \leq\ 1. 
\label{P6b}
\end{equation}
We note in passing that this choice reduces the Martin-Froissart bound by a factor of 2. It also leads to 
one of our asymptotic sum rules \cite{Pancheri:2005jr,Pancheri:2004xc}
\begin{equation}
 \tilde{M}(s, b = 0) \to\ 1\ {\rm as}\ s\to \infty . 
\label{P6c}
\end{equation} 
Using a phenomenological model such as the  Phillips  and Barger (PB) model of Ref. \cite{Grau:2012wy} for the elastic amplitude, 
one can see that experimental data at $7\ {\rm TeV}$ by the TOTEM group at the LHC support Eq.(\ref{P6c}).
The value is $ 0.95 \pm 0.01$ \cite{Grau:2012wy} at $7\ {\rm TeV}$ corresponding to $G_{in}(s,b=0)\to 1$. 
The other mathematically allowed possibility $G_{in}(s,b=0)\to 0$ as $s\to \infty$ that leads to 
${\tilde M}(s,0)\to 2$ is ruled out by  data  at present LHC energies.

The result of Eq.~(\ref{P6a}) however is incomplete. As discussed at the { beginning of this section,}
  if the eikonal function $\Omega(s,b)$ is constructed through a randomly 
distributed Poisson sum of incoherent scatterings, then the elastic ratio ${\cal R}_{el}(s)$
turns out to be larger than its experimental value at LHC, which is still  $\simeq 1/4$, hence quite far from the 
black disk limit of Eq. (\ref{eq:blackdisk}). The reason behind this generic fact is that there is a non-negligible fraction of the
inelastic cross-section, called diffractive which is not  truly random but which maintains quite a bit of coherence with the scattered
particles. 

One type of contribution to  single diffractive events occurs when one of the two scattering particles ends up in a state with 
the same internal quantum numbers with a close by mass but perhaps with a spin-flip (e.g., $p \to\ p^*$) \cite{Pumplin:1982na}, 
while simultaneous break up of both  scattering particles into diffractive states
as defined above  contributes to 
double diffraction. Even at high energies, 
such as at LHC7 for example,
the contribution of the summed diffractive 
cross-section to the total is  $(10\div15) \%$ and 
hence it needs to be properly understood and formulated \cite{Miettinen:1978jb,Pumplin:1968bi}. References to other work on this 
subject can be found in \cite{Pumplin:1982na,Lipari:2009rm}.
The analysis below follows these references in outline. Other references can be found  in the Durham and Tel Aviv analyses, 
which include contributions coming  from high mass diffraction.


The underlying essential physics of diffraction can be incorporated by supposing that the incident particles
are in a superposition of ``diffractive eigenstates'' defined as
\begin{equation}
| A > = \sum_k C_k(A) | \psi_k >,
\label{P7}
\end{equation} 
with $P_k(A) = \sum_k |C_k(A)|^2$ giving the probability of finding the diffractive eigenstate $k$ in $A$ 
and $\sum_k P_k(A) = 1$. The interaction with the other particle produces a mixture of diffractive and
non diffractive states 
\begin{equation}
T| A > = \sum_k C_k(A) T_k | \Psi_k > + {\rm non\ diffractive\ states}. 
\label{P8}
\end{equation} 
Hence with the breakup of only one particle $A$ of the initial state, the elastic amplitude becomes
\begin{equation}
< A | T | A > = \sum_k | C_k(A)|^2 T_k = \sum_k P_k(A) T_k \equiv\ < T >,
\label{P9}
\end{equation} 
where the average $<. >$ denotes an average over the diffractive state probabilities. As before neglecting 
the real part of the diffractive amplitudes, we would have the following expressions:
\begin{eqnarray}
{\cal A}:\   S_T(s,b) \equiv\  \frac{d^2\sigma_{tot}}{d^2{\bf b}} = 4 \Im m <A|T|A> \nonumber\\
= 2 \sum_k P_k(A) {\tilde M}(s,b) = 2 <{\tilde M}(s,b)>\nonumber\\
\label{P10a}
\end{eqnarray} 
and
\begin{eqnarray}
{\cal B }:\  S_{el}(s,b) \equiv\ \frac{d^2\sigma_{el}}{d^2{\bf b}} = 4 |<A|T|A>|^2 \nonumber\\
= 4 |\sum_k P_k(A) T_k|^2 =  |<{\tilde M}(s,b)>|^2.\nonumber\\
\label{P10b}
\end{eqnarray} 
The sum of the elastic and diffractive differential cross-section (in b-space) reads
\begin{eqnarray}
{\cal C}:\    \frac{d^2\sigma_{el+diff}}{d^2{\bf b}} = 4 \sum_k | < \Psi_k|T|A>|^2 \nonumber\\
 =  < {\tilde M}(s,b)^2>.\nonumber\\
\label{P10c}
\end{eqnarray} 
 The basic result for the diffractive part of the differential cross-section in  impact parameter space
is obtained through Eqs. (\ref{P10b}) and (\ref{P10c})
\begin{equation}
{\cal D}:\  S_{diff} \equiv\   \frac{d^2\sigma_{diff}}{d^2{\bf b}}  =  < {\tilde M}(s,b)^2> - < {\tilde M}(s,b)>^2.
\label{P11}
\end{equation} 
In words, $S_{diff}$ is given by the dispersion $< (\Delta{\tilde M})^2 >$ in the absorption probabilities and hence it 
would vanish identically were all components of the initial state absorbed equally. If averages are taken over
both incident particles, $d\sigma_{diff}$ would include both single and double diffraction dissociation.
  
Now let us see how to obtain   the Pumplin upper bound on $S_{diff}$. Since, by virtue of Eq.(\ref{P6b}), the absorption probabilities
$0 \leq {\tilde M}_k \leq 1$,  and ${\tilde M}^2_k \leq {\tilde M}_k$, we have also that
\begin{eqnarray}
<{\tilde M}(s,b)> = \sum_k P_k M_k \leq\ 1;\nonumber\\ 
{\rm and} <{\tilde M}^2(s,b)> \leq\ <{\tilde M}(s,b)>.\nonumber\\
\label{P12}
\end{eqnarray} 
This leads to the Pumplin inequality
\begin{equation}
S_{diff}(s,b) \leq\  [\frac{1}{2} S_T(s,b) - S_{el}(s,b)].
\label{P13}
\end{equation} 
and the integrated version of Eq.(\ref{P13})  leads to Eq.~(\ref{P14}) for the ratio of the elastic + diffractive cross-section  to the total cross-section.
{For $pp$ scattering at ISR, $\sigma_{diff}\approx (8.5\ \pm 1.5) \ mb$ which is over a half of the limit $\approx 14 \ mb$,
predicted by the above Eq.(\ref{P13}).
Presently,  even at the highest energies, the limiting value of $1/2$, as given by ${\cal R}_{el+diff}(s)$ in Eq.(\ref{P14}) is closer to 
the experimental results than the black disk limit ($1/2$) for ${\cal R}_{el}(s)$ in Eq.(\ref{eq:blackdisk}), as we have seen from  Fig.~\ref{fig:sigeltosigtotnewxifae}.

Pumplin \cite{Pumplin:1982na} further observed that while $s$-channel helicity conservation is a good approximation for the elastic 
amplitude such is not the case for diffraction dissociation.

To proceed further, one needs to formulate a model  for the probability distribution for diffraction dissociation at high energies, 
which must  respect several requirements:
\begin{itemize} 
\item (i) { for diffraction dissociation into a continuous mass spectrum, an analytic dependence on mass} ; 
\item (ii) the constraints that all ${\tilde M}(s,b)< 1$;  
\item (iii) ${\tilde M}(s,b) \to 1$ as $b\to 0$  (asymptotic constraint);
\item (iv)  ${\tilde M}(s,b) \to 0$ as $b \to \infty$ (asymptotic constraint). 
\end{itemize}
As discussed next,  a simple probability distribution meeting these requirements, which is amenable to
an analytic treatment, is given by \cite{Miettinen:1979ns,Pumplin:1982na}
\begin{equation}
\frac{dP}{dz} = \frac{z^{\lambda - 1}}{\Gamma(\lambda)} e^{-z}
\label{P15}
\end{equation} 
with $0\leq z\leq \infty$ and the dispersion $<(\Delta z)^2> = <z> = \lambda$. 

\subsubsection{Specific models 
with a continuos distribution
}
In 1979 Miettinen and Thomas \cite{Miettinen:1979ns} discussed  how  to modify the Glauber eikonal formalism to 
include diffraction within a picture in which  quarks and gluons  are - with different  spatial distributions- the  
constituents of a nucleon or a hadron. 
Since  earlier analyses, in particular Good and Walker \cite{Good:1960ba}, it had been known that if  
the different components of a composite system have different distributions 
(absorption strengths), then inelastic states are excited and the elastic formula for the amplitude needs to be modified. 
The aim in \cite{Miettinen:1979ns} was to show  that, taking into account fluctuations in the wave function, 
one can find the matter distribution. This is analogous  to the distinction between
a ``charge distribution'' extracted from $pp$ elastic scattering 
and the proton charge distribution 
obtained through the proton electromagnetic form factor, extracted from $ep$ scattering. 

Their first step is to substitute
the expression for the elastic amplitude between two hadrons $A$ and $B$ scattering at a distance $b$
\be
T_{el}^{AB}=1-e^{-<\Omega(b)>_{AB}}\label{eq:simple}
\ee
with a different amplitude,  corresponding to the different possible configurations $i$ and $j$ in which the 
two hadrons find each other  at the point of collisions,, i.e.
\be
t^{ij}(b)=1-e^{-\Omega_{ij}(b)}
\ee
Calling $p_i$ the probability that the hadron is in a given configuration $i$, the complete amplitude is 
obtained by averaging over all the possible configurations
\be
t_{el}^{AB}(b)=\sum_{ij}p_i^A p_j^B t_{ij}^{AB}=1-<e^{-\Omega(b)}>_{AB}\label{eq:composite}
\ee
The connection between the simple eikonal of Eq~(\ref{eq:simple}) and this description can be seen by writing Eq.~(\ref{eq:composite}) as
\be
t_{el}^{AB}(b)=1-e^{-<\Omega(b)>_{AB}}H_{AB}(b)
\ee
with the corrections to the simple eikonal  embedded into the moments $\mu_k^{AB}(b)$ of the eikonal spectrum, i.e.
\bea
H_{AB}=\sum_{k=0}\frac{(-1)^k}{k!}\mu_k^{AB}(b)\\
\mu_k^{AB}(b)=\big{<} ( \Omega(b)-<\Omega(b)_{AB}>)^k \big{>}_{AB}
\eea
{ The connection to matter distribution in $b$-space enters the eikonal function $<\Omega(b,s)>_{AB}$  
from the overlap of the average matter densities of the incident particles, i.e. the s-dependence is factorized 
into a constant $K_{AB}$ and }
\be
<\Omega(b,s)>=K_{AB}\int d^2\vecb '  \rho(\vecb ') \rho (\vecb -\vecb ')
\ee
To exemplify this model,  Eq.~(\ref{eq:composite}) is written as 
\be
t_{el}^{AB}(b)=\int_0^\infty d\Omega P_{AB}(\Omega,b)(1-e^{-\Omega})
\ee
with the function $P_{AB}(\Omega,b)$ defined by the probabilities $p_i^A,\ p_j^B$. Now, the model becomes easy to solve 
under the hypothesis that the $b-$dependence of  $P_{AB}(b)$ is only a function of the scaling variable 
$z=\Omega(b,s)/<\Omega(b,s)>$. 
A  simple function in the variable $z$ is considered, namely  
\be
P_{AB}(z)=N z^a e^{-\lambda z}
\ee
which obeys the constraints 
\be
\int dz P_{AB}(z)=1;\  \int z dz P_{AB}=1,
\ee
as requested by normalization and the first moment condition. 
One has 
\be
N=\lambda^\lambda/\Gamma(\lambda);\ {\rm and}\ a= (\lambda -1). 
\ee
The integration is then  done immediately, and 
the differential cross-sections in $\vecb$-space now become
\begin{eqnarray}
\frac{1}{2} S_T(s,b) = <{\tilde M}(s,b)> = \int dP(z) [1 - e^{-z \Omega(s,b)/\lambda}]\nonumber\\
= 1 - [1 + \Omega(s,b)/\lambda]^{- \lambda}\nonumber\\;
\label{P16}
\end{eqnarray} 
\begin{equation}
S_{el}(s,b) = 1 - 2[1 + \Omega(s,b)/\lambda]^{- \lambda} + [1 + \Omega(s,b)/\lambda]^{- 2\lambda};
\label{P17}
\end{equation} 
and 
\begin{equation}
S_{diff}(s,b) = [1 + 2\Omega(s,b)/\lambda]^{-\lambda} - [1 + \Omega(s,b)/\lambda]^{-2\lambda}.
\label{P18}
\end{equation} 
A few observations about this model are in order:
\begin{itemize}
\item (i) As $\lambda \to \infty$, the earlier eikonal limits are reached
since $S_{diff} \to 0$. 
\item (ii) For $\lambda = 1$, Eq.(\ref{P16}) reduces to a ``Fermi'' distribution
\begin{equation}
\frac{1}{2} S_T(s,b) = <{\tilde M}(s,b)> \to\ \frac{1}{\Omega(s,b)^{-1} + 1}; [\lambda\ =\ 1].
\label{P19}
\end{equation}
\end{itemize} 
Further extensions of this model can be found in the work by Lipari and Lusignoli \cite{Lipari:2009rm}, who use the mini-jet formalism to study 
 the contributions to the inelastic cross-section from different hadronic configurations 
participating to the scattering. The authors start with a careful discussion of the relation between $\sigma_{jet}$, the cross-section 
for producing a mini-jet pair, and $\sigma_{inel}$, the total inelastic cross-section. Many authors have had difficulty in understanding 
why  $\sigma_{jet}$ can be larger than  $\sigma_{inel}$: this is simply a reflection of the fact that the inelastic cross-section measures 
the probability of inelastic interactions between hadrons, whereas $\sigma_{jet}$ measures the probability of parton-parton collisions, 
so that, since $\sigma_{inel}$ includes at least  one parton-parton  collision,  the ratio $\sigma_{jet}/\sigma_{inel}$   in fact describes the 
average number of mini-jet pairs produced in one inelastic collision and this can be quite large.

The authors next consider  the average number of collisions $<n_{jet}(b,s,p_\perp^{min})>$, taking place at  impact parameter $\vecb$, 
\begin{align}
<n_{jet}(b,s,p_\perp^{min})>=
\int d^2b_1\int d^2b_2 P_{int}(\vecb-\vecb_1+\vecb_2)\times\nonumber\\
\int dp_\perp\int dx_1dx_2\sum_{j,k,j',k'}F_j^{h_1}(x_1,b_1,\mu^2) F_j^{h_2}(x_2,b_2,\mu^2)\times \nonumber\\
\frac{d{\hat \sigma}_{jk\rightarrow j'k'}}{dp_\perp}\ \ \ \ \ \ \ \ \ \ \ \ \ \ \ \ \ \ \ \ \ \ \ \ \ \ \ \ \ \ \ \ \ \ \ \ \ \ \ \ \ \ \ \ \ \ \ \ \ \ \ \ \ \ \ \ \ \ 
\end{align}
 obtained  from  the relevant $x$ and $b$-space distributions   for each  colliding parton,  $j$ going into a hadron 
 $h$, $F_j^{h}(x,b,\mu^2)$, where $\mu$ is a hard scale defining the applicability of the mini-jet perturbative description. 
 Since partons have a spatial distribution, this number depends on  the probability $P_{int}(\vecb)$ that two partons, 
 separated by a  distance   $\vecb$, interact with each other, a standard practice in mini jet models.\\ 
 
Defining parton configurations in the colliding hadrons as ${\mathbb C}_1$  and ${\mathbb C}_2$,  
the average number of mini-jet collisions is obtained by summing over all possible configurations, i.e.
\be
\int d{\mathbb C}_1 
\int d{\mathbb C}_2 P_{h_1} ({\mathbb C}_1)P_{h_2}({\mathbb C}_2) n_{jet}(b,{\mathbb C}_1,{\mathbb C}_2)=<n_{jet}(b,s)>\label{eq:lipari1}
\ee
Upon assumption of the   factorization hypothesis
\be
n_{jet}(b,{\mathbb C}_1,{\mathbb C}_2)=<n_{jet}(b,s)>\alpha({\mathbb C}_1,{\mathbb C}_2)
\ee
with $\alpha({\mathbb C}_1,{\mathbb C}_2)$ a constant, real parameter independent of energy, the final result will be obtained 
by integrating over all values of the parameter $\alpha$, i.e. over all possible configurations, with a distribution given by a probability 
function $P(\alpha)$ which must satisfy the two conditions
\bea
\int d\alpha \ P(\alpha)=1 \label{eq:lipari2}\\
\int d\alpha \ \alpha \ P(\alpha)=1\label{eq:lipari3}
\eea
where the second condition follows from Eq.~(\ref{eq:lipari1}). 
In this  generalized Good and Walker structure  the proposed final expression
for   the elastic amplitude is 
\be
F_{el}(q,s)=\int \frac{d^2b}
{2\pi}
\eiqb
\int d\alpha 
\ p(\alpha)
 \big{[}
  1-e^{-\frac{<n(b,s)>\alpha}{2}}
  \big{]}
\ee
with the simple eikonal case corresponding  to   $p(\alpha)\rightarrow \delta(\alpha-1)$. One can see that in such case   
the  elastic cross-section  
 includes  both  elastic and diffractive processes, so that in the simple eikonal  case the inelastic diffraction contribution 
 vanishes. The cross-sections in momentum  space can now be explicitly written as
\bea
\frac{d\sigma_{el}}{dt}=\pi \big{[} \int db\ b\ J_0(b\sqrt{-t})\int d \alpha\ p(\alpha)\times \\
 \big{(} 1-exp[-\frac{<n(b,s)>\alpha}{2}\big {)} \big{]}^2
\eea
The differential cross-section inclusive of elastic and  diffraction processes is calculated to be
\begin{align}
\frac{d\sigma_{diff+el}}{dt}=\pi \int d\alpha \ p(\alpha) \big{[} \int db\ b\ J_0(b\sqrt{-t})\times \nonumber \\
 \big{(} 1-exp[-\frac{<n(b,s)>\alpha}{2}\big {)} \big{]}^2
\end{align}
An explicit model is built, with the   probability function  $p(\alpha)$ such as to satisfy the two  conditions given by 
Eqs.~(\ref{eq:lipari2}) and (\ref{eq:lipari3}),  and chosen to be 
\be
p(\alpha)=\frac{1}{w\Gamma(\frac{1}{w})}\big{(} 
 \frac{\alpha}{w}
 \big{)}^{\frac{1}{w}-1}exp[-\frac{\alpha}{w}] \label{eq:liparipalpha}
\ee
as in \cite{Miettinen:1979ns}.
The model   depends upon the parameter $w=<\alpha^2>-1$. Upon performing the integrations in $\alpha$, one now has 
\begin{align}
\frac{d^2\sigma_{diff}}{d^2b}=(1+<n(b,s)>w)^{-1/w}-\\
\big{(}1+\frac{<n(b,s)>w}{2} \big{)}^{-2/w}\\
\frac{d^2\sigtot}{d^2b}=2-2\big{(}1+\frac{<n(b,s)>w}{2} \big{)}^{-1/w}\\
\frac{d^2\sigel}{d^2b}=
\big{(}
 1-\big{(}1+\frac{<n(b,s)>w}{2} \big{)}^{-1/w}
 \big{)}^2
\end{align}
The authors apply a parametrization of the mini-jet model to evaluate and predict the diffraction cross-section as a function of energy. 
Using  the approximation
\begin{equation}
\frac{\sigma_{TD}}{\sigma_{el}}
\equiv 
\frac{\sigma_{BD}}{\sigma_{el}}
\simeq \frac{\sigma_{DD}}{\sigma_{TD}}\equiv \frac{\sigma_{DD}}{\sigma_{BD}}  
\end{equation}
  where  $T/B$ refers to Target or  Beam  diffraction (D) and $DD$ stands for double diffraction,
 one obtains
\begin{equation}
\sigma_{diff}=\sigma_{SD}+\sigma_{DD}\simeq \sigma_{SD}(1+\frac{\sigma_{SD}}{4\sigma_{el}}).
\label{eq:maurizio}
\end{equation}
 The last term in  Eq. (\ref{eq:maurizio}) includes an extra factor of 2 at the denominator, which  corrects Eq. (85) of  
 Ref. \cite{Lipari:2009rm}  . \footnote{Private communication by the authors.}
The  results, for two different choices of the model mini-jet parametrizations, are   shown in Fig.~\ref{fig:LL} 
in comparison with, then available, experimental data for single diffraction.
\begin{figure}
\hspace{-1.5cm}
\centering
\resizebox{0.5\textwidth}{!}{
\includegraphics{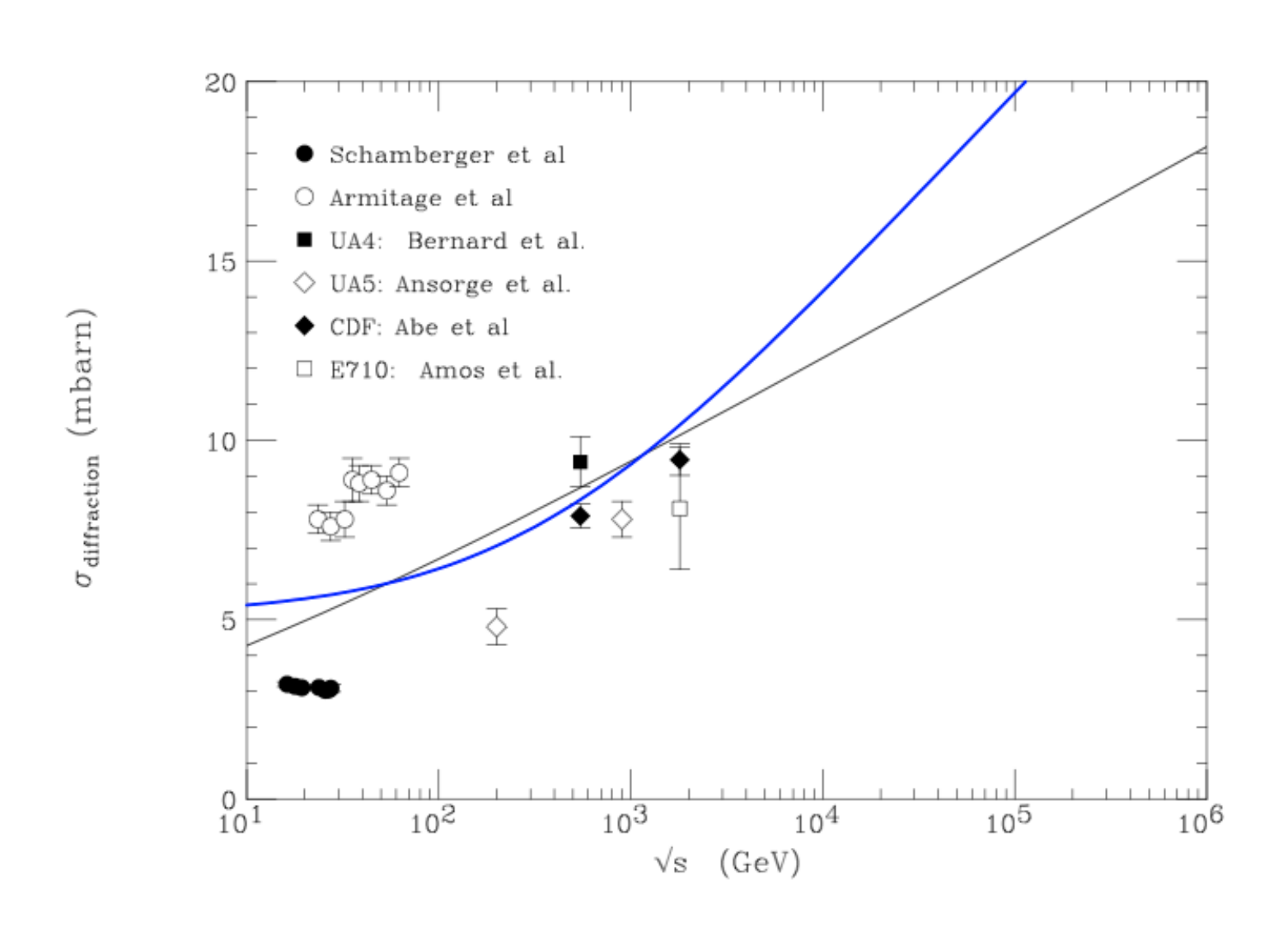}}
\caption{Inelastic diffraction cross-section evaluated in an eikonal mini jet model within a generalized Good and Walker 
mechanism, and different mini-jet cross-section parametrizations. The experimental results correspond to single diffraction 
only. 
Reprinted  figure   with permission  from \cite{Lipari:2009rm}, \copyright (2009) by the American Physical Society.}
\label{fig:LL}
\end{figure}



\subsubsection{The  Durham model by Khoze, Martin and Ryskin}\label{sss:KMRel}
The Durham model by Valery Khoze, Alan Martin  and Misha Ryskin (KMR) offers   a unified description of   both soft and semi-hard interactions. 
 This model,  developed and continuously refined over many years, also with other collaborators, presents  a QCD based Pomeron 
and parton phenomenology .
 
 Recent papers by the Durham group 
 \cite{Ryskin:2012ry,Martin:2012mq,Ryskin:2011qe,Martin:2011gi} 
 describe data on multiparticle production and present differential cross-section analyses. The
  most recent version of the
  model uses a two and  three-channel eikonal formalism and multipomeron exchange diagrams, incorporating both unitarity and Regge behaviour.  In \cite{Martin:2011gi}, the authors compared their original predictions, for the total cross-section and   diffractive quantities, with TOTEM results. Adjustments of their earlier parameters  appears
in  \cite{Martin:2012mq}. 
  
The essential ingredients in their model can be summarized as follows:
\begin{itemize}
\item 
Input Pomeron trajectory for soft and hard processes
\item 
Eikonalization of the amplitudes
\item 
Inclusion of diffractive processes
\item 
Pion loop corrections to the Pomeron trajectory and small $|t|$ slope of the elastic differential cross-section 
\end{itemize}
We shall  briefly describe each of the  first three items and  then  present their most recent phenomenology 
for the elastic differential cross-section. The inclusion of pion loop corrections will be treated separately in 
 \ref{KMRsmallt} 
\par\noindent 
\begin{itemize}
\item KMR1: Pomeron trajectory
\end{itemize}
This approach uses Reggeon Field Theory 
with a phenomenological soft Pomeron, while for hard  interactions a QCD partonic approach is employed \cite{Martin:2012nm}.
In the hard domain, where perturbative QCD and the standard partonic approach can be used, their Pomeron is 
associated with the BFKL singularity. In this perturbative domain, there is a single hard Pomeron exchanged with
\be
\alpha_P^{bare}(t) = (1 + \Delta) + t \alpha^{'bare};\ \Delta =\ 0.3;\ \alpha^{'bare}\lesssim 0.05\ {\rm GeV}^{-2}
\label{eq:2comPom}
\ee 
It is noted that, although the BFKL equation should be written for gluons away from the infrared region, 
after resummation and stabilization, the intercept of the BFKL Pomeron depends only weakly on the scale 
for reasonably small scales. We reproduce in Fig.~\ref{fig:BFKLpom} their description of the connection 
between the intercept of the BFKL Pomeron and the value for $\alpha_s$.
The figure shows how the intercept $\Delta$ goes to a smooth almost constant behaviour as $\alpha_s$ increases.

It is useful to note here that in  mini-jet pictures, the value of $\Delta=\ 0.3$ corresponds with the 
mini-jet cross-section growing as $\simeq\ s^{0.3}$. Thus, the bare KMR Pomeron  plays a role  
similar to that of parton-parton scattering  folded in with parton densities and summed over all 
parton momenta in the mini-jet model. Also, the bare slope is interpreted by KMR to be associated with
 the size of the Pomeron: $\alpha'^{bare}\sim\ 0.05\ {\rm GeV}^{-2} \propto\ 1/<k_t^2>$. It is related to a hard 
 scale, of the order of a few {\rm GeV}, which can find its counterpart in $p_{tmin}$ of the mini-jet models. 
 However, a precise correspondence still needs to be worked out. 

On the other hand, the transition from the hard to the soft domain in KMR requires multi-pomeron exchanges 
through re-summation of soft $k_t$-processes, thus lowering the scale of $<k_t^2>$ from the earlier hard to a 
soft scale. As a result, the BFKL  Pomeron (i) bare intercept decreases and at the same time (ii) the slope
 increases by a factor $\sim 5$ from its bare value. Hence, one has an  
 effective linear Pomeron trajectory such as the one given by the Donnachie and Landshoff parametrization, i.e. 
\be
\alpha_P^{eff}(t) \simeq\ 1.08 + 0.25 t.
\label{Y2}
\ee
KMR found empirical evidence for the above trajectory also in virtual photo-production of vector mesons at HERA. 

Once again, we may find a correspondence of the change in the effective slope with the mini-jet model parameter 
$p_{tmin}\sim 1.1\ {\rm GeV}$ through the observation that  $(p_{tmin}/\Lambda)^2\sim 5$ with $\Lambda\sim 500\ MeV$. 

%

We add here some further
details regarding the ``BFKL multi-Pomeron'' 
 approach. In the Durham model, 
the semi-hard particle production due to a single BFKL  Pomeron is shown in the left plot of Fig.~\ref{fig:abcKMR} 
from \cite{Ryskin:2011qh}.
 In this figure, Y is the rapidity interval. In  (a), the $k_t$ of the partons are not ordered. 
The multiplicity of partons grows as $x^{-\Delta}$ with $\Delta\equiv \alpha_P(0)-1$, and partons drift to lower 
$k_t$ values because of the running of the strong coupling constant. Notice the imposition of a cut-off $k_0$, 
below which the cascade is forbidden to develop. In the central figure, (b), the structure of standard 
DGLAP-based MonteCarlo cascade  is strongly ordered in $k_t$: there is a driving process, in the central rapidity region,  
with the highest parton momentum and then, from the initial parton a cascade ordered from  larger to 
smaller momenta. In this description, because of the singularity of the hard 
parton-parton cross-section $\sim 1/k_t^4 $, an energy dependent cut-off $k_{min}$, called an infrared cut-off, needs to be 
introduced, as can be seen in  PYTHIA 8.1 \cite{Sjostrand:2007gs}.  This cut-off is applied to the hard matrix elements, not  to the parton cascade, 
which is stopped when $k_t=k_0$. This  cut-off depends on the chosen PDF set. The multiple interaction possibility 
is included through eikonalization, both for the DGLAP and BFKL description.
 Finally, the third figure, (c),   includes absorption of low $k_t$ partons. An effective ``infrared'' cut-off', $k_{sat}(x)$  
 limits the low-$k_t$ partons. This parameter which depends on $x$, hence on the energy, is dynamically generated by the 
 enhanced multi-Pomeron diagrams.  Notice that the variable conjugate to  impact parameter $b$ 
is the momentum transfer exchanged through the Pomeron ladder, while the variable $k_t$ is the transverse 
momentum of the intermediate parton.   

 We observe   that this description  follows from the classic GLR paper \cite{Gribov:1984tu},  where all gluons, those involved in hard scattering 
as well as the soft ones from radiation before the scattering, are considered on the same footing and described through evolution equations. 
This differs from  an approach such as
the soft-$k_t$ resummation model (BN mini-jet model), where soft gluons are treated as a separate factorized term from the hard mini-jet cross-section and resummed through a semi-classical procedure a' la Bloch and Nordsieck (thus the name BN) \cite{Bloch:1937pw}. 
 \begin{figure}
\resizebox{0.5\textwidth}{!}{
\includegraphics{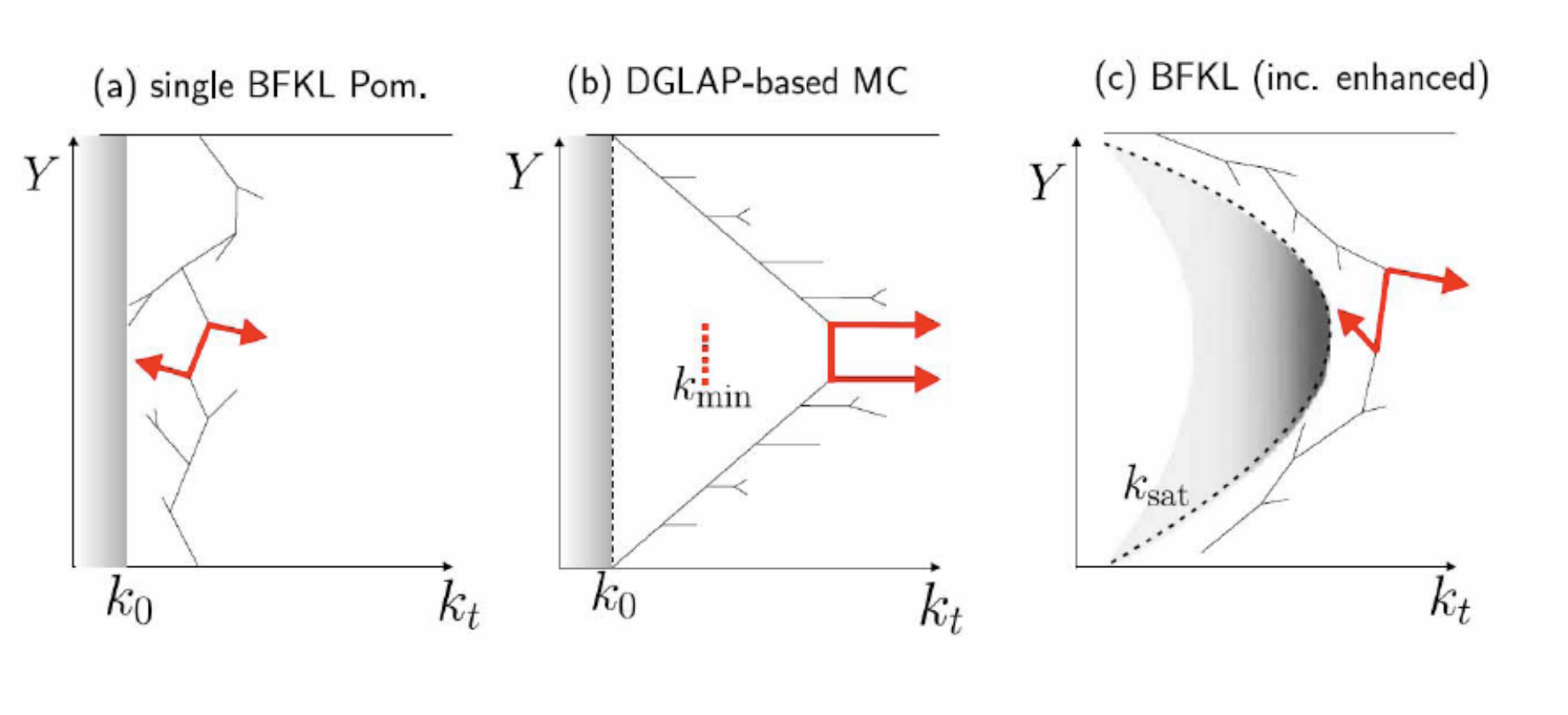}}
\caption{Graphical description of mechanism of semi-hard interactions in $pp$ collisions from
 \cite{Ryskin:2011qh}. Reprinted with permission from \cite{Ryskin:2011qh}, \copyright (2011) by IOP.}
\label{fig:abcKMR}
\end{figure}  
\par\noindent 
\begin{itemize}
\item  KMR2: Eikonalization of the scattering amplitufde 
\end{itemize}
A simplified description of the model can be found in Appendix A of  \cite{Khoze:2000wk}. 
Starting with the usual  expressions for 
the total and elastic cross-section in impact parameter space 
\bea
 \sigtot=2\int d^2\vecb_t A_{el}(b_t)\\
 \sigel=\int d^2 \vecb_t |A_{el}(b_t)|^2
\eea
the scattering amplitude  $A_{el}$ is approximated as  to be completely imaginary,. i.e. there is no real part.

With an effective (for illustration)   Pomeron trajectory written as $\alpha_P(t)=\alpha_P(0)+\alpha'_P t=1+\Delta +\alpha'_P t $ 
and vertex with exponential t-dependence $\beta_p exp(B_0 t)$, the starting point is 
 \begin{align}
 \Im mA_{el}(s,t)&=&\beta_P^2(t)\left( \frac{s}{s_0}\right)^{\alpha_P(t)-1}\ \ \ \   \ \ \ \ \ \ \ \ \ \ \\
 &=&\beta_P^2(t)\left( \frac{s}{s_0}\right)^{\alpha_P(0)-1} e^{\alpha'_P\ t\  \log \frac{s}{s_0}}
 \label{eq:KMRamplel}
 \end{align}

  Writing the amplitude in $b-$space as 
  the Fourier transform of Eq. ~(\ref{eq:KMRamplel})  with $t=-q^2 $
 \begin{align}
 {\cal F}[A(s,t)]=\frac{1}{(2\pi)^2}\int d^2\vecq e^{i\vecb \cdot \vecq} A(s,t)\ \ \ \ \ \ \ \ \ \ \\
= \left( \frac{s}{s_0}\right)^{\Delta} e^{(B_0 /2+\alpha'_P\  \log \frac{s}{s_0}) b^2)/4}.
 \end{align}
this will be the input for the successive eikonalization procedure.
In the case of a single channel the amplitude is written in term of the opacity function $\Omega(b)$, i.e. 
\be
\Im m A_{el}=[1-e^{-\Omega(b)/2}]
\ee 
and then  eikonalizing it, one   obtains
 \be
 \sigtot=4\pi \Im m A(s,0)=2\int d^2\vecb
  [
 1-e^{
 -\Omega(b,s)/2)}
 ]
 \ee
with the {\it opacity } 
\begin{align}
 \Omega(b,s)=\frac{ 
 \beta_P ^2 (s/s_0)^{\alpha_P(0)-1} 
 }
 {4\pi B_P}
 e^{-b^2/4B_P}\\
  B_P=\frac{1}{2}B_0+\alpha'_P \log (s/s_0)
\end{align}
  A  Good and Walker model to include low-mass  diffractive dissociation is then developed as we shall now   describe.
\par\noindent 
\begin{itemize}
\item KMR3: Inclusion of diffractive processes
\end{itemize}
The KMR approach to diffraction has evolved through the years,  as more precise and higher energy  data allowed further  
understanding. We summarize here some of the main ingredients of this model, whose most recent application to LHC 
data can be found in \ \cite{Martin:2012nm}.
The authors begin with $s-$channel unitarity,
\be
2\Imo T_{el}(s,b)=|T_{el}(s,b)|^2+G_{inel}(s,b)
\ee  
which, for $\Omega$ real  is satisfied  by 
\be
T_{el}(s,b)= i(1-e^{-\Omega/2})
\ee
One then writes the total elastic and inelastic cross-sections in impact parameter space 
as
\be
\frac{d^2\siginel}{d^2b}=\frac{d^2\sigtot}{d^2b}-\frac{d^2\sigel}{d^2b}=2 \Imo T_{el}-|T_{el}|^2
\label{eq:inel}
\ee
which coincides with $1-e^{-\Imo \Omega}$ and brings  the interpretation of $e^{-\Omega}$ as the probability 
for inelastic interactions. One notices that this interpretation  is valid also  in case of a non-negligible  real part of $\Omega$. 


Diffraction is defined as elastic scattering and low-mass proton dissociation, to distinguish it from 
high mass dissociation,  the two types of processes being  graphically described in Fig.~\ref{fig:KMRdiff} from \cite{Khoze:2010by}. 
 Apparently the multi-Pomeron vertex controls both the saturation scale and high-mass dissociation. 
\begin{figure}
\resizebox{0.5\textwidth}{!}{
\includegraphics{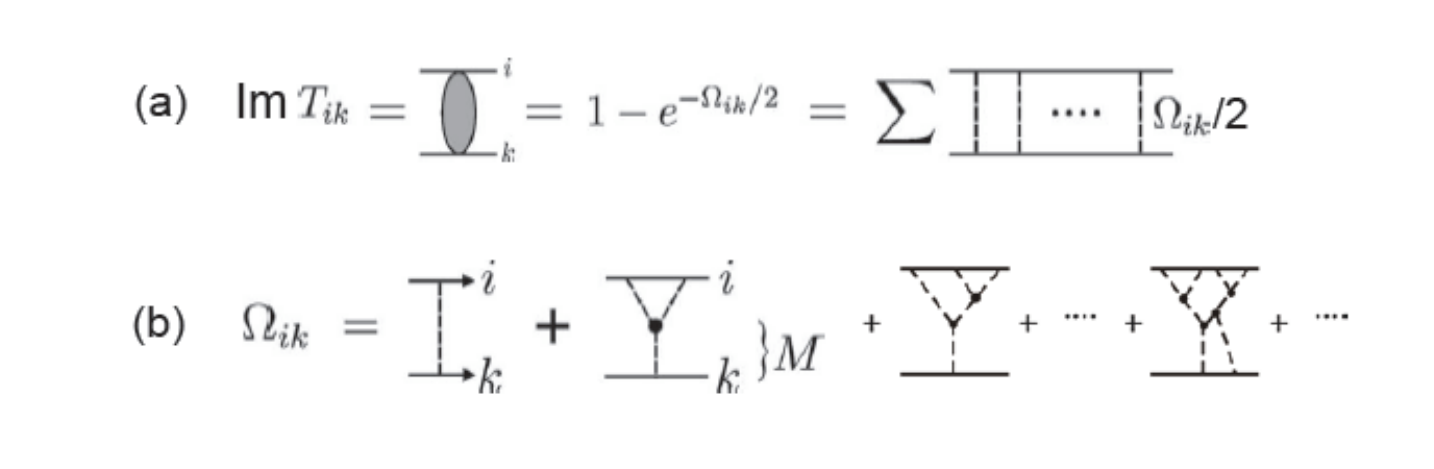}}
\caption{Graphical description of low-mass proton dissociation (a) and high mass dissociation 
(b) through triple Pomeron and multi-Pomeron corrections,  from \cite{Khoze:2010by}. Reprinted with permission  from \cite{Khoze:2010by},
\copyright (2010) by Springer.}
\label{fig:KMRdiff}
\end{figure}
The elastic differential cross-section is written as
\be
\frac{d\sigma_{el}}{dt}=\frac{1}{4\pi}|
\int d^2
 \vecb e^{i\vecq_t\cdot \vecb}
 \sum_{i,k}  |a_i |^2 |a_k|^2  (1-e^{-\Omega_{ik}(b)/2})
  |^2
\ee
In \cite{Ryskin:2012az} a three channel formalism  is used to describe the diffractive final states. Here we shall describe 
their two-channel formalism. In \cite{Khoze:2000wk}, the  GW formalism is applied to the states   $pp,\ pN^*,\ N^*N^*$, 
and the relative  processes are  introduced through a parameter $\gamma$ which describes the effective coupling 
of the proton to the excited state $N^*$, which will then decay into the observed low mass diffractive products. 
This is done by modifying the one channel description into a two channel one, i.e. by the substitution
\bea
\beta_p\rightarrow \begin{pmatrix} \beta(p\rightarrow p)&\beta(p\rightarrow N^*) \\\beta(N^*\rightarrow p) & \beta(N^*\rightarrow N^*)\end{pmatrix}\\
\simeq \beta(p\rightarrow p) \begin{pmatrix} 1&\gamma\\ \gamma & 1\end{pmatrix}
\eea
The expression for the elastic amplitude  thus modified to take into account the other channels, is now 
\begin{align}
\Imo A_{el}(b_t)=1-\frac{1}{4}e^{-(1+\gamma)^2\Omega(b_t)/2}-\nonumber\\
\frac{1}{2}e^{-(1-\gamma^2)\Omega(b_t)/2}-\frac{1}{4}e^{-(1-\gamma)^2\Omega(b_t)/2}]\\
\Imo A(pp\rightarrow N^*p)(b_t)=\ \ \ \ \ \ \ \ \ \                   \ \ \ \ \ \    \ \nonumber\\
\frac{1}{4}[e^{-(1-\gamma)^2\Omega(b_t)/2}-e^{-(1+\gamma)^2\Omega(b_t)/2}]\\
\Imo A(pp\rightarrow N^*N^*)(b_t)= \frac{1}{4}[-e^{-(1-\gamma)^2\Omega(b_t)/2}+\nonumber\\
2 e^{-(1-\gamma^2)\Omega(b_t)/2}-e^{-(1+\gamma)^2\Omega(b_t)/2}]
\end{align}
In this model the opacity function is real and the amplitude acquires a real part  
through 
\bea
\frac{\Reo A}{\Imo A}=\tan (\frac{\pi \lambda}{2})
\eea
where
\begin{equation}
\lambda=\frac{\partial \log(Im A)}{\partial \log s}
\end{equation}
 The elastic proton-Pomeron vertex is parametrized as 
   \be
  V{p\rightarrow P}=\frac{\beta_p}{(1-t/a_1)(1-t/a-2 )}\\
  \ee
with $\beta^2_p$ to be obtained from  $pp$ total cross-section.

\par\noindent
\par\noindent
\begin{itemize}
\item {KMR high mass diffractive dissociation}
\end{itemize}
The previously described decomposition of elastic scattering into GW type states cannot be applied to other types of 
inelastic processes {where the final state has a continuous mass distribution,
 and  does not have the same quantum numbers as the initial proton. As we shall also see in the case of the Tel Aviv model, 
 triple-Pomeron exchanges are invoked for this type of events, with the cross-sections written as \cite{Khoze:2000wk}
\begin{align}
\frac{M^2 d\sigma_{SD}}
{dt dM^2}=\frac{1}{16\pi^2}
g_{3P}(t) \beta(0) \beta^2(t)
(
\frac{s}{M^2}
)^{2\alpha(t)-2}(\frac{M^2}{s_0})^{\alpha(0)-1}
\\
\frac{M^2 d\sigma_{DD}}{dt dy_1 dy_2}=\frac{1}{16\pi^3}g_{3P}^2(t) \beta^2(0) 
exp[(
1+\alpha_{\mathbb P}(0)-2\alpha_{\mathbb P}(t)
)\Delta y] \times  \nonumber\\
\times (\frac{M^2}{s_0})^{\alpha(0)-1}
\end{align}
where $\beta(t)$ is the coupling of the Pomeron to the proton and $g_{3P}(t)$ is the triple Pomeron vertex, 
obtained from a fit to low energy ISR data in  \cite{Ryskin:2012az}. $M$ and $y$ are the diffractive mass and 
the rapidity regions for double diffractive dissociation. To avoid these contributions to grow too rapidly and violate 
unitarity, saturation and screening  must be imposed and survival probability factors are introduced, as described 
for instance in \cite{Khoze:2000wk}.

\begin{itemize}
\item { KMR5 Recent phenomenology}
\end{itemize}
After the general overview of the model, let us see how KMR apply it to elastic scattering at LHC. The basic building blocks of 
this model are now the  three-channel eikonal formalism
and 
multipomeron exchange diagrams, incorporating both unitarity and Regge behaviour. These physical requirements are parametrized through 
 the following 
 procedure:
\begin{itemize}
\item the $s-$dependent  bare Pomeron intercept $\Delta=\alpha_P(0)-1$
\item the bare Pomeron slope $\alpha'\simeq 0$
\item a parameter $d$, which controls the BFKL diffusion in $k_t$
\item the strength 
 of the triple-Pomeron vertex
\item the relative weight of the diffractive states $\gamma_i$, determined by low mass diffractive dissociation
\item the absolute value $N$ of the initial gluon density.   
\end{itemize}
In \cite{Martin:2011gi}, the authors compare their original predictions, for the total cross-section and   diffractive quantities, with the 
TOTEM results, and, as stated earlier, they find the need to do new adjustments of the parameters,  acknowledging  7 lessons, 
among which i) the fact that  
   Totem result, $\sigma_{inel}=\sigma_{total}-\sigma_{el}=73.5 \ mb$,
 is higher than the others, CMS and ATLAS extrapolated -KMR believe it is due to low mass diffraction being higher than expected- and that  ii)
 the ratio
 $\sigma_{el}/\sigma_{total}\simeq 1/4$ according to TOTEM and it cannot grow further. It is also acknowledged that  to describe 
 both the forward peak and the dip in $d\sigma_{el}/dt$ the Durham model would need to introduce more parameters.

Adjustment of the parameters leads to new results, which 
we present in Table \ref{tab:KMR0} from \cite{Martin:2012mq}.
In this table,   some KMR results for total cross-sections both prior to  and including recent LHC data are shown.
\begin{table*}
\caption{Values for various total cross-section components, in two different models: at left, in the original KMR model, prior to the LHC data 
\cite{Ryskin:2011qe} , at right in the KMR 3-channel eikonal from \cite{Ryskin:2012ry}, 
inclusive of LHC TOTEM data at $\sqrt{s}=7\ {\rm TeV}$. }
\label{tab:KMR0}
\centering
\begin{tabular}{||c||c|c|c|c||c|c|c|c|c||}
\hline
\multicolumn{5}{|c|}{KMR pre-LHC7}&\multicolumn{5}{|c|}{3-channel KMR post-LHC7}\\
\hline
energy& $\sigtot$& $\sigma_{el}$&$\sigma^{SD}_{lowM}$ &$\sigma^{DD}_{lowM}$ &$\sigtot$&$\sigma_{el}$ &$B_{el}$ 
&$\sigma^{DD}_{lowM}$ &$\sigma^{DD}_{lowM}$  \\
TeV&mb&mb&mb&mb&mb&mb&${\rm GeV}^-2$&mb&mb\\ \hline
1.8 &72.7 &16.6 &4.8 &0.4 &79.3 &17.9 &18.0 &5.9 &0.7 \\ \hline
7    &87.9 &21.8 &6.1 &0.6 &97.4 &23.8 &20.3 &7.3 &0.9 \\ \hline
14  &96.5 &24.7 &7.8 &0.8 &107.5 &27.2 &21.6 &8.1 &1.1 \\ \hline
100&122.3 &33.3 &9.0 &1.3 &138.8 &38.1 &25.8 &10.4 &1.6 \\ \hline
\hline
\end{tabular}
\end{table*}
This  adjustment  does ameliorate the situation, bringing their prediction quite close to  most recent TOTEM result 
(at this writing) of  $\sigma_{ total}=98.58\pm 2.23\ mb$ \cite{Antchev:1472948}. 

Finally,  recent  descriptions of the elastic differential cross-section in the small $|t|$ range in this model is shown in 
 Figs.~\ref{fig:KMRdsigdt} from
\cite{Khoze:2014nia}, where a good description of how the various choices of formalism and parameters lead to the final results.
\begin{figure}
\centering
\resizebox{0.5\textwidth}{!}{
\includegraphics{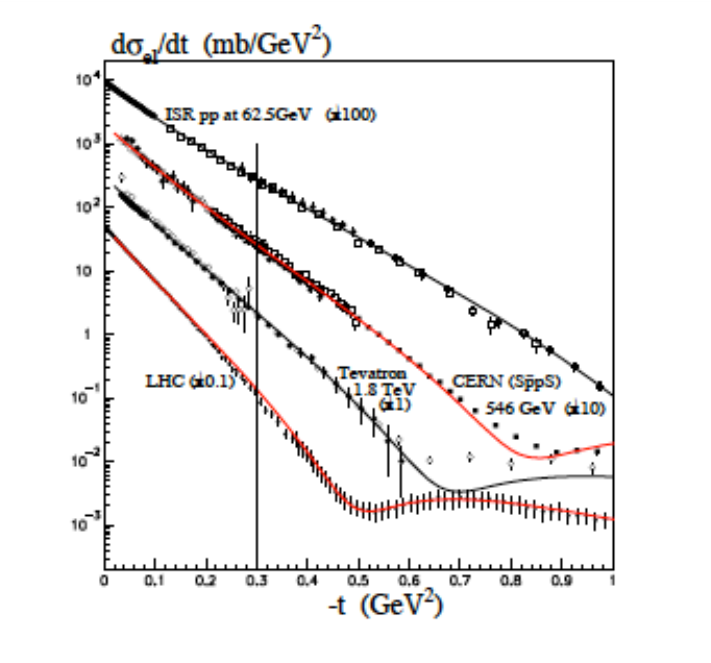}}
\caption{The elastic differential cross-section for $pp$ or $p{\bar p}$ using 
a  two channel eikonal model, which includes  the pion loop  contribution to the pomeron trajectory,   from \cite{Khoze:2014nia}.
Reprinted with permission from \cite{Khoze:2014nia}, \copyright (2014) by IOP.}
\label{fig:KMRdsigdt}
\end{figure}
 \subsubsection{Very small $t$ behaviour\label{KMRsmallt}}
 
As already seen when discussing early measurements
at ISR, various experiments reported an increase in the effective slope as $-t$ approached 
 zero. 
 Such an effect is also seen 
 when comparing   LHC7 data with a phenomenological 
 application such as the  PB model revisited in \cite{Grau:2012wy,Fagundes:2013aja}. In this application, the two exponential model 
 with a constant phase describes  very well the region $0.2<|t|<2.5 \ 
  {\rm GeV}^2$, but misses the optical point by some $ 10\%$, signaling that the slope has a more complicated $t$-dependence 
  than that  given by two exponentials.  In general, attempts to describe the elastic differential cross-section near 
 the optical point as well as the region past the dip
encounter difficulty in describing the very small $t$ behavior. 
Presently, for LHC, a complete description  
 is still lacking. From the theoretical point of view, it can be argued that this region should have contributions 
 from the nearby  thresholds in the unphysical region.   The presence of this behavior at very small $t$ was advocated 
 long time ago  by Anselm and Gribov \cite{Anselm:1972ir} as well as   by Cohen-Tannoudji, Ilyin and Jenkovszky 
 \cite{CohenTannoudji:1972gd} and has been  discussed by the Durham  \cite{Khoze:2000wk}  and  the Cosenza group 
 \cite{Fiore:2008tp}. In the following,    we shall describe two attempts to include such effects, the one  in  the Durham model, 
 and then a model   proposed by   Pumplin.

In the KMR model  \cite{Khoze:2000wk},  the Pomeron trajectory of Eq.~(\ref{eq:2comPom}) is modified  so as to include a very small $|t|$ effect.
 It is then called a two component Pomeron, consisting of three terms: the 
 first two correspond to the usual linear trajectory  describing  the  large scale, small impact parameter space $b_t$, while  
 the third term  corresponds to pion-loop insertions and contributes a correction to the small $t$-behaviour of the local slope. 
Such  improvement over the linear Pomeron trajectory consists in introducing the pion loop corrections as prescribed by 
Anselm and Gribov\cite{Anselm:1972ir}. The underlying physical idea is that pions being the lightest hadrons and the 
massless pion limit expected to be neither divergent nor negligible may provide observable corrections to the high 
energy slope of the elastic differential cross-section in the low momentum transfer limit.
The expression proposed in  \cite{Khoze:2000wk} for the modified Pomeron trajectory is
\be
\alpha_\mathbb{P}(t)=\alpha(0)+\alpha' t -\frac{\beta^2_\pi m^2_\pi}{32\pi^3}h(\tau);\ [\tau = \frac{4 m_\pi^2}{|t|}],
\label{Z1}
\ee
where
\be
h(\tau)=\frac{4}{\tau}F^2_\pi(t)[2\tau - (1+\tau)^{3/2}\log(\frac{\sqrt{1+\tau}+1}{\sqrt{1+\tau}-1})+\log\frac{m^2}{m^2_\pi}]
\label{eq:htau}
\ee
 with $m = 1\ {\rm GeV}$ a semi-hard scale.
 The coefficient $\beta^2_\pi$ specifies the Pomeron residue to the $\pi \pi$ total cross-section
  and
 \be
  F_\pi(t)=\frac{1}{1-t/a_2}.
 \ee
 The above modification  should then 
 explain the very small-$t$ dependence of the effective slope of the differential elastic cross-section in the KMR model.


Reflecting the same type of singularity in low-$t$ elastic scattering, a model was constructed by Pumplin
for the elastic amplitude  in  impact parameter space \cite{Pumplin:1991ea}  } which has the virtue that its Bessel transform, i.e., the elastic amplitude 
as a function of the momentum transfer $t\ =\ -q^2$, has branch points at the ``right thresholds'': $t\ =\ \mu^2; (2\mu)^2; 
(3\mu)^2,....$. Or,  it can be modeled as the two gluon exchanges with a threshold at $t\ =\ 4m_0^2$, where $m_0$ is the
effective gluon mass.

Defining the elastic amplitude as
\begin{equation}
\label{p1}
\sigma_{total}(s) = 4 \pi \Im m F(s, t=0),
\end{equation}
with
\begin{equation}
\label{p2}
F(s, t) = i \int_o^\infty b db J_o(qb) \tilde{F}(s, b)
\end{equation}
and
\begin{equation}
\label{p3}
\tilde{F}(s, b) = 1 - e^{-\Omega(s,b)}
\end{equation}
in the limit of a purely imaginary amplitude, which is a good approximation at large $s$ and small $t$, 
Pumplin models the $b$-dependence of the real $\Omega$, as follows 
\begin{equation}
\label{p4}
\Omega_o(s, b) = \eta(s) e^{\mu[b_o(s) - \sqrt{(b^2 + b_o(s)^2)} ]}. 
\end{equation}
Using the decomposition
\begin{equation}
\label{p5}
\frac{e^{- [k \sqrt{q^2 + \mu^2}]}}{\sqrt{q^2 + \mu^2}} = \int_o^\infty bdb J_o(qb) \frac{e^{- [\mu \sqrt{k^2 + b^2}]}}{\sqrt{k^2 + b^2}},
\end{equation}
it is possible to invert Eq.(\ref{p2}) as a convergent power series given by
\begin{equation}
\label{p6}
F_o(s, t) = ib_o(s)^2 \eta(s) \sum_{n=1} \frac{(-\eta)^{n-1}}{n!} [\frac{y_o(1+y)}{y^3}]e^{y_o-y},
\end{equation}
where $y_o =\ n\mu b_o(s)$ and $y =\ \sqrt{y_o^2 - t b_o^2(s)}$.

The series  expansion for the elastic amplitude obtained in Eq.(\ref{p6}) converges rapidly and hence was proposed as quite useful for numerical
computations. Also, the resulting slopes and curvature as a function of the momentum transfer are quite smooth. Remaining always 
in the small $t$ range, the model provides clear insights into the intricacies involved in obtaining accurate estimates of the slopes
and curvatures as a function of $t$. Let $B(0)$ and $C(0) = B^{'}(0)/2$ be the forward slope and the forward curvature respectively.
Below we quote their numerical values at $\sqrt{s} = 19.4\  $and $\ 546\ {\rm GeV}$ obtained in this model.
\begin{eqnarray}
\sqrt{s} = 19.4 {\rm GeV}:B(0) = 12.44 {\rm GeV}^{-2}; C(0) = 7.72 {\rm GeV}^{-4}\nonumber\\
\sqrt{s} = 546 {\rm GeV}:B(0) =  16.82 {\rm GeV}^{-2}; C(0) = 13.65 {\rm GeV}^{-4}\nonumber\\
\label{p7}
\end{eqnarray}
However, the sizable forward curvature cautions  against accepting forward slopes obtained through single exponential fits. Another important 
point is that the curvature decreases strongly as a function of $t$ and changes sign at larger values  of $t$
\cite{Grau:2012wy,Pumplin:1991ea}. 
\vskip 0.3cm

\subsubsection{Elastic diffraction in AdS/CFT}
A strong emphasis  on the 
embodiment of N=4 Super Yang Mills (SYM) physics \cite{Gotsman:2010nw} is part of the work by the Tel Aviv group, 
Gotsman, Levin and Maor (GLM). Because of this as well as of  the intrinsic interest of the results, 
we now briefly discuss the approach to this problem in the context of the string/gauge theory developed in a series of papers by the Brown 
University group \cite{Brower:2007xg,Brower:2007qh,Brower:2006ea}. In these papers  string/gauge
duality has been employed to obtain interesting results for the Pomeron and the physics of elastic 
amplitudes for small $t$. 

 A brief summary of results can be found in \cite{Tan:2011zzd}.
We can not give here a detailed description of  the elegant formalism developed in \cite{Brower:2007xg,Brower:2007qh,Brower:2006ea}, 
but present below results regarding the Pomeron intercept and a modified eikonal
expression derived from $AdS_5$. 
In their string/gauge theory, a scalar kernel ${\cal K}(s,t,z,z')$ as a function of  the 4-dimensional invariants ($s,t$) and 
two {\it bulk} coordinates ($z, z'$) from the fifth dimension, describes  the Green's function for the Pomeron. 
At $t =0$ its imaginary part reads
\begin{equation}
\label{b1}
\Im m {\cal K}(s,t =0,z,z') \approx\ \frac{s^{j_o}}{\sqrt{\pi {\cal D} ln(s)}} e^{ln^2(z/z')/{\cal D} ln(s)},
\end{equation}
where the Pomeron intercept $j_o$ and ${\cal D}$ are given by
\begin{equation}
\label{b2}
j_o = 2 - 2/\sqrt{\lambda};\ \  {\cal D} = 2/\sqrt{\lambda},
\end{equation}
with $\lambda$ denoting the (large) 't Hooft  coupling constant. Three loop universal anomalous dimension of the Wilson operators in N=4 SUSY Yang-Mills model were first obtained in \cite{Kotikov:2004er}.

The expression for the imaginary part of the Pomeron propagator in Eq.(\ref{b1})  is similar to its corresponding form (in 
the weak limit) with its BFKL  expression
\begin{equation}
\label{b3}
\Im m {\cal K}(s,t =0,p_\perp,p_\perp') \approx\ \frac{s^{j_o}}{\sqrt{\pi {\cal D} ln(s)}} e^{ln^2(p_\perp/p_\perp')/{\cal D} ln(s)},
\end{equation}
where $j_o =  1 + \alpha N (\frac{4 ln(2)}{\pi}), \alpha = g_{YM}^2/4\pi$ is the $SU(N)$ coupling constant for $N$ colours
and ${\cal D} = (14 \zeta(3)/\pi) \alpha N$. The above allows one to identify the correspondence between diffusion in 
the gluon mass virtuality given by $ln(p_\perp^2)$ in Eq.(\ref{b3}) with diffusion in the radial coordinate $ln(z^2)$ in the dual
$AdS_5$ given by Eq.(\ref{b1}).

A reduction from $AdS_5$ to the needed $AdS_3$ propagator is then made. The latter depends on the $AdS_3$ chordal distance
\begin{equation}
\label{b4}
v = \frac{(x_\perp - x_\perp')^2 + (z - z')^2}{2 z z'}
\end{equation}
and a dimension $\Delta_+(j) - 1$ obtained for the physical spin $j$ operators occurring in BFKL/DGLAP:
\begin{equation}
\label{b5}
\Delta_+(j) = 2 + \sqrt{4 + 2\sqrt{\lambda} (j - 2)} = 2 + \sqrt{2\sqrt{\lambda}(j - j_o)},
\end{equation}
where Eq.(\ref{b2}) has been used to obtain the last part of the above equation. Rewriting Eq.(\ref{b5}) as 
\begin{equation}
\label{b6}
j(\Delta) = 2 + \frac{(\Delta -2)^2}{2\sqrt{\lambda}} = j(4 - \Delta),
\end{equation}
one sees the symmetry $\Delta \Leftrightarrow (4 - \Delta)$. Hence, the function $\Delta_+$ interpolates
correctly (i) the value $j = j_o$ at $\Delta = 2$ giving the BFKL exponent as well as (ii) $j = 2$ at $\Delta = 4$
corresponding to the energy-momentum tensor, the first DGLAP operator. A schematic representation of the relationship 
between $j$ and$\Delta $ is shown in Fig.~\ref{fig:ads}.
\begin{figure}[htbp]
\begin{center}
\resizebox{0.5\textwidth}{!}{
\includegraphics{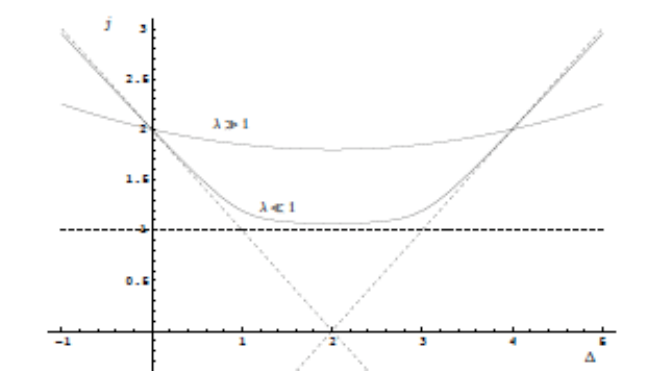}}
\caption{Schematic curve from \cite{Brower:2007xg} showing spin $j$ versus dimension  $\Delta$ as given in Eq.(\ref{b6}) 
both for $\lambda<< 1$ and $\lambda >> 1$. All curves pass through the points $j =2; \Delta = 0$ and $j = 2, \Delta = 4$
where the anomalous dimension vanishes. The dashed lines show $\lambda =0$ DGLAP branch(slope 1);
BFKL branch (slope 0) and inverted DGLAP branch (slope $-1$). Reprinted from \cite{Brower:2007xg}, \copyright (2007)
by Springer. }
\label{fig:ads}
\end{center}
\end{figure}
In the eikonal limit - for its domain of validity see \cite{Brower:2007qh}- the amplitude for scattering $1+2\to 3+4$ is written
as
\begin{equation}
\label{b7}
-2is \int (d^2b)(dzdz') P_{13}(z) P_{24}(z') e^{ib_\perp\cdot q_\perp} [e^{i\chi} -1],
\end{equation}
where $P_{13}(z),  P_{24}(z')$ are wave functions in the bulk coordinates and for Pomeron exchange, the generalized eikonal 
$\chi$ which is a function of ($s,b,z,z'$) is given in terms of the Pomeron kernel by
\begin{equation}
\label{b8}
\chi(s, b, z, z') = \frac{g_o^2 R^4}{2 (zz')^2 s} {\cal K} (s,b,z,z').
\end{equation}
In the limit where Pomeron exchange dominates, for $s\to\infty$ and $\lambda$ fixed, the eikonal reads
\begin{equation}
\label{b9}
\chi \sim \frac{e^{i\pi(1 -j_o/2)} (zz's)^{j_o -1}}{\sqrt{v(2 + v)}}
\end{equation}
We may pause here to note the similarity and differences between this approach [for $j_o >1 $ and moderate values of $z,z'$] and 
the minjet model for total hadronic cross-sections. Also in the minijet model, the eikonal is proportional to $s^\epsilon$ with 
$\epsilon\sim 0.3$, a power of the total energy. However, in the minijet model, there is an exponential suppression for large $b$ 
which softens the power law growth in energy of the eikonal to powers in logarithms of energy of the elastic amplitude. Eq.(\ref{b9}) 
by contrast shows that the eikonal decreases only as a power of $v$ (or $b$) for large $v$.    

\subsubsection{Gotsman, Levin and Maor: the Tel Aviv model}

Recent descriptions of the model, which  embody both Reggeon field theory and features from   N=4 SYM gauge theory,  
can be found in \cite{Gotsman:2012rm}
 and in \cite{Gotsman:2013lya} where  the diffractive peak  at LHC and a summary of experimental and various model results are presented.
 High energy predictions   for the mass distribution for  both low and high mass diffraction can be found in \cite{Gotsman:2013uca}.

For what concerns the elastic differential cross-section, from the optical point to past the dip, 
this model, 
similarly to KMR, describes the diffraction peak region
where non-perturbative effects are dominant, and  is not yet extended
to the dip region.

 To describe elastic scattering and include the different 
components of diffraction, single (SD), double (DD), and central (CD),  low-mass 
and high mass diffraction, GLM
introduce both a Good and Walker (GW) formalism 
with GW diffractive 
states, as well as    non-GW processes. 

In previous papers, whose reference can be found in  \cite{Gotsman:2012rq}, GLM had obtained a good description 
of the  total and the elastic differential cross-section in the small $-t$ region, with a 
given choice of the parameters defining the model. However, TOTEM data, in particular the rather high total cross-section value, 
were seen to require some changes in the parametrization \cite{Gotsman:2012rm}.  Only one parameter seems however 
to be in need of a change, namely the Pomeron intercept $\Delta_\mathbb{P}$. Other parameters need not to be changed,  
at least for a good description of high energy  data, namely $\sqrt{s}\gtrsim 500\ {\rm GeV}$, where the present interest lies for this model.
Before giving some more details on this model, we note the following input elements:
\begin{itemize}
\item both GW states and non GW states contribute to diffraction 
\item GW states contribute to both low and high mass mass diffraction, non-GW states only to high mass diffraction
\item the amplitude in the eikonal is purely imaginary (the model is basically applied to the low $t$-region)
\item  dipole form factors inspired by  the proton form factors  give the factorizable impact parameter dependence in the eikonal functions 
\item the s-dependence of the amplitudes comes from  a  single  Pomeron with $\alpha'_\mathbb{P}=0$ and intercept $\Delta_\mathbb{P}=0.23$
\item  multi-Pomeron interactions described by {\it enhanced} and {\it semi enhanced }  diagrams contribute to both GW and non-GW states 
\end{itemize}
From the above one notices once more that the s-dependence of this model is the same as in mini-jet models, with a term $s^\epsilon$, 
which is not associated with any explicit  $t$-dependence.

To take into account the whole spectrum of  diffraction, GLM starts with  a simple
 Good and Walker model with two eigenwave functions $\psi_1$ and $\psi_2$. 
Upon diagonalization of the interaction matrix,  the wave functions of the  two observed states, a hadron and a diffractive state, the latter  with mass 
small compared to the energy of the process,   respectively $\psi_h$ and $\psi_D$, are written as
\be
\psi_h=\alpha\psi_1+\beta\psi_2, \ \ \ \ \ \ \ \ \ \ \psi_D=-\beta\psi_1+\alpha\psi_2
\ee
with $\alpha^2+\beta^2=1.$
One then constructs the scattering amplitude $A_{i,k}(s,b)$ in impact parameter space $\vecb$, with $i=h,D$  by solving the unitarity condition 
\be
2\Im m A_{i,k}(s,b)=|A_{i,k}(s,b)|^2+G_{i,k}(s,b)
\ee
i.e.
\be
A_{i,k}(s,b)=i (1-e^{-\Omega_{i,k}(s,b)/2}) \label{eq:glm-aik}
\ee
The eikonal function $\Omega_{i,k}$ is the imaginary part of the scattering amplitude for a single Pomeron exchange, i.e.  
\be
\Omega_{i,k}=g_i(b)g_k(b)P(s)=g_i(b)g_k(b)s^\Delta
\ee
where  the Pomeron-proton vertex $g_i(b)=g_i S_i(b)$ and  $S_i(b)$ is the Fourier transform of the proton-like  form factor, 
a dipole with a scale $m_i$. The zero value for the Pomeron slope, i.e. $\alpha'=0$  in GLM is understood to be in agreement 
with the results from N=4 SY as described previously, and is crucial as it allows for resummation of all 
pomeron interactions 
 One sees that the impact space 
dependence in each eikonal $\Omega_{i,k}$ depends on 4 parameters, the scales $m_i$ and $m_k$ in the form factors and 
the proportionality factors $g_i,\ g_k$. The amplitude in Eq.({\ref{eq:glm-aik}}) gives the 
multipomeron exchange contribution to the elastic and 
GW
diffractive states. 

The contribution of GW states is obtained using the formulae
\begin{align}
a_{el}(b)=i(\alpha^4 A_{1,1}+2\alpha^2\beta^2A_{1,2}+
\beta^4A_{2,2})\\
a_{sd}^{GW}=i \{\alpha\beta \{ -\alpha^2A_{1,1}+(\alpha^2-\beta^2)A_{1,2}+\beta^2 A_{2,2}\}\\
a_{dd}^{GW}=i \alpha^2\beta^2 \{A_{1,1}-2A_{1,2}+ A_{2,2}\}
\end{align}
with the corresponding cross-sections  
\begin{align}
\sigma_{tot}=2\int d^2\vecb \ a_{el}(s,b)\\
 \sigma_{i}=\int d^2\vecb \ |a_{i}(s,b)|^2\ \ \ \ \ \ \ i=el,sd,dd\\
\end{align}

To these contributions, one needs to add non-GW terms, which are produced by  Pomeron interactions, in this model only considering triple Pomeron interactions. 
These contributions 
involve the triple Pomeron coupling $G_{3\mathbb{P}}$. 
For single diffraction into a mass $M$, with $Y=\log (M^2/s)$ the authors obtain
\bea
A^{sd}_{i;k,l}&=&\int d^2 \vecb\  2 \Delta (\frac{G_{3 {\mathbb P}}}{\gamma}
 \frac{1}{\gamma^2})\\
&\times &g_i(\vecb-\vecbp,m_i)
g_l(\vecbp,m_l) g_k(\vecbp,m_k)\\
 &\times& Q(g_i,m_i,\vecb-\vecbp,Y_m)\\
 &\times &Q(g_k,m_k,\vecbp,Y-Y_m) \\
 &\times &Q(g_l,m_l,\vecbp,Y-Y_m)
\eea
where $G_{3 {\mathbb P}}$ is the triple Pomeron vertex proportional to $\alpha_s^2$ and 
\begin{align}
Q(g,m,\vecb,Y)=\frac{G_{\mathbb P}(Y)}
{1+G_{3 {\mathbb P}}/\gamma) g G_ {\mathbb P}(Y)S(b,m)}\\
G_{\mathbb P}(Y)=1-exp[\frac{1}{T(Y)}]
\frac{1}{T(Y)}
\Gamma(0,\frac{1}{T(Y)}
)\\
T(Y)=\gamma e^{
\Delta_{\mathbb P}Y}\\
\gamma^2=\int\frac{d^2k_t}{4\pi^2} G^2_{3 {\mathbb P}}
\end{align}
Similarly, one can obtain the amplitudes ${\tilde A}^{dd}_{i,k}$ for double diffraction in terms of the functions $Q$ written above, 
details can be found in   \cite{Gotsman:2012rm}. The final result is that the integrated cross-section for single diffraction is the sum 
of two terms, the GW and non-GW, likewise for double diffraction. To obtain the integrated cross-sections however, a further step 
is required, i.e. the amplitudes $A^{sd}_{i;j,k}$ and $A^{dd}_{j,k}$ are multiplied by the survival probability factors $e^{-\Omega_{j,k}}$.

The model cannot be solved from first principles, and needs phenomenological inputs such as the Pomeron intercept, the interaction vertices between  
the Pomeron with the two GW states, the low-energy amplitude of dipole-target interactions and one constant, for the GW states,  $\beta$. 

The connection with $N=4\ SYM$ is considered in \cite{Gotsman:2010nw} where the model is discussed in light of satisfaction of two ingredients: i) 
the need to deal with a large coupling constant, and ii) the requirement to match with high energy QCD. Thus, the Pomeron intercept can be large,  
but the slope of the Pomeron trajectory is very small. These  results, already introduced in the previous papers, match with the fact that, in the strong 
coupling limit, $N=4\ SYM$  theory has a soft Pomeron, i.e. the reggeized graviton with a large intercept, $\alpha_{\mathbb P}=2-2/\sqrt{\lambda}$ 
with $\lambda=4\pi N_c\alpha_s^{YM}$, $\alpha_s^{YM}$, the QCD coupling constant. The other   important ingredient of the GLM model is   the natural 
matching with perturbative QCD, where the only vertex that contributes is the Pomeron vertex:  this is also understood  in $N=4$  SYM  since  the Pomerons 
(gravitons) interact by means of the triple Pomeron vertex, which infact is small,  at least $\propto 2/\sqrt{\lambda}$.  Another matching result between GLM 
and $N=4\ SYM$ is that  at large $\lambda$, only processes from diffraction dissociation contribute to the scattering amplitude. In GLM model, diffraction
 indeed plays also a large role, as discussed before. 
 
 The main ideas of this model can be summarized as follows:
 \begin{itemize}
\item a double face Pomeron, with
\begin{enumerate} 
\item a large intercept, $\alpha_{\mathbb P}(0)-1$, a direct consequence of $\Delta=1-2/\sqrt{\lambda}$ when $ \lambda$ 
becomes large
\item a short distance behavior, indicated by a small slope $\alpha'_\mathbb{ P} \simeq 0$, as a QCD matching prescription
\end{enumerate}
\item Good Walker mechanism and triple Pomeron vertex
 \end{itemize}
{To the above, one needs to add an important crucial component of GLM description of diffraction, i.e.   
the survival probabilities for large rapidity gaps.} From the eikonal 
 formulation, the survival probabilities are obtained from the quantities 
 \be
 P_{ik}^S=exp[-\Omega_{ik}(s,b)], 
 \ee
 which represent the probability that the initial state $|i,k >$ does not break up after the scattering.
 

After  the publication of the TOTEM actual data for the differential cross-section and confirmation of earlier results for total elastic and inelastic, 
GLM were able to consolidate their parameters, and present definite expectations for this model \cite{Gotsman:2012rm}.  
The detailed results 
for all the quantities of interest appear in the paper. These include $\sigma_{tot}, \sigma_{el}$ and $\sigma_{inel}$ at 
$\sqrt{s}=1.8,\ 7, 8, 57\ {\rm TeV}$ in addition to $ \sigma_{sd}, \ \sigma_{dd}$, for which the separate values obtained for the GW and 
non-GW contributions are also listed. 

We reproduce in 
Fig. ~\ref{fig:glmdiff} 
the fit to data in the small t-range for different energies up to LHC14. Since the model is only to be applied at vary small $-t$ values, 
 as one approaches the dip
the model starts  not to reproduce well the data.

\begin{figure}
\centering
\resizebox{0.5\textwidth}{!}{
\includegraphics{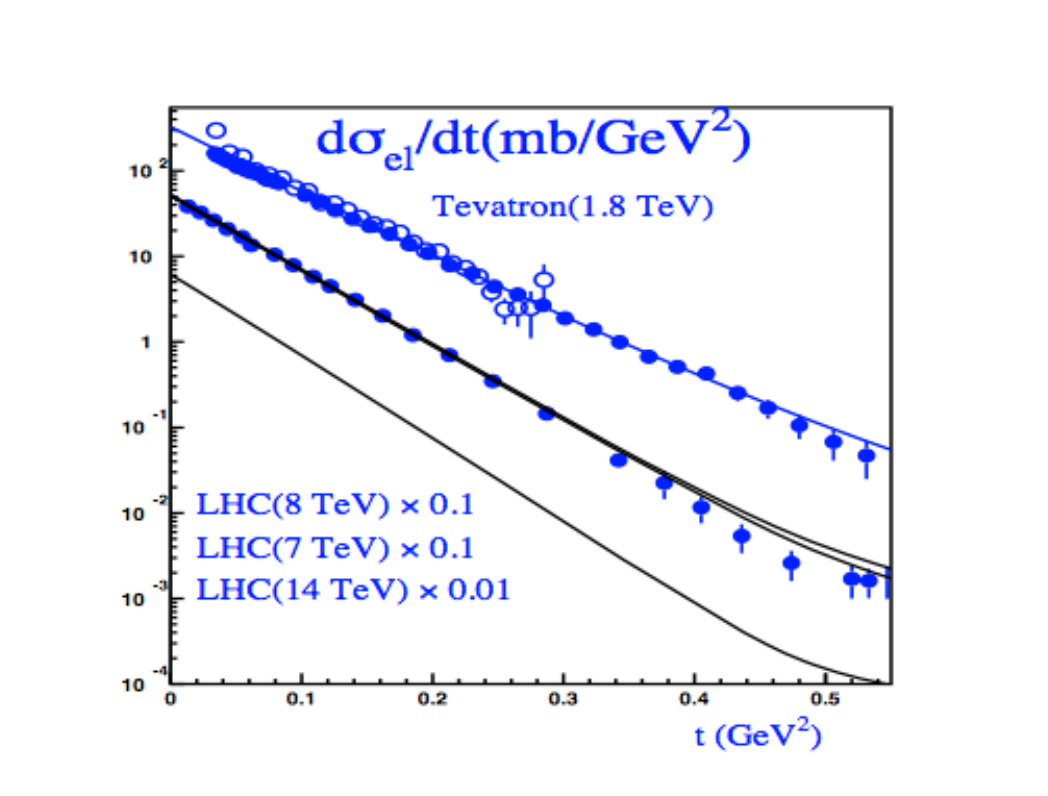}}
\caption{Description of low-$t$  elastic scattering in GLM from  \cite{Gotsman:2012rm}.
Reprinted from \cite{Gotsman:2012rm}, \copyright (2012) with permission by Elsevier.}
\label{fig:glmdiff}
\end{figure}

 Because of the importance of diffraction in this model, a warning is pronounced about the black disk limit in \cite{Gotsman:2012rq}. 
 The model results indicate a very slow approach of the elastic amplitude to saturation of this limit, in good qualitative agreement  
 with what  we have noted elsewhere in this review, namely that  the AUGER results for the inelastic cross-section indicates  
 $\sigma_{el}(57 \ {\rm GeV})\ne 0.5 \sigma_{total}$. 
 To summarize, it is observed that present Reggeon models all introduce at least two different 
 mechanisms for diffraction. In GLM, these 
 two mechanisms are : i) the GW production of diffractive states with unspecified finite mass, independent of energy, and 
 (ii) diffraction due to 
 Pomeron interactions, with a dependence on the Pomeron parameter $\Delta_{\mathbb P}$. In other models, there are distinctions between 
 low-mass and high-mass diffraction, and central diffraction. A discussion of this point can be found at the end of Ref.\cite{Gotsman:2012rm}. 
 
We note that   a model along very similar lines to GLM (and KMR) has also been  developed by Ostapchenko \cite{Ostapchenko:2010gt}. This model is compared with GLM and other similar models for diffraction in 
 \cite{Gotsman:2013lya}.   Ref.  \cite{Gotsman:2013lya}  presents  a good recent comparative discussion of diffraction amplitudes up to LHC7 energies. 

\subsubsection{A comment about soft gluons and diffraction. 
}\label{sss:diffandsoftguons}
Before ending  this brief review of  selected contributions   present in the literature, we shall comment on the possible interpretation of diffraction in terms of soft gluon emission. In the BN model of \cite{Grau:1999em,Godbole:2004kx}, the contribution to diffraction 
is not present. What drives the increase of the cross-section is the result of parton-parton scattering (mini-jets) dressed with soft gluon emission from the initial valence quarks. Eikonalization then re-sums  multiple interactions, each one of them coming with its own cloud of soft gluons. 
The present version of this model does not explicitly  include soft gluon emission from the spectator quarks. However, diffraction along the initial state protons in such a view should include both emission from the quarks participating to the scattering, but also from the so-called spectator quarks. When  the proton is hit by the other hadron, 
quarks {\it also}  undergo intra-beam scattering. Such emission is along the colliding particles. This is in agreement with the original view of diffraction a' la Good and Walker where the target and/or the projectile  dissociates  into a state with no change of quantum numbers: the only  gluon emission process which, by definition,  does not change the proton quantum numbers is resummed soft gluon emission. This is the dynamical picture which should be addressed. If nothing else but diffraction a' la Good and Walker is present, then this emission 
{\it could} be accompanied by a factor inhibiting parton-parton scattering, namely a survival probability factor.  
\subsection{One-channel mini-jet model for total, elastic and inelastic cross-sections \label{ss:Onechannel}}
As just discussed in  \ref{sss:diffandsoftguons},
our BN mini-jet model has not yet dealt with diffraction, being so far a one-channel eikonal model. However, this model can provide interesting insight in the distinction between correlated  and uncorrelated inelastic processes, and what in included in $\sigel$ in one-channel eikonal models.

Here we shall discuss the mini-jet contribution to total, elastic and inelastic cross sections, using a one-channel eikonal formulation 
as described in greater detail in \cite{Fagundes:2015vba}. \\
To construct the total cross section, mini-jets are embedded into the eikonal formulation. Starting with
 \be
\sigma_{total}=2 \int d^2 {\vec b} [1-\Re e( e^{i\chi(b,s)})]\label{eq:sigtot}
\ee
and neglecting the real part in the eikonal at very high energy, the above expression further simplifies into
\be \sigma_{total} =2 \int d^2 {\vec b} [1-e^{-\chi_I(b,s)}] \label{eq:onechtot}
\ee
where $\chi_I(b,s)=\Im m\chi(b,s)$.  Notice that $\Re e\chi(b,s)\simeq 0$ is a reasonable  approximation 
for the scattering amplitude in $\vecb$-space at $t=0$, where very large values of the impact parameter dominate 
and  phenomenologically     the ratio of the real to the imaginary part of the forward scattering amplitude $\rho(s) << 1$. 
By properly choosing a function $\chi_I(b,s)$, all  total hadronic cross sections, $pp$, $p{\bar p}$, $\pi p$, etc.,   can  be described  up to 
currently available data \cite{Grau:2010ju}. In the vast majority of  models, new data have often  required  an adjustment of the 
parameters which give $\chi_I(b,s)$. 

In previous publications, we had proposed a band whose upper border gave a good prediction for LHC results.  
By updating the model and anchoring the parameter set to LHC results, one can now proceed to refine our predictions 
for higher energies, LHC13 and beyond to the cosmic rays region.

The eikonal function of the mini-jet model of \cite{Grau:1999em,Godbole:2004kx} 
is given by
\bea
2\chi_I(b,s)=n_{soft}^{pp}(b,s)+n_{jet}^{pp}(b,s)\nonumber\\
=A_{FF}(b)\sigma_{soft}^{pp}(s)+A_{BN}^{pp}(p,PDF;b,s)
\sigma_{jet}(PDF,p_{tmin};s)\label{eq:chi}\nonumber\\
\eea
The first term  includes collision with $p_t \le p_{tmin}\sim (1\div 1.5 )\ {\rm GeV}$, the second is
obtained from the mini-jet cross section. The term $n_{soft}^{pp}(b,s)$ is not predicted by our model so far 
and we parametrize it here with $\sigma_{soft}^{pp}(s)$,  obtained  with a constant and one or more decreasing 
terms, and $A_{FF}$, the impact parameter distribution  in the non perturbative term, obtained through a  
convolution of two proton form factors.

As expected,  the second term  in Eq.(\ref{eq:chi}) is numerically negligible at energies $\sqrt{s}\lesssim 10 \ {\rm GeV}$. 
The perturbative, mini-jet, part discussed previously is  defined  with $p_t^{parton}\ge p_{tmin}$ and is determined 
through a set of perturbative parameters for the jet cross section, namely a choice of  PDFs  and the appropriate  
$p_{tmin}$. Since  soft gluon re-summation includes all order terms in soft gluon emission,  our model   uses  LO,   
library distributed,  PDFs.

{The results of the LHC updated analysis of the one-channel BN model have been presented in Fig.~\ref{fig:pptot-onech}, where, as mentioned, 
  the    ${\bar p} p $ points are shown, but  have  not been used for the phenomenological fit, and 
values for $pp$ extracted from cosmic ray experiments have not been used either.
In this figure both "old" densities such as GRV  \cite{Gluck:1998xa} and "newer" ones such as MSTW \cite{Martin:2009iq} have been included and compared   with   other models  \cite{Block:2011vz}
 and one-channel model predictions   such as in 
\cite{Khoze:2014nia,Fagundes:2012rr}. Table \ref{tab:total} contains the points corresponding to our model 
results for both GRV and MSTW densities.}\footnote
{An error in the arXiv posted paper  \cite{Fagundes:2015vba} 
in the table for MSTW predictions at 13 and 57 TeV has now been corrected.}
Results for MRSTdensities can be found in \cite{Fagundes:2014fza}, together with  details of  different parameter sets used for the different PDFs.
\begin{table}[htb]
\caption{Total cross section values in mb, from the mini-jet model with two different PDFs sets. }\label{tab:total}
\centering 
\begin{tabular}{|c|c|c|}
\hline
$\sqrt{s}$ {\rm GeV}&$\sigma_{total}^{GRV}$ mb &$\sigma_{total}^{MSTW}$ mb\\ \hline
5		&	39.9	   &39.2	          	 \\
10		&	38.2	   &	38.6          \\
50		&	41.9	   &	 42.2         	 \\
500		&	63.2	   &	62.0	          \\
1800	&	79.5	   &	 77.5        \\
2760      &     85.4   &    83.6         \\
7000	&	98.9	   &	  98.3       \\
8000	&	100.9 & 101.3		        \\
13000	& 108.3	& {111.7}\\
14000	&	109.3  &113.7         \\
57000	&	131.1	&{149.2}\\
\hline
\end{tabular}
\end{table}

We notice that our model is able to describe very well all the total cross section  accelerator data,  and gives  a good agreement 
with cosmic ray  data.  The AUGER point falls within the two different parametrizations we are using, full line for MSTW and dashes 
for GRV. By construction, both parameterizations remain very close up to  LHC7 and LHC8  energies, and start diverging as the energy 
increases, as a consequence of  the uncertainty on the very low-x behavior of the densities.

To summarize 
in the model we have proposed, past ISR energies, mini-jets appear as  hard gluon-gluon collisions accompanied by 
soft gluon emission $k_t$-resummed down into the infrared region. In this language, we have a {\it dressed} hard 
scattering process, with the mini-jet cross section giving the same energy behavior as the hard Pomeron,  and soft 
gluon resummation providing {\it the dressing},  in which the hard interaction is embedded. The eikonal formulation 
then transforms this {\it dressed hard gluon } interaction into a unitary ladder.  The main difference with other mini-jet 
models such for instance in  \cite{Giannini:2013jla}, is the taming mechanism  ascribed to soft gluon resummation in the infrared region.

We now turn to discuss the inelastic cross section. The inelastic total cross section is defined by subtraction  from  
the total and the elastic cross sections.   However, experimentally, it is usually defined only in specific phase space regions, 
and eventually extrapolated via MC simulation programs, which also include parameters and choice of models in the diffractive region. 
{One exception is TOTEM which covers a large rapidity range.} 

Here, we shall focus on one,  theoretically well defined, part of the inelastic cross section, what we define as 
{\it uncorrelated}, which is appropriately described in the mini-jet context and through the one-channel mode. 
In the following we shall see how. 

Since our study \cite{Achilli:2011sw} on the inelastic cross section at LHC, soon followed by the first 
experimental results \cite{Aad:2011eu}, data related to measurements in different kinematic regions have appeared. 
Extensive and detailed measurements have been obtained  for  the inelastic proton-proton cross section by  
CMS \cite{Chatrchyan:2012nj}, ATLAS \cite{Aad:2011eu}, TOTEM \cite{Antchev:2013haa,Antchev:2013iaa}, 
ALICE \cite{Abelev:2012sea}  and LHCb \cite{Aaij:2014vfa} Collaborations. These measurements cover different regions, 
central and mid-rapidity, large rapidity, high and low mass diffractive states. 
Extensive QCD modeling, including minijets \cite{Ostapchenko:2014mna,Ostapchenko:2005nj,Kohara:2014cra,Goulianos:2014hqa}, 
goes in describing the different regions.

Here, we concentrate on the implication of any given one-channel eikonal model. Thus, we  repeat the argument about the relation  
between the Poisson distribution of independent collisions and diffractive processes given in \cite{Achilli:2011sw},
where we stressed that the inelastic cross section in a one-channel eikonal model coincides with the 
sum of independently  (Poisson) distributed collisions  in b-space. Namely, with
\be
\sigma_{total}=\sigma_{elastic}+\sigma_{inel}\label{eq:sigonech}
\ee
then, in a one-channel ({\it one-ch})  mode,  
\be
\sigma_{inel}^{one-ch}\equiv \sigma_{tot}-\sigma_{elastic}^{one-ch}=\int d^2 {\vecb} [1-e^{-2 \chi_I(b,s)}]\label{eq:siginelonech}
\ee
But since 
\be
\sum_1^{\infty} \frac{({\bar n})^n e^{-{\bar n}(b,s)}}{n!}=1-e^{-{\bar n}(b,s)}\label{eq:poisson}
\ee
one can  identify the integrand at the right hand side of Eq.~(\ref{eq:siginelonech}) with a sum of  totally independent 
collisions, with $2 \chi_I(b,s)={\bar n}(b,s)$.  We suggest that this means that in so doing one  excludes diffraction and 
other quasi-elastic processes from the integration in Eq.~(\ref{eq:siginelonech}). Hence, the  simple splitting of the total 
cross section as in  Eq.~(\ref{eq:sigonech}) needs to be better qualified when a one-channel  eikonal is used. In such a case, 
the ``elastic'' cross section 
\be
\sigma_{elastic}^{one-ch}=
\int d^2 {\vec b} |1-e^{-\chi_I(b,s)}|^2 \label{eq:sigel-onech}
\ee
must be including  part of the inelastic contribution, i.e. 
\bea
\sigma_{elastic}^{one-ch}=\sigma_{elastic}+\nonumber\\ 
\ diffractive\  or \ otherwise \ correlated \ processes\nonumber\\
\eea
and $\sigma_{inel}^{one-ch}$  is only the non-diffractive part.  Within this approach, we can compare Eq.~(\ref{eq:siginelonech}) with data.

This comparison is shown in  the left hand panel of Fig.~\ref{fig:olga-inel-danielppall},  from \cite{Fagundes:2015vba}
where   
 inelastic cross section  data up to AUGER energies 
\cite{Collaboration:2012wt} are plotted. The blue band corresponds to the expectations from     Eq.~(\ref{eq:siginelonech}) 
where the same eikonal function $\chi_I(b,s)$  which gives the total cross section as seen in  the right hand plot of Fig. ~\ref{fig:olga-inel-danielppall} is used. Having anchored 
the eikonal $\chi_I(b,s)$ to the LHC total cross section,  the band indicates  the  spread of predictions  due to the different 
asymptotic low-x behavior of the employed densities,  as the energy increases beyond LHC8. The top curve corresponds to 
MSTW, the lower one to GRV.

\begin{figure*}[htb]
\resizebox{0.4\textwidth}{!}{
\includegraphics{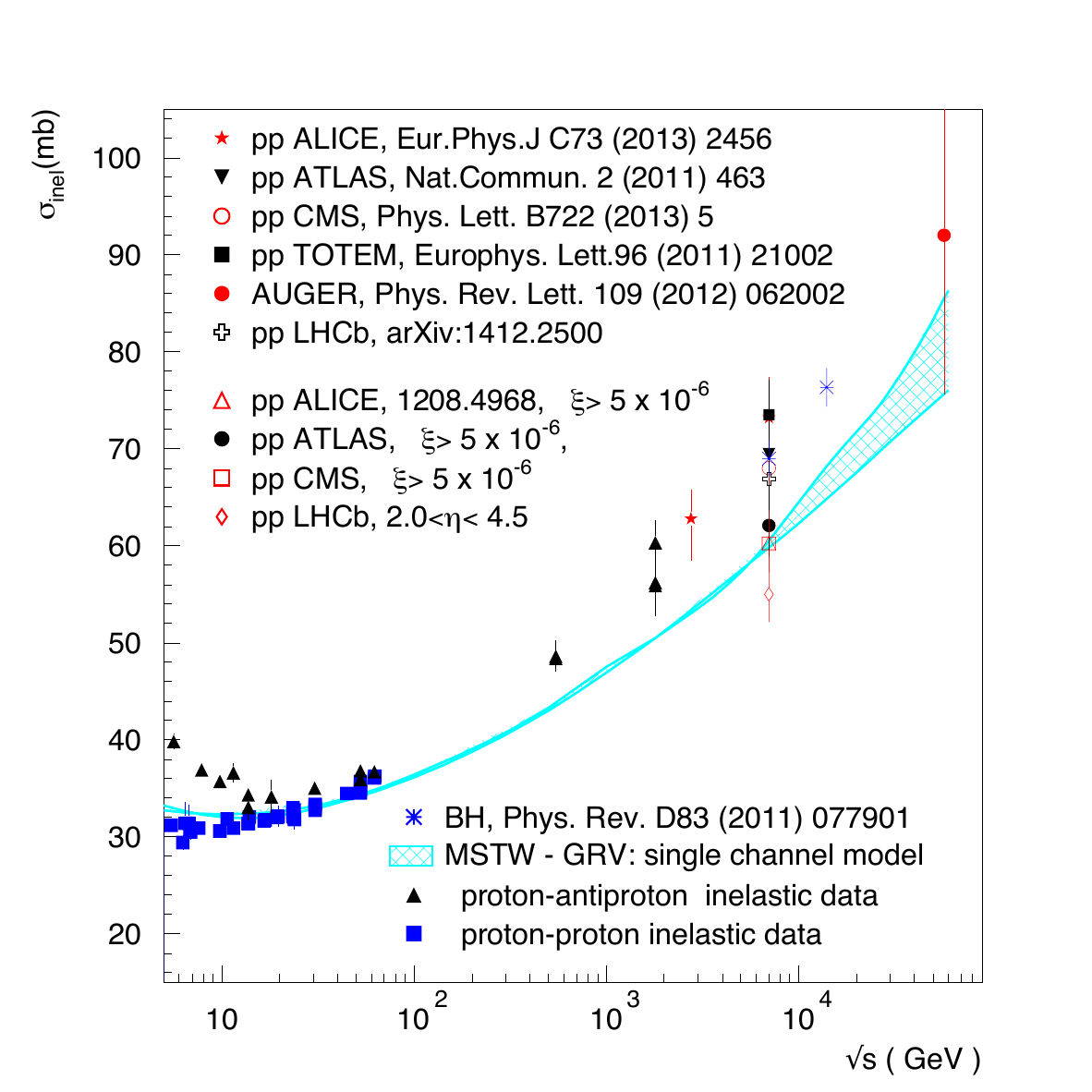}}
\resizebox{0.5\textwidth}{!}{
\includegraphics{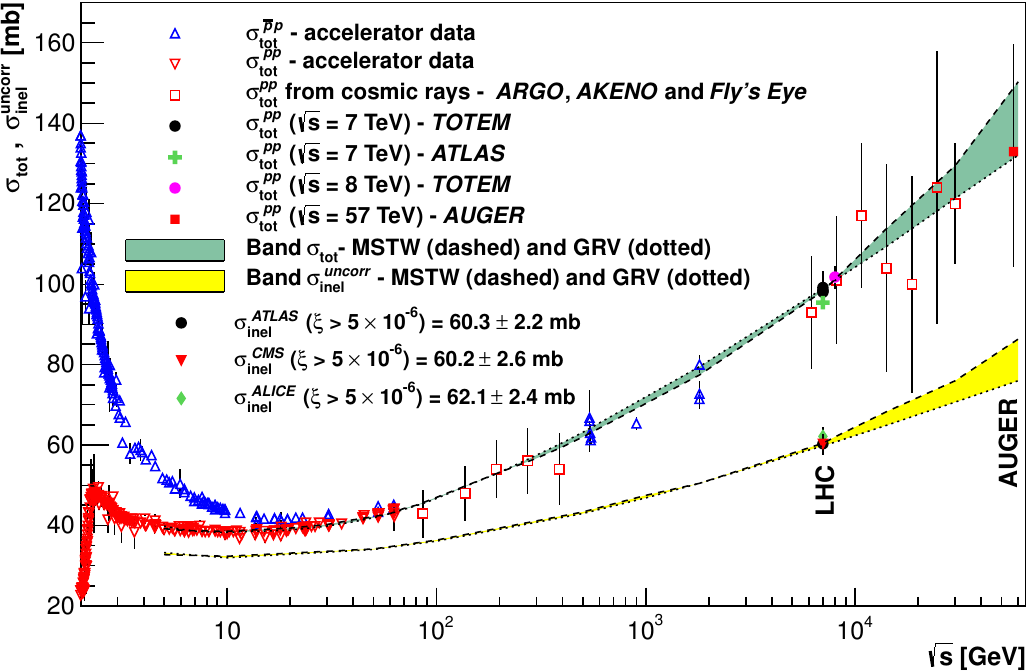}}
\caption{In these plots, the bands correspond to uncertainties  related to  the very low-x behavior of the PDFs used in the calculation of the total and inelastic cross-sections.  At left, data for the inelastic cross section are compared with  GRV and MSTW densities in the eikonalized QCD mini-jet with soft gluon resummation model, 
called BN model in the text and discussed therein. The inelastic uncorrelated cross section is compared with inelastic processes 
for $M^2_X/s > 5 \times 10^{-6}$ as measured  by ATLAS \cite{Aad:2011eu}, CMS \cite{Chatrchyan:2012nj}  and 
ALICE \cite{Abelev:2012sea}. We also show comparison with  Bloch and Halzen (BH) results  \cite{Block:2011uy}. In the right panel, results for our BN-model  are shown for  
$pp$ total cross section together with   the inelastic uncorrelated part of the inelastic cross section, obtained from the one-channel mini-jet model.  
Accelerator data at LHC include TOTEM \cite{Antchev:2013paa,Antchev:2013iaa} and 
ATLAS measurements \cite{Aad:2014dca}.
Panels are reprinted   Fig. (4) and Fig. (5)  from \cite{Fagundes:2015vba} \copyright
(2015) by the American Physical Society.}
\label{fig:olga-inel-danielppall}
\end{figure*}

The comparison with experimental data is very interesting. While the present LHC inelastic cross section data  span a 
range of values corresponding to different kinematic regions,  Eq.~(\ref{eq:siginelonech}) identifies  the region where 
uncorrelated events described by mini-jet collisions, parton-parton collision with $p_t>p_{tmin}$, play the main role. 
From the comparison with data, we can identify it with  the region  $\xi=  M^2_X/s \ge 5\times  10^{-6}$ where  
three LHC experiments, ATLAS \cite{Aad:2011eu}, CMS \cite{Chatrchyan:2012nj}  and ALICE \cite{Abelev:2012sea}, agree to a 
common value within  a small  error. This measurement is in the  high mass region (for instance, at LHC7 the lower 
bound gives $M_{X}=15.7 \ {\rm GeV}$).  LHCb results correspond to a lower cross section, but they  do not cover the same 
region of phase space. 

The above results 
are summarized in  Fig.~\ref{fig:olga-inel-danielppall} where the bands  
correspond to different PDFs used in the calculation of mini-jets and to their different extrapolation to very low-x 
at the cosmic ray energies.  


The dashed yellow band is the one-channel inelastic cross section that only includes Poisson-distributed
independent scatterings. That is, once the parameters of the eikonal $\chi_I(b,s)$ are chosen to give an optimal reproduction of 
the  the total cross section, the computed inelastic cross section immediately gives the uncorrelated part of the total inelastic cross section.
The importance of this fact for cosmic ray deduced $pp$ cross sections has been noticed in \cite{Fagundes:2014fza} 
and discussed in \ref{sss:ourcosmics}. 

\subsubsection{
A phenomenological proposal for 
isolating the diffractive component.} 

The total cross section, which our BN  model successfully describes, includes different components, but only  one 
of them 
is well defined  experimentally as well as theoretically,  that is the elastic cross section.
It is  well known that  one-channel eikonal models fail to simultaneously describe the total and the elastic cross section 
through the entire available CM energy range, with the same parameter set. In the last sub-section, we have delineated this 
shortcoming through the 
observation \cite{Fagundes:2015vba} that once mini-jets become operative  past the {\it soft edge} identified by Block 
and collaborators in \cite{Block:2014lna},
the computed elastic 
cross section includes correlated inelastic collisions  and the computed inelastic lacks the same (i.e., its correlated inelastic part). 
We now discuss this matter in detail so as to make these statements quantitative. We shall do so through  the one-  channel 
 mini-jet model with a suitable parametrization of diffractive data.

In one- channel eikonal models, with the inelastic part given by  Eq.~(\ref{eq:siginelonech}), the elastic  part of the total 
cross section is given by Eq.~(\ref{eq:sigel-onech}). Notice that  
whereas Eq.~(\ref{eq:siginelonech}) is exact, in Eq.~(\ref{eq:sigel-onech}) the real part of the eikonal function 
has been neglected, as in Eq.~(\ref{eq:sigtot}). 

Eq.~(\ref{eq:sigel-onech}) reproduces with a good approximation the elastic cross section data up to the onset of minijets, 
deviating significantly from the data already at energies around 100 {\rm GeV}. In particular, at the  Tevatron, 
Eq.~(\ref{eq:sigel-onech}) gives an elastic cross section roughly 30 \% higher than the data. This 
is shown in  the left hand plot of Fig.~\ref{fig:elastic-onech-diff}, where the one-channel result from 
Eq.~(\ref{eq:sigel-onech})  is plotted together with elastic scattering data and an empirical parametrization 
of all elastic differential cross section  $pp$ data from ISR to LHC7 \cite{Fagundes:2013aja}. 

The analysis  of \cite{Fagundes:2013aja} is based on the   Phillips and Barger model for the elastic differential cross section 
\cite{Phillips:1974vt}, described in \ref{sss:PB},
implemented by a form factor term to fully reproduce the optical point, and 
hence the total cross section, as well as the forward slope. Through suitable predictions for the high energy behavior 
of the parameters, the parameterization of \cite{Fagundes:2013aja}  provides  a model independent prediction  
both for  elastic and total cross sections at very high energies, and hence can  be used as a good test of different models 
in the high energy region beyond present accelerator data.

\begin{figure*}[htb]
\centering
\resizebox{0.8 \textwidth}{!}{
\includegraphics{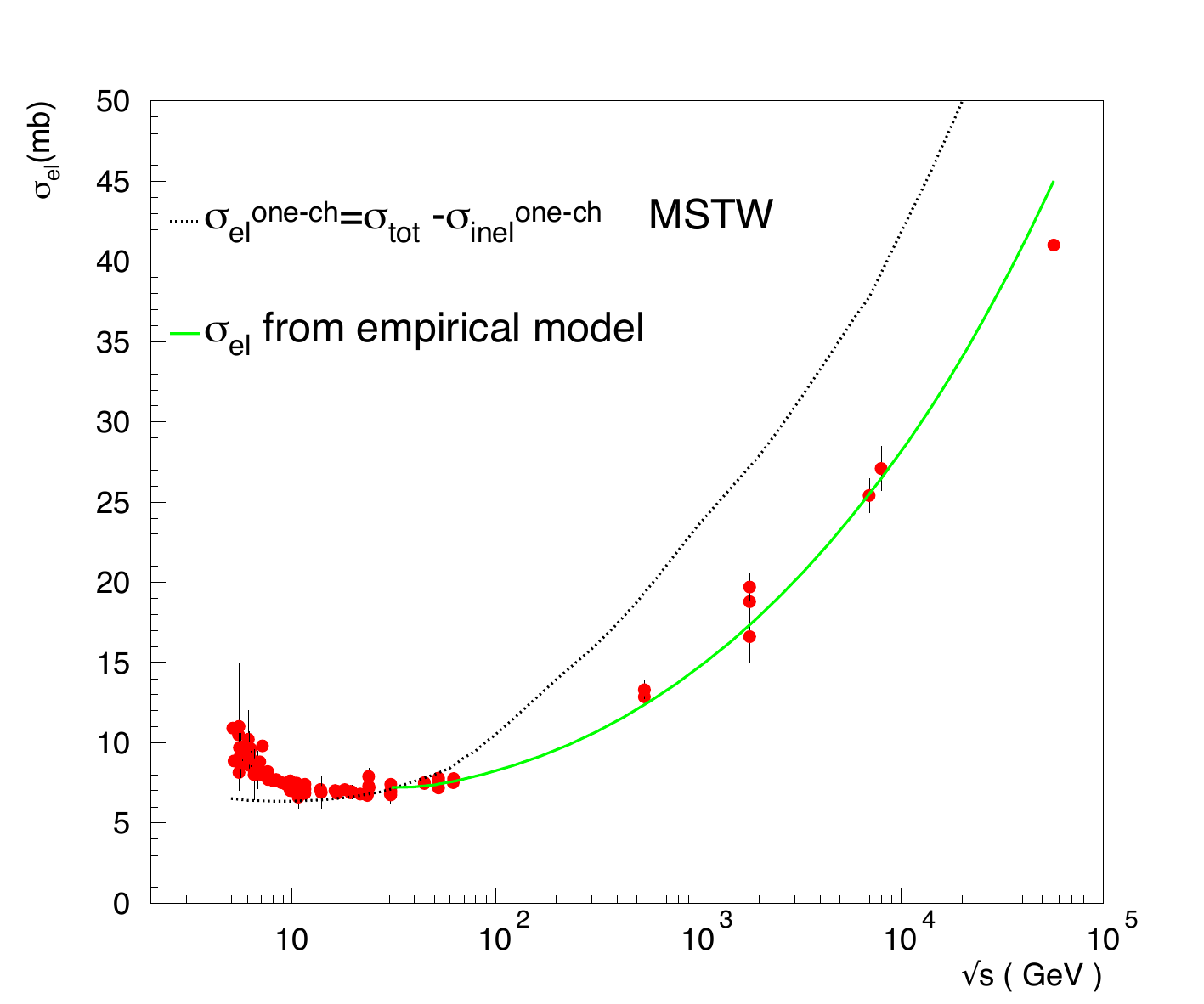}
\includegraphics{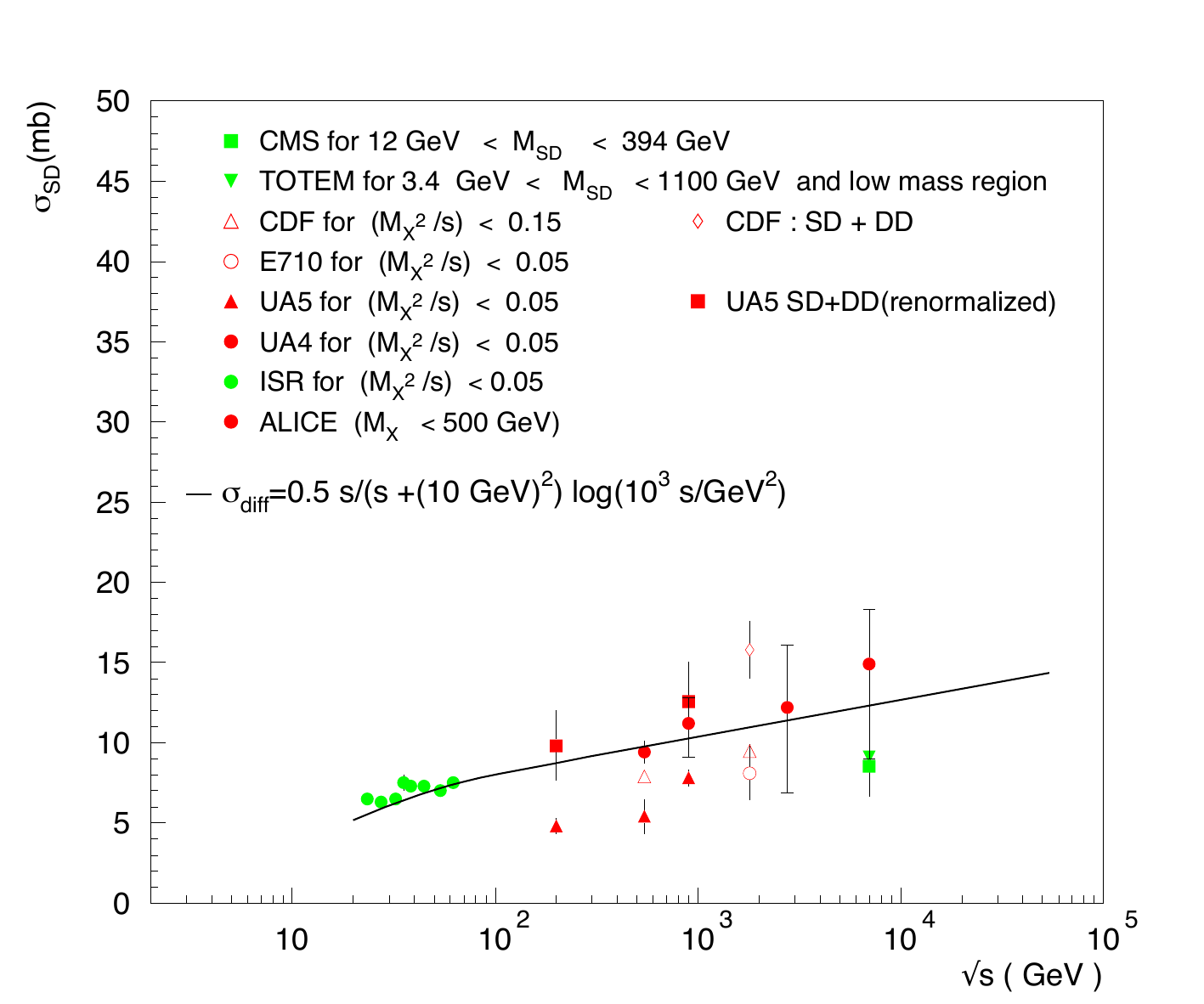}}
\caption{
At left, we show the elastic $pp$ cross section from  the one-channel model given by the top  curve, with  choice of MSTW PDF as in the upper cure of the right hand plot  of Fig.~\ref{fig:olga-inel-danielppall}. The green curve corresponds to the  empirical parametrization of all differential elastic $pp$ data \cite{Fagundes:2013aja} up to$ \sqrt{s}=7$ TeV. Comparison is done with both $pp$ and $p{\bar p}$ data. The right hand panel  shows  diffraction data from E710 \cite{Amos:1992jw}, 
UA5 \cite{Ansorge:1986xq,Alner:1987wb},  UA4 \cite{Bernard:1986yh}, ISR \cite{Armitage:1981zp}, 
CDF \cite{Affolder:2001vx}, CMS \cite{Dutta:2014wda}, TOTEM \cite{Antchev:2013iaa}  and 
ALICE \cite{Abelev:2012sea}  compared with  the  parametrization given by  Eq.~(\ref{eq:sigdiff-EU})  mentioned in the text.
Reprinted Fig.(6) 
with permission from \cite{Fagundes:2015vba}
 \copyright (2015) by the American Physical Society. }
\label{fig:elastic-onech-diff}
\end{figure*}
The left hand plot of Fig.~\ref{fig:elastic-onech-diff}  
from \cite{Fagundes:2015vba} shows that  at low energies, before the onset of mini-jets, 
one- channel models may be used to describe both elastic and total cross sections. However,  past  ISR energies 
the  threshold  of perturbative QCD, reflected in the appearance of the  {\it soft edge}, is crossed, 
and  one-channel models fail.  One-channel models are also unable to reproduce the behaviour of the 
differential elastic cross section, and multichannel models with added parameters are then 
needed to describe diffraction. The difficulty with proper descriptions of diffraction is that at different energies, 
different parts of the phase space are accessed by different experimental set-ups, as we show in the right hand 
plot of Fig.~\ref{fig:elastic-onech-diff}. 
 
For the argument to follow, we consider an estimate of $\sigma_{Diff}$  given by  Eq.~(36) of \cite{Engel:2012pa},
 which provides a good interpolation  of Single Diffractive (SD)  data, from ISR to the LHC results from ALICE, CMS 
 and  TOTEM, as we shown in Fig.~\ref{fig:elastic-onech-diff}.  i.e. 
\be
\sigma_{Diff}(s) = [\frac{(0.5 mb)\ s}{s + (10\ {\rm GeV})^2}] \log (\frac{10^3 s}{{\rm GeV}^2}), \label{eq:sigdiff-EU}
\ee
We have adopted this parameterization for the full diffractive component at high energy. This is an approximation, justified 
at very high energy by the TOTEM result for Double Diffraction(DD) \cite{Antchev:2013any}, namely $\sigma_{DD}\simeq 0.1\ mb$, 
although this result was obtained in a narrow range of pseudo rapidity and more data are needed to conclude that DD does 
not play a significant role at LHC energies. At lower energy the definitions vary, as we show in this figure. 

We shall now show how the one-channel mini-jet model presented here can be used to predict  the full inelastic cross section 
at higher energies. 

We start with the elastic cross section, and consider  now  the difference
\be
\sigma_{elastic}^{one-ch}=\sigma_{tot}-\sigma_{inel}^{one-ch}
\ee   
which includes diffractive (otherwise said, correlated inelastic) contribution, as also discussed in general terms 
in \cite{Kopeliovich:2003tz}, among others. If
\be
\sigma_{inel}=\sigma_{inel}^{one-ch}+\sigma_{Diff}\label{eq:sigineltrue}
\ee
then, we should be able to obtain the  measured  elastic cross section from 
\be
\sigma_{elastic}=\sigma_{elastic}^{one-ch}-\sigma_{Diff} \label{eq:sigeltrue}
\ee

We compare the procedure outlined through Eqs.(\ref{eq:sigdiff-EU}) and (\ref{eq:sigeltrue})    
with experimental data and with the   empirical parametrization for the elastic cross-section data of \cite{Fagundes:2013aja}.
This is shown in 
Fig.~\ref{fig:sigeltrue} 
from \cite{Fagundes:2015vba}. We see that 
 such a procedure gives a good description of the elastic cross section at high energy, basically past the CERN $Sp{\bar p}S$.   
\begin{figure}[htb]
\resizebox{0.5\textwidth}{!}{
\includegraphics{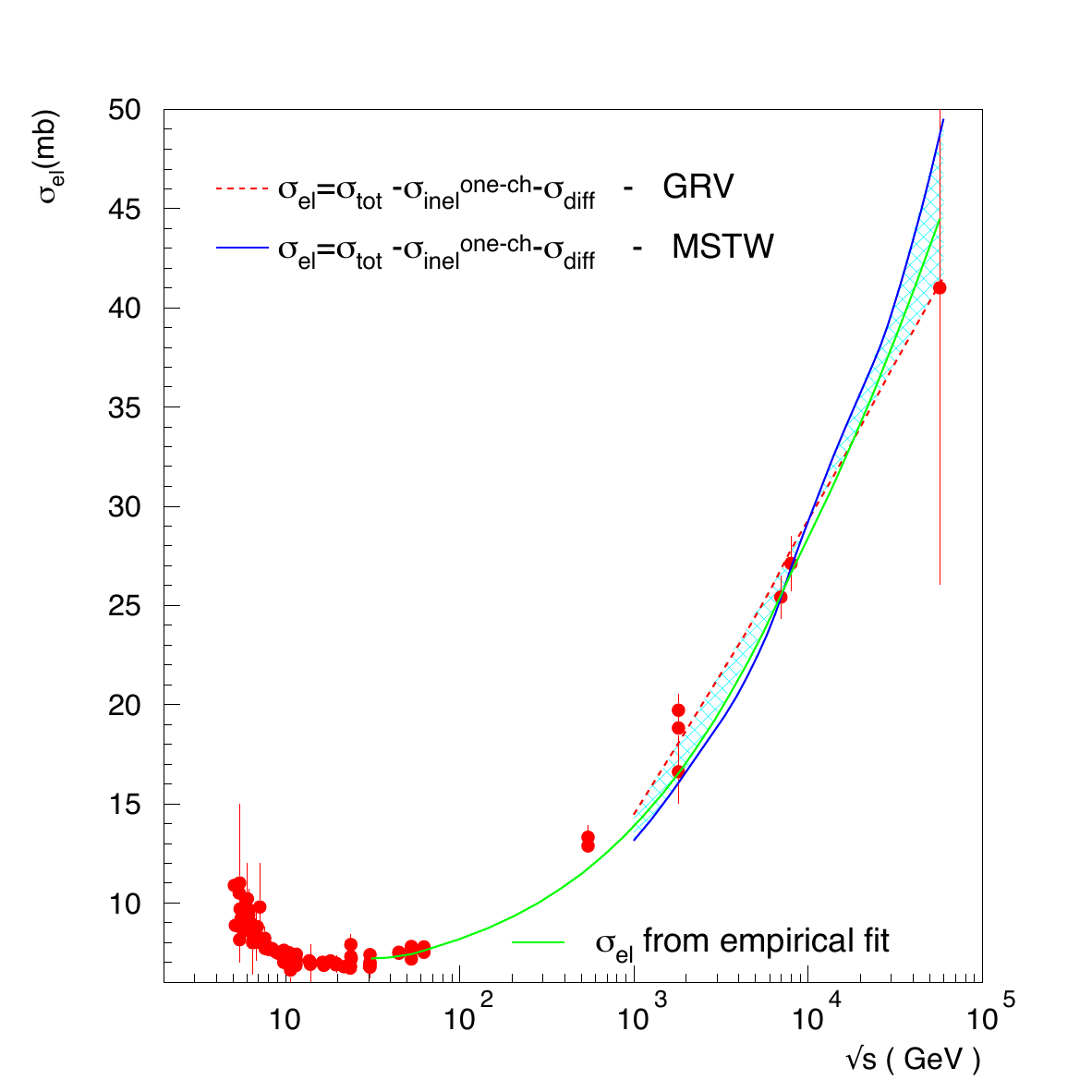}}
\caption{The total elastic cross section obtained by subtracting Single  Diffractive contributions,  indicated as $\sigma_{diff}$,  from the one-  channel model result.  The resulting curve is compared with $pp$ and $p {\bar p}$  data and  the empirical parametrization of \cite{Fagundes:2013aja}
which is seen to fall within the two model predictions. Reprinted left hand plot of Fig.~(7) wiith permission from \cite{Fagundes:2015vba}, \copyright (2015) by the American Physical Society.}
\label{fig:sigeltrue}
\end{figure}
\begin{figure}[htb]
\resizebox{0.5\textwidth}{!}{
\includegraphics{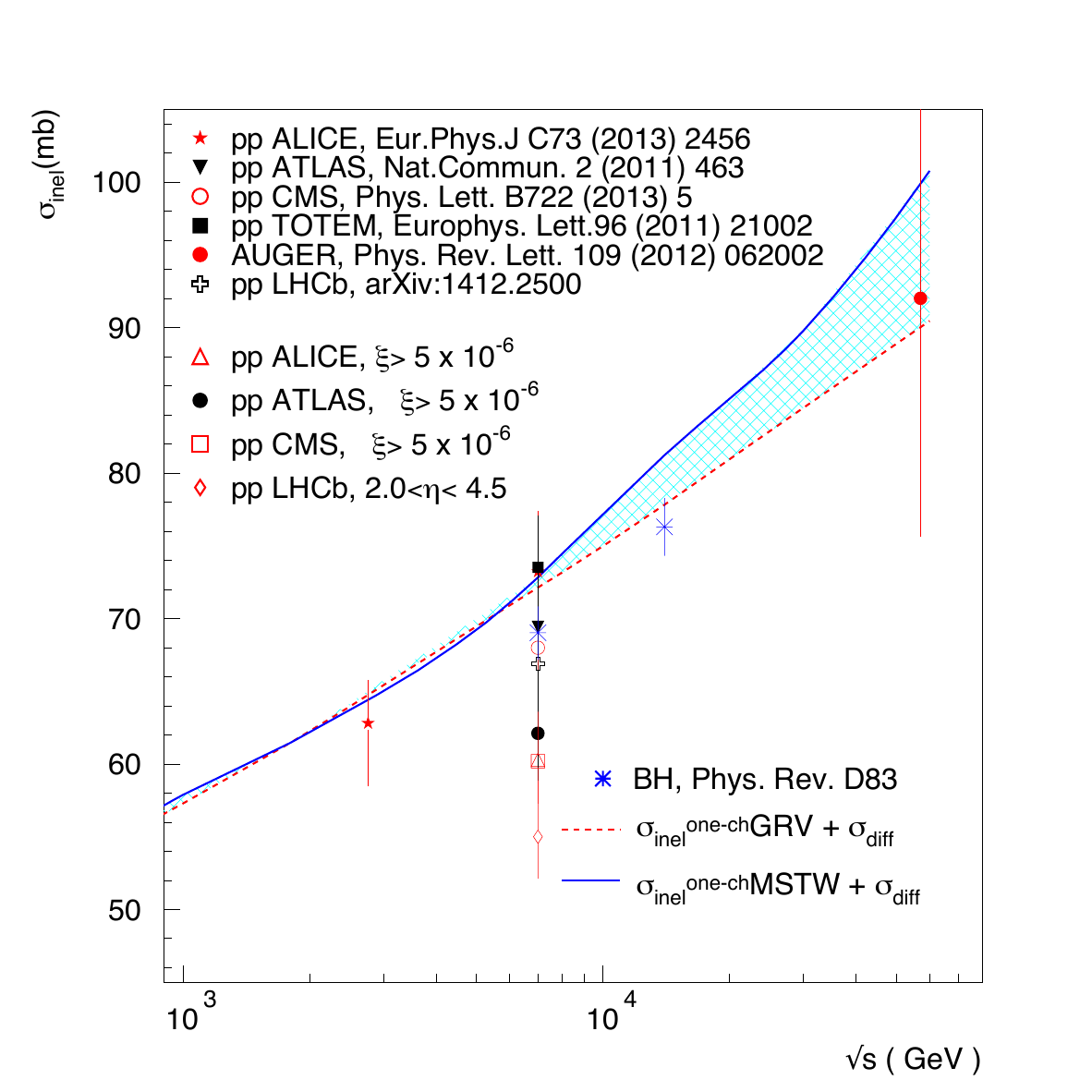}}
\caption{The inelastic cross section: at high energies, adding diffraction brings the one-channel result in agreement to data. Reprinted right-hand plot of Fig.(7) with permission from \cite{Fagundes:2015vba}, \copyright (2015) by the American Physical Society.}
\label{fig:sigineltrue}
\end{figure}
Likewise, from Eq.~(\ref{eq:sigineltrue}),  we can see that  by adding the diffractive part, parametrized as in 
Eq.~(\ref{eq:sigdiff-EU}), to the prediction from the one-channel model, it is possible to obtain a good description 
of  the high energy behavior of the inelastic cross section. This is shown in  Fig.~\ref{fig:sigineltrue}. 
It must be noticed that this procedure shows agreement with data only past ISR  energies (in fact from $Sp{\bar p}S$ onwards) 
 and that a   model  describing both the low and the high energy will have to go beyond the one-channel exercise 
described here.  In Table \ref{tab:13tev}, we show the predictions from this model for  the inelastic cross section at LHC13, 
$\sqrt{s}=13\ {\rm TeV}$.\footnote{
 In the published PRD version of \cite{Fagundes:2015vba}, this table had an error in the MSTW predictions at $\sqrt{s}=13\ TeV$.It has now been corrected in the arXiv version.}
\\
\begin{table}[htb]
\centering
\caption{Minijet model predictions for the inelastic cross section  at $\sqrt{s}=13\ {\rm TeV}$. Predictions of
$\sigma_{inel}$ in the full phase-space were obtained by adding  $\sigma_{diff} (13\
\text{{\rm TeV}})= 12.9$ mb to  $\sigma_{inel}^{uncorr}\equiv \sigma_{inel}^{one-ch}$.\label{tab:13tev}}
\vspace*{.2cm}
\begin{tabular}{c|c|c}
\hline \hline
$PDF$ & $\sigma_{inel}^{uncorr}$ (mb) &
$\sigma_{inel}$ (mb)\\
\hline
$GRV$ &  64.3  & 77.2 \\
\hline
$MSTW$ & {66.9} &{79.8} \\
\hline \hline
\end{tabular}
\end{table}
 
The result of this subsection confirms the interpretation  that  at high energies, past the beginning of the 
rise and the onset of mini-jets, 
 the one-channel inelastic cross section  is  devoid of most of the diffractive contribution.
We have shown that the onset and rise of the mini-jet cross section provide the dynamical mechanism
behind the appearance of a {\it soft-edge} \cite{Block:2014lna}, i.e., a threshold in the total cross section around 
$\sqrt{s}\simeq (10\div 20)\ {\rm GeV}$. Thus, our model for the total $pp$ cross section that utilizes mini-jets with 
soft-gluon re-summation has a built in soft-edge. It has been updated with recent PDFs for LHC 
at $\sqrt{s} = 7, 8\ {\rm TeV}$ and predictions made for higher energy LHC data and cosmic rays. 

We have also discussed in detail the reasons behind failures to obtain correct values for the elastic cross sections 
from a one-channel eikonal that obtains the total cross section correctly. It has been shown, through the
use of phenomenological descriptions of diffractive (otherwise said, correlated inelastic) cross sections, that
one-channel elastic cross section is indeed a sum of the true elastic plus correlated inelastic cross sections.   
An application of this fact to cosmic ray data analysis for the extraction of $pp$ uncorrelated-inelastic 
cross sections shall be presented elsewhere.

\subsection{Conclusions} \label{ss:conclusion-elastic}
 As we have seen, the description of the fundamental dynamics of hadronic scattering is still proceeding along 
 different lines, a Regge-Pomeron interpretation, a microscopic description of the scattering, or analytical constraints 
 and asymptotic theorems. These different ways are not incompatible, and may ultimately come together. We conclude 
 this section with two comments, one  on the differential elastic cross-section, and one on the integrated total cross-sections.
 \subsubsection{The differential elastic cross-section before and soon after the LHC started}
A comparison of the state-of-the-art of theoretical predictions  before and soon after  the LHC started operating,  can be glimpsed from  Fig.~\ref{fig:diffel-islamall-eggertall}.   
In the left panel,  we show 
 a compilation of different model predictions at $\sqrt{s}=14\ {\rm TeV}$   from \cite{Islam:2007nr}, done in 2007, and  at right   we see how predictions at 
$\sqrt{s}=7\ {\rm TeV}$ compared with actual  LHC data, as shown from K. Eggert's talk at  \href{<https://indico.in2p3.fr/event/6004/session/7/contribution/116/material/slides/0.pdf>}{Hadron Collider Physics Symposium},  November 2011, Paris, France.
  \begin{figure*}[htb]
\resizebox{0.4\textwidth}{!}{
\hspace{+0.5cm}
\includegraphics{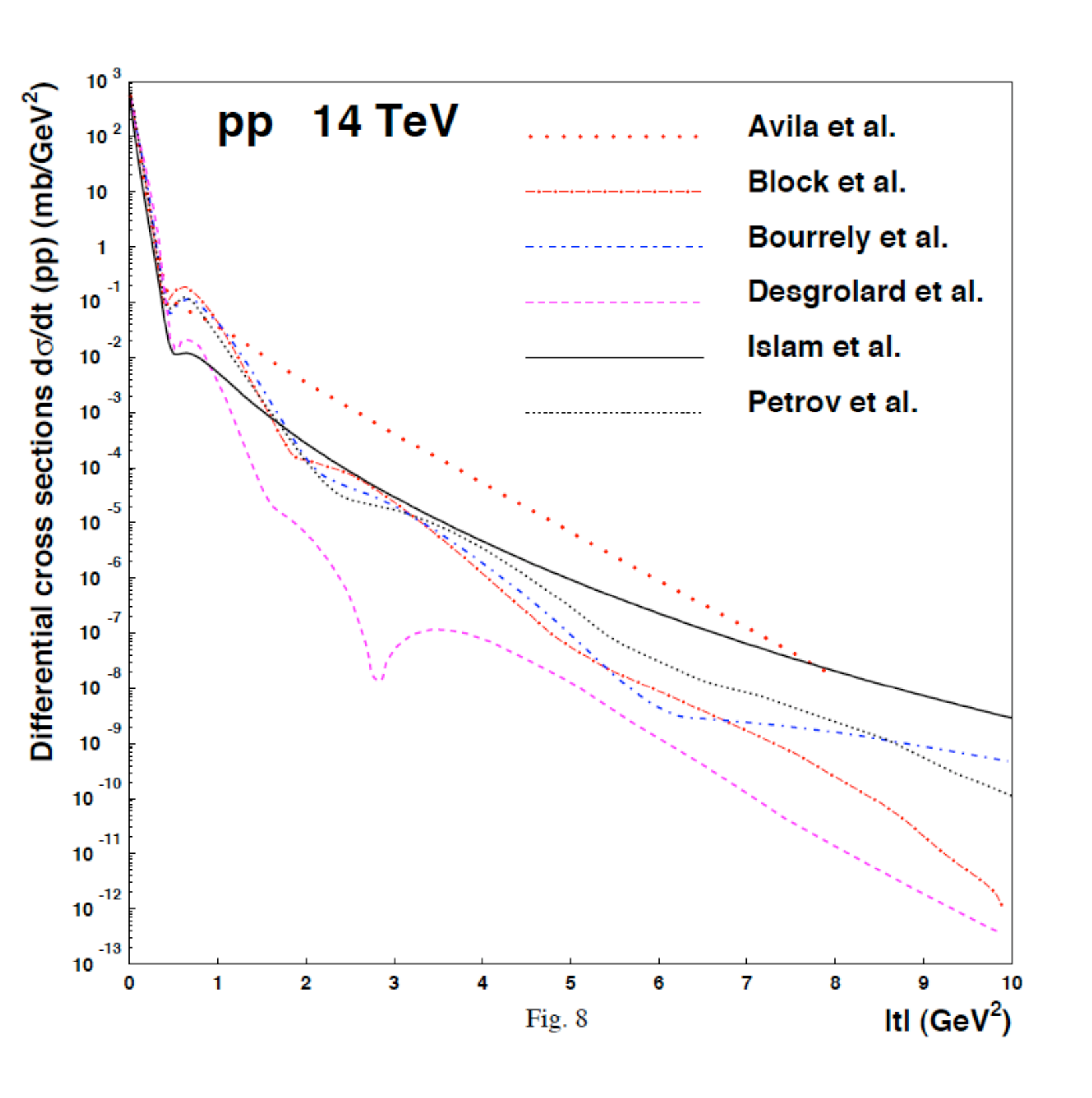}}
\hspace{+.1cm}
\resizebox{0.6\textwidth}{!}{
\includegraphics{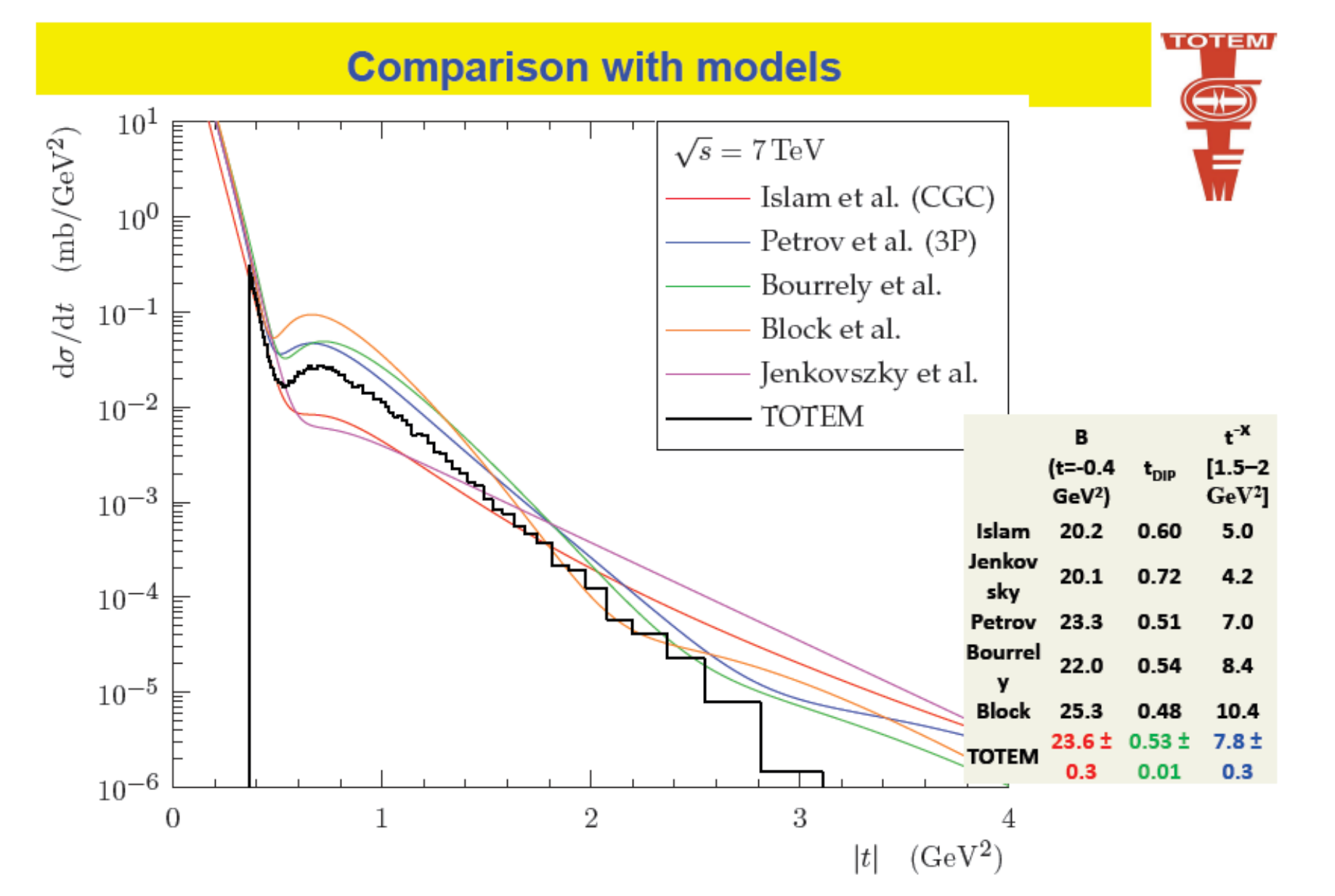}}
\caption{ Left  panel: a compilation of various models for the elastic differential cross-section at LHC14, prepared in 2007, before the LHC started, from 
\cite{Islam:2007nr}.
Right panel: a compilation of  model predictions  for the elastic differential cross-section at $\sqrt{s}=7$ TeV and first comparison with TOTEM data, from K. Eggert's talk  
 at  \href{<https://indico.in2p3.fr/event/6004/session/7/contribution/116/material/slides/0.pdf>}{Hadron Collider Physics Symposium},  November 2011, Paris, France.  Reproduced  with permission of the authors.}
\label{fig:diffel-islamall-eggertall}
\end{figure*}
As new results from LHC  appeared  at $\sqrt{s}=7\ TeV$ 
 the parameters of some models 
 had  to be  updated, and agreement with the new data was easily obtained. This however is not a satisfactory 
 situation,
 since the parameters should remain stable or at least have a predictable energy dependence
 leading to further understanding of the dynamics.
 It is to be hoped that with the new results from LHC which  will appear  at $\sqrt{s}=13 $ and $14\ {\rm TeV}$ such understanding may become closer. 
\subsubsection{A fit to the future imposing Froissart limit and the Black Disk picture} 
As a commentary to this and previous sections, we present in Fig.~\ref{fig:lastblock}
a recent analysis by Block, Durand, Ha and Halzen  \cite{Block:2015mjw}, in which  the high energy behavior of all 
three cross-sections, $\sigtot, \siginel,\sigel$ has been constrained  by  the Froissart limit, $\sigtot \sim [\ln s]^2$,  and  
the black disk  behavior, $\sigel/\sigtot\rightarrow 1/2,  \ 8 \pi B/\sigtot \rightarrow 1$.
\begin{figure}[htb]
\resizebox{0.6\textwidth}{!}{
\includegraphics{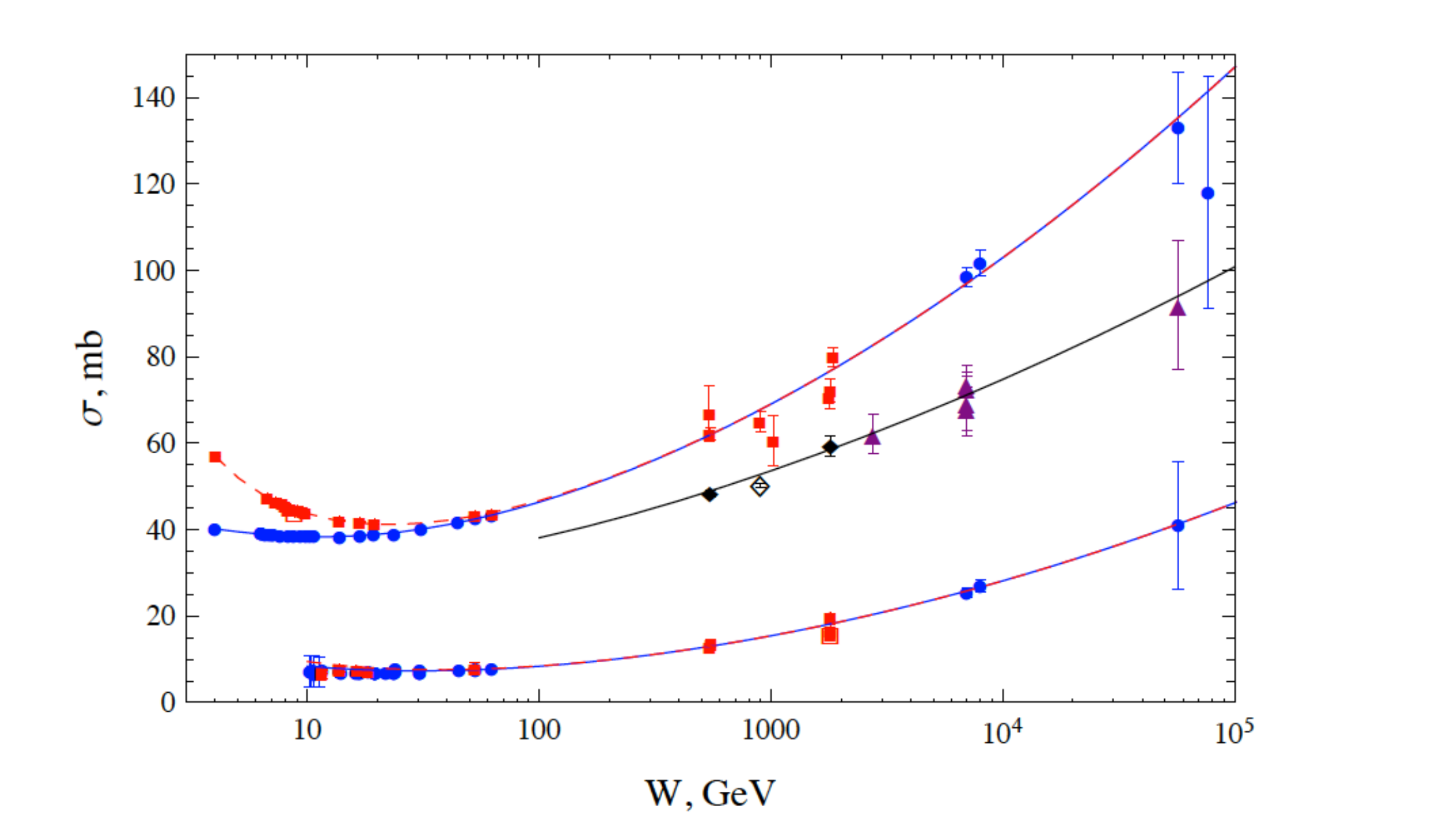}}
\caption{Total $pp$ and $p{\bar p}$ cross-sections from Fig. 1 of \cite{Block:2015mjw}, where the curve is a fit to the data, with high energy constraints from the Froissart bound and the black disk limit. 
Reprinted with permission from \cite{Block:2015mjw}, Fig.(1), \copyright (2015) by the American Physical Society}
\label{fig:lastblock}
\end{figure}

\section{ Photon processes}
\label{sec:photons}

Measurements of the total hadronic cross-section  
are made  with 
different {projectiles and targets} 
involving altogether different techniques. 
The   list includes:
\begin{itemize}
\item heavy ion collisions, most recently LHC experiments for $pA$ scattering, 
\item collisions between primary particles in cosmic rays with the nuclei of the atmosphere, which have been discussed in the section on cosmic ray measurements, Sect. \ref{sec:cosmic}, 
\item photon processes, which include real and virtual photon scattering on nucleons or nuclei, both in motion, as in  HERA, 
or with fixed target, or photons against photons as in $electron-positron$ collisions.   
\end{itemize}
As of 2015,  LHC plans to measure $\gamma p$ and  $\gamma \gamma$  collisions 
but no results are yet available for what concerns total cross-sections. LHC can also study $\pi p$ and $\pi \pi $ 
cross-section, as we describe 
in the next section. 

This section will draw from the extensive set of measurements at HERA, performed by the two 
experiments ZEUS \cite{Chekanov:2001gw} and H1 \cite{Aid:1995bz}, which have measured the  
total cross-section $\sigma^{\gamma p}_{tot}$ at $\sqrt{s}_{\gamma p}\equiv W =209$ and $ 200\ {\rm {\rm GeV}}$ 
respectively. Recently, the ZEUS collaboration has  presented measurements  of the energy variation of 
the cross--section in the range $ 194\ {\rm GeV} \le W \le 296\ {\rm GeV}$ \cite{Collaboration:2010wxa}. Total 
cross-sections at HERA
have been measured   also with virtual photons, {in  a wide range  of the virtual photon squared momentum 
$Q^2$, } including the transition from  
$\gamma^* p$ to $\gamma p$ with the ZEUS Beam Pipe Calorimeter 
\cite{Bornheim:1998gn,Haidt:2001nr,Haidt:2003zz}. {The HERA measurements  
include vector meson production and  are of interest  for QCD studies,  adding  an important kinematic variable to the cross-section modeling.} 
 
We shall review some representative models for photon initiated processes and discuss the transition from virtual to real photons. We touch upon 
an extensive theoretical literature on the subject through Sakurai's VMD model; the Gribov picture; Haidt's phenomenology; 
applications of the  Balitsky-Kovchegov (BK) equation in its various formulations; saturation and geometric scaling; the mini jet models and factorization schemes.
Various items of interest can be found in the next subsections  as follows:
\begin{itemize}
\item kinematics is defined in \ref{ss:kinematics},
\item  Vector Meson Dominance Model proposals are presented in \ref{ss:VMD},
\item the BK evolution equations are introduced in \ref{ss:BKgen},
\item the transition from virtual to real photons and analyses by Haidt and collaborators can be found in \ref{ss:gamstar},
\item specificic models for $\gamma p$ scattering are in  \ref{ss:gamp},
\item vector meson production from real and virtual photon scattering is discussed in \ref{ss:gampVector} and \ref{ss:gamstarpVectors},
\item the total $\gamma^* p$ cross-section can be found in \ref{ss:gamstar-total},
\item data and some models for  real and virtual photon-photon scattering  are  presented in \ref{ss:gamgam} and \ref{ss:gamstargamstar}.
\end{itemize}

\subsection{Data and kinematics for $e p  \rightarrow e X$}\label{ss:kinematics}
The standard process  to be studied, 
\begin{equation}
e + p\rightarrow  e'  + X,
\end{equation}
 is shown in Fig.~\ref{fig:ep}.
\begin{center}
\begin{figure}
\resizebox{0.6\textwidth}{!}{%
  \includegraphics{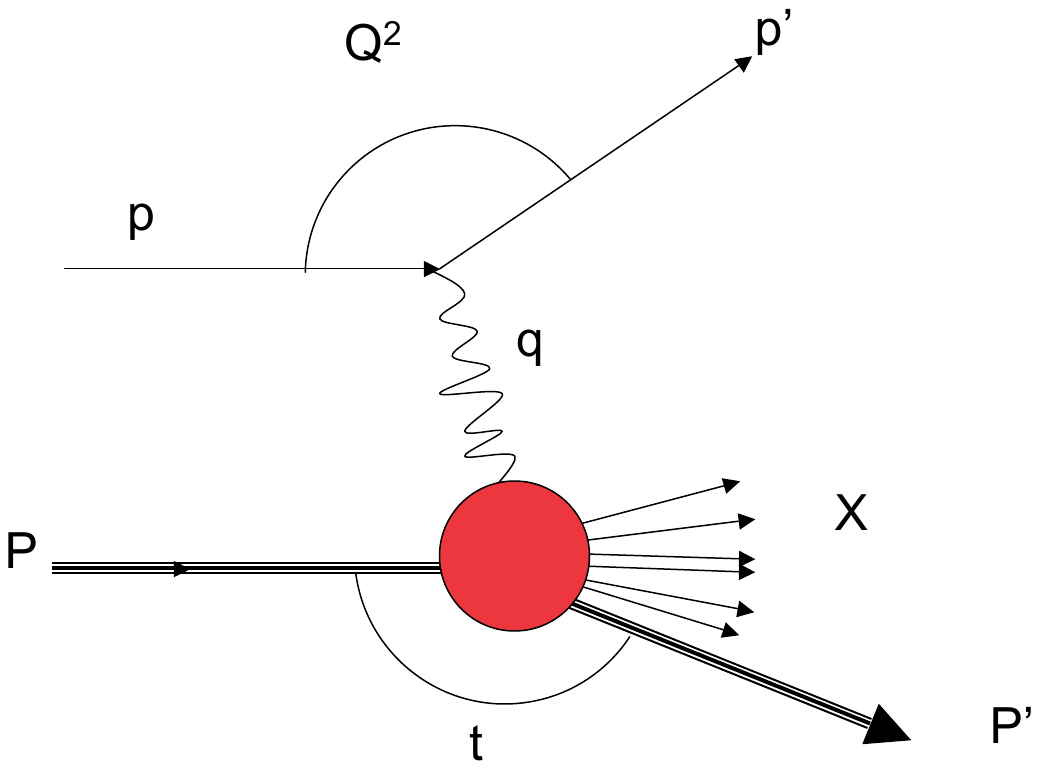}
}
\caption{Electron/positron-proton scattering}
\label{fig:ep}
\end{figure}
\end{center}
For this process  one defines the following  kinematic variables:
\begin{eqnarray}
q=p-p'\\
Q^2=-q^2\\
W^2=(q+P)^2\\
y=\frac{P\cdot q} {P\cdot p}\\
t=(P-P')^2\\
s = (p + P)^2 = (\frac{Q^2}{xy}) + M^2 - m_e^2
\end{eqnarray}
where $q,p,p',P,P'$ are four-momenta. 
These measurements probe a vast kinematic region for the scattering of photons, which can be generally divided as :
\begin{itemize}
\item Photoproduction (PHP) with real photons, with $q^2 \approx 0$ 
\item Deep Inelastic Scattering (DIS) with virtual photons $Q^2=-q^2 \approx (10\div 10^5)\ {\rm GeV}^2/c^2$
\item The transition region of quasi-real photons,  $Q^2 \sim m^2_\rho$
\end{itemize}
Both the PHP and the DIS have been studied extensively and are reasonably well described by various theoretical models. 
The third region has received less attention, but its kinematic range provides valuable $\gamma p$ measurements in a 
continuous range of values for  $\sigtotgamp$ in the HERA energy region \cite{Bornheim:1998gn}.  
We shall discuss these measurements in a separate subsection.

The hadronic cross-section for photons on protons is obtained from electron or positron scattering on protons. 
The protons can be at rest, as in the early measurements, 
or in motion as in the measurements taken with HERA at DESY. 
The leptons in the incoming  beam were electrons in the early measurements,  positrons at HERA.  
From 1992 until 2007,  
at HERA, the lepton energy was $E_e=27.6\ {\rm GeV}$ and the proton energy ranged from $E_p=460\ {\rm GeV}$ to 
$E_p=920 \ {\rm GeV} $ 

In  Table~\ref{tab:heraxsection}, we reproduce the data on $\sigtotgamp$ from HERA.  
Notice that the earlier experiments \cite{Ahmed:1992qc,Derrick:1992kk} spanned  through a $\gamma p$ c.m. energy range, 
and what is reproduced in the table is the average value as given in HEPDATA Reaction Database:\ \  
http://www.slac.stanford.edu/cgi-hepdata/.  

Also, ZEUS 
indicates that, because of various improvements, their  latest value \cite{Chekanov:2001gw} for $\sigtotgamp$ supersedes 
the first ones \cite{Derrick:1992kk,Derrick:1994dt}.  
All these measurements correspond to a photon 4-momentum squared $Q^2< (0.01\div 0.02)\ {\rm GeV}^2$.
 \begin{table}[htb]
\caption{Results of measurements of $\sigtotgamp$ at   HERA}
\label{tab:heraxsection}       
\begin{tabular}{|c|c|c|}
\hline\noalign{\smallskip}
Experiment               & $\sqrt{s} $          & $\sigtot$        \\
                                    & $( {\rm GeV})  $         &  (mb)   \\
\noalign{\smallskip}\hline\noalign{\smallskip}
ZEUS\protect\cite{Chekanov:2001gw} & 209                       & 174    $\pm 1$ (DSYS=13) \\
 H1\protect\cite{Aid:1995bz}                   &200                        &165.3 $\pm$ 2 .3 (DSYS=10.9 ) \\
  ZEUS \protect\cite{Derrick:1994dt}     &167 $\div$ 194	&143 $\pm$ 4  (DSYS= 17)\\
H1 \protect\cite{Ahmed:1992qc}&$<195>$&159 $\pm$ 7     (DSYS=20)\\
ZEUS \protect\cite{Derrick:1992kk} &$<210>$&154 $\pm$ 16 (DSYS=32)\\ 
 \noalign{\smallskip}\hline
\end{tabular}
\end{table}

In Table \ref{tab:zeusvirtualx}, we also reproduce total cross section measurements  by the ZEUS experiment 
as a function of the $\gamma p$ invariant mass W and 
virtual photon polarization, EPS in the table, obtained using  an extrapolation of General Vector Meson Dominance (GVMD) 
 and the assumption $\sigma_L = 0$,  from \cite{Breitweg:1998dz}, where details of the extrapolation can be found.
\begin{center}
\begin{table}[htb]
\caption{$\gamma + p \rightarrow X$}
\label{tab:zeusvirtualx}
\begin{tabular}{|c|c|c|}   
\hline\noalign{\smallskip}
W(GEV)	&  EPS	&$\sigtot  (\mu b)$                            \\
\noalign{\smallskip}\hline\noalign{\smallskip}
  104	          &0.99	&156.2 $\pm$ 5.3     (DSYS=16.1)\\
  134	          &0.98	&166.1 $\pm$ 5.2     (DSYS=11.0)\\
  153	          &0.96	&174.7 $\pm $4.9     (DSYS=12.9)\\
  173	          &0.92	&175.5 $\pm$ 5.0     (DSYS=11.7)\\
  190	          &0.88	&181.8 $\pm$ 4.7     (DSYS=12.8)\\
  212	          &0.80	&186.8 $\pm$ 4.8     (DSYS=13.5)\\
  233	          &0.69	&192.5  $\pm$ 4.7     (DSYS=13.3)\\
  251	          &0.55	&204.8 $\pm$ 5.6     (DSYS=17.0)\\
\noalign{\smallskip}\hline
\end{tabular}
\end{table}
\end{center}

\subsubsection{Kinematics for photoproduction}\label{sss:kinphotoprod}
Let us now  set the kinematics and the relevant definitions for photoproduction processes, 
i.e. when $Q^2\approx 0$. The $\gamma p$ \x\ is extracted from the process shown in 
Fig.~\ref{fig:ep}.  In the forward direction, the kinematic variables are related 
to the measurable quantities, energy and lepton scattering angle in the laboratory frame, through
\begin{eqnarray}
Q^2=2E_eE'_e(1-\cos\theta_e)\approx E_eE_e'\theta_e^2\\
y=1-\frac{E_e'}{2E_e}(1+\cos\theta_e)\approx 1-\frac{E_e'}{E_e}
\end{eqnarray}
The relation between  $ep$  and  $\gamma p$ \x s can then be expressed   through 
\begin{eqnarray}
\frac{d\sigma^{ep}(y)}{dy}=\sigma_{tot}^{\gamma p}\times \nonumber\\
\frac{\alpha}{\pi}\big[ \frac{1+(1-y)^2}{y}\ln \frac{Q^2_{max}}{Q^2_{min}}-2\frac{1-y}{y}(1-\frac{Q^2_{min}}{Q^2_{max}})
\big]=\nonumber \\
\sigma_{tot}^{\gamma p}\times {\bf \cal F}\nonumber\\ 
\label{eq:phpx}
\end{eqnarray}
where $Q^2_{min}=\frac{m^2_ey^2}{1-y}$.
The $\gamma p$ \x\ can be extracted after integrating the above expression in the variable $y$ with the integration limits
\begin{equation}
y_{min/max}=1-\frac{E'_{e\ max/min}}{E_e}
\end{equation}
Eq.~(\ref{eq:phpx}) defines the flux ${\cal F}$ whose determination depends upon the experimental  resolution on the incoming 
and outgoing positrons $\Delta E_e$ and $\Delta E'_e$. 

\subsubsection{Parton model variables}\label{sss:partons}
 The quantities measured in $e-p$ scattering can be    
related to the parton model underlying the scattering process. 
By proper choice of  the scattering frame and in  the very large momentum limit, the variables relating 
the parton model description to  the process shown in Fig.~\ref{fig:epx} in  
Deep Inelastic Scattering  can be  related to  measurable quantities \cite{Bjorken:1968dy}. 
\begin{center}
\begin{figure}
\resizebox{0.6\textwidth}{!}{%
\includegraphics{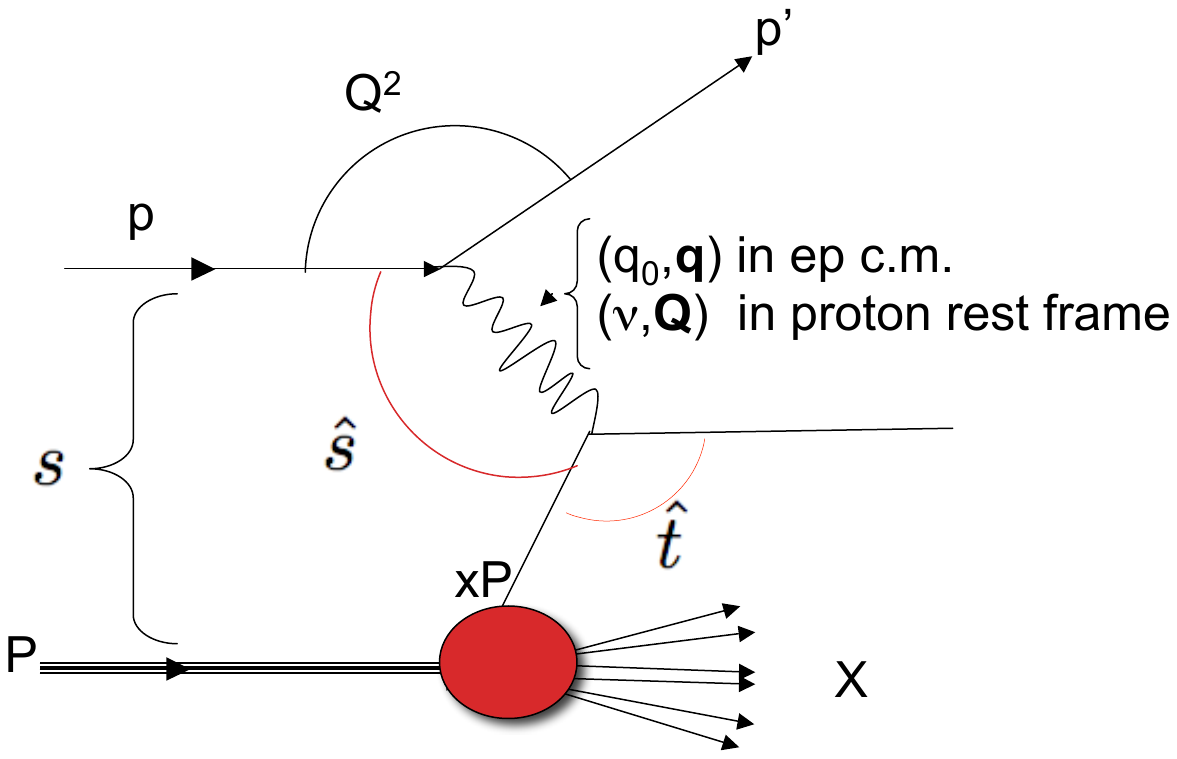}}
\vspace{-2cm}
\caption{Kinematic variables for parton scattering through ep collisions.}
\label{fig:epx}
\end{figure}
\end{center}
Following  here ref.\cite{Bornheim:1998gn}, with
\begin{itemize}
\item {$x$} the fraction of proton momentum carried by the struck quark
\item {$y$} the relative energy transfer from the electron to the proton in the proton rest frame
\item {$W$} the c.m. energy of the photon and proton system
\end{itemize}
one has
\begin{equation}
y=\frac{Q^2}{sx}, \ \ \ \ \  \ W^2=m^2_p+Q^2(\frac{1}{x}-1) \ \ \ \ \ 
\end{equation}
If we neglect the proton mass, $W^2 \approx\ s y(1 -x)$. For small $Q^2$ being discussed here, $x<< 1$, and we have 
$W^2 \approx\ s y$. 

Various kinematic regions of interest 
can now be discussed.
For  total \x \  measurements in photoproduction $\sigtotep$ \ one has
\begin{eqnarray}
0.4<y<0.6\ \ \ \  ZEUS \ \ \ \ \ Experiment,  \\
0.3<y<0.7 \ \ \ \ H1 \ \ \ \ \ \ \ \ \ \ \ \   Experiment
\end{eqnarray}
and $Q^2_{min}\sim 10^{-8}\ {\rm GeV}^2$.  

In terms of the longitudinal and transversely polarized photon cross-section, and neglecting terms 
of order $m^2_p/s$, the   electron-proton \x\  is given by 
\begin{eqnarray}
\frac{d^2\sigma_{ep}}{dydQ^2}=\frac{\alpha}{2\pi}\frac{1 - x}{Q^2}{ \Large [}\left(\frac{1+(1-y)^2}{y}-\frac{2(1-y)}{y}
\frac{Q^2_{min}}{Q^2}\right)\sigma_T\nonumber\\
+ \frac{2(1-y)}{y}\sigma_L {\Large ]} \nonumber\\
= (\frac{x}{y}) \frac{d^2\sigma_{ep}}{dxdQ^2},\ \ \ \ \ \ \ \ \
\end{eqnarray}
where 
\begin{equation}
Q^2_{min}\approx m^2_e\frac{y^2}{1-y}
\end{equation}

In DIS, where $Q^2>0$, in the region of interest  in the variable $y$, the $Q^2_{min}$ can be 
neglected and the expression for the DIS \x  \ , in the parton variable $x$, becomes
\begin{equation}
\frac{d^2\sigma_{ep}}{dxdQ^2}
=\frac{\alpha}{2\pi}
\frac{1-x}{xQ^2}
(1+(1-y)^2)
(\sigma_T+
\frac{2(1-y)}
{1+(1-y)^2}\sigma_L)
\end{equation}
The longitudinal and transverse \x s are related to the  structure functions $F_{2}$ as
\begin{equation}
F_2=\frac{Q^2}{4\pi^2\alpha}(1-x)(\sigma_T+\sigma_L)
\label{F12}
\end{equation}
and $F_2$ is seen to represent the sum over quark and antiquark densities in the proton.

Now one can relate the total cross-section for scattering of a virtual photon on a proton to 
 this sum, i.e.
 \begin{equation}
\sigma^{tot}_{\gamma*p}(x,W)\approx \frac{4\pi^2 \alpha}{Q^2}F_2(x,Q^2) \label{eq:sig*F2}
\end{equation}
and since $x$ is proportional to the cm energy in the photon-proton system, one can thus obtain the total 
cross-section for a range of energies.

On the other hand, in photo-production, $\sigma_L<< \sigma_T$ and one obtains for the 
total photo-production cross-section
\begin{align}
\frac{d\sigma_{ep}}{dy}=\frac{\alpha}
{2\pi}\frac{1+(1-y)^2}{y} \times\nonumber\\
\left[\ln(\frac{Q^2_{max}}{Q^2_{min}}) -\frac{2(1-y)}{1+(1-y)^2}(1-\frac{Q^2_{min}}{Q^2_{max}}) 
\right]\sigtotgamp(W_{\gamma p})
\end{align}

In this review we are basically interested in models for total cross-section. 
Before discussing the models currently used, 
since most models do use Vector Meson Dominance in some fashion, it is useful to recall how it was first proposed.

\subsection{Photons and  Vector Meson Dominance}
\label{ss:VMD}
Gribov, in his description of the interaction of quanta with nuclei, which will be summarized in \ref{sss:gribov},
refers  to  the idea proposed by Bell \cite{Bell:1964eu} that, at high energy,  Vector Meson Dominance could result in the amplitude 
for $\pi-nucleus $ to be proportional to surface terms  rather than to volume terms. 
 
 The idea arose when Bell recalls   Adler's study of neutrino scattering  on nuclei, 
 \begin{equation}
\nu+\alpha \rightarrow l+\alpha^*
\label{eq:adlerneutrino}
\end{equation}
where $\alpha^*$ is a group of strogly interacting particles. 
Adler noticed that, using Partially Conserved Axial Current (PCAC) and Conserved Vector Current (CVC), 
incident virtual pions can actually describe the interactions of neutrinos on nuclei by obtaining for the 
cross-section of process  (\ref{eq:adlerneutrino})
\begin{equation}
\frac{\partial^2\sigma}{\partial q^2 \partial W^2}\propto \sigma(W,-q^2)
\end{equation}
where $q^2$ is the momentum transfer between the incoming neutrino and the outgoing lepton $l$ and 
$W$ is the mass of the hadronic system $\alpha^*$, and $\sigma(W,m^2_\pi)$ would be the total 
cross-section for the reaction $\pi \alpha\rightarrow \alpha^* $. This result was something of a paradox, 
because neutrino's should be sensitive to the entire nucleus, in his language the nucleus should be transparent to the 
neutrino,  and not just to the surface, which is  what happens to pions. Following this line of reasoning, 
Stodolsky \cite{Stodolsky:1966am} produced, what at the time appeared as, a similar paradox for photo-reactions, 
by using the {\it $\rho-photon$} analogy . The discussion as to how the cross-section for photon-nucleus is 
not proportional to the atomic number $A$, but is more similar to surface effects, is interesting and we shall 
reproduce here the main ingredients of Stodolsky's argument.

The usual result that the cross-section for $\gamma-A$ should be proportional to $A$, follows from the optical theorem. 
Taking only the first scattering of the photon, order $\alpha$, the scattered waves are summed up for all the nucleons 
and then from $\Im m F_{nucleus}=A\Im m f_{nucleon}$, the optical theorem gives $\sigma_{\gamma A}=A \sigma_{\gamma-nucleon}$. 
But things are complicated by the fact that {\it quasi-elastic} channels may only apparently contribute to the 
elastic amplitude, and they really should be included as multiple scattering processes. for instance, in $\pi d$ scattering, 
such quasi-elastic processes are  $\pi^-+p\rightarrow \pi^0+n $ 
followed by $\pi^0+n\rightarrow \pi^-p$. This process should be considered at the same level as $\pi^-+p\rightarrow \pi^-+p$ followed by a 
second scattering $\pi^-+n\rightarrow \pi^-+n$. 
Thus Stodolsky is led to consider that the photon and the $\rho$-meson have the same quantum numbers 
and that one can consider $\rho$ -production as a quasi-elastic process, in such a way that the photon will 
fluctuate into a $\rho$-meson with amplitude proportional to $e$, and then reconvert into a photon, and this process will give 
a contribution of order $e^2$ to the cross-section. 
We are repeating this here 
since it shows once more that when dealing with complex systems, a straightforward application of the optical theorem may not work.

\subsubsection{Sakurai's VMD }\label{sss:sakurai}
 In 1969 Sakurai \cite{Sakurai:1969ss} proposed the Vector 
 Meson Dominance (VMD) 
 for high energy electron proton inelastic scattering. 
 Following the conjecture   \cite{Stodolsky:1966am,Sakurai:1968mz} that the total photo-absorption hadronic cross-section could be 
 calculated from diffractive production of $\rho, \omega$ and $\phi$ mesons, Sakurai went on to show that, when both longitudinal and transversely 
 polarized  photon contributions are included in the calculation of the total $ep$ \x, then  the VMD model and experimental results are fully compatible.     
 
 In this paper, the following kinematics is defined: $q=(\vecq, \nu)$ is as usual the  momentum transfer between electrons,    
 $\sqrt{s}$ the {\it missing hadronic } mass. 
 Also notice that he uses a metric such that $q^2>0$ corresponds to space-like photons. The   inelastic differential $ep$-\x \  
 is written in terms of the transverse 
 and longitudinal \x 's $\sigma_T$  and $\sigma_L$ as
 \begin{align}
\frac{d^2\sigma}{dq^2}=\frac{E'}{E} \frac{4\pi \alpha^2}{q^4} \big[ W_2(q^2,\nu) \cos^2\frac{\theta}{2} +
2 W_1(q^2,\nu)\sin^2\frac{\theta}{2}
\big]\\
W_2=\frac{K}{4\pi^2\alpha} \frac{q^2}{q^2+\nu^2} (\sigma_T+\sigma_S),\ \ \ \  W_1= \frac{K}{4\pi^2\alpha}\sigma_T\\
K=\nu - \frac{q^2}{2m_p}=\frac{s-m^2_p}{2m_p}
\end{align}
$\sigma_T$ and $\sigma_S$ are obtained from the transverse and longitudinal 
components of the electromagnetic current. The VMD hypothesis  then relates the electromagnetic matrix element $<A| j_\mu|p>$ 
between a given final hadronic state $|A>$ to the vector meson dominated one as 
\begin{equation}
<A| j_\mu|p>=\frac{m^2_\rho}{f_{\rho}} \frac{1}{q^2+m^2_\rho} <A| j_\mu^{\rho}|p>
\label{eq:vmd}
\end{equation}
where $ j_\mu^{\rho}$ stands for the source density of the neutral $\rho$-meson field and Eq.(\ref{eq:vmd}) defines 
the coupling between the photon 
and the $\rho$-meson. The transverse and longitudinal cross-sections are then given by
\begin{eqnarray}
\sigma_T=(\frac{e}{f_\rho})^2 F^2(q^2) \sigma_{\rho p}^\perp(K)\\
\sigma_S=(\frac{e}{f_\rho})^2 F^2(q^2) \frac{q^2}{m^2_{\rho}}(\frac{K}{\nu})^2\xi (K) \sigma_{\rho p}^\perp(K)
\end{eqnarray}
with
\begin{eqnarray}
F^2(q^2)=(\frac{m^2_\rho}{q^2+m^2_\rho})^2\\
\xi=\frac{\sigma_{\rho p}^\parallel}{\sigma_{\rho p}^\perp}
\end{eqnarray}
Accordingly Sakurai obtains for the structure function $W_2$
\begin{equation}
\nu W_2 (q^2,\nu)= \frac{m^2_\rho}{4\pi^2\alpha}\frac{K}{\nu}\frac{1}{1+m^2_\rho/q^2}\\
{\cal F}
\end{equation}
with
\begin{equation}
{\cal F}=
 [\frac{1}{1+m^2_\rho/q^2}]^2 \xi(K)(\frac{K}{\nu})^2+\frac{m^2_\rho}{q^2}\sigma_{\gamma p}(K)
\label{eq:vw2}
\end{equation}
$\nu W_2 (q^2,\nu)$ is then shown to become a universal function of $\nu/q^2$ in the Bjorken limit $q^2\rightarrow \infty$ 
and fixed $ q^2/\nu $ \cite{Bjorken:1968dy}.

Adding the other vector mesons is easily done by considering their isospin properties so that the overall contribution can be 
written by the simple substitution
\begin{equation}
\frac{1}{f_\rho}\rightarrow \frac{1}{f_\rho}[1+\frac{1}{\sqrt{2}}+ \frac{1}{3}]
\end{equation}
where one can make the approximation  $m^2_\rho\approx m^2_\omega\approx m^2_\phi$. 
Sakurai derives from VMD, a relationship between the Bjorken scaling function $F_2(x)$ and asymptotic $\sigma_{\gamma p}$
\be
F_2(x) \to\ [\frac{\xi(\infty) m_\rho^2}{4 \pi^2 \alpha}](1 - x)^2 \sigma_{\gamma p}(\infty).
\ee

\subsubsection{Gribov's model}
\label{sss:gribov}
We shall now summarize 
the model in which Gribov  first described  the interactions of  photons with matter. 
In \cite{Gribov:1968gs}, Gribov advances the idea that the character of the interaction of photons with nuclei and the development 
of surface effects [$A^{2/3}$ -dependence] at high energies have no connections with $\rho$-mesons or $\pi$-mesons, but are solely determined by 
distances which are significant in those interactions. 
In the above paper, Gribov is referring to  the idea proposed by Bell \cite{Bell:1964eu} about Vector Meson 
Dominance which was   discussed in \ref{ss:VMD}.

In fact, the expression proposed by Gribov for the cross-section of photons on nuclei, which includes only hadronic processes,  is
\begin{equation}
\sigma_{\gamma}=2\pi R^2 (1-Z_3)
\label{eq:sigribov}
\end{equation}
where $R$ is the nuclear radius, and $Z_3$ is the charge renormalization constant due to hadrons,  
which can be written in terms of the cross-section for electron-positron annihilation into hadrons,
\begin{equation}
1-Z_3=\frac{e^2}{\pi} 
\int \rho(x^2)\frac{dx^2}{x^2}
\end{equation}
Gribov's explanation of Eq. ~(\ref{eq:sigribov}) is that $2\pi R^2$ is the geometrical 
\x\  for the interaction of hadrons with nuclei and the other factor is related to the length of time that the photons spend in the hadron state. 
To estimate this time, one first needs to establish the region of validity of Eq.~(\ref{eq:sigribov}). This is obtained by first considering 
 a photon of momentum P in the Laboratory frame, and write the  relevant longitudinal  scale  as $\delta =P/\mu^2$, where
$\mu$ is some characteristic mass. Now, let $l$ be the mean free path length of a hadron in the nucleus, the condition of applicability is 
that   $\delta^2\gg Rl$. If one takes the characteristic mass to be that of  the $\rho$-meson and the path length as defined by the interaction, 
with $ l \sim 1/ m_{\pi}$, then Gribov claims that surface effects will start appearing at energies exceeding $10\ {\rm GeV}$, which
would correspond to a few GeV.

His picture of what happens is  as follows:
first the photon virtually  decays into hadrons (we would now say partons), and then the hadrons 
start interacting with the nucleons in the nucleus.  What matters here is the length of this fluctuation into hadrons, 
which he takes to last for a time $\delta$. He considers  the   two possibilities, $\delta \le l$ and of course $\delta \ge l$. 
Consider the first case and  let $\sigma_{\gamma}$ be proportional to:
\begin{enumerate}
\item  the probability of the photon to hit the nucleus  $\sim \pi R^2$
\item the probability that fluctuations take place inside the nucleus  $\propto \frac{\alpha R}{\delta}$ 
\item the probability that the hadrons forming will have time to complete an interaction with a  nucleon in the nucleus ,  $\propto \delta /l$
\end{enumerate}
Hence $\sigma_{\gamma } \sim \pi R^2 \times \frac {\alpha R}{\delta } \times \delta l\sim \alpha \pi R^3 / l  \propto A \sigma_{\gamma}$. 
But actually as the energy of the photon increases, the duration of the fluctuation will also increase and the probability of interaction 
will increase with energy. 
When the time length of the fluctuation into hadrons exceeds the interaction length, the relevant probability is one and, one 
gets  $\sigma_{\gamma } \sim \pi R^2 \times \frac {\alpha R}{\delta }$ and will decrease as the energy increases. A further effect is due to 
Bell's \cite{Bell:1964eu} {\bf check this ref} observation about the probability that the interaction takes place outside the nucleus is 
$\propto l/\delta$  so that one gets $\sigma_{\gamma}\sim \alpha \pi R^3  l/\delta^2$. When $\delta$ becomes much larger than the 
interaction length, as the energy increases further, the photon will fluctuate into a hadron outside the nucleus and the hadron which 
are thus formed will interact  with a cross-section 
$\pi R^2$. The argument is not full proof, and it appears more as an {\it a posteriori justification}, but the gist of the matter seems 
to be that the \x\ is actually proportional to the nuclear surface and not to the volume. According to Gribov, it is also easy to understand 
the presence of the factor $1-Z_3$. 
To understand it, he then looks at  the forward scattering amplitude, visualized in a figure like Fig.~\ref{fig:gribov1}, which 
we reproduce from \cite{Gribov:1969zy}.
\begin{center}
\begin{figure}
\resizebox{0.6\textwidth}{!}{%
\includegraphics{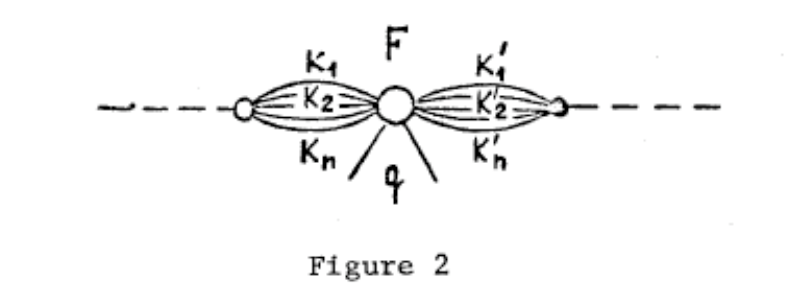}
}
\caption{Cartoon of the forward scattering amplitude from \cite{Gribov:1969zy}. }
\label{fig:gribov1}
\end{figure}
\end{center}
 In the figure, $F$ is the amplitude for scattering of a beam of hadrons on a nucleus of  radius R, with momentum transfer $q$. 
 But in the forward direction, Fig.~(\ref{fig:gribov2}), 
 it is the diagram defining  charge renormalization.
 \begin{center}
\begin{figure}
\resizebox{0.6\textwidth}{!}{%
\includegraphics{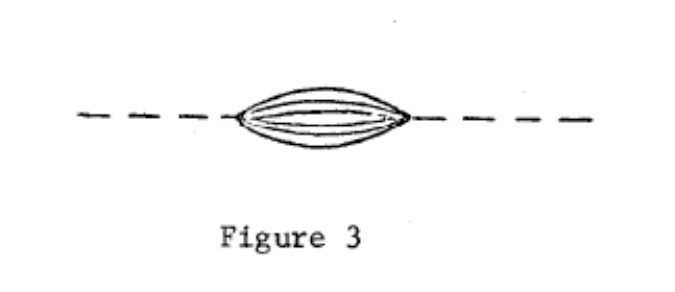}
}
\caption{Charge renormalization,  from \cite{Gribov:1969zy}.}
\label{fig:gribov2}
\end{figure}
\end{center}
For the interaction of electrons with nuclei depicted by Gribov as in Fig. \ref{fig:gribov3}, one has a similar picture, 
except that instead of $1-Z_3$ the cross-section will be determined by the polarization operator from Fig.~\ref{fig:gribov3}.
\begin{center}
\begin{figure}
\resizebox{0.6\textwidth}{!}{%
\includegraphics{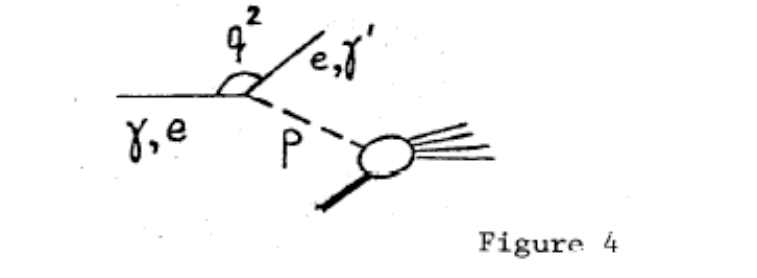}}
\caption{Interaction of electrons with nuclei  from \cite{Gribov:1969zy}.}
\label{fig:gribov3}
\end{figure}
\end{center}

 \subsection{QCD evolution equations for photon processes, 
 BK Equation}
\label{ss:BKgen}


We now turn to the  QCD description of photon processes in terms of evolution equations, addressing 
the transition from the linear BFKL rate equation to the non-linear Balitsky-Kovchegov (BK) equation
\cite{Balitsky:1995ub,Kovchegov:2001ni} and what saturation means. 
The underlying phase transition is also exhibited. We shall then return to  models, including  the transition 
from ($\gamma^{*} p$) to real photon ($\gamma p$) processes and models for (``geometrical'') scaling in the small $Q^2$ region.




\subsubsection{Introduction \label{sss:introBK}}
The BFKL  ``rate'' equation for the density of gluons 
is linear and is expected to break down as the density becomes large [e.g., as in $\gamma^*p$ at low $x$ or in
hadron-nucleus scattering]. A more appropriate equation valid for large densities is provided
by the BK equation\cite{Balitsky:1995ub,Kovchegov:2001ni} that is non-linear and 
incorporates saturation i.e., a maximum steady state value for the density. Similar problems occur in a variety of 
fields of physics, chemistry, biology, logistics, etc. In the following, we shall illustrate the problem and its resolution 
for the important practical case of the photon number for lasers.

\subsubsection{Dynamics behind some simple non-linear rate equations for photons\label{sss:Logistic} }

Here we shall discuss a simple non-linear rate equation and dynamical reasons behind
leading to it. The linear rate equation, where the rate is proportional to the number itself,  of course
leads to an exponential growth or exponential decay as 
\begin{equation}
\label{D0}
\frac{dI(t)}{dt} = +\nu I(t);\ I(t) = I(0) e^{\nu t},
\end{equation} 
depending upon the sign of $\nu$. But in all practical systems, some non-linearity
is bound to be present, giving rise to non exponential behavior in time.

The best studied (and very practical) example is that of the laser. If the rate equation for the mean 
photon number were linear (as above in Eq.(\ref{D0})), the number of laser photons 
would increase exponentially. 
Of course, that can not be, otherwise we would need an infinite source of energy, hence there 
must be some dynamical mechanism to saturate the number. The solution to this problem for lasers
was first given by Lamb. The famous Lamb equation for the light intensity $I(t)$ may be written
as \cite{MW} 
\begin{equation}
\label{D1}
\frac{dI(t)}{dt} = +\nu [ a - I(t)] I(t). 
\end{equation} 
The second term on the right hand side of Eq.(\ref{D1}) arises dynamically through the creation 
\& annihilation of two-photons at a time, just as the first term is related to the creation and annihilation 
of single photons. The parameter $a$ is called the pump parameter and its sign is crucial in determining
the steady state value of $I$. We note here parenthetically that the analog of Eq.(\ref{D1}) written in an 
entirely different context of population and called
the logistics equation  was first written down by Verhulst\cite{Verhulst:1845,Verhulst:1847}.

If $a\leq 0$, the steady state value of $I$ [determined by the vanishing of the left side of Eq.(\ref{D1})],
is $I_{SS}\to 0$. Physically, for negative pump parameter, there is no laser activity. On the other hand,
for $a>0$, $I_{SS} \to a$ and hence the laser intensity increases linearly with $a$.

The innocent looking Eq.(\ref{D1}) has buried in it a (second order) phase transition wherein $a$ acts as 
the order parameter. This is easily seen by considering $I_{SS}$ as a function of $a$. $I_{SS}$ is continuous
at $a=0$ but its derivative is not.

A simple model for a plethora of physical processes such as the mean photon number, intensity, 
mass growth, magnetization etc. is provided by analogues of Eq.(\ref{D1}) where the parameters $\nu$ and $a$
have different physical significance and their signs play a crucial role in determining the fate of that physical
system.   

A partial understanding of the genesis of the quadratic term on the right in Eq.(\ref{D1}) can be obtained
through a consideration of the frequency of a photon mode in a cavity. The frequency of a mode in a 
cavity is inversely proportional to the length of the cavity $L$. Thus, If the geometry of the cavity fluctuates 
via the length scale \begin{math} L \end{math}, then the frequency of the photon oscillator will 
be modulated. Because of such a modulation, the cell cavity will emit or 
absorb two (or more) photons at a time, thus leading to the above rate equation if one truncates
to two photons.

Similar rate equations must exist for any system [depending upon its size for example] where growth may be rapid but 
the growth must cease eventually resulting in a limiting value [such as the maximum size]. In the following section,
we shall discuss the relevant case of the gluon density in QCD where the non-linear direct coupling
$gg\to g$ (absent in QED) automatically provides such a non-linear term. 

But before going on to discuss the case of QCD, let us consider the special case $a = 0$ in Eq.(\ref{D1})
for the number $N(t)$ of photons.
\begin{equation}
\label{D2}
\frac{d N(t)}{dt} = -\nu N^2(t);\ N(t) = \frac{N(0)}{(1 + \nu t)} 
\end{equation} 
The above decay pattern $\sim (1/t)$ for large $t$ is called a hyperbolic decay law and it has 
been observed in certain cases of bio-luminesence. In fact, there are many simple physical systems 
which display hyperbolic decay laws. Typical examples are those which involve the excitation of
pairs in the medium, which then recombine to emit light. This naturally
gives decay laws which one would expect classically to obey 
$dN/dt=-\nu N^2$. Note that this is a purely classical result and
does not require coherent effects between the excited states, which would
also be expected to give the same decay law. 

The important point to remember is that
exponential or hyperbolic behavior can not be theoretically correct for the whole phase space 
even though they may provide good approximations in restricted regions of phase space. Such is
the case in QCD both in hadronic as well as in deep inelastic scatterings.

\subsubsection{Non-linear BK Equations in QCD \label{sss:BKQCD}}
Let us consider the Balitsky approach\cite{Balitsky:1995ub} to the scattering of a virtual photon $\gamma^*(q)$ 
on a hadron of momentum $p$ as summarized by E. de Oliveira \cite{deoliveira:2008}. For the limit of $x_B = [Q^2/2(q.p)]$ small
where $Q^2$ gives a hard scale and $s>> Q^2$, a dipole picture emerges naturally in the limit
of infinite colour $N_c \to \infty$ when planar diagrams become dominant. The photon does not directly
interact with the target hadron but only through an ``onium'' made up of a quark of a given colour accompanied
by an anti-quark of opposite colour to preserve the colour singlet nature of the photon. The onium must
then exchange two gluons with the hadron to preserve the colour singlet nature of the target hadron.

Thus, the photon does not interact directly with the target hadron but through a ``gas'' of non interacting
dipoles. 
Single dipole scattering with the target
hadron leads to BFKL evolution equation and multiple dipole scatterings to the BK equations. To proceed with
the dynamics, Balitsky \cite{Balitsky:2001gj} invokes the general notion that a fast particle in a high energy scattering moves along
its classical trajectory and the quantum effect consists in the acquisition of an eikonal phase along its prescribed 
(classical) path. In QCD, for a fast parton (quark or glue), the eikonal phase is given by the Wilson line that is 
{\it link-ordered} along the straight line collinear to the $4$-velocity $n^\mu$ of the parton. The Wilson line operator may be written as
\begin{equation}
\label{BK1}
U^\eta (x_\perp) = {\mathcal P} exp{\{} ig \int_{-\infty}^\infty du \ n_\mu A^\mu(u n + x_\perp),
{\}}
\end{equation} 
where $A^\mu(x)$ is the gluon field of the target, $x_\perp$ is the transverse position of the target.
In high energy scattering within QCD, Wilson line operators form 
convenient effective degrees of freedom as partons with different rapidities ($\eta$) ``feel'' each other through 
matrix elements of these operators. In the colour dipole model of the photon then, the propagation of 
a quark-antiquark pair 
takes place through the propagation of the colour dipole via the two Wilson lines ordered 
collinear to the quark's velocity. 
Thus the structure function of the hadron becomes proportional to a matrix
element of the colour dipole operator 
which is given by
\begin{equation}
\label{BK2}
{\tilde U}^\eta (x_\perp; y_\perp) = 1 - \frac{1}{N_c} Tr{\{} {\tilde U}^\eta(x_\perp) {\tilde U}^\eta(y_\perp)
{\}},
\end{equation} 
taken between the states of the target hadron. The gluon density is then given approximately by
\begin{equation}
\label{BK3}
x_B G(x_B; \mu^2 = Q^2) \approx\  <p| {\tilde U}^\eta (x_\perp; 0)|p>|_{x_\perp^2 = 1/Q^2}.
\end{equation} 
The energy dependence of the structure function is thus reduced 
to the dependence of the colour dipoles on the slope of the Wilson lines as determined by the rapidity $\eta$. 
A whole hierarchy found by Balitsky emerges as given by equations of the type [valid in LLA for $\alpha_s<<1$ \& 
$\alpha_s (ln x_B) \sim 1$]:  
\begin{eqnarray}
\label{BK4}
\frac{d}{d\eta} <T_{xy}> = \frac{{\bar \alpha}_s}{2\pi} \int (d^2z) {\mathcal M}(x,y;z)\times\nonumber\\
\ [ <T_{xz}> + <T_{yz}> - <T_{xy}> - <T_{xz} T_{yz}> ];\nonumber\\ 
\ {\bar \alpha}_s = \frac{\alpha_s N_c}{\pi}\ \ \ \ \ \ 
\end{eqnarray} 
Once the mean-field approximation (i.e., the factorization) $<T_{xz} T_{yz}> = <T_{xz}> <T_{yz}>$ is made, 
the above becomes a non-linear, but closed, set of  BK evolution equations
In particular, for the Wilson line 
operators it becomes
\begin{align}
\label{BK5}
\frac{d}{d\eta} <{\tilde U}^\eta_{xy}> = \frac{{\bar \alpha}_s}{2\pi} \int (d^2z) \frac{(x - y)^2}{(x -z)^2 (y-z)^2}\times\nonumber\\
\ [ <{\tilde U}^\eta_{xz}> + <{\tilde U}^\eta_{yz}> - <{\tilde U}^\eta_{xy}> - <{\tilde U}^\eta_{xz}><{\tilde U}^\eta_{yz}> ]
\end{align} 
The last (non-linear) term on the right hand side of the BK equation for color dipoles is due to multiple scattering. This
Balitsky-Kovchegov evolution equation is usually written for the dipole hadron cross-section in impact parameter space  
as 
\begin{align}
\label{BK6}
\sigma_{dipole}(x_{01}; Y) = 2 \int (d^2b_{01}) {\mathcal N}(b_{01};x_{01}; Y)\nonumber\\
\ x_{01} = (x_0 - x_1); b_{01} = \frac{x_0 + x_1}{2},
\end{align} 
where ${\mathcal N}(b_{01};x_{01}; Y)$ is the quark-antiquark propagator through the hadron, related to the forward scattering
amplitude of the dipole with the hadron. The BK equation in ``coordinate'' space reads 
\begin{align}
\label{BK7}
\frac{d}{dY} {\mathcal N}(b_{01};x_{01}; Y) = \frac{{\bar \alpha}_s}{2\pi} \int (d^2x_2) \frac{x^2_{01}}{x^2_{02} x^2_{12}}\times\nonumber\\
\ [ {\mathcal N}(b_{01} +\frac{x_{12}}{2} ;x_{02}; Y) + {\mathcal N}(b_{01} +\frac{x_{02}}{2};x_{01}; Y) - \nonumber \\
{\mathcal N}(b_{01};x_{01}; Y)
\ - {\mathcal N}(b_{01} +\frac{x_{12}}{2};x_{02}; Y) {\mathcal N}(b_{01} +\frac{x_{02}}{2};x_{12}; Y) ]\nonumber\\
\end{align} 
The ``time'' here is the rapidity $Y \approx\ 1/x_B$ and the equation has four other variables [two from $x_{01}$ \& two
from $b_{01}$]. The BK equation resums all powers of ($\alpha_s Y$). If the last quadratic term is dropped, then, it reduces to the
linear BFKL equation
\begin{eqnarray}
\label{BK8}
\frac{d}{dY} {\mathcal N}(b_{01};x_{01}; Y) = \frac{{\bar \alpha}_s}{2\pi} \int (d^2x_2) \frac{x^2_{01}}{x^2_{02} x^2_{12}}\times\nonumber\\
\ [ {\mathcal N}(b_{01} +\frac{x_{12}}{2} ;x_{02}; Y) + {\mathcal N}(b_{01} +\frac{x_{02}}{2};x_{01}; Y)\nonumber\\
\ - {\mathcal N}(b_{01};x_{01}; Y)]\nonumber\\
\end{eqnarray} 
Given the complexity of Eq.(\ref{BK7}), it is useful to consider special cases to obtain some familiarity with it.
\subsubsection{Space-independent BK equation in ($0+1$)-dim\label{sss:1dim}}
If one assumes that ${\mathcal N}(b_{01};x_{01}; Y)$ is spatially independent (i.e., independent both of the impact
parameter $b_{01}$ \& the dipole ``size'' $x_{01}$), then the BK equation reduces to the previously discussed 
logistics equation [see Section (\ref{sss:Logistic})]:
\begin{equation}
\label{BK9}
\frac{d}{dY} {\mathcal N}(Y) = \omega [ {\mathcal N}(Y) - {\mathcal N}^2(Y)]; \ \omega > 0.
\end{equation}
As discussed previously, it has two steady-state solutions (or fixed points): an unstable solution ${\mathcal N} = 0$ \&
the other the stable solution ${\mathcal N} = 1$. It should also be clear that the linearized BFKL blow up for large $Y$
has been softened to a maximum value of $1$, i.e., a saturation for small $x_B$, independent of the initial condition.

\subsubsection{Impact-parameter independent BK equation in ($1+1$) dim \label{sss:2dim}}
If we drop only the impact parameter dependence but keep the dipole size, we have 
${\mathcal N}(b;x; Y) \to\ {\mathcal N}(r; Y)$ and the BK equation in ($1+1$) dimension reads
\begin{eqnarray}
\label{BK10}
\frac{d}{dY} {\mathcal N}(|x_{01}|; Y) = \frac{{\bar \alpha}_s}{2\pi} \int (d^2x_2) 
\frac{x^2_{01}}{x^2_{02} x^2_{12}}\times\nonumber\\
\ [ {\mathcal N}(|x_{02}|; Y) + {\mathcal N}(|x_{12}|; Y) - {\mathcal N}(|x_{01}|; Y)\nonumber\\
\ - {\mathcal N}(|x_{02}|; Y) {\mathcal N}(|x_{12}|; Y) ]\nonumber\\
\end{eqnarray} 
Physically, of course, $b$-independence implies an infinite homogeneous hadronic surface 
but where the scattering kernel does depend upon the size of the dipole. Numerical computations
verify general trends already seen in ($0+1$) dimensions\cite{deoliveira:2008} :
\begin{itemize}
\item Saturation {\it occurs} [in contrast to BFKL blowup] for large $Y$;
\item Saturation for large $Y$ is independent of the initial condition [that is, independent of the dipole size];
\item For small $Y$, ${\mathcal N}(r;Y)$ is smaller for smaller dipole size;
\item For small $r$, non-linear corrections are by and large negligible;
\item For large $r$, non-linear corrections are important and ${\mathcal N}(r;Y)\approx\ 1$
\item Saturation scale $Q_s(Y)$: 
\begin{eqnarray}
r< \frac{1}{Q_s(Y)};\ {\mathcal N} << 1;\nonumber\\
\ r> \frac{1}{Q_s(Y)};\ {\mathcal N} \approx\ 1.
\end{eqnarray}
\end{itemize}

\subsubsection{Geometrical scaling in DIS \label{sss:DISGS}}
 The approach to saturation is also discussed
  in a paper by Stasto, Golec-Biernat and Kwiecinski 
through a discussion of 
a geometric scaling \cite{Stasto:2000er} in the low-x region, 
observed at HERA for Deep Inelastic Scattering, $\gamma^* p$ scattering.
 But the result 
 claimed here is not the same as the usual {\it geometric scaling} observed or expected in hadron-hadron scattering. 
 In 
 the purely hadronic case geometric scaling refers to the fact that the scattering amplitude in impact parameter 
 space $G(s,b)$ is only a function of the ratio $\beta=b^2/R^2(s)$, where $R(s) $ is the interaction radius.
 To avoid confusion, one should notice that the interaction radius $R(s)$ in the hadronic case increases with energy, 
 whereas the one in DIS decreases with energy, or with $x\rightarrow 0$.

Such a behavior is understood to represent a unitarity bound, which reflects the fact that the growth with $x$
(as $x \to 0$)  
of the structure functions is tamed by saturation effects. This is also a version of the 
Black Disk model. 

More precisely, the HERA data on the total $\gamma^* p$ scattering cross-section, 
suggest a geometrical scaling of the following form \cite{Stasto:2000er} 
\begin{equation}
\label{gs11a}
\sigma^{\gamma*p}(Q, Y) = \sigma^{\gamma*p}(\tau);\ \tau = \frac{Q^2}{Q^2_s(Y)}.
\end{equation} 
This translates for the scattering amplitude into
\begin{equation}
\label{gs12a}
{\mathcal N}(r;Y) \to\  {\mathcal N}(rQ_s(Y));\ {\rm for\ large\ } Y 
\end{equation} 
Using the form $Q_s(Y) = Q_o e^{{\bar \alpha}_s \lambda Y}$, the scaling form given in Eq.(\ref{gs2})
reduces to 
\begin{equation}
\label{gs13}
{\mathcal N}(r;Y) \to\  {\mathcal N}(Q_oe^{(ln r + {\bar \alpha}_s \lambda Y)}). 
\end{equation}
Eq.(\ref{gs13}) has been interpreted as a {\it traveling wave} with $Y$ as {\it time}, 
(${\bar \alpha}_s \lambda$) as the {\it speed} $v_s$ of the {\it wave} and 
($ln r$) as the {\it spatial coordinate}. Such a wave picture emerges rather naturally 
through a momentum space description as shown  next.

In \cite{Boer:2007ug}, a detailed analysis of 

``extended'' geometrical scaling has been made and its (not at all obvious)
connection with the BK equation investigated. These authors conclude through a numerical analysis
of the BK equation in momentum space that the BK results are qualitatively different from that of the
phenomenological dipole models. In particular, they find that geometrical scaling around the saturation point
is only obtained for asymptotic rapidities.

\subsubsection{Momentum space BK equation \label{sss:MOMBK}}
Let us consider the momentum space amplitude defined through the
Fourier transform\cite{Marquet:2005zf,Ducati:2013wra,Ducati:2013cga}

\begin{equation}
\label{mom1}
{\bar{\mathcal N}}(k;Y) = \int (\frac{d^2r}{2\pi r^2}) e^{i {\bf k}\cdot {\bf r}}
{\mathcal N}(r;Y) . 
\end{equation}
Then the BK equation in momentum space reads
\begin{equation}
\label{mom2}
\frac{d{\bar{\mathcal N}}(k;Y)}{dY} = {\bar \alpha}_s \int (\frac{dk^{'}}{k^{'}}) 
{\mathcal K}(k,k^{'}) {\bar{\mathcal N}}(k^{'};Y) - {\bar \alpha}_s {\bar{\mathcal N}}^2(k;Y). 
\end{equation}
In \cite{Munier:2004xu}, it is shown
that in the saddle point approximation, the BK equation can be mapped into
the FKPP equation\cite{Fisher:1937,Kolmogorov:1937}
of the form
\begin{equation}
\label{mom5}
\partial_t u(\zeta ,t) = \partial_\zeta^2 u(\zeta,t)  + u(\zeta,t) - u^2(\zeta,t), 
\end{equation}
with the dictionary above: $t$ is time, and $\zeta$ is the coordinate. The crucial point is that FKPP equation
does have {\it traveling wave} solutions of the form ($\zeta - v t$), in agreement with the geometrical 
scaling solutions given in Eq.(\ref{gs3}) with $t = Y$, $\zeta = ln (r)$ and $v = {\bar \alpha}_s \lambda$. This
correspondence does provide a window of comfort in the phase space for geometrical scaling. 

So far, we have considered a fixed $\alpha_s$. For a discussion of the results of BK evolution as one changes to 
running $\alpha_s$, we refer the reader to some recent analyses in \cite{Albacete:2009fh} \cite{Lappi:2012vw}.

\subsubsection{Dense hadronic systems \label{sss:CGC}}
For dense hadronic systems, new phenomena in QCD occur and some have been investigated in detail
for heavy ions. For large A nuclei scatterings at high energies, colour glass condensates and colour transparency
have been found through an effective field theory constructed from QCD. It will take us far outside the realm
of this review but we refer the interested reader to excellent expositions by McLerran\cite{McLerran:2003rh}
\cite{McLerran:1993ni}, Venugopalan\cite{Iancu:2003xm} and Mueller\cite{Mueller:1999yb}.

\subsubsection{Beyond BK, fluctuations, Pomeron loops \label{sss:PL}}
That the BK equation does not include fluctuations in the 
gluon (dipole) number 
has been particularly emphasized by Bartels {\it et al.} \cite{Bartels:2012gq}. Thus, if the Pomeron
is considered as a manifestation of the propagation and exchange of two-gluon singlets, then what is missing
in BK becomes the lack of a Pomeron hierarchy as shown in Fig.~\ref{Pomeronloops}.
\begin{figure}[htbp]
\resizebox{0.5\textwidth}{!}
{\includegraphics{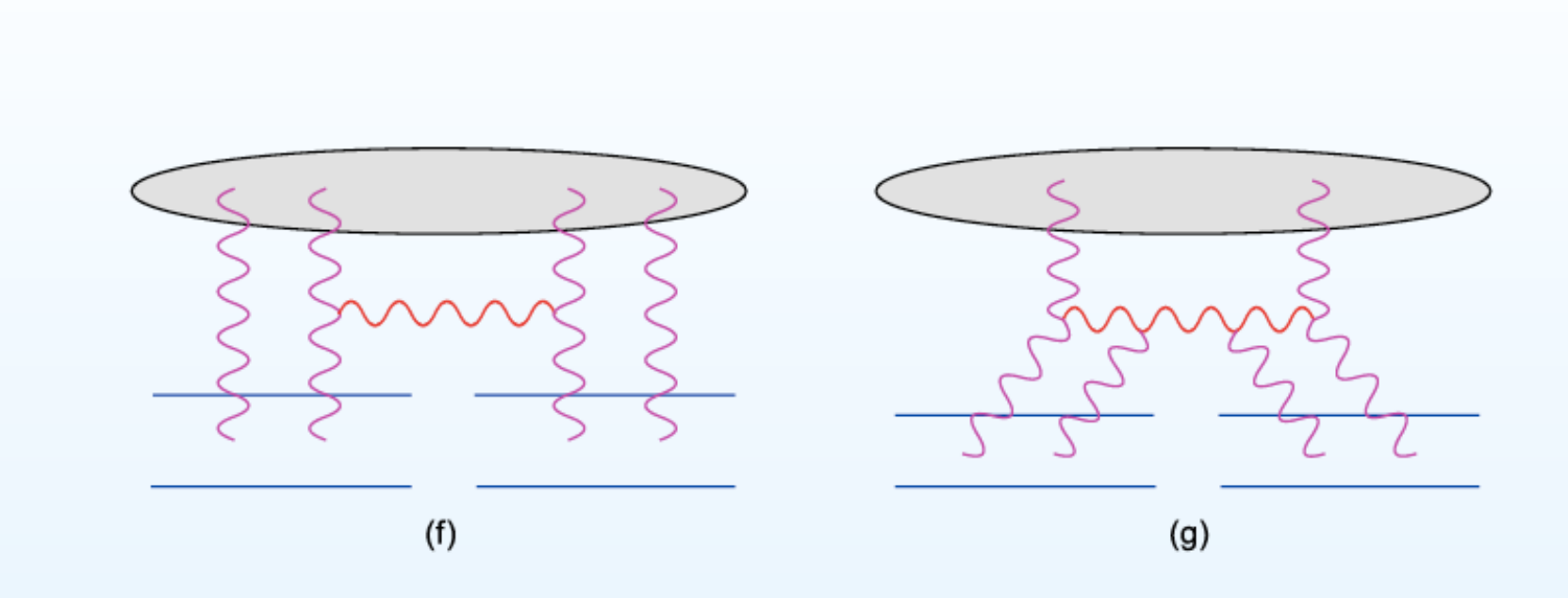}}
\caption{Pomeron loop graphs absent in BK, from \cite{deoliveira:2008}. }
\label{Pomeronloops}
\end{figure}

Work in the past two decades have shown that a Langevin equation
can be formulated to include three Pomeron vertices and we just quote some results. It is claimed \cite{Iancu:2005nj} 
that a BK equation with white noise of the following form -called the stochastic FKPP or $sFKPP$- can mimic 
the Pomeron hierarchy.   
\begin{eqnarray}
\label{PL1}
\partial_t u(\zeta ,t) = \partial_\zeta^2 u(\zeta,t)  + u(\zeta,t) - u^2(\zeta,t)\nonumber\\
\ + \nu (\zeta ,t) \sqrt{(\frac{2}{N_c}) u(\zeta ,t) (u(\zeta ,t) - 1)}, 
\end{eqnarray}
where white noise $\nu (\zeta ,t)$ is defined as
\begin{equation}
\label{PL2}
< \nu(\zeta ,t)> = 0;\ <\nu(\zeta ,t) \nu(\zeta^{'} ,t^{'})> = \delta(\zeta - \zeta^{'}) \delta(t - t^{'}). 
\end{equation}
Strong fluctuations are also discussed in \cite{Marquet:2005ak}.
For further theoretical work on this subject we refer the reader to the literature.

\subsection{Transition from $\sigma(\gamma^*p)$ to real $\sigma(\gamma p)$: Models and phenomenology for low-x physics}\label{ss:gamstar}
After the great successes of Bjorken scaling and the verification of its perturbative QCD calculable violations in 
deep inelastic total cross-section $\sigma(\gamma^*p)$ for large
$Q^2$, came the arduous task of understanding the physics for small $Q^2$ photon masses  
and eventually to bridging the gap to its continuation
to real $\sigma(\gamma p)$ process as $Q^2\to 0$. Excellent quality data exist in this kinematic 
region by the H1 \& Zeus groups from HERA demanding a theoretical and phenomenological
explanation.

Explicitly, the object is to formulate the usual proton EM structure function $F_2(W; Q^2)$ defined as in Eq.~(\ref{eq:sig*F2})
so that it interpolates smoothly to the real photon cross-section $\sigma^{\gamma p}(W^2)$. It should be mentioned
that real photon,  $\sigma^{\gamma p}(W^2)$, \x s  are   obtained through the HERA data in $Q^2 = (0.01\div 0.02) {\rm GeV}^2$ 
region. While such a region lies beyond the realm of perturbative QCD, it does offer the possibility of
extension as well as a challenge to hadronic total cross-section models for its description.

\subsubsection{Phenomenological analyses by Haidt {\it et al.}} \label{sss:Haidt}
Data from HERA on the structure function $F_2(x, Q^2)$  at small (and medium) values of x have been analyzed
in  set of papers by D. Haidt \cite{Buchmuller:1996rz},\cite{Haidt:2000kw},\cite{Haidt:2001hz}
and compared with theoretical expectations. In \cite{Buchmuller:1996rz}, it was shown that the observed rise
at small $x = (Q^2/2 q\cdot p)$ is consistent with a doubly logarithmic increase: a logarithmic increase in $1/x$ along with a logarithmic
growth also with $Q^2$, i.e. 
\be
\label{T1}
F_2(x, Q^2) = a + b [ln(\frac{x_o}{x})] [ln(\frac{Q^2}{Q_o^2})],
\ee
where $a, b, x_o, Q_o^2$ are constants, and the above expression is valid in the perturbative  phase space region at $x < 0.001$.
A stronger increase, which may be incompatible with unitarity when extrapolated to asymptotically §
small values of x, could not be inferred from the data then available.

A few years later, in \cite{Haidt:2000kw},\cite{Haidt:2001hz}, the HERA data for small 
values of the Bjorken variable $x = (Q^2/2 q\cdot p) \leq 0.01$, were described phenomenologically through the expression
\begin{eqnarray}
\label{T2}
F_2(x; Q^2) = m [ln(\frac{x_o}{x})] ln(1 + \frac{Q^2}{Q_o^2})\\
\ m\approx\ 0.4;\ x_o\approx\ 0.04;\ Q_o^2\approx\ 0.5 {\rm GeV}^2.  
\end{eqnarray}
The extension from $ln (Q^2/Q_o^2)$ to $ln (1+Q^2/Q^2_o)$ allows to describe both the perturbative and the nonperturbative regime as long as x is below 0.001. This implies for $F_2$ a behaviour proportional to $Q^2$ for $Q^2 < Q_o^2$ and a logaritmic behaviour above.The strategy adopted by Haidt for a smooth continuation of $\sigma^{\gamma^*p}(W^2; Q^2)$ to very small values of $Q^2$ 
consisted in defining a variable $q = ln(1 + Q^2/Q_o^2)$ and rewriting Eq.(\ref{T2}) as
\begin{equation}
\sigma^{\gamma^*p}(W^2; Q^2) = (\frac{4 \pi^2}{Q_o^2}) [\frac{q}{(Q^2/Q_o^2)}] [\frac{F_2(W, Q^2)}{q}]. \label{eq:gamstarhaidt}
\end{equation}
A virtue of $q$ is that it interpolates smoothly from small $Q^2$ to $ln \ Q^2$ ( for large $Q^2$): since $q \to Q^2/Q_o^2$ as $Q^2 \to 0$, a transition from $\sigma^{\gamma^*p}(W^2; Q^2)$ to $\sigma^{\gamma p}$
becomes amenable. In the region $x < 0.01$, $W^2\sim Q^2/x$ and thus a behavior of $F_2/q \sim [ln(1/x)]$ implies
$F_2/q \sim [ln(W^2)]$. The $q$ dependence of the HERA data were then analyzed through a linear form in $ln(W^2)$:
\begin{equation}
\label{T4}
\frac{F_2(W^2; q)}{q} = u_o(q) + u_1(q) ln(W^2/W_o^2).
\end{equation} 
An almost constant value for the slope $u_1(q) \approx\ 0.4$ was found for large values of $q$. Inclusion of real $\gamma p$
data at $W = 200\  {\rm GeV}$ showed that the transition from the $\gamma^* p$ data available until the lowest value of $Q^2 = 0.05\ {\rm GeV}^2$,
to  real photons in $\gamma p$ seemed to work well.

As Haidt pointed out, for smaller values of q -outside the measured region- Eq.(\ref{eq:gamstarhaidt}) needs to be revised since ($F_2/q$)
is a function of $x$ alone whereas $\sigma^{\gamma p}$ is a function of $W^2$ alone. The suggested replacement to reach real Compton
scattering -so that the $Q^2 \to 0$ limit is reached smoothly- is  
\begin{equation}
\label{T5}
(\frac{x_o}{x}) \to\  [\frac{x_o}{x} (\frac{Q^2}{Q^2 + Q_w^2})],
\end{equation} 
where for consistency $0 \leq Q_w^2 \leq Q_o^2$.  Satisfactory agreement with the HERA data were found for $Q_w^2 = 0.05\ {\rm GeV}^2$.

An attentive reader would note that Haidt's variable $q = ln(1 + Q^2/Q_o^2)$ that becomes linear in $Q^2$ for small $Q^2$, has 
a  parallel in Richardson's proposal of replacing the asymptotic freedom formula for the QCD coupling constant 
$\alpha_{AF}(Q^2)$ to $\alpha_R(Q^2)$ so as to obtain a linearly confining potential \cite{Richardson:1979ri}:
\begin{align}
\label{T6}
\alpha_R(Q^2) = \frac{1}{b\ ln[1 + Q^2/\Lambda^2]};\\ 
\alpha_{R}(Q^2)\to\ [\frac{\Lambda^2}{b\ Q^2}] \ {\rm for}\ Q^2\to 0;\\ 
\alpha_{R}(Q^2)\to\ \alpha_{AF}(Q^2) = \frac{1}{b\ ln[Q^2/\Lambda^2]} \ {\rm for}\ Q^2\to \infty.
\end{align} 
Further discussion and details about singular, confining $\alpha_s(Q^2)$  can be found in Sec. \ref{sec:models} of the present review.  

\subsubsection{Dipole model and Geometrical scaling\label{sss:DMGS}}
As described in Sec(\ref{sss:DISGS}), the phenomenon of saturation and a  geometrical scaling for low-$x$ $\gamma^* p$ processes 
have been obtained from the QCD dipole model. Here we present its essential formulation and phenomenology. 

In this model, the scattering takes place in two steps. First, a virtual (transverse $T$ or longitudinal $L$) photon 
of $4$-momentum $Q$ splits into a $q\bar{q}$ dipole of transverse size ${\bf r}$ that is described through a probability distribution 
$|\Psi(r, z, Q^2)|^2$, where $z$ is the fraction of longitudinal momentum of a quark of mass $m_f$ . Then, a
subsequent scattering of the produced dipole occurs  with the proton that is 
modeled  through a dipole-proton  cross-section $\tilde{\sigma}(r, x)$. Explicitly,
\begin{equation}
\label{GS1}
\sigma_{T, L}(x; Q^2) = \int (d^2{\bf r}) \int_o^1 (dz) |\Psi_{T, L}(r, z; Q^2)|^2 \tilde{\sigma}(x; {\bf r}).  
\end{equation}   
The splitting wave functions for the photon $\Psi$ for a quark of flavour $f$ and charge $e_f$ are given by
\begin{eqnarray}
\label{GS2}
|\Psi_{T}|^2 = [\frac{3 \alpha}{2 \pi^2}] \sum_f e_f^2 \{ [z^2 + (1 -z)^2] (\bar{Q}_f K_1(\bar{Q}_f r))^2\nonumber\\
+ (m_f K_o(\bar{Q}_f r))^2 \}\nonumber\\
|\Psi_{L}|^2 = [\frac{3 \alpha}{2 \pi^2}] \sum_f e_f^2 [2 z(1 - z)\bar{Q}_f K_o(\bar{Q}_f r)]^2,\ \   
\end{eqnarray}   
where $K_{0,1}$ are Macdonald functions and 
\begin{equation}
\label{GS3}
\bar{Q}^2_f = z (1 - z) Q^2 + m_f^2.  
\end{equation}   
It is important to note that the above incorporates the change in the dynamics as $Q^2$ varies from large to
very small values in two ways. Kinematically, as $Q^2$ goes to zero, the effective quark masses $m_f$
begin to set the scale for the process. The important ranges of integration in Eq.(\ref{GS1}) changes with the size of the 
dipoles in two essential ways. The $K$-functions decrease exponentially for large $r$ dipoles whereas
for small size dipoles they provide (inverse) power law dependence. Also, the dipole cross-sections are assumed to ``saturate''
as follows.
\begin{equation}
\label{GS4}
\tilde{\sigma}(r, x) = \sigma_o g(\hat{r});\ \hat{r} = [\frac{r}{R_o(x)}],\   
\end{equation}   
where $\sigma_o$ is taken as a constant (phenomenologically $\sigma_o \sim 23\ mb$) and the function $g$ saturates to $1$
as $\hat{r} \to \infty$:
\begin{equation}
\label{GS5}
g(\hat{r}) = [ 1 - e^{-(\hat{r}^2/4)}],\   
\end{equation}   
The above tames the small $x$ blow up present in the structure functions in DGLAP \& BFKL, as required by unitarity.

Geometrical scaling resides in Eq.(\ref{GS4}) through the fact that $\sigma(\hat{r})$ depends on the dimensionless variable
$\hat{r}$ and thus the saturation radius $R_o(x)$ controls the energy behaviour of the cross-section. Hence, in the region of small
but non-vanishing $x$,  after integration
Eq.(\ref{GS1}) depends only on one dimensionless variable $\tau$:
\begin{equation}
\label{GS6}
\sigma^{\gamma^* p}(x, Q^2) = \sigma_o\ h(\tau); \ \tau = Q^2 R_o^2(x).  
\end{equation}   
Qualitatively, the results -modulo lagarithmic corrections- may be summarized as follows.   
 \begin{eqnarray}
\label{GS7}
\sigma^{\gamma^* p}(x, Q^2) \to\  \sigma_o\ {\rm for}\ \tau \to 0\nonumber\\
\sigma^{\gamma^* p}(x, Q^2) \to\  [\frac{\sigma_o}{\tau}]\ {\rm for}\ \tau >> 1.  
\end{eqnarray}  
A phenomenological form for the saturation radius
\begin{eqnarray}
\label{GS8}
R_o(x) = (\frac{1}{Q_o}) (\frac{x}{x_o})^{\lambda/2};\nonumber\\ 
Q_o = 1 {\rm GeV};\ x_o = 3 \times\ 10^{-4};\ \lambda = 0.29,  
\end{eqnarray}   
seems to work quite well and exhibits scaling for $x < 10^{-2}$\cite{Stasto:2000er,GolecBiernat:1998js} as 
shown in the Fig.~\ref{geoscaling} from \cite{Stasto:2000er}.
\begin{figure}[htbp]
\resizebox{0.5\textwidth}{!}
{\includegraphics{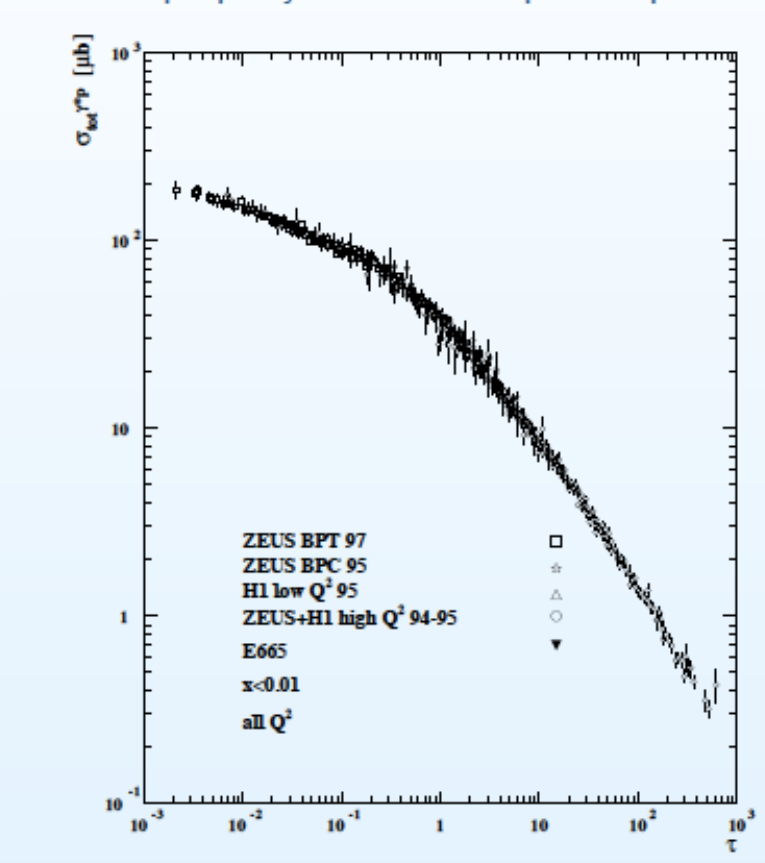}}
\caption{$\sigma^{\gamma^*p}(\tau)$ vs $\tau$, the scaling variable from \cite{Stasto:2000er}.
Reprinted with permission, Fig.(1) from \cite{Stasto:2000er}, \copyright (2001) by the American Physical Society.}
\label{geoscaling}
\end{figure}

For a smooth limit to $Q^2 \to 0$, the Bjorken variable is shifted to 
\begin{equation}
\label{GS9}
x \to\ \tilde{x} = x [1 + \frac{4 m_f^2}{Q^2}],  
\end{equation}  
and a parameter
 \begin{equation}
\label{GS10}
\zeta = (\frac{x}{x_o})^\lambda (\frac{Q^2}{Q_o^2}),  
\end{equation}  
is defined that delineates the ``soft-x'' regime ($\zeta < 1$) from the ``hard-x'' regime where $\zeta> 1$.
A useful interpolation formula that approximately covers both regions has also been given \cite{GolecBiernat:1998js} as
\begin{equation}
\label{GS10a}
\sigma^{\gamma^* p}(x, Q^2) =  \sigma_o\ \{ ln(1 + \frac{1}{\zeta}) + \frac{1}{\zeta} ln(1 + \zeta)\}\  
\end{equation}  
As 
previously discussed 
in the definition of the parameter $q$ 
in  Eq. ~(\ref{eq:gamstarhaidt})
proposed
by Haidt,  a factor $1$ has been added to the argument of the 
logarithms for a smooth limit $Q^2 \to 0$. The above expressions reproduce the change in the slope of the
high $W^2$ cross-section data as $Q^2$ is varied.

 \subsection{Models for $\gamma p$ cross-section}\label{ss:gamp}
 The approaches  to the phenomenological or theoretical description of photon-proton total cross-sections 
 can be roughly divided into some general  categories:
 \begin{itemize}
 \item factorization models, including  the universal Pomeron exchange model by Donnachie and 
 Landshoff description,  which extends very simply from \pp \
 scattering to photon processes and can then be extended, again very simply, to photon-photon processes 
 \item the Reggeon-calculus approach which follows Gribov's picture of the interaction, including the Dual Parton model descriptions
 \item QCD minijet models with photon structure functions 
 \item QCD inspired parametrizations 
 \end{itemize}
 We shall start with Donnachie and Landshoff model, which we have already described in the previous sections.

In 1992, Donnachie and Landshoff  \cite{Donnachie:1992ny} proposed a universal form for all total cross-sections, 
 based on Regge pole behaviour. Their expression, 
based on a simple and economical parametrization of the total cross-section behaviour,  
describes the high energy behavior of all total \x \ with a universal power law. The universality of the slope 
is not always observed, as we have discussed in  \cite{Godbole:2008ex}.  However the DL  expression, 
with slightly different  slopes,    offers  a good  description in the energy range presently reached by accelerators, 
and is still an object of investigation, both theoretically and experimentally.
 
We show in Fig.~\ref{fig:heradl} 
the results from an 
 analysis 
by the ZEUS Collaboration from HERA. The focus of this anaysis is  the slope 
of  $\sigtotgamp$ \cite{Collaboration:2010wxa} as a function of the cm energy W, in the energy range 
spanned by HERA. Parametrizing  $\sigtotgamp$ with $W^{2\epsilon}$ gives $\epsilon=0.111\pm 0.009(stat)\pm 0.036(sys)$.
\begin{figure}
\resizebox{0.5\textwidth}{!}{
\includegraphics{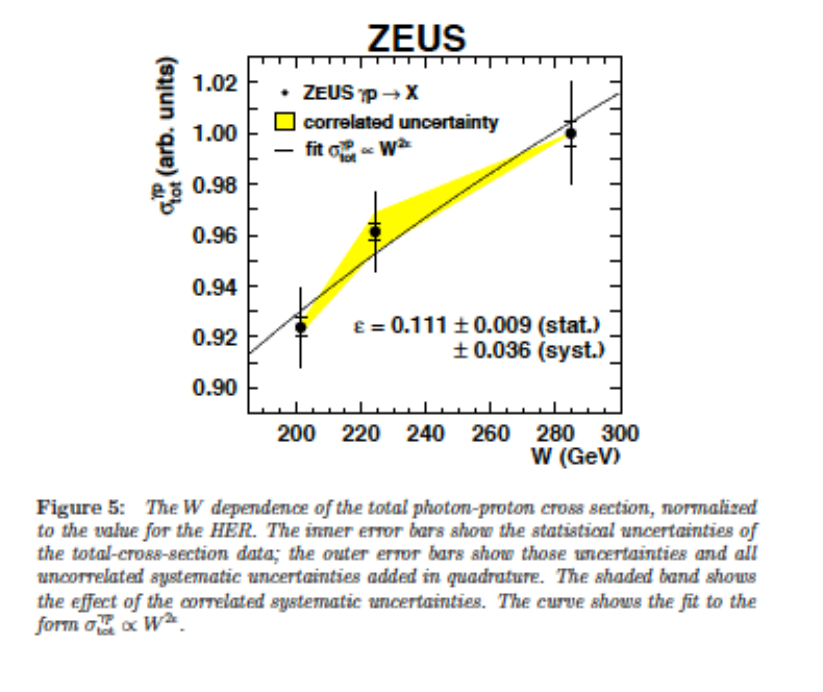}}
\caption{ 
The energy behavior of the photo production cross-section by the ZEUS Collaboration as shown in Fig. (5) from  \cite{Collaboration:2010wxa}. Reprinted with permission from the ZEUS Collaboration from  \cite{Collaboration:2010wxa}. OPEN ACCESS.}
\label{fig:heradl}
\end{figure}

It must be stressed that while a power law behavior is a good parametrization of the energy dependence  
in the HERA region, this is clearly not sustainable at higher energies, as dictated by the Froissart bound.  
The behavior to be expected at the high end of cosmic ray 
energies cannot be gauged from this analysis.
\subsubsection{The Tel Aviv group}\label{sss:GLM}
The  work \cite{Gotsman:1997zw,Gotsman:1999mw} by the Tel Aviv group of Gostman, Levin and 
Maor (GLM) presents a unified description  of DIS total \x \ and  photo-production. 
This work  follows 
Gribov's idea that the scattering of photons on hadrons can be visualized in,  the by now standard,  two stages, i.e.
\begin{enumerate}
\item the virtual photon fluctuates into a $q{\bar q}$ pair (hadron in Gribov's language)
 \item  the $q {\bar q}$ interacts with the hadronic matter
 \end{enumerate}
 In this model one calculates the total cross-section for a generic $Q^2$, and the final expression is written with  
 a contribution from the transverse (T) cross-section as well as one for the longitudinal (L) part. For large $Q^2$,  
 an expression for  contribution from fluctuations of the photon into a heavy quark pair is also given.
  
Following Gribov, the starting expression   for the cross-section, for a photon of mass $Q^2$ scattering off a 
proton, is written  through a dispersion relation in the initial and final hadronic masses  as
 \begin{equation}
\sigtotgampistar=
\frac{\alpha}{3 \pi}
\int
 \frac{\Gamma(M^2)dM^2}
 {M^2+Q^2}
  \sigma(M^2,M'^2,s)
  \frac {\Gamma(M'^2)dM'^2}
  {M'^2+Q^2}
\end{equation}
with 
\begin{equation}
\Gamma^2(M^2)=R(M^2)=\frac{\sigma(e^+e^-\rightarrow hadrons)}{\sigma(e^+e^-\rightarrow \mu^+\mu^-)}
\end{equation}
For large masses  $\Gamma(M^2)\times \Gamma(M'^2)\rightarrow R(M^2)=2$. 
 To describe the hadronic cross-section $\sigma(M^2,M'^2)$, the scattering is first divided according to an  energy scale,  $M_0$, which  
 separates the hard scattering regime where pQCD can be used and the soft region.
 In the soft region, a second scale is needed, because of the difference between gluon and quark sizes. 
 To be more specific, in the soft region, i.e. for $M,M'<M_0$ the following expression is used:
\begin{eqnarray}
\sigma(M^2,M'^2)&=&\sigma_N^{soft}(M^2,s)M^2 \delta(M^2-M'^2)=\nonumber \\
& & [\sigma_{qN}+\sigma_{\bar q N}]M^2\delta(M^2-M'^2)
\end{eqnarray}
and for $M,M'<M_0$,  Gribov's formula is simplified to read
\begin{equation}
\sigma(\gamma^*N)=\frac{\alpha}{3 \pi}\int \frac{R(M^2)M^2dM^2}{(Q^2+M^2)^2} \sigma_N(M^2,s)
\end{equation}

For the soft regime, a Donnachie-Landshoff type expression is used so as to arrive to 
\begin{equation}
\sigma_T^{soft}=\frac{\alpha}{3 \pi}\int_{4m^2_\pi}^{M^2_0} \frac{R(M^2)M^2dM^2}{(Q^2+M^2)^2}\{ A(\frac{s}{M^2})^{\alpha_P-1}+ B(\frac{s}{M^2})^{\alpha_R-1}\}
\label{eq:levinsoft}
\end{equation}
with $A$ and $B$ obtained so as to make the result agree with those from $\rho$-proton interactions. 
In \cite{Gotsman:1997zw}, the constants $A$ 
and $B$ were obtained from DL type fits to $\pi^{\pm}p$.
Since the \x \ thus calculated seem to be higher than the data, some corrections are introduced. 
The calculation for the hard part is done using published 
PDF's for  the gluon distributions inside the proton, and is given by
\begin{eqnarray}
\sigma_T^{hard}=
\frac{2\pi\alpha}{3}
\int_{M_0^2}^{\infty} 
\frac{R(M^2)dM^2}{Q^2+M^2}\times\nonumber \\
\int_0^\infty
\frac{
d{\tilde M}^2
}{
{\tilde M}^4
}
\alpha_s(
\frac{
{\tilde M}^2
}{4
}
)
xG(x,\frac{
{\tilde M}^2}
{4}){\cal I}(M^2,{\tilde M}^2,Q^2)
\end{eqnarray}
with 
\begin{eqnarray}
{\cal I}(M^2,{\tilde M}^2,Q^2)&=& \nonumber \\
 \frac{M^2-Q^2}{M^2+Q^2}
+
\frac{
Q^2+{\tilde M}^2-M^2
}{
\sqrt{
(Q^2+M^2+{\tilde M}^2)^2-4M^2{\tilde M}^2
}
} 
\label{eq:levinhard}
\end{eqnarray}
Notice the lower cutoff for the integration in $M^2$.

Eq.~(\ref{eq:levinsoft}) and  Eq.~(\ref{eq:levinhard}) need to be implemented by the contribution of heavy  quark pairs. 
This is  obtained from Eq.~(\ref{eq:levinhard}) by the substitutions
\begin{equation}
4M^2{\tilde M}^2\rightarrow 4(M^2-4m_Q^2){\tilde M}^2, \ \ \ \ \ \ \  R(Q^2)\rightarrow R^{QQ}(M^2)
\label{eq:levinheavy}
\end{equation}
In both the above equations, $x=x(M^2)=(Q^2+M^2)/W^2$, where $W$ is the energy in the photon-nucleon center of mass system. 
No soft contribution is of course present for the heavy quark term.

 An expression similar to the above is also used to describe a longitudinal component to add to the transverse one. For the {\it soft} contribution, 
 the authors note that {\it a priori} it should be straightforward to replace the  factor $M^2$ with $Q^2$, except they find that, in so doing, 
 the contribution from the soft part  overestimates the experimental data and needs to be reduced. The strategy adopted is to reduce the 
 value of the parameter $M_0$.  For the hard component, the additional degrees of freedom result in an expression proportional 
 to $ Q^2$ (hence going to zero for real photons) . For details, see \cite{Gotsman:1999mw} .
 The overall expression is thus
 \begin{equation}
\sigma(\gamma^*p)=\sigma_T^{soft}+\sigma_T^{hard}+\sigma_{T,QQ}^{hard}+\sigma_L^{soft}+\sigma_L^{hard}
\end{equation}

The resulting fit for the $\gamma p $ \x\ is shown in Fig.~\ref{fig:GLMN}.
\begin{figure}
\resizebox{0.5\textwidth}{!}
{\includegraphics{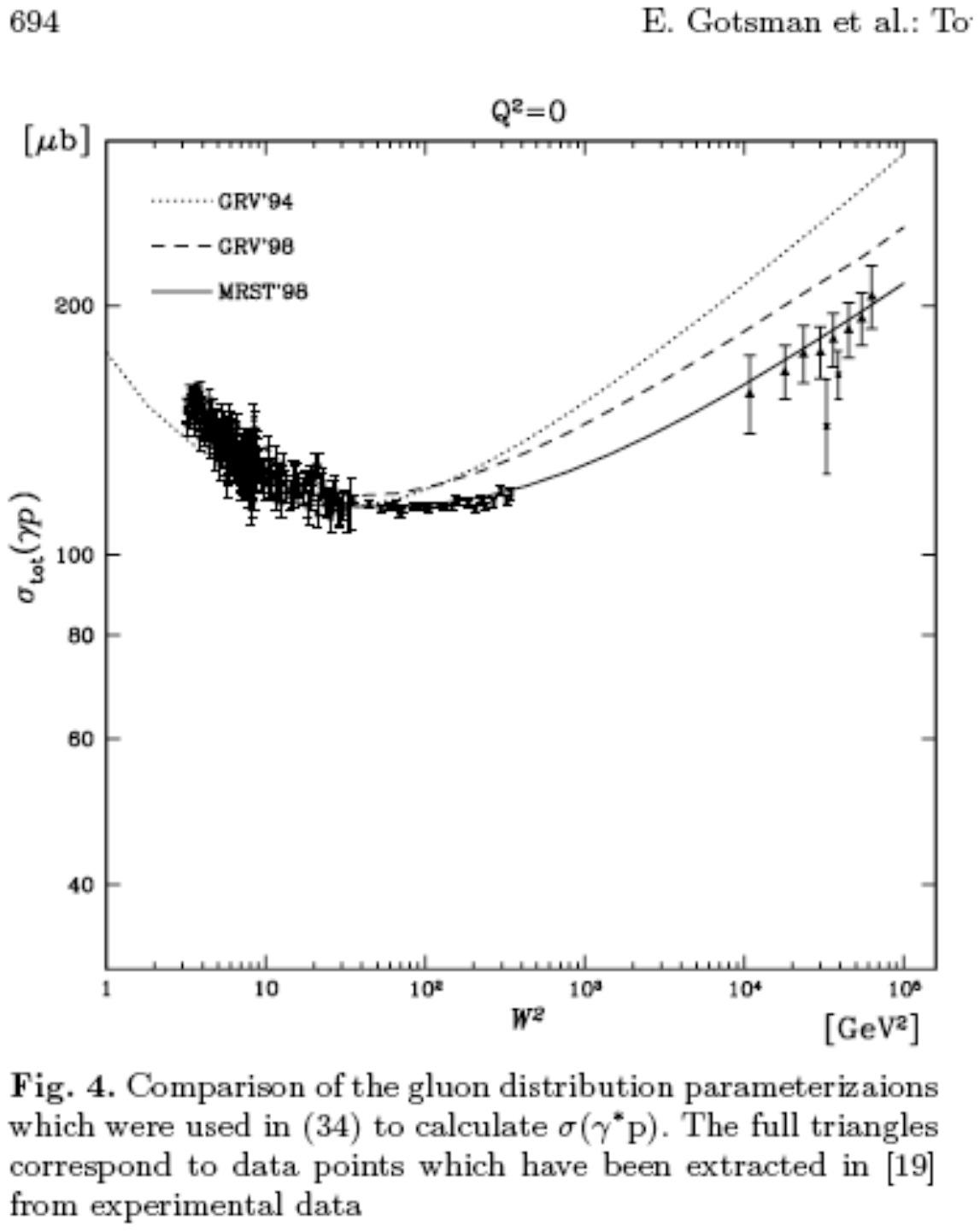}}
\caption{From \protect\cite{Gotsman:1999mw}. Data in the low energy range come from fixed target experiments, 
while \x\ values for $W^2>10^4\ {\rm GeV}^2$ are from \protect\cite{Breitweg:1998dz}. Reprinted with permission  from \cite{Gotsman:1999mw}
\copyright (1999) by Springer.}
\label{fig:GLMN}
\end{figure}
 
 Concerning the pQCD part of this calculation, there are a few  points to  notice:
 \begin{description}
 \item - at low energy, the soft part does use some type of vector meson dominance, being parametrized 
 following $\pi^{\pm}p$, but it needs some adjustments,
 \item - the overall result depends on the gluon densities used for the calculation of the hard part, with  
 MRST\cite{Martin:2002dr} densities for the gluon  giving  a better description than GRV, both GRV94\cite{Gluck:1994uf} or GRV98\cite{Gluck:1998xa},
 \item - a difficulty in the calculation is related to the  low values of $M^2$ integration, where the strategy adopted has been to use
 \begin{equation}
 xG(x,l^2<\mu^2)=\frac{l^2}{\mu^2}xG(x,\mu^2)
 \label{G}
 \end{equation} 
 and, for $Q^2<\mu^2$ to keep fixed  
 the strong coupling constant. 
 \end{description}
In our own QCD calculation of the mini-jet cross-section, the gluon densities have also been extended to very low $x$ values of the gluon 
fractional momenta, 
as discussed  in ~Sect .\ref{sss:BN}, 
and the lower cutoff is given by a phenomenologically determined value $p_{min}\simeq 1\ {\rm GeV}$. 

\subsubsection{Eikonal mini-jet models  for $\gamma p$ scattering}\label{sss:miniets}
We shall now describe how the eikonal mini-jet model  was extended
to photon processes  \cite{Gandhi:1991ve,Fletcher:1991jx,Honjo:1993fi}, and subsequently modified by 
Block {\it et al.} \cite{Block:1998hu} in the QCD inspired model of \cite{Margolis:1988ws}.

In the GLMN approach \cite{Gotsman:1999mw}, the pQCD contribution to the total cross-section was calculated using gluon-gluon scattering 
for the probability of finding a gluon in a proton. For the probability of finding a gluon in a photon, the calculation did not use parton densities, 
but wave functions and various integrations.
A different line of approach to the partonic content of the photon had instead been developed by Drees and Godbole \cite{Drees:1988dk} 
who argued that the hadronic content of the photon consists of quarks and gluons, in a way 
analogous to the partonic content of the proton or the pion. Thus, one could measure and define photon structure functions, which would submit to 
DGLAP evolution just like the hadrons. Such photon densities could  be inserted into a QCD calculation as in the proton-proton case. 
This idea would then allow the calculation of jet \x s and that for production of mini-jets, namely jets with $p_t\ge \approx 1\ {\rm GeV}$. 
To cure the resulting too large number of mini-jets,  a saturation mechanism was invoked in  \cite{Collins:1990bs}, where the 
VMD model was suggested to be used within the eikonal formalism,
in complete analogy with   proton-proton scattering, as discussed in the previous section. 


A  formulation of the calculation of the total $\gamma p$ \x \ was proposed by  Fletcher {\it et al.} \cite{Fletcher:1991jx}, 
following the eikonal mini-jet model for hadronic cross-sections developed earlier by Durand {\it et al.} 
\cite{Durand:1987yv,Durand:1988ax} and extended to photon processes \cite{Gandhi:1991ve}. 
The issues involved, at the time, in correctly extending the model to photon-hadron scattering included  
how to incorporate the photon-hadron coupling into the eikonalization procedure, use of appropriate photon structure 
functions, and gluon shadowing at small $x$.

In 
the mini-jet 
approach, one distinguishes the following steps:
\begin{itemize}
\item the photon interacts with other hadrons ``as a hadron'', namely as an ensemble of quarks and gluons, 
with  a probability $P_{had}$ which is proportional to $\alpha_{QED}$,
\item once the photon has fluctuated into such a hadronic state, one can apply hadronic models for calculation 
of total or inelastic cross-sections, such as  eikonal models with QCD mini-jets to drive the rise,
\item the mini-jet cross-section will be calculated using parton-parton \x s and photon densities, following standard 
parametrizations such as GRV\cite{Gluck:1998xa}, GRS \cite{Gluck:1999ub}, CJKL\cite{Cornet:2002iy}, 
or using QCD inspired parametrizations, or gluon mass models, etc.
\end{itemize}
The proposed expression is
\begin{equation}
\sigma_{inel}^{\gamma p}= P_{had}\int d^2\vecb[1-e^{-n(b,s)}]
\label{eq:halzengampi
}\end{equation}
where
\begin{equation}
n(b,s)=n_0(b,s)+A(b) \frac{\sigma_{parton}}{P_{had}}
\label{eq:nbs}
\end{equation}
In Eq.~(\ref{eq:nbs} ), the first term represents the non-perturbative contribution to the average number of collisions, 
the second is the one which should be calculated perturbatively and which gives the high energy rise  of the \x , through the low-x gluons 
present in the hadronic content of the photon. $n_0(b,s)$ is of order of magnitude of a similar term present in  hadronic interactions, and its 
estimate depends on the low energy modeling of the photons in the hadronic state. The second term has to be calculated using the standard 
parton-parton cross-sections folded in with the photon PDFs.  In many models
\cite{Godbole:2008ex} 
the soft term $n_0(b,s)$ is obtained using the Additive Parton Model (ADM) together with VMD, by putting
\begin{equation}
n_0(b,s)=A_{VMD}\frac{2}{3}\sigma^{nn}_{soft}(s)
\end{equation}
 where  $\sigma^{pp}_{soft}(s)$ would be the same soft \x\  entering the eikonal mini-jet model for proton-proton 
 and/or proton-antiproton scattering. We shall return to this point later. 
 
 The eikonal formulation for this model requires an expression for the impact parameter distribution in the photon. 
 In \cite{Fletcher:1991jx} VMD and the form factor hypothesis  
 are used, and the result is that 
$ A_{VMD}$ is obtained as the Fourier transform of the convolution of two form factors, the proton form factor and the ``photon'' form factor. 
The latter is taken to be the pion form factor, following again a model in which   the number of quarks   
 controls the $b$-distribution during the collision.
For protons, the dipole expression is used, for the pion the monopole expression, so that
 \begin{eqnarray}
A_{VMD}(b)=\frac{\nu^2}{2 \pi}\frac{\mu^2}{\mu^2-\nu^2}\times \nonumber \\
\left [
 \frac{\mu^2}{\mu^2-\nu^2} 
[ K_0(\nu b)-K_0(\mu b)]-\frac{\mu b}{2 }K_1(\mu b)
\right ].
\end{eqnarray}
In most applications of this model, the same expression for $A(b)$ is used  for both the mini-jet term and the soft part. However in general, there is no 
reason to assume that the parton distribution in b-space is the same at very high energy and at low energy. In fact, in the  model to be described next, 
the so called Aspen model \cite{Block:1998hu}, this not so.

Before proceeding further, let us examine the quantity $P_{had}$ which plays a basic role in all the extensions of hadronic models 
to photon total cross-sections.

If the photon, in its interactions with matter, is to be considered just like a hadron, then any model for hadron-hadron scattering should 
be considered extensible to 
photon-hadron scattering. The factor $P_{had}$ represents the probability for a photon to interact like a hadron and  was  
introduced to apply vector meson 
dominance ideas to the eikonalization procedure.  In principle, $P_{had}$ may very well have an energy dependence. 
A possible definition follows the general VMD statement
that the wave function of the photon in its interaction with hadrons can be expressed as \cite{Schuler:1993td} 
\begin{equation}
|\gamma> =Z_3|\gamma_{B}> + \sum_{V=\rho,\omega,\phi}\frac{e}{f_V}|V>+\frac{e}{f_{q{\bar q}}}|q {\bar q}>
\end{equation}
where the first term corresponds to the bare photon, i.e. in its purely electromagnetic interactions, while the second considers 
the non-perturbative component, pictured through VMD,  and the last gives the contribution to the pQCD behaviour at high energy 
from  quarks and gluons.  
 
Given  the general theoretical  uncertainty in total \x \  models, {\it  a phenomenological   strategy}  is   
to ignore this energy dependence and use a VMD model for $P_{had}$, 
or even to use it as a free parameter determined by the normalization of the total $\sigtotgamp$ \x\ at low energy. 

The Aspen model for photons \cite{Block:1998hu} to be described next,  is a  generalization of the Block {\it et al.}  
\cite{Margolis:1988ws} model for protons 
with some differences. The Block model is based on a QCD inspired parametrization 
and uses the eikonal formalism, which guarantees unitarity, namely one starts with 
\begin{equation}
\sigtot=2\int d^2\vecb [1-e^{\chi_I(b,s)}\cos(\chi_R(b,s))]
\end{equation}
In the proton case, $\chi(b,s)$ is a complex function, whose {\it even} component $\chi^{even}$ 
receives contributions from parton-parton interaction through the three separate terms
\begin{eqnarray}
\chi^{even}=\chi_{qq}(b,s)+\chi_{qg}(b,s)+\chi_{gg}(b,s)=\nonumber \\
i\big[
\sigma_{qq}(s)W(b;\mu_{qq})+
\sigma_{qg}(s)W(
b;
\sqrt{\mu_{qq}\mu_{gg}}t
)+\nonumber \\
\sigma_{gg}(s)W(b;\mu_{gg})\big] \label{eq:blockpp}
\ \ \end{eqnarray}
The extension to $\gamma p$ is done as in \cite{Fletcher:1991jx} through 
\begin{equation}
\sigma_{tot}^{\gamma p}(s)= P_{had}\int d^2\vecb[1-e^{-\chi_I^{\gamma p}(b,s)}\cos\chi_R^{\gamma p}(b,s)]
\label{eq:blockgampi}
\end{equation}
In this model the value $P_{had}=1/240$ is used. This value is obtained by fitting the low energy data 
and is very close to the expected VMD value. 
For the cross-sections, $\sigma_{ij}(s)$, and the impact  parameter distribution functions for photons, 
to be used in Eq.~(\ref{eq:blockgampi}), 
the following substitutions are made in Eq.~(\ref{eq:blockpp}):
\begin{eqnarray}
\sigma_{ij}^{pp}\rightarrow \sigma_{ij}^{\gamma p}=\frac{2}{3}\sigma_{ij}^{pp}\\
\mu_i^{pp}\rightarrow \mu_{i}^{\gamma p}=\sqrt{\frac{3}{2}}\mu_i^{pp}
\end{eqnarray}
where the two substitutions are done in the spirit of the Additive Quark Model. We can anticipate that the 
same model will be applied also to $\gamma \gamma $ processes with 
\begin{eqnarray}
\sigma_{ij}^{\gamma \gamma }=\frac{4}{9}\sigma_{ij}^{pp}\\
\mu_{i}^{\gamma \gamma}=\frac{3}{2}\mu_i^{pp}
\end{eqnarray}
The predicted total \x\ in this model is shown in Fig.~\ref{fig:aspengampi}.
\begin{center}
\begin{figure}
\resizebox{0.6\textwidth}{!}{%
  \includegraphics{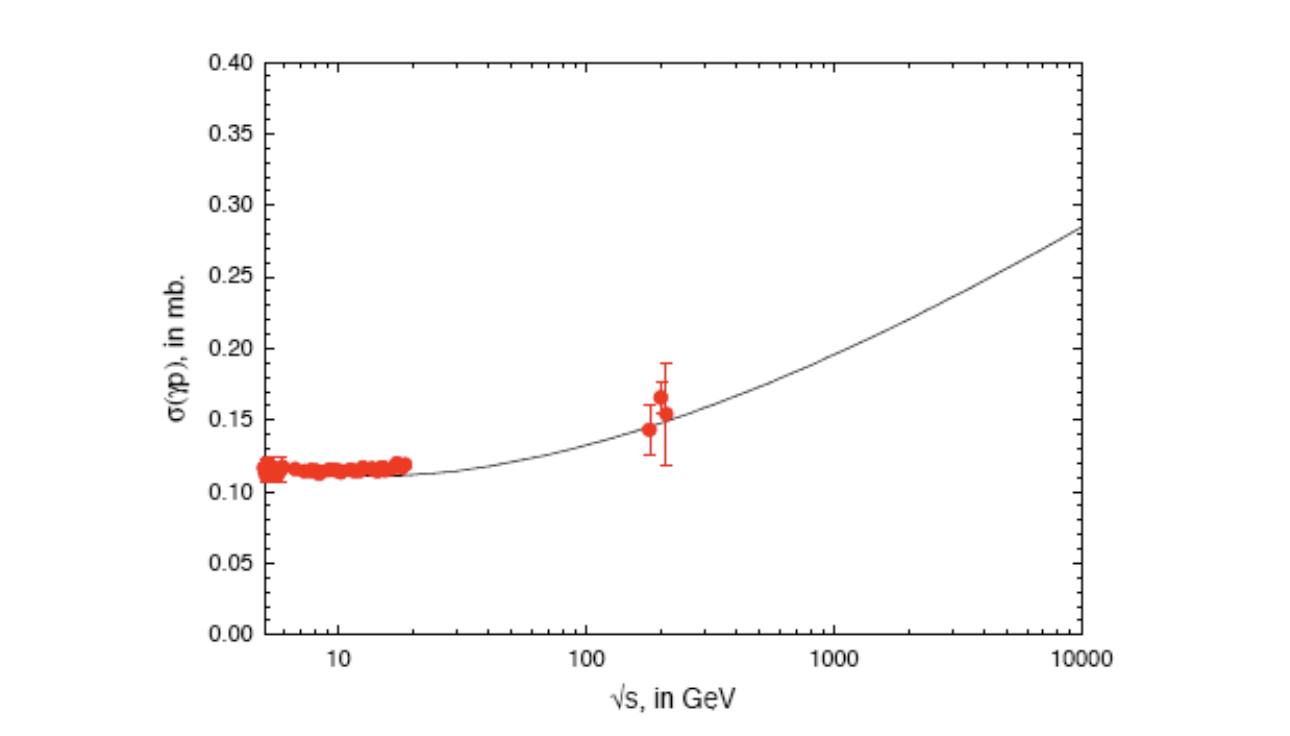}
}
\caption{The total cross section $\sigma_{tot}^{\gamma p}$ in mb vs. $\sqrt{s}$ in GeV, from \protect\cite{Block:1998hu}.
This is  Fig. (7)  from \cite{Block:1998hu}, reprinted with permission, \copyright (1999) by the American Physical Society}
\label{fig:aspengampi}
\end{figure}
\end{center}

Following the QCD inspired model outlined above, Luna and collaborators \cite{Luna:2006sn} 
have also extended their dynamical gluon model to 
photon-proton scattering.

More recently, Block has proposed an analytical amplitude model and has applied it  to both photon and neutrino scattering 
on protons \cite{Block:2014kza}. 

\subsubsection{The BN model : eikonal mini-jet model with soft gluon resummation}\label{sss:bngamma}
In this section we describe our extension of the Eikonal mini-jet model with $k_t$-resummation in the infrared region,
 {\it labeled BN model}, as it is inspired by the 
Bloch and Nordsieck (BN) description of the Infrared catastrophe \cite{Bloch:1937pw}.
As described in the previous section, our aim with this model  is to introduce, together with   
the mini-jet cross-section, a saturation effect which arises from soft gluon emission, 
down into the infrared region, as discussed in the section about the total cross-section.

The model is so far relatively simple, with a limited number of parameters, and thus it can, to a certain extent, be considered 
almost a 
model {\it for testing confinement} through 
a singular quark-gluon coupling below the perturbative QCD expression.

We start with the simplified expression from \cite{Fletcher:1991jx}, namely
\begin{equation}
\sigma^{\gamma p}_{tot} =
2 P_{had}\int d^2 {\vec b}[1-e^{-n^{\gamma p}(b,s)/2}] 
\label{sigtot1}
\end{equation}
with
   \begin{eqnarray}
\label{nbs}
n^{\gamma p}(b,s)=n_{soft}^{\gamma p}(b,s) +n^{\gamma p}_{hard}(b,s)\nonumber \\ 
= n_{soft}^{\gamma p}(b,s) + A(b,s) \sigma_{jet}^{\gamma p}(s)/P_{had}
\end{eqnarray} 
with  $n_{hard}$ including all outgoing parton processes with $p_t>p_{tmin}$. We differ from  other mini-jet models in  approximating 
the eikonal with just the imaginary part \cite{Block:1998hu},  in using a different impact parameter distribution for the soft and the hard 
part \cite{Fletcher:1991jx}, but mostly in our expression and origin of the impact parameter distribution for photons.  In Eq.~(\ref{nbs})  the 
impact parameter dependence has been factored out,  averaging over densities in a manner similar to what was done for the case of the proton  
in \cite{Corsetti:1996wg}. Because the jet cross-sections are calculated using actual  photon densities, 
which themselves give the probability of finding a given quark or gluon in a 
photon, $P_{had}$ needs to be  canceled out in $n_{hard}$. 
 We choose its value, by normalizing the eikonalized cross-section to the data in the low energy region, 
 and we use    $P_{had}=1/240 \approx P_{VMD}$. 
 For the  average number of hard collisions,  we use mini-jets and soft gluon resummation
with  $n_{hard}$ given by:
\begin{equation}
n_{hard}(b,s)={{
A^{AB}_{BN}(b,s) \sigma_{jet}
 } \over{P_{had} }}
\end{equation}  
with the impact distribution function obtained exactly as in the proton-proton case, namely
\begin{eqnarray}
 A^{AB}_{BN}(b,s) =
{\cal N} \int d^2 {\bf K}_{\perp} {{d^2P({\bf K}_\perp)}\over{d^2 {\bf K}_\perp}} 
 e^{-i{\bf K}_\perp\cdot {\bf b}} \nonumber \\
 = {{e^{-h( b,q_{max})}}\over
 {\int d^2{\bf b} e^{-h(b,q_{max})} }}\equiv A^{AB}_{BN}(b,q_{max}(s)).\ \ \
 \label{Eq:abn}
 \end{eqnarray} 
except for the fact that $q_{max}$ the upper limit of integration in the function $h( b,q_{max})$ is to be calculated using proton and 
photon densities. $h( b,q_{max})$ describes the exponentiated, infrared safe, number of single soft gluons of  
all allowed momenta  and is given by, 
\begin{eqnarray}
h( b,q_{max}(s))  = 
\frac{16}{3}\int_0^{q_{max}(s) }
{{dk_t}\over{k_t}} 
 {{ \alpha_s(k_t^2) }\over{\pi}} \nonumber \\  
\times \left(\log{{2q_{\max}(s)}\over{k_t}}\right)\left[1-J_0(k_tb)\right]
\label{hdb}
\end{eqnarray}

We show typical  values taken by $q_{max}$ for different sets of quark densities in Fig.~\ref{fig:qmax_gp_lia_98_lowsc}.
\begin{figure}
\resizebox{0.4\textwidth}{!}{%
  \includegraphics{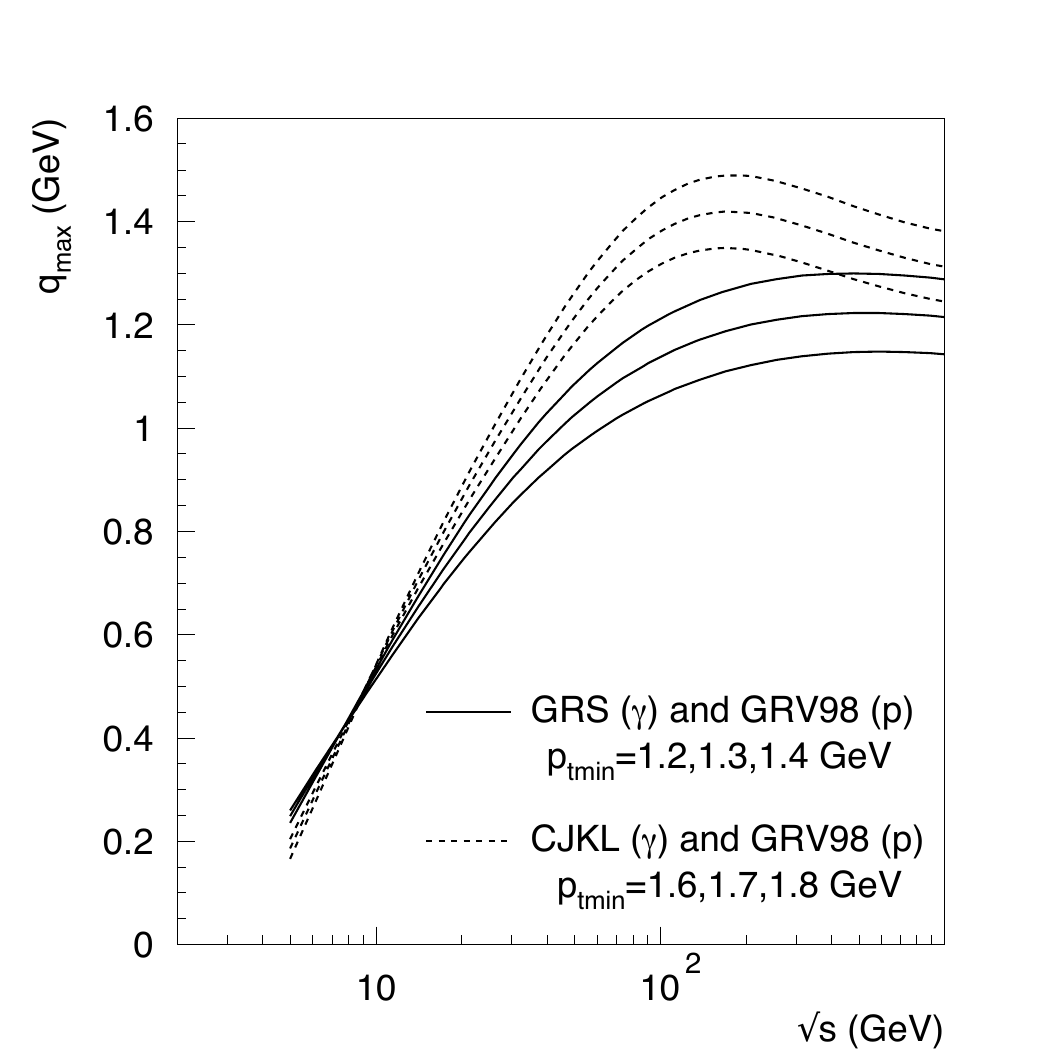}
   }
\caption{The maximum single gluon momentum allowed for soft gluon integration,   $q_{max}$ in GeV,  vs. $\sqrt{s}$ in 
GeV for $\gamma p$ scattering, from \protect\cite{Godbole:2008ex}. Reprinted with permission from \cite{Godbole:2008ex},
\copyright (2008) by Springer.}
\label{fig:qmax_gp_lia_98_lowsc}
\end{figure}
In our model, the expression for $A(b,s)$ for the hard  term in hadron-hadron or hadron-photon scattering remains the same, 
unlike models that use form-factors 
for instance, where the photon needs to be modeled as a meson and then parametrized.

 We show the result of our model in Fig.~\ref{fig:2008ourgamp-comparison} from \cite{Godbole:2008ex}.
\begin{figure}
\resizebox{0.5\textwidth}{!}{%
  \includegraphics{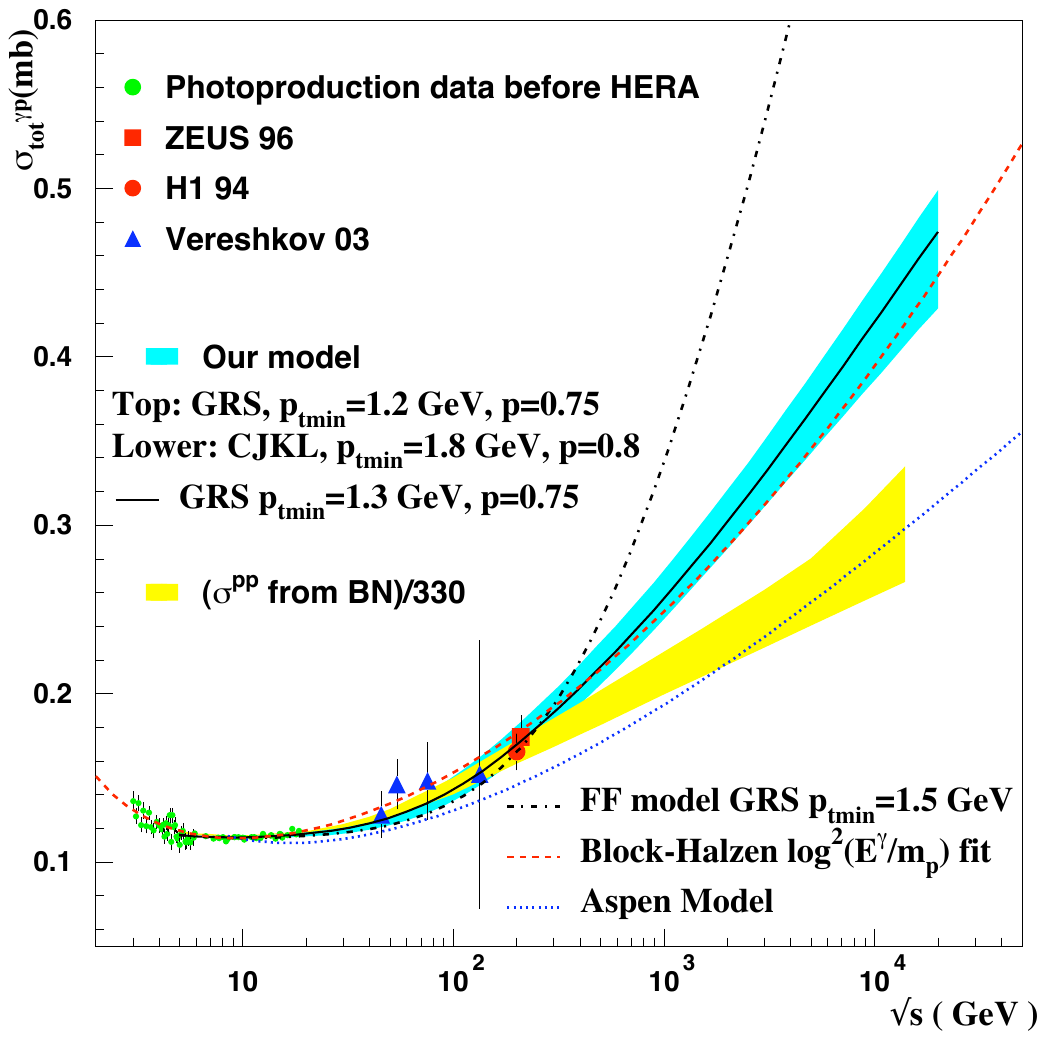}
 }
 \vspace{-2cm}
\caption{The total cross section $\sigtotgamp$ in mb vs. $\sqrt{s}$ in GeV, from \protect\cite{Godbole:2008ex}.
Reprinted from \cite{Godbole:2008ex}, \copyright (2008) by Springer.}
\label{fig:2008ourgamp-comparison}
\end{figure}
In this figure, the high energy parameter set of this description, consisting of the LO PDFs and $p_{tmin}$ value 
used for the mini-jet cross-section calculation, together with the saturation (singularity) parameter $p$, were limited 
to GRV densities for the protons, while two sets of photon PDFs were used.
  A comparison was made with predictions from some available models, such as indicated in the figure and discussed in  \cite{Godbole:2008ex}. 

An updated description of  $\gamma p$, is shown in Fig.~\ref{fig:2014ourgamp-comparison}.  
In this figure, we compare the BN model results obtained with MRST and GRV densities for the proton, GRS for the photon, 
with the recent analysis by Block and collaborators \cite{Block:2014kza}.  
\begin{figure}
\resizebox{0.5\textwidth}{!}{%
  \includegraphics{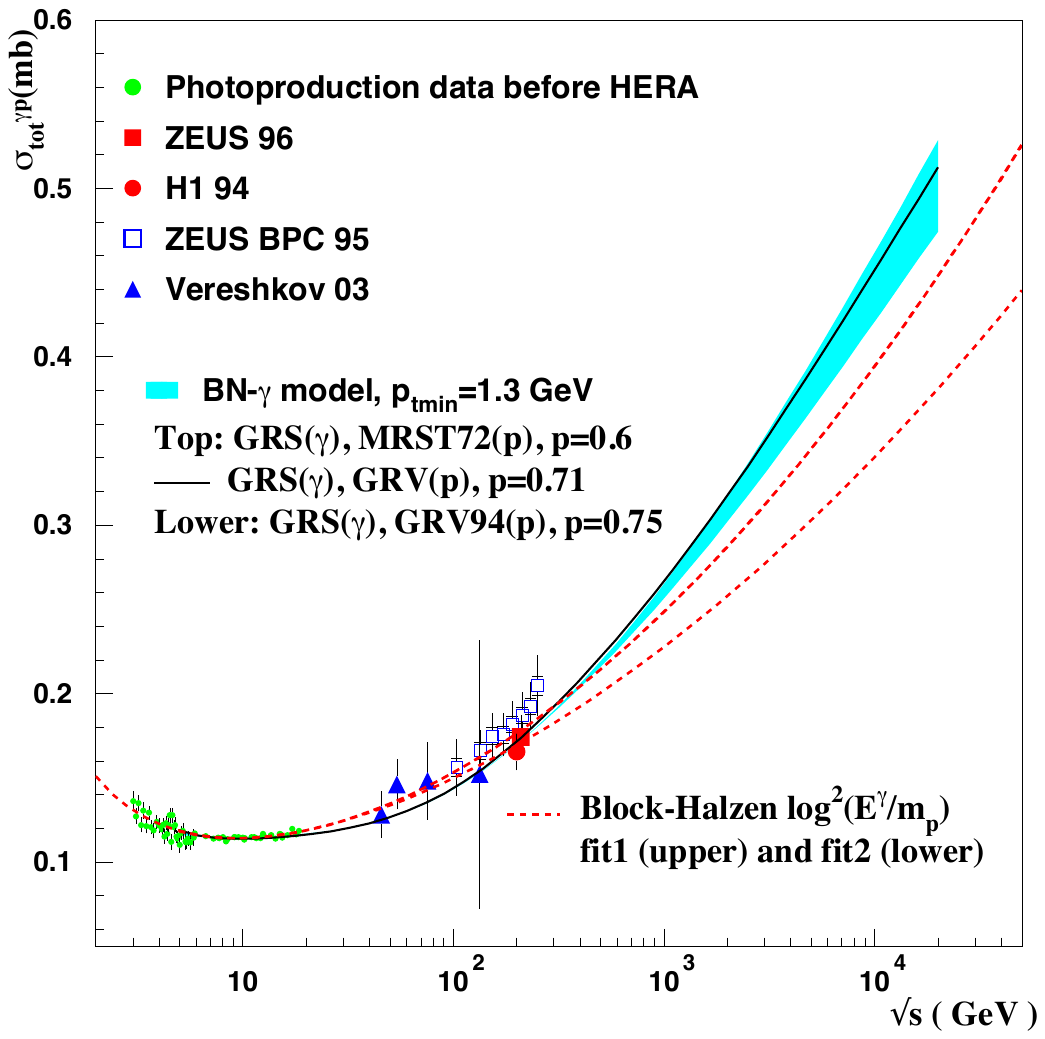}
 }
  \vspace{-2cm}
\caption{The total cross section $\sigtotgamp$ in mb vs. $\sqrt{s}$ in GeV, from  \cite{Cornet:2015qda}.
This is  Fig.(2) from  \cite{Cornet:2015qda}, \copyright (2015) by the American Physical Society.}
\label{fig:2014ourgamp-comparison}
\end{figure}
The band  in Fig.~\ref{fig:2014ourgamp-comparison} correspond to GRV or MRST densities for the proton. 
The difference with the previous analysis is not large, it depends , as mentioned by now many times, on the  small-x behavior of the densities used. 
From a comparison with accelerator data, we can say only that both curves can be used for cosmic ray extrapolations.
\subsection{$\sigma_{total}(\gamma p)$, and exclusive vector meson production $\sigma(\gamma p \to V p)$}
\label{ss:gampVector}
{In addition to the total $\gamma p$ and $\gamma^* p$ cross-section (which will be discussed in some detail in the next subsection) HERA has provided  interesting data on vector meson exclusive production.}

A compendium of total and exclusive vector meson photo-production data are shown as a function of $W$ in Fig. ~\ref{levy-0907-2178-fig1} from 
Levy's review of HERA experimental results \cite{Levy:2009gy}. Recently, this figure appears in  updated versions, as in   \cite{Caldwell:2016cmw,Favart:2015umi} and is of interest for proposals for future electron-positron colliders \cite{Caldwell:2016cmw}. There is only one variable here, the c.m. energy  $W,$  and the fits are made as a power law 
$\sigma(W) \sim W^\delta$. The parameter $\delta$ rises from $0.16$ for the $\sigma_{total}(\gamma p)$, with the mass 
of the produced vector meson to about $1.2$ for $\sigma(\gamma p \to \Upsilon(1S) p)$.
\begin{figure}
\resizebox{0.5\textwidth}{!}{
\includegraphics{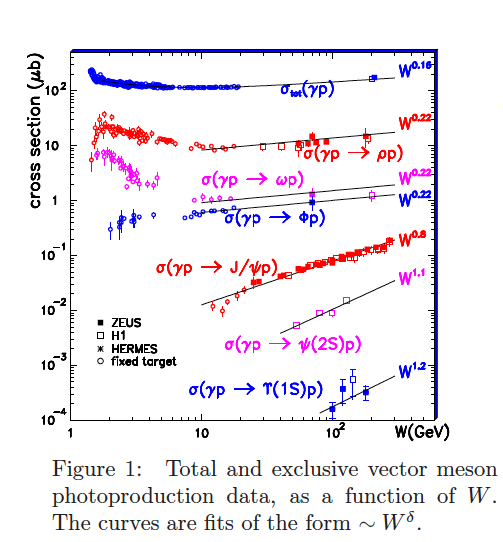}}
\caption{Total and exclusive vector meson photoproduction data, 
 from \cite{Levy:2009gy}. Reprinted with permission from \cite{Levy:2009gy}. OPEN ACCESS.}
\label{levy-0907-2178-fig1}
\end{figure}
In the Regge language, hadron-hadron total cross-section at a CM energy $W = \sqrt{s}$ should grow as $W^{2\epsilon}$,
where $\epsilon = (\alpha_P(0) -1)$ and $\alpha_P(0)$ is the intercept of the Pomeron at momentum transfer $t =0$. {For photo-production, the  value for $\epsilon$ follows the original}
Donnachie-Landshoff power law analysis, discussed at length earlier,  \cite{Donnachie:1992ny}, {i.e.$\epsilon=0.0808$. 
In \cite{Cudell:1996sh}, the value   $\epsilon \approx 0.096$ was  shown to reproduce well pp scattering, while the ZEUS data 
for $\gamma p$ can be fitted with $\epsilon=0.111$ in the HERA energy range, as seen in Fig. ~\ref{fig:heradl}. } 
Thus, $\delta = 2\epsilon \sim 0.192$,
not too far from {either} the HERA value of $\delta \sim\ 0.16$ for $\sigma_{total}(\gamma p)$ {or the ZEUS analysis. At the same time, these differences point to the fact that power law fits, albeit very useful for phenomenological analyses, are often dependent on the energy range and the type of  scattering process.} 

While $\delta = 2\epsilon$ for $\sigma_{total}(\gamma p)$, data show that also the photo-production of light-mass vector mesons 
($\rho^o, \omega, \phi$) are consistent with a soft process. In Levy's review of the data, it is stated that here too there is a large
configuration for the photon to fluctuate into a $q\bar{q}$ pair. On the other hand, as the mass of the vector meson increases the system
is led from the soft to the hard regime: the heavy quarks squeeze the photon into a smaller configuration leading to color screening and
the partonic structure of the proton is resolved. In the hard exclusive regime, the cross-section should be proportional 
to the square of the gluon density and hence there should be a strong dependence on $W$. This is clearly manifested by the HERA data, as discussed and summarized in \cite{Levy:2009gy}. 

\subsection{Electro-production of vector mesons, $\gamma^*p\rightarrow Vp$}
\label{ss:gamstarpVectors}
Virtuality of photons adds another variable $Q^2$ to $W$ that is lacking in purely hadronic cross-sections. HERA data show
interesting results that   can be found in  \cite{Chekanov:2007zr,Levy:2009ek,Levy:1986si}. Here we shall attempt a summary.
\subsubsection{Electro-production of $\rho^o$ meson}
In Fig.(13)  of \cite{Chekanov:2007zr},  reproduced  here as Fig.~\ref{fig:gamstar-rhop-ZEUS-2007},
 $\sigma(\gamma^* p \to \rho^o p)$    is shown as a function of $W$ for different values of $Q^2$. 
\begin{figure}
\resizebox{0.5\textwidth}{!}{
\includegraphics{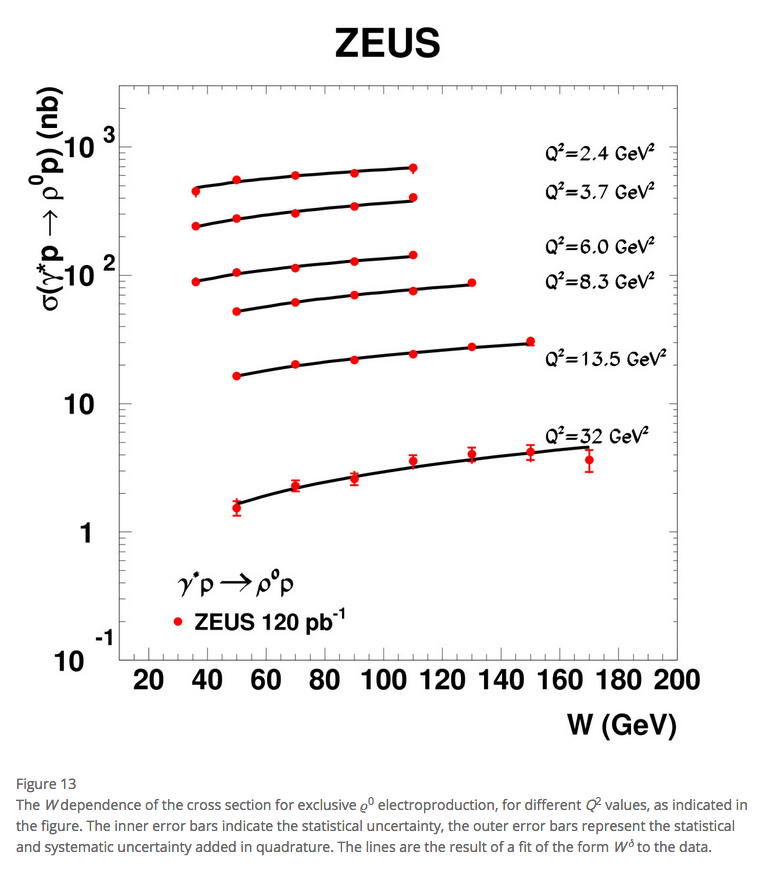}}
\caption{The cross-section for electroproduction of $\rho$-meson, as measured by the ZEUS Collaboration  in Fig. (13) from \cite{Chekanov:2007zr}, 
 as a function of the c.m. energy W, and for different $Q^2$ values. 
Reprinted with permission of  the ZEUS collaboration from \cite{Chekanov:2007zr} \copyright (2007) ZEUS Collaboration. }
\label{fig:gamstar-rhop-ZEUS-2007}
\end{figure}
The data are fitted to a power law $\delta$, which 
 rises from $(0.1\div 0.2$) for low $Q^2$,  as expected for soft processes, to about $0.6$ for large $Q^2$, consistent
with twice the logarithmic derivative of the gluon density, again as expected of a hard process. 

\subsubsection{Electro-production of heavier vector mesons and $\gamma^* p \to \gamma p$}\label{ss:electroproductionVM}

While the general trend of an increase in the cross-section with $Q^2$ is similar for $\phi$, $J/ \psi$ and for deeply-virtual 
Compton scattering $\gamma^* p \to \gamma p$, there is obviously an uncertainty in how to insert the mass $M$ of the produced
vector meson. Quite often, the variable $Q^2 + M^2$ in place of $Q^2$ has been used. In Fig.(8) of Levy, reproduced 
here in Fig.~\ref{fig:levy0907-2178-fig3}, a plot of $\delta$
versus $Q^2+M^2$ is shown.
\begin{figure}
\resizebox{0.5\textwidth}{!}{
\includegraphics{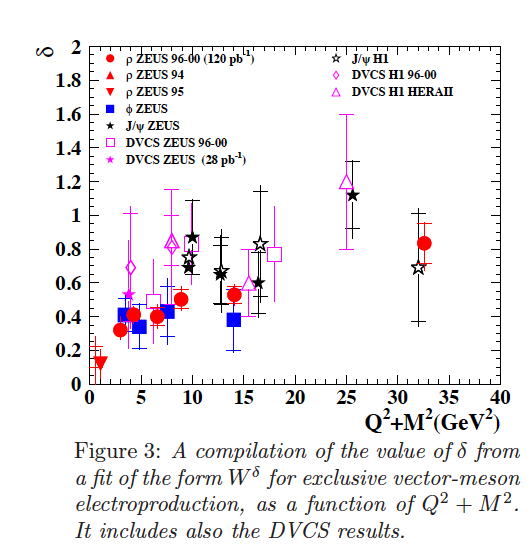}}
\caption{The $Q^2$ dependence of the energy slope for $\gamma^*p\rightarrow Vp$ cross-section, including heavy vector mesons 
electroproduction and Deeply Virtual Compton Scattering results, from \cite{Levy:2009gy} and references therein. Reprinted with permission 
from \cite{Levy:2009gy}, \copyright (2009) by  Science Wise Publ. }
\label{fig:levy0907-2178-fig3}
\end{figure}
 There is an approximate universality showing an increase in $\delta$ as the scale increases.
$\delta$ is found to be small at low scale, consistent with the intercept of a soft Pomeron whereas at larger scales it becomes
close to that expected from the square of the gluon density.  

Further studies to determine 
the best scale to use for vector meson electro-production, led to study the ratio 
$r_V=\sigma(\gamma^*p\rightarrow VP)/\sigma_{tot}(\gamma^* p)$ as a function of $W$. This  ratio can be parametrized 
following Regge arguments, in terms of a Pomeron exchange and of the  slope of the differential cross-section $d\sigma_V/dt$ as
\begin{equation}
r_V\sim W^{\lambda}/b
\end{equation}
More details about the scale dependence of the parameter $\lambda$ and its connections to the $\delta$ parameter 
can be found in \cite{Levy:2009ek}. 
Notice that this analysis  depends on the energy behavior  of $\sigma_{tot}(\gamma*p)$ to which we turn in the next subsection.

\subsection{Total $\gamma^*p$  cross-section}
\label{ss:gamstar-total}
At HERA, extensive measurements in the available phase space have brought a  detailed description of the c.m. energy 
dependence of the total $\gamma ^* p$ cross-section, in a range of values of $Q^2$.
These measurements highlight the transition from real photon scattering to Deep Inelastic Scattering (DIS) region, i.e. 
$0\le Q^2\lesssim 10000\ {\rm GeV}^2$. 
 A comprehensive description, up to $Q^2=2000\ {\rm GeV}^2$,  can be seen in 
 Fig.~ \ref{sigma-levy-Abramovicz} from 
 \cite{Abramowicz:1997ms}. The high energy data have been obtained at HERA, lower energies from  a number of 
 different experiments, and for which we refer the reader to the cited papers. 
\begin{figure}
\resizebox{0.5\textwidth}{!}{
\includegraphics{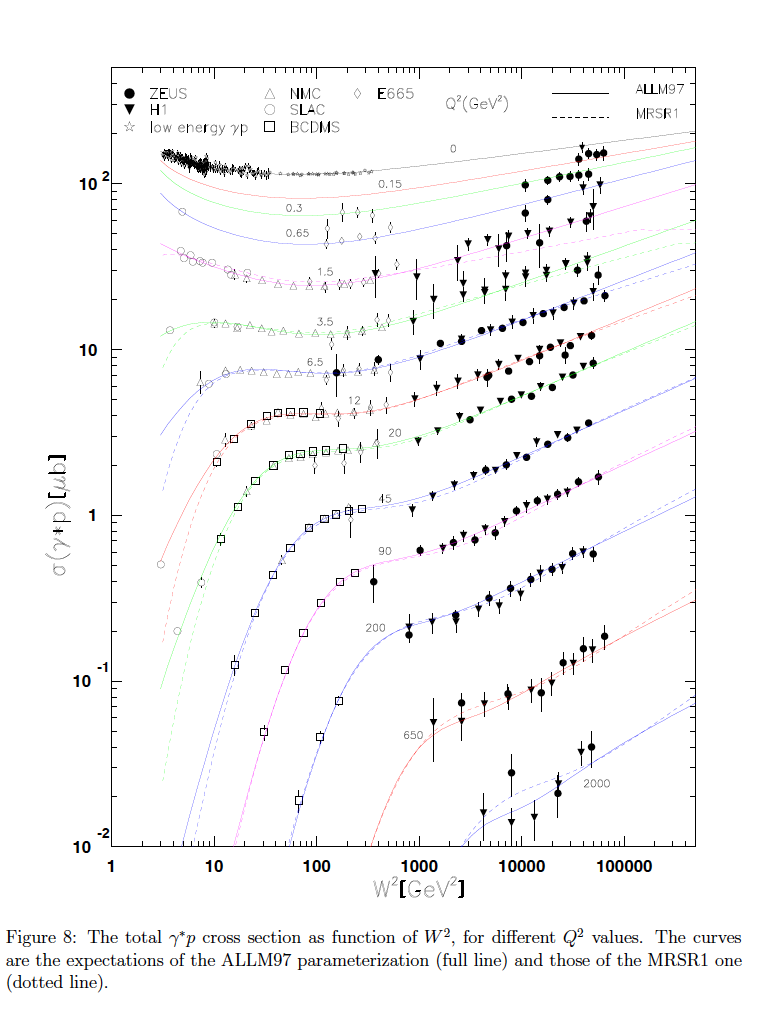}}
\caption{The virtual photon-proton total scattering cross-section for different $Q^2$ values of the virtual photon as a 
function of the squared c.m. energy $W^2$, from \cite{Abramowicz:1997ms}.  Reprinted with permission. }
\label{sigma-levy-Abramovicz}
\end{figure}
One can draw a few 
conclusions from this figure:
\begin{itemize}
\item for real and quasi real photons, the low energy behavior of $\sigma({\gamma^*p})$ exhibits the well known 
initial decrease with energy, followed by an apparent minimum and then a very mild rise. Thus, the  cross-section 
{would follow}  a standard  Donnachie-Landshoff parametrization,
\item 
as $Q^2$ increases beyond $ 20\ {\rm GeV}^2$,  the minimum   disappears and the cross-section  
is everywhere  increasing  with energy
albeit  with different slopes, and the increase with  $W^2$ 
 is steeper for larger   $Q^2$ values,
\item the change in curvature before the high energy rise, moves to higher $W^2$ values as $Q^2$ increases. 
\end{itemize}
In the figure, two different parametrizations are shown,  one described by a continuous line, and labeled as ALLM97, 
which we summarize here, while for the second one, MRSR1, we refer the reader to refs. \cite{Abramowicz:1997ms,Martin:1996as}. 
The ALLM \cite{Abramowicz:1991xz} parametrization {(of which ALLM97 represents an update)} 
describes the proton structure function following the usual split into a Regge and a Pomeron type term, i.e.   
\begin{equation}
F_2(x,Q^2)=\frac{Q^2}{Q^2+m_0^2}\large(      F_2^{\cal P}(x,Q^2)+F_2^{\cal R}(x,Q^2) \large)\\
\end{equation}
with $F_2^{\cal {P},\cal {R}}(x,Q^2)$ a function of a slowly varying variable defined as
\begin{equation}
t=\ln {\large ( }
\frac{\ln \frac{Q^2+Q_0^2}{\Lambda^2}} 
{\ln\frac{Q_0^2}{\Lambda^2}} 
{ \large)}
\end{equation}
The $F_2$ data were then conveyed to $\sigma_{tot}(\gamma^*p)$ using 
\begin{equation}
\sigma_{tot}(\gamma^*p)=\frac{4 \pi^2 \alpha}{Q^2(1-x)}
\frac{Q^2+4M^2x^2}{Q^2}
F_2(W^2,Q^2)
\end{equation}
where $M$ here  is the proton mass.
The ALLM 
$F_2$ is based on 23 parameters, which were updated from pre-Hera data to the nice description shown in Fig. \ref{sigma-levy-Abramovicz}.

The $\gamma^*p$ HERA data have also been studied in terms of Vector Meson Dominance or Color Dipole Picture, 
as shown and discussed in \cite{Schildknecht:2000zt}.


A more recent analysis of HERA data from \cite{Wolf:2009jm} is shown in Fig. ~\ref{gunther-wolf-fig11}, with  the virtual photon cross-section
 \begin{equation}
\sigma^{tot}_{\gamma^*p}=\frac{4\pi^2 \alpha}{Q^2(1-x)}F_2(x,Q^2)
\end{equation}
valid for $4 m_p^2 x^2<< Q^2$, fitted with $F_2(x,Q^2)$ from HERA parametrized according to a  power law, i.e.
\begin{align}
\log_{10}F_2(x,Q^2)=c_1+c_2\cdot \log_{10}(x)+\nonumber \\
+c_3 \cdot \log_{10}(x) \cdot \log_{10}(Q^2/Q_0^2)  +\nonumber \\
+c_4\cdot \log_{10}(x) \cdot ( \log_{10}(Q^2/Q_0^2))^2
\end{align}
\begin{figure}
\resizebox{0.5\textwidth}{!}{
\includegraphics{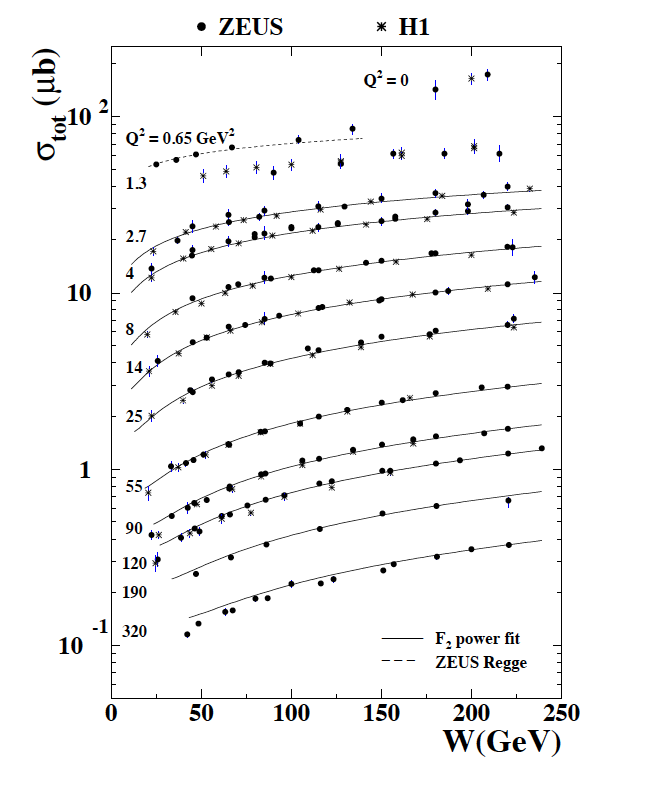}}
\caption{H1 and ZEUS data for real and virtual photon scattering from \cite{Wolf:2009jm}. This is Fig. (11) from  \cite{Wolf:2009jm},  
 \copyright (2010) by IOP, reprinted with permission.}
\label{gunther-wolf-fig11}
\end{figure}
Before moving to briefly discuss two photon processes, we point out that the energy range for photo and 
electroproduction at HERA is limited to values  still far from the asymptotic regime, where purely hadronic cross-sections are 
expected to exhibit a logarithmic  behavior. Thus, the question of a power-law 
vs. a logarithmic 
behaviour is still open, where photon processes are involved.

\subsection{ $\gamma \gamma$ scattering}
\label{ss:gamgam}

Photon-photon scattering  was measured in electron-positron collisions from the very beginning of storage ring colliders 
\cite{Bonolis:2015tos}.  As the available beam energy increased,  data for $\gamma \gamma \rightarrow hadrons$ became 
available.  A compilation of data for the cross-section into hadrons is shown in Fig.~\ref{fig:gamgamdata} from \cite{Godbole:2012jq},
 starting from $\sqrt{s_{\gamma \gamma}} = 1.4 \ {\rm GeV}$ at SPEAR up to LEP measurements, reaching $\sqrt{s}=189\ {\rm GeV}$. 
\begin{figure}[htb]
\resizebox{0.5\textwidth}{!}{
\includegraphics{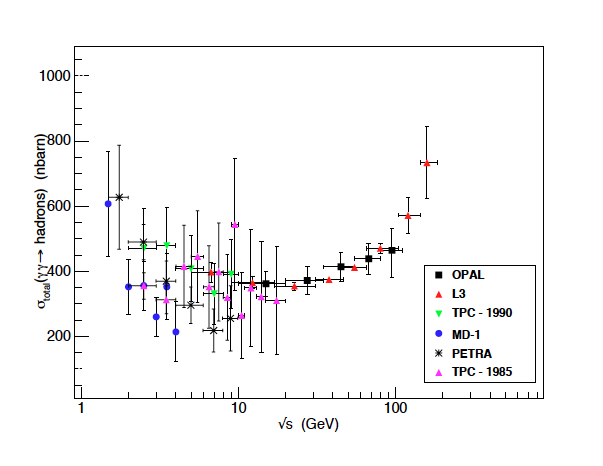}}
\vspace{-1.cm}
\caption{Data for $\gamma \gamma$ total cross-section from \cite{Godbole:2012jq} and bibliographic references therein. Reprinted with permission, \copyright (2012) INFN Frascati Physics Series.  }
\label{fig:gamgamdata}
\end{figure}
The figure indicates that the  trend of the data as a function of the two photon c.m. energy  is 
{consistent with} a  hadronic process, namely it starts with  the usual initial decrease followed by a rise. 
Such behavior is easily obtained in factorization models. 

Indeed 
there are various models which describe photon-hadron scattering through various forms of factorization,  
which would then allow to obtain $\sigtotgamgam$, through the simple statement
\begin{equation}
\sigtotgamgam=\frac{(\sigtotgamp)^2}{\sigtotpn}
\label{eq:bsw}
\end{equation}
where $\sigtotpn$ indicates some combination of $pp$ and ${\bar p}p$ total \x s. 
This is the case of the model by Soffer and collaborators \cite{Bourrely:1999hu}, which follows from their description of $\gamma p$ 
total cross-section effects in \cite{Bourrely:1994bc}.

The Bourelly, Soffer and T.T. Wu ansatz \cite{Bourrely:1994bc} is that $\gamma p $ total \x \ can be obtained from $\pi p$ as
\begin{equation}
\sigma_{tot}(\gamma p)=\frac{1}{3} \alpha\left(\sigma_{tot}(\pi^+p) + \sigma_{tot}(\pi^-p) \right).
\end{equation}
For $\sigma_{tot}(\pi^{\pm}p) $ the authors use,  an early impact picture prediction where a simple 
power-law dependence $s^{0.08}$ was first given.   
From this simple model, one could obtain $\sigtotgamgam$ and compare it with data for  $\gamma \gamma$ extracted  from 
LEP. Through  Eq.~(\ref{eq:bsw}) and  their earlier fit to proton-proton, the authors  \cite{Bourrely:1999hu} obtain the  
results shown in Fig.~\ref{fig:bswgamgam} . 
\begin{figure}
\resizebox{0.5\textwidth}{!}{
\includegraphics{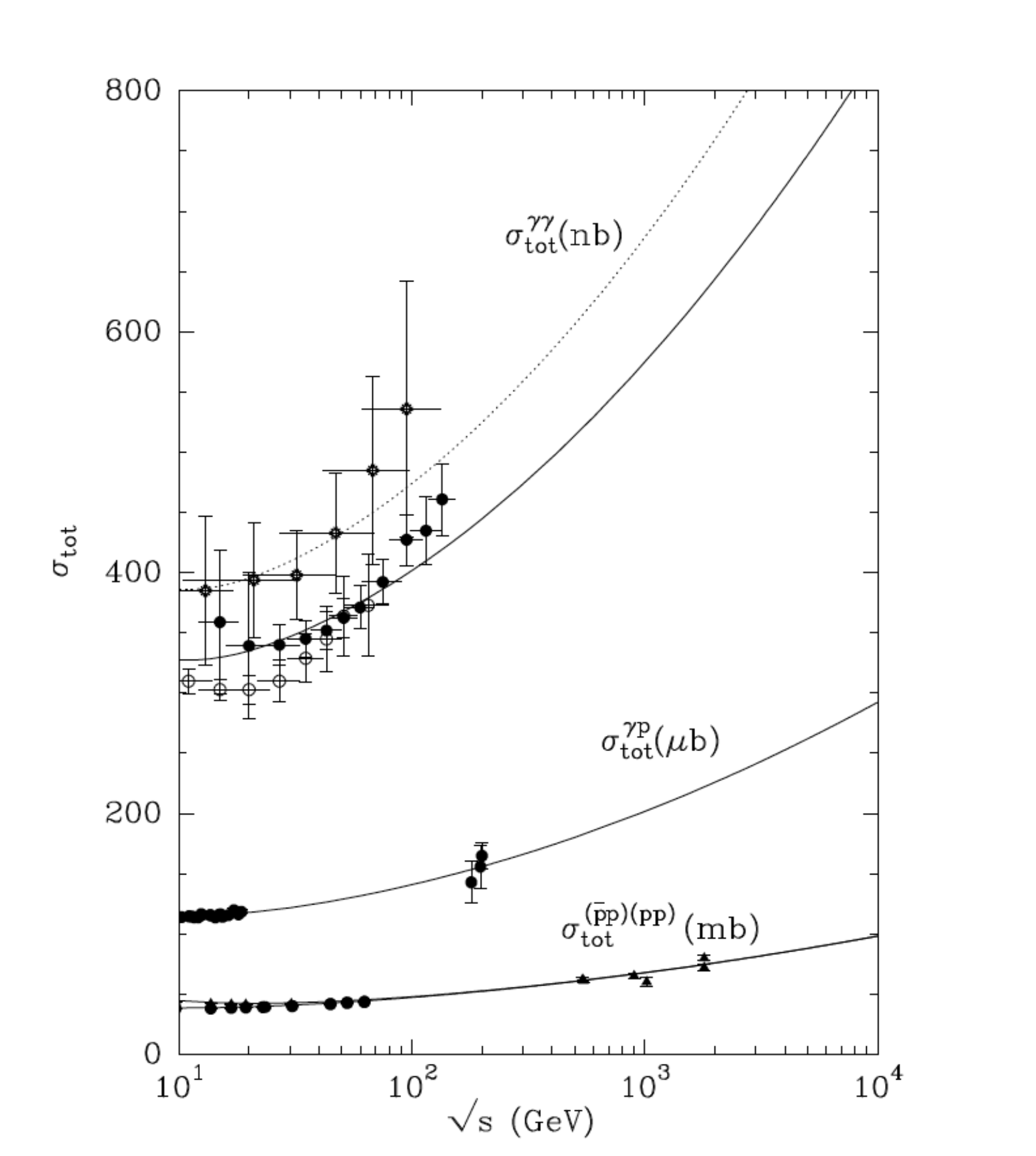}}
\caption{ Description, and comparison with data, of the $\gamma \gamma$ total cross-section in the model by 
Bourrely, Soffer and T.T. Wu, fig. 1 of  \cite{Bourrely:1999hu}. Predictions are also shown, and compared with  
then existing data, for \pp /\pbarp \ and $\gamma p$. Reprinted with permission from \cite{Bourrely:1999hu} 
\copyright (1999) by World Scientific.}
\label{fig:bswgamgam}
\end{figure}

While straightforward factorization models can give a general good description of data up to LEP most recent measurements,    
the limited energy range and the large errors affecting the extrapolation to  full phase-space both at lower and at the highest energies, 
do not provide enough information to  distinguish between  standard power law energy dependence, such as Regge-Pomeron 
exchanges, mini-jets, or QCD driven exchanges.  Such distinction, as is also the case for $\gamma p$,  is left to    future  
colliders or perhaps to LHC. A 2003 compilation of a selection of  models 
 is shown in Fig.~\ref{fig:ourgamgam-0305071} from \cite{Godbole:2003wj}, 
  where the bibliographic references can  be found. Details of models can be found in \cite{Godbole:2003hq}.
\begin{figure}
\resizebox{0.47\textwidth}{!}{
\includegraphics{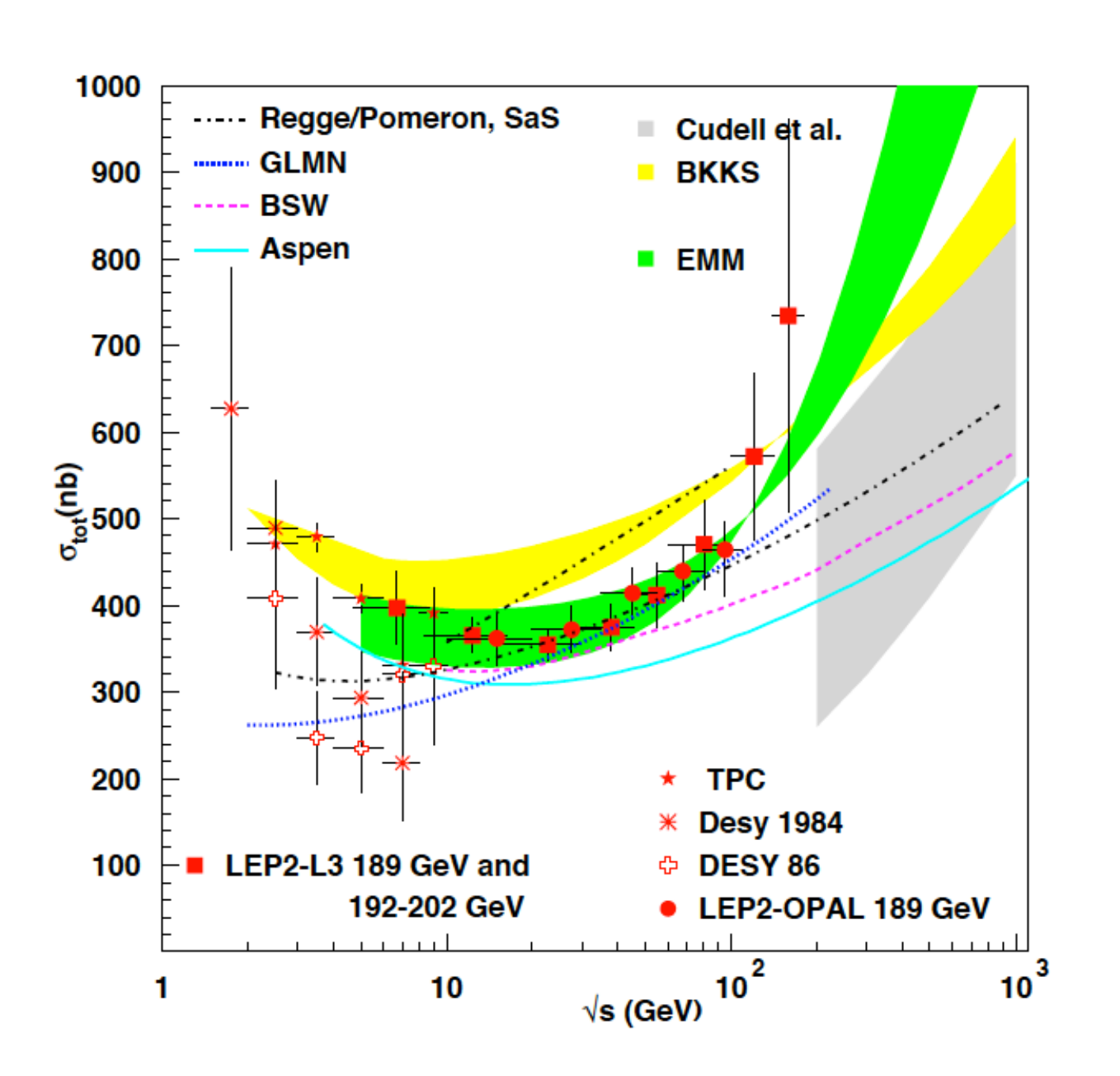}}
\caption{Data and models for $\gamma \gamma \rightarrow hadrons$ from \cite{Godbole:2003wj}.
Reprinted from \cite{Godbole:2003wj} \copyright (2003) by Springer.}
\label{fig:ourgamgam-0305071}
\end{figure}
{
A general fit  to  the LEP data alone, i.e. }
\be
\sigma_{\gamma \gamma}=A s^\epsilon_{\gamma \gamma}+Bs^{-\eta}_{\gamma \gamma} \label{eq:deroeck}
\ee
 {was done in  \cite{Godbole:2003wj} and is shown in Fig.~\ref{fig:deroeckgamgam2003}. }
 
\begin{figure}
\resizebox{0.5\textwidth}{!}{
\includegraphics{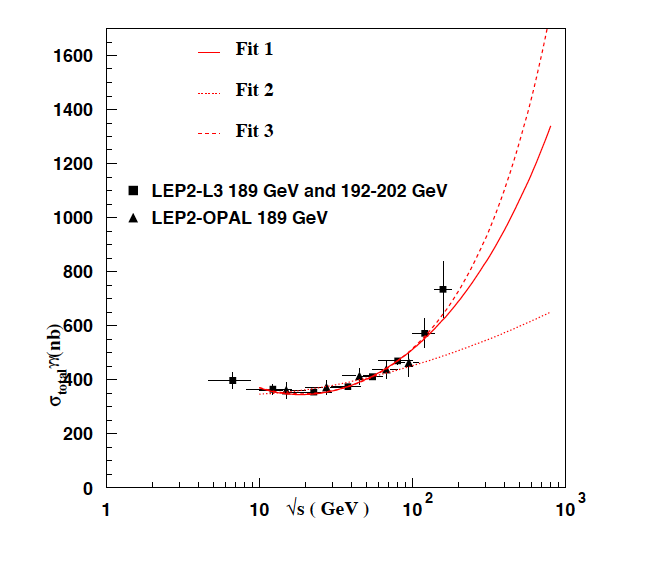}}
\caption{{Data and  power law type  fits to LEP data for $\gamma \gamma$ total cross-section from \cite{Godbole:2003wj}. 
In fit 1, all the parameters of  Eq.~(\ref{eq:deroeck}) are free, in  fit 2, $\epsilon=0.093$, in fit 3 a second Pomeron type term 
is added to Eq.~(\ref{eq:deroeck}) with $ \epsilon_1=0.418$. Reprinted from \cite{Godbole:2003wj} \copyright (2003) with permission by
SISSA.}}
\label{fig:deroeckgamgam2003}
\end{figure}

\subsection{$\gamma^* \gamma^* \to hadrons$}
\label{ss:gamstargamstar}
At LEP, through the measurements of $e^- e^+ \to e^- e^+ + hadrons$, $\sigma(\gamma^* \gamma^* \to hadrons)$ have been
measured by the L3 \cite{Acciarri:1998ix} and OPAL Collaborations \cite{Abbiendi:2001tv}. The kinematics of 
such processes is shown in Fig.~\ref{fig:gamstar}, (fig. (1)  from OPAL).
\begin{figure}
\resizebox{0.5\textwidth}{!}{
\includegraphics{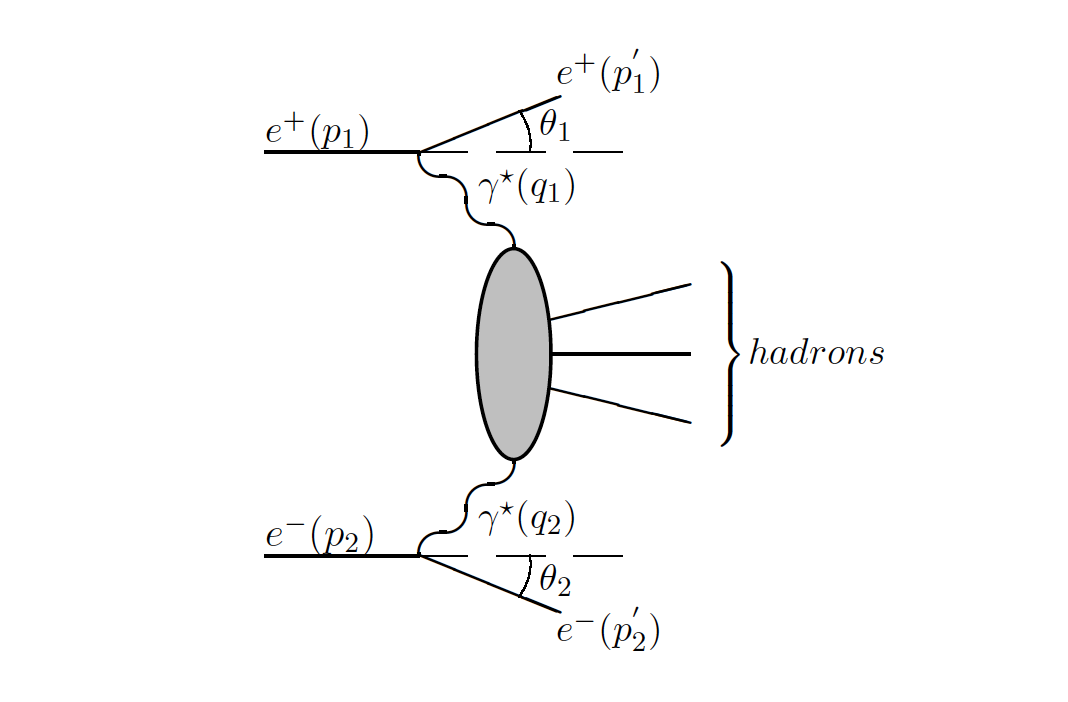}}
\caption{Kinematics of $e^+e^-\rightarrow \gamma^* \gamma^* e^+e^-$, as from Fig.
(1) of \cite{Abbiendi:2001tv}, OPAL Collaboration. Reprinted  with
permission \copyright (2001) by  SPRINGER.}
\label{fig:gamstar}
\end{figure}
{Measurements for the above process for one untagged electron were made  by the ALEPH
Collaboration \cite{Heister:2003an}, who extracted the so-called photon structure functions $F_{2,L}^\gamma$. 
In this case one photon is almost real, and the process is studied as function of a single $Q^2$ value, extracted from   the tagged electron.}

 For the determination of $\gamma^*\gamma^*$ cross-section,
both scattered $e^-$ and $e^+$ have to be tagged at sufficiently large polar angles $\theta_i$, to be observed in the detector. The 
kinematical variables for the process are as follows.
\begin{itemize}
\item the $e^-e^+$ CM energy squared is $s_{e^-e^+} = (p_1 + p_2)^2$
\item the virtualities of the scattered photons are given by $Q_i^2 = (p_i - p_i^{'})^2$;
\item the usual variables of deep inelastic scattering are defined as
\begin{eqnarray}
\label{g^*g^*1}
y_1 = \big{(}\frac{q_1\cdot q_2}{p_1\cdot q_2}\big{)};\ y_2 = \big{(}\frac{q_1\cdot q_2}{p_2\cdot q_1}\big{)};\nonumber\\
x_1 = \big{(}\frac{Q_1^2}{2 q_1\cdot q_2}\big{)};\ x_2 = \big{(}\frac{Q_2^2}{2 q_1\cdot q_2}\big{)};
\end{eqnarray}
\item the hadronic invariant mass squared is $W^2 = (q_1 + q_2)^2$;
\item the Bjorken variables $x_i$ are related to $Q_i^2$ and $W^2$ as
\begin{equation}
\label{g^*g^*2}
x_i = \frac{Q_i^2}{[Q_1^2 + Q_2^2 + W^2]}.
\end{equation}
\end{itemize}
{For comparison with models, an additional variable which incorporates the $W^2$ and $Q_i^2$ dependence, is defined, i.e.}
\be
{\bar Y}=\ln (\frac{W^2}{\sqrt{Q_1^2 Q_2^2}})
\ee
Of course, given three
helicities for each virtual photon with different $Q_i^2$, there are a plethora of physical 
quantities (six in number) that can be measured (4 cross-sections $\sigma_{TT}; \sigma_{TL}; \sigma_{LT}; \sigma_{LL}$) 
and two interference terms   ($\tau_{TT}; \tau_{TL}$), where $T,L$ stand for transverse or longitudinal. Detailed expressions 
for these quantities and discussions can be found in two Phys. Rep.\cite{Budnev:1974de} and \cite{Nisius:1999cv}.

Here, we shall just comment upon salient aspects of the two determinations at LEP of the total 
$\sigma(\gamma^* \gamma^* \to hadrons)$, the ``cleanest'' quantity that can be measured and compared to models. 
Both OPAL and L3 data were taken at 
$\sqrt{s_{e\bar{e}}} = (189\div 209)\ {\rm GeV}$, with similar hadronic mass $W > 5\ {\rm GeV}$ and mean $<Q^2> \sim 18\ {\rm GeV}^2$ ranges. 
{Fig.~\ref{fig:OPAL2002-fig8b} shows the OPAL extracted $\sigma^{\gamma^*\gamma^*}$ as 
a function of the virtual photon c.m. energy W, compared with predictions from PHOJET (solid lines) and a 
Quark Parton Model (QPM) (green dotted lines). }
\begin{figure}
\resizebox{0.5\textwidth}{!}{
\includegraphics{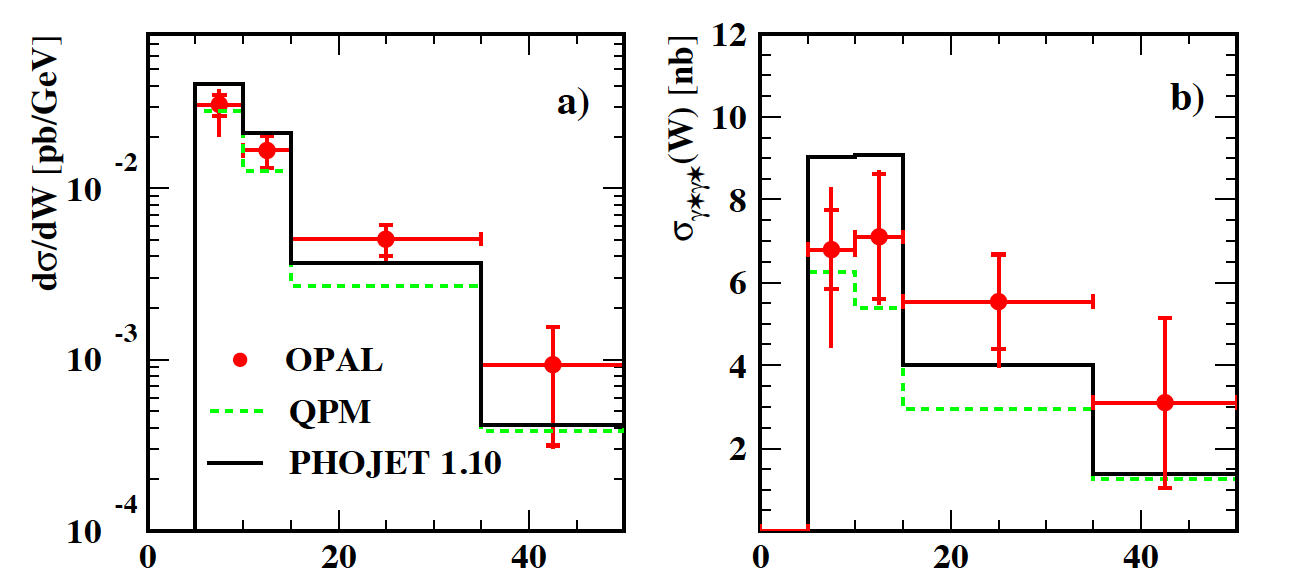}}
\caption{The above Figs. 8a and 8b  from \cite{Abbiendi:2001tv}) of the OPAL Collaboration shows the cross-section for 
$\gamma^*\gamma^*\rightarrow hadrons$ for an average value of the photon's squared momentum $<Q^2>=17.9\ {\rm GeV}^2$.
Reprinted with permission  from \cite{Abbiendi:2001tv} \copyright (2001) by Springer.}
\label{fig:OPAL2002-fig8b}
\end{figure}
The Dual Parton Model (DPM) \cite{Capella:1992yb} is  beneath the PHOJET \cite{Engel:1994vs,Engel:1995yda}  event generator
(PHOJET1.10) used to simulate double-tagged events and obtain the total luminosity $L_{TT}$, through which  the two 
LEP measurements construct $\sigma(\gamma^* \gamma^* \to hadrons)$. DPM contains both hard and soft processes. Hard
processes are incorporated via LO QCD, and soft processes are included  through a phenomenological analysis of $\gamma p$, $pp, p\bar{p}$
data assuming Regge factorization.

{The comparison with QCD models can be seen from the analysis by the L3 collaboration, which we show in 
Fig.~\ref{fig:L3-fig6} from \cite{Acciarri:1998ix}.}
\begin{figure}[htb]
\resizebox{0.5\textwidth}{!}{
\includegraphics{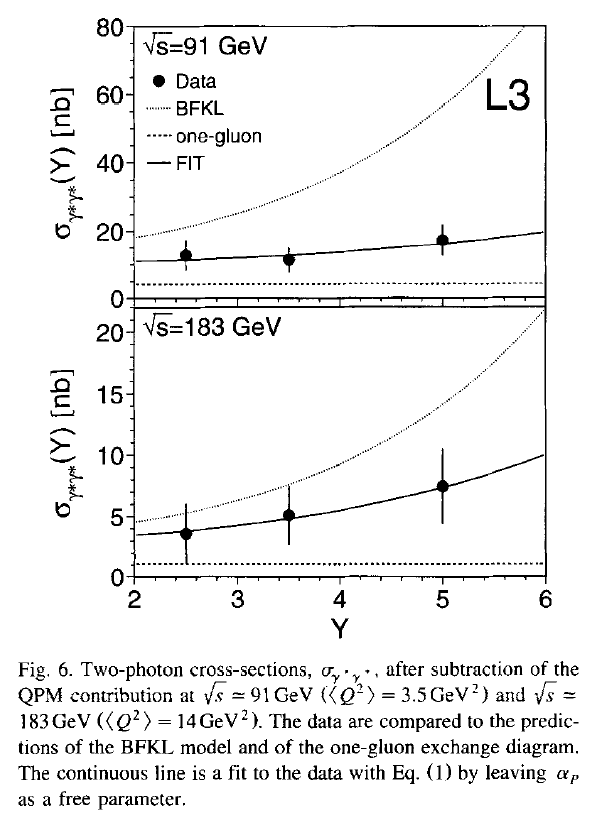}}
\caption{{The above (Fig. 6 from \cite{Acciarri:1998ix}) shows the cross-section measured by the L3 Collaboration  for 
$\gamma^*\gamma^*\rightarrow hadrons$ as a function of the variable Y, for two different $\sqrt{s_{ee}} $ values, 
with corresponding average photon squared momentum, as indicated in the original figure caption. }
Reprinted from \cite{Acciarri:1998ix}, \copyright (1999) with permission by Elsevier.}
\label{fig:L3-fig6}
\end{figure}
{This figure shows that lowest order BFKL predictions for $\sigma(\gamma^* \gamma^* \to hadrons)$ were rather large by a factor of
about 20 or more.}
Subsequent phenomenological results from Next-to-Leading Order
(NLO) have reduced this discrepancy by an order of magnitude. It appears that with theoretical improvements suggested
in \cite{Kwiecinski:1999yx,Boonekamp:1998ve,Altarelli:2000mh,Ciafaloni:1999yw,Brodsky:1998kn}, 
 the BFKL formalism can be reconciled with the two LEP measurements. 
 However, no definite  assessment can be given at present, short of higher energy data becoming available, 
 as also discussed in \cite{Bartels:2004mu}, where further contributions from 
secondary Reggeon exchanges in QCD have been considered.
An interested reader can find further description in the previous references.

\subsection{Conclusions}
Models for photon scattering probe yet another aspect of the total hadronic cross-section, but the absence of data at very high energy,  
for instance $\sqrt{s}\ge 200 \ {\rm GeV}$ into the TeV region,  does not allow for precise tests of model predictions at high energies, 
such as those probed for instance in cosmic ray experiments. The transition from real to virtual photons and from photons to hadrons 
are still rather model dependent. Planned future  measurements, perhaps at LHC, or at future $ep$ or  $e^+e^-$ colliders, would shed 
further light on these transitions in the future.

\section{
LHC program for near forward physics
}
 \label{sec:lhcnow}
 In this section, we  describe the total   and small angle  \x \ measurements  that were programmed to be done at LHC. 
 
Our presentation of  the measurements  at LHC follows the extensive documentation  prepared before the start of  LHC 
\cite{Anelli:2008zza,Aamodt:2008zz,Aad:2008zzm,:2008zzk,Alves:2008zz,Adriani:2008zz}. Since then the LHC has started functioning and a wealth of results has appeared and   updates of the LHC program are planned. 
 An early comprehensive update about various planned experiments can be found in the proceedings of the Blois Workshop, held at CERN in 
 June 2009 \cite{Deile:2010mv}, as well as in presentations at DIFF2010  at Trento Workshop, ECT*. 
 A  recent  review of measurements by TOTEM and ATLAS experiment can be found in \cite{Grafstrom:2015ypa}. 
 For updates as of September 2016, 
  an extensive set of presentations  was done at   the ECT* 2016 Workshop entitled   {\it Forward 
 physics WG: diffraction and heavy ions},  with slides available at :\\
 https://indico.cern.ch/event/568781/timetable/\#al .}\\
 
The experiment dedicated from the outset to measure the total \x \ is the TOTEM experiment \cite{Anelli:2008zza}, 
but other measurements relevant to physics in the forward region have been and will continue to be performed by  
all the LHC experiments: ALICE \cite{Aamodt:2008zz}, ATLAS \cite{Aad:2008zzm}, CMS \cite{:2008zzk}, 
LHC-b \cite{Alves:2008zz} and LHCf \cite{Adriani:2008zz}.
In addition, these experiments have  been providing data  about  the inelastic cross-section, a component of $\sigtot$ \ crucial for a full understanding of the dynamics entering both \pp \ and cosmic ray data. 
Recent results concerning the inelastic total cross-section at the present LHC energy of $\sqrt{s}= 13$ TeV, can be found in  \cite{VanHaevermaet:2016gnh} for  CMS and in  \cite{Aaboud:2016mmw} for ATLAS. 

Various 
experiments 
study particle flows and diffractive physics through a number of  detectors placed at various distances along the 
beam directions, with different physics goals. We show in Fig.~\ref{fig:lhclayout} a schematic drawing of the positions 
of the main experiments around the LHC ring at various Interaction Points (IPs).
\begin{figure}
\begin{center}
\resizebox{0.6 \textwidth}{!}{
\includegraphics{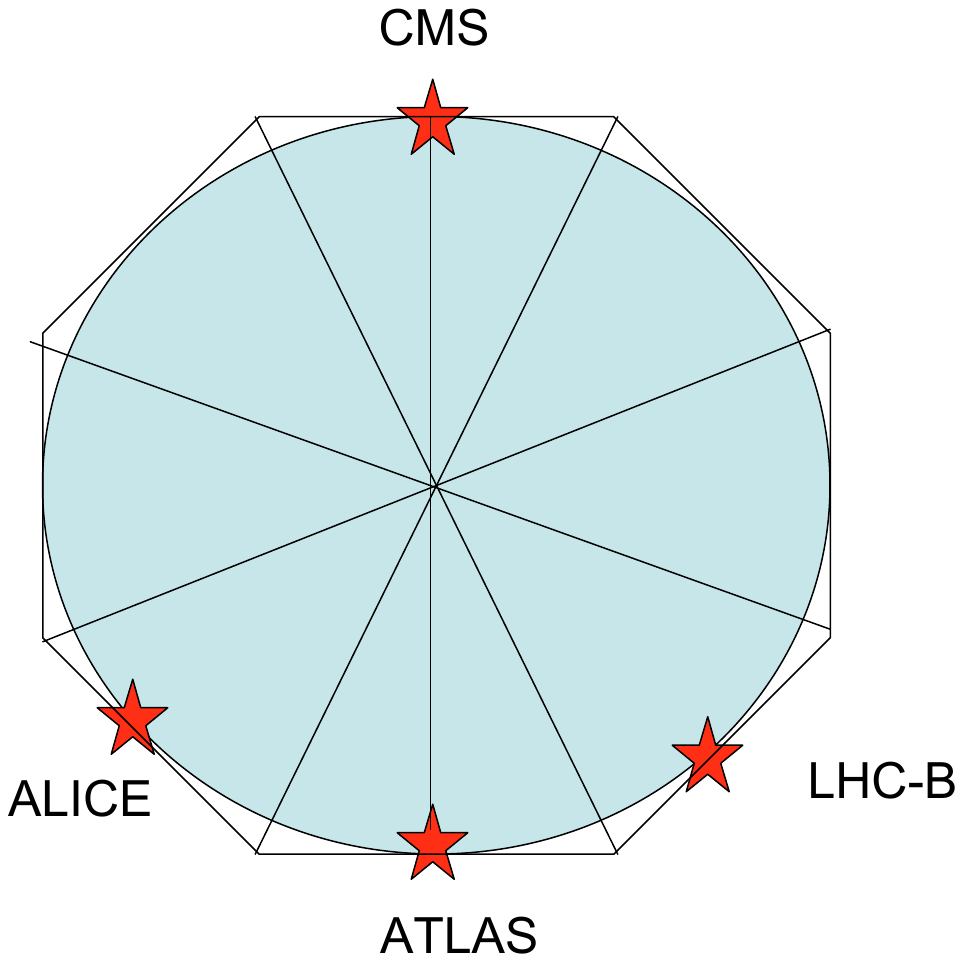}}
\caption{Schematic description of  location of major experimental sites around the LHC ring, ATLAS is at IP1 and CMS at IP5.}
\label{fig:lhclayout}
\end{center}
\end{figure}

 At LHC the phase space range extends to 11 units in rapidity, since in the variable  $y_{max}=\ln\frac{\sqrt{s}}{m}\approx 9.6$. 
 In the variable pseudo-rapidity, $\eta= - \ln \tan\frac{\theta}{2}$, where $\theta$ is the scattering angle of the detected particles, 
 the coverage goes up to 12 or 13 units. The main CMS and ATLAS calorimeters measure energy deposited in the rapidity range 
$|\eta|< 5$, with particle detection and identification  to be performed by the Electromagnetic and Hadronic calorimeters for  
$0<|\eta|<3$ , and  the hadronic  forward (HF for CMS and FCal for ATLAS) calorimeters for $3<|\eta|<5.2$. In this region,
 data can also be collected by ALICE and LHC-B. For forward physics at LHC-B,  see D'Enterria \cite{d'Enterria:2009fa}. 
 The forward calorimeters, however cover only part of the forward region.  With most of the energy deposited in the region 
 $8<|\eta|<9$, other calorimeters are needed and placed  near the beam. In this region, there 
 is the LHCf experiment  measuring particle flows, and the Zero Degree Calorimeters (ZDC)  measuring  neutral particles, 
 while the extreme rapidity region, beyond $|\eta|=9$ will be covered by Roman Pots (RP), with TOTEM in  CMS and ALFA in ATLAS.  

Let us now look 
in more detail to particle detection in  the forward region and to the system of detectors covering the rapidity region 
$|\eta|\simeq 3\div 7$. Up to a distance from the Interaction point (IP) of $(10\div20)\ m$, as we show schematically 
in Fig.~\ref{fig:forward-tracking} the strategy is to surround the beam pipe with tracking calorimeters, as follows:
\begin{description}
\item [ATLAS] with  MBTS, Minimum Bias Trigger Scintillator, at 3.6 m from the interaction point, a 
Hadronic Forward (FCal) calorimeter covering the region $3.1<|\eta|<4.9$ and LUCID,  Luminosity Cerenkov Integrating Detector,   
a luminosity monitor at 17 m; 
\item [CMS] with the Hadronic Forward (HF) calorimeter  placed at 11.2 m from the interaction point, covering the rapidity region 
$|\eta|<5.2 $ (inner part for the region $4.5<|\eta|<5.0$), followed by  CASTOR, Centauro And STrange Objects Research,   
which detects energy flows and  is a Cerenkov calorimeter surrounding the beam pipe $(15\div16.5)\ m$ from the interaction point, 
covering the range $5.2<|\eta|<6.6$ and dedicated to the observation of cascade developments;
\item [TOTEM]  with the two tracking detectors $T_1$ and $T_2$  which cover the region ($3.1\le|\eta|\le 6.5$).
\end{description}

\begin{figure} 
\begin{center}
\resizebox{0.6\textwidth}{!}{%
  \includegraphics{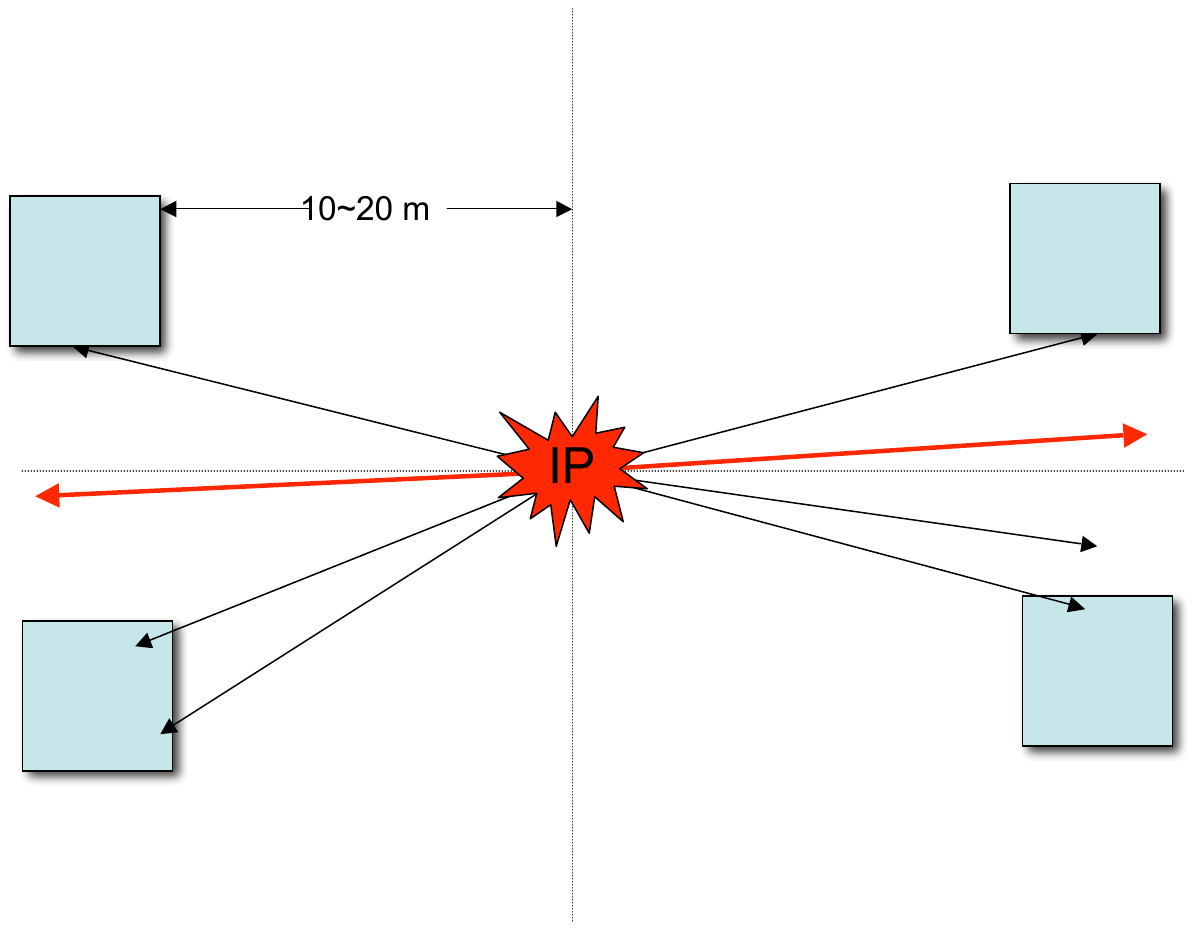}}
\caption{Forward tracking at LHC through calorimeters placed at a distance of $(\sim 10\div 20)\ m$ from the interaction Point (IP).}
\label{fig:forward-tracking}
\end{center}
\end{figure}
These detectors are placed in such a way that it is easy to miss particles scattered in the  very forward direction and they 
are implemented by dedicated set ups like Zero Degree Calorimeters and Roman Pots (RP). The Zero Degree calorimeters 
are placed at 140 meters from the interaction point and cover the rapidity range $|\eta|\ge 8.3$. The ZDC's 
are for the detection of neutral particles such as neutrons, photons and $\pi^0$ and are especially designed for heavy ions 
and diffractive physics. At a distance of 240 meters from the IP3, 
there is ALFA,   Absolute Luminosity For ATLAS,  with Roman Pots, to be placed at an angle  from the beam pipe of 
$3 \ \mu rad$.  In IP5, in the  CMS region, after T1, T2, CASTOR and   ZDC, there is TOTEM with the Roman Pots. 
At even  longer distances, the High Precision Spectrometers at 420 m \cite{Albrow:2008pn}
dedicated to forward Higgs studies \cite{Khoze:2002py,DeRoeck:2008zzg}.

\subsection{The CMS region and cross-section measurements}\label{ ss:CMS}
In Fig.~\ref{fig:layoutCMSforward} we show a schematic view of the layout of various forward physics detectors in and around CMS. 
A similar layout is found also in the ATLAS region.
\begin{figure*} 
\begin{center}
\resizebox{0.8\textwidth}{!}{%
  \includegraphics{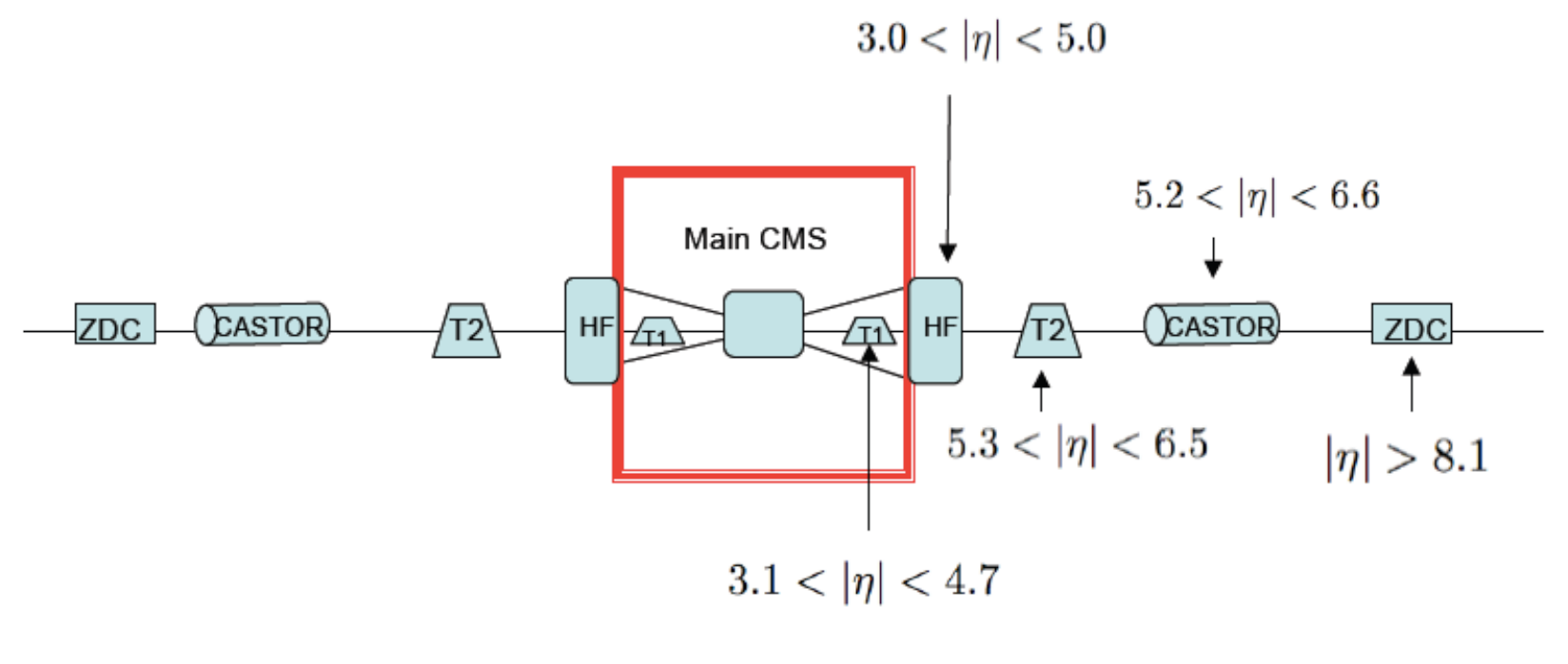}
}
\caption{Schematic view of forward physics detectors in CMS, with T1 and T2 from the TOTEM experiment. }
\label{fig:layoutCMSforward}       
\end{center}
\end{figure*}

We also show  a pictorial  view of the full set up of forward physics detectors in the CMS region in Fig.~\ref{fig:cmsforwardtotem}.
\begin{figure} 
\begin{center}
\resizebox{0.5\textwidth}{!}{%
  \includegraphics{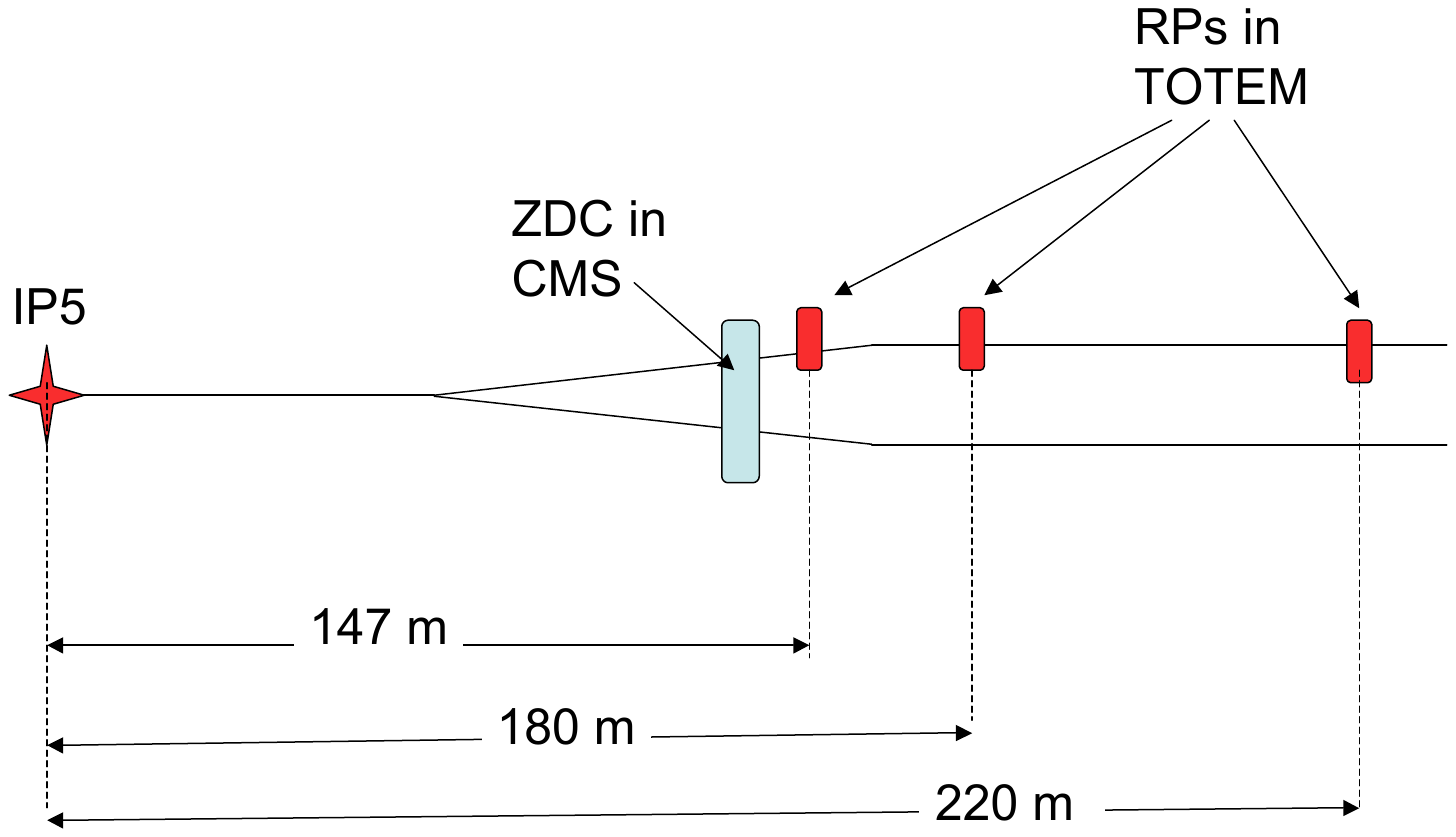}
}
\caption{Interplay between forward detectors between CMS ZDC and TOTEM Roman Pots. Shown are distances in metres.} 
\label{fig:cmsforwardtotem}       
\end{center}
\end{figure}


In the following we shall describe in more detail  the forward physics and experimental layout of interest for total \x \  
and other forward physics measurements. We shall focus on  TOTEM and ZDC, the two experiments and detectors  
in the CMS region where very forward scattering angles  
can be measured and total cross-sections extracted.

\subsubsection{TOTEM}\label{sss:TOTEM}
As stated earlier, TOTEM is the experiment  dedicated to the measurement of the total cross-section \cite{Anelli:2008zza}. 
It is based on the luminosity independent method, which uses both the measurement of the elastic scattering rate at the 
optical point, $t=0$, or as close as possible to it, as well as a measurement of the entire elastic and inelastic events rate 
through the two equations 
\begin{eqnarray}
{\cal L} \sigtot^2&=&\frac{16\pi}{1+\rho^2}\frac{dN_{el}}{d|t|}|_{t=0}\\
{\cal L} \sigtot&=&N_{el}+N_{inel}
\end{eqnarray}
which lead  to 
\begin{equation}
\sigtot=\frac {16\pi}{1+\rho^2} \frac{dN_{el}/d|t|\ |_{t=0}}{N_{el}+N_{inel}}
\end{equation}
The measurement of the elastic and inelastic rate to be done through two detectors, named T1 and T2,  
placed symmetrically with respect to the CMS experiments.  T1 and T2 are trackers embedded into the forward region of the 
CMS calorimeter, within a distance of $10.5$ and $~14\ m$ from IP5 interaction point of the LHC. These detectors provide 
the reconstruction of charged tracks and cover a rapidity interval $3.1 \le |\eta|\le 6.5$, with $T1$ covering the interval 
$3.1<|\eta|<4.7$ and T2 the interval $5.3<|\eta|<6.5$. 

While the measurement of the inelastic rate $N_{inel}$ does not require special machine conditions, measurements in the 
very forward region do. The measurement of the differential elastic cross-section near the optical point 
done through the detection of very  forward protons, with a technique known as Roman Pots (RPs) and used for the first time 
at the ISR\cite{Amaldi:1973yv}. The RPs are placed on the beam-pipe of the outgoing beam at distances between  $147\ m$ 
and $220\ m$ from IP5 and host silicon detectors 
to be moved very close to the beam, inside the vacuum chamber of the accelerator. 


The measurement at the optical point requires special LHC optics, in order to reach the lowest possible value for the momentum 
transfer $t$. For this one needs the beam divergence to be small compared with the scattering angle. We show in 
Fig.~\ref{fig:totemoptics} a schematic description of the relation between beam size and beam divergence,
\begin{figure} 
\begin{center}
\resizebox{0.5\textwidth}{!}{%
  \includegraphics{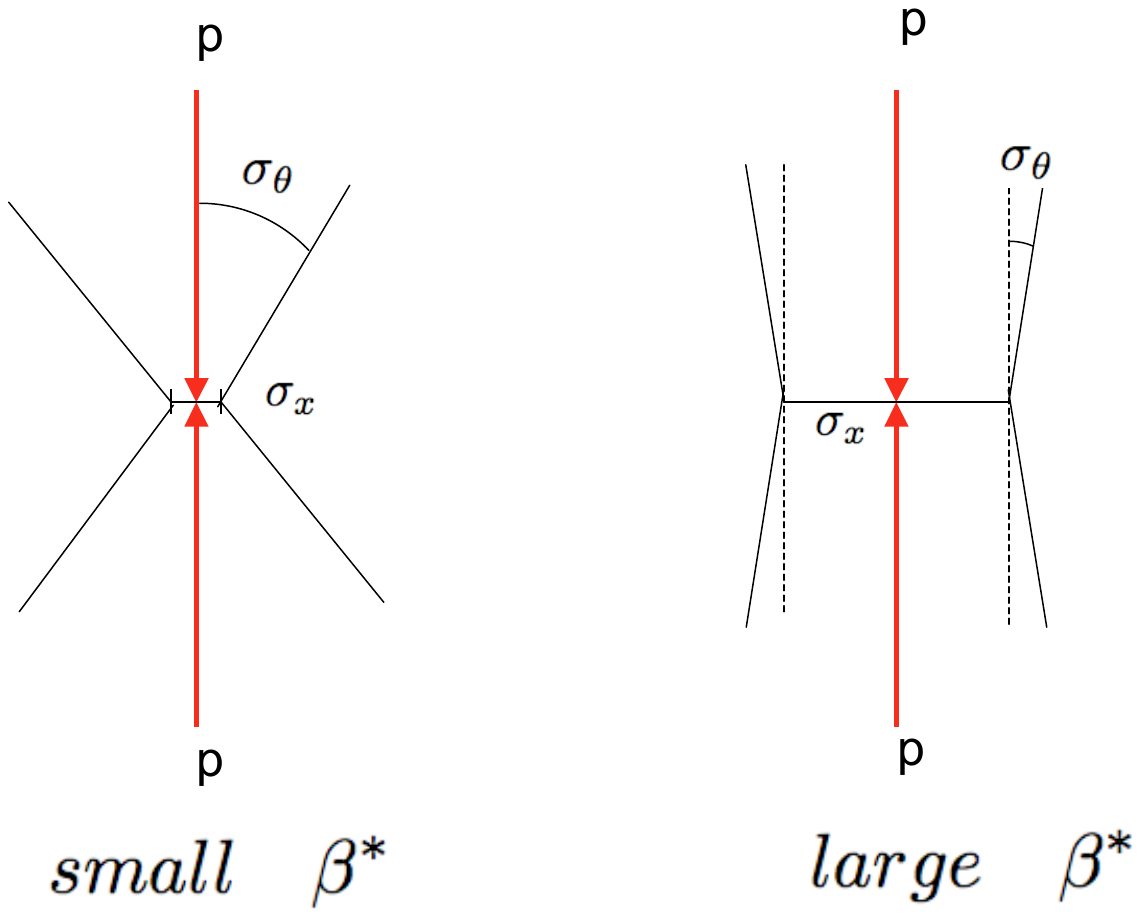}
}
\caption{A  cartoon depicting the meaning of the quantity $\beta^*$ and the relationship between beam size and beam divergence.}
\label{fig:totemoptics}       
\end{center}
\end{figure}
where $\sigma_x$ and $\sigma_{\theta}$ are function of the beam emittance $\epsilon$ and the beam divergence, i.e.
\begin{eqnarray}
\sigma_x=\sqrt{\epsilon \beta_x^*}\\
\sigma_{\theta}=\sqrt{\frac{\epsilon}{\beta_x^*}}.
\end{eqnarray}
Physically, $\beta^*$ is that distance from the focal point where the beam is twice as wide
as at the focal point.  The beam is ``squeezed'' or narrower if $\beta^*$ is low, whereas the beam is ``wide'' and straight for large $\beta^*$.

Thus, the beam divergence $\sim 1/\sqrt{\beta^*}$ is measured by the parameter $\beta^*$, which needs to be as large  as possible. 
This
requires a special value for the parameter 
$\beta^*=1540\ m$. Since such a large value needs a special injection scheme, 
in the early stages of LHC operation (circa 2010), 
a less demanding 
option was planned with $\beta^*=90\ m$. 
At that time , the TOTEM collaboration expected to be able to provide a measurement of the total cross-section with a 5\% 
error within the next three years, with values of the differential elastic cross-section down to values of $|t|> 10^{-2}\ GeV^2$.  
This measurement was to be based on the early optics conditions, $\beta^*=90\ m$ and a luminosity of 
$10^{29}\div 10^{30}\ cm^{-2} sec^{-1}$. Under these conditions, TOTEM Collaboration estimated that about $65\%$ of 
forward protons would be  detected. Later with $\beta^*=1540\ m$, one will be able to reach $|t|> 10^{-3}\ GeV^2$ and, 
with about 90\% of the diffractive protons seen in the detector, with an aim to obtain a measurement at the level of $1\%$.

As for the value of $\rho={\Im m f(t=0)}/{\Re e f(t=0)}$, which we have discussed in earlier sections, 
it was taken to be $\rho=0.14$ following various predictions. This  
was considered adequate, since only the squared value for $\rho$  enters in the equation.
From the analysis of the COMPETE Collaboration \cite{Cudell:2002xe}, we show a compilation of data and best fits as indicated in Fig. ~\ref{fig:competetotem}.
\begin{figure} 
\begin{center}
\resizebox{0.5\textwidth}{!}{%
  \includegraphics{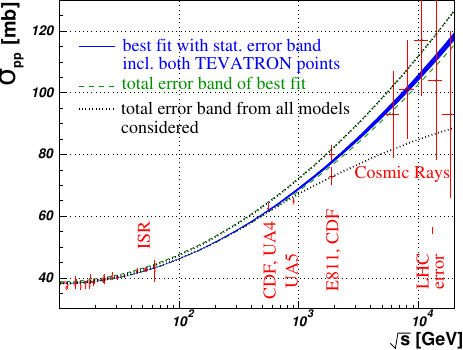}}
\caption{
Fits by the COMPETE collaboration \cite{Cudell:2002xe}  show the wide range of expected results at LHC, due to the tension between data from various experiments at the Tevatron (and at \SpbarpS \ as well). The figure is from \cite{Anelli:2006cz}, with  total cross-section data  compared with various options for the high  energy dependence. 
Figure is reproduced from \cite{Anelli:2006cz}, courtesy  from  J. Kaspar, TOTEM Collaboration.}
\label{fig:competetotem}       
\end{center}
\end{figure}

\subsubsection{ZDC}\label{sss:ZDC}
 
Tuning at zero angle on neutrons, and detecting them with the zero degree calorimeter \cite{Grachov:2008qg} at CMS, 
in addition to a number of diffractive physics measurements, there has also been the hope to measure  
$\pi^+ p$ and $\pi^+ \pi^+$ total cross-sections in an energy range inaccessible so far,
namely in and around 1 TeV \cite{Petrov:2009wr}. Information on diverse initial state particles and their relative rise 
with energy of $\sigtot$ is  crucial for understanding the mechanisms behind the rise of the total cross-section, whether 
or not there is a universal rise, and connections to perturbative QCD. Presently, data for $\pi p$ total cross-section are 
only available in an energy range up to $25\ GeV$\cite{Amsler:2008zzb}. The situation  for $\pi \pi$ is  even less 
favourable.
The mechanism 
proposed to measure these cross-sections in the high energy range is shown in Figs.~\ref{fig:piplhc}  and \ref{fig:pipilhc}, 
namely through detection of neutrons in the very forward direction and production of pions through the charge exchange reaction.
\begin{figure}
\begin{center}
\resizebox{0.4\textwidth}{!}{%
  \includegraphics{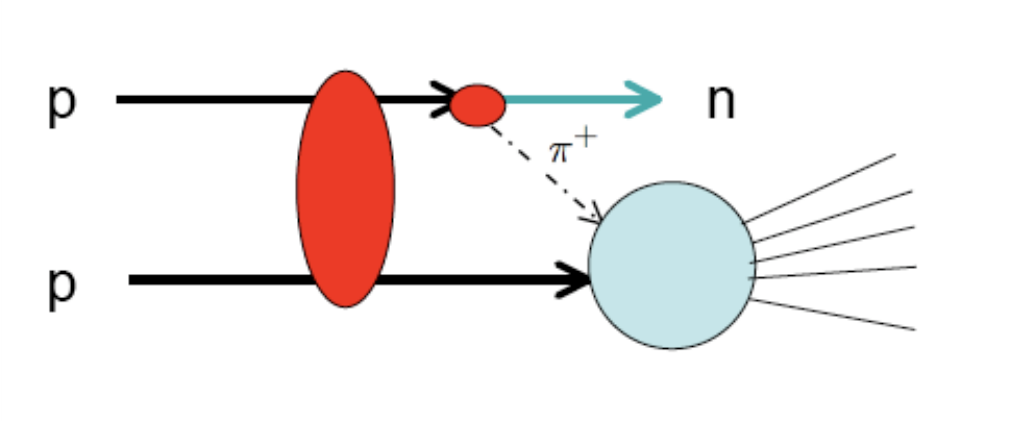}}
\caption{ The charge exchange mechanisms proposed in \cite{Petrov:2009wr} to measure the total $\pi p$ \x at LHC. 
}
  \label{fig:piplhc}
\end{center}
\end{figure}

\begin{figure}
\begin{center}
\resizebox{0.4\textwidth}{!}{%
  \includegraphics{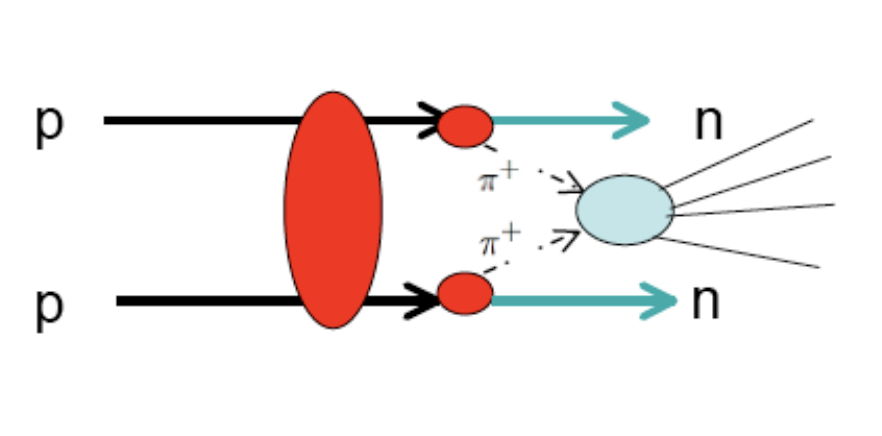}}
\caption{The charge exchange mechanisms proposed in \cite{Petrov:2009wr} to measure the total $\pi \pi$ \x at LHC.}
  \label{fig:pipilhc}
\end{center}
\end{figure}

As proof of the feasibility of such experiments,  Petrov {\it et al.} \cite{Petrov:2009wr}  have extracted data for $\pi^+ p$ \x\ up to 
$50\div 70 \ GeV$ using neutron  and photon 
spectra at previous experiments.  The results are    shown  in Fig. ~\ref{fig:pippetrovfig} from \cite{Petrov:2009wr}, 
where extracted data points are  compared with existing data from the Particle Data Group compilation (PDG)\cite{Amsler:2008zzb}. 
Also shown are two parametrizations, with full line by Donnachie and Landshoff \cite{Donnachie:1992ny} and dashes to indicate  
the  fit by COMPETE  also from \cite{Amsler:2008zzb}.


\begin{figure}
\begin{center}
\resizebox{0.5\textwidth}{!}{
\includegraphics{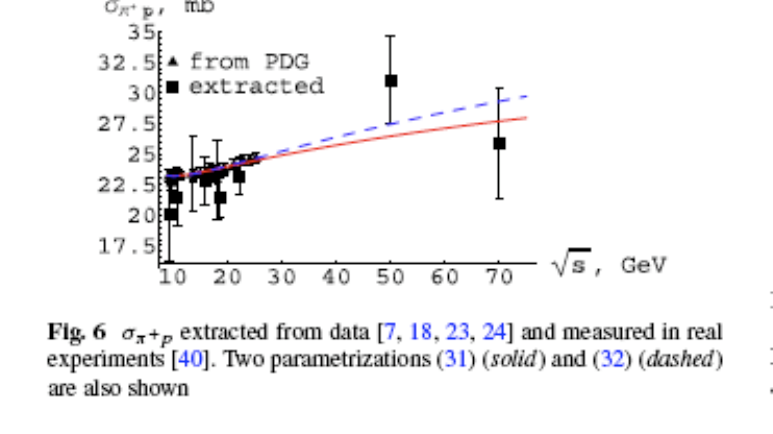}}
\caption{$\pi^+p$ total cross-section from  \cite{Petrov:2009wr}, with both direct and extracted data points extracted  
with two parametrizations \cite{Donnachie:1992ny} (solid) and \cite{Amsler:2008zzb}(dashes). Reprinted with permission from \cite{Petrov:2009wr} \copyright
(2009) by Springer.}
\label{fig:pippetrovfig}
\end{center}
\end{figure}
We also show in Fig~\ref{fig_pipp_low_total_data} a comparison in this energy range between our model \cite{Grau:2010ju} and data, and fits from 
\cite{Petrov:2009wr} as seen in Fig.~(\ref{fig:pippetrovfig}), as well as  comparison  with fits by Block and Halzen \cite{Block:2005pt}.
\begin{figure}
\begin{center}
\resizebox{0.5\textwidth}{!}{
\includegraphics{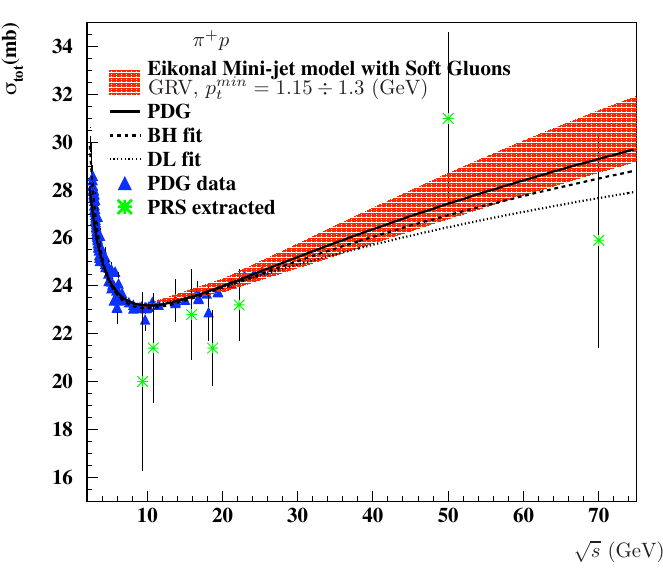}}
\caption{$\pi^+p$ total cross-section data  with both direct and extracted data points  from  \cite{Petrov:2009wr}, compared   
with  parametrizations from DL \cite{Donnachie:1992ny}, COMPETE \cite{Amsler:2008zzb}(full), BH  \cite{Block:2005pt} 
and from our model, as indicated. Figure is reprinted from \cite{Grau:2010ju}, \copyright (2010) with permission by Elsevier.}
\label{fig_pipp_low_total_data}
\end{center}
\end{figure}
The interest of such a measurement can be seen by going to very high energies, where the models differ substantially, as induced in Fig.~\ref{fig_pipp_high_total}.
In the compilation shown in Figs.~\ref{fig_pipp_low_total_data} and \ref{fig_pipp_high_total} (which differ in the energy range)
 we have plotted, together with the existing  data,  four predictions for $\pi^+ p$ total \x  \ as follows:
\begin{itemize}
\item a Regge-Pomeron fit from Donnachie and Landshoff \cite{Donnachie:1992ny}
\begin{equation}
\sigma_{\pi^+p}(mb)=13.63 s^{0.0808}+27.56s^{-0.4525}
\end{equation}
\item the fit from the COMPETE \cite{Amsler:2008zzb} collaboration given as 
\begin{equation}
\sigma_{\pi+p}=Z^{\pi p}+B\ln^2 (\frac{s}{s_0})+Y_1^{\pi+p}(\frac{s_1}{s})^{\eta_1}-Y_2^{\pi^p}(\frac{s_1}{s})^{\eta_2} 
\label{eq:fitcudellpip}
\end{equation}
\item a fit by Halzen and Block \cite{Block:2005pt} of similar functional expression as the one from PDG, 
with  an additional $\ln {s}$ term, i.e.
\begin{eqnarray}
\sigma^{ab}&=&c_0+c_1\ln{(\nu/m_\pi)}+c_2\ln^2{(\nu/m_\pi)}+\nonumber \\
& & \beta (\nu/m_\pi)^{\eta_1}+\delta (\nu/m_\pi)^{\eta_2}
\end{eqnarray}
with numerical coefficients given by  
$c_0=20.11 \  {mb}$, $c_1 = -0.921  \ {mb}$,   
$c_2 = 0.1767  \ {mb}$, 
$\beta=54.4 \  {mb}$,
$\delta = -4.51\  {mb}$, 
$\eta_1 = -0.5$, 
$\eta_2 = -0.34 $
\item the eikonal mini-jet model with initial state soft gluon $k_t$-resummation described in previous section, 
with GRV density functions for  pion and proton and other parameters close to the values used for $\sigtotpp$,
 namely $p_{tmin}=(1.15\div 1.3) \ GeV$, $p=0.75$ and $\Lambda=100\ MeV$ in the soft resummation integral; 
 in this model the low energy data have been independently parametrized with the  expression
\begin{equation}
\sigma_{\pi^+p} = A_0 + A_1[\frac{1 GeV}{E}]^{\alpha_1} - A_2[\frac{1 GeV}{E}]^{\alpha_2}
\end{equation}
with parameters $A_0 = 31.49\ mb$, $A_1 = 58.56\ mb$, $A_2 = 40.52\ mb$, 
 $\alpha_1 = 0.498$, $\alpha_2 = 0.297$ 
 \end{itemize}

\begin{figure}
\begin{center}
\resizebox{0.4\textwidth}{!}{%
\includegraphics{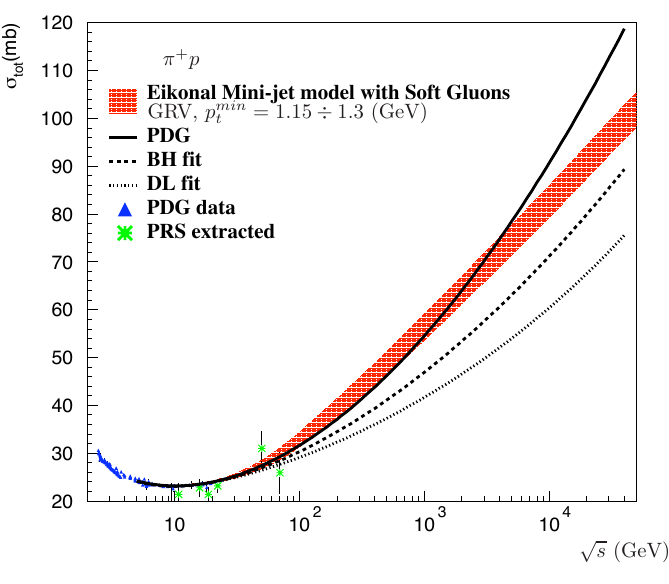}}
\caption{Predictions for $\pi^{+} p$ total \x \ in the LHC energy range  from different models, as described in the text, 
and comparison with  data on $\pi^+ p$ total cross-section.  DL (dots) is from \cite{Donnachie:1992ny}, 
BH (dashes) from \cite{Block:2005pt}, PDG COMPETE (full) \cite{Amsler:2008zzb} and PRS (stars) indicates extracted 
data from \cite{Petrov:2009wr}.  Figure is reprinted from \cite{Grau:2010ju}, \copyright (2010) with permission by Elsevier.}
  \label{fig_pipp_high_total}
\end{center}
\end{figure}

\subsection{ The ATLAS region and forward physics}
\subsubsection{LHCf}

LHCf is the smallest of the LHC experiments and is a detector placed at the ATLAS interaction point, 
with an independent acquisition system,  very easy to correlate with ATLAS.  The experimental set-up 
covers a very  forward kinematics, $\eta> 8.4 $, with   angle from beam axis $\theta<450 \ \mu rad$, 
with detection of very forward $\gamma's$.  Because of radiation problems, LHCf can however take 
data only at low machine luminosity and  needs to be taken out in high luminosity running  conditions.

The LHCf \cite{Bongi:2010zz} experiment will measure the properties of neutral particles produced 
in the very forward region and compare them 
with expectations  from the MonteCarlo simulation programs used in Cosmic Ray Physics. 
The experiment will  use these forward particles from the collision  to simulate cosmic rays of similar energies
in laboratory conditions with particle energies at LHC, at  $\sqrt{s}=14\ TeV$, corresponding  to 
laboratory energies of  $10^{17} \ eV$.
The aim of this  experiment is  to clarify some phenomenological problems encountered in 
extracting physics from cosmic rays, among them  a precise determination of the energy, nature and origin 
of the  particles which initiated the Extensive Air  Showers observed in cosmic ray experiments. 
By observing the energy deposition of controlled particles, like neutrons, $\pi^0$'s and $\gamma$'s, and comparing 
their properties with the two most used MonteCarlo simulation programs, SYBILL\cite{Fletcher:1994bd,Engel:1999db} 
and QGSJET\cite{Ostapchenko:2007qb}, one can hope to resolve some outstanding questions in high energy cosmic 
ray physics \cite{Aloisio:2009sj,Fraschetti:2008if}. In cosmic ray physics,  presently of great interest  is to 
study  the cosmic ray spectra in and  around  the GZK \cite{Greisen:1966jv,Zatsepin:1966jv} cut-off, 
expected to take place at $E_{lab}\approx 10^{19.5} \ eV$. Quite a long time ago, Greisen, Zatsepin and 
Kuzmin (GZK)  predicted that at such energies the flux of cosmic rays could  become too small to be observed.
This effect corresponds to a reduction in  the flux of primary cosmic ray protons once they reach an energy high 
enough  to interact with the  photons from the Cosmic Microwave Background  (CMB) and produce the 
$\Delta(1232)$-resonance, through  $p+\gamma ^{CMB}\rightarrow \Delta\rightarrow \pi p$. Were the cut-off not 
to be observed, the possibility of exotic sources could not be ruled out. While earlier measurements in the GZK cut off 
region had not seen the cut-off, recently 
the observation of the cut-off has been reported by  two experiments,  Auger \cite{Abraham:2008ru} and HiRes\cite{Abbasi:2009nf}. 
They both  observe a decrease of the flux and a change in slope. Some contradictions still exist, 
as one can see from Fig.~\ref{fig:auger-GZK} from \cite{Abraham:2010mj}. This figure shows that,  
even though both HiRES and Auger report the expected GZK flux reduction, there is still a difference in 
normalization between their data.

\begin{figure}
\resizebox{0.4\textwidth}{!}{%
\includegraphics{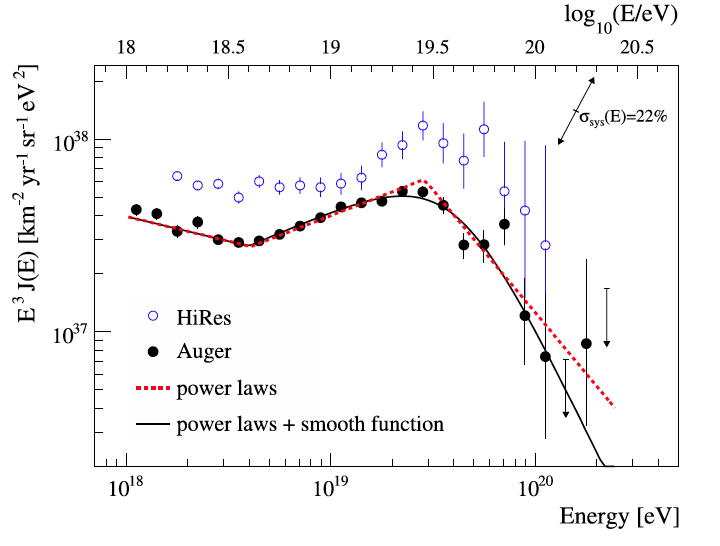}}
%
\caption{Results from measurement of the Ultra High Energy Cosmic Ray flux in the region of the GZK cut-off, from \cite{Abraham:2010mj}.
Reprinted from \cite{Abraham:2010mj}, \copyright (2010) with permission by Elsevier.}
\label{fig:auger-GZK}
\end{figure}

\subsubsection{ATLAS forward detectors}\label{sss:ATLAS}
The positioning of ATLAS forward detectors is shown in Fig.~\ref{fig:atlasforward}, where ALFA indicates the detectors 
for Absolute Luminosity measurement,  ZDC is the Zero Degree Calorimeter for ATLAS , LUCID is the LUminosity 
Cerenkov Integrating Detector. Not shown is MBTS,  the minimum Bias Trigger Scintillator, closest to the IP.

\begin{figure*}
\begin{center}
\resizebox{0.7\textwidth}{!}{%
  \includegraphics{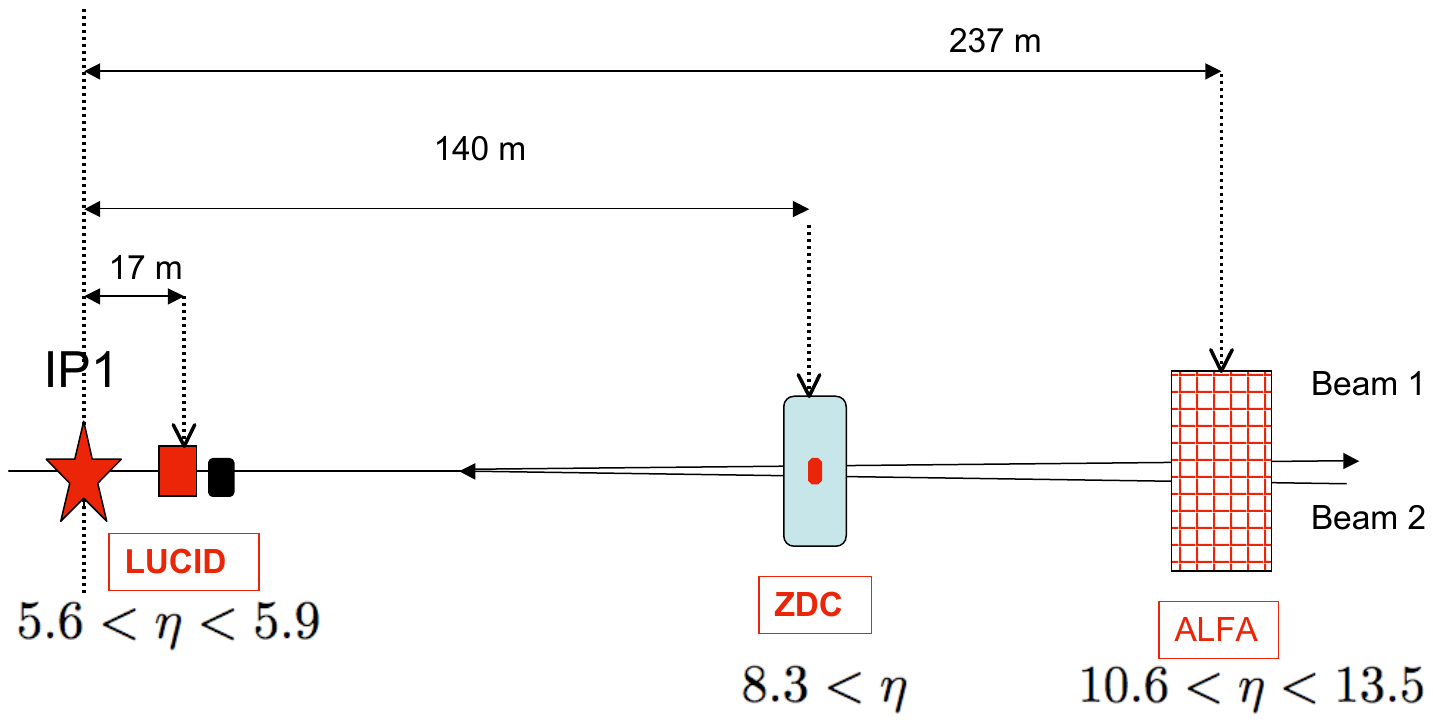}}
\vspace{-3cm}
\caption{The positioning of ATLAS detectors for forward physics.}
\label{fig:atlasforward}
\end{center}
\end{figure*}

One distinguishes  the coverage of pseudo-rapidity in central and forward detectors regions. For ATLAS, in the central region, 
$|\eta|$ coverage  is up to $2.5$ for the inner tracker, $3.2$ for the electromagnetic calorimeters, $4.9$ for the hadronic 
calorimeters, and $2.7$ for the muon spectrometer. The forward detectors cover  rapidity intervals up to $|\eta|< 13.5$ as follows :
\begin{description}
\item[MBTS ] $2.1< |\eta|<3.8$, is the Minimum bias trigger dedicated to  diffractive physics measurements 
\item[LUCID ] covering $ 5.6< |\eta|<5.9$, is  the luminosity monitor, designed to measure luminosity up to 
$10^{33} cm^{-2}sec^{-1}$, with a $3\div 5 \%$ precision, is sensitive to charged particles pointing to the primary 
$pp$ collision, and is needed to provide the minimum bias trigger at high values of pseudorapidity,
\item[ZDC] a Zero Degree Detector,   $ |\eta|>8.3$, will measure production of neutral particles, 
$n,\gamma, \pi^0$, in the forward direction and study both  heavy ions and pp collisions,
\item[ALFA ] $10.6< |\eta|<13.5$ will measure the absolute luminosity and hadronic physics forward parameters. 
\end{description}
The main method  designed to measure the luminosity in ATLAS uses
Roman Pots to make a reference  measurement at low luminosity. This measurement will 
then be used to calibrate  a  monitor when   luminosity is  too high for  use of the RPs.  LUCID, the Beam Condition 
Monitor (BCM) and MBTS  are the three detector systems for luminosity monitoring.

The very forward region in ATLAS is  covered by  Roman Pots (RP) which  measure elastic $pp$ scattering 
at the very small angles needed to extrapolate the differential elastic cross-section to $t=0$,  the optical point for 
total cross-section measurements. As mentioned,  these measurement requires special beam optics (high $\beta^*$) 
and low luminosity, $L=10^{27}cm^{-2}sec^{-1}$. 

\subsubsection{Roman POTS and the ALFA detector}\label{sss:ALFA}

The technique by which one measures the very forward scattering events to  extract the differential elastic cross-section in the very small $t$-region    and thus the total cross-section through the optical method, makes use of the so called {\it Roman Pots}.  Roman Pots do not really look like pots from ancient Rome, where containers were of round "amphora-like" shape \footnote{Comment courtesy of G. Matthiae}, and thus quite different from the cylindrical shape of the actual RPs. They get their name having been used by the Rome-Cern group at the ISR in the early '70 and by their function. The actual detectors are  cylindrical containers which are connected to the vacuum chamber of the accelerator through bellows. While the beam intensity is building up during injection, the RPs are retracted and do not enter the vacuum chamber. After the beams have stabilized  and the collider has reached stable conditions, then the bellows are compressed and the detectors are pushed forward up to a distance of $1 \ mm$ from the beam. We show in Fig. ~ \ref{fig:romanpots} a schematic view of how the detectors will be placed near the beam so as to detect protons scattered at $|t|\approx 6.5 \times 10^{-4}\ GeV^2$. 
\begin{figure}
\begin{center}
\resizebox{0.5 \textwidth}{!}{
\includegraphics{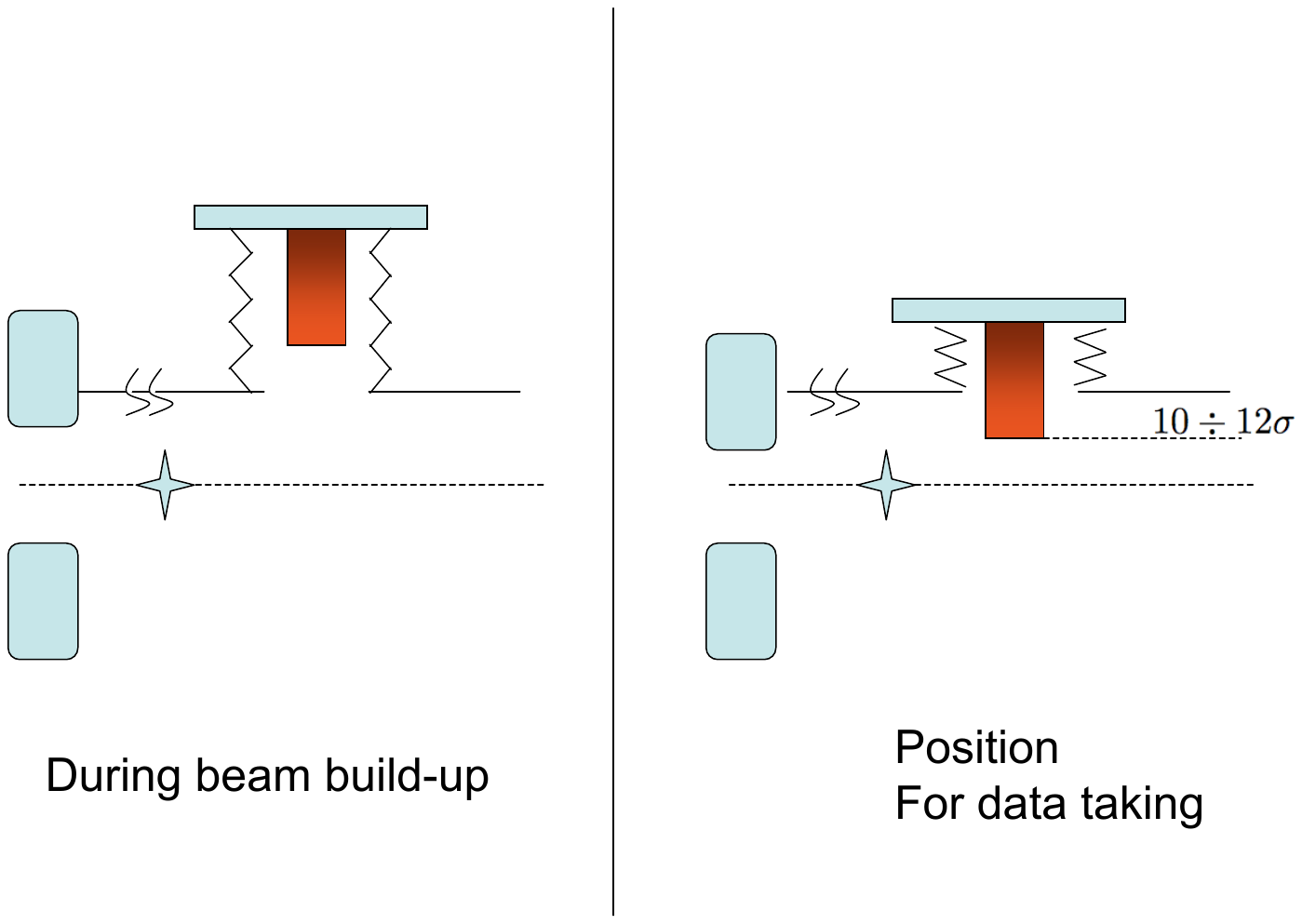}}
\caption{A schematic view of the operation of RPs before and during data taking.}
\label{fig:romanpots}
\end{center}
\end{figure}

For such small values of $t$ one has the following relation between scattering angle and beam parameters:
\begin{equation}
|t_{min}|=\frac{p^2}{\gamma}n_d^2\frac{\epsilon_N}{\beta^*}
\end{equation}
where $n_d$ is the distance from the beam in units of beam size, $\epsilon_N$ is the normalized emittance. ALFA will be used to get an absolute measure of the luminosity by detecting protons in the Coulomb region with a   sought for precision of  3 \%,  an important improvement above the precision obtainable using machine parameters, which is not expected to be better than 20 \%. Such high precision is needed for precise determination of Higgs parameters and Branching Ratioes.

The absolute measurement of the luminosity ${\cal L}$  is extracted from the differential event rate. Up to $|t|\approx 1 \ GeV^2$, the differential rate for elastic scattering, to first order in $\alpha$,  can be written as  
\begin{equation}
\frac{dN}{dt}={\cal L}\large[\frac{4\pi\alpha^2}{|t|^2}+\frac{\alpha \rho \sigtot e^{-Bt^2/2}}{|t|}+\frac{\sigtot^2(1+\rho^2)e^{-B|t|}}{16 \pi} \large]
\end{equation}
By measuring this rate it in the Coulomb region, i.e. below $|t|=10^{-3}$, and after radiative corrections (see section 3) the absolute luminosity can be extracted. In   Fig.~\ref{fig:forward-t}, we show a cartoon representation of the three regions, Coulomb, interference and purely hadronic, which can give information on various hadronic physics quantities of interest. 
\begin{figure}
\begin{center}
\resizebox{0.5 \textwidth}{!}{
\includegraphics{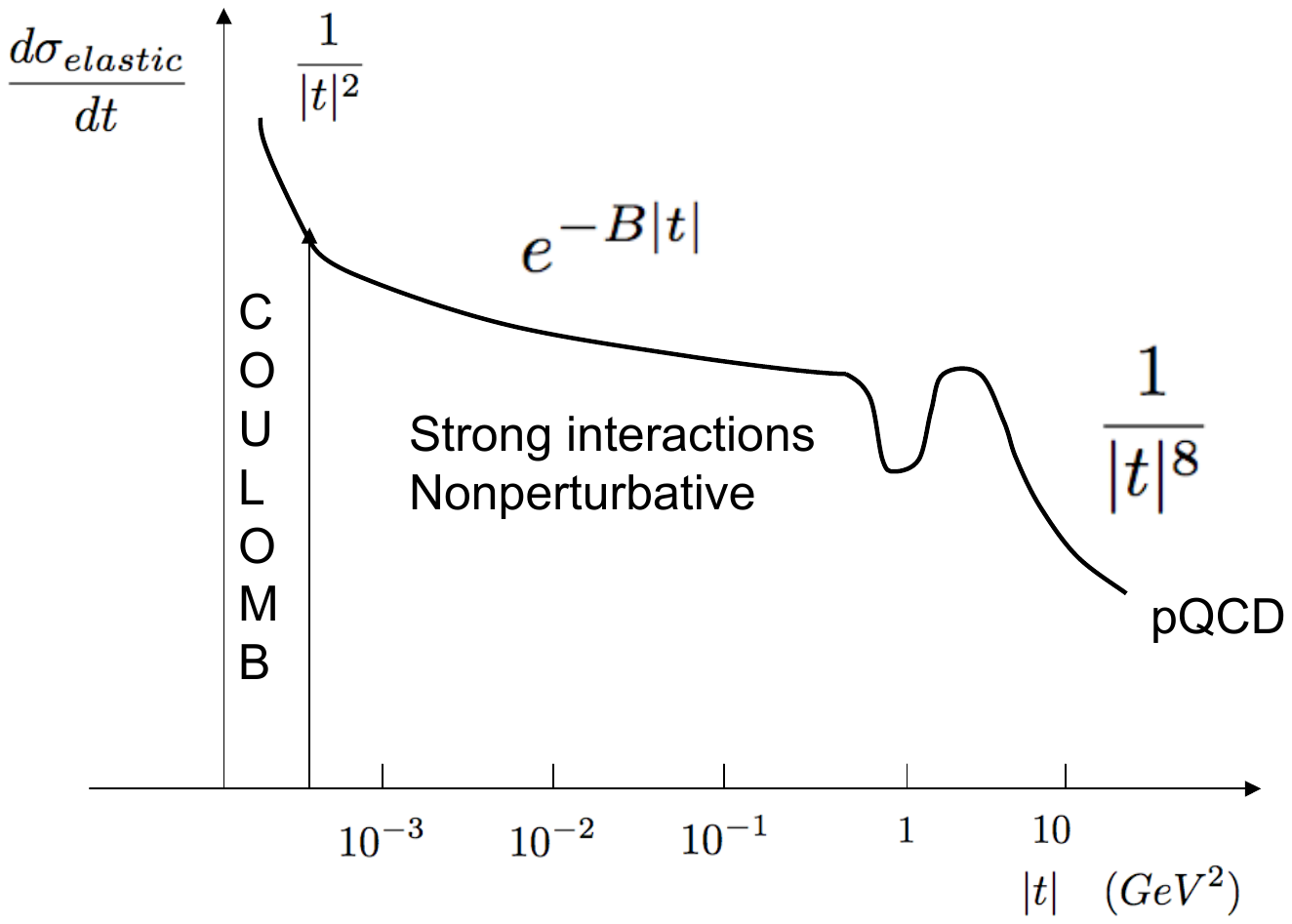}}
\caption{A cartoon  sketching the  differential elastic cross-section as a function of the momentum transfer $|t|$, showing how different $t$ regions will give information on elastic and total scattering parameters. After presentation by  A. Pilkington at Trento Workshop on Diffractive Physics, 4-8 january 2010, ECT*, Italy. Please notice that this figure is purely indicative and it is not in scale. }
\label{fig:forward-t}
\end{center}
\end{figure}
\subsection{Updates about LHC forward physics programs}
In addition to what already mentioned at the beginning of this section, the  interested reader can find  descriptions of updates for LHC forward physics presented by various groups, such  as  a Workshop
on High Energy Scattering at Zero Degree held in March 2013 at Nagoya University in Japan. The slides of all the talks at Nagoya as well 
as presentations at Marseille and Paris in France; at Trento and Reggio Calabria in Italy; CERN, Switzerland; Barcellona, Spain; and at 
Eilat, Israel,  and more, can all be found at:\\
{\it totem.web.cern.ch/Totem/conferences/conf\_tab2013.html,\\
 et sim. for 2014, 2015, 2016.}  By comparing these reports with what we have presented here, and which follow  the plan as of  2008, one can see the great progress  of these years and look with confidence that  future measurements will further reduce errors and clarify many issues.  

Indeed considerable progress has been made in the beam optics and proper functioning of various detectors so much so that now we have
rather precise data on total, elastic and inelastic cross-sections, elastic differential cross-sections and various diffractive results in 
different regions of phase space. Many of these results have been used and discussed throughout this review. For example, 
using dedicated beam optics and the Roman Pots, at $\sqrt{s} = 8\ TeV$, TOTEM at the end of 2012 gives the following values
\cite{Antchev:2013paa}:
\bea
\label{TOTEM8}
\sigma_{tot}(8\ TeV) = (101.7 \pm 2.9)\ mb;\nonumber\\ 
\sigma_{el}(8\ TeV) = (27.1 \pm 1.4)\ mb;\nonumber\\
\sigma_{in}(8\ TeV) = (74.7 \pm 1.7)\ mb.
\eea 
Thus, the total cross-section has been measured with less than $3\%$ error better than the estimated error, after a 3 year run, 
of $5\%$. 
An overview of all the measurements of total, inelastic, elastic and diffractive cross-sections inclusive of data up to 2016 is presented in Fig. ~\ref{fig:alltogether}.
\begin{figure*}
\centering
\resizebox{0.8\textwidth}{!}{
\includegraphics{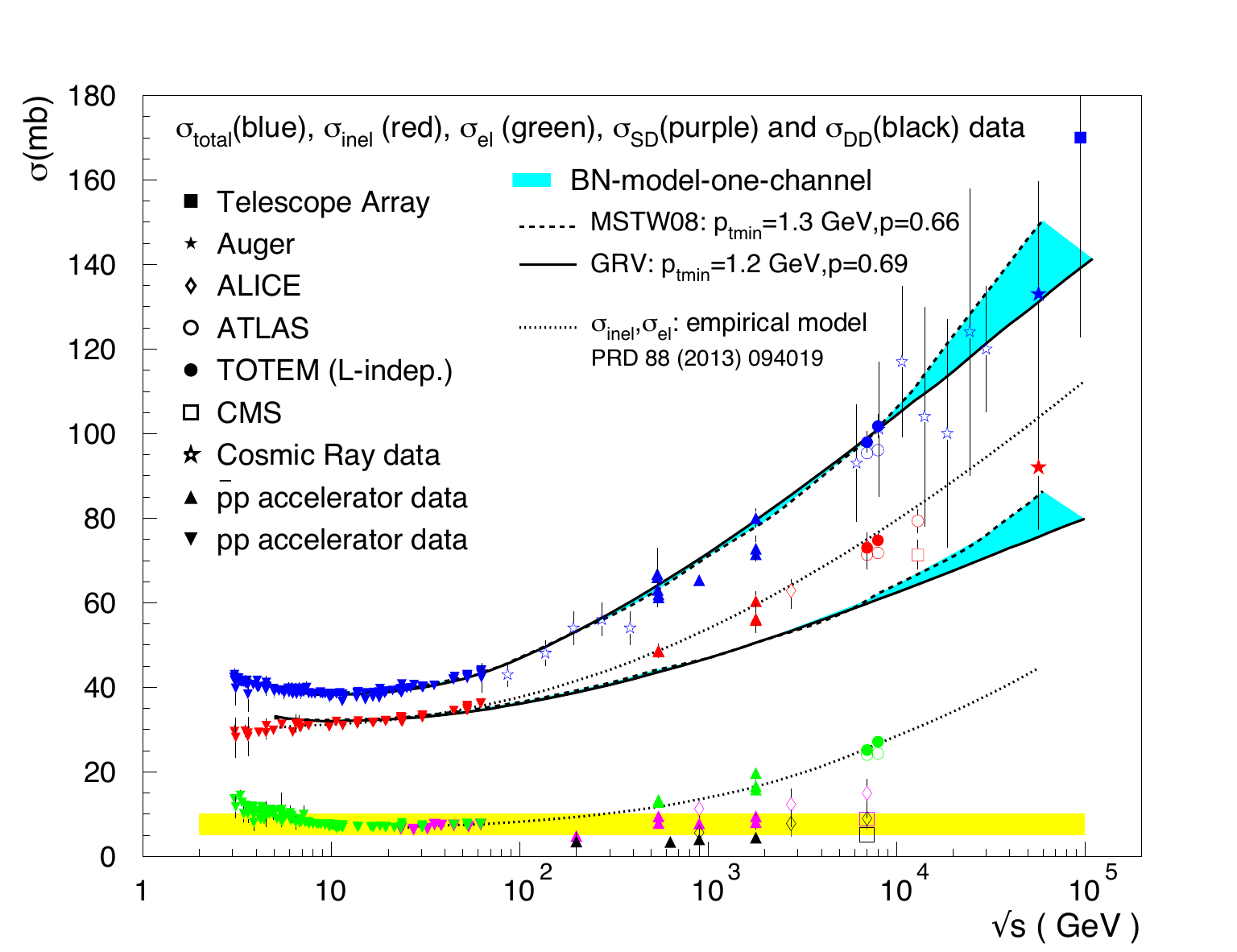}}
\caption{
A compilation of LHC data on the total, inelastic, elastic  and diffractive cross-sections, as of September 2016. Superimposed  curves correspond to the BN model described in Sec. \ref{ss:Onechannel}  for the total  and non-diffractive inelastic cross-section,  and in Sec. \ref{sss:PB} for the elastic cross-section, as in the updated version from  \cite{Fagundes:2013aja}. A band has been drawn to drive the eye for the Single and Double Diffraction data. This figure is courtesy of D. Fagundes, A. Grau and Olga Shekhovtsova.}
\label{fig:alltogether}
\end{figure*}

\section{Conclusions}
Huge progress has been made over the past several decades, both experimentally and theoretically, on the subject of high energy 
total and differential cross-sections. 
 {In this review we have attempted to outline these developments  from early accelerator measurements in the 1950's with fixed target experiments up to proton-proton scattering at the  CERN Large Hadron Collider,  and beyond, where cosmic ray interactions  reach  energies as high as 100 TeV in the proton-proton center of mass.}
 
In proton-proton scattering, two milestones stand out, the first of them  concerning the  energy dependence of the total cross-section. The increase  with energy of the total cross-section is now fully confirmed, and ascribable to the appearance of parton-parton scattering, although  
questions regarding asymptotia and whether the Froissart bound is saturated, are still under debate. The second milestone is the LHC confirmation of the dip in the differential proton-proton elastic cross-section, which had  not been  observed since  the CERN Intersecting Storage Ring experiments in the early '70s. Experiments at the CERN \SpbarpS \ and at the  Tevatron in FermiLab have given  hints that the  presence of the dip  in proton-antiproton scattering may be revealed   as   higher and higher  energies are reached, but  confirmation of the dip in this channel   needs higher energy experiments  which  are not presently planned. During  the same decades, a large set of measurements were performed at HERA in DESY, using both real and virtual photons on nucleons
and nuclei to obtain total and production cross-sections for $\gamma p$, $\gamma^* p$, and through $e^+e^-$ machine at LEP for $\gamma \gamma*$
and $\gamma^* \gamma^*$ final states. These results are mostly complementary to those  from purely hadronic machines and have led to remarkable theoretical developments such as Bjorken scaling, the parton model and various dynamical evolution equations.  

From the theoretical point of view, our review spans from Heisenberg's model  to the rich descriptions which have been developed in more than  60 years in terms of QCD, Reggeon field theory, mini-jets, among others.
The amount of material on the subject is so huge, that some selection was indispensable. Hence, 
we are aware that we could not always acknowledge or survey all the work done during the past 50-60 years 
in a quest of understanding the dynamics underlying the hadronic cross-sections.

We have gathered and presented the material which we could relate to and understand.
Hence, we apologize to those scientists whose work we may not have recognized adequately. 
Many excellent reviews on the subject have been written during the past decades that are 
complementary to our largely historical perspective.



All together, we hope that our work may shed light on the fascination that the subject has held for 
so many scientists for so many years and that shall continue to fascinate in the future through further results
from  LHC and cosmic ray experiments.
  
\section*{Acknowledgments}
One of us, G.P., gratefully 
acknowledges hospitality at the MIT Center for Theoretical Physics and  Laboratory for Nuclear Science,
 as well as at Brown University Physics Department and at Durham University IPPP. 
We  thank  our friends and collaborators for  helping us with this review:    in particular we thank Earle Lomon  for 
continuous suggestions and enlightening conversations, Aron Bernstein of  MIT LNS for support and suggestions, John Negele  for early encouragement,
 Meenakshi Narain and Ulrich Heinz from Brown University for support and hospitality, Michael Murray from CMS for introduction to the physics of the ZDC,   Ruggero Ferrari and  Galileo  Violini 
for suggestions and advice,  L. Bonolis and C. Bernardini for archival work on Heisenberg's work.  We thank
  Giorgio Bellettini for allowing us to use part of the material presented at the meeting on Forward Physics 
at LHC, held at La Biodola, l'Elba, May 27-30 2010. We also thank Marco Bozzo for useful comments. Thanks are due to our 
collaborators Agnes Grau, Olga Shekhovtsova and Daniel Fagundes, who prepared many figures. 
We are  also  grateful to the library services of  INFN Frascati National Laboratories, and in particular to Ms Debora Bifaretti,   
for retrieving many articles, otherwise  not easily available.

Special thanks are due to Dieter Haidt,  
as well as to the EPJC referees, who have  given us  advice and suggestions throughout the  long 
preparation of this review. 

\section*{Note about reproduction of  other author's figures}
As indicated in the figure captions, many figures in this review have been extracted from articles published in  journals or posted on the web site \href{http://inspirehep.net}{inSPIRE}.

 Permission from the publisher was obtained  for all the figures, as indicated in the figure captions. We thank all the authors, whose figures we have reproduced.  We thank many colleagues who provided us with the original files for the figures extracted from their papers. For figures coming from articles with a large number of authors, we secured permission from  the spokesman of the collaboration. In all cases we thank our colleagues  from  these collaboration for their  permission to reproduce these  figures.
 \bibliographystyle{epj}
 \bibliography{general-measure-aug2016,cosmic-models-oct20_2016,elastic_L11,photons_yyy1-oct21,lhcnow-1,extra_dec15-2016}
 \end{document}